# The Detailed Science Case of the Maunakea Spectroscopic Explorer

The MSE Science Team

April 9, 2019

# Contents

# Preface to Version 2 of the Detailed Science Case, 2019

### Purpose and Scope

This is the MSE *Detailed Science Case* (DSC). It is the highest level document in the MSE document hierarchy, and forms the foundation for the Level 0 *Science Requirements Document* (SRD). A snapshot of the entire MSE observatory including technical designs at the end of the Conceptual Design Phase can be found in *The MSE Book 2018*.

The MSE DSC provides the science narrative describing the principal envisioned science goals of MSE and describes its impact on a broad range of science topics. This new version of the DSC builds upon, develops and augments the original science case that was released in 2016. It reaffirms the science capabilities of MSE that form the basis of the SRD, while updating it for recent advances and discoveries (e.g., Data Release 2 of Gaia, GW170817, ...). It includes new contributions and perspectives from the science team, whose membership has grown significantly since 2016 ($>$ 360 members in 2019 compared to $\sim$ 100 members in 2016; for a full list of current science team members, please refer to the Appendix).

### A note on the structure of the DSC

The structure of Version 2 of the DSC is notably different from the original version. The DSC begins with an Executive Summary and an Overview chapter, after which there are seven thematic chapters that describe the science cases for MSE from exoplanetary through to cosmological scales . Each chapter was developed by the relevant thematic Science Working Group (SWGs). Each chapter can be read as a stand-alone document.

### Credit and acknowledgements

This document represents contributions from more than 250 scientists across the international astronomical community over the past 8 years. The author list consists of all those scientists that have actively contributed to the writing of this version of the Detailed Science Case, and/or the 2016 version of the Detailed Science Case, and/or closely related documents.

## Science Working Group leads

*(alphabeticized in author list)*

SWG1 – Exoplanets and stellar astrophysics: Maria Bergemann & Daniel Huber
SWG2 – Chemical nucleosynthesis: Sivarani Thirupathi & David Yong
SWG3 – The Milky Way and resolved stellar populations: Carine Babusiaux & Sarah Martell
SWG4 – Astrophysical tests of dark matter: Manoj Kaplinghat & Ting Li
SWG5 – Galaxy formation and evolution: Aaron Robotham & Kim-Vy Tran
SWG6 – Active Galactic Nuclei and supermassive black holes: Sara Ellison & Yue Shen
SWG7 – Cosmology: Will Percival & Christophe Yeche
SWG8 – Time domain astronomy and the transient Universe: Adam Burgasser & Daryl Haggard
Principal editors of the DSC: Jennifer Marshall & Alan McConnachie

A ninth science working group (SWG9 – Integral Field Unit Development: Guilin Lui) will help develop the science cases and requirements associated with the next-generation integral field unit capability of MSE (see Chapter 2 and the *MSE Book 2018*), that will be published as a supplement to the DSC.

## Principal contributors to the DSC for each SWG

*(alphabeticized in author list)*

SWG1:

Maria Bergemann, Daniel Huber, Vardan Adibekyan, George Angelou, Daniela Barría, Amelia Bayo, Timothy C. Beers, Paul G. Beck, Earl P. Bellinger, Joachim M. Bestenlehner, Bertram Bitsch, Adam Burgasser, Derek Buzasi, Jan Cami, Luca Casagrande, Santi Cassisi, Márcio Catelan, Ana Escorza, Scott W. Fleming, Maksim Gabdeev, Boris T. Gänsicke, Davide Gandolfi, Rafael A. García, Patrick Gaulme, Mark Gieles, Guillaume Guiglion, Gerald Handler, Lynne Hillenbrand, Amanda Karakas, Yveline Lebreton, Nicolas Lodieu, Carl Melis, Thibault Merle, Szabolcs Mészáros, Andrea Miglio, Karan Molaverdikhani, Richard Monier, Thierry Morel, Benoit Mosser, Hilding R. Neilson, Mahmoudreza Oshagh, Adrian M. Price-Whelan, Andrej Prsa, Jan Rybizki, Juan V. Hernández Santisteban, Aldo Serenelli, Victor Silva Aguirre, Rodolfo Smiljanic, Jennifer Sobeck, Dennis Stello, Gyula M. Szabó, Robert Szabo, Silvia Toonen, Pier-Emmanuel Tremblay, Maria Tsantaki, Marica Valentini, Sophie Van Eck, Eva Villaver, Nicolas J. Wright, Siyi Xu, Mutlu Yildiz, Huawei Zhang, Konstanze Zwintz

SWG2:

Sivarani Thirupathi, David Yong, Alexander Ji, Amanda Karakas, Charli Sakari, Drisya Karinkuzhi, R. Michael Rich, Projjwal Banerjee, Terese Hansen, Vinicius Placco

SWG3:

Carine Babusiaux, Sarah Martell, Martin Asplund, Giuseppina Battaglia, Maria Bergemann, Angela Bragaglia, Elisabetta Caffau, Alis Deason, Richard de Grijs, Paola di Matteo, Ken Freeman, Mario Gennaro, Karoline Gilbert, Aruna Goswami, Carl Grillmair, Despina Hatzidimitriou, Misha Haywood, Vincent Hénault-Brunet, Rodrigo Ibata, Pascale Jablonka,

Iraklis Konstantopoulos, Rosine Lallement, Chervin Laporte, Sophia Lianou, Nicolas Martin, Mario Mateo, Carl Melis, David Nataf, Gajendra Pandey, R. Michael Rich, Charli Sakari, Ricardo Schiavon, Arnaud Siebert, Else Starkenburg, Sivarani Thirupathi, Guillaume Thomas, Marica Valentini, Kim Venn, Nicolas Wright

SWG4:

Manoj Kaplinghat, Ting Li, Keith Bechtol, Adam S. Bolton, Jo Bovy, Timothy Carleton, Chihway Chang, Alex Drlica-Wagner, Denis Erkal, Marla Geha, Johnny P. Greco, Carl J. Grillmair, Stacy Y. Kim, Chervin F. P. Laporte, Geraint F. Lewis, Martin Makler, Yao-Yuan Mao, Lina Necib, A. M. Nierenberg, Brian Nord, Andrew B. Pace, Marcel S. Pawlowski, Annika H. G. Peter, Robyn E. Sanderson, Guillaume F. Thomas, Erik Tollerud, Matthew G. Walker

SWG5:

Aaron Robotham, Kim-Vy Tran, Luke J. M. Davies, Nimish Hathi, Joss Bland-Hawthorn, Xu Kong, Linhua Jiang, Luca Cortese, Michael Balogh, Laura Ferrarese, Claudia Lagos, Danilo Marchesini, Khee-Gan Lee, Andreea Petric, Andrew M. Hopkins, Santi Cassisi, Pascale Jablonka, R. Michael Rich, Matthew A. Taylor, Daniel Smith, Jon Loveday, Malgorzata Siudek, Swara Ravindranath, Laura C. Parker, Yjan Gordon, Chris O'Dea, Laurence Tresse, Jennifer Sobeck, Ivana Damjanov, Alessandro Boselli, Sophia Lianou, Yiping Wang

SWG6:

Sara Ellison, Yue Shen, Trystyn Berg, Gisella de Rosa, Laura Ferrarese, Sarah Gallagher, Yjan Gordon, Patrick Hall, Dragana Ilić, Linhua Jiang, Stephanie Juneau, Jens-Kristian Krogager, Claudia Lagos, Xin Liu, Pasquier Noterdaeme, Chris O'Dea, Patrick Petitjean, Andreea Petric, Luka Č. Popović, Daniel Smith, Karun Thanjavur

SWG7:

Will Percival, Christophe Yeche, Maciej Bilicki, Andreu Font-Ribera, Nimish P. Hathi, Cullan Howlett, Michael J. Hudson, Faizan Gohar Mohammad, Jeffrey A. Newman, Nathalie Palanque-Delabrouille, Carlo Schimd, Andrei Variu, Yuting Wang, Michael J. Wilson

SWG8:

Adam Burgasser, Daryl Haggard, Michele Bannister, John Ruan

## Other contributors to the DSC, 2019

*(not previously mentioned)*

Borja Anguiano, Megan Bedell, William Chaplin, Remo Collet, Jean-Charles Cuillandre, Pierre-Alain Duc, Nicolas Flagey, JJ Hermes, Alexis Hill, Devika Kamath, Mary Beth Laychak, Katarzyna Małek, Mark Marley, Andy Sheinis, Doug Simons, Sérgio G. Sousa, Kei Szeto, Yuan-Sen Ting, Simona Vegetti, Lisa Wells

# Preface to Version 1 of the Detailed Science Case, 2016

## Purpose and scope

This is the MSE *Detailed Science Case* (DSC). It is the highest level document in the MSE document hierarchy, and forms the foundation for the Level 0 Science Requirements Document (SRD).

The MSE DSC provides the science narrative describing the principal envisioned science goals of MSE and describes its impact on a broad range of science topics. Science Reference Observations (SROs) describe in detail specific transformational observing programs for MSE that span the range of science described in the DSC. Science Requirements are defined as the science capabilities required to conduct the SROs. The SROs are included as Appendices to the DSC.

## A note on the structure of the DSC

The DSC is an extensive document covering a range of science topics. All chapters and all appendices can be read as stand-alone documents. All chapters begin with a 1–2 page synopsis that summarizes the content of the chapter. The first chapter provides a summary of the entire document.

A 10 page summary of the entire project aimed at the international astronomy community is presented in an accompanying document, *A concise overview of the Maunakea Spectroscopic Explorer*.

## Credit and acknowledgements

This document represents contributions from over 150 scientists across the international astronomical community from the past 5 years. It has been compiled and edited by the MSE Project Scientist in close collaboration with members of the MSE Science Team. It is based on a large number of documents developed by members of the MSE Science Team and its precursor project, the Next Generation Canada-France-Hawaii Telescope (ngCFHT).

The MSE Project Office was established in 2014 after precursor studies led to the ngCFHT Feasibility Studies. The MSE Science Team then undertook a multi-stage process to better define the science capabilities of MSE and develop the associated science case. White Papers were drafted highlighting important science areas not necessarily previously considered by the ngCFHT studies. Phase 1 studies, under the coordination of three leads, were then conducted. The purpose of these studies was to identify the most compelling science areas for further development in three broad areas (Stars and the Milky Way, The Low Redshift Universe, The High Redshift Universe) and to present the science cases for each. Cooperation between these groups was encouraged for areas of overlap. This led to the first set of proposed Science Reference Observations. From this initial set, a shortlist of 12 SROs was selected for further development. These SROs form the basis for the derivation of the Science Requirements.

The main chapters of the Detailed Science Case are based on the compilation of material that has been developed for MSE during this multi-stage process, including the ngCFHT Feasibility Study. The final set of SROs is presented in as Appendices to the DSC.

All original science documents on which the DSC is based are available at
`http://mse.cfht.hawaii.edu/docs/sciencedocs.php`

### Science team leads and coordinators

Carine Babusiaux (Stars and the Milky Way)
Michael Balogh (The low redshift Universe)
Simon Driver (The high redshift Universe)

Pat Côté (Lead author, the ngCFHT Feasibility Study 2012)

### Science Reference Observations leads and contributors

(Full author lists for each SRO are given in the Appendices)

Leads

Carine Babusiaux (SRO-3, Milky Way archaeology and the in situ chemical tagging of the outer Galaxy)
Michael Balogh (SRO-6, Nearby galaxies and their environments)
Helene Courtois (SRO-12, Dynamics of the dark and luminous cosmic web during the last three billion years)
Luke Davies (SRO-11, Connecting high redshift galaxies to their local environment: 3D tomographic mapping of the structure and composition of the IGM, and galaxies embedded within it)
Simon Driver (SRO-9, The chemical evolution of galaxies and AGN over the past 10 billion years ($z<2$)
Laura Ferrarese (SRO-7, Baryonic structures and the dark matter distribution in Virgo and Coma)
Sarah Gallagher (SRO-10, Mapping the inner parsec of quasars with MSE)
Rodrigo Ibata (SRO-4, Stream kinematics as probes of the dark matter mass function around the Milky Way)
Nicolas Martin (SRO-5, Dynamics and chemistry of Local Group galaxies)

Aaron Robotham (SRO-8, Evolution of galaxies, halos and structure over 12Gyrs)
Kim Venn (SRO-2, Rare stellar types and the multi-object time domain)
Eva Villaver (SRO-1 Exoplanets)

Contributors

Jo Bovy, Alessandro Boselli, Matthew Colless, Johan Comparat, Pat Côté, Kelly Denny, Pierre-Alain Duc, Sara Ellison, Richard de Grijs, Mirian Fernandez-Lorenzo, Laura Ferrarese, Ken Freeman, Raja Guhathakurta, Patrick Hall, Andrew Hopkins, Mike Hudson, Andrew Johnson, Nick Kaiser, Jun Koda, Iraklis Konstantopoulos, George Koshy, Andrew Hopkins, Khee-Gan Lee, Adi Nusser, Anna Pancoast, Eric Peng, Celine Peroux, Patrick Petitjean, Christophe Pichon, Bianca Poggianti, Carlo Schimd, Prajval Shastri, Yue Shen, Chris Willot

## DSC editors and principal contributors to new material in the DSC

Michael Balogh (Low mass galaxies)
Jo Bovy (CDM halo substructures)
Pat Côté (Low mass stellar systems)
Scott Croom (IFU studies of galaxy evolution)
Sara Ellison (Galaxy mergers)
Laura Ferrarese (The Virgo Cluster)
Sarah Gallagher (Reverberation mapping)
Rosine Lallement (The Interstellar Medium)
Nicolas Martin (The Local Group)
Patrick Petitjean (The Intercluster Medium)
Carlo Schimd (The Dark Universe)
Dan Smith (SKA synergies)
Matthew Walker (Nearby dwarf galaxies and dark matter)
Jon Willis (Galaxy clusters)

## Phase 1 study report co-authors

Carine Babusiaux, Michael Balogh, Alessandro Bosselli Matthew Colless, Johan Comparat, Helene Courtois, Luke Davies, Richard de Grijs, Simon Driver, Sara Ellison, Laura Ferrarese, Rodrigo Ibata, Sarah Gallagher, Aruna Goswami, Mike Hudson, Andrew Johnson, Matt Jarvis, Eric Jullo, Nick Kaiser, Jean-Paul Kneib, Jun Koda, Iraklis Konstantopoloulous, Rosine Lallement, Khee-Gan Lee, Nicolas Martin, Jeff Newman, Eric Peng, Celine Peroux, Patrick Petitjean, Christophe Pichon, Bianca Poggianti, Johan Richard, Aaron Robotham, Carlo Schimd, Yue Shen, Firoza Sutaria, Edwar Taylor, Kim Venn, Ludovic van Waerbeke, Chris Willot

## Science white papers

Giuseppina Battaglia (The accretion history of the Milky Way halo through chemical tagging)
Pat Cote (Upgrades to the MSE Point Design Specifications)

Aruna Goswami (Study of old metal-poor stars)
Richard de Grijs (Spatially resolved stellar populations and star formation histories of the largest galaxies)
Pierre-Alain Duc (Probing the large scale structures of massive galaxies with dwarfs/GC/UCDs in a variety of environments)
Pat Hall, Charling Tao, Yue Shen, Sarah Gallagher (Opportunistic Transient Targeting)
Misha Haywood, Paola di Matteo (Bulge science with MSE)
Misha Haywood, Paola di Matteo (Galactic archeology: The outskirts of the Milky Way disk)
Matt Jarvis (SKA Synergies)
Iraklis Konstantopoulos (The Redshift Reject Rubric)
Iraklis Konstantopoulos (Dense, Low-Mass Galaxy Groups: A Phase-Space Conundrum)
Iraklis Konstantopoulos (Star Clusters as Chronometers for Galaxy Evolution)
Rosine Lallement (Galactic InterStellar Medium: the 3D era with MSE)
Patrick Petitjean (Three Dimensional Reconstruction of the InterGalactic Medium)
Charli Sakari, Kim Venn et al. (Integrated Light Spectroscopy at High Resolution with MSE)
Carlo Schimd et al.; updated October 2014 (MSE: a velocity machine for cosmology)
Arnaud Seibert (Milky Way structure and evolution, ISM)
Yue Shen (Multi-Object AGN Reverberation Mapping with MSE)
Sivarani Thirupathi (Tracers of Pre-galactic and extra galactic Lithium abundances in Milky Way: towards solving the lithium problem)
Yuting Wang (Optimal tests on gravity or cosmological quantities on large scales)
Yiping Wang (SFH and scale-length evolution of massive disk galaxies towards high-z)

## Other contributions and feedback

(not previously mentioned)

Ferdinand Babas, Steve Bauman, Elisabetta Caffau, Mary Beth Laychak, David Crampton, Daniel Devost, Nicolas Flagey, Zhanwen Han, Clare Higgs, Vanessa Hill, Kevin Ho, Sidik Isani, Shan Mignot, Rick Murowinski, Gajendra Pandey, Derrick Salmon, Arnaud Siebert, Doug Simons, Else Starkenburg, Kei Szeto, Brent Tully, Tom Vermeulen, Kanoa Withington

## MSE Science Team membership (2016)

Nobuo Arimoto
Martin Asplund
Herve Aussel
Carine Babusiaux
Michael Balogh
Michele Bannister
Giuseppina Battaglia
Harish Bhatt
SS Bhargavi
John Blakeslee
Joss Bland-Hawthorn
Alessandro Boselli
Jo Bovy
James Bullock
Denis Burgarella
Elisabetta Caffau
Tzu-Ching Chang
Andrew Cole
Johan Comparat
Jeff Cooke
Andrew Cooper
Pat Côté
Helene Courtois
Scott Croom

Richard de Grijs
Paola Di Matteo
Simon Driver
Pierre-Alain Duc
Sara Ellison
Ginevra Favole
Laura Ferrarese
Hector Flores
Ken Freeman
Bryan Gaensler
Sarah Gallagher
Peter Garnavich
Karoline Gilbert
Rosa Gonzalez-Delgado
Aruna Goswami
Puragra Guhathakurta
Pat Hall
Guenther Hasinger
Misha Haywood
Falk Herwig
Vanessa Hill
Andrew Hopkins
Mike Hudson
Narae Hwang
Rodrigo Ibata
Pascale Jablonka
Matthew Jarvis
Umanath Kamath
Lisa Kewley
Iraklis Konstantopoulos
George Koshy
Rosine Lallement
Damien Le Borgne
Khee-Gan Lee
Geraint Lewis
Robert Lupton
Sarah Martell
Nicolas Martin
Mario Mateo
Olga Mena
David Nataf
Jeffrey Newman
Gajendra Pandey
Eric Peng
Enrique Pérez
Celine Peroux
Patrick Petitjean
Bianca Poggianti
Francisco Prada
Mathieu Puech
Alejandra Recio-Blanco
Annie Robin
Aaron Robotham
Will Saunders
Carlo Schimd
Arnaud Seibert
Prajval Shastri
Yue Shen
Daniel Smith
C.S. Stalin
Else Starkenburg
Firoza Sutaria
Charling Tao
Karun Thanjuvur
Sivarani Thirupathi
Laurence Tresse
Brent Tully
Ludo van Waerbeke
Kim Venn
Eva Villaver
Matthew Walker
Jian-Min Wang
Yiping Wang
Yuting Wang
Jon Willis
David Yong
Gongbo Zhao

## ngCFHT science Feasibility Study membership

(not previously mentioned)

Patrick Boisse, James Bolton, Piercarlo Bonifacio, Francois Bouchy, Len Cowie, David Crampton, Katia Cunha, Magali Deleuil, Ernst de Mooij, Patrick Dufour, Sebastien Foucaud, Karl Glazebrook, John Hutchings, Jean-Paul Kneib, Chiaki Kobayashi, Rolf-Peter Kudritzki, Damien Le Borgne, Yang-Shyang Li, Lihwai Lin, Yen-Ting Lin, Martin Makler, Norio Narita, Changbom Park, Ryan Ransom, Swara Ravindranath, Bacham Eswar Reddy, Marcin Sawicki, Luc Simard, Raghunathan Srianand, Thaisa Storchi-Bergmann, Keiichi Umetsu, Ting-Gui Wang, Jong-Hak Woo, Xue-Bing Wu

# Chapter 1

# Executive Summary

The Maunakea Spectroscopic Explorer (MSE) is an end-to-end science platform for the design, execution and scientific exploitation of spectroscopic surveys. It will unveil the composition and dynamics of the faint Universe and impact nearly every field of astrophysics across all spatial scales, from individual stars to the largest scale structures in the Universe. Major pillars in the science program for MSE include (i) the ultimate Gaia follow-up facility for understanding the chemistry and dynamics of the distant Milky Way, including the outer disk and faint stellar halo at high spectral resolution (ii) galaxy formation and evolution at cosmic noon, via the type of revolutionary surveys that have occurred in the nearby Universe, but now conducted at the peak of the star formation history of the Universe (iii) derivation of the mass of the neutrino and insights into inflationary physics through a cosmological redshift survey that probes a large volume of the Universe with a high galaxy density. MSE is positioned to become a critical hub in the emerging international network of front-line astronomical facilities, with scientific capabilities that naturally complement and extend the scientific power of Gaia, the Large Synoptic Survey Telescope, the Square Kilometer Array, Euclid, WFIRST, the 30m telescopes and many more.

The MSE Observatory has an 11.25m aperture with a 1.5 square degree field of view that will be fully dedicated to multi-object spectroscopy. MSE is designed for transformative, high precision studies of faint astrophysical phenomena. 3249 fibers will feed spectrographs operating at low ($R \sim 3000$) and moderate ($R \sim 6000$) spectral resolution, and 1083 fibers will feed spectrographs operating at high ($R \sim 20/40K$) resolution. All spectrographs are available all the time. The entire optical window from 360–950 nm and the near-infrared $J$ and $H$ bands will be accessible at the lower resolutions, and windows in the optical range will be accessible at the highest resolution.

The entire MSE system is optimized for high throughput, high signal-to-noise observations of the faintest sources in the Universe. High quality calibration and stability is ensured through the dedicated operational mode of the observatory, which ensures that the equivalent of more than 10 million fiber hours of 10m class spectroscopy are available for forefront science every year. The discovery efficiency of MSE is an order of magnitude higher than any other spectroscopic capability currently realized or in development, and it will produce datasets equivalent in number of objects to a SDSS Legacy Survey every 7 weeks, albeit on a telescope with an aperture 20 times larger.

*On stellar and Galactic scales,* MSE will be the most powerful facility available to provide critical stellar spectroscopic observations for stars across the Hertzsprung-Russell diagram. MSE stellar monitoring programs will dramatically improve our understanding of stellar multiplicity, including providing dynamical masses for unprecedented samples of transiting hot Jupiters ($\sim 10^4$). MSE will study the *r*-process element abundances across our Galaxy, and will produce the definitive dataset of the most chemically primitive stars with which to identify the signatures of the very first supernovae and chemical enrichment events in the Universe. MSE will carry out the ultimate spectroscopic follow-up of the Gaia mission, and it is the only facility capable of producing vast high resolution spectroscopic datasets for stars across the full magnitude range of Gaia targets. MSE is critical to our understanding of the faint and distant regimes of the Galaxy, in particular the outer disk, thick disk and stellar halo, and will conduct in-situ chemodynamical analysis of millions of individual stars in all Galactic components. It will play a central role in revolutionary three dimensional ISM mapping experiments that will be boosted by Gaia parallax distances. MSE will also usher in a new era for studies of nearby dwarf galaxies: it will enable accurate and efficient chemo-dynamical measurements across their full luminosity range and will provide spectra for more than an order of magnitude more stars in each system.

*In the dark sector,* MSE will conduct a suite of surveys that provide critical input into determinations of the mass function, phase-space distribution and internal density profiles of dark matter halos across all mass scales. Recent N-body and hydrodynamical simulations of cold, warm, fuzzy and self-interacting dark matter suggest that non-trivial dynamics in the dark sector could have left an imprint on structure formation. Analyzed within these frameworks, the extensive and unprecedented kinematic datasets produced by MSE will be used to search for deviations away from the prevailing model in which the dark matter particle is cold and collisionless.

*On the scales of galaxies and galaxy groups,* MSE will allow the types of revolutionary extragalactic surveys that have been conducted at $z = 0$ to be conducted as a function of redshift out to the peak of cosmic star formation. At low redshift, MSE will probe a representative volume of the local Universe to lower stellar and halo masses then is achievable with current and other upcoming surveys. It will measure the extension of the stellar mass function to masses below $10^8 M_\odot$, for a cosmologically representative, unbiased, spatially complete spectroscopic sample. High redshift extragalactic surveys with MSE will provide a high-completeness, magnitude-limited sample of galaxy redshifts spanning the epoch of peak cosmic star-formation ($1.5 < z < 3.0$). MSE surveys have sufficient areal coverage, depth and temporal character to cover the AGN zoo in this redshift range, and will probe the growth of supermassive black holes (SMBHs) by measuring luminosity functions, clustering, outflows and mergers. A multi-epoch reverberation mapping campaign with MSE will yield $2000 - 3000$ robust time lags, an order of magnitude more than the expected yields from current campaigns, and will enable accurate SMBH mass measurements for the largest sample of quasars to date and unprecedented mapping of the central regions.

*On the largest scales*, MSE can answer two of the most important questions within physics, namely determining the masses of neutrinos and providing insight into the physics of inflation. It can do this by undertaking a cosmological redshift survey that will probe a large volume of the Universe with a high galaxy density. MSE will measure the level of non-Gaussianity as parameterized by the local parameter $f_{NL}$ to a precision $\sigma(f_{NL}) = 1.8$. Combining these data with data from a next generation CMB stage 4 experiment and existing DESI data will provide the first $5\sigma$ confirmation of the neutrino mass hierarchy from astronomical observations.

# Chapter 2

# The scientific landscape of the Maunakea Spectroscopic Explorer

**Abstract**

The Maunakea Spectroscopic Explorer will address key science cases ranging from time-domain searches of sub-stellar mass objects and understanding the statistical properties of exoplanetary hosts to determining the properties of the dark matter particle and measuring the mass of the neutrino. MSE is an end-to-end science platform for the design, execution, and scientific exploitation of spectroscopic surveys. It is positioned to become a critical hub in the emerging international network of front-line astronomical facilities, with scientific capabilities that naturally complement and extend the scientific power of Gaia, the Large Synoptic Survey Telescope, the Square Kilometer Array, Euclid, WFIRST, the 30m telescopes and many more. A continual interaction between its science users and observatory staff will ensure that MSE remains responsive to, and at the forefront of, changing science priorities throughout its many years of operation. The scientific and strategic importance of MSE cannot be overstated, and this is reflected by the strong backing that large aperture, wide field multi-object spectroscopy has on the international scene, and for which MSE is the realization of that ambition.

## 2.1 The composition and dynamics of the faint Universe

MSE will unveil the composition and dynamics of the faint Universe and impact nearly every field of astrophysics across all spatial scales, from individual stars to the largest scale structures in the Universe. Major pillars in the science program for MSE include (i) the ultimate Gaia follow-up facility for understanding the chemistry and dynamics of the distant Milky Way, including the outer disk and faint stellar halo at high spectral resolution (ii) galaxy formation and evolution at cosmic noon, via the type of revolutionary surveys that have occurred in the nearby Universe, but now conducted at the peak of the star formation history of the Universe (iii) derivation of the mass of the neutrino and insights into inflationary physics through a cosmological redshift survey that probes a large volume of the Universe with a high galaxy density.

The scientific capabilities of MSE are summarized in Table 1. This suite of capabilities enables MSE to excel at precision studies of faint astrophysical phenomena that are beyond the reach of 4-m class spectroscopic instruments. A broad range of spectral resolutions – including $R \simeq 2K$, $R \simeq 6$ and $R \simeq 20/40K$ – ensure that the specialized technical capabilities of MSE enable a diverse range of transformational astrophysics from exoplanetary through to cosmic scales. These are discussed in detail in each of the proceeding stand-alone chapters of this document, and we refer to each of the relevant chapters for detailed discussions of the diverse science case for MSE. Here, we summarize a few highlights from each chapter:

- **(Chapter 3) Exoplanets and stellar astrophysics**

  MSE will be the most powerful facility available to provide critical stellar spectroscopic observations from the lowest-mass brown dwarfs to massive, OB-type giants, and including important yet rare stellar types across the Hertzsprung-Russell diagram, such as solar twins, Cepheids, RR Lyrae stars, AGB and post-AGB stars, as well as faint, metal-poor white dwarfs. In the stellar regime, strong synergies exist between MSE and TESS, PLATO, Gaia, eROSITA, LSST and other time domain facilities (see Figure 1). MSE stellar monitoring programs will dramatically improve our understanding of stellar multiplicity, including the interaction and common evolution between companions spanning a vast range of parameter space such as low-mass stars, brown dwarfs and exoplanets, but also pulsating, eclipsing or eruptive stars. MSE will provide dynamical masses for unprecedented samples of transiting hot Jupiters ($\sim 10^4$), allowing the exploration of critical outstanding questions of this intriguing class of planets such as their radius inflation and migration mechanisms. An MSE follow-up campaign of transiting, massive TESS planets will help to disentangle hot Jupiters from brown dwarfs and very low-mass stars, in order to test the mass-radius relation for objects that populate both the high-mass end of the exoplanet regime and the low-mass end of the stellar regime. MSE will be crucial to constrain the poorly-known aspects of stellar physics in the low-mass domain ($0.08 - 0.5$ $M_\odot$), including the equation-of-state of dense gas, opacities, nuclear reaction rates, and magnetic fields.

- **(Chapter 4) Chemical nucleosynthesis**

| Site characteristics | |
|---|---|
| **Observatory latitude** | 19.9 degrees |
| **Accessible Sky** | 30,000 square degrees (airmass < 1.55 i.e., $\delta > -30$ degrees) |
| **Median image quality** | 0.37 arcsec (free atmosphere, zenith, 500 nm) |
| **Av. length of night adjusted for weather** | 8 hours |
| **Observing efficiency (on-sky, on-target)** | 80% |
| **Expected on-target observing hours** | 2336 hours / year |
| **Expected on-target fiber-hours** | 10,119,552 fiber-hours / year (total): 7,589,664 (LR & MR) / 2,529,888 (HR) |

| Telescope architecture | |
|---|---|
| **Structure, focus** | Altitude-azimuth, Prime |
| **M1 aperture / Science field of view** | 80.8 $m^2$ / 1.5 square degrees |
| **Spectrograph system** | 6 x LMR spectrographs (4 channels / spectrograph), all identical, each channel seperately switchable to provide LR and MR modes |

| Fiber positioning system | |
|---|---|
| **Multiplexing** | 4,332 (total): 3,249 (LR & MR) / 1,083 (HR) |
| **Fiber size** | 1 arcsec (LR & MR) / 0.75 arcsec (HR) |
| **Positioner patrol radius** | 90.3 arcsecs |
| **Positioner accuracy** | 0.06 arcsec rms |
| **Positioner closest approach** | Two fibers can approach with 7 arcsecs of each other (three fibers can be placed within 9.9 arcsec diameter circle) |
| **Repositioning time** | < 120 seconds |
| **Typical allocation efficiency** | > 80 % (assuming source density approximately matched to fiber density) |

| Low resolution (LR) spectroscopy | | | | |
|---|---|---|---|---|
| **Wavelength range** | $360 \leqq \lambda \leqq 560$ nm | $540 \leqq \lambda \leqq 740$ nm | $715 \leqq \lambda \leqq 985$ nm | $960 \leqq \lambda \leqq 1320$ nm |
| **Spectral resolution *(center of band)*** | 2,550 | 3,650 | 3,600 | 3,600 |
| **Sensitivity requirement *(pt. source, 1hr, zenith, median seeing, monochromatic magnitude)*** | m = 24.0<br>SNR/res. elem. = 2, $\lambda > 400$ nm<br>SNR/res. elem. = 1, $\lambda \leqq 400$ nm | m = 24.0<br>SNR/resolution element = 2 | m = 24.0<br>SNR/resolution element = 2 | m = 24.0<br>SNR/resolution element = 2 |

| Moderate resolution (MR) spectroscopy | | | | |
|---|---|---|---|---|
| **Wavelength range** | $391 \leqq \lambda \leqq 510$ nm | $576 \leqq \lambda \leqq 700$ nm | $737 \leqq \lambda \leqq 900$ nm | $1457 \leqq \lambda \leqq 1780$ nm |
| **Spectral resolution *(center of band)*** | 4,400 | 6,200 | 6,100 | 6,000 |
| **Sensitivity requirement *(pt. source, 1hr, zenith, median seeing, monochromatic magnitude)*** | m = 23.5<br>SNR/res. elem. = 2, $\lambda > 400$ nm<br>SNR/res. elem. = 1, $\lambda \leqq 400$ nm | m = 23.5<br>SNR/resolution element = 2 | m = 23.5<br>SNR/resolution element = 2 | m = 24.0<br>SNR/resolution element = 2 |

| High resolution (HR) spectroscopy | | | |
|---|---|---|---|
| **Wavelength range** | $360 \leqq \lambda \leqq 440$ nm | $420 \leqq \lambda \leqq 520$ nm | $500 \leqq \lambda \leqq 900$ nm |
| **Wavelength band** | $\lambda$ / 30 | $\lambda$ / 30 | $\lambda$ / 15 |
| **Spectral resolution *(center of band)*** | 40,000 | 40,000 | 20,000 |
| **Sensitivity requirement *(pt. source, 1hr, zenith, median seeing, monochromatic magnitude)*** | m = 20.0<br>SNR/resolution element = 10, $\lambda > 400$ nm<br>SNR/resolution element = 5, $\lambda \leqq 400$ nm | m = 20.0<br>SNR/resolution element = 10 | m = 20.0<br>SNR/resolution element = 10 |

| Science calibration | |
|---|---|
| **Sky subtraction accuracy** | 0.5% requirement (0.1% goal) |
| **Velocity precision** | 100 m/s (HR, SNR/resolution element = 30) |
| **Relative spectrophotometric accuracy** | 3% (LR, SNR/resolution element = 30) |

*Table 1: The detailed science capabilities of MSE*

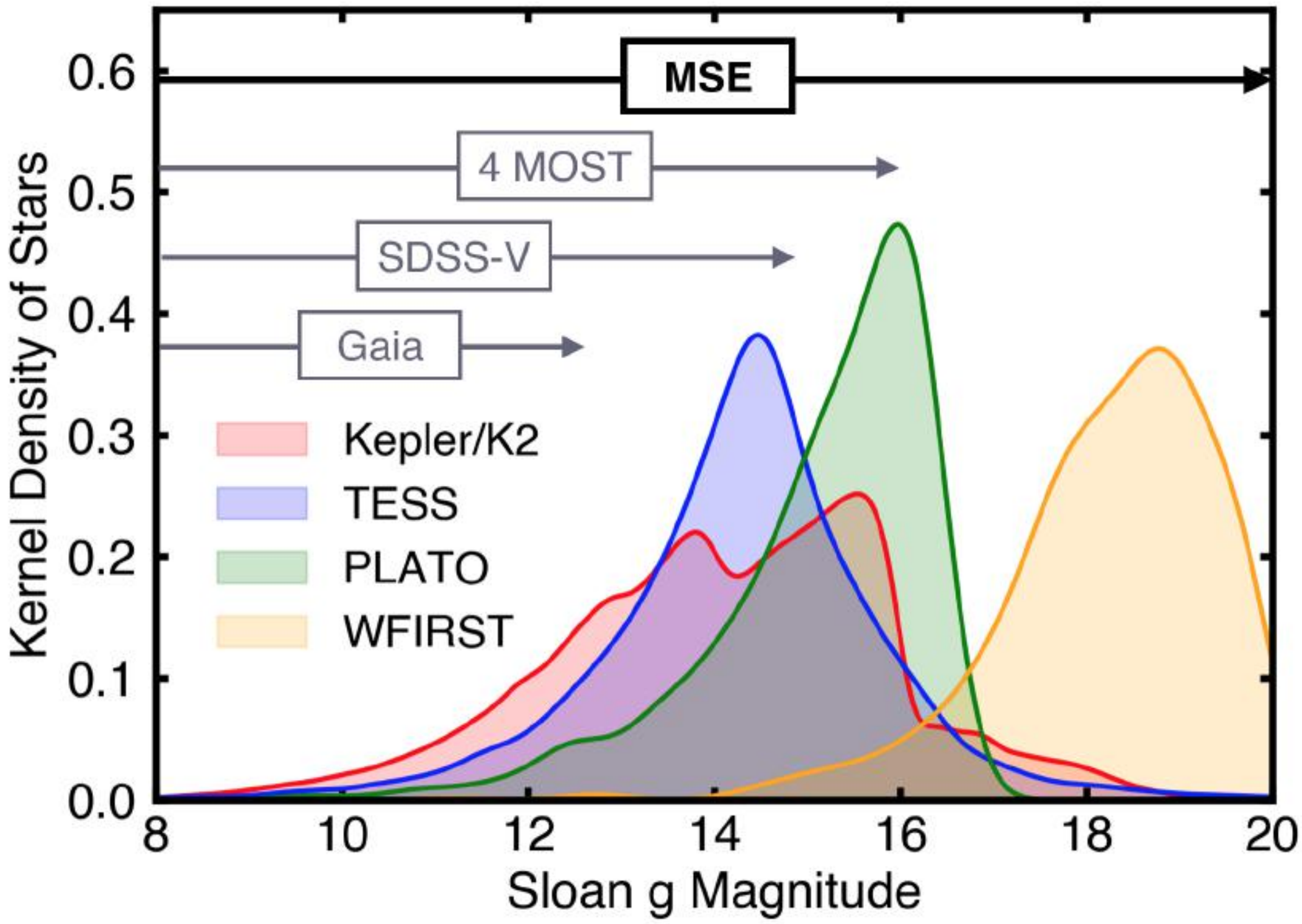

*Figure 1: Figure from Chapter 3: Exoplanets and Stellar Astrophysics. MSE is the only multi-object spectroscopic facility that can provide high-resolution optical spectroscopy for tens of millions of stars with high-precision future space-based photometry. Lines show the g-magnitude distribution for stars with space-based photometry from Kepler/K2 (red, Batalha et al., 2011; Huber et al., 2014, 2016) and predicted yields of stars observed with a photometric precision better than 1 mmag hr$^{-1}$ from an all-sky survey with TESS (blue, Sullivan et al., 2015; Stassun et al., 2018), a typical PLATO field (green, Rauer et al., 2014), and the WFIRST microlensing campaign (orange, Gould et al., 2015). Sensitivity limits of some other MOS facilities that will provide high-resolution ($R > 20K$) spectroscopy at least half of the sky ($> 2\pi$) are shown in grey. Lines are kernel densities with an integrated area of unity.*

MSE is uniquely tailored to understanding the cosmic formation and evolution of the elements of the periodic table. It will trace different nucleosynthetic processes, sites and timescales through the measurement of a large number of chemical species, including in the crucial blue/UV region of the spectrum. It is ideally suited for detecting the EuII 4129Å line, as well as a large number of other neutron-capture elements, covering the full element mass range. MSE will study the *r*-process element abundances in unprecedented numbers of stars across our Galaxy. MSE will produce the definitive dataset of the most chemically primitive stars with which to identify the signatures of the very first supernovae and chemical enrichment events in the Universe. It will enable a large scale study of the lithium abundances down to the lowest metallicities to understand the possible depletion mechanism(s) of lithium due to the first generation of stars. The relative importance of low and intermediate mass stars to the chemical enrichment of the Universe will be quantified by MSE via a systematic and comprehensive chemical abundance study of large samples of evolved stars in diverse metallicity environments, covering a wide range of initial masses. MSE will measure the dimensionality of chemical abundance space using data for millions of stars across all Galactic components and sub-components.

- **(Chapter 5) The Milky Way and resolved stellar populations**

  MSE will carry out the ultimate spectroscopic follow-up of the Gaia mission, and is critical to our understanding of the faint and distant regimes of the Galaxy. It is the only facility capable of producing vast high-resolution spectroscopic datasets for stars across the full magnitude range of Gaia targets. Uniquely, MSE will conduct in situ chemodynamical analysis of individual stars in all Galactic components, searching for inter-relationships between them and for departures from equilibrium. The unprecedented size of the stellar spectroscopic dataset will enable the definitive analysis of the metal-weak tail of the halo metallicity distribution function. MSE will bring about an entirely new era for nearby dwarf galaxy studies, enabling accurate chemo-dynamical measurements to be performed efficiently across the full range of dwarf galaxy luminosities ($10^{3-7}L_{\odot}$), and providing spectra for at least an order of magnitude more stars in each system, reaching well beyond where circular velocity curves are expected to peak. MSE will also provide a comprehensive understanding of the chemodynamics of M31 and M33, essentially enabling a full chemodynamical deconstruction of these galaxies across their entire spatial extent. Finally, MSE will play a central role in revolutionary three dimensional ISM mapping experiments that will be boosted by Gaia parallax distances.

- **(Chapter 6) Astrophysical tests of dark matter**

  MSE will conduct a suite of surveys that provide critical input into determinations of the mass function, phase-space distribution, and internal density profiles of dark matter halos across all mass scales. Importantly, recent N-body and hydrodynamical simulations of cold, warm, fuzzy and self-interacting dark matter suggest that non-trivial dynamics in the dark sector could have left an imprint on structure formation. Analyzed within these frameworks, the extensive and unprecedented kinematic datasets produced by MSE will be used to search for deviations away from the prevailing model in which the dark matter particle is cold and collisionless. MSE will provide an improved estimate of the local density of dark matter, critical for direct detection experiments, and will improve estimates of the J-factor for indirect detection through self-annihilation or decay into Standard Model particles. MSE will determine the impact of low mass substructures on the dynamics of Milky Way stellar streams in velocity space, and will allow for estimates of the density profiles of the dark matter halos of Milky Way dwarf galaxies using more than an order of magnitude more tracers (see Figure 2). In the low redshift Universe, MSE will provide critical redshifts to allow the luminosity functions of vast numbers of satellite systems to be derived, and MSE will be an essential component of future strong lensing measurements to obtain the halo mass function for higher redshift galaxies. Across nearly all mass scales, the improvements offered by MSE in comparison to any other facility are such that the relevant dynamical analyses will become limited by systematics rather than statistics.

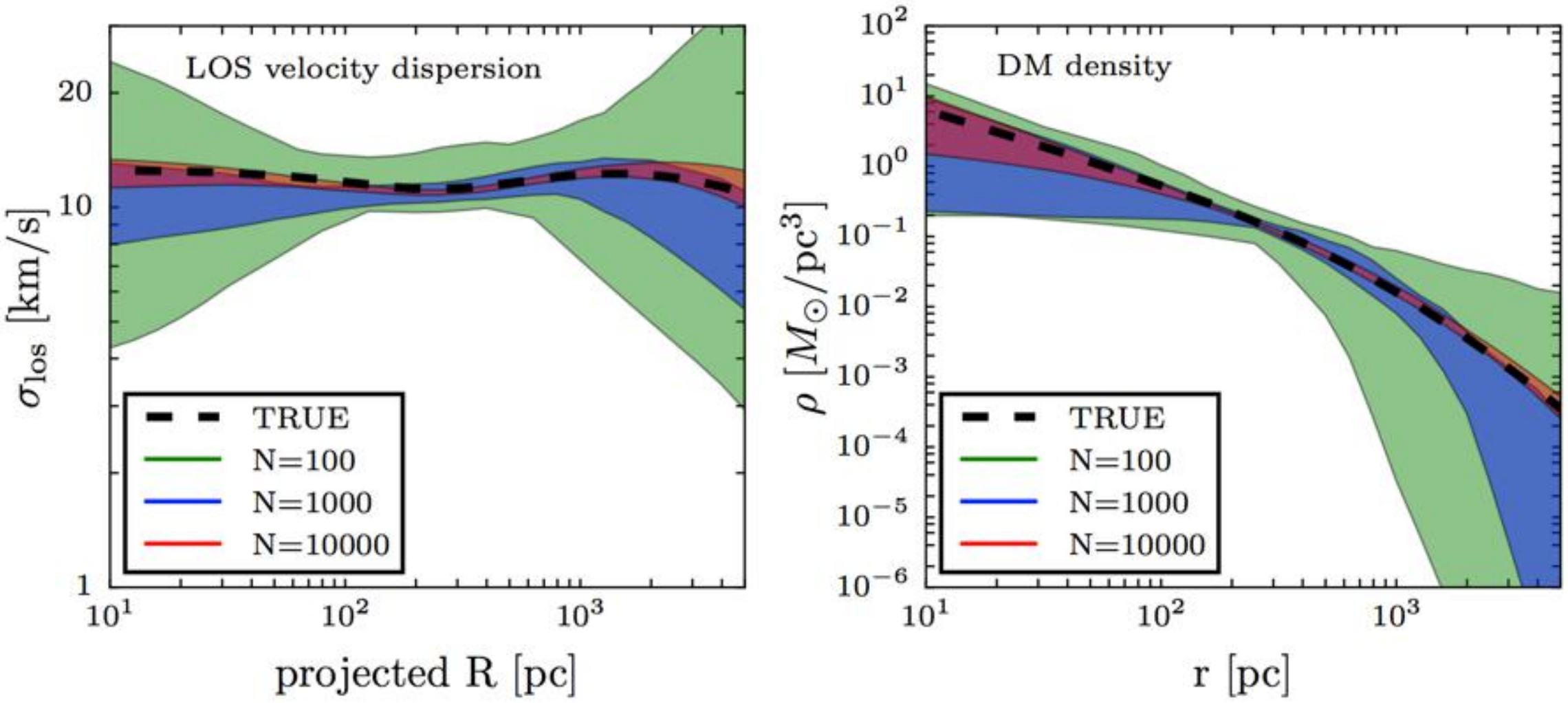

*Figure 2: Figure from Chapter 6: Astrophysical tests of dark matter. Recovery of intrinsic line-of-sight velocity dispersion (left) and inferred dark matter density (right) profiles as a function of spectroscopic sample size. Shaded regions represent 95% credible intervals from a standard analysis (based on the Jeans equation) of mock data sets consisting of line of sight velocities for $N = 10^2$, $10^3$ and $10^4$ stars (median velocity error $2\,{\rm km\,s^{-1}}$), generated from an equilibrium dynamical model for which true profiles are known (thick black lines, which correspond to a model having a cuspy NFW halo with $\rho(r) \propto r^{-1}$ at small radii).*

- **(Chapter 7) Galaxy formation and evolution**

  MSE will allow the types of revolutionary extragalactic surveys that have been conducted at $z = 0$ to be conducted as a function of redshift out to the peak of cosmic star formation. At low redshift, MSE will probe a representative volume of the local Universe to lower stellar and halo masses then is achievable with current and other upcoming surveys (see Figure 3). These surveys will allow a diverse array of science topics from dwarf galaxies, to galaxy interactions in the low stellar mass regime, the environ-mental impact on galaxy evolution and the extension of large-scale structure analyses to low mass groups. A fundamental measurement for MSE will be the extension of the stellar mass function to masses below $10^8 M_\odot$, for a cosmologically representative, unbiased, spatially complete spectroscopic sample. High redshift extragalactic surveys with MSE will provide a high-completeness, magnitude limited sample of galaxy redshifts spanning the epoch of peak cosmic star-formation ($1.5 < z < 3.0$). They will cover the diverse range of environments probed by surveys such as SDSS and GAMA (groups, pairs, mergers, filaments, voids), but at an epoch when the Universe was under half its current age. The scale of insights available from such surveys is difficult to predict given the relatively scarce amount of information we currently have in these regimes (thousands as opposed to millions of targets). As such, MSE will likely have a generation-defining impact on our understanding of how galaxies evolve over 12 billion years of cosmic time.

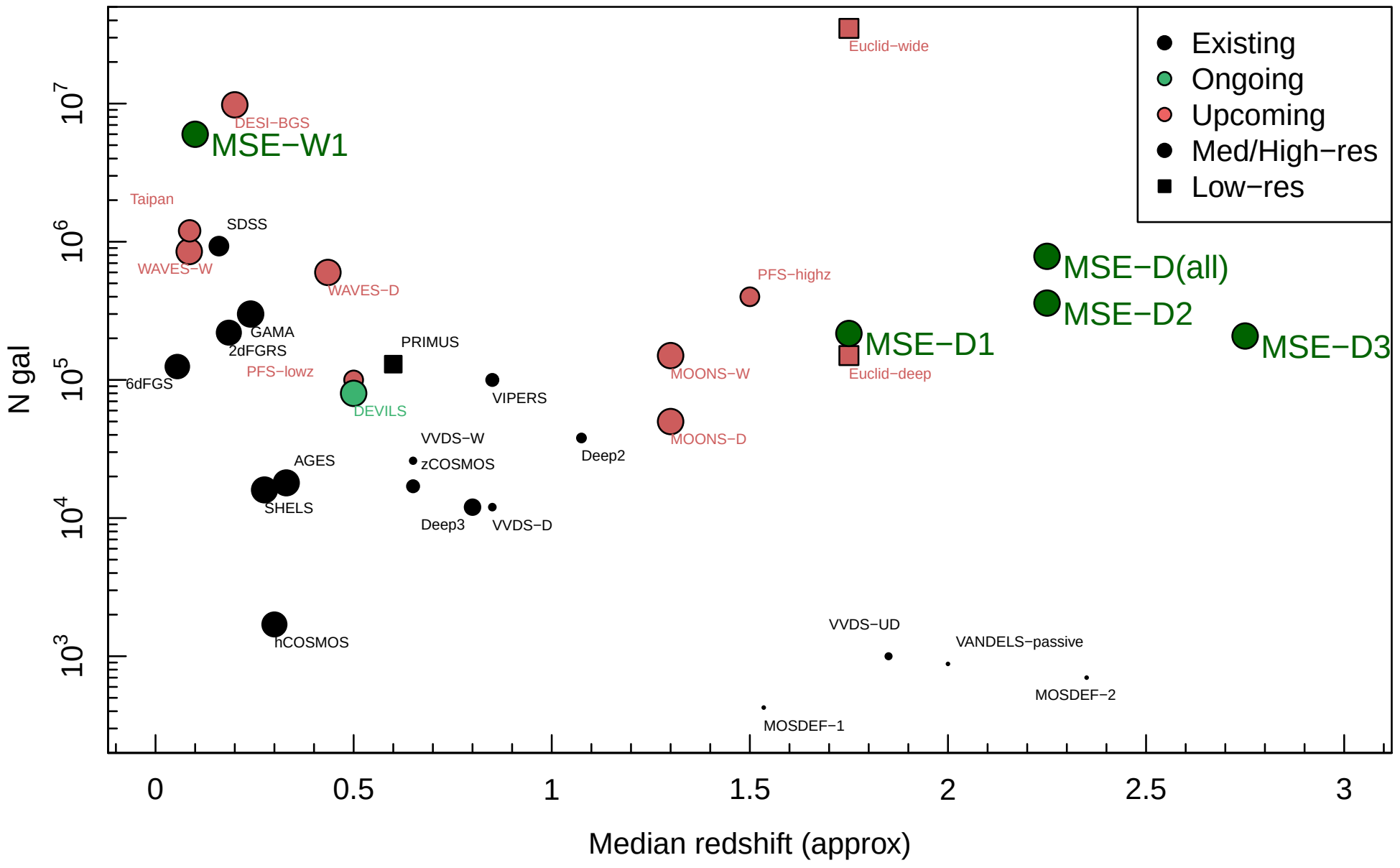

*Figure 3: Figure from Chapter 7: Galaxy formation and evolution. Comparison between proposed MSE extragalactic wide and deep surveys (dark green points) with existing, ongoing and upcoming spectroscopic surveys. Point size approximately scales with survey completeness down to a fixed magnitude limit.*

- **(Chapter 8) Active Galactic Nuclei and supermassive black holes**

  MSE will probe the growth of supermassive black holes (SMBHs), and will characterize their relationship with host galaxies near and far, by measuring luminosity functions, clustering, outflows, variability and mergers. A multi-epoch reverberation mapping campaign with MSE will yield 2000 – 3000 robust time lags of the quasar broad-line region over a broad range of redshift and luminosity. This is an order of magnitude more than the expected yields from current campaigns, and enables accurate SMBH mass measurements for the largest sample of quasars to date and unprecedented mapping of their central regions. MSE will provide large, statistical samples of growing SMBHs with sufficient areal coverage, depth, and temporal character to cover the AGN zoo at $z = 0-3$. It will also build a large sample of very high-$z$ ($z > 7.5$) quasars, and so probe the most distant SMBHs. MSE will simultaneously study the radiation environment close to growing SMBHs and the star formation histories of their host galaxies. MSE will provide better determination of the cosmological density of galaxies that host a binary SMBHs and will constrain the rate of SMBH mergers. Further, MSE will also allow us to better constrain how the cluster environment evolves from one that is conducive to the triggering of efficiently accreting AGN at high-$z$, to one that inhibits (efficient) AGN activity at low-$z$.

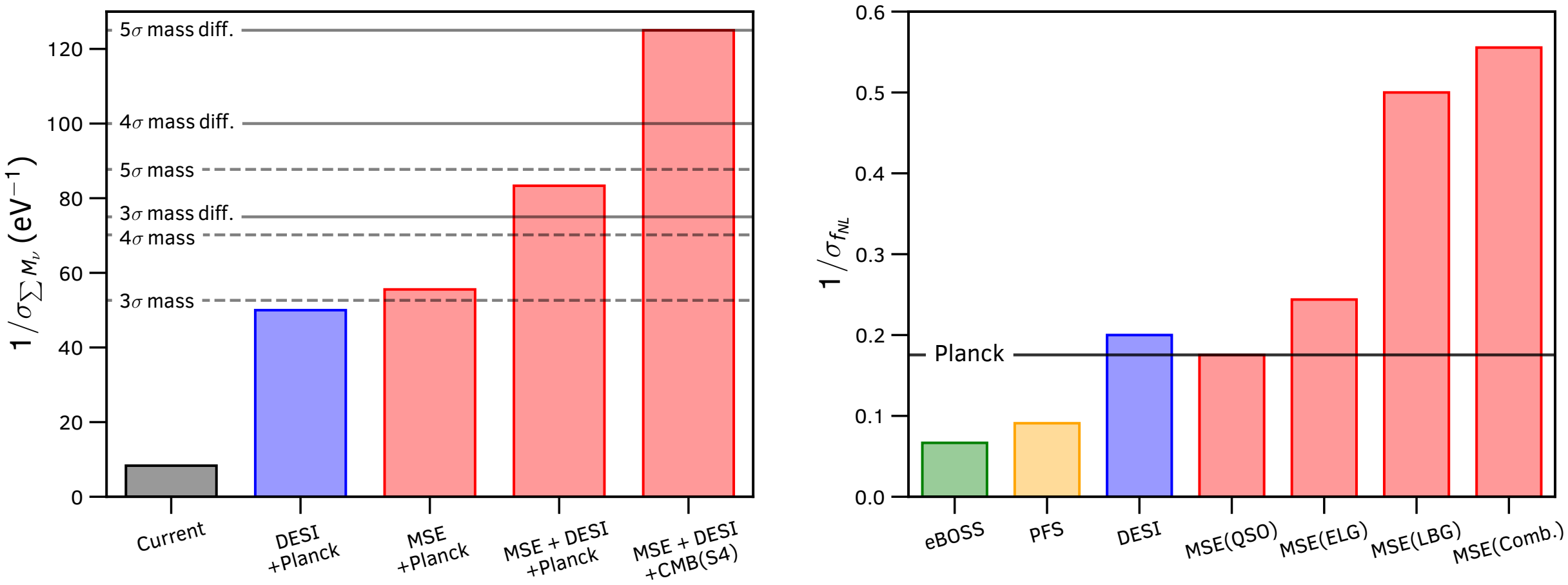

*Figure 4: Figure from Chapter 9: Cosmology. A summary of the neutrino mass (left) and primordial non-Gaussianity (right) constraints achievable with the MSE compared to other surveys. The dashed and solid horizontal lines in the left panel show the requirements for* 3*,* 4 *and* $5-\sigma$ *constraints on the sum of neutrino masses (either hierarchy) and on the mass difference between hierarchies respectively. The horizontal line on the right panel shows the current constraints on* $f_{NL}$ *from Planck CMB data.*

- **(Chapter 9) Cosmology**

  MSE can answer two of the most important remaining questions within physics, namely determining the masses of neutrinos and providing insight into the physics of inflation. It can do this by undertaking a cosmological redshift survey that will probe a large volume of the Universe with a high galaxy density. With such a survey, we expect a measurement of the level of non-Gaussianity as parameterized by the local parameter $f_{NL}$ to a precision $\sigma(f_{NL}) = 1.8$. Combining these data with data from a next generation CMB stage 4 experiment and existing DESI data will provide the first $5\sigma$ confirmation of the neutrino mass hierarchy from astronomical observations (see Figure 4). In addition, the Baryonic Acoustic Oscillations observed within the sample will provide measurements of the distance-redshift relationship in six different redshift bins between $z = 1.6$ and 4.0, each with an accuracy of $\sim 0.6\%$. These high-redshift measurements will provide a probe of the dark matter dominated era and test exotic models where dark energy properties vary at high redshift. The simultaneous measurements of Redshift Space Distortions at redshifts where dark energy has not yet become important directly constrain the amplitude of the fluctuations parameterized by $\sigma_8$, at a level ranging from 1.9% to 3.6% for the same redshift bins. In addition to this major program, MSE is able to address many other areas of cosmological interest, for example a deep survey for LSST photometric redshift training and pointed observations of galaxy clusters to $z = 1$.

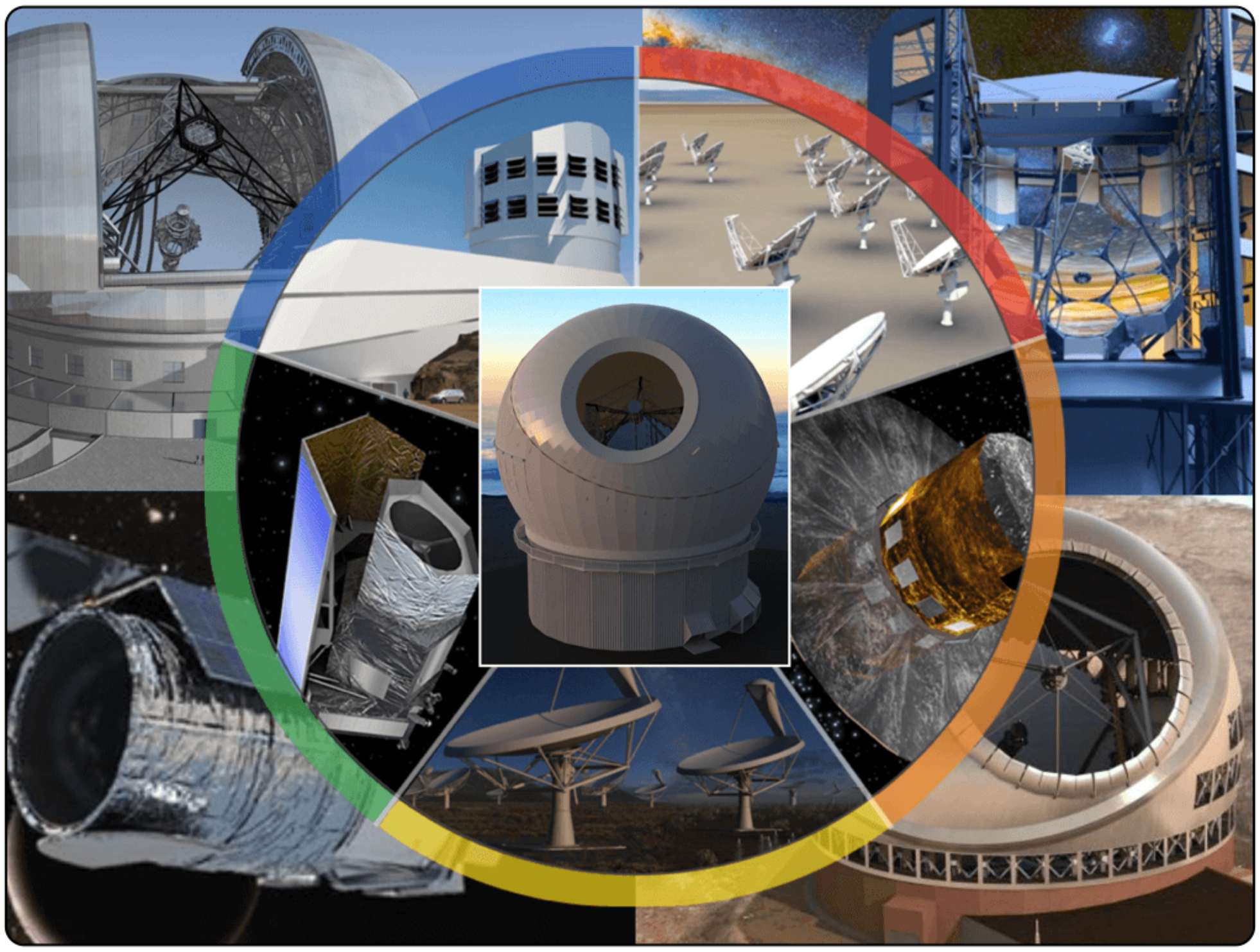

*Figure 5: Some of the most notable next generation facilities that, together with MSE, will help define the international network of astronomical facilities operational beyond the 2020s. Outer ring (clockwise from top left): the Giant Magellan Telescope, the Thirty Meter Telescope, WFIRST, the Extremely Large Telescope. Inner ring (clockwise from top left): the next generation Very Large Array, Gaia, the Square Kilometer Array, Euclid, the Large Synoptic Survey Telescope.*

### 2.2 MSE and the international network of astronomical facilities

Astronomy has entered a new, multi-wavelength realm of big facilities. Figure 5 shows just some of the most notable of these new observatories. Collectively, they represent many billions of dollars of investment and many decades of development by large teams from the international community. They probe questions as diverse and fundamental as the origin of life and the composition of our Universe. The Second Data Release of Gaia is causing a revolution in our understanding of the stellar populations of the Milky Way and new insights into its formation. Multiple Square Kilometer Array (SKA) pathfinders are in operation, the Large Synoptic Survey Telescope (LSST) is nearing completion of construction, and construction of the 30m-class telescopes is underway. These are no longer “next generation” ambitions, but instead are the cutting edge of what is possible now. Perhaps most excitingly, our view of the Universe is no longer limited only to the electromagnetic spectrum, such that the term “multiwavelength” is no longer sufficient to completely describe the windows into the Universe that we have at our disposal. The detection of the first gravity waves in February 2016 by LIGO (Abbott et al., 2016) opened up a fundamentally new window on the Universe. The subsequent panchromatic, multi-facility, multi-messenger study of the LIGO detection GW170817 in 2017 (Drout et al. 2017; Shappee et al. 2017) was a watershed moment in astronomy that demonstrated the power of coordinated observations using multiple eyes on the sky.

Each of the facilities shown in Figure 5 has a well-defined science case and can operate as a stand-alone facility. However, the recent history of astronomy demonstrates that it is through the combination of data from these new facilities and extant telescopes that many of the major advances will be made. The science that will ultimately emerge from this collaboration is far more diverse than we can currently anticipate.

The development of MSE explicitly recognizes the existence of this international network of astronomical facilities that will define the future of frontline astrophysics. MSE is positioned to be a critical hub in this network, with scientific capabilities that naturally complement and extend the capabilities of the facilities shown in Figure 5. Many of the synergies are obvious, many cannot yet be anticipated, but all are important to the future health of astronomy. Here, we discuss MSE in this global context of astronomical capabilities.

### 2.2.1 Optical imaging of the Universe

The most powerful ground-based imaging capability in the next decade will be the Large Synoptic Survey Telescope (LSST). Currently in construction on Cerrro Pachon, LSST is a 6.2m (effective aperture) telescope with a field of view of 9.6 square degrees that will enter into full science operations in 2023[1]. Its primary mission is to undertake a 10-year program to monitor $\sim 25\,000$ square degrees, building up deep *ugriz* images of the sky through the co-addition of $\sim 1000$ exposures per filter at each position. “Deep fields” will also be obtained over considerably smaller areas. Each individual exposure lasts $\sim 15$s and has a single-pass depth of $r \simeq 24.5$. The final, co-added, LSST main survey depth is $r \simeq 27.5$ (LSST Science Collaboration et al., 2009).

The discovery space of LSST is enormous. The design of the surveys has been guided by four main science themes, specifically the search for new objects in the solar system, mapping of the Milky Way, the search for transient astrophysical phenomena, and probing dark energy (see LSST Science Collaboration et al. 2009). In practice, however, LSST can be expected to impact every area of astronomy. Over the 10 year baseline, LSST will take $\sim 5.5$ million images and will have measured $\sim 7$ trillion single epoch sources, for a total of $\sim 37$ billion distinct objects. It is a requirement that photometric zeropoints are stable to $\sim 10$ millimags; with nearly 1000 visits per patch of sky, LSST will probe the variable and transient optical

[1]https://www.lsst.org/about/timeline

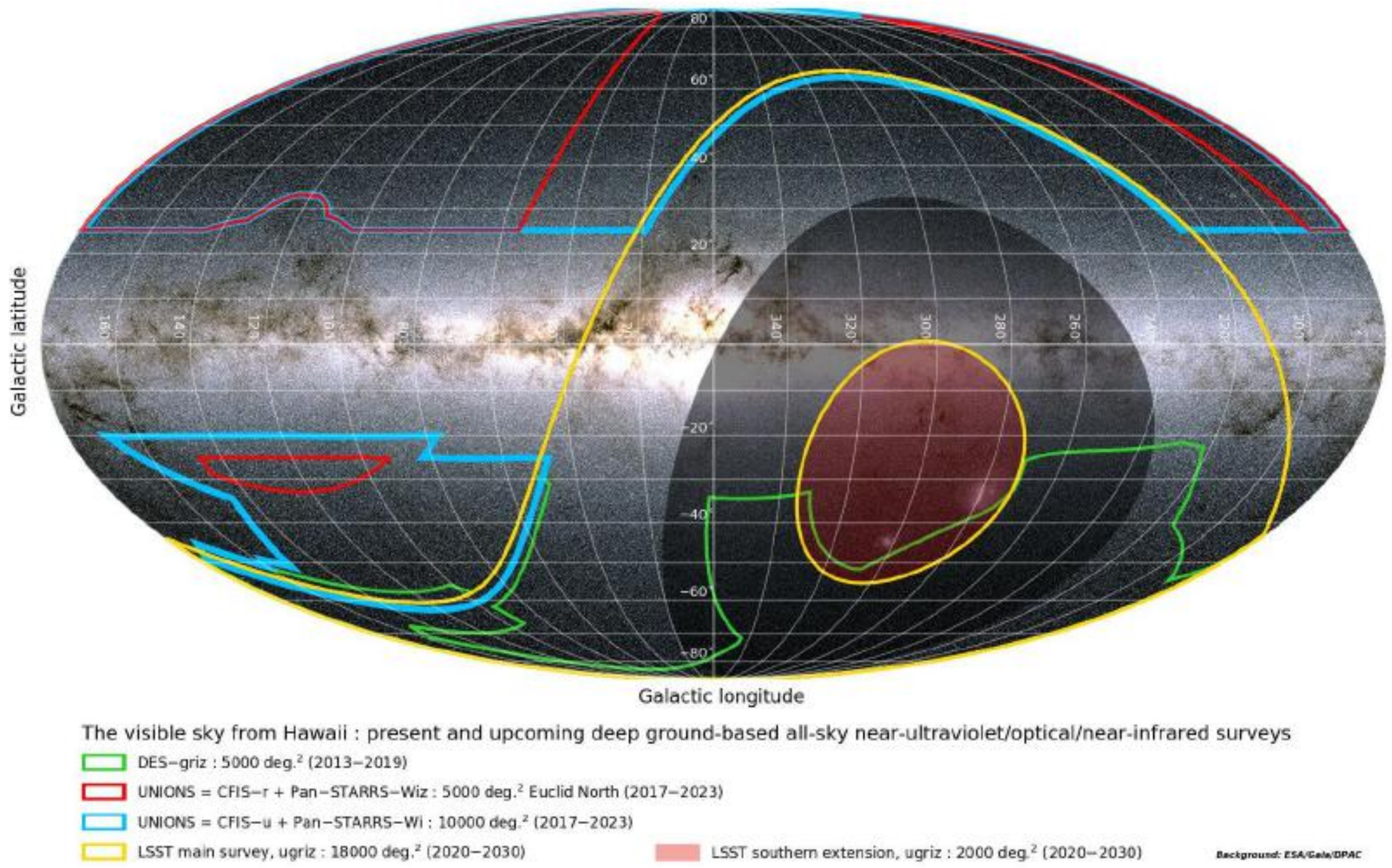

*Figure 6: Mollweide projection in Galactic coordinates showing the sky coverage of current and future deep ground-based optical surveys. The dark area is the part of the sky not visible from Hawaii. The background image is from Gaia.*

sky with unprecedented accuracy and cadence.

In the northern hemisphere, the entire $3\pi$ steradians of sky visible from Hawaii has been mapped in $grizY$ by the Pan-STARRS1 telescope as part of the PS1 survey (Flewelling et al., 2016) ($5-\sigma$ point-source depth of $r \simeq 21.8$). Pan-STARRS2 was commissioned in 2018. Among various surveys that the two Pan-STARRS telescopes are conducting, they are targeting approximately 5000 square degrees of the extragalactic sky, concentrating around the North Galactic Cap, to reach $i = 24.8$ and $z = 24.6$ ($5-\sigma$ point-source depths). This complements CFHT/MegaCam $u-$ and $r-$band photometry in the same area taken as part of the Canada-France Imaging Survey (CFIS; Ibata et al. 2017), that reaches $u = 24.7$ and $r = 25.1$ ($5-\sigma$ point-source depths). This multi-band, multi-telescope effort intends to incorporate $g-$band in the future for complete $ugriz$ imaging, and is called the "Ultraviolet Near-Infrared Optical Northern Survey", or UNIONS. It is a stand-alone survey program that is a component of the ground-based imaging campaign for the Euclid space satellite. Figure 6 shows the sky coverage in Galactic coordinates provided by current and upcoming deep ground-based optical surveys, specifically UNIONS, LSST, and the Dark Energy Survey.

Many other extensive wide field imaging surveys on 4–8 m telescopes already exist, both at optical wavelengths (e.g., the CFHT Legacy Survey and other CFHT programs; the Subaru SuprimeCam imaging archive) and at NIR wavelengths (e.g. the UKIDSS northern sky survey using UKIRT; various surveys on VISTA). Hyper-Suprime-Cam (HSC) on the Subaru telescope on Maunakea is currently conducting a 5 year strategic survey to map $\sim$ 1500 square degrees along the celestial equator to unprecedented depth (Aihara et al., 2018). In the south, the Dark Energy Survey (DES; Abbott et al., 2018b) has mapped 5000 square degrees of sky in $grizY$ using the Dark Energy Camera on the 4m Blanco telescope to a depth of $r = 25$ ($5-\sigma$ point-source depth; see Morganson et al., 2018). Observations were completed in January 2019. This area also contributes towards the Dark Energy Camera Legacy Survey (DECaLS), which is approximately 1 magnitude shallower than DES. These surveys, in combination with the Beijing-Arizona Sky Survey (BASS) and the Mayall z-band Legacy Survey (MzLS), are producing a 14,000 square degree survey of the sky in $grz$, and provides pre-imaging for DESI; see Dey et al. 2018.

MSE is situated at an equatorial location ($l = 19.9°$), with the entire sky north of $\delta = -30°$ accessible at airmass less than 1.55 (30,000 square degrees) The scientific synergies between MSE and deep imaging surveys such as UNIONS, LSST (where more than half of the primary survey area is accessible), Subaru/HSC, PS1 and others, are extensive. It is worth emphasizing that the majority of the billions of sources identified is these surveys are far outside the capabilities of 4-m class spectroscopic follow-up. As an 11.25m aperture facility, MSE can obtain spectroscopic data for sources identified in a single pass of LSST. Smaller aperture spectroscopic facilities are unable to exploit the discovery potential of these surveys to the same degree as MSE, and emphasizes the strong need for large aperture, dedicated MOS.

### 2.2.2 Wide field optical and infrared science from space

Imaging from space is dominated by two missions: Euclid and WFIRST. Euclid, scheduled for launch in 2022[2], is an ESA-led mission to understand the nature of dark energy and dark matter through weak lensing and galaxy clustering. It is a 1.2m NIR/optical telescope that will image a minimum area of 15,000 square degrees at $|b| > 30$ degrees during its anticipated 5-year lifetime. Limiting AB magnitudes are expected to be 24 in YJH, and 25.8 in a single, broad optical filter (RIZ; $5-\sigma$ point source depths). Slitless spectroscopy will also be obtained. Overall, more than a billion galaxies will be observed (Laureijs et al., 2011). WFIRST is a complementary, NASA-led mission planned for a mid-2020's launch[3] that has a considerably larger aperture than Euclid, at 2.4m. In terms of imaging, one of its primary missions will be to map $\sim 2000$ square degrees of sky to depths of $J \sim 26.7$, in addition to a set of ultra-deep degree-scale fields (Spergel et al., 2015). Precision photometry and astrometry will be obtained for all sources. Importantly, up to 25% of WFIRST time is likely going to be P.I.-time, ensuring that deep pointed surveys of various fields will occur. Both Euclid and WFIRST will also obtain several deep fields as part of their primary science, resulting in tens of square degrees with exceptionally deep photometry.

Examples of the sort of scientific opportunities produced by Euclid and WFIRST for MSE are numerous. A driving theme of extragalactic science with MSE is linking galaxies to their surrounding large scale structure. As such, the emergence of structure on kiloparsec scales is of great interest. Driver et al. (2013) propose that galaxies form via two stages, firstly bulge formation via some dynamical hot process (i.e., collapse, rapid merging, disc instabilities and/or clump migration), and secondly disc formation via a more quiescent dynamically cool process (i.e., gas in-fall and minor-merger accretion events). The pivotal redshift for these processes is $z \sim 1.5$. Euclid and WFIRST will provide extremely deep high-spatial resolution imaging ($\sim 0.2$") and will discern structure to sub-kpc scales out to $z > 2.5$ at rest-optical wavebands over extremely large areas. This will allow a direct measurement of the epochs at which the various structures emerge (e.g., bulge, disc formation) and how they evolve (i.e, growth of spheroids, bulges, and discs). However, to fill in the void between very nearby surveys such as SDSS and GAMA (Driver et al. 2011) and very distant surveys with HST and JWST requires the combination of Euclid/WFIRST imaging with photometric-redshifts and MSE spectra. Indeed, the combination of MSE spectral analysis with Euclid/WFIRST NIR imaging, ground-based optical (LSST, UNIONS) and radio (SKA, ngVLA) will provide a complete blueprint of galaxy evolution from the present epoch to the peak of the cosmic star-formation era ($z = 0$ to $z \sim 2.5$). The crucial element is the ability to obtain reasonable signal-to-noise ratio (SNR) spectra ($\sim 30$) for high-z ($z \sim 2.5$), faint ($i \sim 25$ mag) systems with a high-level of completeness. MSE is the only spectroscopic facility that provides these capabilities.

---
[2] `https://www.euclid-ec.org`
[3] `https://wfirst.gsfc.nasa.gov/index.html`

### 2.2.3 Multi-messenger and time-domain astronomy

MSE will operate in a scientific era with far greater synoptic coverage of the sky through wide-field, panchromatic surveys such as LSST, ZTF, SKA-1 and Euclid, which are expected to greatly increase the yield of transient events (e.g.,Ridgway et al. 2014). The dedicated spectroscopic capabilities of MSE enables key science in the time domain through multiple types of observations. In addition to *multiplexed* science, there is also *multi-threaded* science; that is, tackling science cases that on a per field basis do not require the full multiplexing capability of MSE - and so can be conducted in parallel to other programs - but which integrated over multiple fields produce extensive datasets of relatively rare targets.

Major multiplexed time-domain programs are already being planned in both the stellar and extragalactic regimes. As an example of the former, multi-epoch stellar radial velocities to a per-epoch precision of 100 m/s will enable the discovery/confirmation of stellar and substellar companions down to Jovian-mass planets, with orbit periods of up to several years. Such observations are particularly important for follow-up of pre-survey transit detections made by Kepler, TESS and Plato, providing in principle dynamical masses for every hot Jupiter uncovered in that survey. The combined mass and radius measurements will be a powerful probe of the physics of hydrogen-degenerate matter by mapping the mass-radius relationship across the maximum-degeneracy inflection (see Chapter 3). As an example of a multiplexed extragalactic time domain program, a ground-breaking reverberation mapping campaign will reveal the inner structures of AGN. Hundreds of epochs of observations of thousands of high- z quasars over a period of several years will accurately measure a large sample of SMBH masses through time and map the evolution of their inner regions (see Chapter 8).

Multi-threaded programs are particularly relevant for time-domain science, including the investigation of periodic (e.g., binaries/exoplanets, pulsation), evolutionary (e.g., post explosion supernovae) and bursting behavior (e.g., flares, CV novae), solar system moving objects (e.g., main-belt and more distant asteroids, trans-Neptunian objects), as well as astrophysical transients (e.g., supernovae, kilonovae, neutron star mergers, gravitational wave electromagnetic counterparts, fast radio bursts), on time-scales from minutes to years. MSE is not envisioned to operate primarily as a "prompt Target of Opportunity" facility, but by dynamically positioning even only 1% of fibers per field on recently reported transients (alerts from within the past few hours or nights), then a database of $> 100,000$ spectra of faint transients will be amassed *per year*. Rare extraordinary triggers (e.g., highly localized GW signals, extreme bursts, near-field supernovae) might also be considered for follow-up on shorter timescales if the potential scientific yield is of sufficient value.

The LIGO-Virgo discovery and localization of neutron star merger GW170817 spectacularly ushered in the era of multi-messenger astrophysics with gravitational waves. Perhaps no other event has had a greater impact on the prospects for time domain astrophysics in recent years, a science whose future prospects hinges crucially on access to rapid, wide field spectroscopy. Key to understanding binary neutron star mergers and their associated kilonovae is early detection and monitoring of their spectrum. The objectives of upcoming spectroscopic campaigns will be to detect the blue spectra of GW counterparts, to characterize the shock breakout, to place constraints on heavy element production ($r$-process elements in particular), and to characterize the host galaxies and environments of the binary neutron star mergers. In addition to kilonovae follow-up, another area of particular scientific opportunity for MSE is the rapid response spectroscopic follow-up of other explosive variables and transients; novae, supernovae, hypernovae, gamma ray bursts, X-ray flashes, fast radio bursts, and others, on timescales of hours and days.

MSE is preparing for success in this area through the early identification and adoption of key operational procedures tuned to enable time domain observations. Coordination of MSE observations with time domain imaging surveys such as LSST is a critical element of this planning. One excellent possibility being considered is to plan for wide-field survey follow-up of Euclid, ZTF, and/or LSST deep drilling fields, with a set of fibers held in reserve and placed in "real time" on new transients that are triggered during planned observations ("planning to be lucky"). Another high-risk, high-reward strategy is for MSE to target the highest-sensitivity regions of the LIGO-Virgo footprint in real time. This will allow for very rapid dynamic placement of fibers on new candidate electromagnetic counterparts to GW sources, and could be combined with a planned survey of AGN or galaxies in similar regions of the sky.

### 2.2.4 The era of Gaia

Gaia is a landmark astrometric space mission whose primary focus is a detailed understanding of the structure and composition of our Galaxy. It was launched in 2014 and had its second data release (DR2) in April 2018, based on the first 2 years of science observations (see Figure 7). Gaia is conducting an all-sky survey that is measuring the positions, parallaxes and proper motions of $\sim 1.3$ billion stars between $3 \lesssim G \lesssim 21$, approximately 1% of the entire stellar content of the Milky Way. Objects brighter than $G \sim 15$ mag typically have parallax uncertainties better than 0.04 milliarcseconds, and the precision for the brighter sources by the end of the mission are expected to be of order 20 microarcseconds (the approximate width of a human hair viewed at a distance of 1000 km). Beyond astrometry, Gaia is obtaining multi-band photometry for all sources, and it is additionally equipped with a Radial Velocity Spectrometer (RVS) that is measuring velocities for objects brighter than $G \sim 17$ mag (roughly 150 million stars) to an accuracy of 1–10kms$^{-1}$. Basic astrophysical information – including interstellar reddening and atmospheric parameters – is being acquired for the brightest ∼ 5 million stars. Chemical abundance information will also be provided for a few elements (i.e., Mg, Si, Ca, Ti and Fe for stars of spectral type F-G-K) for stars brighter than 12th magnitude $G \simeq 12$.

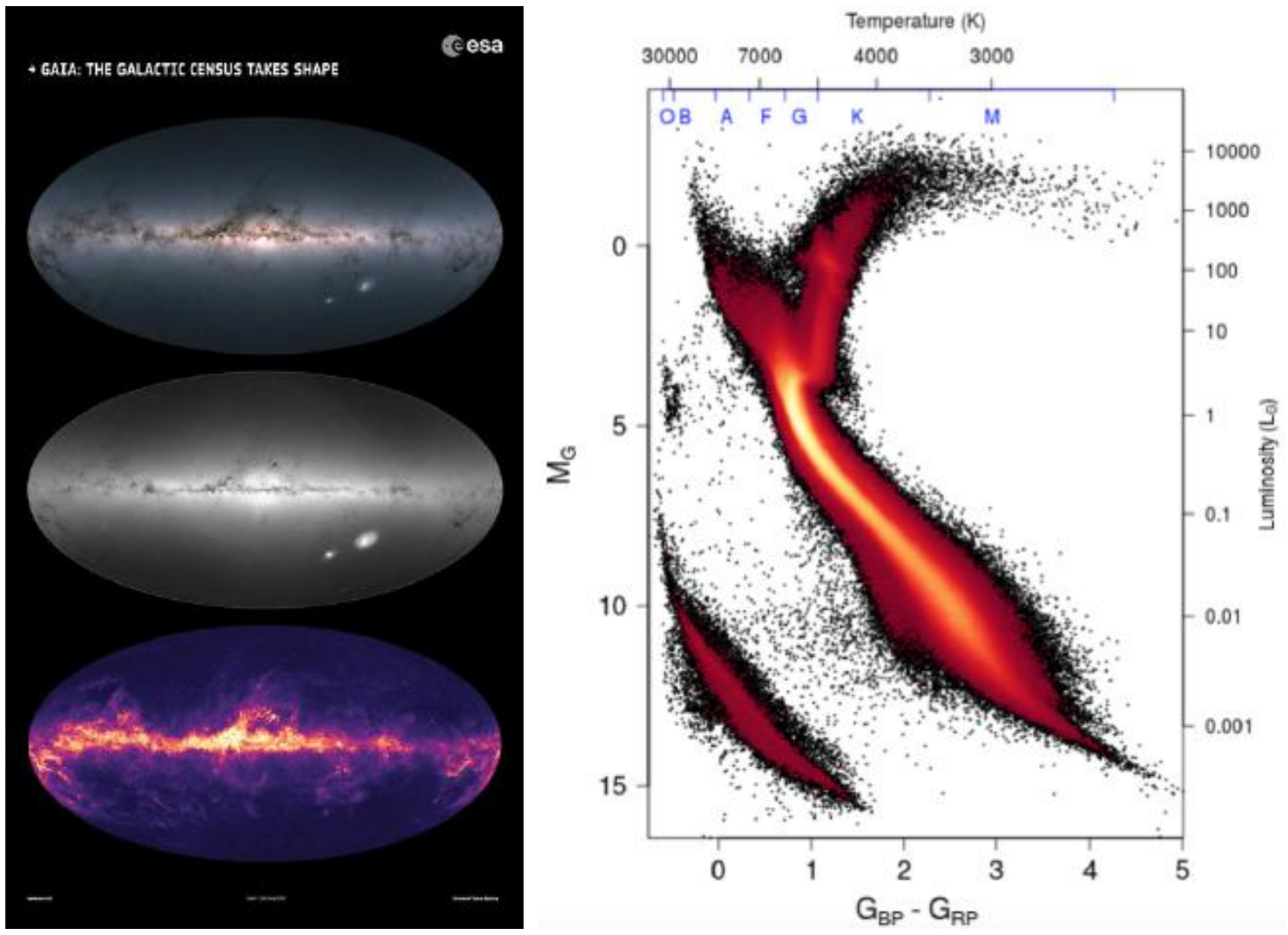

*Figure 7: Left panels: The all-sky view of our Milky Way Galaxy and nearby galaxies from Gaia DR2. The maps show the total brightness and colour of stars (top), the total density of stars (middle) and the distribution of interstellar dust (bottom)*[4]*. Right panel: A Gaia colour magnitude diagram (Hertzsprung Russell Diagram) of 4,276,690 stars with low foreground extinction,* $E(B–V) < 0.015$ *mag, shown with square-root scaling. Figure from Gaia Collaboration et al. (2018a).*

Gaia is revolutionizing our vision of the Milky Way and its local environment. The Gaia-Enceladus merger remnant has been discovered, showing that our Galaxy had a major (4 : 1) merger around 10 Gyrs ago. Remnants of Gaia-Enceladus are prevalent in the inner halo, and its accretion helped shape the thick disk (Belokurov et al., 2018b; Myeong et al., 2018c; Helmi et al., 2018b). This reinforces earlier ideas of the inner stellar halo's formation (e.g. Meza et al., 2005; Navarro et al., 2011) through an ancient merger (see review by Freeman & Bland-Hawthorn, 2002). The thin disk has been shown to be locally in a strongly perturbed state (Antoja et al., 2018), reinforcing earlier signs of disequilibrium in the disc (e.g. Minchev et al., 2009; Widrow et al., 2012) including a dynamical warp (Poggio et al., 2018) and a strong flare in the outer disk (Thomas et al., 2019). This result confirms many of the predictions of pre-Gaia DR2 models of Laporte et al. (2018b) of the interaction of the Milky Way disc with the Sagittarius dwarf galaxy (see Laporte et al., 2018d), which has long been suspected to play a role in shaping the structure of the Galactic disc (Ibata & Razoumov, 1998; Quillen et al., 2009; Purcell et al., 2011; Gómez et al., 2013). New open clusters (Castro-Ginard et al., 2018), halo streams (Malhan et al., 2018) and dwarf satellites (Torrealba et al., 2018) are being found. Hypervelocity stars, globular clusters, streams and dwarf spheroidals are being used to derive the Milky Way potential (e.g. Eadie & Jurić, 2018). Those are only a few examples. Overall, Gaia is confirming that the Galaxy is not an equilibrium figure and that the different components are not trivially separated. Instead, there is a strong interplay between them. External events and internal dynamics have blurred out the different components with cosmic time, at least at some level.

[4]Acknowledgement: Gaia Data Processing and Analysis Consortium (DPAC); Top and middle: A. Moitinho / A. F. Silva / M. Barros / C. Barata, University of Lisbon, Portugal; H. Savietto, Fork Research, Portugal; Bottom: Gaia Coordination Unit 8; M. Fouesneau / C. Bailer-Jones, Max Planck Institute for Astronomy, Heidelberg, Germany.

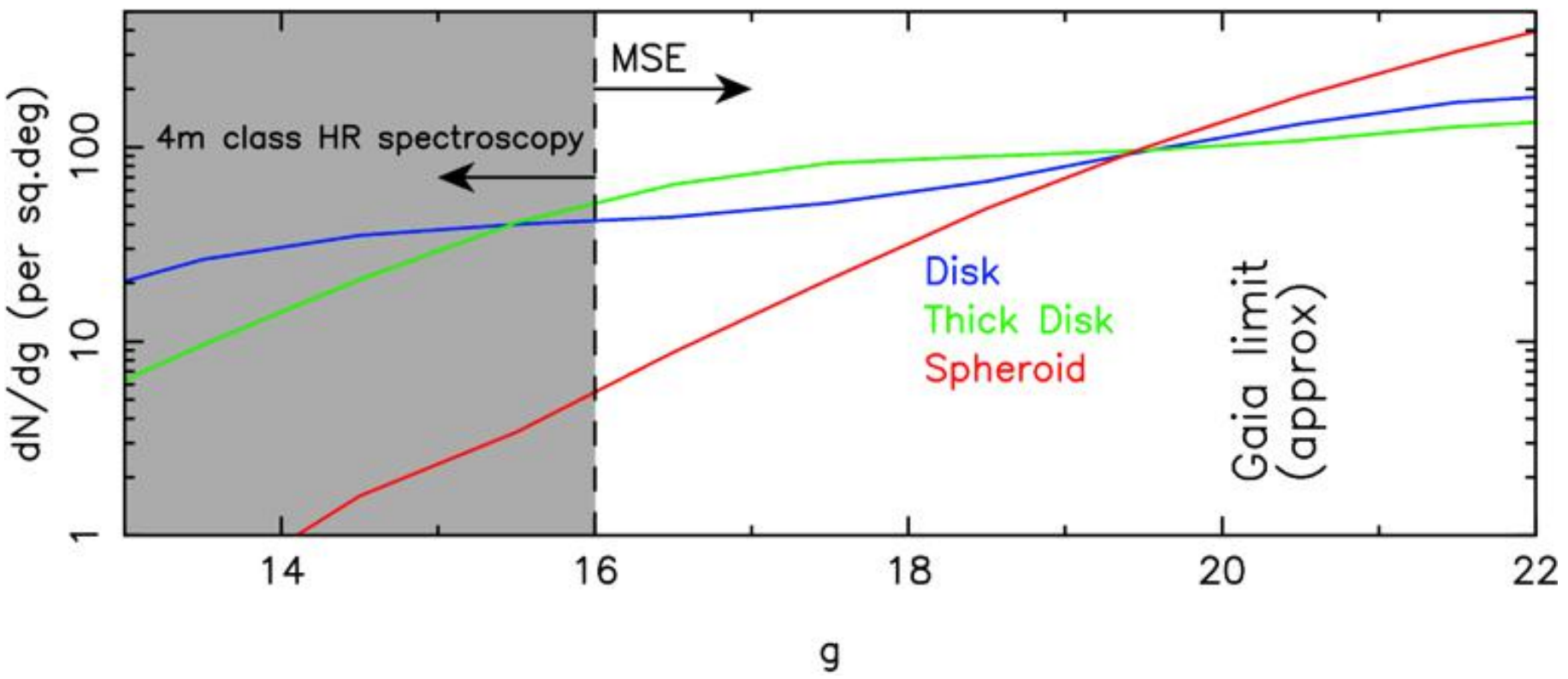

*Figure 8: Differential star counts as a function of magnitude for the three main Galactic components, based on the Bescancon model of the Galaxy (Robin et al. 2003) for a 100 square degree region in the vicinity of the north Galactic cap. The shaded region indicates the magnitude range accessible at high resolution to 4m class spectrographs (typically operating at* $R \sim 20,000$*–*$40,000$*). MSE is the only facility able to access the thick disk and spheroid at high resolution in the regions of the Galaxy in which they are the dominant components. MSE targets at high resolution span the full luminosity range of targets that will be identified with Gaia.*

The importance of ground-based spectroscopy to supplement Gaia data cannot be overstated. Of the numerous recent Galactic Archaeology papers using Gaia DR2 data - including those listed above - more than 20% use spectroscopic complements. MSE is the only survey spectrograph planned that will be able to observe millions of the faintest Gaia stars at high resolution. In terms of data on the dynamics of stars, the astrometric accuracy of Gaia is matched by similarly accurate radial velocities only for the very brightest subset of its targets. MSE radial velocities will give access to the full 6D position/velocity space for Gaia stars, and MSE data will give spectroscopic distances for stars in the range not covered by accurate Gaia parallaxes. In terms of chemistry, these spectra will carry information on the abundance of 20 to 30 elements from various nucleosynthetic families. This means that MSE can access the detailed chemodynamical signatures of every Galactic component and sub-component using *in situ* analysis of individual stars, as demonstrated in Figure 8. For comparison, AAT/GALAH (which has a magnitude limit of V = 14) estimate that only 0.2% of their targets will be halo stars[5] and WHT/WEAVE's (G<16) estimate is 3%

[5] https://galah-survey.org/survey_design

(C. Babusiaux, *private communication*). These surveys provide essential insight into the nearby Galaxy and the main disk in particular. From Figure 8, the thick disk becomes the dominant component at higher latitudes for $g \geq 16$. This is easily accessible with MSE even with minimal pre-selection of targets. As an 11.25m facility, MSE will obtain good SNR at high resolution in reasonable exposure times across the full magnitude range of Gaia targets, including where the stellar halo is dominant ($g \geq 19.5$ at high Galactic latitude).

The Gaia dataset is a unique and comprehensive resource for astronomy, given the unprecedented accuracy with which it will measure the 3-D positions of Galactic stars. There is a near perfect synergy between MSE and Gaia on multiple fronts, and the science that the combination of these facilities enables features heavily throughout this document.

### 2.2.5 Synergies at long wavelengths

At radio wavelengths, considerably development activity is underway and, in the near-future, SKA-1 is expected to have a profound impact. Among the many science goals of this transformational telescope, SKA-1 has the capability of detecting Milky Way-type galaxies via synchrotron radiation into the epoch of reionization, finding AGN of all types and luminosity, and tracing the neutral hydrogen content of galaxies to $z \sim 2$. On longer timescales, the Next Generation Very Large Array (ngVLA) is planned for the 2030s[6] and has an extensive science case, complementary to that of SKA-1. Key science topics that will be addressed by ngVLA include probing interstellar medium and star formation physics, and conducting large cosmological surveys looking at the cold molecular gas fuelling star formation in galaxies all the way back to the epoch of reionization.

Optical spectroscopy is crucial to maximize the scientific output of SKA-1 and ngVLA, and to gain the biggest leap in our understanding of galaxy formation and cosmology. For example, the extreme sensitivity of SKA-1 means that it will be able to detect normal star-forming galaxies and radio-quiet active galactic nuclei in radio continuum emission out to the highest redshifts beyond the so-called cosmic noon ($z > 1$) and back to the earliest times. However, the lack of spectral features in the radio continuum means that additional redshift information is required if we wish to harness this immense sensitivity. While photometric redshifts are often sufficient to calculate luminosity functions, they do not allow robust source classification, or accretion mode diagnostics, and are insufficient for most other science. Only with spectroscopy are we able to quantify a galaxy's environment, measure ages and metallicity, and unlock the true potential of the billions of dollars that have been invested in surveying the radio sky.

A good example of the type of synergy we can expect between SKA and MSE in this arena can be seen with the upcoming WEAVE-LOFAR survey (Smith et al., 2016), to be conducted using the 4-m class WHT/WEAVE spectroscopic facility (e.g Dalton et al., 2012), with first light expected in early 2020. WEAVE-LOFAR will target sources selected from each tier of the LOFAR Surveys Key Science Project (Röttgering et al., 2011), and Figure 9 shows the expected star formation rate sensitivity of the three tiers of the LOFAR Surveys Key Science Project compared to other existing surveys. This figure highlights the already mighty star formation rate sensitivity of radio surveys in comparison to the deepest possible, confusion-limited, surveys with the *Herschel Space Observatory* (Pilbratt, 2011) and with the SCUBA-2 array on the James Clerk Maxwell Telescope (Holland et al., 2013). The sensitivity of Tier 3 of the LOFAR survey is extreme and will allow the routine detection of sub-millimeter-like galaxies at $z > 5$. However, the critical redshift information on the x-axis of Figure 9 is lacking without optical spectroscopy. WEAVE-LOFAR will provide this essential data for complete samples at $z \leq 1$, and for large but incomplete samples of rare luminous objects beyond this. However, only with a larger mirror, and dedicated massively-multiplexed survey instrument such as MSE, can we begin to fully understand the diversity in the radio source population, and use it to study the key processes shaping galaxy formation (namely star formation and accretion) using complete samples, at the key epoch when activity in the Universe was at its peak. Optical spectroscopy of radio selected targets is very efficient due to the profusion of emission-lines in radio sources, which renders it unnecessary to detect optical continuum emission for much of the science. Even so, for faint radio sources at $z > 1$, there is no substitute for a large mirror with a massively multiplexed spectrograph.

[6] `http://ngvla.nrao.edu`

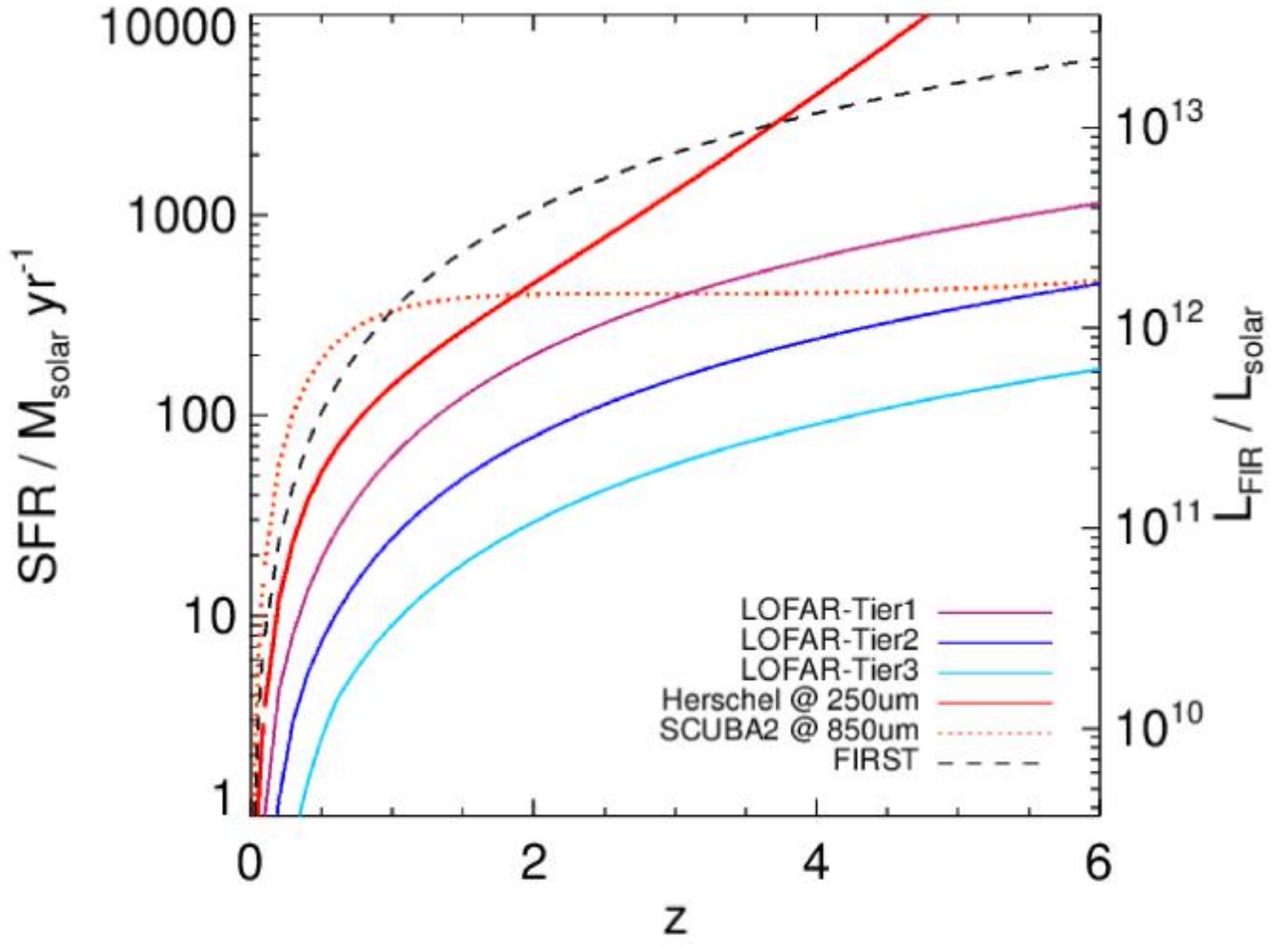

*Figure 9: Star formation rate sensitivity of selected far-infrared observatories and radio continuum surveys as a function of redshift. The red lines show the confusion-limited SFR sensitivity of the Herschel Space Observatory at 250 μm and the SCUBA-2 instrument on the JCMT at 850 μm (solid and dashed, respectively). The black dashed curve shows the SFR sensitivity of radio surveys from the J-VLA over Stripe 82 (Heywood et al. in prep; solid line), and the colored lines show the three tiers of the LOFAR Surveys KSP; Tier 1 (dotted line), Tier 2 (dot-dashed line), and Tier 3 (dot-dot-dot-dashed). Spectroscopic redshifts are necessary to provide the x-axis for real data in this plot, and will be provided for LOFAR by WHT/WEAVE. MSE and SKA-1 can expect similar synergies in this field. Figure from Smith et al. (2016).*

The challenge posed by the sensitivity of SKA-1 for optical astronomy is summarized in Figure 10, which shows the results of simulations presented in Meyer et al. (2015), detailing the fraction of HI sources that will be identified in SKA-1 that will also be detected in the optical, as a function of $r$-band limiting magnitude. Several different configurations for SKA-1 (left panel) are considered (including the now defunct SKA-SUR mode). Clearly, to match to SKA-1 detection thresholds for HI sources requires optical data extending to $r \sim 24$ at least (note that radio continuum – as opposed to HI – selected population of galaxies in the already existing LOFAR surveys do not have complete cross-identifications to this depth). While this is readily achievable with imaging, it is at the limit of 4-m spectroscopic capabilities; for example, the detection limits for one of the most ambitious extragalactic surveys to be conducted on VISTA/4MOST, called WAVES, is marked in Figure 10, and even this is below the 90% completeness threshold for several of the possible configurations. Preferable for synergy with the radio surveys is to ignore the optical magnitudes altogether, and focus on exploiting the large fraction of emission-line dominated sources in radio-selected samples, leveraging the exquisite emission line sensitivity of MSE to directly harness the star formation and accretion sensitivity of SKA-era radio surveys. The construction of large statistical datasets to capitalize on the SKA-1 data requires the survey speed of the spectroscopic facilities to be high, and this scales with the aperture of the facility, therefore very large aperture facilities such as MSE are essential.

Numerous other synergies exist between MSE and SKA-1 and ngVLA. For example, even these facilities will not be sensitive enough to directly detect HI in all but the gas rich galaxies out to $z \sim 2$. Therefore, a key focus is on stacking galaxies with known redshifts. Here, spectroscopic redshifts are essential, since the uncertainty of even the best photometric redshifts for radio sources (e.g. Duncan et al., 2018) is prohibitively large. MSE will provide the ideal spectroscopic sample to carry out HI stacking analyses of galaxies, subdivided by various galaxy properties such as age, morphology, redshift etc.

MSE, SKA-1 and the ngVLA naturally provide measurements of distinct tracers of the underlying density distribution of the Universe via galaxy redshifts catalogues (e.g., precise redshifts for luminous red galaxies from MSE and precise redshifts for lower mass, gas- rich

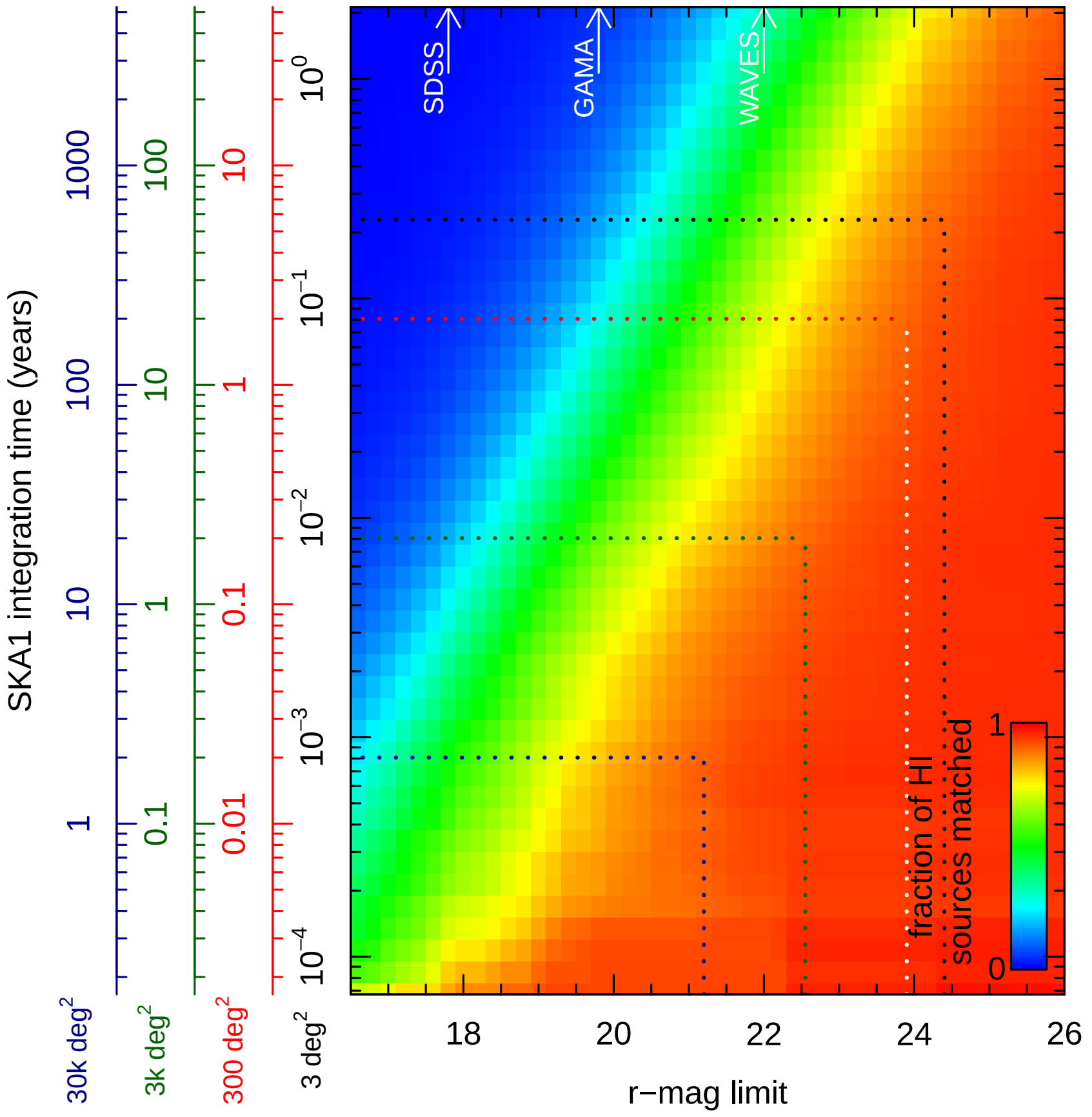

*Figure 10: Fraction of HI SKA-1 survey sources that are detected in an $r$-band apparent magnitude-limited sample. Four surveys are considered (including the now defunct SKA-SUR mode): 3 deg$^2$ using SKA1-MID (black), 300 deg$^2$ using SKA1-SUR (red), 3000 deg$^2$ using SKA1-SUR (green), and 30,000 deg$^2$ using SKA1-SUR (blue). Dotted lines indicate the r-magnitude limit needed to achieve matches for 90% of sources in each survey area (given 2000 hrs of integration for SKA1-MID surveys, and 2 years of telescope time for SKA1-SUR surveys). Dashed lines indicate 90% completion thresholds. Figure from Meyer et al. (2015)*

galaxies detected in HI with radio surveys). These galaxies trace the underlying dark-matter distribution with vastly different bias, thus allowing the so-called "multi-tracer" technique (Seljak, 2009) to be used in order to overcome cosmic variance effects (e.g., Ferramacho et al., 2014). Such an approach is necessary since it is increasingly apparent that cosmology is becoming limited by systematics not statistics.

A further area of direct synergy between MSE and future radio facilities is to obtain spectroscopic redshifts of distant galaxies that are used for weak lensing. In much the same way as for optical surveys, redshift information adds significant power to weak lensing analyses, allowing the growth of structure to be traced. Radio weak lensing is in itself extremely complementary to optical weak lensing surveys, as the systematic uncertainties are different, e.g., the wavelength-dependent PSF is known analytically in radio interferometry but the source density is generally lower. Thus the same MSE spectroscopic surveys that will be used to supplement optical weak lensing surveys will play the same role in radio weak lensing with future facilities.

It is worth pointing out that synergies with non-survey telescopes will also be important. Such studies amount to only a small number of fibers per pointing of MSE, and provide excellent science return for relatively little investment. For example, the ALMA archive is rapidly filling with observations of fields accessible to MSE. Although many of the highly-obscured star forming galaxies that constitute the bright sub-millimeter galaxy population are so heavily reddened that they will be too faint for spectroscopy with MSE, surveys with MSE will be able to characterize their environments. Studies of the faint ($< 1$ mJy at 0.85 mm) sub-millimeter galaxy population is now possible with ALMA, and many of these galaxies are not as heavily obscured in the optical (e.g. Fujimoto et al., 2016) and could be targeted for spectroscopy with MSE. Studying these objects at high redshift can provide a link between normal star forming galaxies on the main sequence and the hyperluminous obscured starbursts that dominate sub-millimeter surveys.

### 2.2.6 30m-class telescopes and MSE

The Thirty Meter Telescope (TMT), the European Extremely Large Telescope (ELT) and the Giant Magellan Telescope (GMT) will become the premier astronomical OIR facilities for detailed, high spatial resolution views of the faintest astronomical targets when they see first light in the 2020s (ELT, mid-2020s; TMT & GMT, late 2020s, although science operations with GMT are planned earlier with a reduced number of segments). Together, each of these ~USD1B facilities can access the entire sky with unprecedented collecting areas and with fields of view of order a few arcminutes.

Essential to the efficient scientific exploitation of these forefront observatories is target identification. Ideally, this will use a coordinated suite of supporting facilities to ensure these forefront facilities maximize their science impact by targeting the most scientifically compelling phenomena.

MSE will occupy an important role in the era of 30m-class telescopes through its ability to provide statistically significant samples of OIR spectra of relatively faint sources identified in wide field surveys at a range of wavelengths. Targets can be selected from this sample based on criteria specific to the individual science case, using spectral information derived over the same wavelength to which the 30m-class telescopes are sensitive. Observations with these giant facilities can then focus on higher SNR, higher spectral resolution and/or higher spatial resolution (in the case of spatially resolved sources). Given the plethora of sources identified by current and future wide field surveys at faint magnitudes, this type of filtering is essential for nearly all science programs.

The wide field perspective of MSE and the high precision small field perspectives of the 30m-class telescopes naturally encourages the development of combined science programs between the facilities. For example, precision measurements of the dark matter mass profiles of Milky Way satellites requires spatially complete radial velocity surveys extending out to very large radius. The 30m-class telescopes can provide critical astrometric data in the central regions of these dark matter halo, by measurement of the tangential velocities of stars at the center of the dark matter halos. This allows the derivation of tangential velocities for a subset of stars that provides significant leverage in breaking degeneracies in the dark matter profile in the central regions (see discussion in Chapter 6, and also Evslin 2015, 2016). On larger astrophysical scales, mass measurements of galaxy clusters using complete kinematic data for cluster members can be obtained using MSE, and compared to precision lensing mass measurements for the same clusters using the 30m-class telescopes.

## 2.3 From science cases to facility requirements

The following chapters discuss in detail some of the most anticipated, high profile, science objectives of MSE, with the important caveat that it is impossible to predict how astrophysics will evolve over the next few decades in response to discoveries and realizations that have not yet been made. Nevertheless, it is clear that diverse fields of research have converged in requiring dedicated large aperture multi-object spectroscopy resources. It is also clear that the capabilities of MSE will be a critical component of future lines of astronomical enquiry. But how exactly are the necessary capabilities of MSE identified from the science cases?

The Appendices of Version 1 of the *Detailed Science Case* describe a suite of Science Reference Observations (SROs). These were identified by the international science team as science programs that are *transformative in their fields and which are uniquely possible with MSE*. This last point is particularly important given the large number of spectroscopic resources that are available and which are discussed in more detail in Section 2.6. The SROs were selected to span the range of anticipated fields in which MSE is expected to contribute, and were developed in considerable detail for Version 1 of the *Detailed Science Case* published in 2016. The new, revised, *Detailed Science Case* for MSE has built on the original document and through this process the science team has reaffirmed the continued importance of the SROs to the design of MSE. As such, the traceability of the SROs to the overall science case is maintained by highlighting in each chapter those SROs that are most relevant to the science objectives being discussed.

The science requirements for MSE – i.e., the highest level design requirements for the facility – are defined as the suite of capabilities necessary for MSE to carry out the science objectives described in the suite of SROs. The MSE architecture and technical requirements flow directly from the science requirements and deliver the capabilities that they describe. Table 2 cross references each SRO to each of the high level MSE science requirements. The science requirements are described and discussed in detail in the *MSE Science Requirements Document*[7]. Detailed discussion of how these science requirements impact the entire design of MSE is given in the *MSE Book 2018*[8].

| | | Resolved stellar sources | | | | | Extragalactic sources | | | | | | |
|---|---|---|---|---|---|---|---|---|---|---|---|---|---|
| | | *Exoplanet hosts* | *Time domain stellar astrophysics* | *Chemical tagging in the outer galaxy* | *CDM subhalos and stellar streams* | *Local Group Galaxies* | *Nearby galaxies* | *Virgo and Coma* | *Halo occupation* | *Galaxies and AGN* | *The interGalactic Medium* | *Reverberation mapping* | *Peculiar velocities* |
| Spectral resolution | *Low spectral resolution* | | R~2000 (white dwarfs) | | | | R~3000 | R~3000 | R~2000-3000 | R~3000 | | R~3000 | R~1000-2000 |
| | *Intermediate spectral resolution* | | Any repeat observations | Essential, R~6500 | Essential, R~6500 | Essential, R~6500 | Velocities of low mass galaxies | Velocities of low mass galaxies | | | R~5000 | | |
| | *High spectral resolution* | R>=40000 | R>=50000 | Essential, R~20-40k | Essential | Young stars | | Bright globular clusters | | | | | |
| Focal plane input | *Science field of view* | ~2000 sq. deg | all-sky | 1000s sq. deg | 1000s sq. deg | 100s sq. deg | 3200 (100) sq. deg | ~100 sq. deg | ~1000s sq. deg | ~300 sq. deg | 40 sq. deg | 7 sq. deg | all-sky |
| | *Multiplexing at lower resolution* | | | | | | ~5000 galaxies/sq. deg | 100s target (galaxies/GCs) /sq.deg | >5000 galaxies/sq. deg | 770 galaxies/sq. deg | | 600AGN/ deg | 1000s galaxies/sq. deg |
| | *Multiplexing at moderate resolution* | | | 1000s stars/sq.deg to g~23 | 1000s stars/sq.deg to g~23 | few-thousands stars/sq.deg | ~5000 galaxies/sq. deg | | | | 500 galaxies/sq.deg | | |
| | *Multiplexing at high resolution* | ~100 stars/sq.deg @ g=16 | ~1000 stars/sq.deg to g=20.5 | ~1000 stars/sq.deg to g=20.5 | ~1000s stars/sq.deg to g~23 | | | | | | | | |
| | *Spatially resolved spectra* | | | | | | Goal | Goal | | | | | Yes |
| Sensitivity | *Spectral coverage at low resolution* | | | | | | 0.37 - 1.5um | 0.37 - 1.5um | 0.36 - 1.8um | 0.36 - 1.8um | 0.36 - 1.8um | 0.36 - 1.8um | Optical emission lines |
| | *Spectral coverage at moderate resolution* | | | Strong line diagnostics in optical | CaT essential | CaT essential | Goal: Complete | Goal: Complete | | | Goal: Complete | | |
| | *Spectral coverage at high resolution* | Strong lines for velocities; tagging | Strong lines for velocities | Chemical tagging | Strong lines for velocities | | | | | | | | |
| | *Sensitivity at low resolution* | | | | | | i=24.5 | i=24.5 | i=25.3 | i=25 / H=24 | | i=23.25 | i=24.5 |
| | *Sensitivity at moderate resolution* | | | g>20.5 | g~23 | i=24 | i=24.5 | i=24.5 | | | r~24 | | |
| | *Sensitivity at high resolution* | g=16 @high SNR | g=20.5 | g=20.5 | g~22 | | | | | | | | |
| Calibration | *Velocities at low resolution* | | | | | | v~20km/s | v~20km/s | v~100km/s | v~20km/s | | v~20km/s | v~20km/s |
| | *Velocities at moderate resolution* | | | v~1km/s | v~1km/s | v~5km/s | v~9km/s | v~9km/s | | | v~20km/s (10km/s goal) | | |
| | *Velocities at high resolution* | v<100m/s | v~100m/s | v~100m/s | v<1km/s | | | | | | | | |
| | *Relative spectrophotometry* | | | | | | ~4% | | | | | Critical: 3% | |
| | *Sky subtraction, continuum* | few % | few % | few % | few % | <1% | <1% | <1% | <0.5% | <0.5% | <1% | <1% | <1% |
| | *Sky subtraction, emission lines* | | | | important (CaT region) | important (CaT region) | critical | critical | critical | critical | critical | critical | |
| Operations | *Accessible sky* | Plato footprint (ecliptic) | Gaia footprint (all sky) | Gaia footprint (all sky) | Gaia, PS1, HSC footprint | Northern hemisphere (M31, M33, dec~+40) | LSST overlap useful (10000 sq. Deg) | NGVS footprint dec+12 | LSST overlap useful (10000 sq. Deg); Euclid (all sky) | LSST overlap useful (10000 sq. Deg); Euclid (all sky) | LSST overlap useful (10000 sq. Deg); Euclid (all sky) | all sky target distribution | all sky target distribution |
| | *Observing efficiency* | maximize | maximize | maximize | maximize | maximize | maximize | maximize | maximize | maximize | maximize | maximize | maximize |
| | *Observatory lifetime* | Monitoring >= years | Monitoring >= years | Survey >= 5 years | Survey >= 5 years | Survey ~100 nights | Survey ~few years | Survey ~100 nights | Survey ~7 years | Survey ~100 nights | Survey ~100 nights | Monitoring ~5 years | Survey ~years |

*Table 2: Cross-references between Science Reference Observations (in the columns) and Science Requirements (grouped in rows). Blue borders indicate that the SRO is used in the derivation of the requirement; blue boxes indicate that the requirement is highly relevant; grey boxes indicate that the requirement has some relevance to the SRO.*

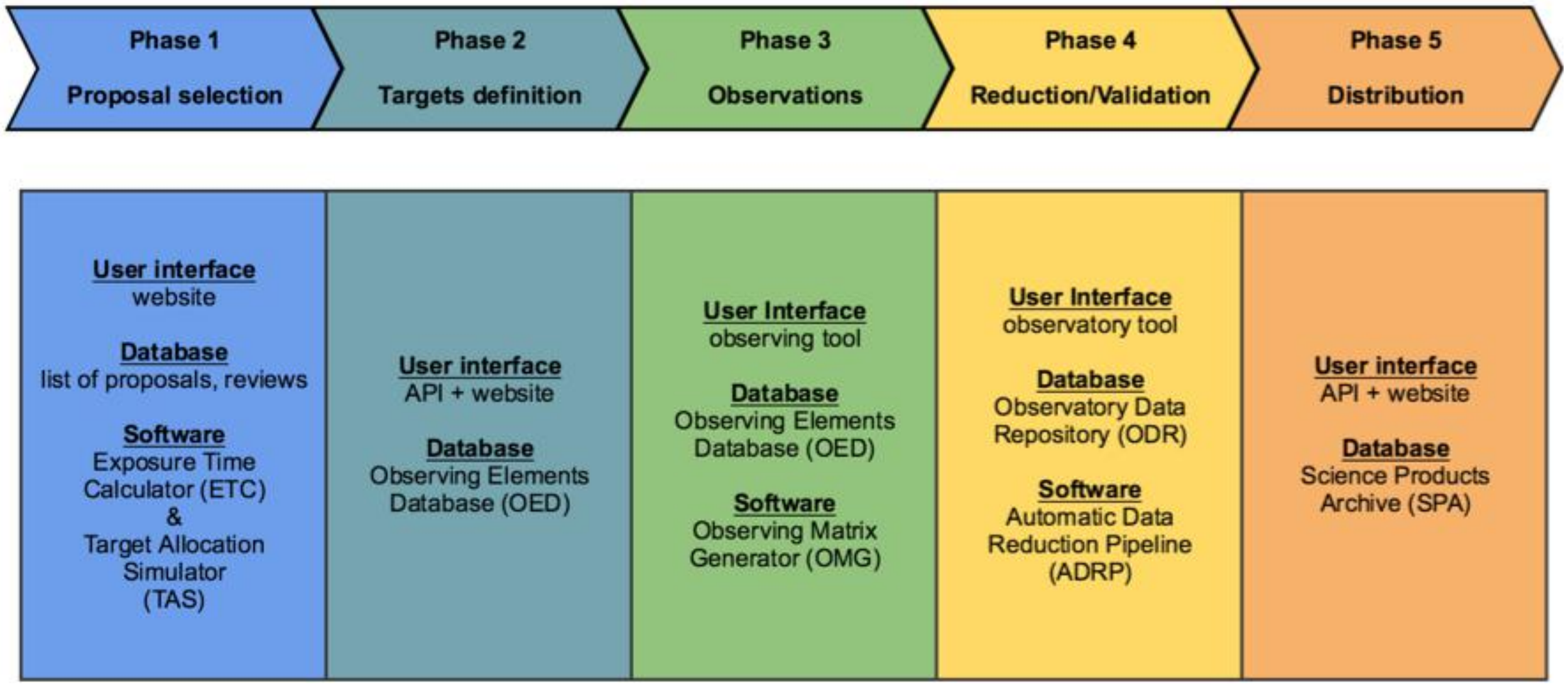

*Figure 11: MSE science operation phases*

### 2.4 From science cases to a science platform

Many of the SROs listed in Table 2 may be precursors to observing programs that MSE will carry out. However, it is not necessarily the case that this is so. At first light of the facility, a new set of forefront science topics might have emerged that the MSE community will want to address. Nevertheless, the suite of capabilities that MSE will have, in response to the SROs, will ensure that MSE is at the cutting edge of astronomical capabilities, whatever the science themes to be explored.

However, MSE is more than a facility consisting of a telescope and instrumentation suite. Fundamentally, MSE is an end-to-end science platform for the design, execution and scientific exploitation of spectroscopic surveys. Such a platform requires not only the main science hardware shown in Figure 13, but a suite of supporting infrastructure, including software and databases, that interface with the science user community. In addition, it requires operational methodologies that enable surveys which match the scope, ambition and science goals of the user community. To deliver all of this requires continual interaction between the science users and observatory staff to develop practices that ensures MSE remains responsive to, and at the forefront of, changing science priorities throughout its many years of operation.

Ongoing science development in MSE is building on the science cases described in this docu-ment in order to identify all those elements required to deliver to users the science platform that will best enable their research in the next decade. In particular, a suite of surveys are being designed that each have as their primary focus a single transformative science goal, but which, by their extensive nature, will also enable a vast array of ancillary science. These Design Reference Surveys (which we collectively refer to as the DRS) will become the primary means by which the scientific impact of all future MSE design decisions and choices (ranging from hardware, to software, to operational, etc) are judged and quantified.

[7] `https://mse.cfht.hawaii.edu/?page_id=15`

[8] `https://ui.adsabs.harvard.edu/#abs/2018arXiv181008695H/`

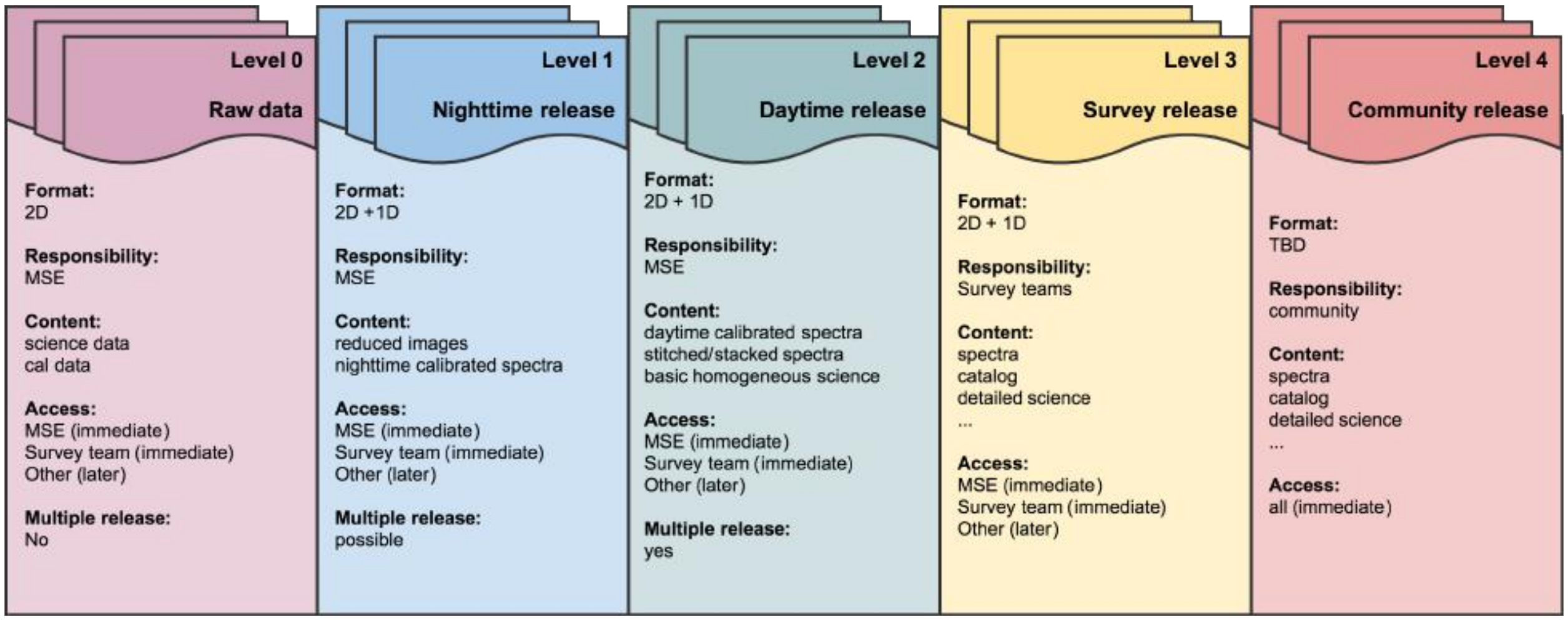

*Figure 12: Primary MSE data products.*

DRS development is an ongoing task, and the DRS will be updated at regular intervals as technical specifications are updated and as the science context evolves. In the first instance, the DRS will involve target selection, fiber allocations, queue scheduling and simulation of each survey, based on the historical weather conditions at the MSE site and the facility design parameters. Future iterations of the DRS will likely eventually involve full simulation of the photon light path through the MSE system and the production of mock images at the detector level. Such simulations will provide essential information to both the observatory teams and the science community on critical design elements of the facility and the surveys, not just limited to hardware, and judged via quantifiable science-driven metrics. This includes how best to schedule multiple large surveys to most efficiently reach specified science goals.

MSE's anticipated survey model, that will be tested through the development of the DRS, will allow for both large scale, multi-year, surveys, as well as smaller scale surveys that will be completed on shorter timescales. Only a small number of large scale programs are expected to be scheduled at any one time (e.g., a major dark time survey and a major bright time survey), with the remaining time given over to the smaller programs, that are expected to lead to publications on more rapid timescales. There will be frequent calls for MSE survey programs, and the procedures for these calls are anticipated to bear some similarities to the procedures used in ESO Public Surveys. That is, letters of intent will be solicited and large community-based teams will respond.

Partners in MSE contribute towards and lead the development of these legacy and strategic surveys. MSE will work closely with the selected survey teams (those astronomers named on the proposal) to progress through each phase of the operations required to realise the observations. These phases are outlined in Figure 11. Once observations have been conducted, high quality data and basic data products will be made rapidly available to the survey team and to the entire MSE team (all astronomers in partner communities). Standard, homogeneous, derived-data products will be released by the MSE Observatory at regular intervals. Specialized (i.e., survey-specific) derived-data products will generally be the responsibility of the survey team, working in collaboration with the MSE Observatory. The anticipated levels of data products are described in Figure 12. All raw data, homogeneously derived basic data products, and possibly advanced data products, will eventually be made available to the international community, after a proprietary period set by the MSE partnership, and expected to be a few years.

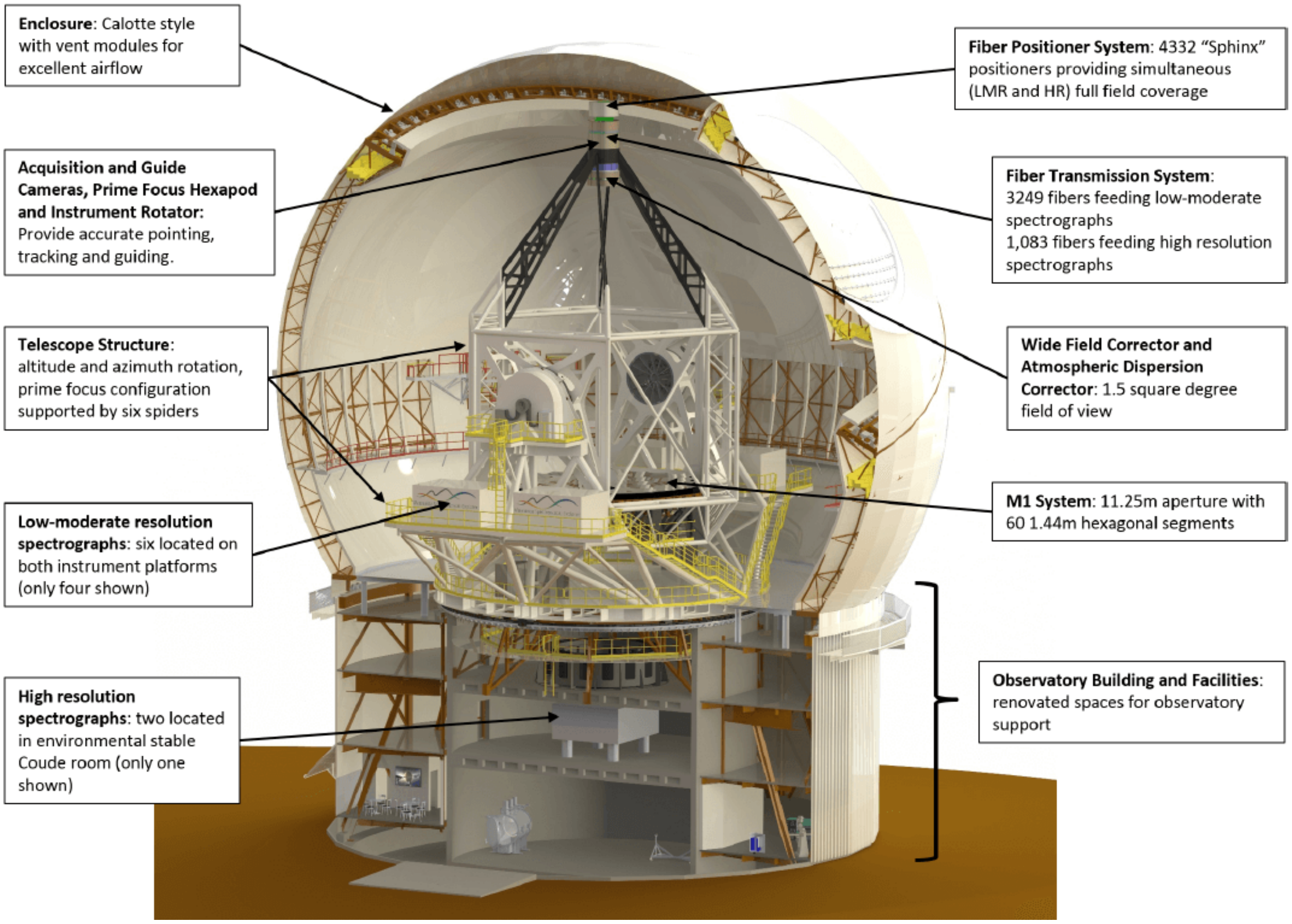

*Figure 13: A cut-away of the MSE Observatory.*

### 2.5 The science capabilities of MSE

Table 1 summarizes the scientific capabilities of MSE, and Figure 13 shows a cut-away of the layout of the Observatory. The scientific impact of MSE will be made possible and affordable by upgrading the existing Canada-France-Hawaii Telescope (CFHT) infrastructure on the Maunakea summit, Hawaii. CFHT is located at a world-class astronomical site with excellent free-atmosphere seeing (0.4 arcseconds median seeing at 500 nm). The Mauna Kea Science Reserve Comprehensive Management Plan for the Astronomy Precinct[9] explicitly recognizes CFHT as one of the sites that can be redeveloped. CFHT is an iconic 3.6-m telescope with four decades of operational experience and a legacy of discovery on Maunakea. MSE will build on the experience of CFHT and incorporate the latest technical advancements made by other top astronomical facilities.

MSE will replace CFHT with an 11.25 m aperture telescope, while retaining the current summit facility footprint. The rotating CFHT enclosure will be replaced by a Calotte enclosure that is only 10% larger than the current size, leaving the foundation and much of the remaining infrastructure intact. Building renovations and structural upgrades will be internal, so the outward appearance of MSE will remain very much unchanged from that of CFHT. Inside, however, a modern observatory will perform cutting-edge science at one of the best astronomical sites in the world, with access to three quarters of the entire night sky. We refer the reader to the *MSE Book 2018*[10] for a detailed discussion of the technical design of MSE.

MSE will have an 11.25 m segmented primary mirror and a 1.52 sq. degree field of view. Three different spectral resolution settings are possible. At the lowest resolution, 3,249 spectra, spanning the entire optical spectrum in the J-band, can be obtained in a single pointing. At moderate resolution, the same number of spectra can be obtained for approximately half the optical waveband and for the full H-band. The low and moderate resolutions are provided by the same spectrograph system (six identical spectrographs, with four channels each), and each channel can switch settings independently. At the highest resolution, 1,083 spectra will be obtained per pointing for three windows in the optical region of the spectrum. At all resolutions, MSE can encompass the faintest targets, and will remain well calibrated and stable over its lifetime. Located at the equatorial site of Maunakea ($l = 19.9°$), MSE will access the entire northern hemisphere and half of the southern sky, making it an ideal follow-up and feeder facility for a large number of both existing and planned ground- and space-based facilities.

The diverse science enabled by MSE and discussed in this document spans all astronomy, from exoplanets, the microphysics of stars and the interstellar medium through to the dynamics of dark matter, the physics of black holes and the mass of neutrinos. What are the key science-enabling capabilities of MSE that facilitate this impact and diversity?

### 2.5.1 Key capability 1: survey speed and sensitivity

MSE must have a high survey speed. This is a function of sensitivity, field of view, multiplexing, and observing efficiency. Many of the surveys MSE is expected to undertake cover large areas of sky. MSE therefore has a 1.52 square degree field of view, hexagonal for ease of tessellation. MSE has an observing efficiency of 80%, defined as the fraction of night time spent collecting science photons (excluding losses due to weather). This will make it one of the most efficient optical observatories in the world.

[9] `http://www.malamamaunakea.org/management/master-plan`
[10] `https://ui.adsabs.harvard.edu/#abs/2018arXiv181008695H/`

MSE must have a large aperture, to provide sufficient sensitivity to follow up faint sources identified by imaging surveys conducted by large-aperture telescopes, such as LSST. As a filtering facility for the 30m telescopes, MSE must also provide excellent sensitivity to faint sources, which can then be followed up using higher spatial resolution (e.g., Integral Field Units), higher spectral resolution, and/or higher SNR by these giant facilities. Consequently, MSE will be the largest optical facility after the 30m telescopes. At a low spectral resolution, MSE will obtain a SNR per resolution element of two for magnitude 24 sources at all wavelengths (point sources, monochromatic AB magnitude; i.e., the approximate depth of a single LSST visit) over the course of an hour-long observation. At high spectral resolution, MSE will obtain a SNR per resolution element of 10 for magnitude 20 sources at all wavelengths (point sources, monochromatic magnitude; i.e., covering the full luminosity range of Gaia sources) over the course of an hour-long observation.

MSE will utilize the full light-gathering power of its large aperture in pursuit of its driving science goals and is focused towards an understanding of the faint Universe - such as intrinsically faint stars, the distant Galaxy, low mass galaxies, and the high redshift Universe - that is not accessible with smaller apertures. This opens up extensive new areas of research, such as in-situ chemodynamical studies of the distant Milky Way stellar halo, enabling all Galactic components to be studied in an unbiased manner. For extragalactic science, the dynamics and stellar populations of dwarf galaxies that fall below the detection threshold of smaller-aperture facilities at low and moderate redshifts will be easily accessible, and MSE will allow the analysis of sub-L* galaxies out to high redshift. Again, the light gathering power of the large aperture is essential in allowing unbiased analyses of these galaxy populations that could not be achieved with smaller apertures.

### 2.5.2 Key capability 2: spectral performance and multiplexing

A range of spectral resolutions and wavelengths, from UV, through optical, to NIR, are needed to enable a diverse range of scientific investigations. The multiplexing requirements of each mode are determined by a consideration of the expected target densities. The source density of galaxies at $z < 0.2$ brighter than $i = 23$ is $2,100$/sq. degree (or $\sim 3,200/1.5$ sq. degree), which determines the minimum multiplexing for the low and moderate spectral resolution modes (the density of even fainter sources makes it desirable to achieve a much higher fiber density, but here there are many technical limitations). The source density of thick disk and halo stars at high Galactic latitudes in the critical magnitude range for MSE of $17 < g < 21$ is $\sim 700$ per sq. degree or $\sim 1000$ per 1.5 sq. degree, which determines the minimum multiplexing for the high spectral resolution mode.

MSE is also being designed to incorporate multi-object Integral Field Units after first light. We anticipate that these will feed the low/moderate spectrograph suite, which will be achieved by switching the fiber positioning and fiber transmission systems.

The wavelength coverage in each mode is determined by the consideration of primary scientific goals. As part of its low/moderate spectral resolution studies, MSE will probe aspects of galaxy evolution in the distant Universe. Sensitivity out to and including H-band ensures that galaxies and AGN can be studied, using the same set of tracers from $z = 0$ to cosmic noon and beyond (see Figure 14). As part of its high spectral resolution studies, a large number of critical nucleosynthetic tracers in stars need to be accessed at blue/UV wavelengths. Many of these features are weak and in crowded spectral regions, which requires a spectral resolution of $R > 20K$.

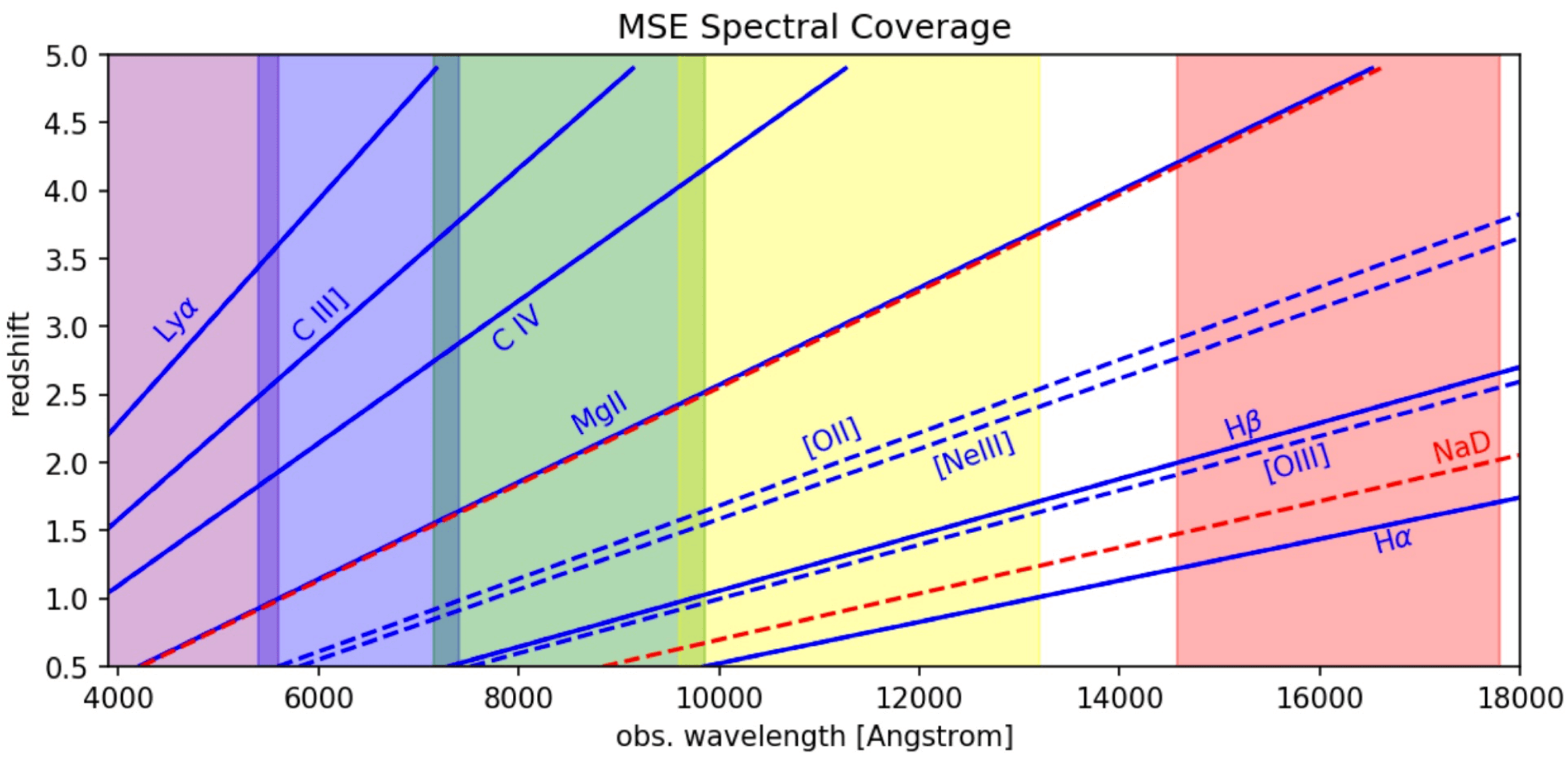

*Figure 14: Observability of spectral features as a function of redshift and observed wavelength for MSE. Vertical colored bands show the available spectral windows. Dashed blue lines mark narrow emission lines, solid blue lines mark emission lines that are potentially broad (in the case of Type 1 AGNs), and red dashed lines mark absorption lines. Based on figure from Chapter 8: AGN and SMBHs.*

### 2.5.3 Key capability 3: dedicated and specialized operations

MSE is specialized for one task: the efficient acquisition of multi-object spectra. This basic operation philosophy enables the production of stable, homogeneously calibrated, well-characterized, high quality data. For instruments that move on and off telescopes at regular intervals, data issues such as calibration, stability and reproducibility can become problematic. MSE can, therefore, address science cases that are very difficult to address using other MOS instruments (for example, time-resolved, high-resolution spectroscopy and quasar reverberation mapping). In addition, multiple avenues will be available to the community to maximise the science output for MSE. Clearly, the dominant avenue is to tackle science cases that require high multiplexing. But in addition to *multiplexed* science there is also *multi-threaded* science; that is, tackling science cases that on a per field basis do not require the full multiplexing capability of MSE - and so can be conducted in parallel to other programs - but which integrated over multiple fields eventually produce large datasets of relatively rare targets.

The impressive power that a large, homogeneous and well characterized dataset can offer has been most successfully demonstrated by the Sloan Digital Sky Survey (SDSS). SDSS has had a profound impact on topics as far ranging as small bodies in the solar system to the reionization of the Universe – a broad appeal that underlies its repeated ranking as the highest-impact telescope of the last decade (e.g. Madrid & Macchetto 2009; Chen et al. 2009). A search of NASA ADS for refereed publications mentioning SDSS in their abstracts returns over 8,600 publications, with more than 436,000 citations[11]. The importance of the spectroscopic component of SDSS can scarcely be overestimated: it is universally recognized that the science impact of SDSS would have been greatly diminished without its spectroscopic element, for the simple reason that broadband photometry alone provides only zeroth-order information on the physical properties of astrophysical sources. The success of SDSS comes despite the fact that it is a relatively small-aperture facility by modern standards, located at a site that cannot compete with Maunakea in terms of median image quality. However, a large part of its success can be traced to the extremely well calibrated and well characterized nature of the data. MSE can be viewed as an evolution of the SDSS concept, using a telescope with a collecting area around twenty times larger, situated at arguably the best optical astronomical site on the planet. The science that emerged from SDSS was far more diverse than originally envisioned at its outset, and it seems fair to anticipate that the same will be true for MSE.

The fiber-positioning technology chosen by MSE ensures that all spectrographs are available at all times, so every MSE observation will use all 4,332 fibers. High spectral resolution observations of (relatively) bright targets will be prioritized during bright times, and low-resolution observations of fainter, extragalactic targets will be prioritized during dark times. Like the multiple generations of SDSS, MSE is designed to be flexible and responsive, and instrument/telescope upgrades will occur (the IFU mode is already envisioned). But like the SDSS over the last few decades, MSE will remain focused on being the world's premier resource for the spectroscopic exploration of the Universe. It is worth noting that, assuming a (conservative) baseline exposure of one hour per field, with eight hours per night available for observations (10.2 hours before weather losses), then MSE will observe around one million astronomical spectra every month: the equivalent of a SDSS Legacy Survey—1,640,960 spectra—every eight weeks. Over the first decade of MSE operations, more than 100 million fiber hours of 10m class spectroscopy will be available to the community for forefront science.

### 2.5.4 Development of the multi-object IFU mode

The design of MSE is modular in order to anticipate the need to upgrade or replace components during its lifetime. Already, MSE is being designed with the capacity to include a future multi-object IFU mode (that is expected to make use of the existing low/moderate resolution spectrograph suite). It is not anticipated that this mode will be available at first light, but rather that it will be incorporated once MSE operations reach a mature level with the MOS fiber mode.

In addition to the science requirements listed in Table 2, a multi-object IFU system has a unique set of science requirements independent of the regular MOS fiber system (number of IFUs, IFU sizes, spaxel sizes, etc). These requirements are determined from analysis of the

[11] Search performed at `https://ui.adsabs.harvard.edu` on January 22 2019

| Class | Facility / Instrument | First light (anticipated) | Aperture (M1 in m) | Field of View ($deg^2$) | Etendue ($m^2\ deg^2$) | Multiplexing | Wavelength coverage (um) | Spectral resolution (approx) | IFU | Dedicated facility |
|---|---|---|---|---|---|---|---|---|---|---|
| Comparison | **SDSS I - IV** | Existing | 2.5 | 1.54 | 7.6 | 640 | 0.38 - 0.92 | 1800 | Yes | Yes |
| 4-m | **Guo Shoujing / LAMOST** | Existing | 4 | 19.6 | 246 | 4000 | 0.37 - 0.90 | 1000 | No | Yes |
| | **AAT / HERMES** | Existing | 3.9 | 3.14 | 37.5 | 392 | windows | 28000, 50000 | No | No |
| | **WHT / WEAVE** | 2019 *a* | 4.2 | 3.14 | 43.5 | 960 | 0.37 - 0.96 | 5000 | Yes | Yes |
| | | | | | | | windows | 20000 | | |
| | **Mayall / DESI** | 2019 *b* | 4 | 7.1 | 89.2 | 5000 | 0.36 - 0.98 | 4000 | No | Yes |
| | **VISTA / 4MOST** | 2022 *c* | 4 | 4.1 | 51.5 | 2436 | 0.39 - 0.95 | 6500 | No | Yes |
| | | | | | | | windows | 20000 | | |
| 8-m | **VLT / MOONS** | 2020 *d* | 8.2 | 0.14 | 7.4 | 1000 | 0.65 - 1.80 | 4000 | No | No |
| | | | | | | | windows | 18000 | | |
| | **Subaru / PFS** | 2021 *e* | 8.2 | 1.25 | 66 | 2394 | 0.38 - 1.26 | 3000 | No | No |
| | | | | | | | 0.71 - 0.89 | 5000 | | |
| 10-m | **MSE** | 2029 | 11.25 | 1.52 | 151 | 4329 | 0.36 - 1.3 | 3000 | Second generation | Yes |
| | | | | | | | 0.36 - 0.95 (50%) 1.5 - 1.85 | 6000 | | |
| | | | | | | | windows | 40000 | | |

*a* http://www.ing.iac.es/Astronomy/telescopes/wht/weavepars.html#dflow
*b* https://www.desi.lbl.gov
*c* https://www.eso.org/sci/facilities/develop/instruments/4MOST.html#status
*d* https://www.eso.org/sci/facilities/develop/instruments/MOONS.html#status
*e* http://pfs.ipmu.jp/schedule.html

*Table 3: Summary of major current and upcoming optical and infrared multi-object spectroscopic instruments and facilities.*

| | 8 - 12 m class facilities | | | | | | |
|---|---|---|---|---|---|---|---|
| | **VLT / MOONS** | | **Subaru / PFS** | | **MSE** | | |
| *Dedicated facility* | No | | No | | Yes | | |
| *Aperture (M1 in m)* | 8.2 | | 8.2 | | 11.25 | | |
| *Field of View (sq. deg)* | 0.14 | | 1.25 | | 1.52 | | |
| *Etendue* | 7.4 | | 66 | | 151 | | |
| *Multiplexing* | 1000 | | 2394 | | 4329 | | |
| *Etendue x Multiplexing* | 7400 *( = 0.01 )* | | 158004 *( = 0.24 )* | | 653679 *( = 1.00 )* | | |
| *Observing fraction* | < 1 ? | | 0.2 (first 5 years) 0.2 - 0.5 afterwards ? | | 1 | | |
| *Spectral resolution (approx)* | 4000 | 18000 | 3000 | 5000 | 3000 | 6500 | 40000 |
| *Wavelength coverage (um)* | 0.65 - 1.80 | windows | 0.38 - 1.26 | 0.71 - 0.89 | 0.36 - 1.3 | 0.36 - 0.95 (50%) 1.5 - 1.8 | windows |
| *IFU* | No | | No | | Second generation | | |

*Table 4: MSE in comparison to other planned MOS instruments on 8-m class telescopes.*

driving science cases. The current document does not provide a description of these science cases, and instead focuses on the extensive science available with the regular MOS fiber mode. A future supplement to the *Detailed Science Case* will describe the driving science of MSE-IFU and highlight its specific science requirements.

### 2.6 Competition and synergies with future MOS

Table 3 compares the core capabilities of a large number of MOS instruments and facilities, including all those that are at advanced stages of design, many of which will be operating on timescales that overlap with MSE. The ESO spectroscopic telescope concept is not listed in these tables, since only a feasibility study has been conducted and detailed technical specifications are not yet available (see Ellis et al. 2017; Pasquini et al. 2018. Table 4 lists only those facilities from Table 3 that have comparable sensitivity to MSE (i.e., 8 -– 10 m

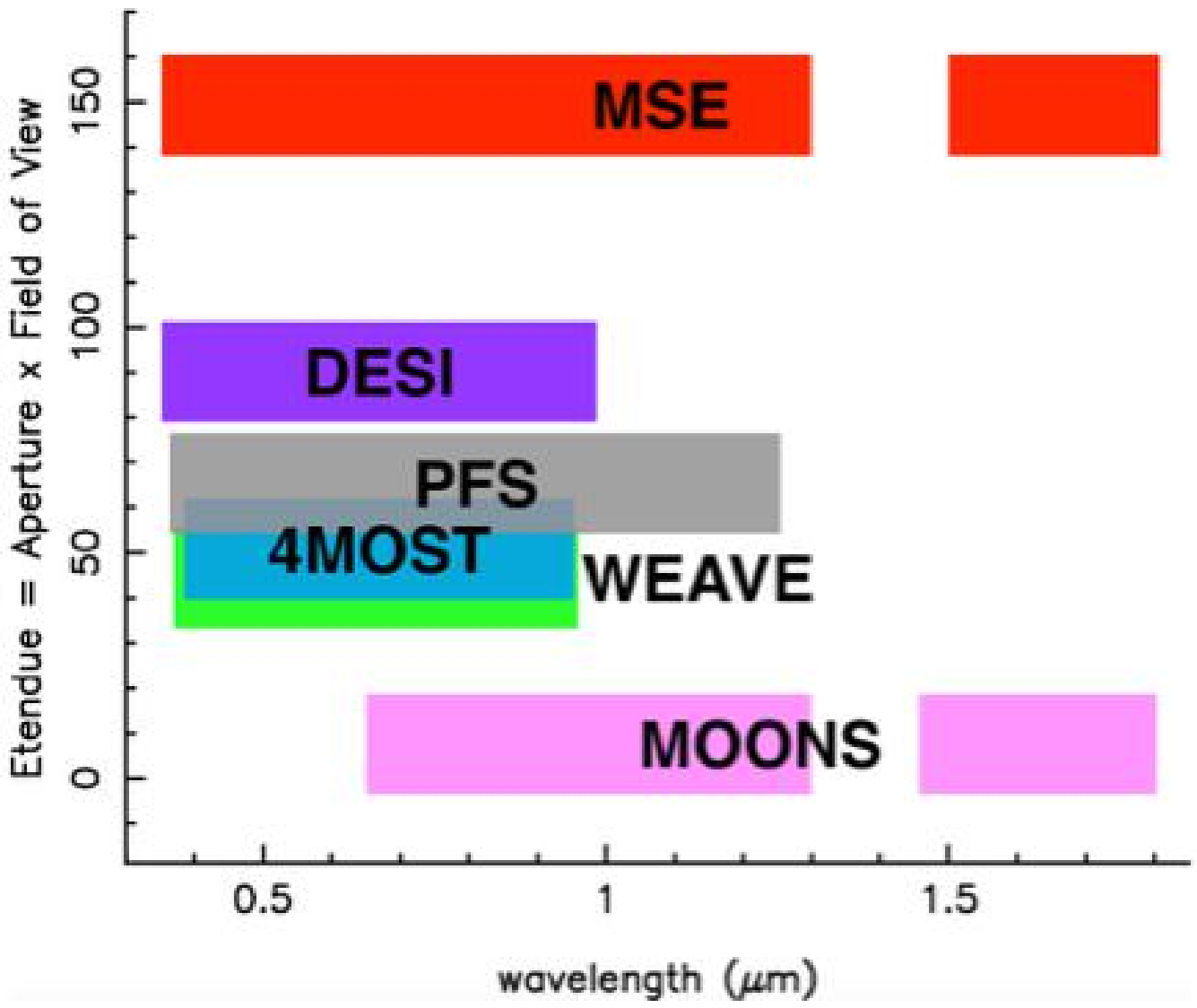

*Figure 15: Etendue versus wavelength coverage for the upcoming* $4-10\,m$ *class wide-field MOS facilities listed in Table 3*

class facilities).

Figure 15 presents a graph showing the wavelength coverage and étendue of the upcoming generation of facilities, listed in Table 3. In addition to its étendue - larger than that of any other upcoming $4-10\,\mathrm{m}$ facility by almost a factor of two - the wavelength coverage of MSE is unmatched by any other wide-field, spectroscopic facility in any aperture class. Figure 16 shows MSE in comparison to the facilities listed in Table 4, that are closest to MSE in terms of sensitivity, but still have collecting areas which are smaller by almost a factor of two. Here, étendue is combined with the multiplexing and the observing fraction, to define a quantity that is essentially the survey speed. With VLT/MOONS, we are assuming that the MSE will remain on the telescope at all times; for Subaru/PFS, the solid blue rectangle reflects our assumption that PFS will account for 20% of telescope time, in line with other Subaru strategic survey programs. The hatched rectangle shows its survey speed were PFS to occupy 100% of telescope time. These considerations make MSE the ultimate spectroscopic facility, that is more than an order of magnitude more efficient at surveys than its closest competitor.

### 2.7 A scientific priority for the coming decade of discovery

The plethora of deep imaging and astrometric surveys at optical wavelengths, and of survey missions at other wavelengths, has resulted in significant focus turning to how to obtain complementary OIR spectral data for the (literally) billions of sources that these surveys will identify. Wide field spectroscopy is a critical “missing link” in the international portfolio of astronomical facilities, especially at large apertures where MSE is the only facility in active development.

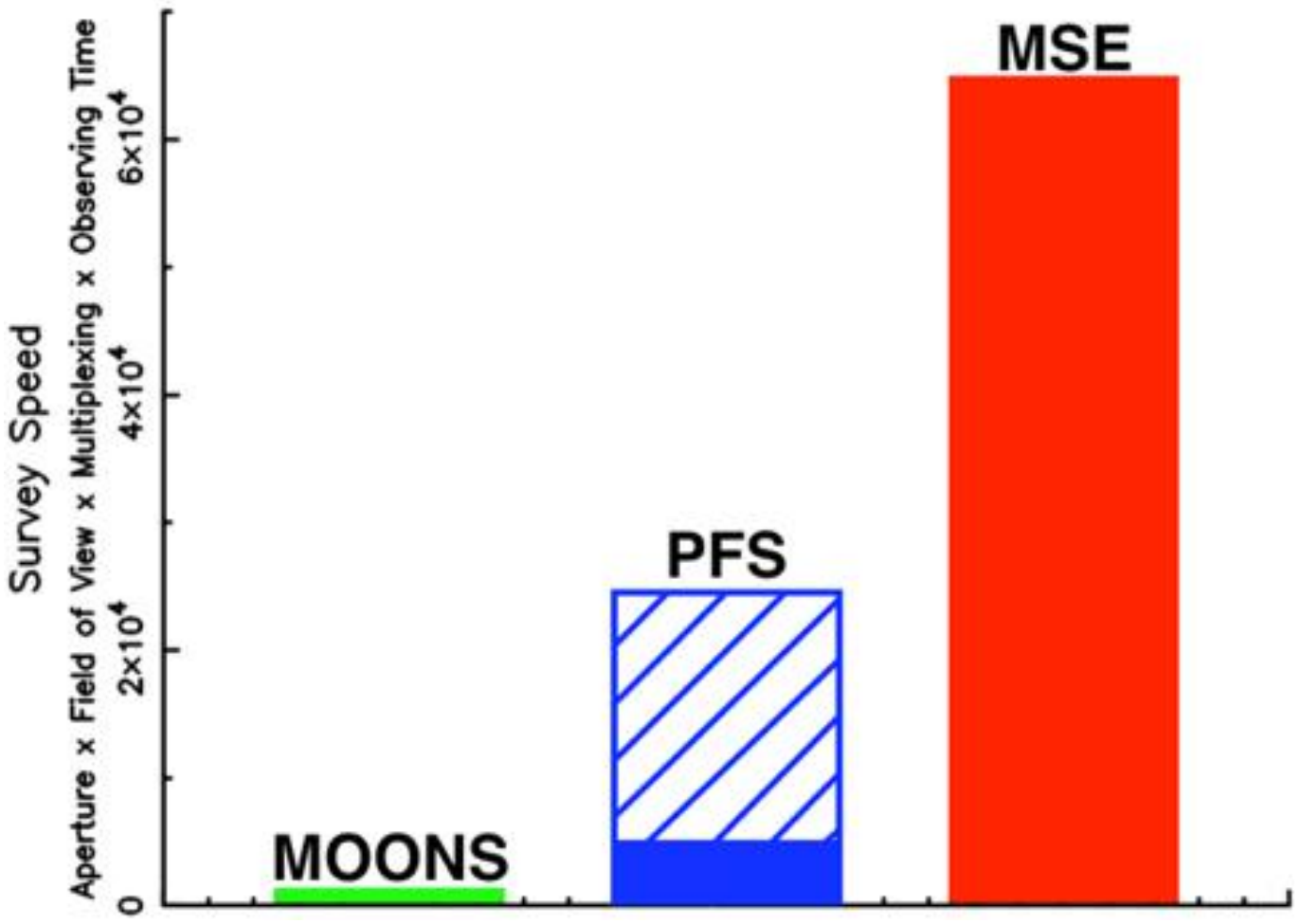

*Figure 16: Comparison of the survey speeds of the three 8–10 m class wide-field MOS capabilities in design or under construction, listed in Table 4*

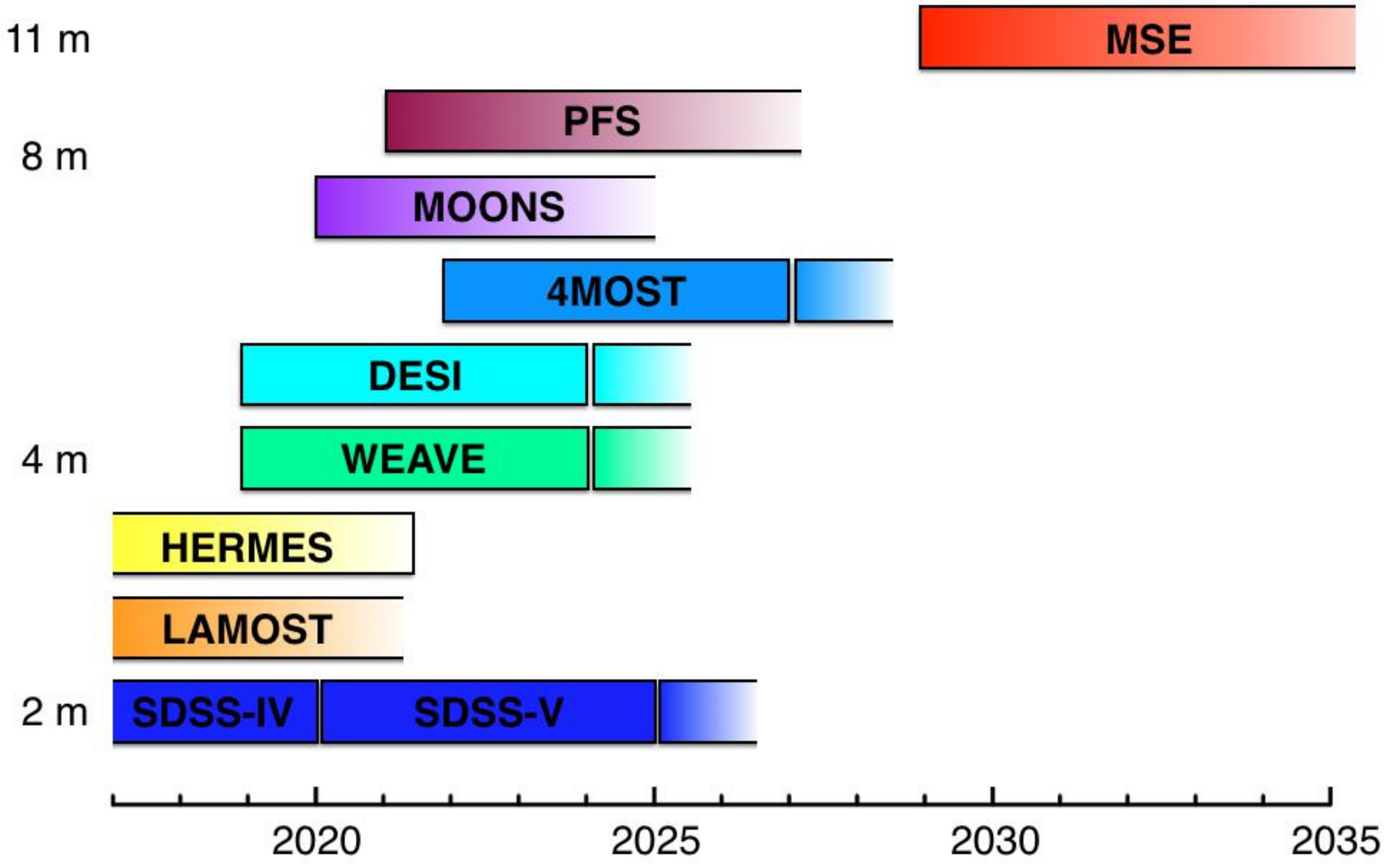

*Figure 17: Current anticipated timelines of the wide-field MOS facilities listed in Table 3. Bounded boxes indicate the duration or lifetime of the survey or facility, and the absence of a vertical solid line indicates that the facility has no clear end date. Several of the facilities will operate initially for a set period of years, with the expectation that their lifespan will be extended beyond the nominal end date. This figure was inspired by a similar figure created by J. Newman.*

In Canada[12], the Long Range Plan 2010, notes that a 10-m class telescope, equipped with an extremely multiplexed spectrograph, "would be a unique resource for follow-up spectroscopy, both for the European Gaia satellite mission, and also for LSST and Euclid/WFIRST". Subsequently, the Mid-Term Review of LRP2010 comments directly on MSE and notes that "The scientific and collaborative opportunities available to MSE partners are a direct indicator of the strategic relevance of MSE to the future astronomy landscape." The Australian Astronomy Decadal Plan 2016 – 2025[13] details the ways in which a dedicated, wide-field spectroscopic facility, integrated into a large telescope, would "provide follow-up spectra of objects identified by the SKA and imaging telescopes like the US-led Large Synoptic Survey Telescope." In Europe, ESO has identified highly multiplexed spectroscopy as a priority, with strong science backing from their community (Primas et al., 2015). An ESO Working Group commissioned to study the potential of wide-field MOS concluded that "such a facility could enable transformational progress in several broad areas of astrophysics, and may constitute an unmatched ESO capability for decades" (Ellis et al. 2017; see also Pasquini et al. 2018). In the US, in response to the National Research Council report, "Optimizing the U.S. Optical and Infrared System in the Era of LSST"[14], NOAO and LSST convened study groups whose top recommendation was that the US community "develop or obtain access to a highly multiplexed, wide-field, optical, multi-object spectroscopic capability on an 8-m class telescope, preferably in the Southern Hemisphere" (Najita et al., 2016).

MSE is located at arguably the premier site on the planet for ground-based optical and infrared astronomy, with access to the entire northern sky and more than half of the southern sky at reasonable airmass. The science presented herein reflects several years of effort by the MSE Science Team in the development of transformational science cases that demonstrate the unique, high impact and exceptionally diverse science enabled by large aperture, wide field MOS. This revised version brings the science case up to date for entering the 2020s. MSE is designed to provide a natural next step beyond what is envisioned for the current and imminent generation of MOS instruments. The stand-alone science potential of MSE is awesome, but moreover the strategic importance of MSE within the international network of astronomical facilities cannot be overstated. This is reflected by the strong backing that large aperture, wide field MOS has on the international scene, and for which MSE is the realization of that ambition.

[12] `https://casca.ca/?page_id=75`

[13] `https://www.science.org.au/supporting-science/science-sector-analysis/reports-and-publications/decadal-plan-australian`

[14] `https://www.noao.edu/meetings/lsst-oir-study/`

# Chapter 3
# Exoplanets and stellar astrophysics

**Abstract**

MSE will be the most powerful facility available to provide critical stellar spectroscopic observations from the lowest-mass brown dwarfs to massive, OB-type giants, and including important yet rare stellar types across the Hertzsprung-Russell diagram, such as solar twins, Cepheids, RR Lyrae stars, AGB and post-AGB stars, as well as faint, metal-poor white dwarfs. In the stellar regime, strong synergies exist between MSE and TESS, PLATO, Gaia, eROSITA, LSST and other time domain facilities. MSE stellar monitoring programs will dramatically improve our understanding of stellar multiplicity, including the interaction and common evolution between companions spanning a vast range of parameter space such as low-mass stars, brown dwarfs and exoplanets, but also pulsating, eclipsing or eruptive stars. MSE will provide dynamical masses for unprecedented samples of transiting hot Jupiters ($\sim 10^4$), allowing the exploration of critical outstanding questions of this intriguing class of planets such as their radius inflation and migration mechanisms. An MSE follow-up campaign of transiting, massive TESS planets will help to disentangle hot Jupiters from brown dwarfs and very low-mass stars, in order to test the mass-radius relation for objects that populate both the high-mass end of the exoplanet regime and the low-mass end of the stellar regime. MSE will be crucial to constrain the poorly-known aspects of stellar physics in the low-mass domain ($0.08-0.5$ M$_\odot$), including the equation-of-state of dense gas, opacities, nuclear reaction rates, and magnetic fields.

**Science Reference Observations** (appendices to the *Detailed Science Case, V1*):
**DSC – SRO – 01** The characterization and environments of exoplanet hosts
**DSC – SRO – 02** Rare stellar types and the multi-object time-domain

## 3.1 Introduction

Stellar astrophysics and exoplanet science are closely connected and rapidly developing fields that form one of the backbones of modern astronomy. Stellar evolution theory, guided by measurements of fundamental parameters of stars from observational techniques such as interferometry, asteroseismology and spectroscopy, underpins models of stellar populations, galaxy evolution, and cosmology. It is now recognized that exoplanets are ubiquitous in our galaxy, and that the formation, evolution and characteristics of planetary systems are closely connected to those of their host stars.

MSE will form a critical component for answering key questions in stellar astrophysics and exoplanet science in the 2020s by covering large fractions of the Galactic volume and surveying many millions of stars per year. Unprecedented datasets for tens of millions of stars such as high-precision space-based photometry from TESS (Ricker et al., 2014) and PLATO (Rauer et al., 2014), astrometry from Gaia (Lindegren et al., 2016), X-ray data from eROSITA (Merloni et al., 2012), and ground-based photometry from transient surveys such as LSST (Jurić et al., 2017) require spectroscopic follow-up observations to be fully exploited. The wide-area, massively multiplexed spectroscopic capabilities of MSE are uniquely suited to provide these critical follow-up observations for all kinds of stellar systems from the lowest-mass brown dwarfs to massive, OB-type giants.

MSE will provide massive spectroscopic follow-up of important yet rare stellar types across the Hertzsprung-Russell diagram, such as solar twins, Cepheids, RR Lyrae stars, AGB and post-AGB stars, but also faint, metal-poor white dwarfs. The detection and characterisation of such objects using high-resolution optical spectroscopy is required to deepen our understanding of stellar structure, fundamental parameters of stars, planetary formation processes and their dependence on environment, internal evolution and dissipation of star clusters, and ultimately the chronology of galaxy formation.

MSE will be uniquely suited for wide-field time-domain stellar spectroscopy. This will dramatically improve our understanding of stellar multiplicity, including the interaction and common evolution between companions spanning a vast range of parameter space such as low-mass stars, brown dwarfs and exoplanets, but also pulsating, eclipsing or eruptive stars.

## 3.2 Information content of MSE stellar spectra

The major science-enabling capabilities of MSE for stellar astrophysics and exoplanet science is its very broad wavelength coverage, from 0.36 to 1.8 micron in the low- and medium-resolution mode. Also in the high-resolution mode, spectra will be taken in three broad, partly overlapping, windows: $360 - 620$ nm at $R \sim 40K$ and $600 - 900$ nm at $R \sim 20K$. This exquisite wavelength coverage, with optimized exposure times to reach a sufficient signal-to-noise ratio (SNR) even in the near-UV, as well as huge multiplexing capacities, put MSE at the top of all available or planned spectroscopic facilities (Figure 18).

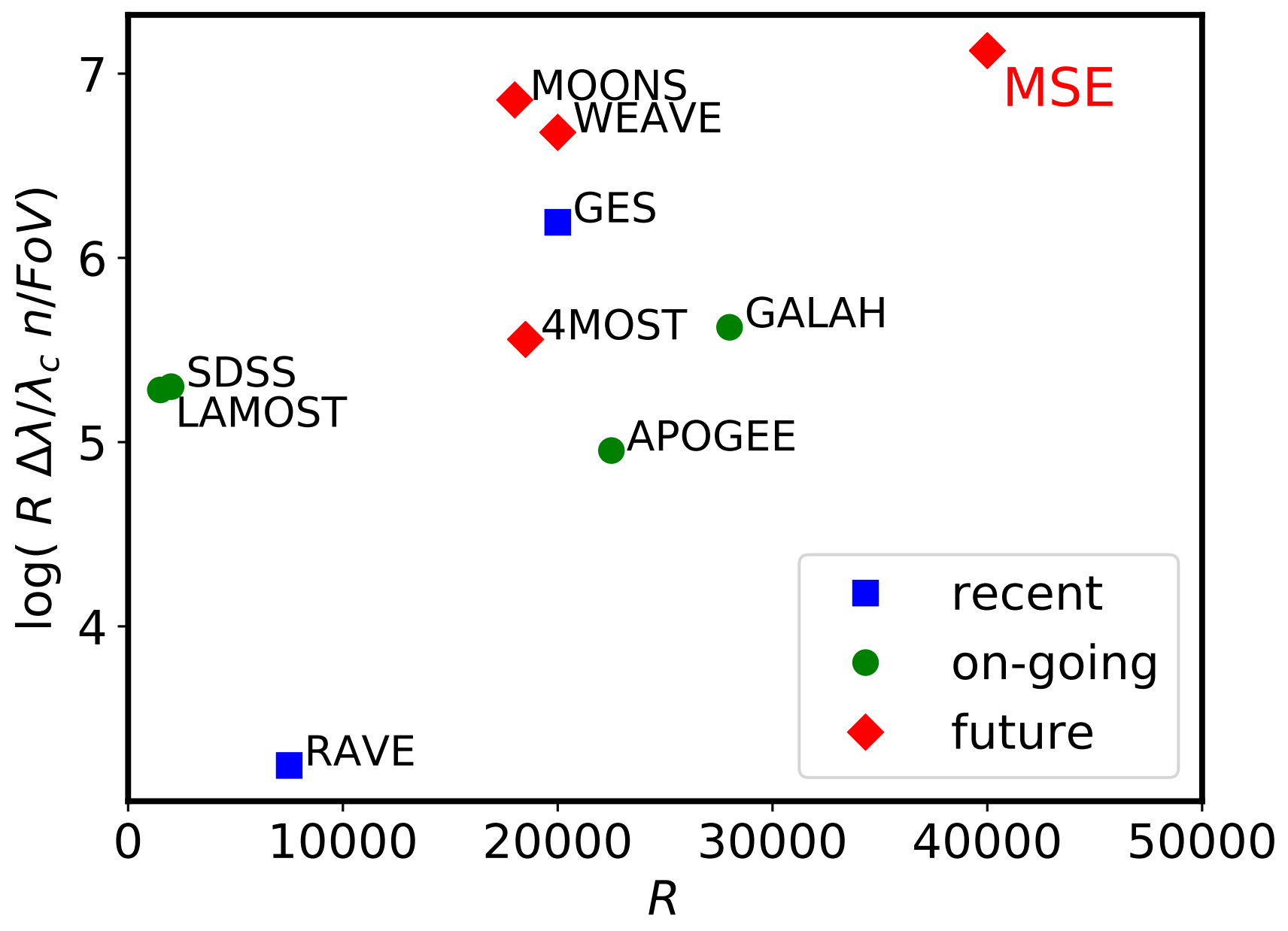

*Figure 18: MSE will dramatically improve the resolving power and sensitivity of past, current and future planned surveys to characterize stars, substellar objects, and exoplanets. Potential of recent, on-going and future ground-based MOS surveys parameterized as a function the resolving power* $\boldsymbol{R}$*, wavelength coverage* $\Delta\lambda$*, central wavelength* $\lambda_c$*, number of fibres for high resolution mode* $\boldsymbol{n}$*, and field of view (Allende Prieto, 2016).*

These wavelength regimes cover not only the critical indicators of stellar surface parameters ($H_\alpha$, Mg triplet lines, over $10^3$ iron lines to determine accurate metallicities), but also *useful diagnostic lines of all major families of chemical elements* (Hansen et al., 2015; Ruchti et al., 2016). These include Li, C, N, O, $\alpha$-elements (Si, Ca, Mg, Ti), odd-Z elements (Na, Al, K, Sc, V), Fe-group elements (V, Cr, Mn, Fe, Co, Ni), Sulphur and Zinc, but also rare-earth and neutron-capture elements (La, Y, Eu, Ce, Th, Nd, Zr, Dy, Ba, Sr, Sm). For instance, one of the heaviest elements (Pb I line at 405.8 nm), a key tracer of s-process, can be systematically targeted. The spectra will cover molecular lines, including the G-band of CH. For high-mass OB type stars, abundances of He, C, N, O, Ne, Mg, Si, and Fe can be determined, as well as terminal wind velocity and mass loss (Nieva & Przybilla, 2012; Bestenlehner et al., 2014). See also discussion in Chapter 4).

The high-resolution spectra will also deliver accurate radial velocities (RVs) with a nominal precision of $100\,\mathrm{m\,s^{-1}}$, projected equatorial rotational velocity ($\upsilon_{\rm e} \sin i$), macroscopic motions, indicators of winds, mass loss, and activity (e.g. Wise et al., 2018), including the Ca H&K lines at ~396.9 and 393.4 nm, the Ca infrared triplet at 849.8, 854.2 and 866.2 nm. The TiO bands at 710 and 886 nm will be particularly useful for M dwarfs and heavily spotted (active) stars. Also the high-resolution MSE mode will permit exploiting the shape of the H$\alpha$ line at 656.28 nm in red giants to measure stellar masses (Bergemann et al., 2016), and hence, accurate distances beyond the Milky Way.

Beyond providing input for physics of stellar structure and exoplanets, MSE stellar spectra will offer a powerful means to test models of stellar atmospheres and spectra. The recent decade has seen breakthrough advances in physical description of stellar spectra, radically influencing the accuracy of diagnostics of fundamental stellar parameters and abundances. Simulations of stellar convection are being developed (Collet et al., 2007; Freytag et al., 2012; Trampedach et al., 2013; Chiavassa et al., 2009, 2011) in cohort with non-local thermodynamic equilibrium modelling of radiative transfer (Bergemann et al., 2010; Bergemann, 2011; Nordlander et al., 2017; Lind et al., 2017; Amarsi & Asplund, 2017; Bergemann et al., 2017). These more sophisticated models will be applied to the MSE spectra, enabling a wealth of constraints on the micro- and macro-physics of stellar atmospheres, including departures from local thermodynamical equilibrium, convection and surface dynamics, influence of pulsations and mass loss on the line profiles.

## 3.3 Exoplanets and substellar mass objects

### 3.3.1 Radial velocity surveys

The systematic discovery of low-mass companions to stars and their connection to the formation and evolution of binaries/triples and planetary systems will be a major focus of astrophysics over the coming decades. MSE, owing to its unique multiplexing and high-resolution spectroscopy capabilities, will be an ideal instrument to probe the statistics of substellar mass objects and massive planets using multi-epoch radial velocities. With a nominal radial velocity precision in high-resolution mode of $100\,\mathrm{m\,s^{-1}}$, MSE will be sensitive to a vast range of parameter space covering low-mass, substellar companions, and high-mass planets for an unprecedented number of stars (Figure 19). Recent results have demonstrated that even a survey with relatively sparse sampling and/or few-epoch radial velocity measurements will provide a powerful tool for detecting companions (e.g., Price-Whelan et al., 2017, 2018) and characterizing binary population statistics (e.g., Badenes et al., 2018).

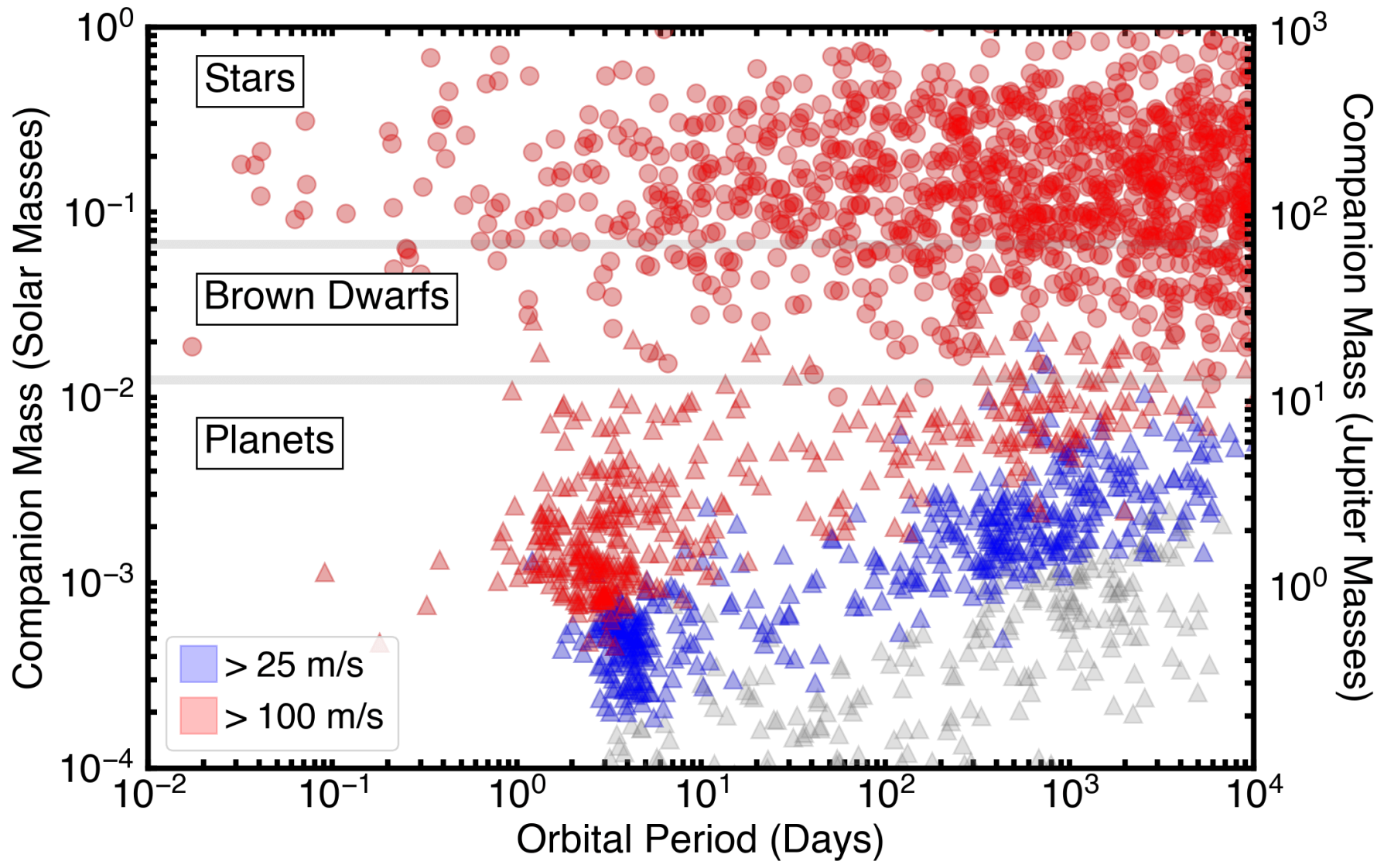

*Figure 19: MSE will enable the detection of stellar and substellar companions to stars spanning a vast range of parameter space. Circles illustrate expected stellar companions drawn from a synthetic population calculated using TRILEGAL for a representative 1.5 square degree field observed by MSE. Triangles show the currently known population of exoplanets. Red and blue symbols highlight companions with an expected radial velocity signal* $> 100\,m\,s^{-1}$ *(nominal MSE performance in high-resolution mode) and* $> 25\,m\,s^{-1}$*. Grey horizontal lines mark canonical mass boundaries between stars, brown dwarfs and exoplanets.*

While the binary statistics at high mass ratios around solar-type stars is relatively well understood (Raghavan et al., 2010), systematic searches for brown dwarfs and high-mass planets around stars of all masses are mostly confined to direct imaging surveys (Chauvin et al., 2010; Brandt et al., 2014; Elliott et al., 2015; Biller et al., 2013; Bowler et al., 2014; Dupuy & Liu, 2017), which are restricted to small field of views and hence sample sizes. This is particularly the case for substellar primaries, whose faint magnitudes have inhibited large-scale searches and studies of close-separation binaries with $< 1$ AU through radial velocity methods, despite evidence that such systems may compose a significant fraction, if not majority, of substellar multiples (Burgasser et al., 2007; Blake et al., 2010; Bardalez Gagliuffi et al., 2014). Multi-epoch radial velocity measurements of the $\sim$1000 known substellar objects within 25 pc of the Sun, made possible by the red sensitivity and resolution of MSE, would enable a detailed assessment of the overall binary fraction of brown dwarfs. At the same time, it would provide new systems for dynamical mass measurements, both critical tests for substellar formation and evolutionary models.

A complete radial velocity survey of the local substellar population would also enable the detection of ”fly-by” stars, which have made (or will make) a close approach to the Sun. Such systems may had a role in shaping the composition and orbits of objects in the outer Solar System (including the hypothesized Planet 9), and even the arrangement of the major Solar planets (Pfalzner et al., 2018). An analysis of Gaia data by Bailer-Jones et al. (2018) indicates an $< 1$ pc encounter rate of 20 stars/Myr, although the authors estimate only 15% of encounters within 5 pc and ±5 Myr have been identified due to the lack of radial velocities for the coolest stars and brown dwarfs. A pertinent example is WISE J072003.20-084651.2 (aka ”Scholz’s star”), a very low-mass star/brown dwarf binary whose kinematics indicate that it passed within 50,000 AU of the Sun 70,000 years ago, but lacks radial velocities in Gaia. Given the high fraction of brown dwarfs in the Solar Neighborhood (20-100% by number, Kirkpatrick et al., 2012; Mužić et al., 2017), MSE measurements would significantly improve our assessment of the incidence of star-Sun interactions in the past/future $50 - 100$ Myr.

Turning to solar-type stars, it is expected that binary companions have a strong influence on the formation of exoplanets, for example by truncating protoplanetary discs (Jang-Condell, 2015), dynamically stirring planetesimals (Quintana et al., 2007), or affecting the orbits of already formed planets through dynamical interactions (Fabrycky & Tremaine, 2007; Naoz et al., 2012). Imaging surveys of the Kepler field have indeed revealed intriguing evidence that exoplanet occurrence is suppressed by the presence of stellar companions (Kraus et al., 2016), emphasizing the need for a census of low-mass and very low-mass companions around planet host stars to better understand the link between binary stars and exoplanets. The high multiplexing capabilities and sensitivity of MSE would allow a complete characterization of the close binary fraction of exoplanet hosts, allowing studies of how and why exoplanet occurrence is shaped by stellar multiplicity, and complementing imaging efforts to detect wider companions (Furlan et al., 2017; Hirsch et al., 2017; Ziegler et al., 2018).

MSE will also be sensitive to planetary-mass objects through RV surveys. Among these systems, planets orbiting stars in clusters, moving groups, and star forming regions (probing the age groups from $> 0.1$ Gyr, for open clusters and moving groups, to $< 10$ Myr, for star-forming regions) are of special interest. They share the same distance, age, and have the same initial chemical composition, and thus represent unique laboratories to study planet formation. The detection of planets in clusters of different ages would shed light on the question if, when and how hot Jupiters migrate to the close orbital distances at which they are observed among old field stars (see Dawson & Johnson, 2018, for a review). So far, only a handful of planets have been discovered in clusters by ground-bound surveys (Lovis & Mayor, 2007; Sato et al., 2007; Quinn et al., 2012, 2014; Brucalassi et al., 2014; van Eyken et al., 2012; Malavolta et al., 2016) and space-based planet searchs (David et al., 2016; Mann et al., 2017; Gaidos et al., 2017; Mann et al., 2018; Livingston et al., 2019). The prospect of expanding this work with MSE is extremely promising: the large number of fibers, their on-sky separation, and the unique sensitivity of MSE are well matched to the densities of stars in typical open clusters, and will provide unique possibilities to probe the population of Jupiter-mass planets and their survival rate depending on mass and metallicity in different environments.

### 3.3.2 Characterization of transiting exoplanets

The need for multi-object spectroscopic (MOS) facilities enabling accurate RV measurements of large samples of stars is particularly acute for characterizing transiting exoplanets. Space-based photometry missions, such as TESS, are expected to discover tens of thousands of transiting Jupiter-mass planets over the coming decade (Barclay et al., 2018). These yields vastly outnumber the available follow-up resources on single-objects spectrographs.

MSE can provide dynamical masses for unprecedented samples of transiting hot Jupiters ($\sim 10^4$), allowing the exploration of critical outstanding questions of this intriguing class of planets such as their radius inflation (Miller & Fortney, 2011) and migration mechanisms (Dawson & Johnson, 2018). The latter can be probed by employing the Rossiter-McLaughlin effect (Rossiter, 1924; McLaughlin, 1924; Triaud, 2017) to measure the projected spin-orbit alignment of the host star and the planet. Since the amplitude of the effect scales linearly with $\omega_{\rm e} \sin i$, even a modest RV precision can be used to increase the current sample of spin-orbit angle measurements, thus providing important clues on the dynamical formation history of hot Jupiters.

An MSE follow-up campaign of transiting, massive TESS planets will also help to disentangle hot Jupiters from brown dwarfs and very low-mass stars, in order to test the mass-radius relation for objects that populate both the high-mass end of the exoplanet regime and the low-mass end of the stellar regime (e.g. Hatzes & Rauer, 2015). The impact of MSE for radial-velocity studies of exoplanets will be even stronger below the nominal $100\,{\rm m\,s^{-1}}$ precision (see Figure 19), which may be achieved using refined wavelength calibrations using telluric lines (e.g. the $\sim 10-30\,{\rm m\,s^{-1}}$ precision with the current CFHT high-resolution spectrograph ESPaDOnS, e.g. Moutou et al., 2007) or new, data-driven methods for the extraction of precise radial velocities (Bedell et al., 2019).

MSE spectroscopy of large samples of transiting planet-host stars will also help to improve the accuracy of exoplanet radii themselves. Our ability to measure transit parameters such as the impact parameter and stellar-to-planet radius ratio is frequenctly limited by our knowledge of stellar limb darkening. Available limb darkening tables can result in a $1-10\%$ bias in planet radius for stars with $T_{\rm eff} > 5000\,{\rm K}$, whereas for cooler main-sequence stars the error can rise up to 20% (Csizmadia et al., 2013). By combining accurate element abundances from MSE with transit light curves, it will be possible to construct improved grids of limb darkening coefficients. Especially valuable will be stars that host multiple planets, since the transit of several planets on different orbits helps to circumvent the degeneracy of limb darkening effects in photometric data.

### 3.3.3 Characterisation of exoplanet and brown dwarf atmospheres

Our understanding of the physics of exoplanets is being revolutionised with the development of new techniques to probe their atmospheres (e.g. Fortney, 2018; Sing, 2018, and references therein). Complementary to space-based photometry, ground-based spectroscopy is a powerful tool that opens up new perspectives on the compositions of exoplanet atmospheres (e.g. Snellen et al. 2010; Kok et al. 2013; Birkby et al. 2013; Nortmann et al. 2018; Brogi & Line 2018). MSE is the only large-aperture MOS facility that can provide multi-epoch, high-resolution spectroscopy of large samples of planetary systems expected to be discovered by TESS. These datasets will come at a low-cost since both the characterization of exoplanet host stars and the exoplanet atmospheres do not interfere with each other, and thus can be achieved simultaneously by optimizing the observations.

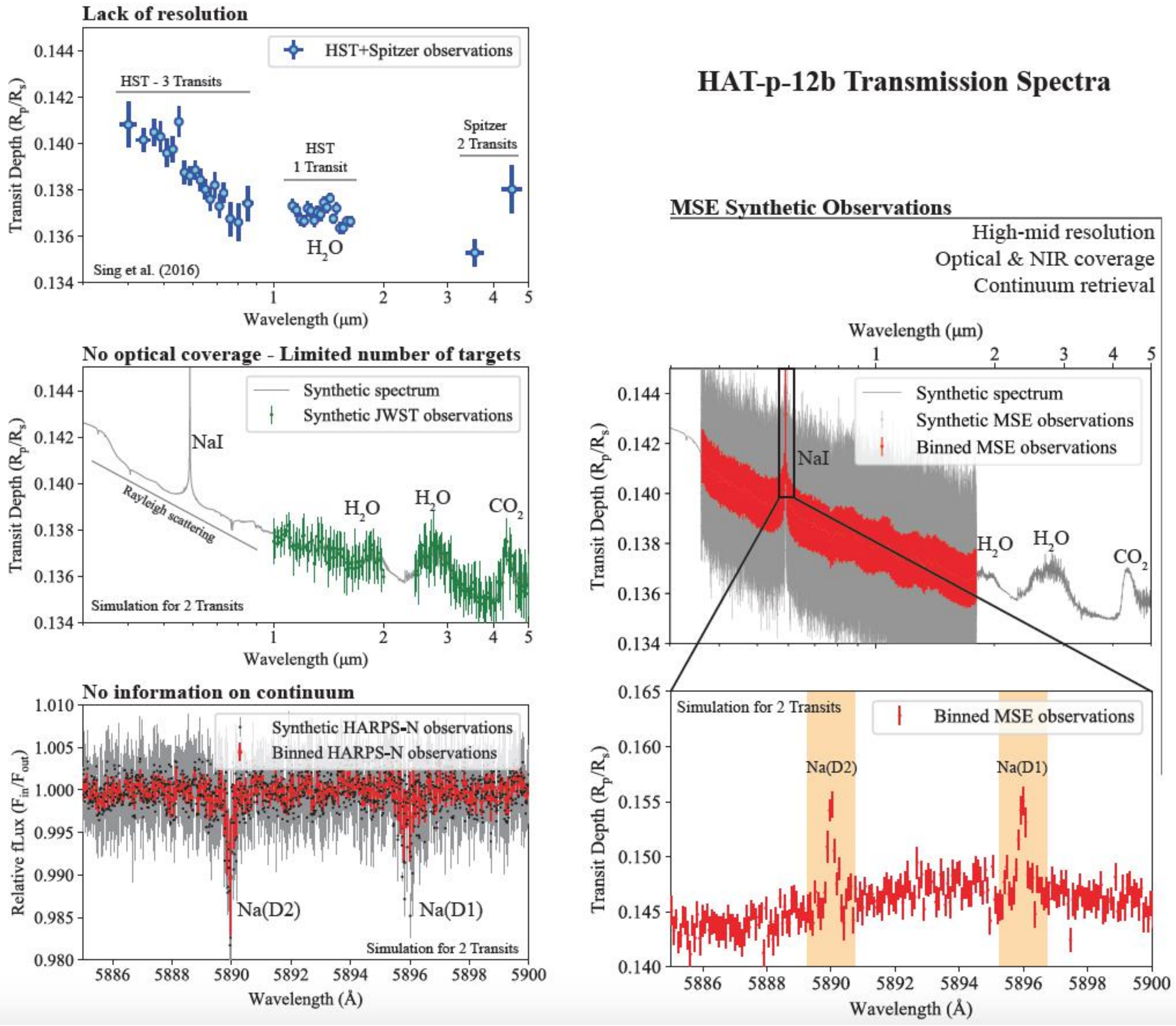

*Figure 20: MSE will provide unique capabilities for the characterization of exoplanet atmospheres. Panels demonstrate the example of HAT-P-12 ($V = 12.8$), a close-in gas giant planet with a transit duration of less than 2.5 hours. Top Left: HST and Spitzer photometry of HAT-P-12, revealing some spectral features but lacking the resolution for a robust detection. Middle Left: Synthetic JWST/NIRCam spectra, following Schlawin et al. 2016. Only a limited number of exo-atmospheres are expected to be characterised by JWST, and the facility has no wavelength coverage shorter than 0.6 $\mu m$. Bottom Left: Synthetic HARPS-N spectra, which lack information on the continuum. Right panels: MSE synthetic spectra after telluric and host star spectra removals by following multi-reference star approach. Both continuum (top right) and resolved sodium lines (right bottom) can be retrieved, resulting in a coherent characterisation of HAT-P-12b spectral features from optical to NIR wavelengths.*

Specifically, the high-resolution mode of MSE (R $\sim 40K$) is well-suited for the application of novel methods such as the cross-correlation technique, resolved-line high-resolution spectroscopy (Fig. 20, lower-left panel), and differential spectro-photometry (e.g. Morris et al. 2015; Boffin et al. 2016). The cross-correlation technique allows to detect complex molecular features, such as H2O, CO, CH4, while resolved-line high-resolution spectroscopy is mostly suitable to detect atomic features, such as Na, K, and He (e.g. Birkby 2018 and references therein). Spectro-photometry, on the other hand, provides information on the continuum, which is crucial to study clouds and hazes on exoplanets (e.g. Sing et al. 2016) and can potentially be applied to spectral lines as well. The combined analysis of lines and continuum puts powerful constraints on the physical structure of the atmosphere of the planet at different pressure levels, as well as its absolute mass and orbital inclination (Brogi et al., 2012). For the former, longitude-dependent T/P profiles, chemical composition (including isotopologues, like $^{13}$CO, heavy water HDO, and $CH_3D$ (Mollière & Snellen, 2018)), atmospheric dynamics and escape, and the formation of clouds and hazes may be inferred. Nearby, cool main-sequence stars are the best targets to map chemistry of rocky planet atmospheres with the accuracy that is required to make a first step towards understanding bio-signatures.

Spectrophotometry cannot be achieved through single-object high-resolution spectroscopy. Thus single-object spectroscopy, which has been the most commonly used ground-based method to study exo-atmospheres, is unable to retrieve the continuum. Although challenging, the technical capabilities of MSE allow high-resolution spectroscopy and spectophotometry for transiting and non-transiting exoplanets. For close-in transiting planets, spectral monitoring on timescales of $3-5$ hours would be sufficient (e.g. Sing 2018). Strong spectral features of transiting exoplanets with extended atmospheres are typically detectable by one or a few visits, and usually achieve S/N ratios of a few hundred in the continuum. On the other hand, observing the atmospheric spectra of non-transiting planets requires longer observational time, in order to span over a significant portion of their orbits. However the latter is not time-critical and can be sparse in the time domain, as long as the spectra probe different orbital phases.

Telluric lines and sky emission corrections are the key to explore exo-atmospheres from the ground (e.g. Bedell et al., 2019). Currently, high-resolution studies iteratively fit the modelled telluric spectra to absorption features in the science spectra. Employing the modelling approach is mostly due to the lack of high-quality correction frames. However, such correction frames can be obtained by simultaneously observing a handful of reference stars and the plain sky. The wide field of view of MSE allows the identification of the most suitable reference stars to ensure high-quality spectral contamination removal, with calibration exposures obtained in a configuration that is as close as possible to that of the science observations.

MSE will also enable new detailed studies of the atmospheres of exoplanet analogues: isolated low-temperature L and T dwarfs in the vicinity of the Sun. At these temperatures, liquid and solid condensates are present at the photosphere, shaping both the emergent spectra and (through surface inhomogeneities in cloud structures) driving photometric variability of up to 1%. Spectrophotometric monitoring studies from HST (Apai et al., 2013) and the ground (Schlawin et al., 2017) have enabled detailed exploration of the vertical stratification of cloud layers and particle grain size distribution of these systems (Buenzli et al., 2012; Lew et al., 2016; Apai et al., 2017). These few measurements have provided necessary clues for interpreting the evolution of brown dwarfs through the L dwarf/T dwarf transition (where clouds may be dynamically disrupted; Burgasser et al., 2002) and constraining global climate models of brown dwarf and exoplanet atmospheres (Showman & Kaspi, 2013). The low-resolution mode of MSE would provide both the scale and sensitivity to measure panchromatic light curves for hundreds of variable brown dwarfs (whose rapid rotations require monitoring periods of hours) to fully explore the diversity of cloud behaviors in these objects, as well as dependencies on mass, metallicity, rotation rate and magnetic activity. Indeed, the broad spectral coverage of MSE's low-resolution mode would permit simultaneous investigation of the weather-activity connection (Littlefair et al., 2008) .

#### 3.3.4 Exoplanet host stars and proto-planetary disks

Fundamental parameters of stars are paramount to understand the formation and physical properties of exoplanets. Both theory (Ida & Lin, 2008; Bitsch & Johansen, 2017; Nayakshin, 2017) and observations (Santos et al., 2004; Fischer & Valenti, 2005; Johnson et al., 2010) demonstrated that the probability of giant planet formation increases with host star metallicity. However, the shape of this relationship, as well as its dependence on the detailed abundance pattern, is still not understood (Udry & Santos, 2007; Johnson et al., 2010; Mortier et al., 2013). Large, homogeneous, and unbiased samples of stars across the full mass and metallicity range are needed to investigate the planet occurrence rates in different environments (e.g. Santos et al., 2004; Sousa et al., 2008; Buchhave et al., 2014; Schlaufman, 2015; Zhu et al., 2016; Mulders et al., 2018).

MSE will be an ideal next-generation facility to address these outstanding puzzles by mapping the detailed chemical composition of the planet-host stars of all ages, including the systems around pre-MS stars (T Tau and Herbig objects). With a wide wavelength coverage and large aperture, MSE will enable accurate measurements of abundances of volatile (O, C, N) and refractory elements (Fe, Si, Mg) for large samples of stars with and without detected planets, allowing quantitative constraints on planet formation scenarios. Recent simulations show that the key parameter is water ($H_2O$) to Si ratio, with large $H_2O$ fractions favoring the birth of giant planets, while low $H_2O$ support the growth of super-earths (Figure 21, Bitsch & Johansen, 2016). Some studies suggest that the formation of CO leads to water depletion (Madhusudhan et al., 2017) that could potentially inhibit efficient gas giant formation. Hence, a detailed chemical mapping of the volatiles and refractories in the atmospheres of the host stars will help to understand the chemical composition of the building blocks of planets (e.g. Booth et al., 2017; Maldonado et al., 2018).

In addition, MSE can provide unique constraints on the interaction between the star and its proto-planetary disc. This will require high-resolution spectra of stars in multiple systems, i.e., those that can be assumed to share the same age and initial composition. Stars usually accrete their discs, except for special cases when processes like external photo-evaporation may take place (e.g. in clusters). However, during planet formation, the forming planet removes material from the proto-planetary disc which is consequently not accreted onto the central star (e.g. Bitsch et al., 2018). This may give rise to the abundance differences between the stars in a binary systems (Tucci Maia et al., 2014; Ramírez et al., 2015; Teske et al., 2016), potentially revealing the planet formation location, although different explanations are also possible (e.g. Adibekyan et al., 2017). The direct engulfment of planetary companions also creates large observable abundance differences that appear to have trends that are distinct from disc consumption (e.g., Oh et al., 2018). Hence, by detailed mapping of abundances in binary systems, MSE will place valuable constraints on planet formation and destruction pathways.

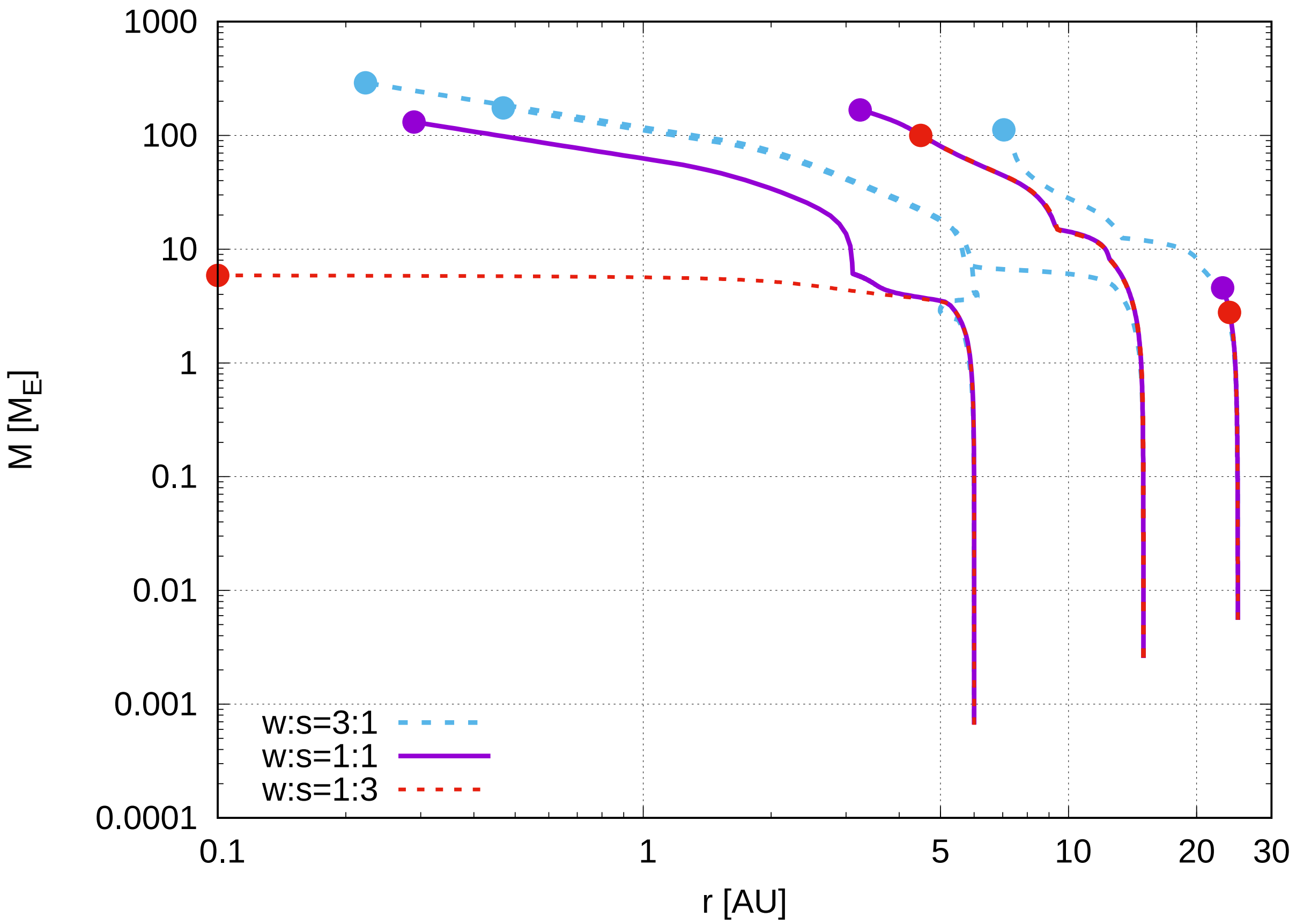

*Figure 21: Detailed chemical abundances of large numbers of planet hosts obtained by MSE will constrain the building blocks of planet formation. Lines shows growth tracks of planets in discs with different water-to-silicate ratios (w:s), starting with pebble accretion followed by the accretion of a gaseous envelope after reaching* $5-10$ *Earth masses. The total metallicity is 1% for all simulations, only the composition of the build material is varied. Planets growing in water poor discs grow slower and to smaller masses due to the reduced water content, while planets growing in discs with large water content grow easier to gas giants, especially in the outer disc where building material is rare. Figure adapted from Bitsch & Johansen (2016).*

### 3.3.5 Planetary systems around white dwarfs

White dwarfs are a common end stage of stellar evolution, and almost all exoplanets detected today are orbiting stars that will eventually evolve into white dwarfs. What happens to the asteroids, comets, and planets when the host star evolves off the main sequence? Recent studies propose engulfment of close-in planets by evolved stars (Schröder & Connon Smith, 2008; Villaver & Livio, 2009). In particular, whether a planet would survive the AGB phase depends on the competition of tidal forces arising from the star's large convective envelope and of the planets' orbital expansion due to stellar mass loss (Mustill & Villaver, 2012). Intriguingly, no single planet orbiting a white dwarf has been detected yet, but the presence of planets can be inferred by the detection of material that has been most likely disrupted into the Roche lobe of the star (e.g. Mustill et al., 2018).

The most direct example is WD 1145+017, which displays long, deep, asymmetric transits with periods between 4.5 – 5.0 hours first discovered by the K2 Mission (Howell et al., 2014; Vanderburg et al., 2015; Gänsicke et al., 2016; Rappaport et al., 2016; Gary et al., 2017). The transits are believed to be produced by fragments from an actively disintegrating asteroid in orbit around the white dwarf. In addition, WD 1145+017 belongs to a small group of white dwarfs with infrared excesses from a circumstellar dust disk (Farihi, 2016). It is widely accepted that these hot compact disks are a result of asteroid tidal disruption (Jura, 2003). Infrared observations of white dwarf disks show that they can be variable on a few year timescale (Xu & Jura, 2014), further demonstrating the dynamic nature of these systems.

All white dwarfs with dust disks are also heavily "polluted"; that is, elements heavier than helium are present in their atmospheres from accretion of planetary debris. In a pioneering paper, Zuckerman et al. (2007) demonstrated that the bulk composition of exo-planetary debris can be accurately measured from high-resolution spectroscopy of polluted white dwarfs (Figure 22), including rock-forming elements such as refractory lithophiles (Si, Mg, Al, Sc, Ca, Ti), siderophiles (Fe, Ni, Mn), and volatiles (C, O, S, N) (Jura & Young, 2014). To zeroth order, exo-planetary debris has a composition similar to that of bulk earth, with O, Fe, Si, and Mg being the dominant four elements (Xu et al., 2014) with a small amount of C and N (Jura et al., 2012; Gänsicke et al., 2012). From the large variations of Si to Fe ratio observed in polluted white dwarfs, it has been suggested that differentiation and collision must be widespread in extrasolar planetary systems (Jura et al., 2013; Xu et al., 2013).

Recent results suggest that white dwarfs in some special cases are accreting from specific layers of a massive, differentiated rocky object (see, e.g. Melis et al., 2011; Raddi et al., 2015;

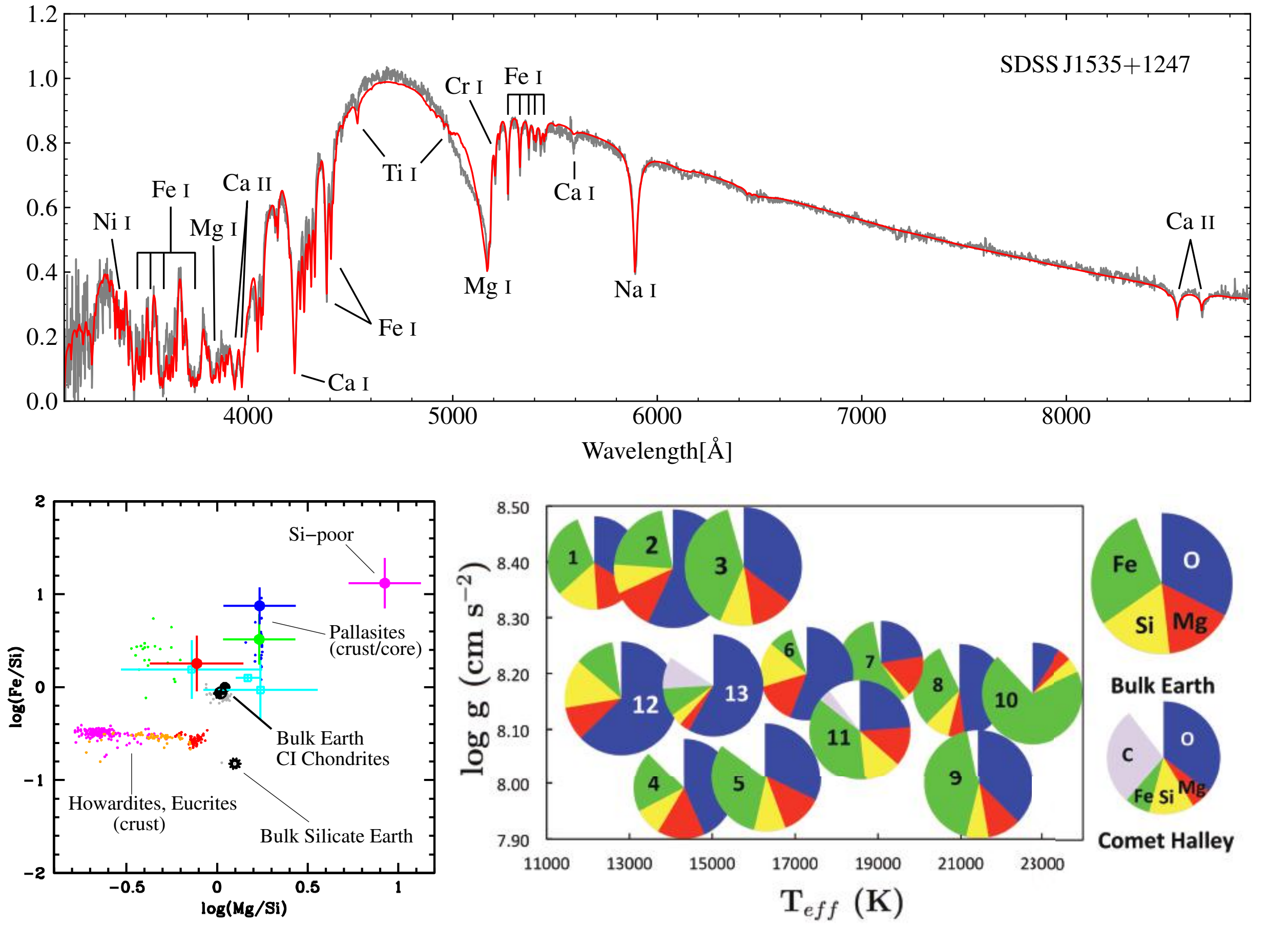

*Figure 22: MSE will provide direct measurements of the bulk composition of thousands of exo-planetesimals through spectroscopy of white dwarfs that were polluted by planetary debris. The abundances of Fe, Si, Mg, and trace elements (such as Sc, V, Ti, and Ni) are consistent with rocky planetesimals (Zuckerman et al., 2007; Gänsicke et al., 2012), though there is evidence for water-rich planetesimals (Farihi et al., 2013; Raddi et al., 2015), and Kuiper belt-like objects (Xu et al., 2017). Detailed abundances are currently measured only for a few dozen exo-planetesimals (bottom right, Xu et al. 2014), limited by the small number of strongly metal-polluted white dwarfs known. MSE will particularly excel at targeting cool, old white dwarfs, which provide insight into the formation of rocky planets early in the history of the Milky Way.*

Melis & Dufour, 2017). White dwarfs with pollution having significant enhancements of iron and deficient silicon and magnesium could be accreting the remains of a differentiated body's core (e.g. Melis et al., 2011), while white dwarfs that are iron-poor could have material originating in the crust-mantle region (i.e., the surface) of a differentiated body (e.g. Zuckerman et al., 2011). Water-rich exo-asteroids (Farihi et al., 2013; Raddi et al., 2015; Gentile Fusillo et al., 2017) and volatile-rich asteroids, similar to the composition of comet Halley (Xu et al., 2017), have been detected. Exotic compositions with no solar system analog, such as carbon-dominated chemistry, appear to be rare, if they exists at all (Wilson et al., 2016). Polluted white dwarfs can thus deliver information directly applicable to the study of rocky planets inside and outside our Solar system and break degeneracies on surface and interior composition that cannot be addressed with other available techniques (Dorn et al., 2015; Rogers, 2015; Zeng & Jacobsen, 2017). These measurements provide important inputs into planet formation models (Carter-Bond et al., 2012; Rubie et al., 2015).

MSE is ideally suited to rapidly advance the study of exo-planetesimal abundances in polluted white dwarfs. Gaia Data Release 2 (Gaia Collaboration et al., 2018b) recently uncovered an all-sky sample of $\sim 260,000$ white dwarfs that is homogeneous and nearly complete down to $G \lesssim 20$ (Gentile Fusillo et al., 2019). By extending the follow-up to $G \sim 20-21$ compared to smaller MOS facilities, MSE will dramatically increase the number of old white dwarfs with evolved planetary systems. An example is vMa2, the third-closest white dwarf ($d = 4.4$ pc, $V = 12.4$, van Maanen 1917) with a cooling age of $\sim 3.3$ Gyr and strong Ca, Mg, and Fe contamination (Wolff et al., 2002), indicating that it is accreting planetary debris. MSE spectroscopy of the Gaia white dwarfs will identify vMa2 analogs (Fig. 22, top panel) out to several 100 pc, and result in $\sim 1000$ strongly debris-polluted systems. MSE is, indeed, the only MOS project that can perform high-spectral resolution observations of white dwarfs. Hot white dwarfs can be heavily polluted but yet have weak enough lines such that R $\lesssim 5\,000$ spectra would not be able to detect them. Only R $> 20K$ spectra will show the weak CaII or MgII lines heralding the dramatic pollution present for these objects.

The detailed abundance studies of these systems will take the statistics of exo-planetesimal taxonomy to a level akin to that of solar system meteorite samples. The progenitors of the Gaia white dwarfs span masses of $M_{\rm ZAMS} \sim 1-8\,{\rm M}_{\odot}$, and the ages of these systems will range from a few 100 Myr to many Gyr, providing deep insight into the planet formation efficiency as a function of host mass and into the signatures of galactic chemical evolution on the formation of planetary systems.

### 3.4 Stellar physics with star clusters

The statistics and dynamical properties of ensembles of stars in clusters are being revolutionised by the ESA's flagship facility Gaia (Gaia Collaboration et al., 2018a) which provides positions, parallaxes and kinematics for huge samples of clusters and associations across the full range of ages. Complementary to this, numerous ground- and space- facilities, such as VVV (Minniti et al., 2010), HST, and J-PLUS (Cenarro et al., 2018), are used to obtain high-precision and deep photometry of cluster members (e.g. Borissova et al., 2011; Dotter et al., 2011; Soto et al., 2013). Future photometric time-series facilities like LSST (LSST Science Collaboration et al., 2017), using its wide-fast-deep (WFD) observing strategy (Pris-inzano et al., 2018), will probe most distant and faints regions in the Milky Way providing up to 2000 new clusters.

However, detailed physical insights into the physics of these systems is hampered by lack of the critical component - a detailed spectroscopic characterisation that provides accurate line-of-sight velocities, fundamental parameters and chemical composition of stars. Current instruments, such as UVES@VLT and Keck can only observe a handful of stars at a time with high-resolution, while the quality of data with fiber instruments (Giraffe@VLT) if often compromised by narrow-filter observations. This is the area where the technical capabilities of MSE will be un-matched: MSE will be the only facility in 2025s to provide massive spectroscopic follow-up of clusters detected with Gaia and LSST.

With its large FoV, large aperture, and broad wavelength coverage, MSE will map stellar clusters out to 100s of kpc (Fig. 23), providing critical information on the evolution of coeval ensembles of stars in different environments.

### 3.4.1 Pre-main sequence stars

Pre-MS stars with ~1 to $6M_{\odot}$ share the same location in the HR-diagram as their evolved counterparts in the post-MS phase. Hence, it often not possible to constrain the evolutionary stage of stars, i.e. before or after the MS, only by their position in the H-R diagram. The main difference between stars in the two evolutionary phases lies in their inner structures. Using asteroseismology, the frequencies of pressure and gravity modes can be observed as periodic variations in luminosity and temperature or as Doppler shifts of spectral lines, providing critical information about stellar interiors that remove this ambiguity (e.g. Zwintz, 2016).

With MSE, large sample of young clusters and star forming regions can be targeted to obtain high-resolution, high SNR time-series spectroscopy to study line profile variations for pulsating young stars. Good candidates are the MYSTIX sample of clusters (e.g Kuhn et al. 2015) or associations such as Cygnus OB2 (Wright et al. 2015). Typical pulsation periods of different classes of pre-MS pulsators lie between ~0.5 days and 3 days for slowly pulsating B and $\gamma$ Doradus type stars, between ~18 minutes and 6 hours for $\delta$ Scuti type objects and between ~five minutes to 20 minutes for the currently only predicted solar-like oscillators. The analysis of time dependent variations of spectral absorption line profiles will provide sensitive probes of the pulsation modes.

Simultaneous to the time series observations of pulsating young cluster members and candidates, the less massive and, hence, fainter cluster members - typically T Tauri like objects - can be targeted with the remaining MSE fibers. Using high-resolution spectroscopy of T Tauri stars, the activity of these low-mass pre-MS objects can be studied through the time-dependent properties of the chromospheric $H_{\alpha}$ and Ca II lines. Emission lines originating from the circumstellar environment trace infalling and outflowing gas, including broad components of $H_{\alpha}$ and CaII in the accretion flow, and also jet lines such as [OI], [NII], and [SII] as well as other species in the more extreme objects.

Spectroscopic characterisation of the pre-MS stars, especially those in young clusters, will allow measurements of the effective temperature, surface gravity, and detailed element abun-dances. Most readily measured is the increased Li abundance for cool pre-MS stars. Further detailed work on elemental abundance patterns will allow empirical investigation of how material accreted from circumstellar disks changes the chemistry of the pre-MS stars.

MSE will also provide the opportunity to conduct a homogeneous study of the rotation rates of pre-MS stars. With this it is possible to learn how gravitational contraction and the initial stages of out-of-equilibrium nuclear burning influence the stars rotation rates and to test the theoretical assumptions of a first spin-up of rotation during the pre-MS contraction phase and then a deceleration at the final approach to the main-sequence.

### 3.4.2 Open clusters

MSE is the major next-generation facility that will allow an in-depth study of the full stellar mass spectrum in open clusters across the Galaxy (Figure 23). Gaia DR3 (~ late 2021) will provide precision astrometry for the analysis of cluster membership. MSE will complement by allowing the full spectroscopic characterization, probing down to the K and M dwarf domain in clusters out to 2-3 kpc, whereas turn-off regions can be mapped in clusters out to 20-50 kpc, and individual stars on the horizontal branch and tip RGB even to extragalactic distances.

Accurate spectroscopic characterisation by MSE will provide chemical composition for stars of all masses, as well as orbits and masses for a wealth of spectroscopic binaries with accurate Gaia astrometry. In particular, the high-resolution ($R \sim 40\,000$) mode of MSE is ideally suited to determine precision abundances of Li, $^{13}$C and $^{12}$C, N, Na, Al, Fe (e.g. Bertelli Motta et al., 2018; Gao et al., 2018; Smiljanic et al., 2018; Souto et al., 2018), as well as projected rotational velocities, which, combined with rotation periods from $K2$ and TESS, will allow the inclination of the rotation axis to the line of sight to be measured, thereby greatly increasing the accuracy of stellar radii estimates. Accurate rotation velocities and subtle chemical signatures offer unique constraints on the physics of mixing in stellar interiors, including atomic diffusion, radiative acceleration, turbulent convection, and rotational mixing (Richard et al., 2005; Charbonnel & Zahn, 2007; Lagarde et al., 2012; Deal et al., 2018). Recent studies (Marino et al., 2018) suggest that extended MS turn-offs are associated to stellar rotation.

Thanks to the large aperture of MSE, it will be possible to cover a range of evolutionary stages and masses, from the upper main-sequence and turn-off, to the Hertzprung gap and the RGB, reaching the red clump, blue loops and the early-AGB. Also, central stars of planetary nebulae in clusters are excellent candidates to explore the initial-to-final mass relationship, through the opportunity to constrain the age and mass of the progenitor star. Beyond offering a critical test of stellar evolution models, MSE data will give new insights on age indicator diagnostics, such as the main-sequence turnoff, abundances of lithium, and gyro-chronology (e.g. Do Nascimento et al., 2009; Barnes et al., 2016; Beck et al., 2017; Randich et al., 2018).

The MSE data, both low- and high-resolution, will be also crucial to constrain the poorly-known aspects of stellar physics in the low-mass domain (0.08 – 0.5 $M_\odot$, Figure 24), including the equation-of-state of dense gas, opacities, nuclear reaction rates, and magnetic fields, which are thought to be responsible for the inflation of stellar radius observed in very-low-mass (VLM) stars (Feiden & Chaboyer, 2013; Kesseli et al., 2018; Jaehnig et al., 2018). This will ideally complement the Gaia's Ultracool Dwarf Sample, which is expected to contain thousands of ultra-cool L, T, and Y dwarfs in the local neighbourhood (Smart et al., 2017; Reylé, 2018). MSE will, furthermore, probe the transition from structures with radiative cores to the fully convective regime (Jackson et al., 2016) that will enable studies of mechanisms that generate magnetic fields (Feiden & Chaboyer, 2013; Brun & Browning, 2017). Improved stellar radii well test the hypothesis that mass transfer in binary systems of VLM stars could lead to over-massive brown dwarfs (Forbes & Loeb, 2018).

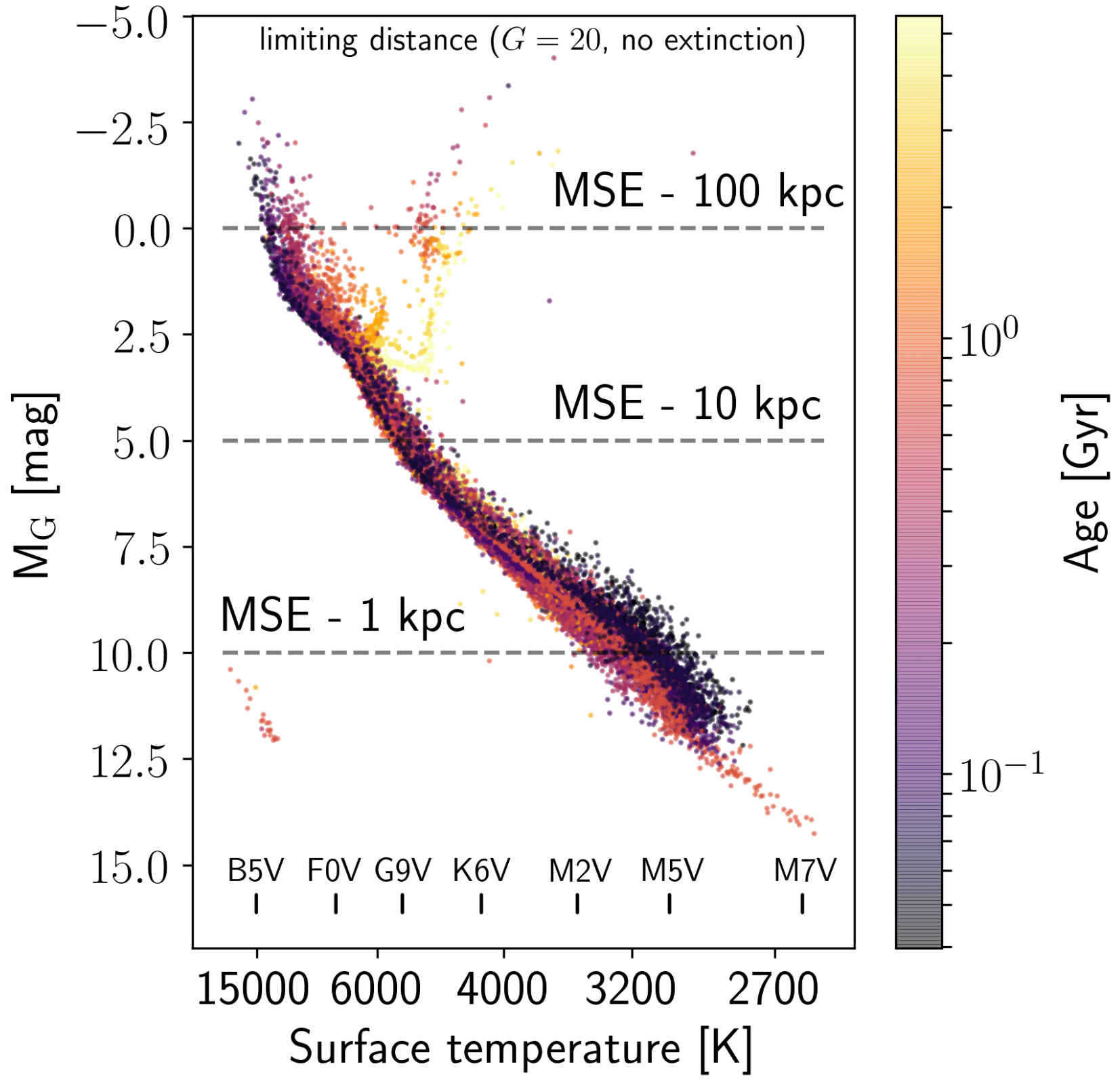

*Figure 23: MSE will allow the full spectroscopic characterization of open clusters, probing down to the K and M dwarf domain in clusters out to 2-3 kpc, whereas turn-off regions can be mapped in clusters out to 20-50 kpc, and individual stars on RGB even to extragalactic distances. A composite colour-magnitude diagram for open clusters colour-coded according to age and metallicity. Horizontal dashed lines show typical distances for a limiting magnitude of* $\boldsymbol{G} = 20$*, corresponding to an approximate faint limit for MSE. Figure adapted from the Gaia DR2 data Gaia Collaboration et al. (2018a).*

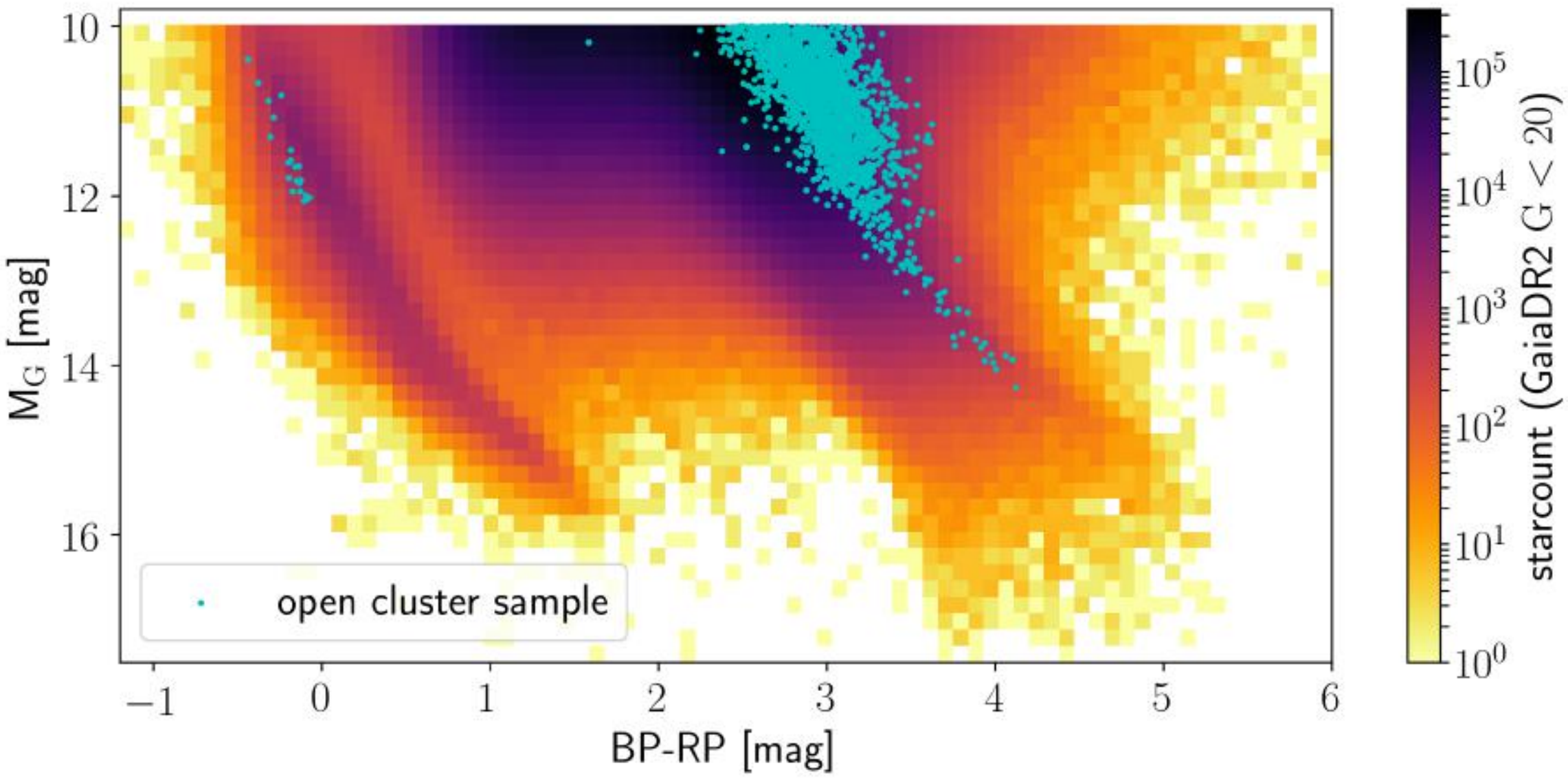

*Figure 24: MSE will be crucial to constrain the poorly-known aspects of stellar physics in the low-mass domain. Color-magnitude diagram of low-mass and very low-mass stars in the Gaia DR2 catalog, with members of open clusters highlighted. Figure adapted from Gaia Collaboration et al. 2018a.*

Ultimately, MSE observations of large samples of solar analogues in clusters (e.g. Fichtinger et al., 2017), which will be possible out to 10 kpc (Fig. 6), will allow their mass-loss rates to be accurately measured. Mass loss is one of the crucial parameters in stellar evolution, as even slow winds of $10^{-13}$ $M_{\odot}$/yr remove the outermost layers of the star at a rate comparable to that of diffusion. This has a subtle, but important effect on the stellar photospheric abundances. Constraining mass loss in Sun-like stars will also offer new insights into the faint young Sun paradox (Gaidos et al., 2000; Feulner, 2012). So far, all attempts to resolve the problem of Earth turning into a state of a global snowball during the first two billion years failed. Yet, strong evidence for abundant liquid water on Mars (Orosei et al. 2018) has sparked renewed interest in this problem. The data that MSE will obtain for solar-like stars at different ages will set new constraints on the hypothesis that the Sun could have been more massive in the past and more luminous than the standard solar models predict (Serenelli et al., 2009; Vinyoles et al., 2017).

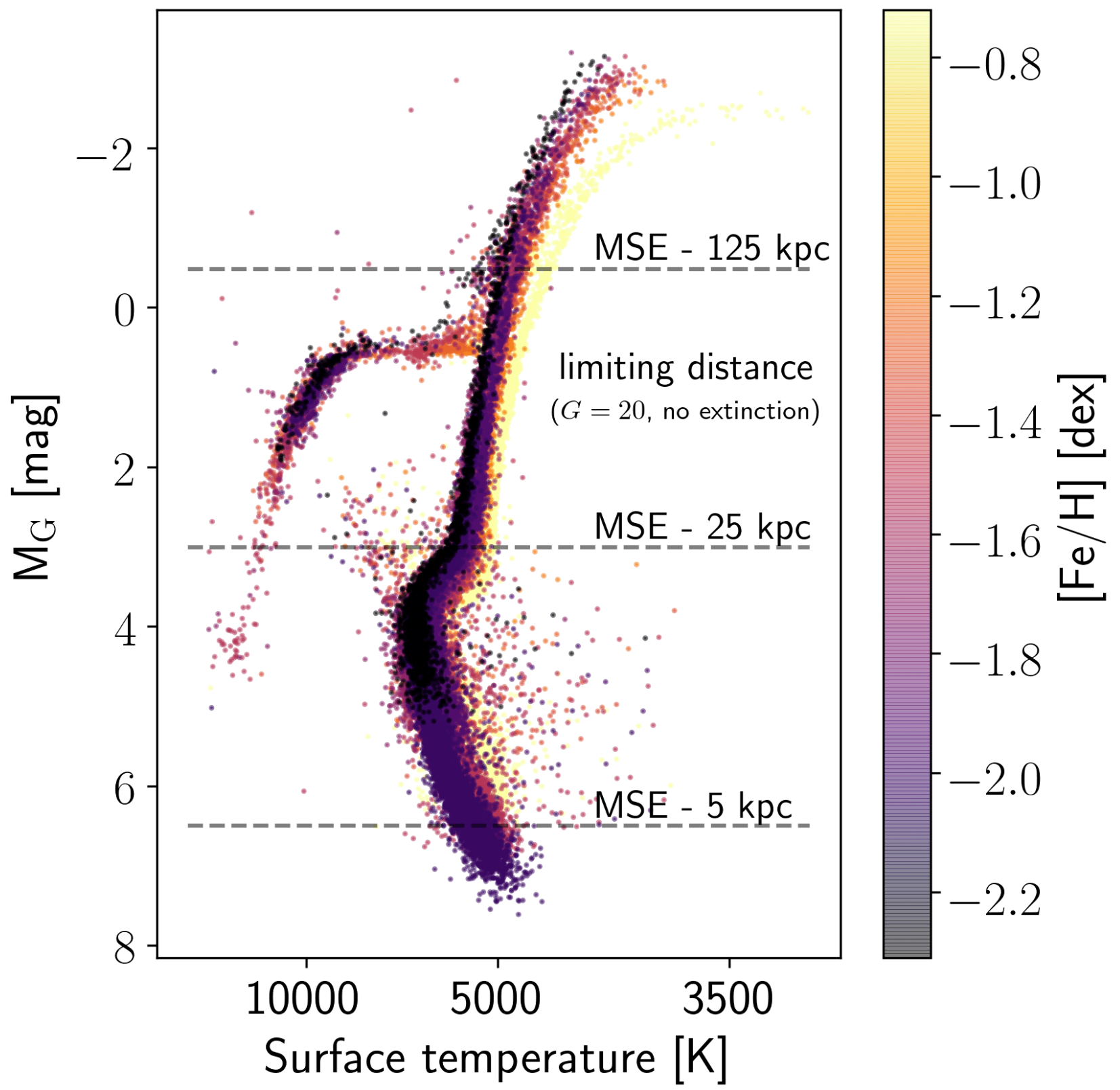

*Figure 25: Capitalising on Gaia, MSE will be the most important next-generation facility to allow a massive spectroscopic census of globular clusters in the Milky Way and its galactic neighborhood out to 130 kpc. A composite colour-magnitude diagrams for globular clusters colour-coded according to age and metallicity. Horizontal dashed lines show typical distances for a limiting magnitude of $G = 20$, corresponding to an approximate faint limit for MSE. Figures adapted from the Gaia DR2 data Gaia Collaboration et al. (2018a).*

### 3.4.3 Globular clusters

The past decade stirred revolution in our understanding of globular clusters (GCs), which were once thought to be simple and coeval stellar populations. Multiple sequences seen in HST photometry (e.g. Sarajedini et al., 2007; Piotto et al., 2015; Marino et al., 2017; Milone et al., 2018), but also strong chemical signatures in form of anti-correlations revealed using high-resolution spectra at the largest 8- ad 10-meter telescopes (e.g. Carretta et al., 2010; Gratton et al., 2012) have been pivotal to prove that most GCs, in contrast to simple OCs, are highly complex entities hosting multiple populations of stars (see Bastian & Lardo, 2018, for a review). The origin of these multiple populations is a major astrophysical problem, that has received considerable attention in theoretical stellar physics seeding a variety of scenarios from fast-rotating massive stars (Decressin et al., 2007; de Mink et al., 2009), to AGB (e.g. Ventura et al., 2001) and supermassive stars ($\gtrsim 10^3\ M_\odot$, Denissenkov & Hartwick, 2014; Gieles et al., 2018), or a combination thereof (Valcarce & Catelan, 2011).

Capitalising on Gaia, MSE will be the most important next-generation facility to allow a massive spectroscopic census of globular clusters in the Milky Way and its local galactic neighborhood, providing new constraints on the evolution and structure of stars in dense environments across the full range of metallicities and ages. The end of mission data from Gaia will be of sufficient quality to provide accurate treament of crowding, in addition to delivering exquisite proper motions and parallaxes accurate to better than 1% out to 15 kpc, as well as precise photometry for brighter stars (Pancino et al., 2017). MSE will complement this with high-quality radial velocities, to obtain accurate kinematics for clusters out to 100 kpc, and, crucially, with detailed chemical composition. These observations will be pivotal to provide constraints on self-enrichment, rotation (Bastian & Lardo, 2018), stellar evolution in multiple systems, mass transfer in binaries, and chemical imprints on surviving members (e.g. Korn et al., 2007; Gruyters et al., 2016; Charbonnel & Chantereau, 2016). It may become possible to detect signatures of internal pollution by neutron star mergers, or nucleosynthesis in accretion discs around black holes (Breen, 2018). Constraining the binary mass function in GCs would also help to improve the models of binary black hole formation (Hong et al., 2018) and test whether the BBH formation, following numerous detections of gravitational waves from mergers with LIGO and Virgo, is facilitated by dynamical encounters in globular clusters (Fragione & Kocsis, 2018).

MSE will provide precise abundances of the key elements that allow tracing the physics of stellar interior: Li, CNO, as well as odd-even pairs of heavier metals, e.g., Na, Mg, Al, and O. Peculiar abundance patterns on the RGB still lack a theoretical explanation in the framework of canonical stellar evolution theory (e.g. Charbonnel & Zahn, 2007; Angelou et al., 2015; Henkel et al., 2017). Thermohaline instability (Angelou et al. 2011, 2012, Lagarde et al. 2011, 2012), rotation (Palacios et al., 2006; Denissenkov & Tout, 2000), magnetic buoyancy (Palmerini et al., 2011; Hubbard & Dearborn, 1980), internal gravity waves or combinations there of (Denissenkov et al., 2009) have been proposed as potential mechanisms that trigger the mixing. Beyond, there is a debated problem of missing AGB stars in second population of massive clusters (see MacLean et al. 2018 for a review).

With MSE, the key evolutionary stages - the main-sequence, turn-off and subgaints, the RGB bump up to the RGB tip and horizontal branch - will be homogeneously and systematically mapped in clusters out to 30 kpc, expanding the previous high-resolution samples by orders of magnitude, and hence potentially providing new strong constraints on the physical and dynamical evolution of stars in globular clusters and their ages (VandenBerg & Denissenkov, 2018; Catelan, 2018).

### 3.4.4 White dwarfs

White dwarfs offer a unique opportunity to constrain ages of stellar populations (Winget et al., 1987; Richer et al., 1997; Fontaine et al., 2001; Tremblay et al., 2014). The total age of stellar remnants with the mass slighly above 0.6 $M_{\odot}$ is dominated by the white dwarf cooling age, allowing to accurately pin down the age of the system. Despite the prospects, limitations to this technique still remain, owing to the complexity of the cooling physics of the models. Also very deep observations are required to probe the cool and faint remnants in old systems, such as the Galactic halo (Kilic et al., 2019) and globular clusters (Hansen

et al., 2002).

Gaia detected ≃ 260 000 white dwarfs, an unprecendeted sample, homogeneous to the magnitude of $G \lesssim 20$ (Gentile Fusillo et al., 2019). LSST and Euclid will push the faint limit to 22 − 23 mag. MSE will be the ideal facility to obtain spectroscopic follow-up of these faint old white dwarfs, and to determine acccurate $T_{\rm eff}$, masses, and cooling ages. Pushing to a fainter magnitude limit, compared to current MOS facilities, is essential to define the initial-to-final mass relation at lower stellar masses (see Section 3.6.3), but also to constrain the exotic physics of cool dense WDs: crystallisation (Tremblay et al., 2019), convective coupling between the envelope and the core (Fontaine et al., 2001; García-Berro et al., 2010; Obertas et al., 2018), as well as non-ideal gas physics (Blouin et al., 2018).

## 3.5 Asteroseismology, rotation, and stellar activity

Continuous, high-precision photometry from space-based telescopes such as CoRoT (Baglin et al., 2006) and Kepler/K2 (Borucki et al., 2011; Howell et al., 2014) have recently initiated a revolution in stellar astrophysics, with highlights including the application of asteroseismology across the H-R diagram, and the investigation of rotation-activity relationships across a range of stellar masses and ages. Current and future missions planned over the coming decade such as TESS (Ricker et al., 2014), PLATO (Rauer et al., 2014) and WFIRST (Spergel et al., 2013) will continue this revolution by extending the coverage of high-precision space-based time-domain data to nearly the entire sky. At the same time, Gaia data releases (Gaia Collaboration et al., 2018c) and ground-based transient surveys such as Pan-STARRS (Chambers et al., 2016), ATLAS (Tonry et al., 2018; Heinze et al., 2018), ASAS-SN (Shappee et al., 2014; Jayasinghe et al., 2018), ZTF (Bellm et al., 2019), and LSST (Jurić et al., 2017) will provide more sparsely sampled light curves revealing variability in millions of stars across our galaxy.

A notorious problem for the interpretation of this wealth of time-domain photometry is that the majority of targets are faint, thus making systematic spectroscopic follow-up time consuming and expensive. Dedicated high-resolution spectroscopic surveys of the Kepler field, for example, have so far covered less than 20% of all stars for which light curves are available (Mathur et al., 2017). Furthermore, currently planned spectroscopic surveys capable of surveying large regions of the sky will only cover a small fraction of all stars for which high-precision space-based photometry will be avaiable (Figure 26). MSE is the only MOS facility that will make it possible to break this bottleneck and fully complement space-based photometry with abundance information, enabling investigations across a wide range of long-standing problems in stellar astrophysics.

### 3.5.1 Solar-like oscillations

The *Kepler* mission detected oscillations excited by near-surface convection (solar-like oscillations) in approximately 100 solar-type main-sequence stars (Davies et al., 2016a; Lund et al., 2017). This sample enabled the first systematic asteroseismic determinations of the ages, masses, radii, and other properties of solar-type stars (e.g., Silva Aguirre et al., 2015, 2017; Angelou et al., 2017; Bellinger et al., 2016, 2018). For the best of these targets it was possible to use asteroseismology to infer their internal structure in a manner that is essentially independent of stellar models (Bellinger et al., 2017).

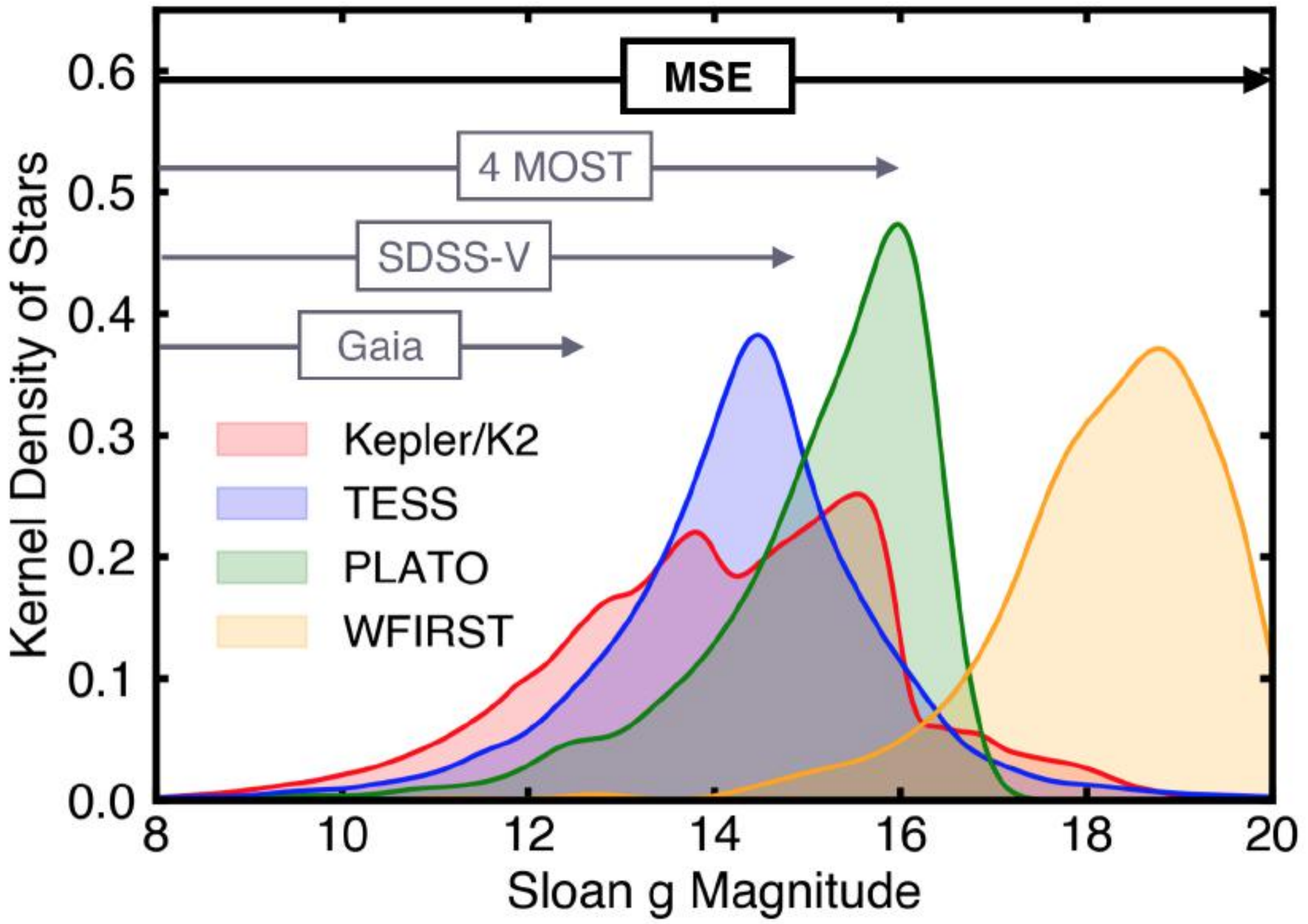

*Figure 26: MSE is the only MOS facility that can provide high-resolution optical spectroscopy for tens of millions of stars with high-precision future space-based photometry. Lines show the g-magnitude distribution for stars with space-based photometry from Kepler/K2 (red, Batalha et al., 2011; Huber et al., 2014, 2016) and predicted yields of stars observed with a photometric precision better than 1 mmag hr$^{-1}$ from an all-sky survey with TESS (blue, Sullivan et al., 2015; Stassun et al., 2018), a typical PLATO field (green, Rauer et al., 2014), and the WFIRST microlensing campaign (orange, Gould et al., 2015). Sensitivity limits of other MOS facilities that will provide high-resolution ($R > 20000$) spectroscopy at least half of the sky ($> 2\pi$) are shown in grey. Lines are kernel densities with an integrated area of unity.*

The TESS and PLATO missions are expected to increase the number of solar-like oscillators across the H-R diagram by several orders of magnitude (Schofield et al., 2019; Rauer et al., 2014). MSE, with its large aperture, unique multiplexing capabilities and broad-band wavlength coverage, will provide high-resolution and high SNR spectra for all of these targets, providing critical complementary information for asteroseismic analyses such as detailed chemical composition, surface rotation rates, and multiplicity.

Combining spectroscopic follow-up from MSE with asteroseismic data will allow the identification of best-fit evolutionary models, revealing accurate ages, masses, radii, and other properties for an unprecedented number of stars. The combined information will also provide constraints on the role of convection and turbulence in stellar evolution (Salaris & Cassisi, 2015; Tayar et al., 2017; Salaris et al., 2018; Mosumgaard et al., 2018). Inversions of asteroseismic frequencies and measurements of frequency separations will permit comparisons with the best-fit stellar models, placing an ensemble of constraints on stellar structure across a large range of ages, masses, and metallicities and allowing strong tests of the physics of stellar interiors.

Asteroseismology also provides precise stellar surface gravities, with accuracies better than 0.05 dex (Morel & Miglio, 2012; Huber, 2015). Asteroseismic $\log \mathrm{g}$ values depend only mildly on $T_{\mathrm{eff}}$, and thus have been widely used to $\log g$ values measured by spectroscopic pipelines. For example, CoRoT targets have been observed by the Gaia ESO Survey as calibrators (Pancino et al., 2012), *Kepler* targets have been used for calibrating APOGEE (Pinsonneault et al., 2014) and LAMOST (Wang et al., 2016), and K2 targets have been used for constructing the training sample for GALAH (Buder et al., 2018) and for calibrating atmospheric parameters in RAVE (Valentini et al., 2017). Observations of PLATO and TESS asteroseismic targets at every resolution will thus provide a powerful calibration set for MSE, in addition to providing powerful constraints on galactic stellar populations (galactic archeology) enabled by combining asteroseismology and spectroscopy (e.g. Miglio et al., 2017).

While oscillation frequencies are valuable diagnostics of stellar interiors, the amplitudes of solar-like oscillations are poorly understood, owing to large uncertainties when modeling the convection that stochastically drives and damps the oscillations (Houdek, 2006). Empirical relations between stellar properties and oscillation amplitudes have been established (Huber et al., 2011; Corsaro et al., 2013; Epstein et al., 2014), but the scatter of these relations exceeds the measurement uncertainties by a factor of 2, indicating a missing dependency on metallicity that has yet to be established. The large number of solar-like stars observed by PLATO and by MSE will provide unprecedented insights into how chemical compositions affect the pulsation properties of stars across the low-mass H-R diagram.

### 3.5.2 Stellar activity and rotation

Stellar rotation is a fundamental diagnostic for the structure, evolution, and ages of stars. Rotation has long been associated with stellar magnetic activity, but the dependence of age-activity-rotation relationships on spectral types are still poorly understood.

Strikingly, recent Kepler results indicate a fundamental shift in the magnetic field properties of stars near solar age, sparking several follow-up efforts to study the variation in magnetic activity for solar-type stars (van Saders et al., 2016; Metcalfe et al., 2016). The wide wavelength range of MSE optical spectra enables diagnostics using multiple activity-sensitive spectral indices (e.g. Wise et al., 2018), including the Ca H&K lines at ~396.9 and 393.4 nm, the $H_\alpha$ line at 656.28 nm or the Ca infrared triplet at 849.8, 854.2 and 866.2 nm., as well as the estimates of projected rotation velocities. For a sub-sample, rotation periods of thousands of stars from time-domain surveys will be available. Thus, MSE will explore rotation-activity relationships across the full HRD on an unprecedented level.

High-cadence photometry from Kepler has also provided new insights into high-energy environment of stars by detecting flares and probing their dependence on flare energies and spectral type (Davenport, 2016). Understanding stellar flares is tightly connected to stellar activity cycles and prospects of habitability of exoplanets, in particular for low-mass stars for which the habitable zones are close to their host stars (Shields et al., 2016). Recent studies attempting to link chemical abundance patterns such as the production of Li to super-flares observed in Kepler stars have yielded ambiguous results (Honda et al., 2015), highlighting the need of systematic spectroscopic follow-up that can be provided with MSE.

MSE will be operating during the PLATO mission, and thus allow the possibility of simultaneous high-precision photometry and high-resolution spectroscopy for a large sample of stars for the first time. Depending on the achievable RV precision, this would allow investigations of the correlation between photometric variability and RV jitter (e.g., Saar et al., 1998; Boisse et al., 2009; Haywood et al., 2014; Cegla et al., 2014; Bastien et al., 2014; Oshagh et al., 2017; Tayar et al., 2018; Yu et al., 2018), including dependencies on stellar parameters (such as spectral type, age, metallicity), providing comprehensive insight about stellar magnetic activity.

### 3.5.3 Opacity-driven pulsators

Space-based missions such as TESS and PLATO will reveal a vast number of pulsating variable stars in the classical instability strip over the coming decades, including $\delta$ Scuti variables, RR Lyrae stars, and Cepheids. CoRoT, Kepler, and OGLE have discovered several intriguing dynamical effects in these classical pulsators, including modulations (Blazhko effect), resonances, additional (nonradial) modes, and period doubling (Szabó et al., 2014; Anderson, 2016; Smolec, 2017). Spectroscopic characterizations of all these pulsating stars with MSE will shed new light on the critical dependence of their physical and dynamical parameters on metallicity and detailed chemical composition.

One of the major challenges with the application of the Leavitt law (Leavitt, 1908) for measuring cosmic distances with Cepheids and, thereby, constraining the Hubble constant, $H_0$, is calibrating its metallicity dependence (Freedman et al., 2012; Riess et al., 2016, 2018a). Empirical studies show that metal-poor Cepheids appear to be more luminous for the same pulsation period, at least at optical wavelengths (Freedman & Madore, 2011; Gieren et al., 2018). At infrared wavelengths the dependency flips, and the nature of the metallicity dependency of the period-luminosity relationship is unknown. Predictions from theoretical models suggest that the metallicity dependency at optical wavelengths should be opposite to what is found from empirical studies (Bono et al., 2008, and references therein).

MSE will allow a systematic spectroscopic characterization of large samples of Cepheids detected with transient surveys, Gaia and space-based photometry missions. Current samples are small (Romaniello et al., 2008; Genovali et al., 2013, 2014) and biased, especially at long pulsation periods. MSE will provide high-resolution spectra for statistically significant samples of both types of Cepheids (I, II) across a large range of pulsation phases and pulsation periods. This will provide critically important information to finally pin down the impact of metallicity on the Leavitt law, thereby improving the precision of the local value of the Hubble constant $H_0$ and probing the origin of the controversial results from the direct measurement of $H_0$ and that based on Planck combined with a concordance ΛCDM model.

Likewise, the number of high-resolution spectroscopic studies of RR Lyrae stars is currently very limited, requiring photometric techniques that are difficult to calibrate owing to the lack of spectroscopic data (Hajdu et al., 2018), thus jeopardizing their use as distances indicators and tracers of structures. MSE will make an important step forward in this direction.

High-resolution spectroscopy is also critical to interpret the pulsations of lower luminosity stars in the instability strip such as $\delta$ Scuti and $\gamma$ Doradus variables. In particular, assigning mode identifications to observed pulsation frequencies has been a major bottleneck for modeling the interior structure and deriving fundamental parameters these stars, although some progress on identifying regular frequency patterns similar to solar-like oscillators has been made (Antoci et al., 2011; Breger et al., 2011; García Hernández et al., 2015). Additionally, a novel method using frequency shifts of $\delta$ Scuti pulsations (Murphy et al., 2014; Shibahashi et al., 2015) has been used to discovery a vast number of wide binary stars (Murphy et al., 2018) and exoplanets (Murphy et al., 2016) around these intermediate-mass stars. High-resolution spectroscopy with MSE of a large number of $\delta$ Scuti and $\gamma$ Doradus pulsators will be critical to narrow down the parameter space for modeling observed pulsation frequencies, and provide follow-up RV measurements for phase-modulation binaries identified from light curves.

### 3.6 Stars in multiple systems

Stellar multiplicity is an inherent characteristic to the formation and evolution of single and multiple stars because the stars are born in clusters and associations. In the solar neighborhood, the fraction of multiple systems is estimated at 40% for solar-type stars and can reach 90% for O-type stars (Moe & Di Stefano, 2017). Beyond this region, samples suffer from complete biases and selection effects of observing techniques.

MSE and Gaia are poised to revolutionize the field thanks to the possibility to monitor the radial velocities (RVs) of many thousands of spectroscopic binaries (SBs) on time scales of days to years down to very faint magnitudes. This will provide a unique facility to study stellar multiplicity from a statistical point of view on an unprecedenting scale (Sect. 3.6.1), facilitated by enormous progress in theoretical tools that enable using few-epoch spectroscopy to identify and characterize binaries and multiple systems (e.g. Price-Whelan et al., 2018).

In addition, SBs that show eclipses (EBs) are the gold standards for accurate masses and radii (Eker et al., 2018), fundamental to calibrate distance scales to the Local Group galaxies (Sect. 3.6.2). One of the simplest outcomes of the binary evolution occurs for wide binaries with WD that can provide new insights in the initial-to-final mass relation (Sect. 3.6.3).

MSE will also play a pivotal role in driving forward our understanding of the complex evolution of stellar interactions. Tidal interactions between stars can alter the birth eccentricity and period distributions of these systems and also provides the opportunity to constrain the internal structure of the member stars through measurements of the tidal circularization rate (e.g., Verbunt & Phinney, 1995; Goodman & Dickson, 1998; Price-Whelan & Goodman, 2018). The end result of strong interactions are binaries containing at least one compact stellar remnant – which are key objects across a wide range of astrophysics: all confirmed Galactic stellar mass black holes reside in binaries (Corral-Santana et al., 2016), and the most precise tests of gravitation come from binary pulsars (Antoniadis et al., 2013). Compact binaries also include the progenitors of some of the most energetic events in the Universe, supernovae Type Ia (SN Ia) and short gamma-ray bursts (GRBs), and the progenitors of all gravitational wave events detected to date (Abbott et al., 2016; Abbott et al., 2017a,b). MSE will be the key to observationally characterise large samples of compact binaries emerging from time-domain and X-ray surveys (Sect. 3.6.4) and the progenitors of gravitational wave events (Sect. 3.6.5), providing critically important tests and calibrations to binary evolution theory.

### 3.6.1 The binary census in the Milky Way and Local Group galaxies

A homogeneous census of multiple systems in different environments, from dense star forming regions to faint globular clusters, is fundamental to infer multiplicity rates and to provide strong constraints on formation and evolutionary pathways for single and multiple stars. MSE, combined with Gaia, will allow deep and wide spectroscopic monitoring to discover, characterize, and classify millions of Galactic binaries and thousands of binaries in Local Group galaxies.

Ground-based spectroscopic surveys such as RAVE (Steinmetz, 2019), GES (Gilmore et al., 2012), APOGEE (Majewski et al., 2016, 2017), LAMOST (Luo et al., 2015) and GALAH (De Silva et al., 2015) have identified thousands of single-lined (SB1 El-Badry et al. *e.g.* 2018), hundreds of double-lined (SB2 Fernandez et al. *e.g.* 2017) and tens of multiple-lined (SB3 Merle et al. *e.g.* 2017) candidates. MSE will improve upon these facilities (Fig. 27, left panel), owing to (i) an increased resolving power, (ii) large multiplexing, that will allow a simultaneous follow-up of numerous binaries in dense and faint clusters and (iii) a higher RV precision essential for an SB detection (e.g. velocity precisions for current MOS surveys and Gaia reach a few km/s for late-type and tens km/s for early-type stars).

Gaia DR3 is expected to identify tens of millions spectroscopic binaries. Yet, orbital solutions will be available only for $G \leq 16$ stars with periods less than 10 years (right panel of Fig. 27). A 10-year survey of MSE will allow monitoring of SBs with orbital periods up to 20 years. In addition, MSE will discover and characterize a wealth of new SB candidates around fainter stars ($16 \leq G \leq 23$). Compared to current state-of-the art studies (Raghavan et al., 2010; Duchêne & Kraus, 2013; Moe & Di Stefano, 2017), MSE samples of stars in multiple systems will provide an unprecendeted view of their population statistics: multiplicity frequencies and fractions, period, mass ratio, and eccentricity distributions in different environments. These data will also offer new insights into the controversial dependence of the binary fraction on metallicity (Badenes et al., 2018).

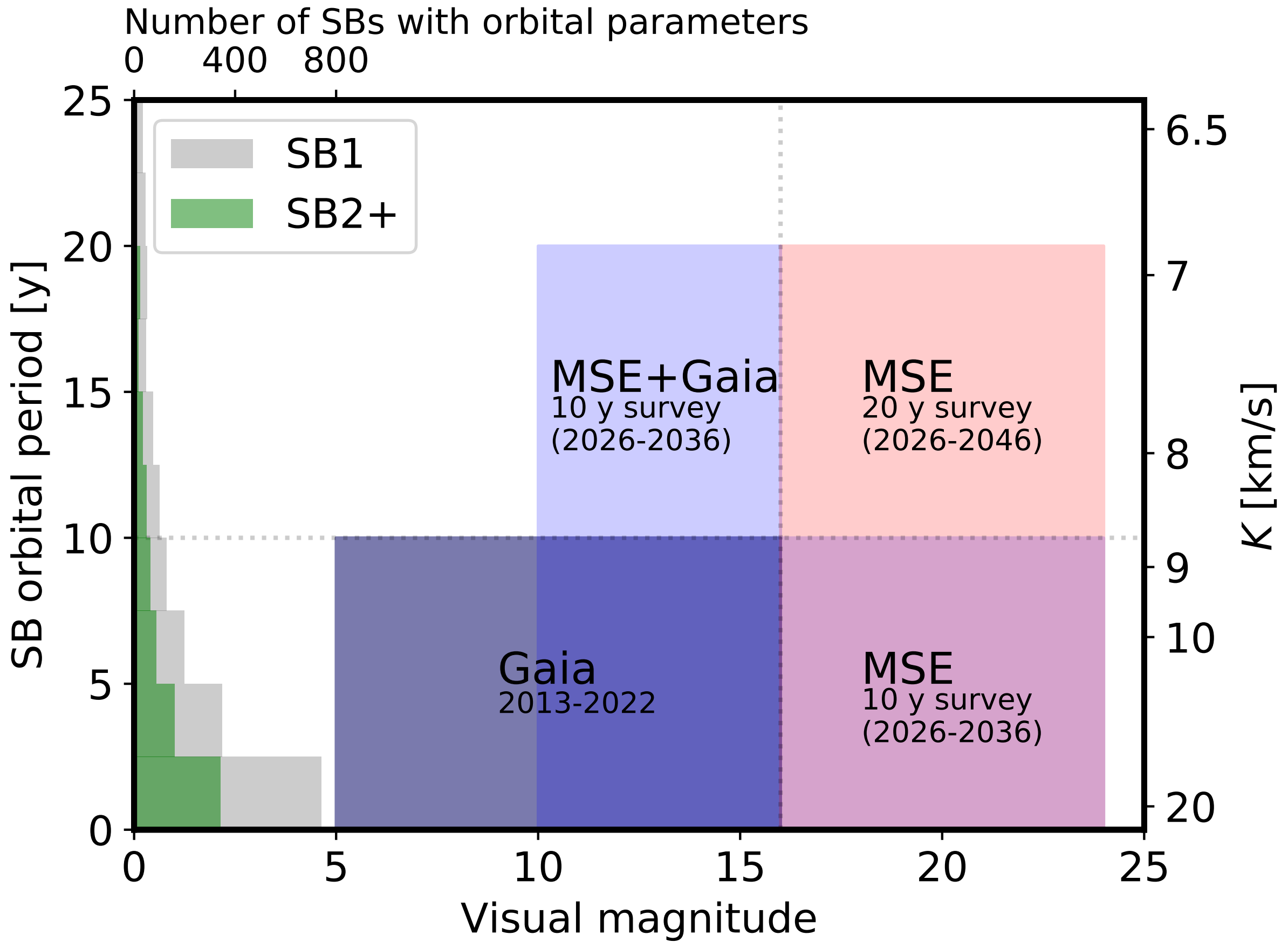

*Figure 27: Potential spectroscopic binaries (SB) discovered and characterized by Gaia, Gaia+MSE and MSE alone assuming 10 and 20 year surveys as a function of the visual magnitude. The left horizontal histograms show the periods distribution of known spectroscopic binaries with one (SB1, grey) and two or more components (SB2+, green) from the 9th catalogue of spectroscopic binary orbits (Pourbaix et al., 2004), based on decades of observations. The right vertical right scale gives the RV semi-amplitude* $K$ *for a twin binary with solar mass components on a zero eccentricity orbit seen edge-on.*

MSE will also characterise binary systems containing pulsating variable stars, which are otherwise challenging to follow-up with existing facilities due to phase smearing. Such systems are of key importance to constrain evolutionary and pulsation models of pulsating variables (e.g. Pietrzyński et al., 2010).

### 3.6.2 Eclipsing binaries

Eclipsing binaries (EBs) are fundamental calibrators for distances and stellar parameters, such as radii and masses. The CoRoT and Kepler missions discovered thousands of EBs, as well as other interacting binaries such as hearbeat stars and Doppler beaming EBs (Kirk et al., 2016; Deleuil et al., 2018). These yields are expected to increase by orders of magnitude with current and future space-based missions, such as TESS, WFIRST and PLATO. Many of these EBs will be too faint for Gaia (Figure 26), but MSE will be perfectly suited to obtain the masses and distances to these systems.

EBs have been fundamental to determine distances to the Magellanic Clouds, M31, and M33 (*e.g.* Guinan, 2004; North et al., 2010; Pietrzyński et al., 2013; Graczyk et al., 2014). An increasing number of extragalactic binaries are being found as members of dwarf galaxy members of the Local Group (*e.g.* Bonanos, 2013). LSST is expected to detect and characterise ∼ 6.7 million EBs, of which 25% will likely be double-lined binaries (Prsa et al. 2011). The MSE follow-up of these systems will allow studies of the properties of EBs that have formed in galaxies with dynamical and star formation histories different from that of the Milky Way and to greately improve the accuracy of extragalactic distance indicators.

### 3.6.3 Wide binaries as probes of post-main sequence mass loss

The stellar initial-to-final mass relationship (e.g. El-Badry et al., 2018) is a critical diagnostic of the evolution of asymptotic giant branch (AGB) stars, since the final mass of a star is determined by the combined action of mixing processes and mass loss. However, this relationship is still poorly understood, owing to significant systematic discrepancies between theoretical predictions and semi-empirical results (Salaris & Bedin, 2019).

The upcoming Gaia DR3 will discover a large number of long-period binaries containing a WD and a main-sequence star, which will offer an exquisite opportunity to improve constraints on the initial-to-final mass relationship. The total age of the system can be determined by combining MSE spectroscopy and Gaia astrometry for the un-evolved primary star. The mass of the WD progenitor is then constrained by making use of the WD cooling age and chemical abundances for the companion. With a large sample statistics, MSE will hence provide a number of powerful constraints on the initial-to-final mass relationship.

### 3.6.4 Compact white dwarf binaries

Compact binaries containing at least one white dwarf (CWDBs) are the most common outcome of close binary interactions, and are also easily characterised in terms of their physical properties. They, therefore, play a critical role in advancing our understanding of the complex physical processes involved in the evolution of binaries that undergo interactions. SDSS has demonstrated the enormous potential that observational population of large samples of CWDBs has for testing predictions of compact binary evolution theory (e.g. Gänsicke et al. 2009), providing constraints on the progenitors of SN Ia (Maoz et al., 2018), and calibrating empirical parameters on which binary population models are based, such as the common envelope efficiency (Zorotovic et al., 2010).

MSE will play a pivotal role characterizing large samples of several sub-classes of CWBDs, overcoming three major limitations of current studies: (1) CWBDs are intrinsically faint, and require a much larger aperture than ongoing MOS spectroscopic facilities can provide; (2) the SDSS samples of CWDBs were serendipitous identifications, hence incomplete and subject to biases that are difficult to quantify; (3) measuring key properties, in particular, orbital periods, required follow-up of individual CWBDs. The large aperture of MSE, access to large and well defined CWBD target samples, and the ability of mulit-epoch spectroscopy will address all three issues.

**Interacting CWDBs:**Interacting CWDBs exhibit extremely diverse observational characterstics, and were historically serendipitously identified via X-ray emission, optical colours, variability and emission lines. Consequently, the known population of CWDBs is very heterogeneous. Recent systematic time-domain surveys have started to produce well-defined samples of CWDBs (Drake et al., 2014; Breedt et al., 2014), which will be significantly augmented by the ZTF and LSST. The eROSITA mission (Predehl et al., 2018) will provide the first all-sky X-ray survey since more than two decades, and lead to the detection of intrinsically faint CWDBs with low accretion rates. The majority of these CWDBs will be fainter than 19th magnitude, and MSE will be uniquely suited to provide the spectroscopic follow-up to determine their fundamental properties. Gaining a comprehensive insight into the properties of interacting CBWDs is critically important to the development and testing of a holistic theoretical framework for the evolution of all types of compact binaries.

**Detached post-common envelope binaries (PCEBs):** Binaries which are sufficiently close to interact, once the more massive component leaves the main sequence, usually enter a common envelope phase. During this phase, the orbital separation shrinks by orders of magnitudes, leading to compact binaries with periods of hours to days (Ivanova et al., 2013). Our understanding of this phase is still fragmentary, and it is often modelled based on empirical fudge factors, which require observational calibration (Zorotovic et al., 2010).

SDSS demonstrated the potential of multi-object, multi-epoch spectroscopy to identify PCEBs (Rebassa-Mansergas et al., 2007), yet expensive individual follow-up of these systems was necessary to determine their binary parameters (Nebot Gómez-Morán et al., 2011). MSE will obtain radial velocity follow-up of several thousand PCEBs. identified in a homogenous way using Gaia parallaxes, variability information, and deep pan-chromatic imaging survey that are rapidly emerging, such as Pan-STARRS and LSST. Large aperture and high spectral resolution of MSE will permit the characterisation of systems spanning a much wider range of orbital separations and mass ratios. This will provide crucial tests on the theory of common envelope evolution (Zorotovic et al., 2010) and the binary populations models built on it (Schreiber et al., 2010).

**Double-degenerates:** Binaries in which both components have initial masses $\gtrsim 1\,M_\odot$ may go through two common envelope phases, resulting in short-period double-degenerates (DDs), which are key objects both in the context of SN Ia and gravitational waves. To date, only $\simeq 200$ DDs have been identified, largely due to the fact that medium to high resolution time-series spectroscopy is required to distinguish them from single white dwarfs. SPY (Napiwotzki et al. 2001) is the only high-resolution survey for DDs, yet this is a heterogeneous sample of only $\simeq 1000$ white dwarfs. SDSS identified several 10 000 white dwarfs, but it was only sensitive to the systems with the largest radial velocity amplitudes – $\simeq 200-300$ km/s – inherently resulting in a strong bias towards the shortest-period system and unequal mass ratio binaries with extremely low-mass companions (Brown et al., 2016). Combined, SPY and SDSS demonstrated that the fraction of DDs among the white dwarf population is $\simeq 5\%$, provided some constraints on the DDs as SN Ia progenitors (Maoz & Hallakoun, 2017; Maoz et al., 2018), and discovered a handful of ultra-compact DDs, which show orbital decay due to gravitational wave emission on time scales of years (Hermes et al., 2012).

The white dwarf sample identified with Gaia (Gentile Fusillo et al., 2019) finally provides the opportunity for a systematic and unbiased characterisation of the entire population of DDs. The large aperture and high spectral resolution of MSE will be critical to obtain precision spectroscopy for $\simeq 150\,000$ white dwarfs, which will result in $\simeq 10\,000$ DDs – sufficiently large a sample to quantitatively test the evolutionary channel that includes SN Ia progenitors. The DD sample assembled by MSE will also provide comprehensive and timely insight into the low frequency gravitational foreground signal (Nissanke et al., 2012; Korol et al., 2017) that has the potential to set the sensitivity threshold for LISA (to be launched in $\sim 2035$) in this frequency range.

### 3.6.5 Massive stars as progenitors of compact object mergers

Massive stars in multiple systems are the progenitor of compact object mergers such as binary black hole (BBH), neutron star and black hole (NSBH) and binary neutron star (BNS) systems. LIGO and Virgo gravitational wave observatories have confidently detected 10 BBH and 1 BNS mergers in two observation runs since September 2015 (e.g. The LIGO Scientific Collaboration et al., 2018). The statistics of those events will significantly improve with future upgrades and more detectors going online in the near future. However, the evolution of the progenitor massive star systems is poorly understood, and is unclear how such compact object systems can form in the first place. In addition to the complex evolutionary path of a single massive star, a companion in a close orbit induces additional poorly understood physical processes such as mass transfer and common envelope evolution.

Multiplicity properties of massive stars are well studied in the Galaxy ($Z \approx Z_\odot$) and in the Large Magellanic Cloud ($Z \approx 0.5Z_\odot$) (e.g. Sana, 2017, for an overview). While population synthesis calculation favour a metallicity upper limit of $Z \lessapprox 0.1Z_\odot$ (e.g. Belczynski et al., 2016; Marchant et al., 2016) to form such compact binary systems, recent studies suggest the upper limit can be as high as the metallicity of the Small Magellanic Cloud ($Z \approx 0.2Z_\odot$, e.g. Kruckow et al., 2018; Hainich et al., 2018).

The high sensitivity of MSE and its capability of wide field time-domain stellar spectroscopy will allow stringent tests on these predictions by efficiently observing and monitoring the massive star population in low redshift galaxies in our Local Group for the first time. MSE will provide homogeneous and high quality spectra of massive star multiple systems with a large variety of orbital properties over all evolutionary stages and a wide range of metallicities, including dwarf galaxies in the Local Group with the highest star formation rates as determined from $H_\alpha$ luminosity and oxygen abundances (Figure 28). In particular, adopting a completeness down to a $15\,M_\odot$ star on the main sequence (O9.5V) with a typical absolute magnitude $M_V \approx -4$ mag demonstrates that MSE will open a new window to the stellar physics of massive stars in low metallicity environments. Depending on the distance of the dwarf galaxy, the adopted resolution, and the number of available fibers it is possible to be even complete down to late B dwarfs ($\sim 5\,M_\odot$).

In addition to metallicities, time-domain stellar spectroscopy will provide us an additional independent method to derive stellar parameters from their orbital solutions and allow us to test in more detail the physics at low metallicity in state of the art stellar structure calculations. The mass ratio and spin distributions of compact binary mergers from gravitational wave observations will probe the prediction from population synthesis modelling and their predicted characteristics of BBH, NSBH or BNS systems before they merge (e.g. Eldridge & Stanway, 2016; de Mink & Mandel, 2016; Marchant et al., 2016). With MSE we will be able to probe evolutionary paths of massive binary system and discover new and unexpected evolutionary channels and massive stellar systems. In addition, at low metallicity lies the key in the understanding of the nature of pulsational pair-instability supernovae and long-duration Gamma-Ray Bursts, which might be a result of close binary evolution as well (e.g. Marchant et al., 2018; Aguilera-Dena et al., 2018).

### 3.7 Asymptotic giant branch evolution

All stars with initial masses between about $0.8M_\odot$ and about $10M_\odot$ evolve through the AGB phase of evolution. Stars in this mass range are important contributors to dust and chemical evolution in galaxies and are largely responsible for the production of the *slow* neutron capture process (Karakas & Lattanzio, 2014). The physics of stars in this mass range is highly uncertain owing to the lack of understanding of convective mixing and mass-loss. AGB stars are bright, long-period pulsators and can be seen at large distances out to 1 Mpc and beyond (Menzies et al., 2019). They are hence useful probes of young to intermediate-age stellar populations in different physical environments and of on-going nucleosyntheis (e.g. Shetye et al., 2018; Karinkuzhi et al., 2018b).

MSE's exquisite high-resolution capabilities and wavelength coverage will make it possible to provide abundances of a wide range of elements heavier than iron, and hence bring new constraints on nucleosynthesis in low-mass stars. Currently, the quality of AGB spectra taken at 4- or 8-m facilities in the energy range required for precision abundance diagnostics, is compromised owing to the faintness of these stars in the blue. With its wide aperture, MSE will overcome these limitations.

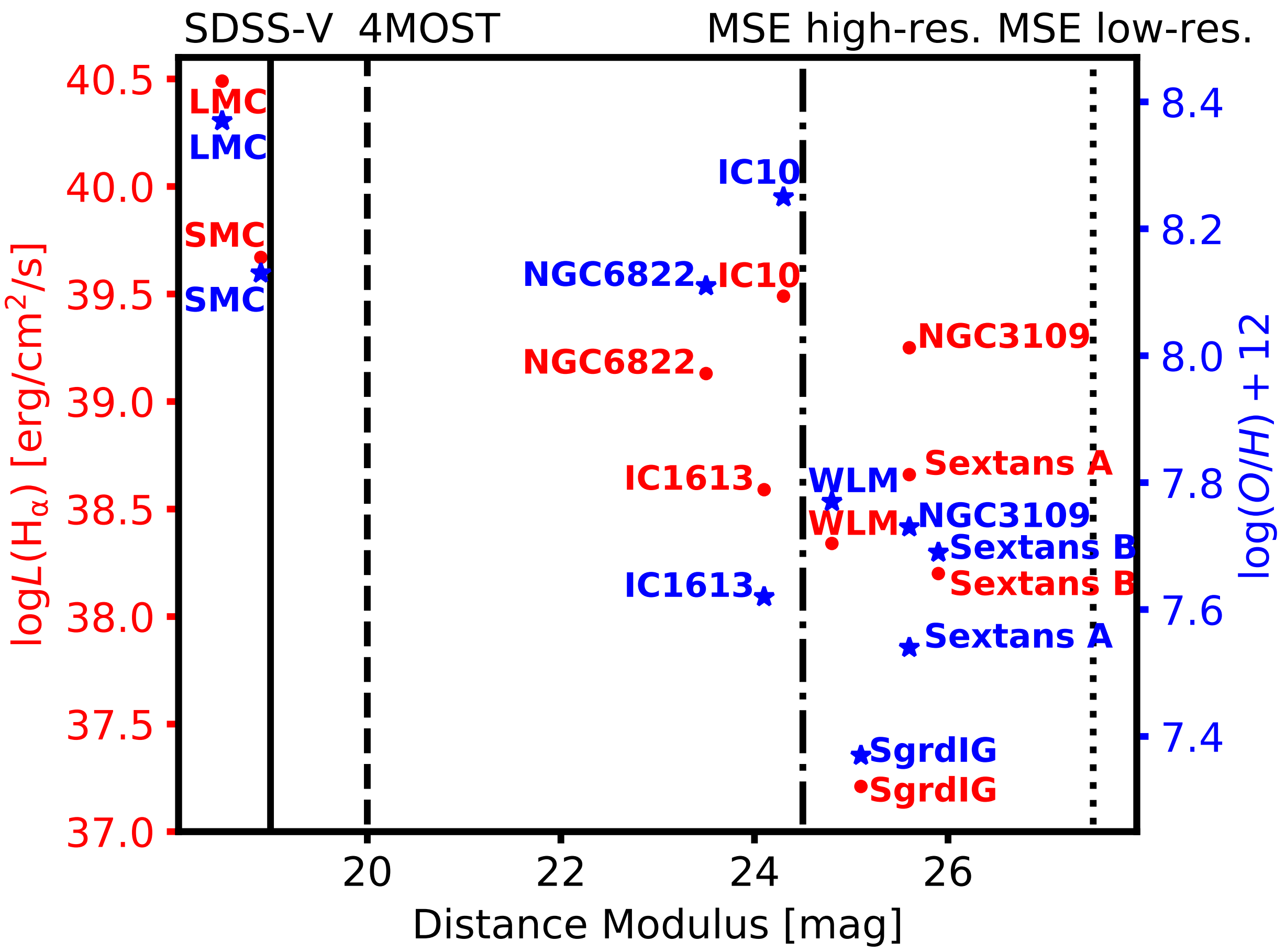

*Figure 28: MSE will enable the spectroscopic characterization of massive stars in local group dwarf galaxies with unprecedented completeness, providing new insights into the progenitors of compact object mergers. Vertical black lines indicate the distance limits of different surveys for a* $15M_{\odot}$ *main sequence star (O9.5V,* $M_{\rm V} \approx -4$ *mag). Red circles and blue stars show the* $\rm H_\alpha$ *luminosity (*$L(\rm H_\alpha)$*, a proxy for the number of expected massive stars) and Oxygen abundance (*$\log(O/H)+12$*, an indicator for metallicity) as a function of distance modulus. Distances and* $\rm H_\alpha$ *luminosities are adopted from Kennicutt et al. (2008), Oxygen abundances are taken from van Zee et al. (2006) (LMC, SMC, WLM, NGC 6822, NGC 3109, Sextans A, Sextans B and IC 1613), Tehrani et al. (2017) (IC 10) and Saviane et al. (2002) (Sagittarius dwarf irregular galaxy, SgrdIG).*

Also, post-AGB stars, the progeny of AGB stars, are exquisite tracers of AGB evolution and nucleosynthesis. During the brief post-AGB phase, the warm stellar photosphere makes it possible to quantify photospheric abundances for a very wide range of elements from CNO up to the heaviest *s*-process elements that are brought to the stellar surface during the AGB phase. Pilot studies of post-AGB stars (Kamath et al., 2014) have revealed that the objects display a much larger chemical diversity than anticipated (van Aarle et al., 2013; De Smedt et al., 2016; Kamath et al., 2017). Yet, the samples are small and heterogeneous, hence the element production in low-mass stars remains shrouded in mystery.

MSE will have the depth and sensitivity to collect large samples of the Galactic, LMC and SMC post-AGB stars, and to enable massive spectroscopic diagnostics of the key elements, including C/O, N, iron-peak and s-process in these rare sources. These data will constrain critical poorly-understood physics, such as binary evolution through Roche Lobe overflow, that can truncate evolution along the AGB (Kamath et al., 2015, 2016) or result in stellar wind accretion, affecting the surface composition of the companion star (e.g., produce a barium star or CH-type star) while leaving the AGB star intact. The sample will provide key insights into the physical properties of post-AGB stars in diverse environments, therefore constraining their role in chemical enrichment.

### 3.8 Very metal-poor stars

Stars in the halo system of the Galaxy with metallicities one thousand times lower than the Sun provide a direct touchstone with the nucleosynthesis products of the very first generations of stars. Such objects have been found in increasing numbers over the past few decades using surveys such as the HK survey, the Hamburg-ESO survey, SDSS/SEGUE, SkyMapper, and LAMOST (see, e.g. Beers & Christlieb, 2005; Yanny et al., 2009; Howes et al., 2015; Li et al., 2018). Ongoing and forthcoming surveys such as the Pristine Survey and the 4MOST surveys of the halo (Starkenburg et al. 2018, Christlieb et al. 2019, Helmi & Irwin et al. 2019 in press) promise to identify many more such stars (Youakim et al., 2017).

MSE will be the key next-generation facility to greately extend the areal coverage and the depth of the ongoing surveys and provide a high-resolution follow-up of the available candidates found in low-resolution. Of particular importance are the frequencies of the various known subsets of metal-poor stars, including the *r*- and *s*-process-enhanced stars, and the carbon-enhanced metal-poor (CEMP) stars, as a function of metallicity. Ongoing survey efforts (Hansen et al., 2018a; Yoon et al., 2018) have provided some information, but detailed understanding requires enlarging the samples by at least an order of magnitude, which can be readily accomplished by MSE.

Large number statistics of ultra-metal-poor stars with accurate chemical-abundance patterns is essential to reveal the range of nucleosynthesis pathways that were available in the early Universe, but also to provide a direct test of the importance of binary evolution of these systems (e.g. Arentsen et al., 2019). This will be a unique opportunity to probe the physical properties and mass distribution of the very first generations of massive stars (e.g. Kobayashi et al., 2014), precious information that will not be revealed in any other way.

# Chapter 4

# Chemical nucleosynthesis

**Abstract**

MSE is uniquely tailored to understanding the cosmic formation and evolution of the elements of the periodic table. It will trace different nucleosynthetic processes, sites and timescales through the measurement of a large number of chemical species, including in the crucial blue/UV region of the spectrum. It is ideally suited for detecting the EuII 4129Å line, as well as a large number of other neutron-capture elements, covering the full element mass range. MSE will study the *r*-process element abundances in unprecedented numbers of stars across our Galaxy. MSE will produce the definitive dataset of the most chemically primitive stars with which to identify the signatures of the very first supernovae and chemical enrichment events in the Universe. It will enable a large scale study of the lithium abundances down to the lowest metallicities to understand the possible depletion mechanism(s) of lithium due to the first generation of stars. The relative importance of low and intermediate mass stars to the chemical enrichment of the Universe will be quantified by MSE via a systematic and comprehensive chemical abundance study of large samples of evolved stars in diverse metallicity environments, covering a wide range of initial masses. MSE will measure the dimensionality of chemical abundance space using data for millions of stars across all Galactic components and sub-components.

**Science Reference Observations** (appendices to the *Detailed Science Case, V1*):
**DSC – SRO – 03** Milky Way archaeology and in situ chemical tagging of the outer Galaxy

### 4.1 Motivation: From BBN to B2FH

Some 13.8 billion years ago, the Universe began with the Big Bang. A few minutes later, all of the hydrogen in the Universe was formed, and it remains the most abundant atom. Big Bang Nucleosynthesis (BBN) then produced helium, the second most abundant atom in the Universe, along with traces of lithium. Big Bang Nucleosynthesis ended after 20 minutes.

Where, when, and how, were the remaining elements of the periodic table produced? The answer is in the stars. Those chemical elements heavier than lithium include the essential ingredients for planets and indeed life as we know it. As Carl Sagan famously wrote:

> The nitrogen in our DNA, the calcium in our teeth, the iron in our blood, the carbon in our apple pies were made in the interiors of collapsing stars. We are made of star stuff.

The basic framework for the synthesis of the chemical elements in stars was established by Hoyle (1946), Burbidge et al. (1957, hereafter B2FH), and Cameron (1957). B2FH identified eight nucleosynthetic processes: (1) hydrogen burning (2) helium burning (3) the $\alpha$ process (4) the $e$ (equilibrium) process (5) the $s$ (slow neutron capture) process (6) the $r$ (rapid neutron capture) process (7) the $p$ (proton) process (8) the $x$ process.

The exact environments (e.g., low-mass stars, high-mass stars, merging neutron stars) and when those events occur (e.g., during the main sequence evolution, in the dying stages etc.), however, continue to be debated. Figure 29 illustrates our current knowledge of the emergence of the Periodic Table. For each element, the relative contributions from different objects are indicated by distinct colors. What this picture lacks, however, is the time-dependence of the yields of the chemical elements. For example, while the majority of the carbon and barium atoms in the Sun are produced from dying low-mass stars, are we certain that this is also the case in the first 500 Myr after the Big Bang? Probably not. Indeed, many of the key details concerning the emergence of the elements of the Periodic Table remain unknown. Moreover, if we had both a more precise constraint on the origin of elements, and significantly, measurements of the composition of stars made to much greater precision, we might be able to discern with greater detail the dominant processes of the formation and assembly of the major populations of the Milky Way.

These broad studies of chemical and chemo-dynamical evolution continue to demand measurements of the highest precision on ever increasing samples of stars. Unfortunately, most of the large-scale spectroscopic surveys that are planned or contemplated are fundamentally limited, either by utilizing the infrared H-band (as is the case for SDSSV) or by using relatively low spectral resolution and SNR (e.g., 4MOST). For the former, heroic work is underway to derive detailed compositions of late-type stars for the iron-peak and $\alpha$-elements, but there are few, if any, suitable lines of heavy ($r$- and $s$-process) elements available in the infrared. Hence, optical surveys are required to answer one of the burning questions of our time: what is the origin of $r$-process elements? The MSE facility presently stands alone as a large aperture multi-object spectroscopic facility with sufficiently high spectral resolution in the optical passband to address this and other key questions relating to the origin and evolution of the chemical elements. In this chapter, we highlight key areas in which MSE will enable transformative progress in our understanding chemical nucleosynthesis.

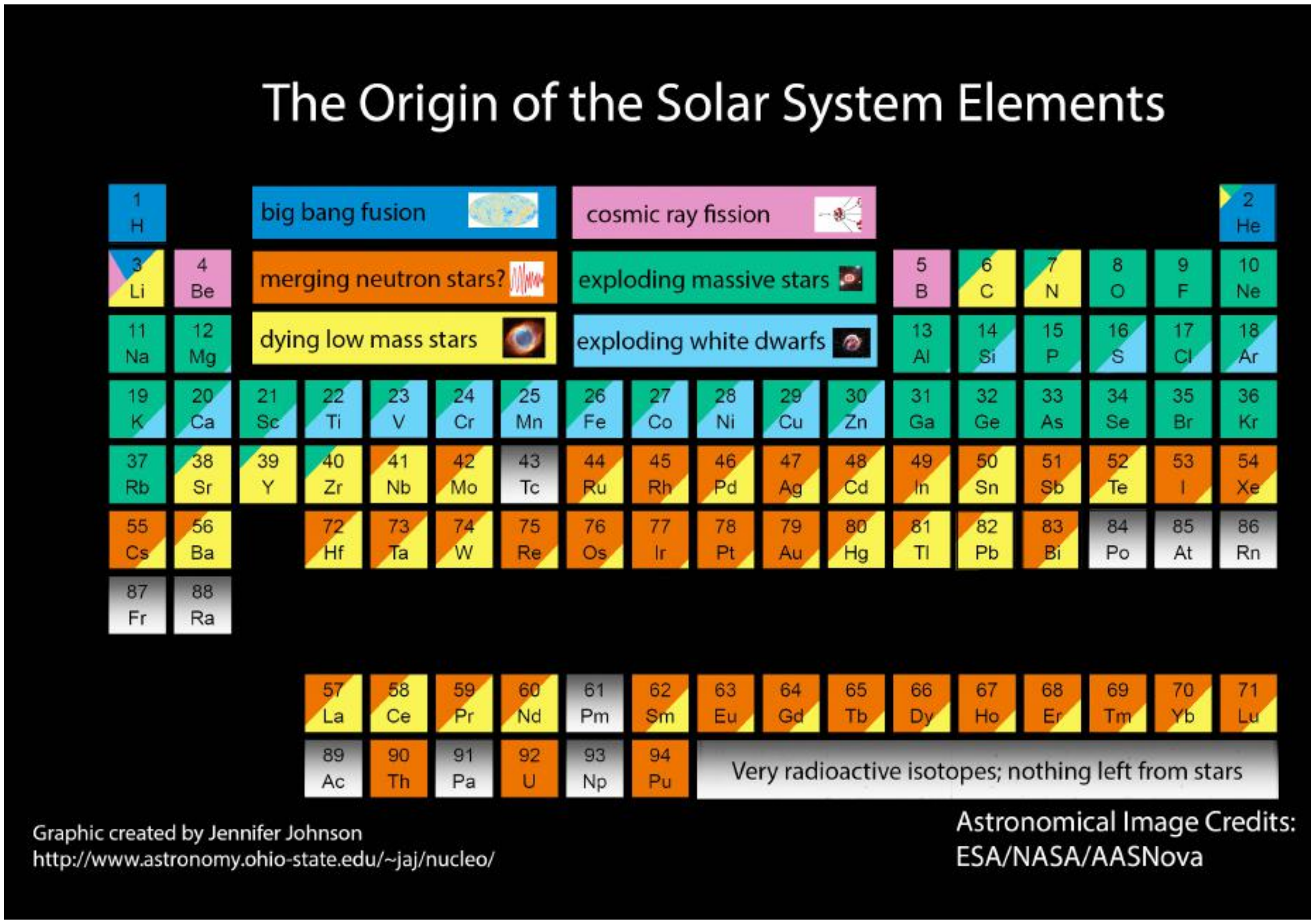

*Figure 29: The periodic table, color-coded to indicate the currently favored pathways of element production. (Figure from Jennifer Johnson,* `http://www.astronomy.ohio-state.edu/~jaj/nucleo/`*).*

### 4.2 Metal-poor stars and the first chemical enrichment events in the Universe

The very first generation of stars formed from the primordial material produced by Big Bang Nucleosynthesis, namely hydrogen, helium, and traces of lithium. Those first stars were likely much more massive than the sun ($M_\star > 10\,\mathrm{M}_\odot$), and have long-since died (Hirano et al., 2014). The chemical elements that they produced, and in particular the relative amounts of each element, depend upon the properties of the first stars, e.g., mass, explosion energy, rotation, mass cut, explosion mechanism. Given that they no longer exist, what is the most promising avenue to study the nature of the first stars?

The chemical elements produced by the first stars were returned to the interstellar medium from which a second generation of stars formed. Those second generation stars include some low-mass objects ($M_\star < 1\,\mathrm{M}_\odot$) which still live today and their atmospheres retain the chemical composition of the interstellar gas at the time and place of their birth. That is, the oldest and most chemically primitive stars are fossils, which contain the nuclear ashes of the first stars to be born in the universe.

These second generation stars, however, are rare. One way to identify such objects is by the paucity of "metals" (elements heavier than helium) in their atmosphere, using iron as a proxy. Only eight of these so-called metal-poor stars are known with less than 1/100,000th the solar iron-to-hydrogen ratio, [Fe/H] $< -5$, (Frebel & Norris, 2015); these are the most chemically ancient objects known. They exhibit an enormous range in their relative chemical abundance ratios, which demand a wide variety in the properties of the first supernovae, e.g., mass, explosion energy, rotation, mass cut, explosion mechanism (Nomoto et al., 2013). More data at the lowest metallicities are needed to better constrain the diversity of properties, and relative frequency, amongst the first supernovae.

With a large aperture, high spectral resolution, blue wavelength coverage, and enormous multiplexing capability, MSE provides an unparalleled competitive advantage in the discovery and analysis of metal-poor stars. By studying the chemical compositions and dynamics of unprecedented numbers of the most chemically primitive stars, we can probe the details and properties of the very first supernovae and chemical enrichment events in the Universe.

### 4.3 The cosmological lithium problem

Spite & Spite (1982) first noted that metal-poor stars show a constant lithium abundance, with a narrow scatter. Such a plateau in lithium abundance was expected if these abundances represented the primordial lithium that was produced during the Big Bang. However, the lithium abundance of the plateau was found to be $2-4$ times lower than the lithium abundance that was predicted from baryonic densities from CMB experiments (Fields, 2011). This is the so-called "Cosmological lithium problem".

Proposed solutions to this problem include:

1. Uncertainties in the nuclear cross-section and non-standard BBN (Jedamzik, 2004);
2. Stellar depletion of lithium due to diffusion and mixing process (Korn et al., 2006);
3. Uncertainties in the stellar parameters and stellar modelling (Meléndez & Ramírez, 2004);
4. Pre-galactic chemical enrichment and mixing, leading to lithium depletion (Piau et al., 2006).

Recently, experiments that measured the nuclear cross-sections at Big Bang energies (Anders et al., 2014) are consistent with the theoretical predictions.

In order to distinguish between depletion due to stellar processes (diffusion, convection, binarity) and depletion of ISM lithium abundances due to Population-III stars, it is crucial to have observations of pre-galactic or extra-galactic lithium abundances. The prospects for observing directly the lithium abundance in the proto-galaxy or in an external galaxy are extremely limited. Howk et al. (2012) measured interstellar lithium in the Small Magellanic Cloud (SMC). However, the metallicity of the SMC is not very low when compared to Galactic halo stars in which lithium has already been measured. MSE will enable us to solve the cosmological lithium problem by studying lithium abundances in halo stars across the entire Galaxy as well as pre-galactic lithium abundance in high velocity clouds.

**Lithium in high velocity clouds with MSE:** High Velocity Clouds (HVC) are neutral condensations of intergalactic gas that are falling towards the galaxy. HVCs are metal poor, and their deuterium abundances were found to be consistent with primordial value. Thus, these are pristine and primitive objects where one can observe deuterium and lithium in the same system (Prodanović & Fields, 2004).

Currently, such observations are challenging since a vast area of sky needs to covered and large numbers of objects need to be observed to identify stars which are behind HVCs. MSE is ideal for large sky coverage and unique with its R > 20K mode for this study. Ongoing and planned spectroscopic surveys on 4m class facilities with similar spectral resolution simply cannot reach the faint magnitudes of these stars. With precise distances from Gaia for stars in the HVC field, MSE will be the ideal instrument to survey the field of HVC clouds, and select suitable stars (e.g., BHB and hot stars free of stellar lines), that lie behind these clouds to derive the lithium abundances of the clouds.

**Lithium in outer halo stars with MSE:** Current observations of Galactic halo stars reveal a narrow plateau in lithium abundance down to [Fe/H] $\simeq -2.5$. Below this value, the scatter increases dramatically (Sbordone et al., 2010). This so-called "meltdown of the Spite plateau" represents another facet to the cosmological lithium problem.

In the context of lithium abundances in halo stars, it is worth noting that our understanding of the Galactic halo has evolved from (i) Eggen et al. (1962a) versus Searle & Zinn (1978) to (ii) inner and outer components (Carollo et al., 2007) to (iii) high- and low-$\alpha$ populations with distinct kinematics (Nissen & Schuster, 2010) to (iv) a major (early) merger with Gaia-Enceladus and multiple halo components (Helmi et al., 2018b; Myeong et al., 2018b). So it is important to recognise that halo stars of comparable metallicity may have been born in very different environments with different chemical enrichment histories.

With the availability of Gaia data and chemical abundances from MSE, all halo stars can potentially be classified into various populations. Of particular interest would be to trace metal-poor halo stars to individual progenitors (e.g., stripped dwarf galaxies) and look for differences in lithium abundance as a function of kinematics, i.e., environment. The combination of Gaia kinematics and MSE abundances will enable a study of Galactic halo stars (of unprecedented scope and scale) to disentangle (i) lithium depletion from stellar processes (diffusion, convection, binarity), (ii) lower ISM lithium abundances due to environment (accretion from dwarf galaxies vs. in-situ halo formation) and (iii) Li depletion due to Pop III stars.

### 4.4 The promise and potential of chemical tagging to probe the origin of the elements

Several recent works have demonstrated the potential of chemical tagging as a probe to study the nuclear origin of the elements and also the possible sites. Chemical tagging was originally envisaged as a tool to identify groups of stars dispersed across the sky but with common chemical compositions indicating common formation sites. With sufficient numbers of stars and element abundance measurements, chemical tagging will retrospectively identify those Galactic building blocks and enable "temporal sequencing of a large fraction of stars in a manner analogous to building a family tree through DNA sequencing" (Freeman & Bland-Hawthorn, 2002). This method, however, also provides an exceptionally powerful probe of the various possible nuclear processes and sites of element production in the Galaxy.

The application of chemical tagging requires the measurement of a large number of chemical species per star. However, exactly how many measurements are required, and which species are the most effective tracers, is a complex question. For example, two different species that are found to vary in lock-step with each other as a function of any other variable will provide considerably less discriminating power than two species whose behavior over a large sample of stars is less correlated. Clearly, the dimensionality of chemical abundance space is linked to the number of unique pathways by which chemical enrichment can take place.

Ting et al. (2012) presented a detailed study of the structure of chemical abundance space as a pathfinder study for the AAT/HERMES spectrograph and the GALAH survey (De Silva et al., 2015). They undertook a principal component analysis of the abundance space of stars observed in several different environments ranging from the solar neighbourhood to metal-poor halo stars. For the metal-poor stars ($-3.5 <$ [Fe/H] $< -1.5$ dex), they used a dataset of nearly 300 stars from Barklem et al. (2005) and the First Stars Survey (Cayrel et al., 2004; François et al., 2007; Bonifacio et al., 2009) for which there are measurements for 17 different chemical elements.

Ting et al. (2012) found seven to nine independent dimensions, and Figure 30 illustrates the first four principle components. Here, neutron capture elements are different shades of red (dark red represents lighter $s$-process elements, red represents heavier $s$-process elements and orange are mostly $r$-process elements), light odd-Z elements are dark brown, $\alpha$ elements are blue, iron-peak elements are green, and Cr and Mn are black. In full, they find that there are typically seven to nine dimensions in chemical abundance space for the solar neighbourhood and metal-poor stars, and that dwarf galaxies may have extra dimensions, perhaps associated with their longer star formation timescales:

- *The first principal component* in Figure 30 reveals the presence of a site that produces both $r$-process elements and $\alpha$ process elements (perhaps $r$-process core collapse SNe);

- *The second principal component* reveals the presence of a process that produces an anti-correlation between $\alpha$ elements with iron-peak and neutron-capture elements (perhaps normal core collapse SNe);

- *The third principal component* reveals a process producing an anti-correlation between

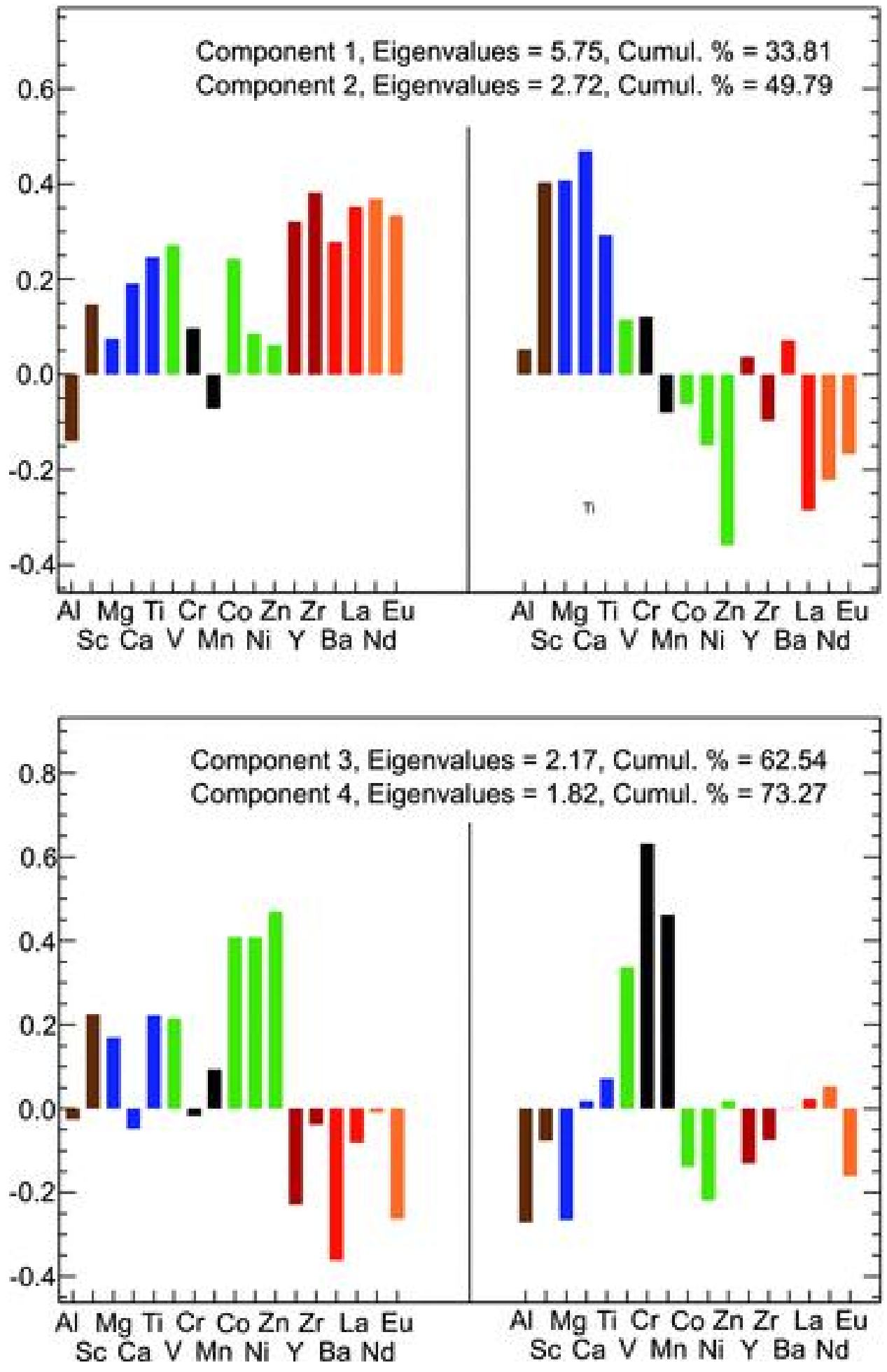

*Figure 30: The normalized principal components of 17 elements for the low-metallicity Barklem et al. (2005) sample. The upper plot and lower plots show the first two principal components and the third and fourth principal components, respectively. Figure from Ting et al. (2012).*

$\alpha$ elements and iron-peak elements with neutron-capture elements (perhaps hypernovae);

- *The fourth principal component* shows a strong contribution to Cr and Mn (both of which are synthesized in the incomplete Si-burning region).

In addition to providing insight into the chemical space that MSE will explore, the analysis of Ting et al. (2012) also shows the power of principal component analysis to chemical abundance studies. Hogg et al. (2016a) have also made important progress in chemical tagging by recovering known objects simply based on similarities in chemical abundance ratios. Ultimately, the dimensionality of chemical abundance space will be explored more fully by MSE using data for millions of stars across all components and sub-components, in contrast to the 300 stars in Figure 30. We return to discussion of chemical tagging in Chapter 5.

## 4.5 Sites of i-process and the origin of the CEMP-r/s stars

With increasing numbers of very metal-poor stars ([Fe/H] < −2), it has become clear that a large fraction exhibit high abundances of carbon relative to iron ([C/Fe] > 0.7), the so-called carbon enhanced metal-poor (CEMP) stars (Beers & Christlieb, 2005; Placco et al., 2014). Moreover, CEMP stars exhibit a large scatter in their abundance patterns, particularly for the heavy elements, atomic numbers Z > 30, (Beers & Christlieb, 2005).

Among the CEMP stars is a subclass that exhibit enhancements for both the rapid neutron-capture process elements and the slow neutron-capture process elements, *r*- and *s*-process, respectively. The origin of these CEMP-r/s stars, is particularly puzzling. The most common scenario to explain the peculiar abundance pattern of CEMP-r/s stars is that these objects started as *r*-process enhanced stars. Later, *s*-process enriched material was added by a binary companion during the asymptotic giant branch (AGB) phase. While this simple model may be able to explain some of the CEMP-r/s stars, it has now become clear that it cannot explain the chemical abundance patterns of many of the CEMP-r/s stars (Jonsell et al., 2006; Lugaro et al., 2012; Abate et al., 2015).

On the other hand, the intermediate neutron-capture process (*i*-process), with typical neutron densities in between that of the $s-$ and *r*-process (Cowan & Rose, 1977) has been recently shown to provide a promising explanation for CEMP-r/s stars. One-zone parametric studies have shown (e.g., see Figure 31) that the *i*-process leads to an excellent match to the observed heavy element abundance pattern in most CEMP-r/s stars (Hampel et al., 2016). This has led to the suggestion that CEMP-r/s stars should in fact be re-classified as CEMP-i stars.

The site of the *i*-process, however, is still under debate. Within the binary mass transfer scenario for CEMP-r/s stars, several sites have been proposed. Most of these sites involve ingestion of protons in He-rich layers in low- to intermediate-mass stars during the late stages of stellar evolution (Fujimoto et al., 2000; Campbell et al., 2010; Herwig et al., 2011; Jones et al., 2016) or during rapid accretion on a white dwarf (Denissenkov et al., 2017). Alternatively, the *i*-process can also occur in convective He shells of massive stars ($M_\star \sim 20-30\,\mathrm{M}_\odot$ which can pollute the ISM at early times (Banerjee et al., 2018). In this scenario, the low-mass star observed today is born directly as a CEMP-r/s star which need not be in a binary configuration. This is in sharp contrast to all other scenarios that involve surface pollution via mass transfer from the binary companion. MSE will enable large-scale long-term radial velocity monitoring campaigns to check for radial velocity variations (i.e., binarity) of CEMP-r/s stars which is essential for shedding more light on their possible origin(s). More details on science with long-term stellar radial velocity monitoring campaigns can be found in Chapter 3.

Recently, a very interesting CEMP-r/s star RAVE J094921.8-161722 (Gull et al., 2018) has been discovered. The simple scenario of a superposition of *r*- and *s*-process mentioned gives a perfect fit for the observed abundance pattern in this star (see Figure 32). This star is in fact the first bonafide *r*+*s* star. The detection of key elements that confirm this are Th, Os, and Ir that are derived from the *r*-process, while elements such as Ba and Pb are due to the *s*-process. In particular, Th can only be made via the *r*-process and its abundance in this star agrees very well with the scaled solar ***r***-process abundance (when normalized to Eu along with the 2nd ***r***-process peak elements Os and Ir). Future detection of ***r***-process elements such as Os, Ir, and Th will be crucial in discriminating CEMP-r/s from CEMP-i stars.

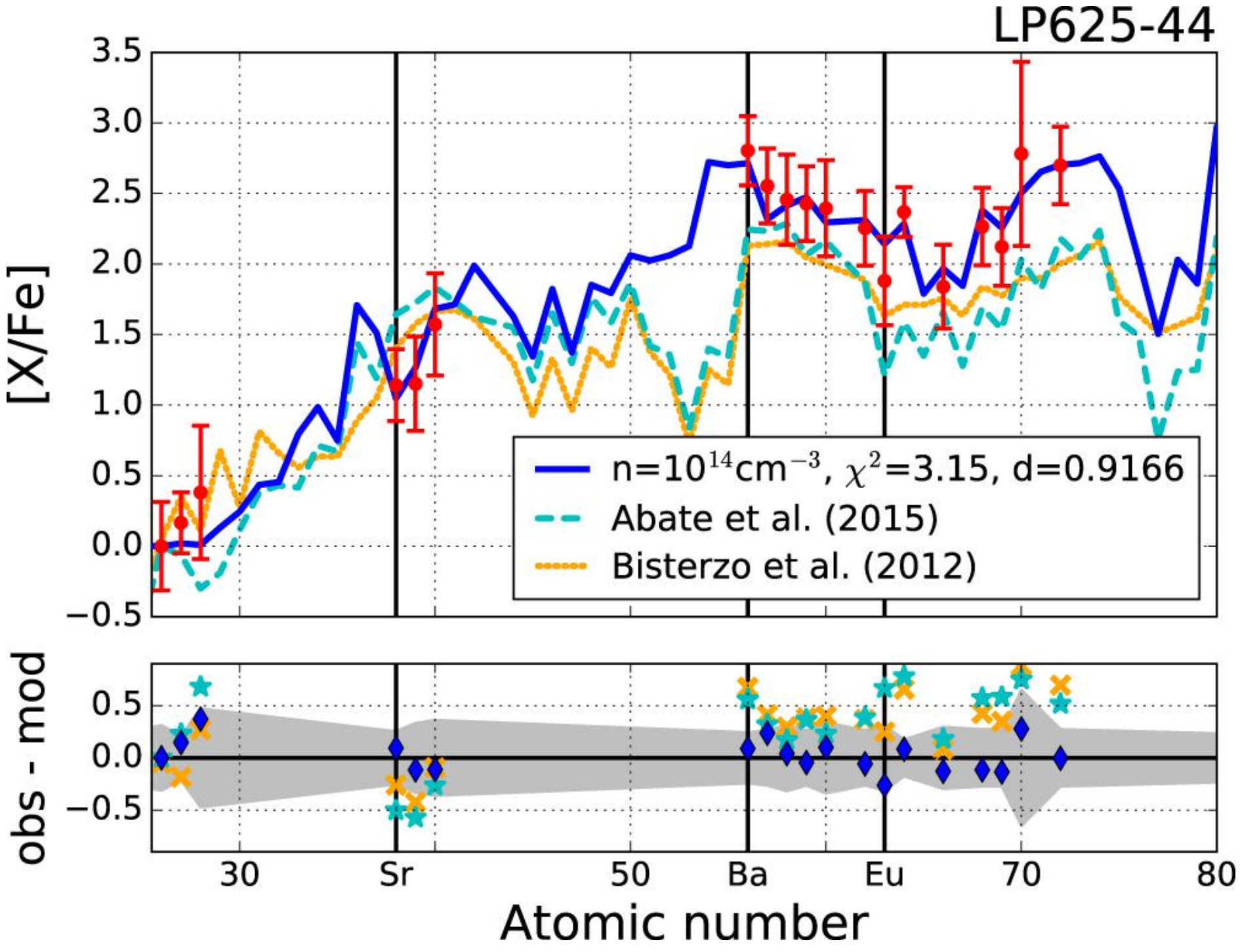

*Figure 31: Best-fitting model (blue) for CEMP-s/r star LP625-44 (red dots) using the parametric nucleosynthesis calculations with a neutron density of $n = 10^{14}$ cm$^{-3}$. Also shown is the best fitting model from Abate et al. (2015) with AGB nucleosynthesis (cyan), and the* **s***-process model from Bisterzo et al. (2012) with an initial* **r***-process foundation of [r/Fe] 1.5 (orange). The models from Abate et al. (2015) are an excellent match to most of the CEMP-s stars, which are the product of mass transfer with a previous AGB companion. Figure from Hampel et al. (2016).*

### 4.6 Studying AGBs and their progeny

Stars with the initial mass from $0.8 - 8\,M_{\odot}$ evolve through the asymptotic giant branch (AGB) phase. The AGB population in the Galaxy and Magellanic Clouds is rich in variety, due to their range in initial mass which leads to an enormous range in age and therefore metallicity (e.g., from 12 billion year old AGB stars in globular clusters to young, relatively massive OH-IR stars in the disk of the Galaxy). The progeny of AGB stars include post-AGB stars and planetary nebulae, which can be observed out to great distances. Furthermore, binary mass transfer with an AGB star can lead to a zoo of different stellar types including CEMP-***s*** stars, CH and barium stars, R-type stars, and R Coronae Borealis stars.

AGB stars are present in the halo, thick disk, thin disk, bulge and also among star clusters.

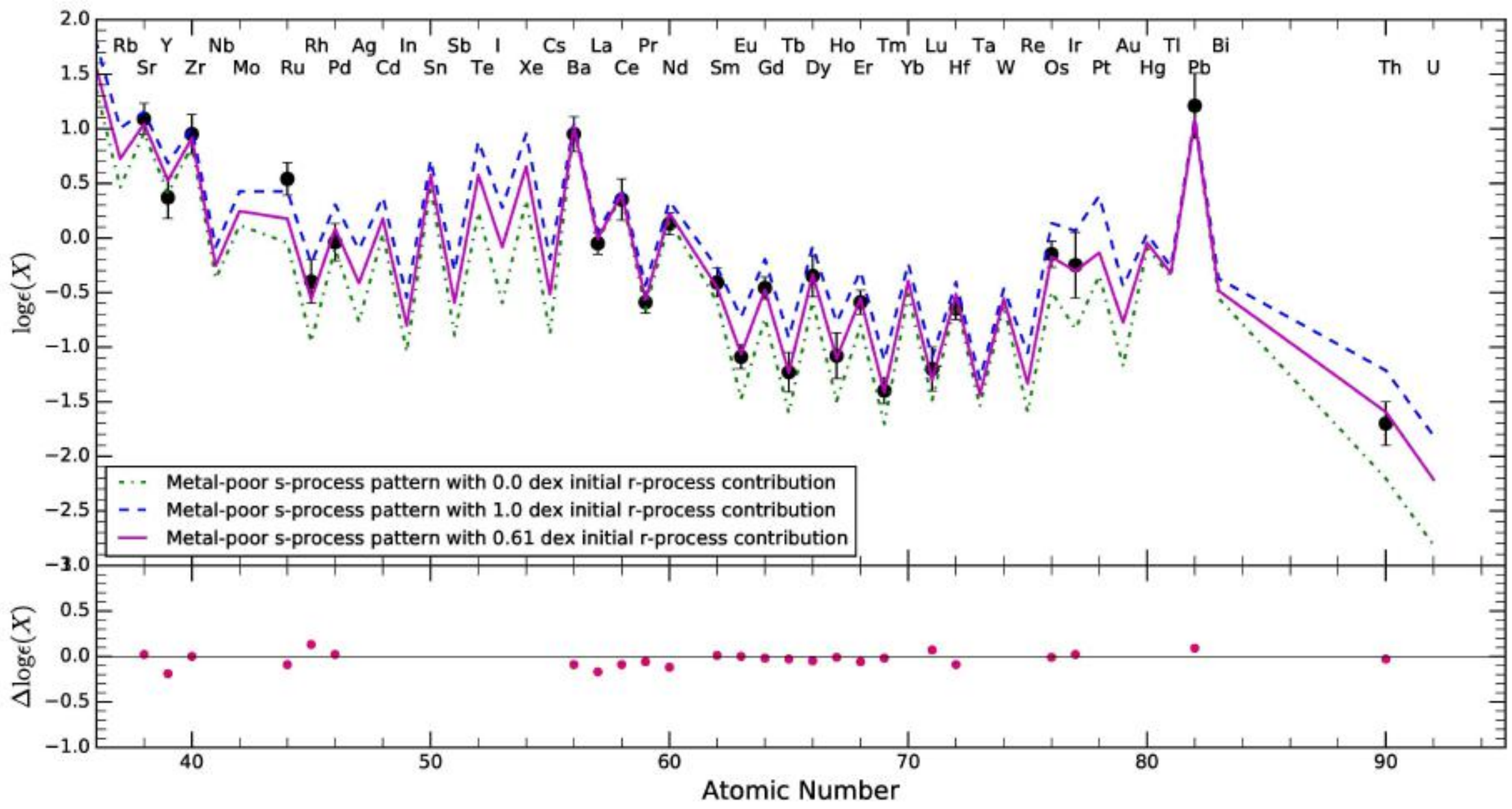

*Figure 32: RAVE J094921.8-161722 abundances in comparison with results from three metal-poor s-process models. The best fit (magenta line) is achieved with an s-process model combined with an initial r-process component of [Eu/Fe] = +0.6. The other models have r-process contributions of [Eu/Fe] = +0.0 (blue line) and [Eu/Fe] = +1.0 (green line). Residuals (i.e., the difference between observations and the best-fit model) are shown in the bottom panel. Figure from Gull et al. (2018).*

They go through rich nucleosynthesis during shell burning phases of hydrogen and helium along with complex mixing processes and mass loss to thereby contribute to the chemical enrichment of the Galaxy (see review by Karakas & Lattanzio, 2014).

Although the last few years have seen tremendous progress as far as Galactic chemical evolution is concerned, large gaps still remain in our understanding of the astrophysical sites and the production mechanism of many elements and their isotopes, in particular the neutron-capture elements (*s*-process and *r*-process). This is partially because it is difficult to directly observe elemental abundances in the stars making heavy elements. AGB stars are notoriously difficult to study in detail, with their cool extended dynamic atmospheres dominated by molecules.

Recent progress has come about in two ways. First, the computation of large grids of model atmospheres of cool giants, combined with accurate atomic data had led to more precise abundances of chemical elements. Second, large surveys of post-AGB stars in the Magellanic Clouds (Kamath et al., 2015), the progeny of AGB stars, and follow-up high-resolution observations have revealed some of the most heavily enriched objects known to date (e.g., De Smedt et al., 2012). While post-AGB stars are rare, the warm stellar photosphere makes it possible to quantify photospheric abundances for a very wide range of elements from CNO up to some of the heaviest *s*-process elements including Pb that are produced during the AGB phase. Therefore post-AGB stars can provide direct and stringent constraints on the parameters governing stellar evolution and nucleosynthesis, especially during the uncertain AGB phase.

Recent examples that showcase the improvements in model atmospheres include the study of S-type stars using the latest MARCS model atmospheres by Van Eck et al. (2017); and the dynamical model atmospheres used to provide more accurate Rb abundance determinations by Zamora et al. (2014). Examples that showcase the confrontation between theory and observations have mostly come from using the progeny of AGB stars: post-AGB stars or barium/CH-type stars. Barium and CH type stars are a higher metallicity link to the CEMP-*s* stars, which are found in the Galactic halo. The studies by Neyskens et al. (2015) and Karinkuzhi et al. (2018a) use the Zr – Nb ratio to test thermodynamic conditions of the *s*-process after mass transfer has occurred, while Cseh et al. (2018) compare a large sample of barium star data to AGB theoretical predictions from different groups. The main conclusion is that current AGB models with rotation are a poor match to the observations of the ratio of light-*s* (Y, Sr, Zr) to heavy *s* elements (La, Ce, Nd).

To clarify element production in low- and intermediate-mass stars requires a systematic and comprehensive chemical abundance study of large samples of evolved stars in diverse metallicity environments, covering a wide range of luminosities (or initial masses). MSE's high-resolution spectroscopic mode will provide the required spectral resolution and spectral coverage to investigate the C/O ratio, abundances of C, N, $\alpha$-elements, iron-peak elements and neutron-capture element in Galactic AGB and post-AGB stars. MSE + Gaia will complement and optimise current studies with UVES/VLT, MIKE/Magellan, and GALAH + Gaia and will allow for an efficient exploitation of the Galactic and close-by extragalactic sources. The ultimate goal of this study is to quantify the contribution of low and intermediate-mass stars to the chemical enrichment of the Universe.

### 4.7 Survey of *r*-process elements

Half of the neutron-capture elements in the Sun are primarily made in the rapid neutron-capture process (*r*-process). The astrophysical site (or sites) producing these elements is still a subject of intense debate. Theoretically, the high neutron fluxes required to synthesize the full *r*-process, up to and including the actinides, limit sites to the births or deaths of neutron stars. The expected sites include rare types of core-collapse supernovae (e.g., Winteler et al. 2012; Siegel et al. 2018) and neutron star binary or neutron star-black hole mergers (e.g., Lattimer et al. 1977; Metzger et al. 2010). Additionally, sites with lower neutron fluxes such as normal core-collapse supernovae may contribute in the light *r*-process element range, up to Ba (e.g., Arcones & Montes, 2011; Wanajo, 2013).

The detailed study of stellar abundances has provided several key insights about the *r*-process site: the *r*-process occurs relatively early, enriching metal-poor stars with [Fe/H] $\sim -3$ (Sneden et al., 1994; Hill et al., 2002; Frebel et al., 2007); *r*-process nucleosynthesis appears to be ubiquitous in many environments (e.g., Roederer et al., 2010; Roederer, 2013, 2017); *r*-process element production must be rare compared to core-collapse supernovae (e.g., Ji et al., 2016; Macias & Ramirez-Ruiz, 2018).

The recent direct detection of a kilonova afterglow following a gravitational wave signal of a binary neutron star merger has clearly established that neutron star mergers *do* produce copious amounts of *r*-process elements (Abbott et al., 2017c,b). However, neutron star mergers still face chemical evolution challenges, likely requiring additional prolific *r*-process sites or significant adjustments to our understanding of compact object formation or merger physics (e.g., Argast et al., 2004; Cescutti et al., 2015; Wehmeyer et al., 2015; Belczynski et al., 2018; Côté et al., 2018; Safarzadeh et al., 2018; Siegel et al., 2018).

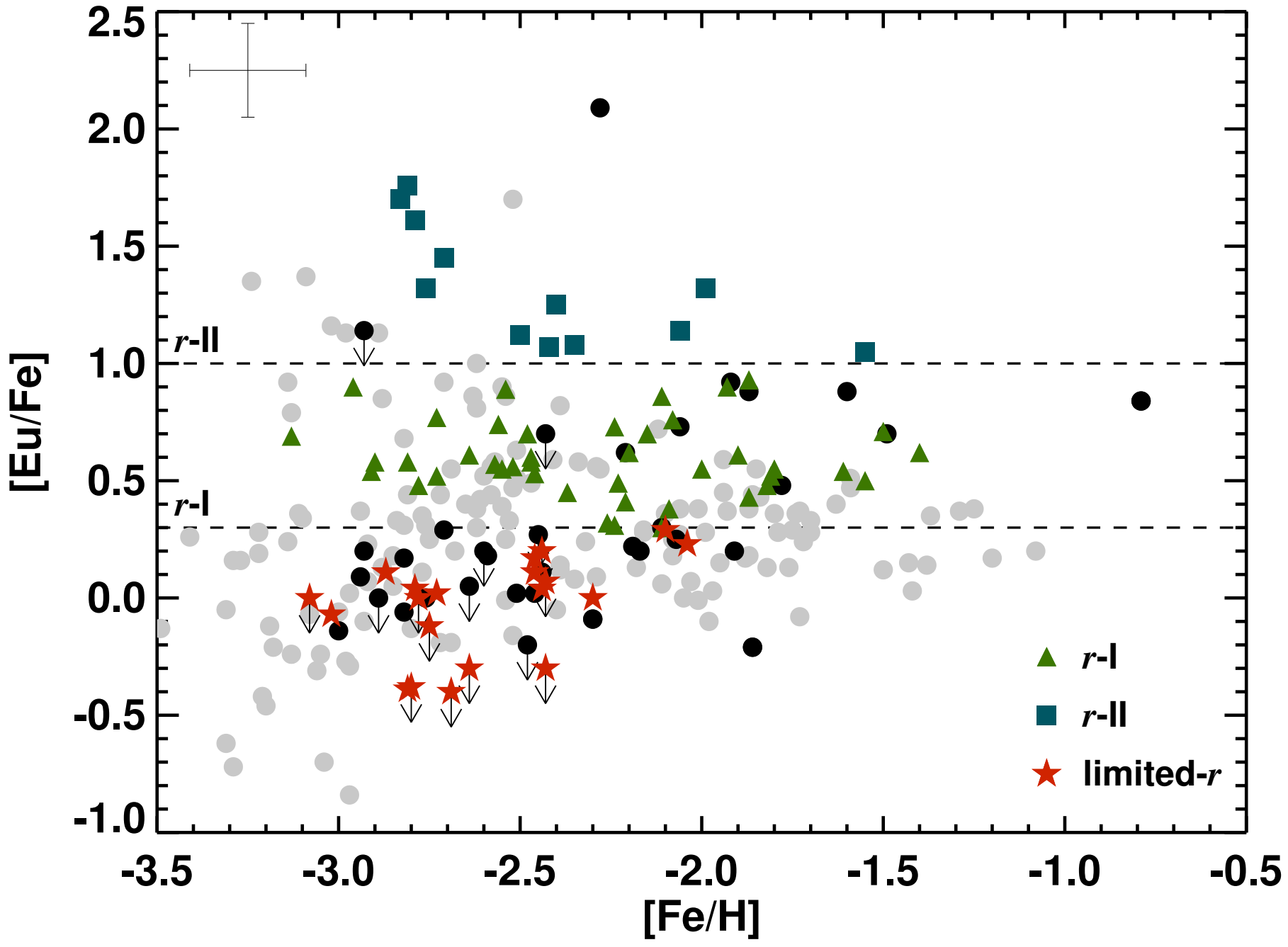

*Figure 33:* [Eu/Fe] *as function of metallicity for halo stars that highlights the large spread in Eu abundances observed at low metallicity. Figure from Hansen et al. (2018b).*

Furthermore, the variations in the observed *r*-process pattern of stars needs to be accounted for. Different types of *r*-process element enhancement have been detected in stars (see Figure 33). One group (*r*-I and *r*-II stars) exhibit enhancement in the heavy *r*-process elements (like Eu) while another (limited-*r* stars) exhibit larger abundances for the light neutron-capture elements like Sr, compared to the heavier species (Barklem et al., 2005; Hansen et al., 2018b). For the *r*-I and *r*-II stars the abundance pattern from barium to iridium, spanning the second through third *r*-process peaks including the rare earth elements/lanthanides, has been found to be Universal across metallicity and environment (e.g., Hill et al., 2002; Sneden et al., 2003, 2008; Hansen et al., 2017) (see Figure 34). Simultaneously a large scatter is observed in these stars for the first peak *r*-process elements (e.g., Sr, Y, and Zr) relative to the lanthanides (e.g., McWilliam, 1998; Ishimaru & Wanajo, 1999; Hansen et al., 2012; Siqueira Mello et al., 2014). Also, the abundances for the radioactive actinide elements Th and U vary greatly from star to star (Hill et al., 2002, 2017; Holmbeck et al., 2018; Ji & Frebel, 2018). Of the limited-*r* stars only a couple have been analysed in detail (e.g., Honda et al., 2006; Roederer et al., 2014). Thus the detection of common abundance features for this group of stars, which can constrain the *r*-process responsible for their abundance signature, still awaits homogeneous analysis of larger samples. Understanding the origin of the *r*-process not only helps complete our understanding of the periodic table, but *r*-process elements provide key constraints on subjects such as stellar age dating with radioactive Th and U, and galaxy formation across different environments.

The key challenge in our understanding of the *r*-process elements is the still relatively small sample sizes of stars with a significant number of *r*-process element measurements. The strongest *r*-process element lines are only identifiable with high-resolution spectroscopy at blue wavelengths (e.g., the EuII 4129Å line). Past and current surveys have identified a number of stars with detectable *r*-process elements (e.g., Barklem et al., 2005; Hansen et al., 2018b; Sakari et al., 2018). To date, stars have been individually targeted for followup. However, disentangling the many *r*-process sources, the extent of variation within those sources, and understanding the role of potential contaminants will ultimately require growing the samples by at least an order of magnitude. With an 11m aperture and spectral resolution of $R$ >20K, MSE is ideally suited for detecting the EuII 4129Å line, as well as a large number of other neutron-capture elements, covering the full *r*-process element mass range including the actinide element Thorium. MSE will study the *r*-process element abundances in unprecedented numbers of stars across our Galaxy.

### 4.8 Nucleosynthesis and chemical evolution in dwarf galaxies

Many questions about nucleosynthesis are studied with stellar abundances of Milky Way stars. But with a few notable exceptions (e.g., post-AGB stars, the most iron-poor stars), observed stellar abundances convolve nucleosynthetic sources with galaxy formation and evolution processes. Fortunately, the Milky Way's dwarf satellite galaxies provide collections of stars with independent chemical enrichment histories, allowing study of how nucleosynthesis varies with galaxy formation history and environment (e.g., Tolstoy et al., 2009). Most importantly, the lower star formation efficiencies in these systems amplify the impact of time-delayed nucleosynthetic sources, such as AGB stars (e.g., Venn et al., 2004, 2012), Type Ia supernovae (e.g., Kirby et al., 2011; McWilliam et al., 2018; Hill et al., 2018), and neutron star mergers (e.g., Ji et al., 2016; Duggan et al., 2018). The most metal-poor stars in dwarf galaxies are an ideal place to search for signatures of the first Pop III stars (Ji et al., 2015; Salvadori et al., 2015; Magg et al., 2018), and provide a coherent environment to understand the building blocks of the Milky Way's metal-poor stellar halo (Frebel & Norris, 2015). Dwarf galaxy stars also provide a unique window into variations in the initial mass function with environment (McWilliam et al., 2013; Geha et al., 2013; Carlin et al., 2018).

The key challenge for studying nucleosynthesis in dwarf galaxies has been low statistics due to their relatively large distances ($30 - 300$ kpc). The brightest red giants in these galaxies typically have $V > 17$, requiring hours of integration of the largest telescopes and significant multiplexing for statistical coverage of these systems. Moderate resolution spectroscopy ($R > 6000$) is sufficient to obtain many abundances of relatively metal-rich stars ([Fe/H] $\gtrsim -2$). As metallicity decreases and lines get weaker, higher resolution is needed to obtain abundance information about the stars, and many usable lines shift to bluer wavelengths (e.g., Kirby et al., 2011; Duggan et al., 2018; Hill et al., 2018). Wide field coverage of dwarf galaxies is also important: most spectroscopic studies to date have focused on the innermost regions of dwarf galaxies, but there appear to be abundance gradients and different chemodynamical populations in dwarf galaxies (e.g., Koposov et al., 2011; Kordopatis et al., 2016). The ideal facility to study dwarf galaxies are thus wide field, high-resolution, multi-object spectrographs with blue coverage on large aperture telescopes, exactly following the specifications of MSE's high-resolution mode.

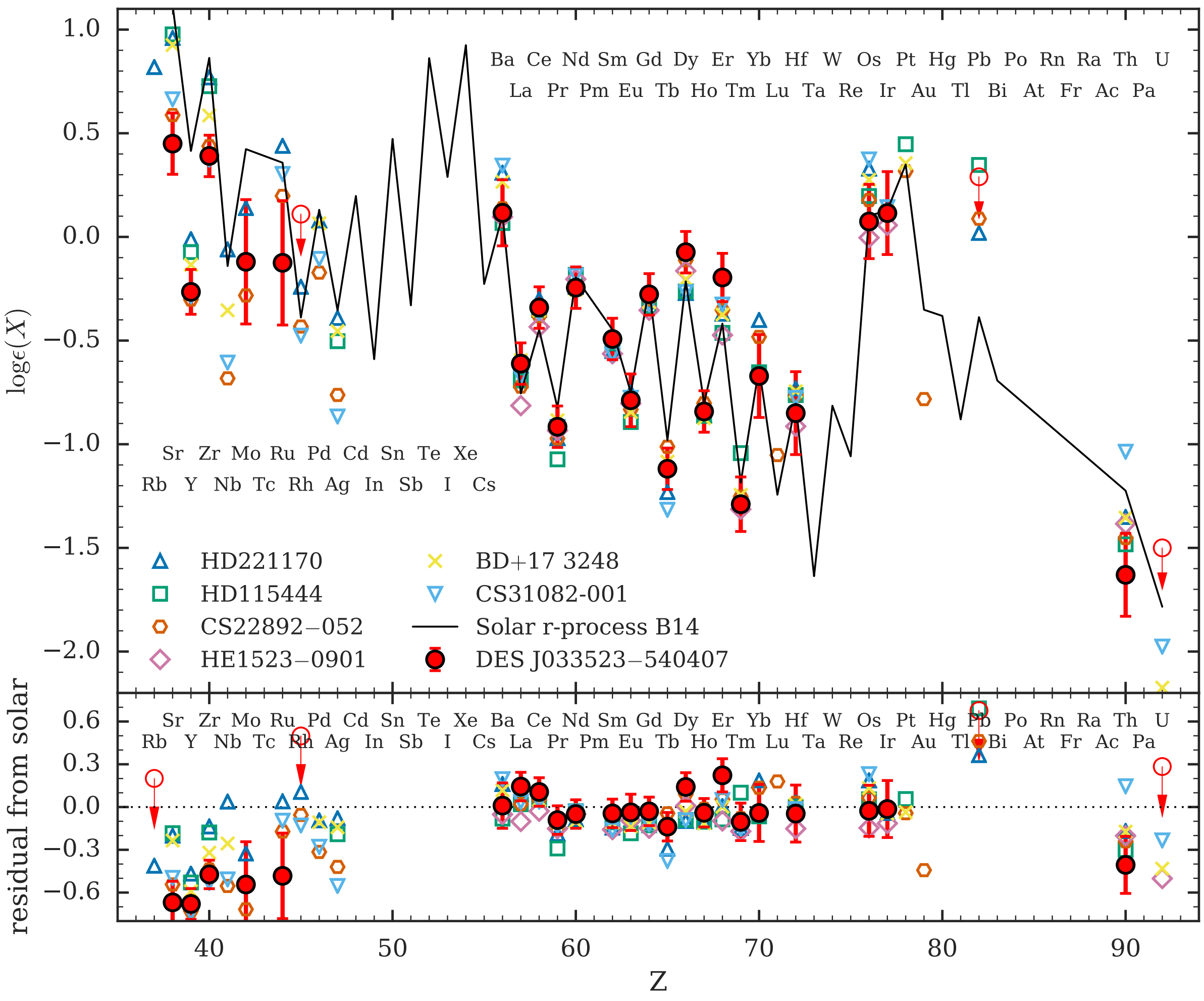

*Figure 34: Top panel: Neutron-capture element abundances in seven r-process-enhanced stars compared to solar r-process component (B14, Bisterzo et al., 2014) and six well-studied r-process enhanced stars (Sneden et al., 2008). Bottom panel: residuals relative to the solar r-process component. The abundance scatter in the first peak and actinides vary significantly more than the universal (main) r-process pattern from Ba-Ir. Figure from Ji & Frebel (2018).*

# Chapter 5

# The Milky Way and resolved stellar populations

**Abstract**

MSE will carry out the ultimate spectroscopic follow-up of the Gaia mission, and is critical to our understanding of the faint and distant regimes of the Galaxy. It is the only facility capable of producing vast high resolution spectroscopic datasets for stars across the full magnitude range of Gaia targets. Uniquely, MSE will conduct in situ chemodynamical analysis of individual stars in all Galactic components, searching for inter-relationships between them and for departures from equilibrium. The unprecedented size of the stellar spectroscopic dataset will enable the definitive analysis of the metal-weak tail of the halo metallicity distribution function. MSE will bring about an entirely new era for nearby dwarf galaxy studies, enabling accurate chemo-dynamical measurements to be performed efficiently across the full range of dwarf galaxy luminosities ($10^{3-7}L_{\odot}$), and providing spectra for at least an order of magnitude more stars in each system, reaching well beyond where circular velocity curves are expected to peak. MSE will also provide a comprehensive understanding of the chemodynamics of M31 and M33, essentially enabling a full chemodynamical deconstruction of these galaxies across their entire spatial extent. Finally, MSE will play a central role in revolutionary three dimensional ISM mapping experiments that will be boosted by Gaia parallax distances.

**Science Reference Observations** (appendices to the *Detailed Science Case, V1*):
**DSC – SRO – 03** Milky Way archaeology and in situ chemical tagging of the outer Galaxy
**DSC – SRO – 04** Stream kinematics as probes of the dark matter mass function around

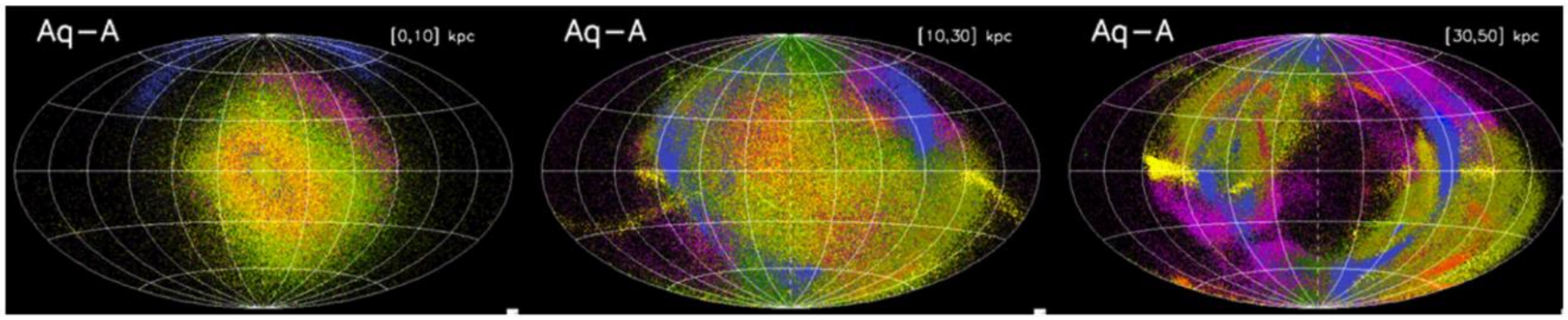

*Figure 35: Simulation of the distribution of field red giant branch stars on the sky at various distances from the Sun for the stellar halo of the simulation Aq-A. The different colours correspond to stars originating in different progenitors. Different progenitors cannot be distinguished between 10 and 30 kpc in photometric surveys alone, and velocities and chemical abundances are essential to reveal the individual events. Figure from Helmi et al. (2011).*

the Milky Way
**DSC – SRO – 05** Dynamics and chemistry of Local Group galaxies

### 5.1 Context: Galactic archaeology in the era of Gaia

Our Galaxy provides the most important laboratory available for exploring galaxy chemo-dynamical evolution. Eggen et al. (1962b) were the first to show that stellar abundances and kinematics could be used to understand the formation and evolution of our Galaxy, and thus guide ideas of galaxy formation in general. Indeed, their analysis of 221 very nearby stars remains, arguably, the single most influential observational paper on galaxy formation. In it, they proposed that metal-poor stars in the halo of the galaxy were formed during the rapid collapse of the protocloud that eventually became the Milky Way. An alternative proposal was offered by Searle & Zinn (1978), whose analysis of the stellar populations of a number of Galactic globular clusters led them to infer that they were formed in independent "protogalactic fragments" that later assembled the outer parts of the Galaxy. Key aspects from both of these original scenarios are naturally contained in the current paradigm of hierarchical structure formation (Davis et al., 1985), in which galaxy formation takes place through the condensation of gas in dark matter halos to form proto-galaxies eventually merging to form larger systems (White & Rees, 1978).

These fundamental papers demonstrated that the chemical and dynamical characteristics of the early proto-Galaxy, its constituent building blocks, and the subsequent evolutionary processes that have produced the Milky Way (and its subcomponents) can be explored through the chemodynamical studies of their present-day stellar populations. Contemporary studies of the abundances and dynamics of stars in the Galaxy show that this approach is powerful (see Freeman & Bland-Hawthorn, 2002). Stars which were long accreted into the Milky Way and are now fully phase-mixed can readily be identified in the space of integrals such as energy and angular momentum (e.g. Helmi & de Zeeuw, 2000).

Figure 35 shows a realization of a simulated stellar halo in a Milky Way-like Galactic halo taken from the Aquarius simulations (Springel et al., 2008). Depending on the distance regime being examined, orbital timescales can vary from a few 100 Myr to Gyrs. There

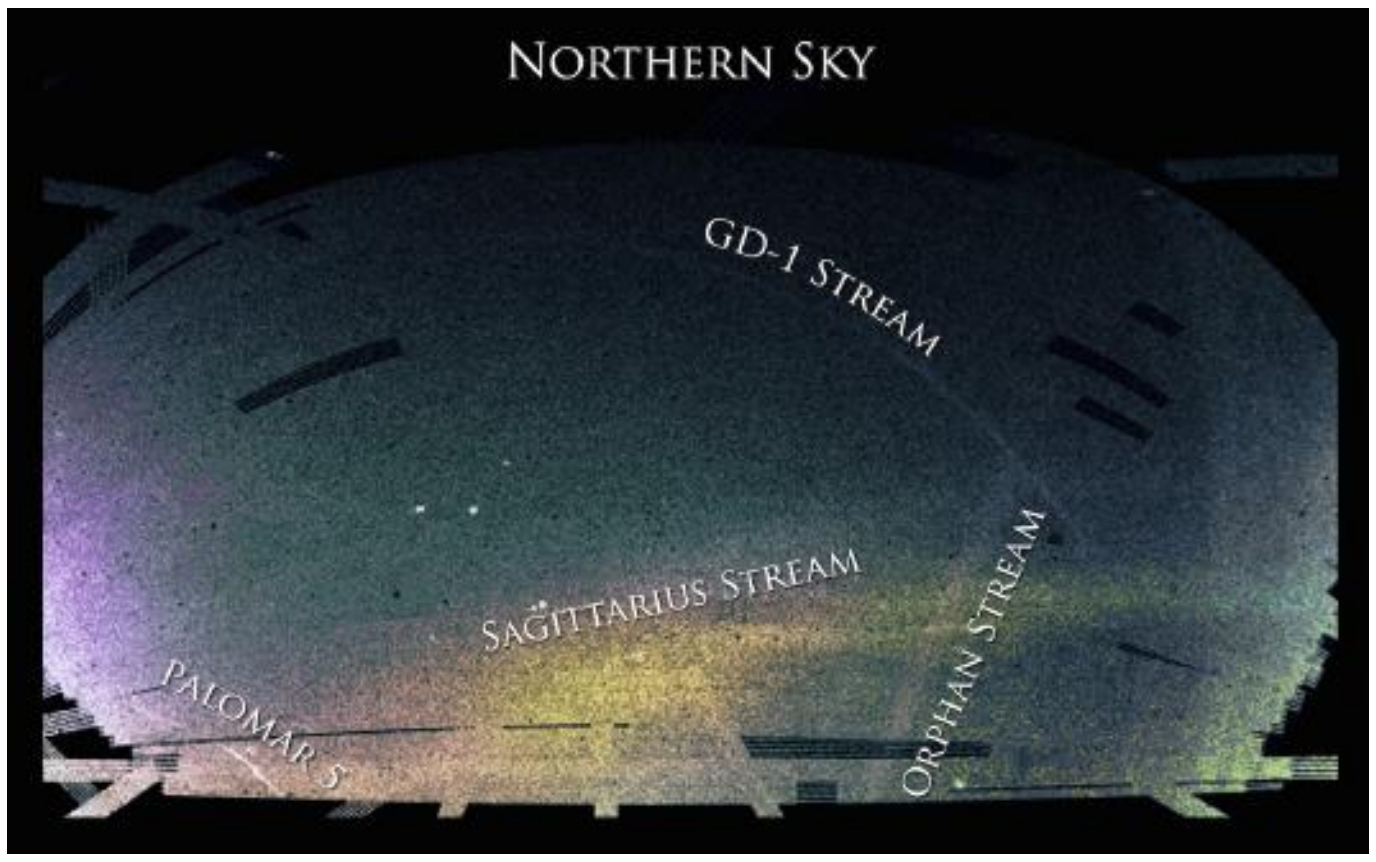

*Figure 36: Stellar density map of the Milky Way halo in Celestial coordinates (right ascension increases to the left, declination increases to the top) in the footprint of the SDSS imaging survey in the Northern sky. A matched filter analysis reveals a large number of stellar streams and other substructures, the most prominent of which are labeled. Figure from Bonaca et al. (2012).*

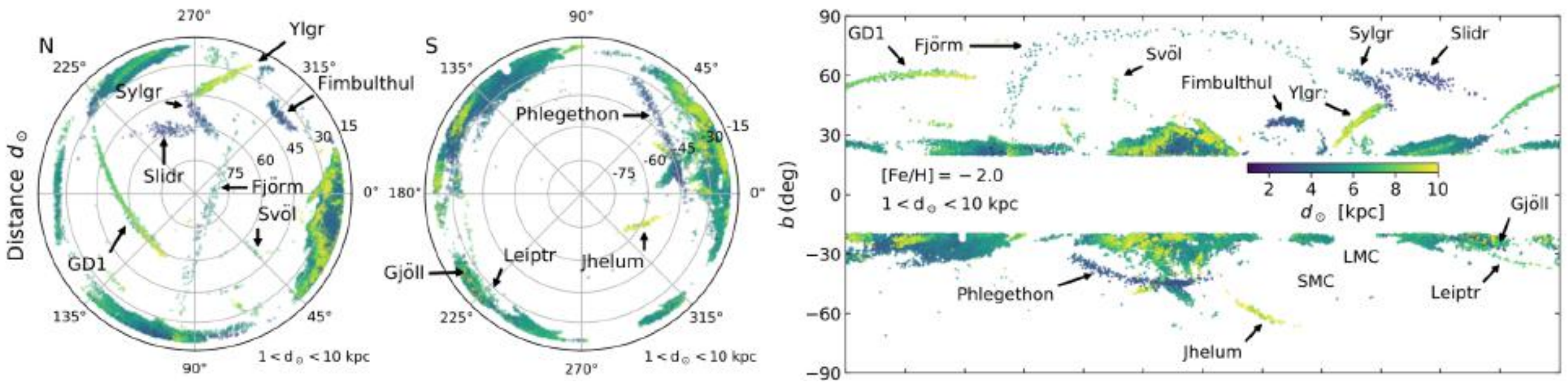

*Figure 37: Spatial maps of a large number of stellar streams detected in Gaia DR2 through combined analysis of spatial and proper motion data, by Ibata et al. (2019), and color coded by distance. The left and center panels show Zenithal Equal Area projections centered on the north and soth Galactic poles, respectively. Distance solutions assume a metal-poor template with an age of 12.5 Gyr and [Fe/H] = −2.0. Figure adapted from Ibata et al. (2019)*

has been previous success in using only positional information to identify substructures in the halo, not least of which is the iconic "field of streams" (Belokurov et al. 2006a; see Figure 36). However, it is clear that a complete unveiling of the succession of individual merger events that contributed to the formation of the stellar halo will require information beyond three spatial coordinates; recent results from Gaia such as the many streams found by Ibata et al. (2019) (see Figure 37) are a dramatic example of the success of using dynamical as well as spatial information. Extending to further dimensions such as that of chemical space, it may be possible to identify stars formed from similar birth places by measuring the abundances of different chemical species. This procedure, commonly known as "chemical tagging" (Freeman & Bland-Hawthorn, 2002), forms a fundamental and complementary approach to the identification of different Galactic "building blocks" by means of chemistry alone.

The combined multi-dimensional information associated with every star (chemical abundances and phase-space information) in the Galaxy thus encode its formation and evolutionary history, providing us a window into understanding the fundamental processes through which the Milky Way was assembled. MSE will provide spectra for millions of stars throughout every component of the Galaxy, and contribute to the creation of the most detailed dataset ever assembled for a single galaxy.

Existing or future high resolution large scale spectroscopic surveys (including AAT/GALAH, WHT/WEAVE, VISTA/4MOST and SDSS-V) are still probing the very nearby Galaxy where the thin disk component is dominant (Figure 38). While their survey strategies try to include as many halo stars as possible, the stars they will be able to observe at high spectral resolution are limited in distance (Figure 39). For example, AAT/GALAH (which has a magnitude limit of $V = 14$) estimate that only 0.2% of their targets will be halo stars and WHT/WEAVE's ($G < 16$) estimate is 3% (C. Babusiaux, *private communication*).

Fainter than $g \geqslant 16$, the thick disk becomes the dominant component at higher latitudes and is easily accessible with MSE even with minimal pre-selection of targets. As an 11m aperture facility, MSE will obtain good SNR at high resolution in reasonable exposure times even in the magnitude range where the stellar halo is dominant (i.e., $g \geq 19.5$ at high Galactic latitude). A critical component of Galactic science with MSE in comparison to other existing or proposed spectroscopic facilities is the ability to access the detailed chemodynamical signatures throughout every Galactic component and sub-component using *in situ* analysis of individual stars.

Gaia is revolutionising our vision of the Milky Way and its local environment. The Gaia-Enceladus merger remnant has been discovered, showing that our Galaxy had a major $(4:1)$ merger around 10 Gyrs ago. Remnants of Gaia-Enceladus are prevalent in the inner halo, and its accretion helped shape the thick disk (Belokurov et al., 2018b; Myeong et al., 2018c; Helmi et al., 2018a). This reinforces earlier ideas of the inner stellar halo's formation (e.g. Meza et al., 2005; Navarro et al., 2011) through an ancient merger (see review by Freeman & Bland-Hawthorn, 2002). The thin disk has been shown to be locally in a strongly perturbed state (Antoja et al., 2018), reinforcing earlier signs of disequilibrium in the disc (e.g. Minchev et al., 2009; Widrow et al., 2012) including a dynamical warp (Poggio et al., 2018) and a strong flare in the outer disk (Thomas et al., 2019). This result confirms many of the predictions of pre-Gaia DR2 models of Laporte et al. (2018c) of the interaction of the Milky Way disc with the Sagittarius dwarf galaxy (see Laporte et al., 2018d), which has long been suspected to play a role in shaping the structure of the Galactic disc (Ibata & Razoumov, 1998; Quillen et al., 2009; Purcell et al., 2011; Gómez et al., 2013). New open clusters (Castro-Ginard et al., 2018), halo streams (Malhan et al., 2018) and dwarf satellites (Torrealba et al., 2018) are being found. Hypervelocity stars, globular clusters, streams and dSphs are being used to derive the Milky Way potential (e.g. Eadie & Jurić, 2018). Those are only a few examples. Overall, Gaia is confirming that the Galaxy is not an equilibrium figure and that the different components are not trivially separated. Instead, there is a strong interplay between them. External events and internal dynamics have blurred out the different components with cosmic time, at least at some level.

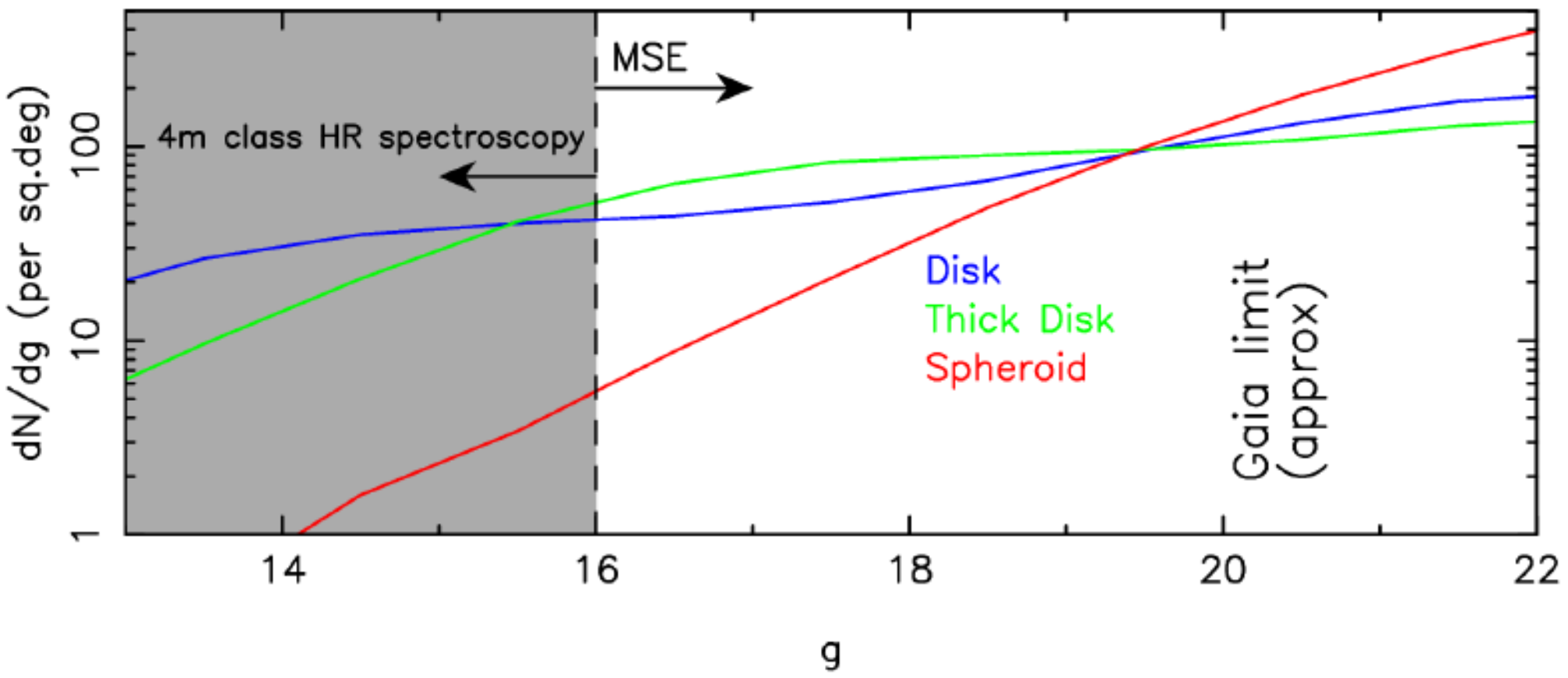

*Figure 38: Differential star counts as a function of magnitude for the three main Galactic components, based on the Besançon model of the Galaxy (Robin et al., 2003) for a 100 square degree region in the vicinity of the north Galactic cap. The shaded region indicates the magnitude range accessible at high resolution to 4m class spectrographs (typically operating at R ∼ 20 000 – 40 000). MSE is the only facility able to access the thick disk and spheroid at high resolution in the regions of the Galaxy in which they are the dominant components. MSE targets at high resolution span the full luminosity range of targets that will be identified with Gaia.*

High resolution spectroscopy is an essential complement to Gaia in order to provide critical detailed abundance information. Of the numerous recent Galactic Archaeology papers using Gaia DR2 data, more than 20% use spectroscopic complements. MSE is the only survey spectrograph planned that will be able to observe millions of the faintest Gaia stars at high resolution. These spectra will carry information on the abundance of 20 to 30 elements from various nucleosynthetic families. Detailed abundances allow us to carry out chemical tagging (see Section 5.3), to identify stars with common origins and to distinguish stars from environments with different star-formation histories. In addition, MSE radial velocities will give access to the full 6D position/velocity space for Gaia stars, and MSE data will give spectroscopic distances for stars in the range not covered by Gaia parallax (Figure 39).

MSE is also an ideal companion to Milky Way science program of LSST. Specifically, LSST will produce highly accurate photometry and astrometry for stars at distances of tens to > 100 kpc (LSST Science Collaboration, 2009). The wealth of data on Milky Way stars that will come from the deep time-domain, multi-band observations of LSST will be crucial input for some of the target selection of MSE, and the spectroscopic data from MSE will return detail on kinematics and abundances that are not otherwise accessible for LSST data.

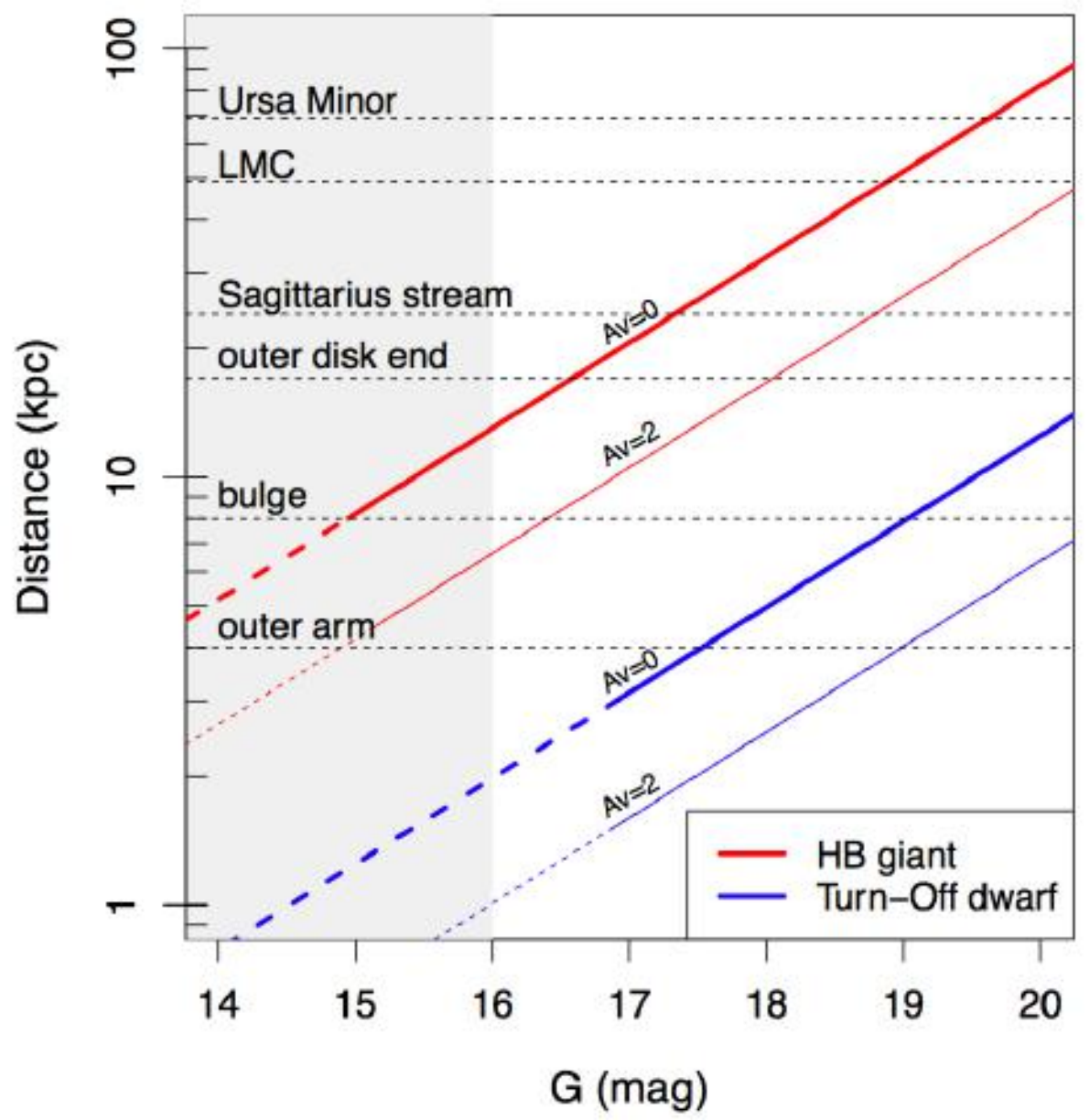

*Figure 39: Distance range probed as a function of magnitude by using two different stellar tracers: horizontal Branch stars (in red) probe to greater distances at a fixed magnitude than turn-off stars (in blue). Tip of the red giant branch stars (not shown) probe beyond the Milky Way virial radius. The shaded grey area corresponds to the magnitude range covered by the 4m-class spectroscopic surveys, and are not a priority for MSE. Horizontal dashed lines indicate the distance threshold required to probe some prominent Galactic structures and satellites. Extinction shifts the red and blue lines to the right as illustrated with the $A_V$ = 2 mag presented here. For each colored line, the dashed section represents the distance – magnitude range where Gaia parallax accuracies are better than 20%. The solid sections represents the distance – magnitude range where distances derived through spectroscopy will be essential to complement Gaia.*

Chemical abundances from MSE are also important to fully exploit asteroseismic data from space-based missions. For example, the combination of metallicities from spectroscopy with asteroseismic parameters are required to derive the masses and therefore ages of red giant stars, which serve as powerful probes of galactic stellar populations (Miglio et al., 2013). This will be particularly critical for space-based missions that will extend the era of Galactic Archaeology into the 2020s and 2030s using red giant asteroseismology, including TESS, PLATO (Miglio et al., 2017) and WFIRST (Gould et al., 2015). The unmatched sensivity of MSE will enable spectroscopic follow-up of rare stellar types such as halo stars with space-based photometry at an unprecedented level, thus completing the efforts of shallower surveys which will probe significant fractions of the thin and thick disc (Figure 39).

The source density of Gaia targets accessible to MSE is shown in Figure 40, showing the need for the MSE high multiplex. Another important characteristic of MSE is its ability to perform multiple visits of large numbers of stars, in particular to find and characterize binary stars (see Chapter 3) . This capability is of critical importance for resolved stellar population studies, since undetected binaries can bias both the abundance determinations and the radial velocity dispersion.

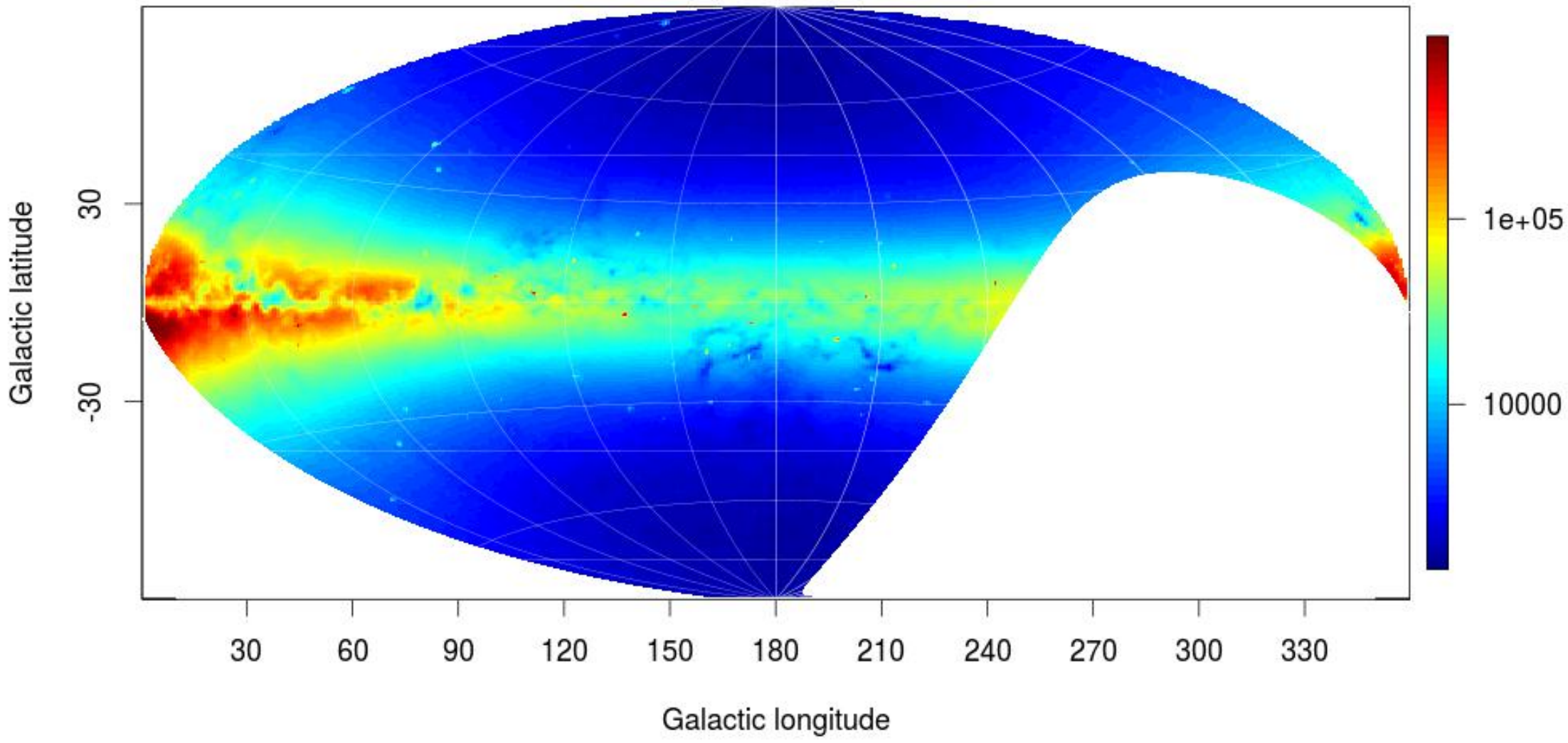

*Figure 40: Aitoff projection of stellar density (per deg$^2$) to G=20 in Gaia DR2 that will be accessible to MSE ($\delta > -30°$). The high multiplexing of MSE is ideally suited to this very large set of potential targets.*

In practice, the potential role for MSE in furthering our understanding of the Galaxy in the post-Gaia era is extensive and covers all components and sub-components of the Galaxy. It is not our intent to detail all possible science programs, but rather to focus on those areas for which MSE is uniquely powerful.

## 5.2 The chemodynamical evolution of the Milky Way

Examination of Figure 39 and Figure 40 emphasizes the crucial role of MSE for understanding the otherwise least accessible parts of our Galaxy, *the faint and distant regimes where the outer disk, thick disk and stellar halo are dominant.* MSE will undertake a range of related analyses that aim to determine the inter-relation between these components, and what continuities/discontinuities exist between their stellar populations.

### 5.2.1 The Galactic disk

Extragalactic observations have revealed all the complexity of the outer regions of galaxy disks: truncated/anti-truncated surface brightness profiles, breaks in metallicity profiles, U-shaped age profiles, complex star formation histories, and more (e.g. Zhang et al., 2018b; Barker et al., 2012, 2011). For the Milky Way, the outer regions are still largely uncharted in 6D phase space. The Galactic Anticenter shows successions of overdensities above and below the midplane (e.g. Newberg et al., 2002; Rocha-Pinto et al., 2004; Sharma et al., 2010) which have been commonly been interpreted as stellar debris from disrupting dwarf galaxies (e.g. Peñarrubia et al., 2005; Sheffield et al., 2014). These overdensities have recently been shown through the study of their stellar populations to be composed of disk stars (e.g. Price-Whelan et al., 2015; Sheffield et al., 2018) and chemically tagged as thin disk material (Bergemann et al., 2018). These observations support the predictions of recent models of the Sagittarius dwarf galaxy's interaction with the Galaxy (Laporte et al., 2018b). Other features in the Anticenter include large overarching thin streams such as the Anticenter stream (Grillmair, 2006). These streams have been re-interpreted as remnants of tidal tails from the Milky Way disk, excited from past satellite interactions (Laporte et al., 2019b), and could be used to constrain the flattening of the Galactic potential and strength of satellite interactions.

MSE is crucial for mapping the 3 dimensional velocity field of the disk beyond the extended solar neighbourhood, and indeed beyond the edge of the outer disk. Such a detailed 3 dimensional map of the velocity field will contain information on the mass distribution of the Galaxy. This will allow us to quantitatively map the boundary and substructures in the outer disk, provide a detailed chemical description, and trace its formation history and its link with the inner disk. We will be able to exploit non-axisymmetric features to decipher its dynamical history. These datasets will capture the ongoing perturbations of the disk driven by spiral arms (e.g. Monari et al., 2016), the bar (e.g. Dehnen, 2000) and the coupling between the two, as well as perturbations from massive satellites interacting with the Milky Way, as discussed above. It will allow measurement of the various pattern speeds at play in the Galaxy and the location of their resonances (e.g. Sellwood, 2010).

For younger disk stars, MSE will be able to track dissolving open cluster trails among field stars, and identify stars from unbound clusters that are actively dispersing, associations, and moving groups using velocities and chemical tagging. Open clusters are the building blocks of the young disk, and MSE will be able to follow up previously unknown clusters discovered by Gaia. MSE will study clusters across a range of Galactic environments down to their faintest stars. This data will be crucial for investigating the stellar initial mass function and the initial-to-final mass relation in different environments, and to test models of stellar structure and evolution (see more details in Chapter 3).

### 5.2.2 The Galactic bulge

The turn-off for stars in the bulge is at $V \sim 20$ ($H \sim 17.5$) in Baade's Window, which means that the sensitivity and field of view of MSE makes it is well suited to studies of the Galactic bulge. VLT/MOONS, with its high resolution and near infrared capabilties, will do some extraordinary science in the bulge in areas of higher obscuration. A key theme of MSE science for the bulge is to determine the link between the inner galactic substructures to the outer Galaxy. Specifically, the detailed abundance distribution of the bulge and its outskirts, and its comparison with those homogeneously obtained for the halo and the disk studies, will reveal the continuity/ discontinuity between these populations. The bulge is densely populated and different populations dominate at different Galactic longitude and latitude (e.g. Ness et al., 2013). Reconstructing the distribution of its chemistry, kinematics and age profiles therefore requires good statistics over a large area, up to $|b| > 10$-$20°$.

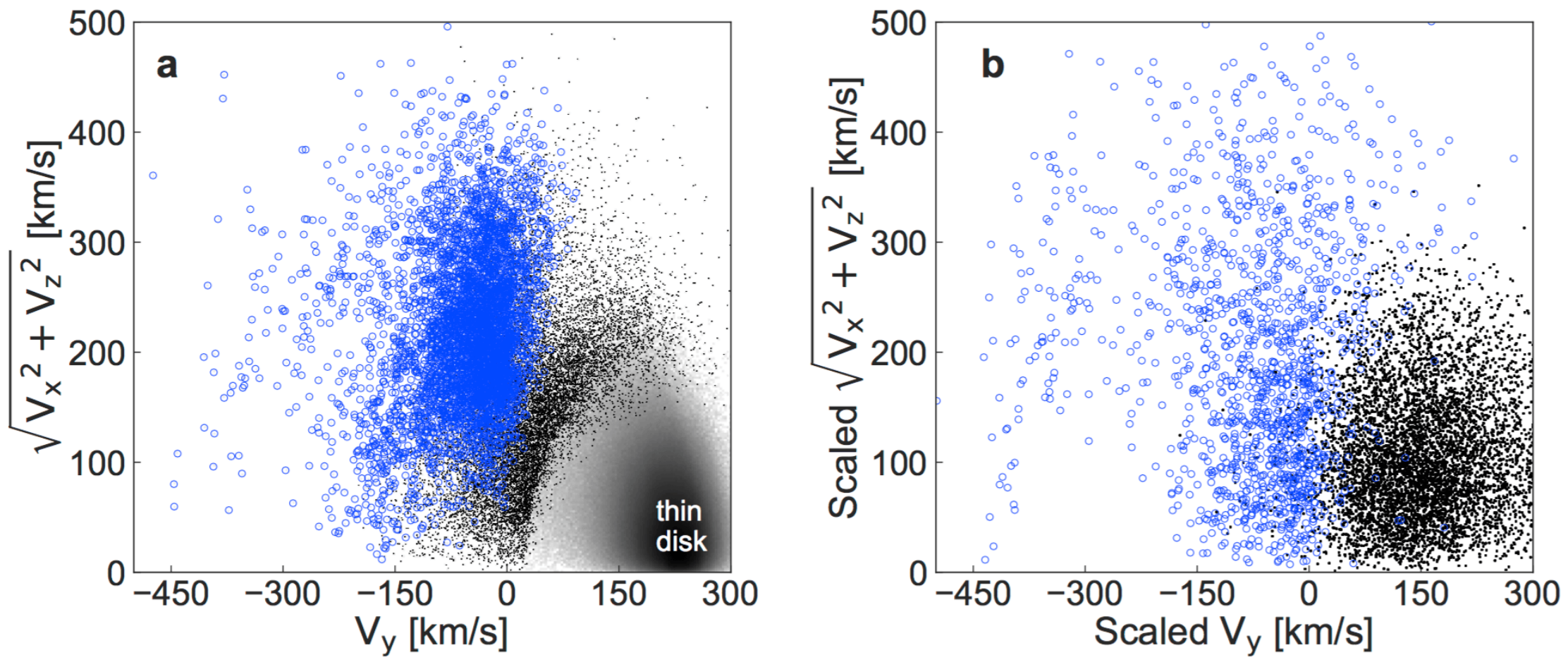

*Figure 41: Left panel: velocity distribution of stars in the solar vicinity. Disk stars are plotted as grey contours, halo stars are shown as points. The blue points show halo stars belonging to a structure with a slight retrograde motion. Right panel: velocity distribution of star particles in a simulation of a 5:1 merger, that triggers the formation of a thick disk. Results for the Milky Way demonstrate that it is qualitatively consistent with having undergone a similar merger around 10Gyrs ago. The minor galaxy in that merger has been called Gaia Enceladus, the remnants of which are traced by the blue stars in the left panel. Gaia Enceladus is potentially responsible for the formation of the thick disk. Figure from Helmi et al. (2018a).*

It is also possible that the bulge contains the remains of the most ancient mergers that shaped the core of our Galaxy (e.g., Howes et al., 2015). Stars in the inner Milky Way are kinematically well mixed, and so chemical abundances are required to reveal their origins. The structure and kinematics of the bulge is that of a barred system presenting a boxy/peanut shape. However, there are also indications of the presence of a primordial structure within the inner galactic regions (Hill et al., 2011; Schiavon et al., 2017b), but its relative mass and connection with the local old structures is unknown. We require detailed abundance studies for many elements in order to identify the presence of a separate primordial bulge component and distinguish it from, or link it to, with the structures we identify in the solar neighbourhood.

### 5.2.3 The stellar halo

The role of accretion in the formation history of the halo has been shown to be highly significant even within the local sphere through the discovery of Gaia-Enceladus (Helmi et al. 2018b; see Figure 41). Gaia data for nearby stars show that our previous understanding of the boundary between the in-situ halo and the thick disk needs to be revised (Haywood et al., 2018). Full homogenous coverage of the halo must be extended from local stars, which will be massively targeted by WEAVE/4MOST, to more distant regions. As such, MSE will concentrate on the outer halo, and will extend the high resolution surveys of WEAVE and 4MOST that focus of the inner halo. MSE is the best facility to target the outer halo, which shows significantly different structure, metallicity and kinematic properties than the inner halo.

MSE will definitively quantify the ratio of halo stars formed *in situ* versus those that were accreted throughout the halo. It will measure abundance gradients in the *in situ* population and record the chemical imprint of the accreted progenitors' formation histories (see Section 5.3). The observation of halo white dwarfs will also provide strong constraints on the age and IMF of the halo components (see detailed discussion in Chapter 3).

The Milky Way stellar halo is rich in substructure (e.g. Belokurov et al., 2006b; Malhan et al., 2018). Kinematics of the halo out to the virial radius will provide a better mapping of its distant streams and substructures, including globular clusters and their outskirts. From the spatial and kinematic shape and deformation of those substructures, a detailed mapping of the overall mass distribution of the Milky Way will be drawn, including dark sub-halos (see detailed discussion in Chapter 6). The detailed study of the halo and its streams will allow the reconstruction of the history of the infall of satellites and its impact on halo density features (e.g. Vesperini & Weinberg, 2000). Those studies need huge observational datasets due to the large number of accretion events and the interplay between the different structures, as demonstrated by the perturbation of the Orphan Stream by the LMC (Erkal et al., 2018b).

## 5.3 In-situ chemical tagging of the outer Galaxy

A critical tool available to MSE lies not just in dynamical analyses, but in the derivation of detailed chemical abundances. Recognising the suggestion by Freeman & Bland-Hawthorn (2002) that "the major goal of near-field cosmology is to tag individual stars with elements of the protocloud", MSE will provide abundance ratios for 20 to 30 elements formed through multiple nucleosynthetic channels for many millions of targets. Of course, this applies not just to the stellar halo but also to other Milky Way components, including the thin disk, thick disk and bulge. As previously discussed, the study of the chemical properties of the halo has relied, with a few exceptions, on relatively local samples of halo stars that pass near enough to the Sun to be observable at high spectral resolution. MSE shifts the paradigm towards *in situ* analyses of the outer parts of the halo.

Figure 42 provides an important example of the power of large chemical abundance datasets in helping to unravel the formation of Milky Way and its satellites. A stellar population follows a well defined route in this diagram, with roughly constant $\alpha$ abundance at lower metallicity giving way to monotonically decreasing $\alpha$ abundance at higher metallicity. The transition between the two regimes is referred to as the "knee" in the distribution, and the metallicity at which the knee occurs is influenced by early star formation and self-enrichment in a galaxy. Type II supernovae are $\alpha$-rich, and explode only a few-to-tens of Myr after a star formation event. Type Ia supernovae are iron-rich in comparison, and are delayed by at least a few hundred Myr prior to contributing to the chemical abundances. A galaxy with a knee at higher metallicity must have had more intense star formation at early times (i.e., more Type II supernovae contributed to both the $\alpha$ and iron abundance prior to Type Ia supernovae contributing to the iron abundance only).

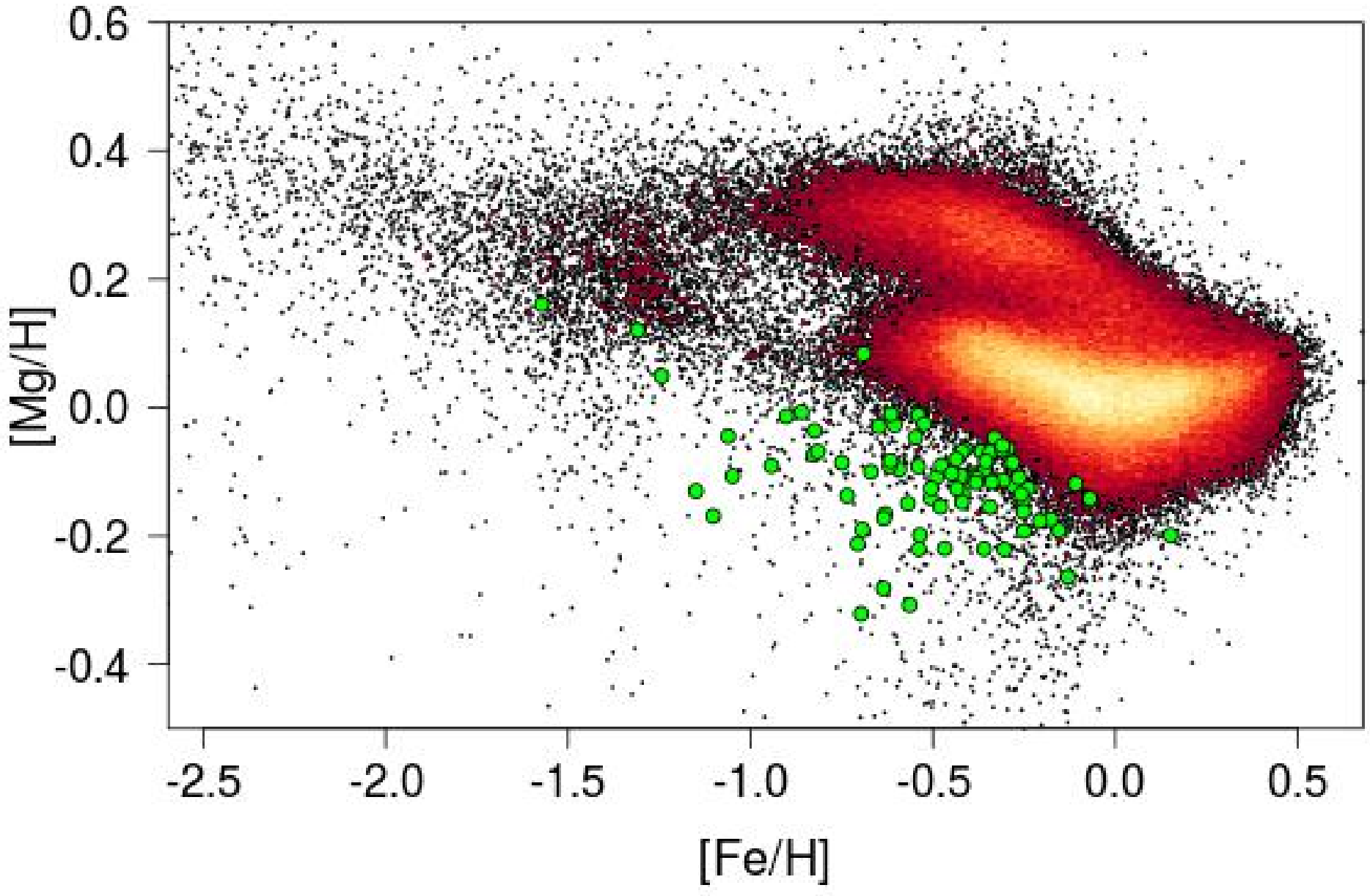

*Figure 42: APOGEE DR14 α element [Mg/H] as a function of [Fe/H]. The colour scale represents the square root of the stellar density. The thick disk and the thin disk form the 2 most prominent sequences in this diagram. The slight overdensity at [Fe/H] -1.5 corresponds to the Gaia Enceladus stars. Green points highlight the Sagittarius dwarf galaxy stars.*

The insights available through chemical abundance studies are powerful, and large statistical datasets of high resolution chemical abundances from survey projects such as SDSS/APOGEE, AAT/GALAH and VLT/GES have become publicly available in recent years. For any spectroscopic survey, the higher the number of distinct nucleosynthetic pathways sampled by the abundances that can be measured, the better we can trace and distinguish the distinct evolutionary paths of the different stellar populations (see Chapters 4 and 3 for a detailed discussion of the different abundances that MSE will be able to track). MSE will be particularly well suited to studies of the r-process in comparison to many previous large scale surveys, and will learn from many important considerations that have been highlighted by recent analyses of these datasets:

**The chemical homogeneity of star formation events:** A fundamental premise of chemical tagging is that stars formed in an individual cluster (which henceforth we adopt as shorthand for "a star formation event", regardless of whether the event forms a recognizable stellar cluster) share a single chemical "fingerprint", such that there is a chemical signature that can be used to relate the stars even if they do not cluster strongly in phase space. This is predicted by theoretical studies of turbulent mixing in gas clouds (Feng & Krumholz, 2014).

Some observational tests for abundance homogeneity in open clusters (e.g., De Silva et al., 2007, 2009; Friel et al., 2014) have been positive, finding that open clusters are typically homogeneous in abundance within the observational abundance errors. Other work has aimed to shrink those error bars in search of low-level abundance complexity. As one example,

Liu et al. (2016) used differential analysis to determine high-precision abundances for stars in the Hyades, and found correlated star-to-star variations in many elements at the 0.02-dex level.

**Abundance changes following star formation:** Not all elements are equally useful for chemical tagging. Effects such as diffusion (e.g., Önehag et al., 2014; Bertelli Motta et al., 2018; Gao et al., 2018), first dredge-up (Iben, 1965; Salaris & Cassisi, 2005), and extra mixing (e.g., Sweigart & Mengel, 1979; Lagarde et al., 2019) all affect particular surface abundances during specific evolutionary phases. There is also some correspondence between the presence of planetary systems and the abundances of refractory elements in stellar atmospheres (e.g., Meléndez et al., 2014; Spina et al., 2015; Oh et al., 2018), though the mechanism that produces this effect is not yet clear.

These effects can be corrected for, inasmuch as we understand them observationally and theoretically, to convert the observed abundance pattern into an "initial" abundance pattern. However, there is a risk of adding uncertainty through these corrections, and it may be more straightforward to carry out chemical tagging only with those elements that are not affected by these processes.

**Uncertainties in analysis:** Traditional techniques to determine stellar abundances have an accuracy that, empirically, appears to be limited to 0.1 dex for whatever element is being measured. Generally, this accuracy reflects systematic uncertainties in the underlying stellar atmosphere models used to interpret the data and subtle physical effects such as microturbulence and non-local thermodynamic equilibrium.

Already the volume of data being acquired by current surveys is large enough to make standard spectroscopic analysis impractical, and with the step to larger projects like MSE the data volume will only increase. New methods for determining stellar parameters and abundances have been developed in recent years using regression (Ness et al., 2015) and neural networks (Fabbro et al., 2018; Ting et al., 2018). These methods identify the optimal set of stellar parameters and abundances to reproduce an observed spectrum quickly and self-consistently. The errors in the abundances produced by these methods are fundamentally different from the errors in traditional abundance determination methods, and may have systematic dependences on the details of the method, especially on the "training set" of well studied stars used for regression, or the architecture of the neural network.

**Sampling rate and practical cluster mass limits for chemical tagging:** Ting et al. (2015) examine some of the survey considerations for successful chemical tagging. Key parameters they consider include:

- Sampling rate; that is, the number of targets in the survey compared to the possible number of targets (Ntarget/Nsample). Higher sampling is required in order to be sensitive to lower mass clusters;

- The number of "distinct" regions of chemical space in which the clusters are distributed, Ncells. The relevant region of chemical space to be explored will have a finite volume, in which each of the clusters will occupy a unique point. However, once measurement errors are considered, it is clear that some clusters may become indistinguishable from other clusters in chemical space, particularly if the number of clusters is high, the volume of chemical space is small, and/or measurement errors are large, i.e., the dimensionality of chemical space, and typical measurement uncertainties, are both critical parameters in defining Ncells;

- The number and mass distribution of individual star clusters ("star formation events") contributing to the signal. Clearly, measurement of these quantities is a fundamental goal of chemical tagging experiments.

Ting et al. (2015) conclude that, through strategic observational programs, it is possible to statistically reconstruct the slope, high mass cut-off and evolution of the cluster mass function for studies that focus on the stellar disk of the Milky Way for surveys that sample of order 1 million stars. The short dynamical times in the disk presumably make this one of the most challenging environments for this type of study. It is worth noting that earlier work by the same authpr suggests that different observational strategies may need to be developed for different Galactic components (Ting et al. 2012). That earlier study uses principal component analysis on abundances for stars in the thin and thick disk, the halo, and open and globular clusters, and finds that the dominant abundance patterns are different in each environment. This implies that these components may be best traced through different sets of chemical species.

Because of the intrinsic complexity of the problem of chemical tagging, there is not a single best method to use. A number of authors have carried out studies of the effectiveness of various chemical tagging approaches. To list a few examples, Mitschang et al. (2013, 2014) used the Manhattan distance (a quadratic sum of normalized distances in each abundance) to identify stars with similar abundance patterns in the disk near the Sun. Kos et al. (2018) used the dimensionality reduction method tSNE (t-distributed stochastic neighbour embedding) to separate members of open and globular clusters from nearby field stars using only abundance information. Blanco-Cuaresma & Fraix-Burnet (2018) use a phylogenetic tree to create a hierarchical classification for open cluster stars based on abundances, and find that they can successfully separate the clusters with more or less success depending on the elements used in the classification.

The analysis by Hogg et al. (2016b) is a nice demonstration of chemical tagging in action (Figure 43): sub-structures have been found using a k-means clustering in the abundance space only. From the phase space coordinates of the stars, which were not used in the identification, it is clear that the stars do form coherent structures.

These recent analyses of different aspects of chemical tagging emphasize the importance of carefully planned observing programs to optimally (i) sample the structures being probed (ii) probe the numerous dimensions of chemical abundance space (iii) use phase-space information in addition to chemical abundance information, and (iv) enable precision determination of individual abundances.

### 5.4 First stars and the progenitors of the Milky Way

The unprecedented size of the stellar spectroscopic dataset for MSE will enable the definitive analysis of the metal-weak tail of the halo metallicity distribution function (MDF). This key observable has a direct bearing on models for the formation of the first stars, and on the dark baryonic content of galaxies. The first stars to be formed after the Big Bang were formed with the "primordial" chemical composition: i.e., hydrogen and helium, plus traces of lithium. A protogalactic cloud consisting of such a gas may have had difficulty in providing cooling mechanisms efficient enough to allow the formation of low-mass stars. Several theories on star formation postulate the existence of a "critical metallicity" below which only extremely massive stars can form (Bromm & Loeb, 2003; Schneider et al., 2012, and references therein). Other theories invoke fragmentation to produce low-mass stars at any metallicity (Nakamura & Umemura, 2001; Clark et al., 2011; Greif et al., 2011).

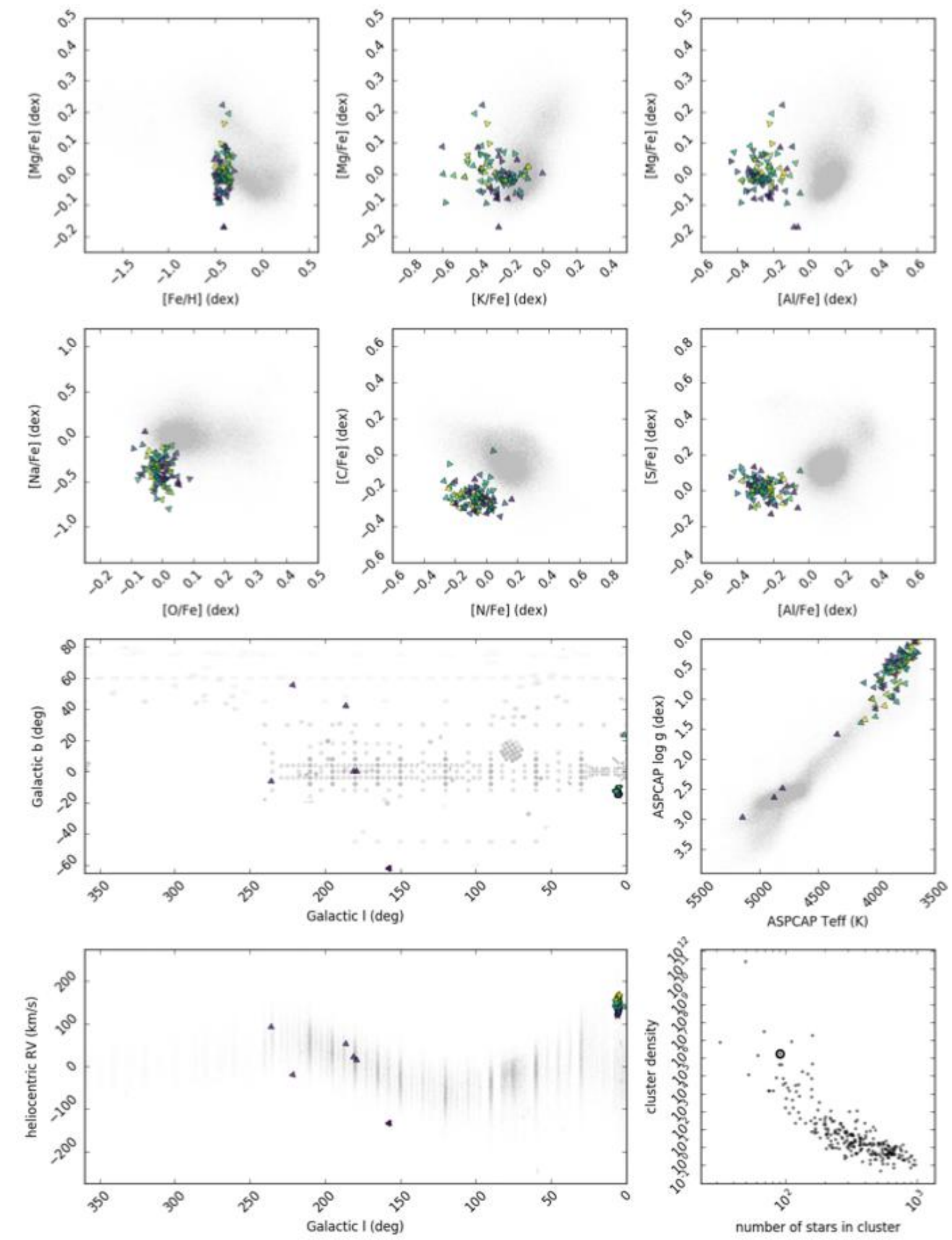

*Figure 43: Greyscale points show projections of approximately* $10^5$ *individual stars from the SDSS/APOGEE survey. The top set of panels show projections in abundance space only. Clustering analysis of this (15-dimensional) space reveals substructures. One of these most prominent substructures is shown as the large points. The lower panels show the structure of this feature in phase space coordinates that were not used for its identification. In fact, this feature is due to the Sagittarius dwarf spheroidal galaxy. Figure from Hogg et al. (2016b).*

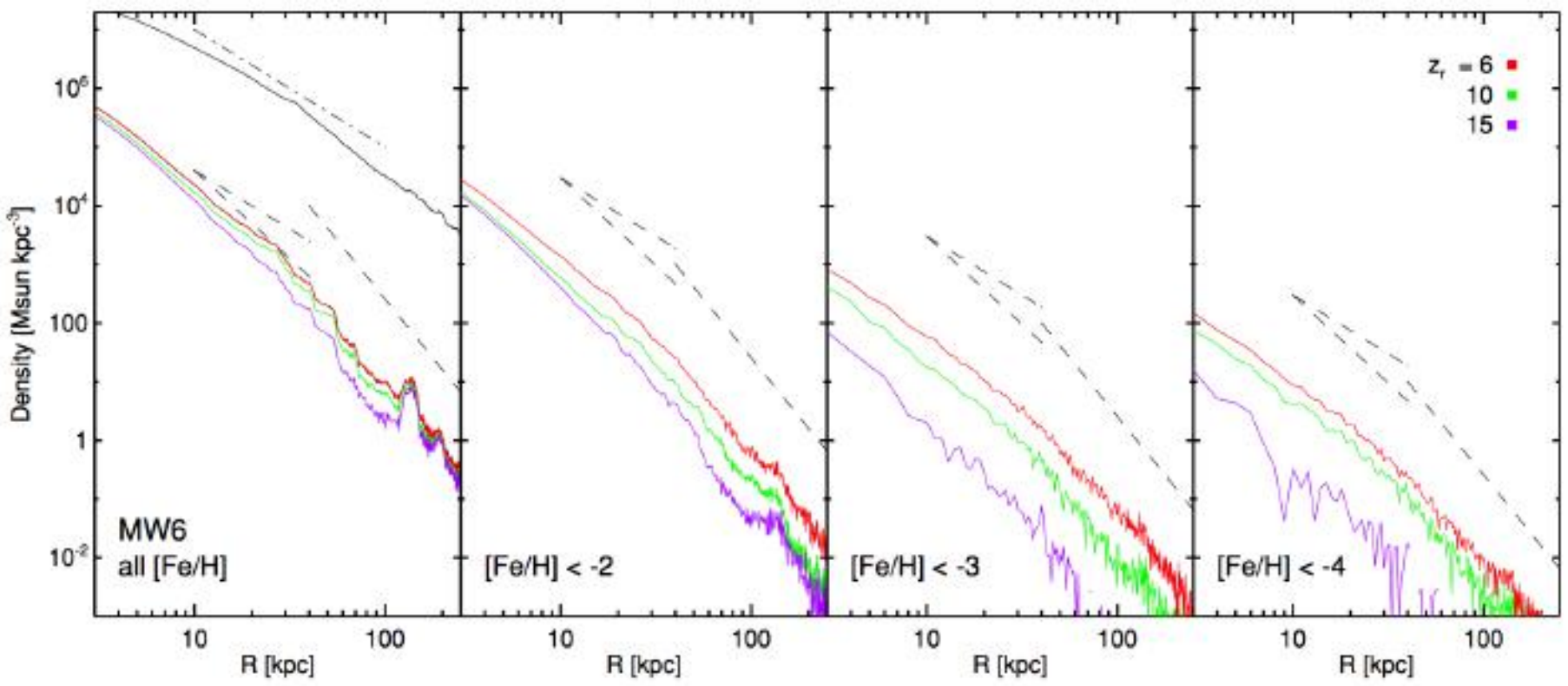

*Figure 44: Stellar mass density profiles for a realization of a cosmologically-consistent Milky Way mass halo for all stars (left panel) and three metallicity cuts (from left to right), [Fe/H] < -2, -3, and -4, and three redshifts of re-ionization in each panel (as described by the inset text). The dashed lines mark power laws with slopes -2, -3, and -4. Figure from Tumlinson (2010).*

The implication of these considerations for the baryonic content of galaxies is obvious: if the first generation of massive stars that reionized the universe formed along with low-mass stars, a significant fraction of these would now be present as old, cool white dwarfs, while a small fraction of them (essentially those of mass less than 0.8 $M_{\odot}$, which have main-sequence lifetimes comparable to the Hubble time) would still be shining today and can in principle be observed. Further, if there is a critical metallicity, then the metal-weak tail of the MDF ought to show a sharp drop at this value. MSE can examine the chemical abundance distributions of old stars to search for nucleosynthetic signatures of these first stars, and can also construct a precise MDF to search for evidence for a critical metallicity.

Detailed modeling of the formation of a Milky Way mass galaxy in which the distribution of these first stars are monitored (e.g., Brook et al., 2007) reveals fascinating insights into the expected growth of the Milky Way with time. Tumlinson (2010) simulated a suite of six Milky Way analogues and incorporate baryonic processes on top of the dark matter "scaffolding" to allow analysis of the evolution of the implied stellar populations. They find that the fraction of stars dating from the oldest epochs is a strong function of radius within the Galaxy and with metallicity. The former trend reflects the inside out growth of dark matter halos, and implies the oldest stars are some of the most tightly bound to the Galaxy. The latter trend reflects the trend of increasing metallicity with time.

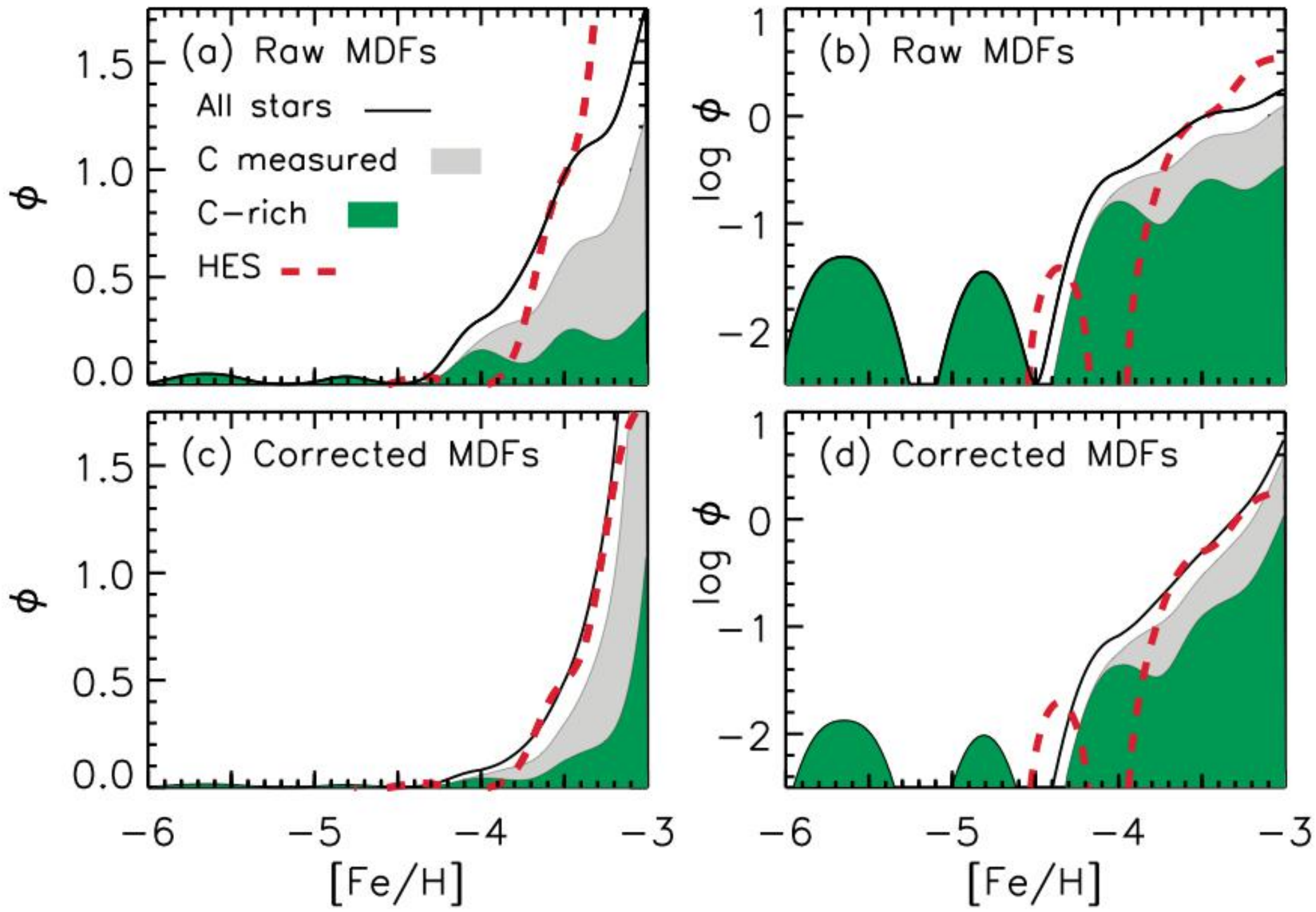

*Figure 45: The metallicity distribution function, [Fe/H], based on the high-resolution abundance analysis of Yong et al. (2013a) at the metal-weak end, [Fe/H]<-3 dex. Left panels show a linear scaling, and right panels show a logarithmic scaling. The upper and lower panels refer to the raw data and those following completeness corrections on the range -4.0 < [Fe/H] < -3.0, as described by Yong et al. (2013b). Green and grey color-coding is used to present the contribution of C-rich and C-normal stars for which measurement was possible, respectively. The dashed line shows the Hamburg-ESO MDF based on the data of Schörck et al. (2009). Figure from Frebel & Norris (2015).*

Figure 44 shows the radial distribution of stars in a halo realisation from Tumlinson (2010) as a function of metallicity. This study showed that stars that formed at redshifts z = 6 - 10 likely have metallicities more metal poor than [Fe/H] = -3 dex. To test the validity of these models, and to probe the formation of the Galaxy and its stars at the earliest times, it is therefore important not just to find metal poor stars, but to develop a detailed understanding of the shape of the full MDF and its spatial variation in the Galaxy, including out to very large radius. Such a program requires *in situ* analysis of large numbers of metal poor stars, and is a core science goal of MSE.

To put the potential of MSE studies of the metal-weak tail of the Galaxy in context, Figure 45 shows the state-of-the-art MDF for stars more metal-poor than [Fe/H]= -3 dex. For homogeneity, this is based on a high resolution analysis by Yong et al. (2013a). Also shown as a dashed line is the MDF derived from the Hamburg-ESO Survey (Schörck et al., 2009). The Hamburg-ESO survey is recognized as a landmark study of the metallicity of the Galactic halo, and is based on a total of 1638 stars. The Yong et al. (2013a) sample has 86 stars with [Fe/H] < -3 dex, of which 32 have [Fe/H] < -3.5 dex. Current surveys will increase the known sample of metal-poor stars significantly over the next few years, and will produce important lists of candidate metal poor stars for spectroscopic follow-up e.g., the SkyMapper project (which found a star with [Fe/H] ~ -7, Keller et al. 2014), and the Pristine project on CFHT (Starkenburg et al., 2018). However, full characterization of the MDF through *in situ* spectroscopic analysis requires the large aperture, high resolution, highly multiplexed capabilities of MSE.

MSE will provide the definitive study of the metal-weak structure of the Galaxy, by providing metallicities for a sample of several million stars at resolution R > 20 000, allowing a complete, homogeneous characterization of the halo MDF down to [Fe/H] ~ -7. Connection of the observed abundance patterns of these extremely metal-poor stars with those found in the ultra-faint dwarf galaxies (e.g. Frebel, 2018) will provide direct tests of the environments and mass distributions of what are likely to be the earliest mini-halos in which star formation took place. Refined numerical chemical evolution models will be able to compare the predicted frequencies of such stars with these observations, and thereby constrain the range of possible progenitors responsible for their abundance patterns.

### 5.5 The Local Group as a time machine for galaxy evolution

The collection of galaxies in the immediate vicinity of the Milky Way – the Magellanic Clouds, the Andromeda Galaxy, Triangulum, and the ~ 100 currently known dwarf galaxies that constitute the "Local Group" – are the nearest examples of galaxies spanning a wide range of morphological types. They are the only galaxies in the Universe, aside from the Milky Way itself, for which large numbers of individual stars can be spectroscopically observed from the ground. As such, these nearby systems offer unique insights into galaxy formation and evolution.

The size of the Local Group is of order a few Mpc in diameter, as defined by the zero-velocity surface where the gravitational attraction of the galaxies of the Local Group exactly balances the Hubble flow (McConnachie 2012). Analysis of the formation of Local Group-like structures in cosmological simulations (Boylan-Kolchin et al. 2015) shows that, at $z = 7$, the co-moving linear size of the Local group is 7 Mpc (i.e., a volume of 350Mpc). At early epochs, the Local Group therefore probes a cosmologically representative volume of the Universe. At z $\leqslant$ 3, the co-moving size of the Local Group exceeds that of the Hubble Ultra-Deep Field, and is similar to the volumes that will be probed by JWST (see Figure 46).

The complementarity of studies of the Local Group population of galaxies to high redshift studies of galaxy evolution with JWST, WFIRST and other future facilities is clear. At early times, the Local Group and JWST explore similar volumes, with crucial differences. Specifically, the Local Group contains representative numbers of faint galaxies that will never be detectable at high redshift with JWST, and whose formation can be explored through their stellar fossil record. In contrast, JWST is sensitive to galaxies that are intrinsically brighter and, therefore, rarer. This is shown explicitly in Figure 47.

Local Group galaxies subtend large angles on the sky and contain large numbers of stars that are, individually, quite faint. A complete exploration of the deep field of the Local Group will require a wide field of view, high sensitivity, and significant multiplexing. MSE is therefore essential to enable the detailed study of chemodynamics and galaxy formation in the Local Group.

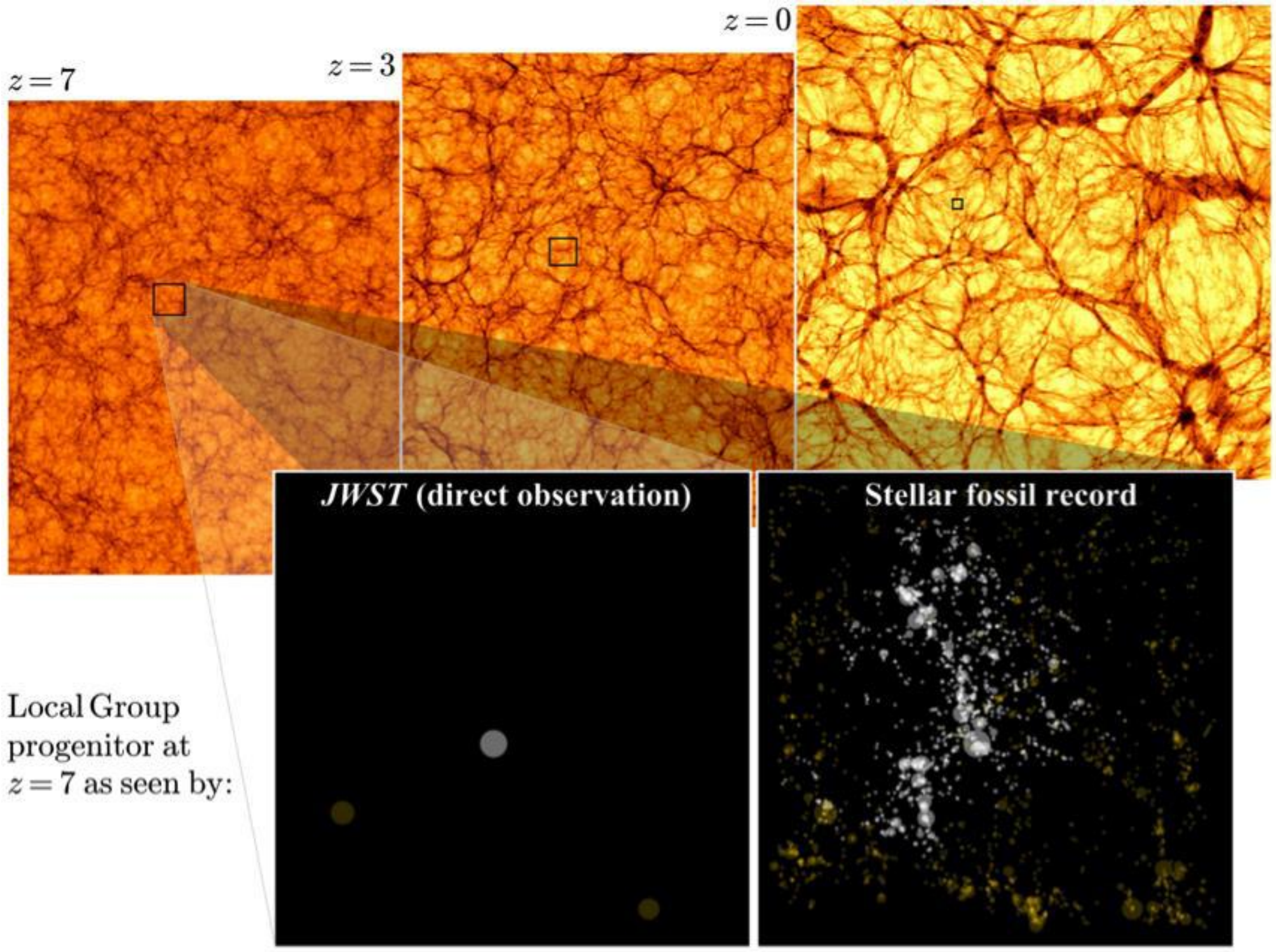

*Figure 46: Density slices of the Illustris simulation at different redshifts, each with a side length of 106.5 co-moving Mpc. The square in each panel indicates the co-moving size of a Local Group-type structure at each redshift. These are comparable in size to the HST Ultra-Deep Field. The zoom-in panels shows an expected image of the galaxies visible at* $z = 7 \pm 0.02$ *using JWST (left panel), and through the stellar fossil record in the Local Group (right panel; white symbols indicate structures still present in the Local Group at* $z = 0$*, whereas gold symbols represent objects that are no longer present in the Local Group at* $z = 0$*). Figure from Boylan-Kolchin et al. (2015).*

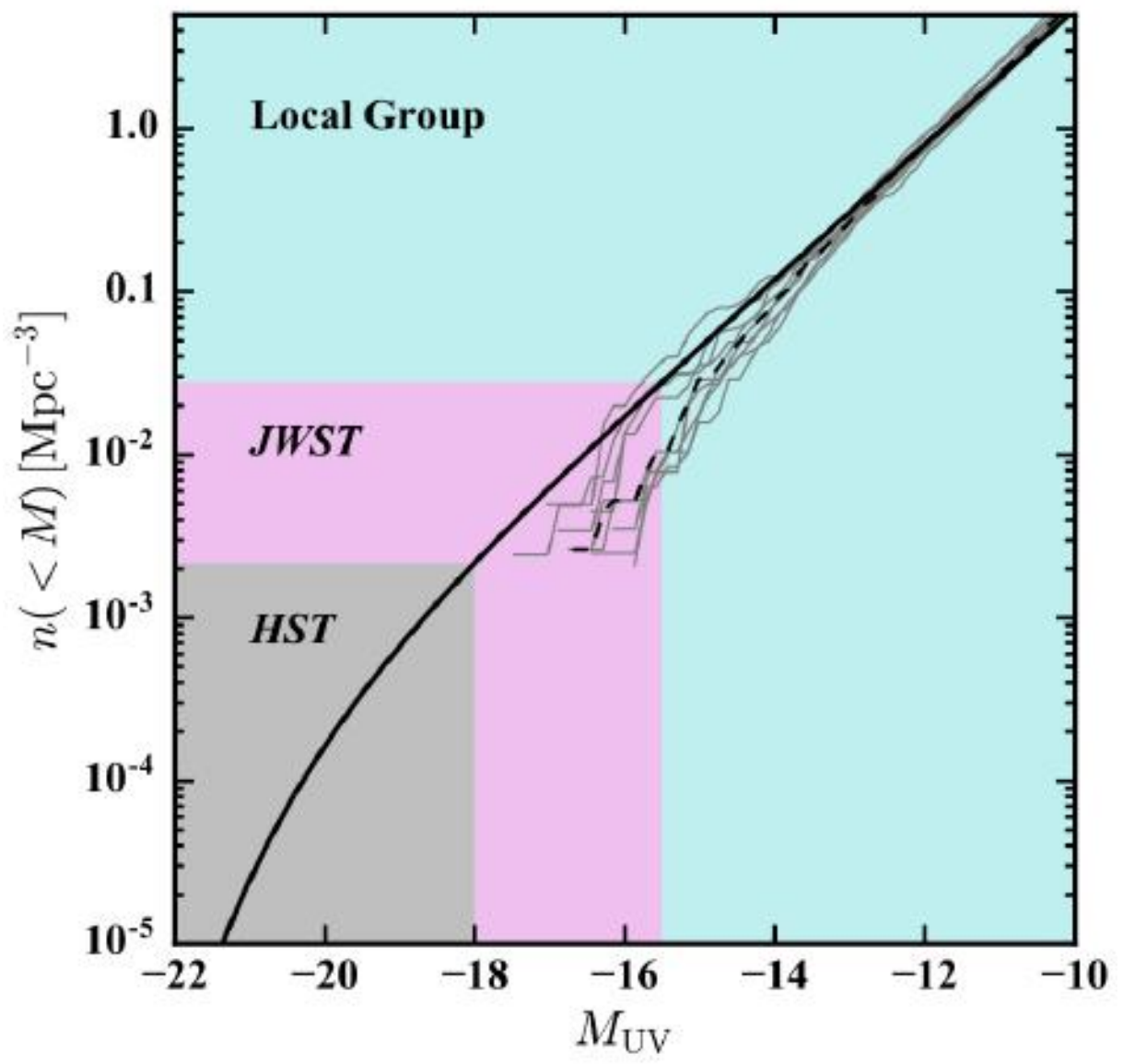

*Figure 47: The solid black line represents the cumulative luminosity function of galaxies at $z = 7$ (Finkelstein et al., 2015), extrapolated beyond $M_{\rm UV} = -18$ to faint magnitudes. The thin grey lines represent the cumulative luminosity functions in simulations of Local Group environments (the ELVIS simulations, see Garrison-Kimmel et al., 2014), and the dashed black line represents the mean. The approximate detection limits of HST, JWST and the Local Group are shown, highlighting the complementarity of the near and far-field regimes for a full understanding of galaxy evolution across the luminosity function. Figure from Boylan-Kolchin et al. (2016).*

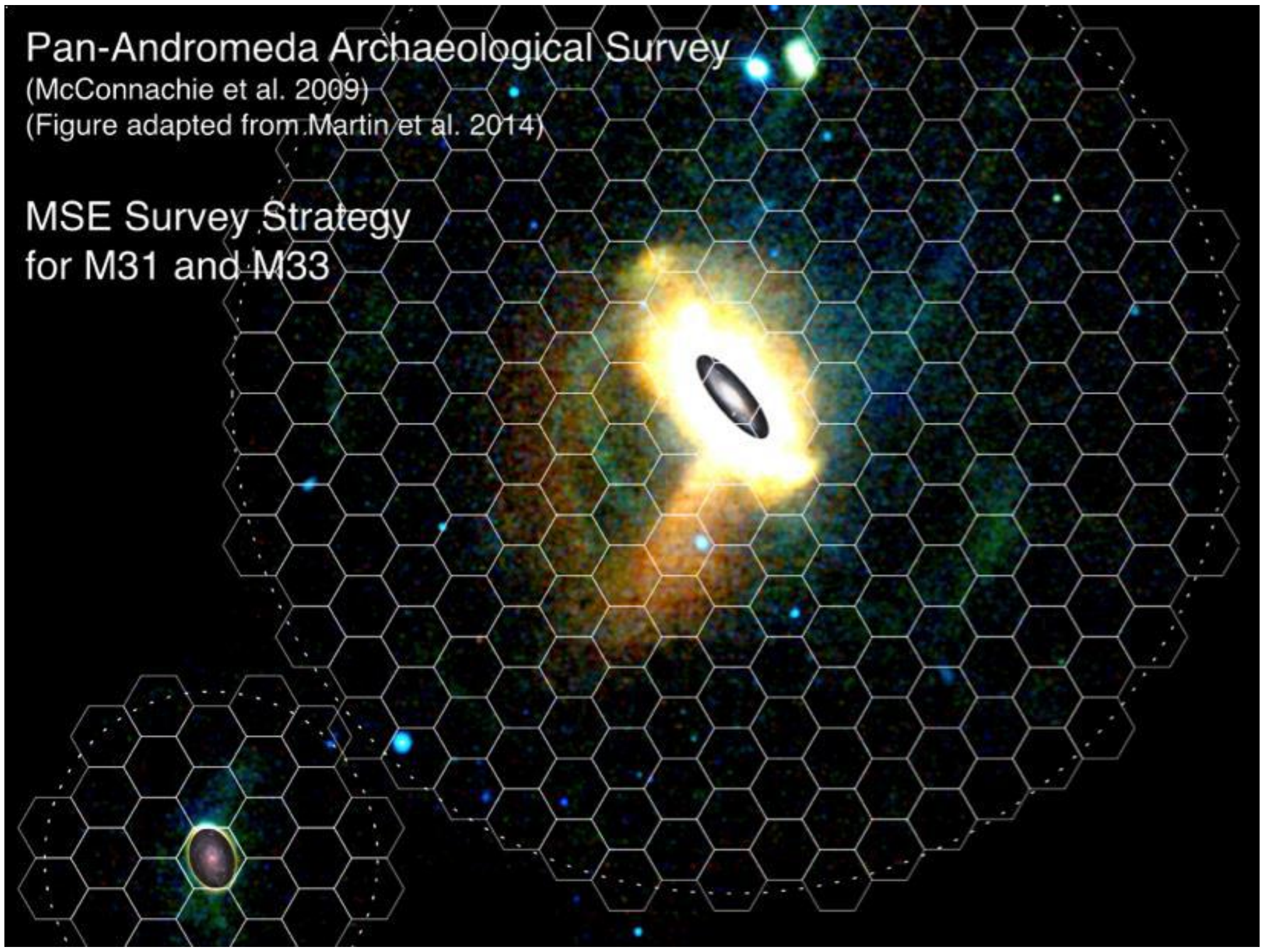

*Figure 48: Spatial distribution of candidate red giant branch stars in the environs of M31 and M33, as identified from colour-magnitude cuts from the Pan-Andromeda Archaeological Survey. Colour-coding corresponds to the colour of the RGB stars, such that redder RGB stars (likely higher metallicity) appear red, and bluer RGB stars (likely metal-poor) appear blue. Dashed circles highlight maximum projected radii of 150 kpc and 50 kpc from M31 and M33 respectively. The tiling strategy for an MSE survey of this region is overlaid. Figure adapted from Martin et al. (2014).*

### 5.5.1 The chemodynamical deconstruction of the nearest L* galaxy

MSE will conduct a systematic survey of the kinematics and chemistry of Local Group galaxies within 1 Mpc at a level of detail that has never been achieved before and cannot be achieved with any other current or planned facility. Gaia is producing an unprecedented data set, but detailed information on the properties of the resolved stellar populations of other galaxies is necessary to better understand the context in which to interpret results for the Milky Way. It is in this context that the capability of MSE to provide the ultimate chemodynamical decomposition of the two closest Local Group spirals (M31 and M33) is most important. These galaxies represent the obvious connection between our highly detailed description of the Milky Way and the low-resolution studies of more distant galaxies in the realm of a few Mpc and beyond, where we can obtain significant samples of galaxies (as a function of galaxy type, environment, mass, and so on). MSE will make a pivotal contribution by undertaking a large spectroscopic survey of these two Local Group galaxies. We note that Subaru/PFS intends to observe approximately 66 square degrees of the halo of M31 (50 individual exposures; Takada et al., 2014a). However, the outer regions of M31 have a complex morphology generated by its active minor merger history. A comprehensive understanding of the chemodynamics of M31 and M33 requires full spatial coverage and high completeness, especially in the target-rich inner regions of these galaxies.

Figure 48 shows the spatial distribution of candidate RGB stars in the environs of M31 and M33, as identified by colour-magnitude selection from 400 square degrees of contiguous gi imaging with CFHT/MegaCam as part of PAndAS (McConnachie et al. 2009, 2018b). PAndAS resolves point sources at the distance of M31 (D ~ 780 kpc; McConnachie et al., 2005) to $g \simeq 25.5$ and $i \simeq 24.5$ at SNR ~ 10. More than $10^7$ stellar sources are shown in Figure 48. The typical colour of tip of the RGB stars in M31 is $(g - i) \simeq 1.3$, and $g_{TRGB} \sim 22.5$. PAndAS therefore reaches to (nearly) the horizontal branch level, providing photometry of sufficient depth to provide spectroscopic targets for a 10m-class facility.

In Figure 48, the effective surface brightness of the faintest visible features is of order 33 mag arcsec$^{-2}$. This corresponds to just a few RGB stars per square degree. Note that the disk of the Milky Way is located to the North so there is increasing contamination in the colour-magnitude of the RGB locus by foreground dwarfs; the reddest RGB stars are particularly affected by this source of contamination. Young, blue stellar populations — and even intermediate-age populations such as asymptotic giant branch (AGB) stars — are not present in the outer regions of M31 in any significant numbers. Nevertheless, Figure 48 demonstrates the complexity of the outer halo of M31. Figure 49 is an attempt to more rigorously quantify the relationships between all the various substructures that can be seen in the PAndAS map, and which uses only the projected stellar positions (see McConnachie et al. (2018b) for details). Clearly, to make further progress in understanding the hierarchy of strucutre present in this galaxy's halo requires the addition of physical parameters estimated through spectroscopy.

Any spectroscopic study of the outer regions of the M31 halo will necessarily concentrate on the older, evolved, RGB population. For this reason, we consider separately surveys of the outer halo (characterized by a low surface density of evolved giant star candidates) and the inner galaxy (with a high surface density of targets from a mixture of stellar populations).

- **An outer halo survey of M31/M33**: This program aims at obtaining a complete, magnitude-limited, spectroscopic census of every star in the outer regions (40−150 kpc) of an L* galaxy halo to provide complete kinematics for every star, supplemented by metallicity estimates for most stars and $\alpha$ abundances for the brightest ones. Ultimately, such a survey will allow us to chemodynamically deconstruct a nearby galactic halo. Such a survey will provide a crucial testbed for the hierarchical formation of L* galaxies, and yield the optimal data set to constrain the dark matter content of M31 and M33.

- **Stellar populations in the inner regions of M31/M33**: This program focuses on the thin disk/thick disk/halo transition region to measure the extent of the disks, characterize the relationship between these components, and to determine the role of

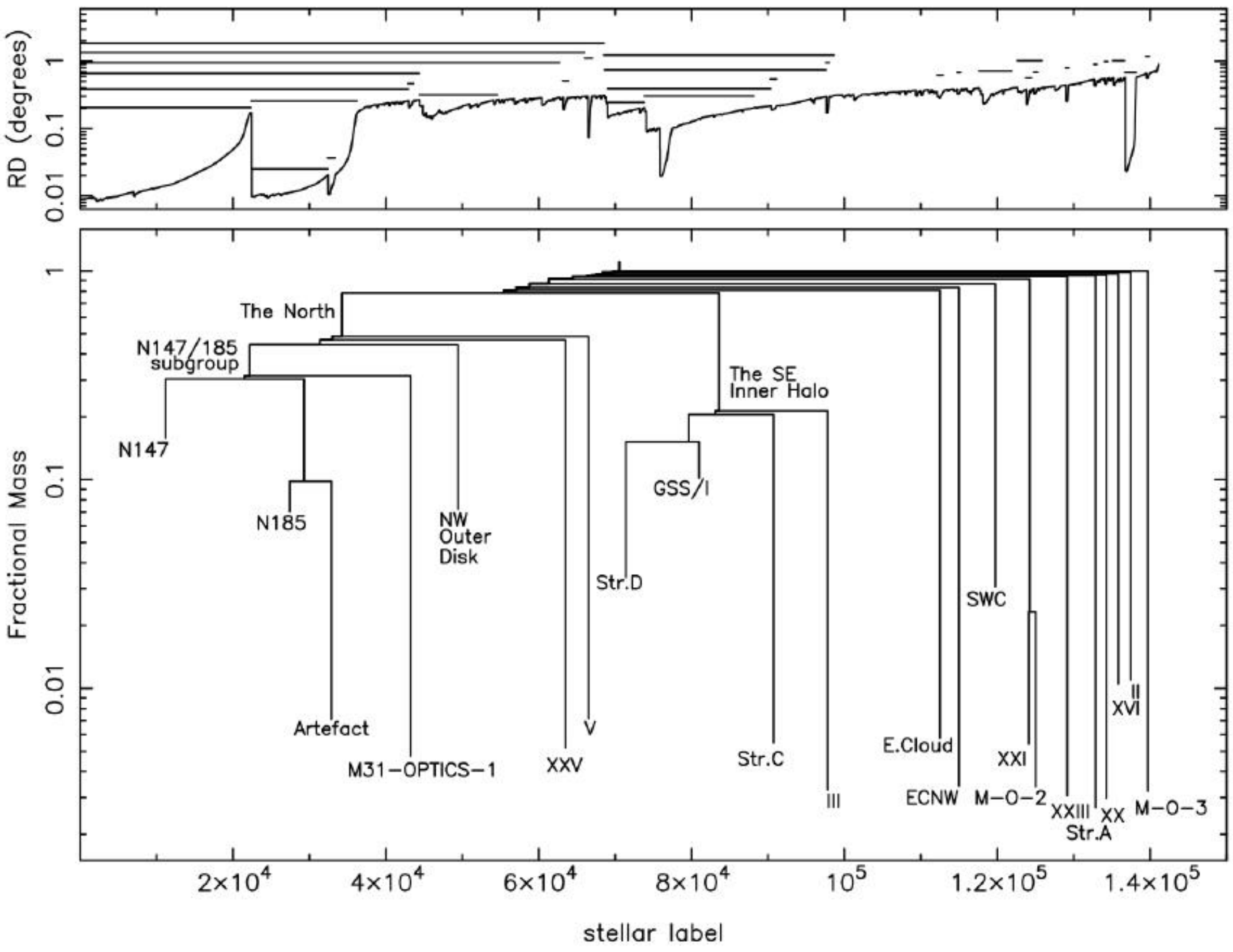

*Figure 49: Top panel: the M31 stellar halo "reachability diagram", that quantifies the clustering of stars around M31 (see McConnachie et al. 2018b for details). Clusters of stars appear as valleys in this diagram, and automatically identified clusters are indicated with horizontal lines. Bottom panel: tree diagram, where each cluster is represented by a vertical line centered on the middle of the cluster on the x-axis. The starting (lower) position on the y-axis is the fractional mass the cluster contains relative to the total mass. Horizontal lines connecting clusters indicate the merging of multiple clusters into their parent cluster. Figure from McConnachie et al. (2018b).*

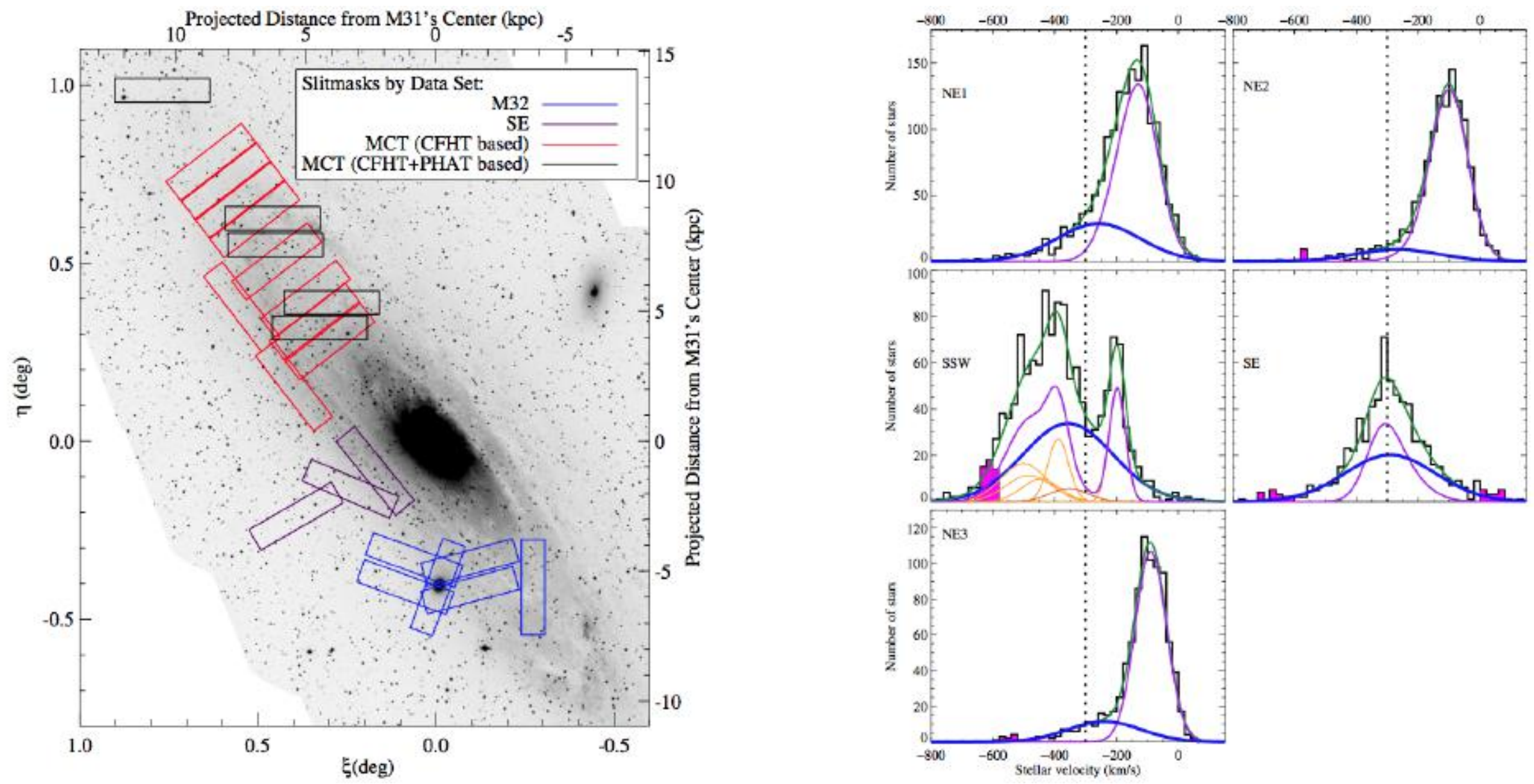

*Figure 50: Left panel: The positions of twenty-four Keck/DEIMOS multiobject slitmasks from the SPLASH survey of M31 overlaid on the Choi et al. (2002) KPNO Burrell Schmidt B-band mosaic image of M31. Note that this image covers approximately 1.5 degrees on a side, to be compared with the ~ 25 x 25 degree image shown in Figure 48. Right panels: Maximum likelihood fits of a kinematically hot spheroid to each of five sub-regions of the survey, after excluding the velocity range encompassing the giant stellar stream (Ibata et al., 2001) and its associated tidal debris (shaded pink). Violet lines show the cumulative region cold component; blue show the best-fit spheroid Gaussian; green show the sum of these two components. The dashed lines show the systemic velocity of M31 relative to the Milky Way. Individual subregion cold components are shown in orange in the SSW region panel, but left out of the other panels for clarity. The complexity of the disk region of M31 is clear. Figures from Dorman et al. (2012).*

mergers in the evolution of the disk and inner halo (see also work by the Panchromatic Hubble Andromeda Treasury, PHAT, Dalcanton et al. 2012)

The richness of this dataset can be anticipated by considering what has already been achieved for M31 and M33 using Keck/DEIMOS (hundreds of pointings in the vicinity of M31 targeting RGB star candidates, for a total of $> 10,000$ stars, primarily by the SPLASH and PAndAS collaborations (e.g., Collins et al. 2011; Dorman et al. 2012 and referneces therein). Keck/DEIMOS has a $5 \times 16$ arcmin field of view and is able to observe of order 100 targets per pointing, making it the most well suited large aperture spectrograph currently available for studies of M31.

Figure 50 show velocity histograms for RGB stars in the vicinity of the disk of M31, from spectroscopic studies of this region by Collins et al. (2011) and Dorman et al. (2012). These authors show that there are significant contributions to these histograms from the inner spheroid of M31 (the locations of their fields and the resulting velocity histograms are shown in Figure 50). It is clear that this region is complex, and maximum likelihood analysis suggests that multiple components are present in each field. They find evidence that the spheroid rotates, and more closely resembles that of an elliptical galaxy than a typical spiral galaxy bulge.

All of the analyses of the dynamics of M31 stars are based on kinematic subsamples that are very sparse and highly incomplete. More than 10 million RGB candidates are present in the PAndAS map of M31 shown in Figure 48, and yet more than a decade of research using Keck/DEIMOS has barely sampled 1% of the possible candidates. The capabilities of MSE are essential for collecting a comprehensive data set for both the inner and outer regions of M31. The high stellar density in the inner regions requires extreme multiplexing, and the large extent and significant foreground contamination on the outskirts of M31 require a wide field of view and a high multiplex.

To quote Dorman et al. (2012): "the literature is full of vocabulary such as 'bulge,' 'spheroid,' 'inner spheroid,' 'outer spheroid,' 'disk,' 'thin disk,' 'thick disk,' 'extended disk,' and so on. There is not yet a consensus on the best combination of these nouns to represent M31". M31 exhibits extreme dynamical complexity, due in part to its vigorous history of minor mergers. The next major advance in our understanding of its structure and history requires a coherent and holistic view of stellar kinematics and abundances across its entire spatial extent, which MSE is uniquely able to provide.

### 5.5.2 Dwarf galaxies

The Local Group contains dwarf galaxies across a wide range of mass and morphology. Intensive spectroscopic surveys are required to advance our understanding of their dynamics so that we can unveil the properties of the dark matter subhalos they inhabit. Recent efforts to systematically gather large radial velocity samples with sub-km/s uncertainties for nearby dwarf galaxies have shown the power of data sets of a few thousand spectra in a single system (Walker et al., 2009). However, as the dynamical modeling of these systems improves, these data sets are also now showing their own limitations, and it is necessary to extend the information gathered from velocities to the realm of chemical abundances (see Walker & Peñarrubia 2011; Kirby et al. 2008, 2011; Strigari 2013). At the moment, large uncertainties remain on the properties of dwarf galaxy dark matter halos (total halo mass, inner dark matter density profile, dynamical state, etc.), and some have gone so far as to question whether these objects contain dark matter at all (Hammer et al., 2018). We refer the reader to Chapter 6 for an extensive discussion of using dwarf galaxies to better understand the properties of the dark matter particle. The presence of stellar substructures within some dwarf galaxies, as well as the presence of extratidal stars in the outskirts of others, all hint at complexity that can only be addressed using large data sets compiled via high-precision spectroscopy. The recent discovery of a past Galactic merger with the Gaia-Enceladus system further highlights the importance of considering dwarf galaxies as remnants from the population of primordial systems that built the Galactic halo.

MSE will bring about an entirely new era for nearby dwarf galaxy studies, enabling accurate chemo-dynamical measurements to be performed efficiently across the full range of dwarf galaxy luminosities ($10^{3-7}L_{\odot}$). With respect to current studies and for all Local Group dwarf galaxies with $\delta > -30°$, MSE will provide spectra for at least an order of magnitude more stars in each system, reaching well beyond where circular velocity curves are expected to peak. Furthermore, the high multiplexing and large field of view of MSE will enable efficient spectroscopic variability surveys of the fainter half of the Local Group dwarf galaxy sample ($< 10^5 L_\odot$). This capability will provide crucial information about unresolved stellar binary systems, long suspected of systematically inflating the velocity dispersions measured for the coldest ultra-faint dwarfs (e.g., McConnachie & Côté 2010 and references therein). Crucially, MSE would also be well-positioned to understand how the dynamics of dwarf galaxies respond to evolutionary effects, in particular tidal stripping, due to its ability to explore sparse outer fields while ensuring high completeness in the presence of strong contamination. The samples gathered with MSE will have the power to constrain the formation and evolution of the sample of ~70 dwarf galaxies within the Local Group, the only satellite systems that can be observed in such detail. Most of these satellites will not have been studied systematically before MSE comes online, owing either to the faintness of the target stars beyond ~100 kpc and the inability of upcoming 4m-telescope surveys to observe them, or to the lack of survey facilities on 10m-class telescopes, necessary to conduct a systematic survey.

Connecting the dynamics of dwarf galaxies to their metallicities and chemical abundances is a key area of current research and will remain so well into the era of MSE. There are crucial open questions regarding the chemistries, metallicities, and star formation histories of dwarf galaxies that MSE will address:

- The full metallicity distribution, including spatial gradients in metallicity and stellar populations. This would provide important information on whether star formation propagated inwards or outwards and whether dwarf-dwarf interactions have a significant role in their formation and evolution (e.g., see discussion in Stierwalt et al. 2015). These data would also indicate if there was significant metallicity evolution over time, and would make it possible to break the age-metallicity-reddening degeneracy definitively in color-magnitude diagram studies.

- The dispersion in abundance ratios at fixed metallicity: e.g., the patterns of [$\alpha$/Fe] vs. [Fe/H] and the relative proportions of other element groups in relation to the iron-peak elements (i.e., light elements, odd-Z elements, light and heavy s-process, r-process). These abundances hold important clues to the history of star formation in these galaxies and about the links between accreted dwarf galaxies and the Milky Way's globular cluster content (see also Chapter 3 for a detailed discussion of the stellar populations in globular clusters).

- A census of rare stellar species (extremely metal-poor stars, carbon stars, etc), including their overall numbers and spatial distributions.

As an illustration of the sheer power of MSE for chemodynamical studies of nearby dwarf galaxies we consider NGC 6822. It is one of the nearest dwarf irregular galaxies, and lies at an ”intermediate” distance for Local Group galaxies (~500 kpc). It is one of the more intriguing targets for detailed study because of ongoing disturbances in its HI velocity field and very active star formation. There is some evidence for young stellar populations associated with infalling HI clouds, and for deviations from circular disk rotation. However, the large angular

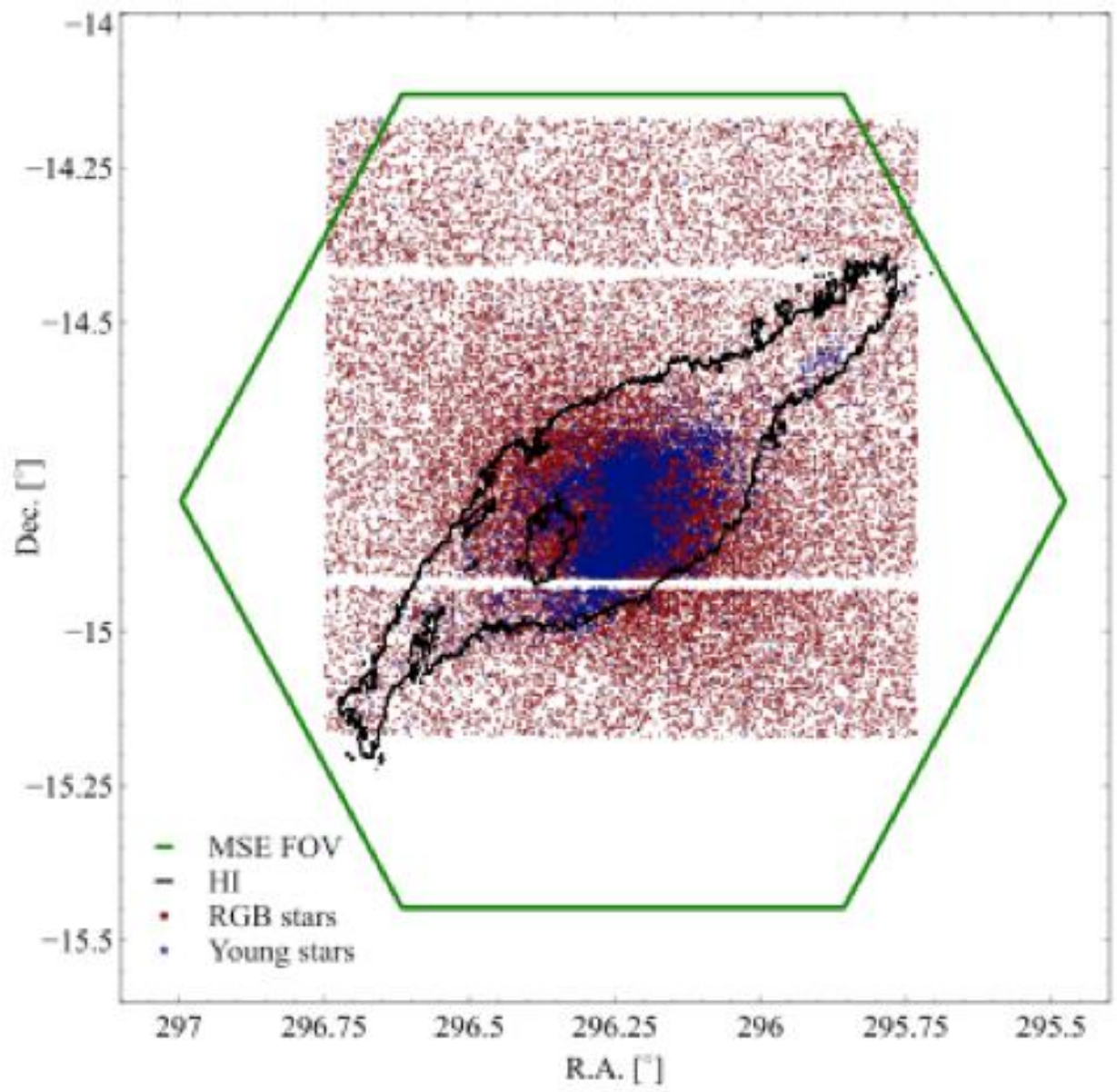

*Figure 51: Spatial distribution of red giant branch stars (red dots) in the barred irregular galaxy NGC6822 from CFHT/MegaCam observations (see Higgs et al. 2016). The contours show the boundary of the HI disk from de Blok & Walter (2006). The MSE field of view is shown for comparison; clearly, MSE is ideal for wide field spectroscopic studies of Local Group dwarf galaxies*

scale of the system (∼ 1 degree across) and the likelihood that the substructures contain only a small fraction of the stars means that the system remains poorly understood, although the sample of stellar spectra from VLT/FLAMES and Keck/DEIMOS is steadily growing. Nevertheless, these have not been enough to identify the population substructures or true dynamical state this galaxy or galaxies similar in mass (Kirby et al., 2012, 2014). Even smaller (and, apparently "simpler") galaxies like WLM still have large uncertainties (e.g., Leaman et al., 2013). Indeed, it is really only for the Small Magellanic Cloud that the structure is now becoming apparent based on spectroscopic samples of several thousand stars (e.g., Dobbie et al., 2014a,b). Obtaining thousands to tens of thousands of spectra in these dwarf galaxies is therefore necessary for understanding dwarf galaxy structure and evolution.

The number of potential targets in NGC6822 available for spectroscopy with MSE is far beyond the capabilities of existing multi-object spectrographs. For illustration:

- the number of potential "red star" targets (i.e., stars with ages > 500 Myr, most of which are likely older than 1 Gyr) is:
  - $\sim$ 2500 stars above the tip of the red giant branch (TRGB), including $\sim 10^3$ carbon stars ($18 < I < 19.5$);
  - $\sim$ 7500 stars within 0.5 mag of the TRGB ($19.5 < I < 20$);
  - $\sim 30,000$ stars between 0.5 and 1.5 mag of the TRGB ($20 < I < 21$).

- the number of potential “blue star” targets is:
  - $\sim$ 1000 stars with ages < 50 Myr ($15.5 < g < 18.5$);
  - $\sim$ 1000 stars with ages 50–200 Myr ($18.5 < g < 20.5$);
  - $\sim$ 2000 stars with ages 200–400 Myr ($20.5 < g < 21.5$).

With MSE, tens of thousands of member stars spanning all ages could easily be observed with multiple fiber set ups in a single pointing. It would therefore be possible, for example, to measure radial velocities accurate to better than 5 $\mathrm{km\,s^{-1}}$ for every AGB and red supergiant star, and nearly all RGB stars within 1.5 mag of the TRGB, in just a handful of MSE nights at medium resolution and with SNR $\sim 10-20$. For bright member stars, repeated observations over a period of several years would allow unprecedented studies of variability for stars in the late phases of evolution.

Closer to us, all observable stars in recently identified ultra-faint Milky Way dwarf galaxies could be targeted with a single fiber configuration and a monthly to yearly cadence to consistently study spectroscopic variability (e.g., Koposov et al., 2010; Martinez et al., 2011a). For more extended and brighter systems, the possibility to probe down to the oldest main sequence turn-off will allow for an exquisite modeling of the gravitational potential with thousands of targets. It would also enable a systematic mapping of the stellar extent of the dwarf galaxies, combined with a search for extratidal stars on a much larger scale than previously.

### 5.5.3 Globular clusters

The multiplexing capability of MSE will allow us to capture detailed information about globular clusters and their member stars across many different environments within the Milky Way and the Local Group dwarfs. Metal-poor globular clusters may have formed in dark matter mini-halos ($< 10^8$ $\mathrm{M_\odot}$) in the earliest phases of galaxy evolution (Saitoh et al., 2006). Indeed, extended stellar halos have been discovered around several Milky Way globular clusters (e.g. Kuzma et al., 2018; Marino et al., 2014), which closely resemble theoretical predictions of globular clusters evolving in their own low-mass dark matter halo (Peñarrubia et al., 2017). However, a similar signal in the stellar density profile is expected from debris of a progenitor dwarf galaxy, and from the interaction between a dark matter-free globular cluster and Galactic tides. Kinematics and abundances are required to disentangle these three scenarios.

The large field of view of MSE is ideally suited for exploring the kinematics and abundances of stars in the outskirts of globular clusters out to the dynamical radius ($\sim 40-60$ arcmin). The large number of fibres, combined with membership information from Gaia, makes it possible to confidently identify outlying member stars that can be used to study the interaction between the cluster, the Galactic tidal field, and any dark matter halo belonging to the cluster. The kinematics of stars in the outskirts of globular clusters is highly complementary to the emerging kinematics of stars in cluster centres from integral field unit studies (e.g., Kamann et al., 2018) and Hubble Space Telescope (HST) proper motions (e.g., Watkins et al., 2015) and will be important for mass modelling efforts. Globular clusters are typically mono-metallic while dwarf galaxies experience metallicity evolution over time. As a result, we can use metallicities from the low-resolution MSE fibres to firmly classify these low-mass systems as “true” globular clusters or dwarf galaxy remnants.

Studies of cluster formation, evaporation and destruction by the Galactic field require a thorough mapping of cluster dynamics, based on radial velocity measurements for as many stars as possible. The global velocity field can reveal the presence of rotation, warped structure or kinematic subgroups due to mergers, while the velocity dispersion profile gives insight into the dynamical state of the cluster. The peak projected rotational velocity near the cluster core is in the order of, or below, the internal velocity dispersion. Therefore, a velocity precision below 100 m/s is needed for many cluster members at different projected positions, and probabilistic global model fitting is required to evaluate these data. MSE is especially well suited to study the rotation of globular clusters, as its radial velocity precision will be below 100 m/s, and it will observe a large number of stars in each cluster.

Global rotation has been observed in a few globular clusters (Lane et al., 2010; Bianchini et al., 2018). The commonly accepted explanation involves the cluster having undergone a merger with another cluster in the past, which could provide an explanation for the presence of multiple stellar populations in globular clusters. These multiple populations are seen in color-magnitude space, and in the light-element abundance anomalies that are seen to some degree in all Galactic globular clusters and some clusters in the Magellanic Clouds (e.g., Hénault-Brunet et al., 2015).

These anomalies take the form of anticorrelations between C and N, O and Na, and Mg and Al, a pattern that resembles the equilibrium abundances produced by the CNO, NeNa, and MgAl hydrogen burning cycles. The most extensive spectroscopic survey of the result of NeNa and MgAl cycles in globular clusters (Carretta et al., 2009c,b,a) revealed that these abundance inhomogeneities are ubiquitous in Galactic globular clusters, though they do not appear to occur in other star formation environments. Large-scale high-resolution spectroscopic surveys have expanded the available data on globular cluster abundances in recent years, including a study by Mészáros et al. (2015) investigating the C-N and Mg-Al anticorrelations in globular clusters in the APOGEE Majewski et al. (2017) survey. This was extended by Masseron et al. (2018) to include APOGEE data for more stars in the same clusters, with an updated analysis method that made it possible to study finer details of the C-N and Mg-Al anticorrelations. Pancino et al. (2017) used data from the Gaia-ESO survey Gilmore et al. (2012) to carry out a similar study in Southern globular clusters, mainly focusing on Mg and Al abundances.

The magnitude limits of the current large-scale spectroscopic surveys mean that they only observe the most luminous evolved stars in each cluster. The ability of MSE to observe down to the main sequence in Galactic globular clusters will result in two key improvements over current survey capabilities. Including lower-mass targets will significantly increase the sample size, and make it possible to study the abundance behavior across a range of stellar evolutionary phases, allowing a clear separation of primordial and evolutionary abundance effects. The increased spectroscopic resolution and S/N will also make it possible to study even finer details of the abundance inhomogeneity, particularly in Na-O and Mg-Al.

Martell & Grebel (2010) combined the abundance anomalies in globular clusters with the concept of chemical tagging (Freeman & Bland-Hawthorn, 2002) to identify red giants in the halo field with a high probability of having formed in a globular cluster, using the SDSS-II/SEGUE survey (Yanny et al., 2009). Further studies by Martell et al. (2011), Ramírez et al. (2012), Lind et al. (2015), Martell et al. (2016a) and Martell et al. (2016b) confirmed the presence of this subpopulation of halo stars, using a variety of data sources and abundance tags. Later work by Schiavon et al. (2017a) and Fernández-Trincado et al. (2017) used the APOGEE survey to uncover a large and unexpected population of N-rich stars in the inner galaxy, and Mg-poor and Al-rich stars in the Milky Way disk, respectively. By collecting detailed abundances for stars throughout the Galaxy, MSE will drastically increase the number of known field stars with globular cluster origins. Chapter 3 discusses other potential uses for stars in Galactic globular clusters, including detailed studies of stellar evolution.

## 5.6 The interstellar medium

### 5.6.1 3D mapping the Galactic ISM

Multi-wavelength observations of the Galactic ISM have never been so detailed nor so abundant as now. High-quality emission maps of the gas and dust have been produced thanks to a wide range of ground- and space-based facilities that operate at almost all wavelengths, including $\gamma$ rays, X-rays, UV, IR, sub-mm/mm, mm and radio. Datasets come in the form of two-dimensional images in specific spectral bands, or as data cubes (i.e., two spatial dimensions plus a spectral, or a polarimetric, dimension). Yet, paradoxically, our understanding of the ISM structure remains surprisingly limited, even at the largest spatial scales, primarily because we lack accurate distances to the structures that are responsible for the observed emission and their three-dimensional (3D) distribution.

Our understanding of the origin, evolution, interplay of the various phases of the Galactic ISM and of the links with star formation and stellar populations in general, is seriously hampered by the lack of realistic 3D density and velocity distributions, and this lack prevents the construction of quantitative, physically motivated models. A very caricatural example is the one of the so-called North Polar Spur/Loop 1 features, the most prominent features of the X ray, sub-mm and radio continuum full-sky maps, generated in the Solar neighborhood or at the Galactic Centre depending on authors. This lack of 3D perspective impacts negatively on a large number of research fields: influence of turbulence, magnetic field, metallicity or dust properties on star formation, models of cavities blown by stellar winds and supernovae, mixing of stellar ejecta, large scale evolutionary modeling of the Milky Way bulge, bar, and spiral arms, respective roles of accretion and internal evolution on the large scales, etc. For all these fields, the 3D structure of the ISM is an invaluable ingredient to make full profit of 2D photometric, spectroscopic, and polarimetric datasets.

Moreover, some of the major ground or space projects require a highly detailed description of the Galactic ISM. For instance, interpreting observations dedicated to the cosmic microwave background (CMB) polarization requires an accurate description of the Galactic dust emission spectra and polarization properties and, subsequently, of the spatial distribution of grains, their size distribution and temperature. To produce realistic models of the radiation field that heats the dust, it is necessary to compute the propagation of photons, and subsequently, again, the 3D distribution of the interstellar matter. Similarly, understanding the energy spectrum and spatial variability of Galactic cosmic rays (GCRs) requires a detailed knowledge of the 3D ISM distribution through which they propagate, and modeling of the diffuse $\gamma$ ray emission doubly requires this knowledge because the emission is generated by interaction of cosmic rays with ISM nuclei and by up scattering of the interstellar radiation field by cosmic electrons. Finally, obscuration by dust poses a significant obstacle to our understanding of stellar populations within the Milky Way, and, in order to characterize the properties of distant stars one has to know the reddening amplitude and the reddening spectral law very accurately. In the case of Gaia, star and IS dust studies must be conducted simultaneously.

Studies of specific astrophysical targets also benefit from detailed three-dimensional maps of the ISM. Knowing foregrounds, environments and backgrounds helps identifications and disentangling of features, as well as studies of interactions with the ambient medium. There is an increasing number of structures discovered around stars in the UV, optical or infra-red (bow-shocks, trails, ionization cavities) and their understanding depends on the physical and dynamical properties of the ambient medium. Finally, searches for particular categories of objects could be more efficiently conducted if one may use maps to identify the most favorable sky regions.

Building accurate 3D maps of the ISM requires (i) massive amounts of distance limited absorption data, i.e., stellar surveys at high spectral and spatial resolution covering large fractions of the sky, and (ii) accompanying information on the target distances. MSE and Gaia will provide this powerful and unprecedented synergy. Gaia and photometric surveys can be used to build 3D dust maps, however, only high resolution spectra provide full information on the interstellar absorbers, including on their kinematics, and only high resolution spectra provide full information on the stellar parameters, and, in turn, accurate estimates of the reddening and reddening law associated with the sightline. The first set of information can be used to compare different species and deduce physical and chemical properties of the absorbing medium, and, importantly, can be combined with emission data (especially CO, HI radio spectral cubes) by means of velocity matches, allowing to assign distances to the emissive structures previously identified in position-position-velocity space. The second series of informations can bring invaluable input on the reddening law and the dust properties. Therefore, one can distinguish several levels of mapping studies:

- An initial distance assignment for intervening clouds based on all types of absorption data, including dust reddening, gaseous lines or diffuse interstellar bands. The resulting 3D maps will be a general tool of wide use.

- Differential mapping, i.e., comparison of the distributions of the various absorbers, e.g., diffuse interstellar bands (DIBs) and dust. This enables a variety of studies on the multi-phase structure of the ISM.

- Combination with emission data through Doppler shift matches.

### 5.6.2 Three-dimensional ISM mapping: MSE perspectives

Several mapping efforts have been done based on various techniques and data types (for a more complete list, see Lallement et al. (2019). Most techniques reconstruct the column of dust or IS species in narrow beams and treat angular bins separately. Marshall et al. (2006) produced the first large scale reddening map by adjusting the 3D Galactic dust distribution to 2MASS photometric data and the Besancon stellar population model. Green et al. (2015, 2018) developed a Bayesian reconstruction of reddening profiles using Pan-STARRS and 2MASS data. Kos et al. (2014) and Zasowski et al. (2015) built 3D maps of integrated columns of a diffuse band carrier based on RAVE and SDSS/APOGEE respectively. A full threedimensional Bayesian method has been developed by Vergely et al. (2001) and applied to various datasets, including composite reddening-DIB data (Lallement et al., 2014; Capitanio et al., 2017; Lallement et al., 2018). Sale et al. (2014) and Rezaei Kh. et al. (2018) developed a 3D multi-scale inversion technique and applied it to data from the IPHAS survey and APOGEE respectively.

Recently the situation has evolved very rapidly thanks to the Gaia data. Chen et al. (2018) and Lallement et al. (2019) presented 3D maps based on Gaia-DR2 distances and photometry, in combination with IR survey photometric data. Figure 52 shows the dust distribution in the midplane.

It is clear that this situation will change dramatically with the combination of Gaia and MSE. Future updated Gaia parallaxes and future Gaia spectrophotometric and spectroscopic data will be available for combination with all new informations on absorption lines and reddening from optical/infrared spectroscopy, allowing to revolutionize studies of the ISM and its 3D structure. Reciprocally, Gaia data analyses will in turn benefit from the increasingly accurate and detailed Galactic extinction maps by allowing to break degeneracies between reddening and temperature for very faint targets.

Such progresses require massive, all-sky data and high resolution (to disentangle, extract absorption lines and bands), and MSE, with wide-field, high multiplexing, high resolution, high efficiency and high spectral coverage fulfills ideally all requirements. In addition to unprecedented 3D mapping, it would also address long-standing questions about the ISM such like the formation and destruction sites of many interstellar species, including DIB carriers.

- The large aperture and high throughput increases the number of weak targets that can be individually studied and the distance coverage. In the case of large-scale ISM mapping, MSE s high observing efficiency reduces survey durations to tractable levels, putting it far beyond the reach of any other existing or planned facility/instrument.

- The wide field of view facilitates the required coverage over large fractions of the Galaxy.

- The high multiplexing increases the number of targets along a given sightline and the spatial resolution of the maps. For detailed study of individual structures, MSE s wide field and dense fibre coverage is well matched to many Galactic features, such as stellar streams, astrospheres, and individual interstellar clouds, targets that would otherwise be accessible in the radio region alone; columns and/or stellar types used to derive the reddening, and it allows us to couple absorptions with emission spectra by means of velocity identification.

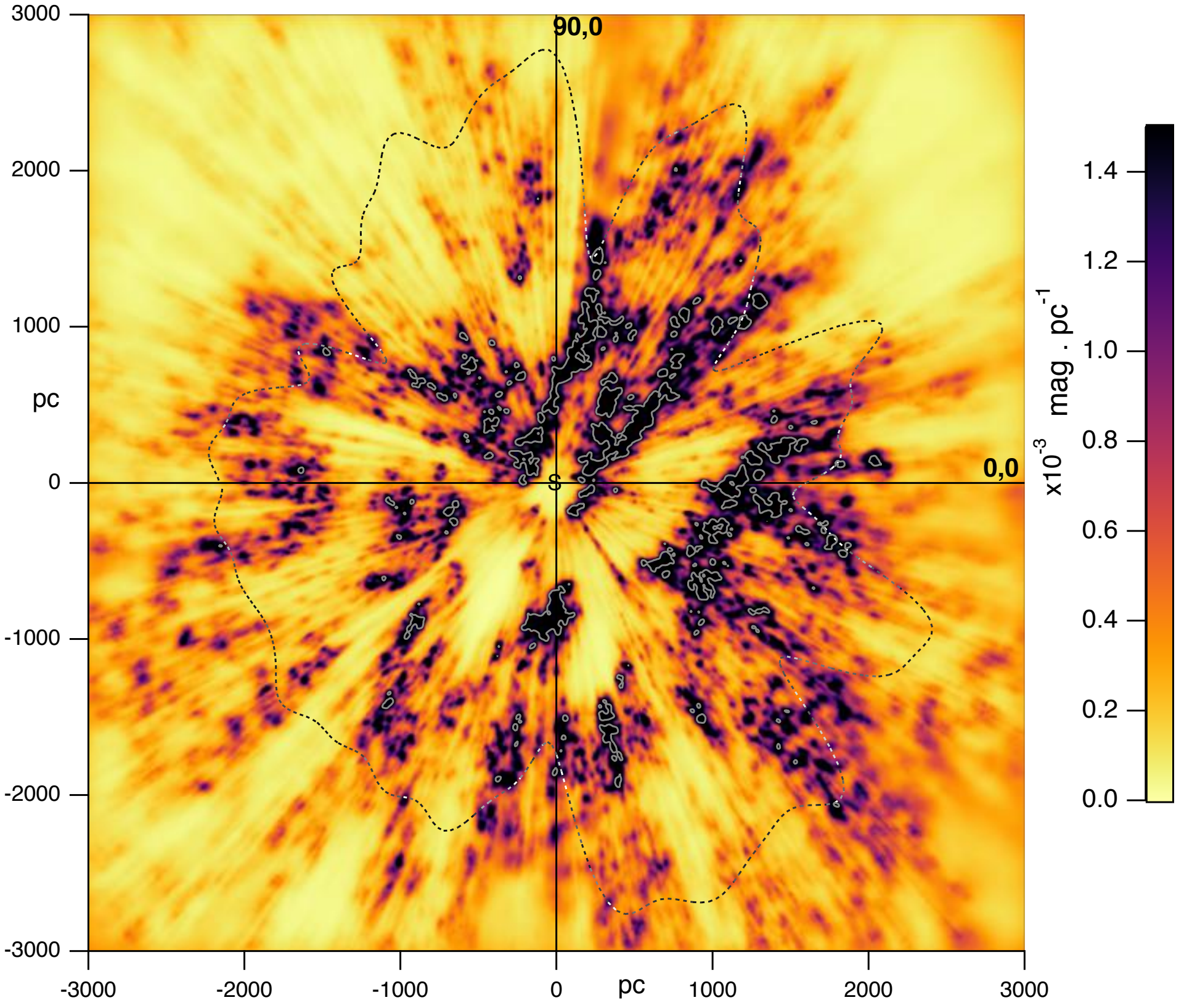

*Figure 52: Dust density along the Galactic plane (Sun at the center; the Galactic center to the right) based on Gaia DR2 and 2MASS. The color scale represents the differential extinction in units of mag per parsec. The dashed black line represents the distance beyond which the final resolution of 25 pc is not achievable due to target scarcity. The dotted white contours correspond to a differential extinction of 0.003 mag/pc and delimit the dense areas. MSE will be able to extend the scale of this map up to the outer disc borders and add extra dimensions with the ISM velocities and physical properties. Figure and caption from Lallement et al. (2019).*

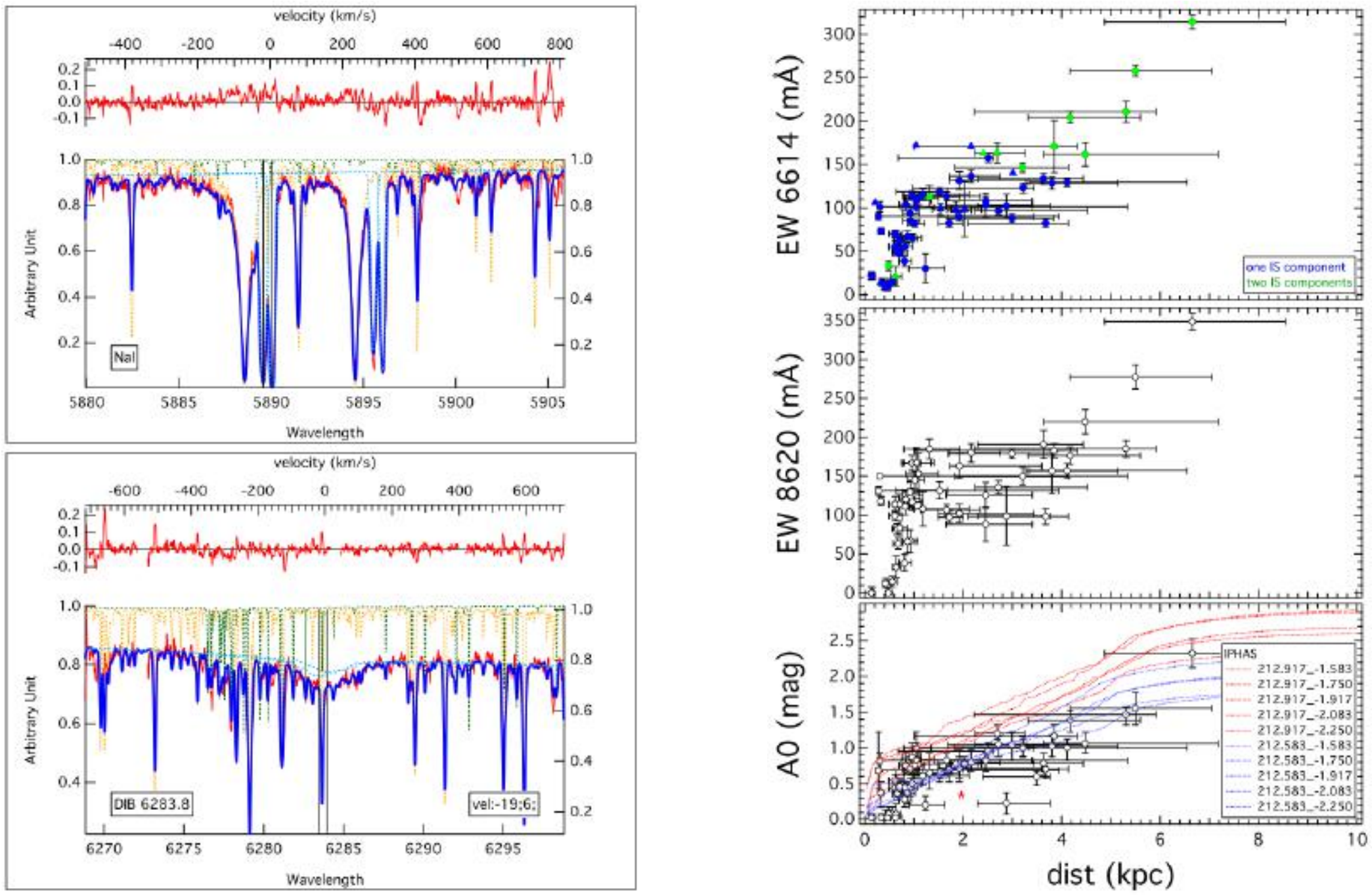

*Figure 53: Illustration of ISM mapping based on spectroscopy. Left: Interstellar NaI doublet (top) and the 6284 Å diffuse band (bottom) extracted from a V=14.7 target star at D=2.8 kpc. The $R = 48K$ VLT/UVES spectrum (red) is modeled with the product of a synthetic stellar spectrum (yellow), an interstellar profile/DIB model with two velocity components (light blue) and a synthetic telluric absorption (green). Right: application of such fitting methods to radial cloud mapping based on the strengths of two DIBs (6284 Å, 8620 Å; top and middle panels, respectively) and extinction (A0; bottom panel). Distances are derived from the spectroscopic stellar parameters and photometric data. DIBs and extinction vary in a similar way and trace the local Arm and Perseus. Individual spectroscopic data show a stronger variability that may correspond to different physical properties of the clouds or smallscale structure, illustrating the potential of high resolution spectroscopy compared to photometry. MSE and Gaia distances should produce massively increased and improved data of this kind. Figures from Puspitarini et al. (2015).*

The wide spectral interval of MSE offers the possibility to extract a large number of gaseous species absorptions. We list in Table 5 the major gaseous lines and the interstellar gas phases they are tracing. While the ionized and diffuse atomic ISM fraction is best traced in the optical, the molecular and dense atomic phase is best traced at blue wavelengths. Molecular clouds are detected in the blue with CH (4300Å), $CH^+$ (4230Å) , CN (3870Å) and also metallic lines such as TiI (3630Å), FeI (3860Å), MnI, NiI, AlI (3940Å), and some information on the molecular phase can be probed with the infrared bands of $C_2$ around 8780Å.

| Feature | Wavelength | Probe | Feature | Wavelength | Probe |
|---|---|---|---|---|---|
| NaI | 5890 | Strong | CH | 4300 | Molecular |
| | 5896 | | CH+ | 4230 | |
| CaII | 3934 | | CN | 3870 | |
| | 3968 | | C2 | 8780 | |
| CaI | 4227 | | TiII | 3230 | Metallic |
| DIB | 5780 | | | 3240 | |
| | 5797 | | | 3380 | |
| | 6284 | | TiI | 3630 | |
| | 6196 | | FeI | 3860 | |
| | 6614 | | MnI | 3940? | |
| NaI | 3300 | Faint | NiI | 3940? | |
| KI | 7699 | | AlI | 3940? | |

*Table 5: Major ISM diagnostics.*

The ISM will be studied as part of the observations that will be planned for the Galactic Archaeology survey, with more than a million sightlines spread across the sky. This ISM survey would be extremely efficient since high-quality spectra (i.e., SNR ~ 100) could be obtained in just ~ 15 min exposures for $g$ = 16 stars of spectral types O, B and A. For reference, a survey of 4950 square degrees covering a swath of 20 degrees centered on the Galactic plane would require ~ 1000 hrs of MSE, or about two weeks per year over the course of a decade-long survey. Such observations could be undertaken in conditions of very high sky brightness.

An illustrative example of the MSE capabilities discussed above is the multicomponent absorption study of one of the fields of the Gaia-ESO Spectroscopic Survey (GES) that is shown in Figure 53. VLT/FLAMES (R = 18 000, 48 000) spectra were modeled as the product of a stellar synthetic spectrum, a DIB or line model and a telluric absorption model, as shown in the left panels. The velocity structure and the absorption strengths evolve with distance, tracing the spiral arms and their kinematics, as shown in the right panels. MSE survey data, by far superior in angular and distance coverage and using Gaia distances, will provide the kinematics of the ISM with unprecedented detail along with its 3D distribution.

### 5.6.3 Targetted studies of the ISM

In addition to the large-scale mapping of the ISM described in the previous section, focused studies in specific regions of the Galaxy can also provide unique ISM science. The list is long and we describe below only several examples, cognizant that other fields of particular interest will arise at the time of MSE first light.

**Small-scale structures seen in polarization, the interface between stars and the ISM:** Detailed information on small-scale (~1”) structures in the ISM is now available from radio polarization data (e.g., Landecker et al., 1999; Ransom et al., 2010; Wolleben, 2007; Wolleben et al., 2010). Some of these structures correspond to shells generated by stellar winds or supernovae. Other structures trace the interfaces between evolved stars or planetary nebulae (PNe) and the ambient ISM, including tails behind fast-moving PNe. Such interfaces, albeit much larger, are also seen at other wavelengths, most notably in the infrared (e.g., Spitzer arcs in Orion) or in the UV (i.e., the spectacular case of Mira's tail; Wareing et al., 2007). In many (and perhaps most) cases, the origins of the detected polarization signatures are far from clear. The lack of information on the distance and size of these objects obscures their true nature and undermines efforts to construct quantitative models.

Observations with MSE are ideal for advancing this area of study. Dedicated observations of target stars located in the field and associated absorption measurements would yield more accurate distances to the radio polarization sources, and thus provide accurate physical dimensions. Moreover, the ability of the new observations to detect variations in the velocity structure at very small angular scales will allow for direct comparison to hydrodynamic simulations of the star – ISM interaction, enable improvement of those models, and provide critical information for the construction of rotation measure models. These models will give important information on the magnetic-field structure in the interaction and tail regions. Emission lines in the bow-shock and tail structures may also be detected and their variability studied as a function of location.

**Supernova Remnants:** Increasingly detailed radio, optical and X ray observations of supernova remnants have raised a number of important questions. Why are there such large differences in the ratios between X ray and radio emissions for supernovae of the same type and with comparable radio emissions? Why are there radiative recombination continua (interpreted as freely expanding and recombining gas) in X ray spectra of mixed-morphology supernova remnants that are known to be interacting with molecular gas (e.g., Miceli et al. (2010))? The nature of the remnant, and the detailed distribution of the ISM surrounding it, are key parameters in understanding these interactions that play a major role in galactic ISM recycling. Observations of the field stars around supernova remnants with MSE will allow the measurement of absorption and emission features that hold clues to the ISM structure, enable correlation studies, and ultimately lead to models that describe the ejecta expansion and shock properties.

**Small-scale structure of diffuse molecular clounds and the CH+ problem:** Whether or not diffuse interstellar clouds are clumpy remains a hotly debated issue. In particular, any structure in the spatial distribution of the major molecule, H2, is very difficult to identify since the column density of this species can be measured only in the far UV, which requires spectral observations of bright background stars with O or B spectral types. The degree of H2 clumpiness is critically important for modeling the abundances within these clouds as it affects the penetration of UV photons and thus photo-destruction processes. Because H2 is closely correlated with CH (Federman, 1982), observations of the blue lines of this radical can be used as a surrogate for H2. By selecting appropriate clouds at intermediate latitudes where the confusion is minimal and the surface density of background stars is still high, one can map the distribution of CH and infer that of H2. The use of background stars makes it possible to probe the spatial structure over a surprisingly broad range of scales, with a dynamic range of about 3000. While the largest separations (~1.5 degrees or ~3 pc for a cloud at 100 pc) can be used to delineate the overall geometry of the cloud and its boundaries, the smallest ones (a few arcseconds, or about 0.001%) would probe the structure at very small scales.

MSE spectra taken at a resolution of $R \geq 20K$ would be sufficient to reach these objectives because, apart from the main transition at 4300Å, several additional weaker features are available around 3890Å, allowing the measurement of line opacities and, hence, CH column densities even if line profiles are unresolved. Such observations will provide simultaneous, and invaluable, information on the CH+ problem . The high abundance of CH+ in the ISM is presently not understood: the reaction presents an energy barrier of 4600K and is insufficient to produce the observed amounts of CH+ in the conditions prevailing in diffuse molecular gas. Similarly, the large abundance of H2 in J > 2 rotational states requires energetic processes that have yet to be identified and which can play an important role in the physics and chemistry of diffuse molecular gas. CH+ is thus an important species in that it can be used to investigate the nature of non-thermal processes that may play a key role in the physics of interstellar clouds. Several scenarios involving either shocks (Pineau des Forêts et al., 1986), vortices (Godard et al., 2009) or cloud/intercloud interfaces have been suggested as the additional energy source required to overcome the CH+ formation energy barrier. All these scenarios imply the presence of localized regions heated to higher temperatures but with very different geometries (i.e., shocks and interfaces are essentially 2-D structures) while in the scenario involving turbulence, vortices are distributed over the whole cloud volume. Thus, a detailed study of the spatial distribution of CH+ with MSE will allow us to identify conclusively the process at work.

**The carriers of the diffuse interstellar bands (DIBs):**

The carriers of DIBs are still unknown, excepted for a few bands associated with the fullerene cation C60+, but it is widely accepted that they are large organic molecules in the gas phase, linear carbon chains, polycyclic aromatic hydrocarbons (PAHs), and fullerene-like compounds being the most favored candidates. Such macro-molecules are key elements of the life cycle of interstellar matter from stellar ejecta to star formation and understanding their role and evolution would help elucidating unclear aspects of star and planet formation. Being likely the largest reservoir of organic matter in the Universe, they additionally deserve a particular interest in the frame of the origin of living systems. There is a large number of DIBs between 4200Å and 9000Å (see, e.g., Hobbs et al., 2009; Friedman et al., 2011). Some DIBs are very tightly correlated with color excess, some with atomic gas, and some with molecular gas. Recent studies have shown that the strength of most DIBs correlates with the radiation field and the molecular fraction and that the carriers disappear in the dense cores, showing that the cloud history and physical state (i.e., shocks, ionization, cooling, condensation, shielding against UV photons, etc.) influences the DIB absolute and relative strengths along with all other species. Through detailed mapping and ISM-phase assignment of the hundreds of diffuse bands detectable in the optical, MSE will bring new constraints on the conditions of formation and disappearance of the large carbonaceous molecules producing the absorption bands. Despite being generally broader than gaseous lines (the narrowest DIBs are about 0.05 nm wide), DIBs also contain important kinematic information as they are Doppler-shifted according to the radial velocity of their carriers, and this information can be combined and compared with the one from the gaseous lines. Depending on their relationship to the extinction or the gas columns, their detection may require high SNR and/or observing highly extinguished stars, and, in this respect, MSE offers optimal capabilities. It would provide for the first time detailed properties of the clouds to which DIB carriers belong and detailed spatial distributions of the DIB carriers within clouds with high spatial resolution, based on spectroscopic measurements of multiple targets close to each other in 3D space, and, in turn, unprecedented information on how those carriers evolve within the cloud. Such observations may also disentangle DIBs that originate from the same types of molecules since they should display the same spatial structure. No such studies have yet been attempted because they require not just DIB measurements but also absorption data (needed to constrain the physical and chemical structure of the ISM). MSE will therefore be the ideal facility for such a study. Targets should in this case include cool, evolved stars in order to maximize the spatial resolution, however, DIB extraction from cool star spectra is now an available technique.

# Chapter 6
# Astrophysical tests of dark matter

**Abstract**

MSE will conduct a suite of surveys that provide critical input into determinations of the mass function, phase-space distribution, and internal density profiles of dark matter halos across all mass scales. Importantly, recent N-body and hydrodynamical simulations of cold, warm, fuzzy and self-interacting dark matter suggest that non-trivial dynamics in the dark sector could have left an imprint on structure formation. Analysed within these frameworks, the extensive and unprecedented kinematic datasets produced by MSE will be used to search for deviations away from the prevailing model in which the dark matter particle is cold and collisionless. MSE will provide an improved estimate of the local density of dark matter, critical for direct detection experiments, and will improve estimates of the J-factor for indirect detection through self-annihilation or decay into Standard Model particles. MSE will determine the impact of low mass substructures on the dynamics of Milky Way stellar streams in velocity space, and will allow for estimates of the density profiles of the dark matter halos of Milky Way dwarf galaxies using more than an order of magnitude more tracers. In the low redshift Universe, MSE will provide critical redshifts to allow the luminosity functions of vast numbers of satellite systems to be derived, and MSE will be an essential component of future strong lensing measurements to obtain the halo mass function for higher redshift galaxies. Across nearly all mass scales, the improvements offered by MSE in comparison to any other facility are such that the relevant dynamical analyses will become limited by systematics rather than statistics.

**Science Reference Observations** (appendices to the *Detailed Science Case, V1*):
**DSC − SRO − 04** Stream kinematics as probes of the dark matter mass function around the Milky Way
**DSC − SRO − 05** Dynamics and chemistry of Local Group galaxies

## 6.1 Motivation

Dark matter has been detected through its gravitational influence on galaxies and clusters of galaxies, the large-scale distribution of galaxies and the cosmic microwave background. A cosmological model with particle dark matter convincingly explains a vast array of observations stretching from kiloparsec scales to the horizon and from the present time to the time of last scattering (Davis et al., 1985).

Dark matter density equivalent to about 0.3GeV/cc has been inferred in the solar neighborhood from the motions of disk stars (Kuijken & Gilmore, 1991; Holmberg & Flynn, 2004; Garbari et al., 2012; Bovy & Tremaine, 2012). We know from the orbit of the Milky Way and Andromeda that the two combined have a mass of about $2\times10^{12}M_{\odot}$, far in excess of all the stars and gas (Kahn & Woltjer, 1959; Peñarrubia et al., 2014). The satellite galaxies orbiting the Milky Way provide strong evidence for dark matter, with inferences of 10 to 1000 times more mass in dark matter than stars (Mateo, 1998; Simon & Geha, 2007; Strigari et al., 2008). Away from our Local Group, every galaxy for which we have dynamical information far enough out of the disk of stars has shown evidence for dark matter. Amazingly, all these measurements are consistent within a factor of two with the predictions of hierarchical structure formation models with dark matter.

In larger systems like groups and clusters of galaxies, we see concrete evidence for dark matter through different methods, with densities that scale in the way expected from dark matter models (Navarro et al., 1997). The large-scale distribution of galaxies stretching over hundreds of Mpc is beautifully explained in the context of a model which includes dark matter (e.g. Davis et al., 1985; Springel et al., 2005). On even larger scales and from the time when the Universe was about four hundred thousand years old, we have clear evidence for non-baryonic (dark) matter in the cosmic microwave background (CMB) anisotropies (Komatsu et al., 2009; Planck Collaboration et al., 2018a).

While the model space of dark matter is large (Feng, 2010), the dominant idea in the particle physics community has been that the dark matter particle is the lightest supersymmetric (SUSY) particle (a "neutralino") or the axion. Both candidates have the virtue that they arose in models designed to solve deep problems in particle physics. However, the neutralino or the axion does not have to be the dominant component of dark matter.

Despite enormous progress in mapping the distribution of dark matter in galaxies and a concerted effort to look for certain kinds of dark matter particles in underground laboratories, at colliders and in space, there is no concrete evidence for the identity of the dark matter particle. The lack of detection has ruled out large parts of parameter space (e.g. Cohen et al., 2013; Aprile et al., 2018; Arcadi et al., 2018) and pushed theorists to explore more general models of dark matter (e.g. Schirber, 2018). Many of the models can be broadly classified as dark sector models, i.e., models where the dark matter lives in a secluded sector that is very weakly coupled to the Standard Model of particle physics. Within this theory landscape, there are many ideas being currently explored (Battaglieri et al., 2017).

In the dark sector, there is no compelling reason to expect the dark matter particle mass to be $O(100\,\mathrm{GeV/c^2})$ (i.e., weak scale mass). Just like Standard Model particles can be light and have appreciable interactions via forces other than gravity, the dark matter in the hidden sector can also be light (sterile neutrino dark matter Dodelson & Widrow 1994; Shi & Fuller 1999; fuzzy dark matter, Hu et al. 2000; Hui et al. 2017) and have large interactions, including interactions with itself (for example, like Hydrogen atoms, Kaplan et al. 2010). There may be ways for the dark matter particles to also cool via inelastic interactions (double disk dark matter, Fan & Reece 2013; atomic dark matter, Cyr-Racine & Sigurdson 2013; Foot 2014). Many of these exciting possibilities can only be tested by astrophysical probes (Boddy et al. 2016; Vogelsberger et al. 2018).

Stepping away from the theoretical landscape, there are other reasons to look towards the dark sector. A long standing puzzle in the galaxy formation community has been the presence of spiral and dwarf galaxies with low dark matter densities in the center. This issue is often referred to as the cusp-core problem. Halos formed in N-body simulations with cold collisionless dark matter (CDM) have central "cusps" such that the density increases with decreasing radius ($r$) as $1/r$, whereas rotation curves and stellar kinematics of many galaxies show evidence for "cores" of uniform density (Moore, 1994; Flores & Primack, 1994; de Blok, 2010; Walker et al., 2011). However, there are also galaxies with similar total baryon content that have densities similar to the CDM predictions. Excitingly, this diversity seems to be consistent with predictions of models where dark matter has large self interactions (Kamada et al., 2017).

At the faint end, it becomes harder to decipher cores and cusps but a similar problem, referred to as the "too big to fail" problem, exists (Boylan-Kolchin et al., 2012b; Papastergis et al., 2015). Solutions to these problems in terms of dark matter physics are more varied and can include dark matter self interactions, warm dark matter or fuzzy dark matter.

It is important to note that just the presence of low-density cores in galaxies is not evidence for deviations from the CDM model. It has become clear in recent years that large cores can be created in the context of CDM models with better feedback prescriptions. (e.g. Brooks et al., 2013; Wetzel et al., 2016; Tollet et al., 2016). This progress underscores the point that we cannot talk about dark matter halo properties in isolation from star formation considerations. There is great diversity of dark matter cores across a wide range of galaxies (McGaugh, 2005; Kuzio de Naray et al., 2010; Oman et al., 2015), and the cores are correlated with the stellar distribution (McGaugh, 2005). The richness of the cusp-core issue, coupled with the rapid progress in hydrodynamical simulations, indicates that it should be possible to disentangle feedback physics from dark matter physics.

The case for using astrophysical observables to constrain or measure particle physics models of dark matter is strong. The recent progress in N-body and hydrodynamical simulations of cold, warm, fuzzy and self-interacting dark matter have helped bolster this case, while a wealth of new observations from dwarf galaxies to galaxy cluster scales has opened up the exciting possibility that non-trivial dynamics in the hidden sector could have left an imprint on structure formation. MSE has critical roles to play in this unfolding story and we highlight these below.

## 6.2 How can astrophysics probe the particle nature of dark matter?

### 6.2.1 Dark matter physics

One of the fundamental predictions of CDM model is that structure formation is hierarchical (White & Rees, 1978), with the smallest structures collapsing first and then merging into larger structures (Davis et al., 1985). Indeed, calculations of the matter power spectrum associated with popular "weakly interacting massive particle" (WIMP; e.g., super-symmetric neutralinos) candidates for the dark matter particle imply that the minimum mass of self-bound structures could be as small as an earth mass (Hofmann et al., 2001; Green et al., 2004; Diemand et al., 2005; Loeb & Zaldarriaga, 2005; Bertschinger, 2006). Halos formed in a hierarchical structure formation scenario are predicted to have subhalos, with subhalos hosting their own sub-subhalos down to the scale of the "minimum" mass set by the particle physics. To date, the smallest dark matter halos that have been inferred from observations have mass $\sim 10^5 M_{\odot}$ within their luminous regions (often extrapolated to virial masses $M_{\rm vir} \sim 10^8 M_{\odot}$). These observations thus allow for the possibility that additional physics in the dark sector may inhibit structure formation on smaller scales (Spergel & Steinhardt, 2000; Bode et al., 2001; Dalcanton & Hogan, 2001; Kaplinghat, 2005; Gilmore et al., 2007).

There are two generic ways in which the dark matter particle properties can change the abundance of dark matter halos or their internal density profile. One is an early Universe effect (typically set in place well before matter-radiation equality) and the other a late Universe effect.

In the early Universe, there are two quantities that turn out to be relevant for structure formation. One is the mean velocity of dark matter particles in the early Universe, which sets the free-streaming scale, below which fluctuations are erased. If the dark matter particles have significant interactions with relativistic particles (either dark radiation or interactions with Standard Model leptons or photons), this will lead to an additional suppression of fluctuations through the same mechanism that leads to acoustic oscillations and damping of the cosmic microwave background. The larger of these two effects will determine the suppression scale, $\lambda$ .

The suppression scale, processed through non-linear structure formation, results in lowered abundance for dark matter halos with mass comparable to or below $(4\pi\lambda^3/3)\rho_{\rm matter}$, where $\rho_{\rm matter}$ is the cosmic abundance of matter today. Halos are not just fewer in number, they also grow later because of the suppression in the power spectrum due to these early Universe effects. The slower growth results in a less concentrated halo.

In addition to the physics described above, one can have dark matter properties such as self-interactions and decay that only impact the halos after their formation (late Universe). Both the self interaction of dark matter particles (elastic or inelastic) and dark matter decay impact a wide range of galaxy masses.

It is not necessary to work out the structure formation of each particle physics model sepa-rately. For astrophysical purposes, dark matter particle candidates and associated cosmologies can generally be classified according to whether the dark matter particle's properties (e.g., mass and corresponding free-streaming scale, non-gravitational interactions, etc.) play a role in galaxy formation (Vogelsberger et al., 2016). An example of such a classification is shown in Figure 54.

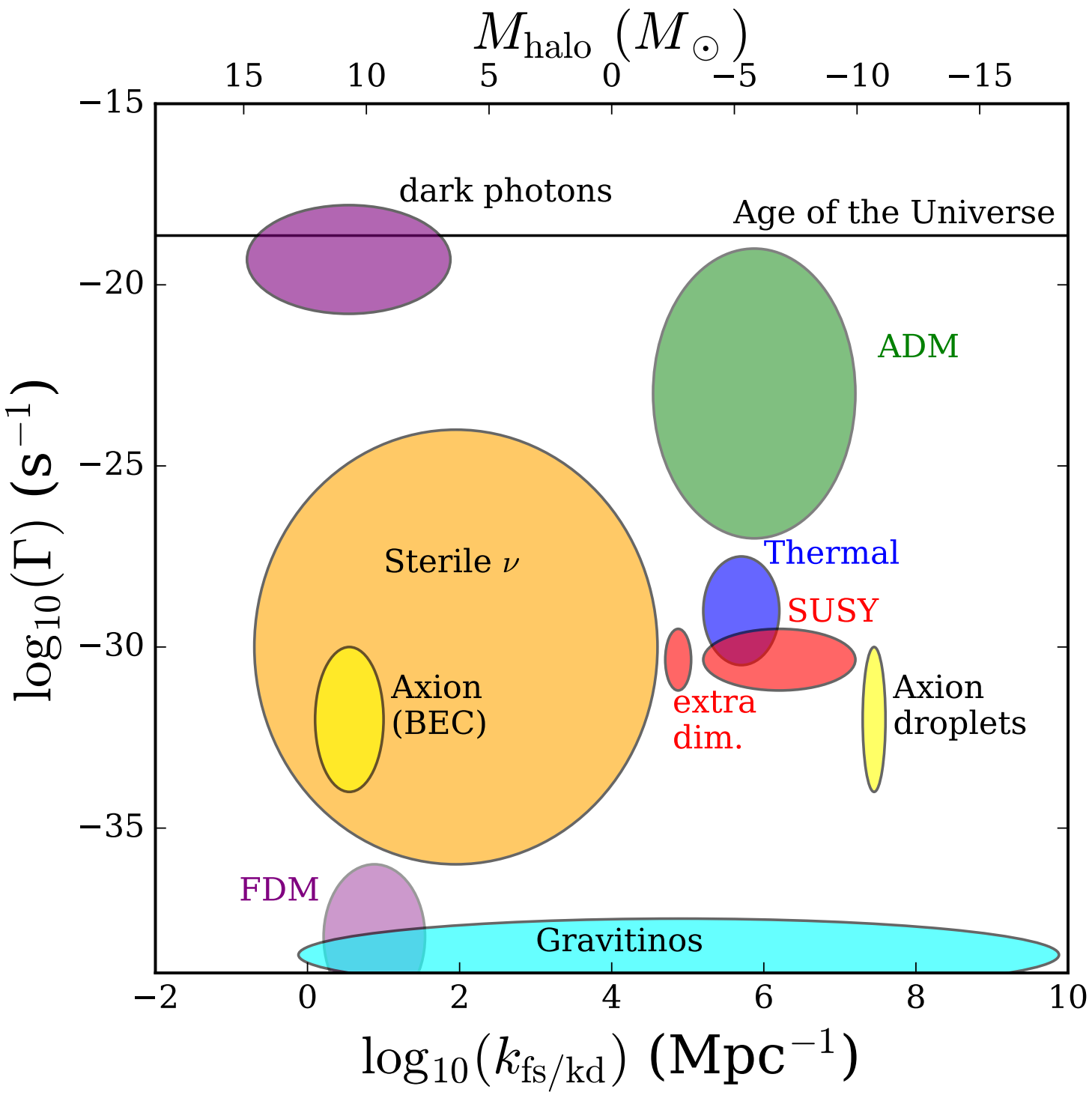

*Figure 54: Various dark matter candidates in a parameter space that directly influences structure formation. The horizontal axis is the characteristic free-streaming wavenumber of the model and the corresponding halo mass is shown on the top. These effects are set in the early Universe, typically well before the epoch of matter-radiation equality. The vertical axis is the characteristic interaction or decay rate, which directly impacts the structure of the halos at late times. Figure from Buckley & Peter (2018).*

For the purposes of delineating the physics, we will use four generic classes of dark matter models, namely CDM, "warm dark matter" (WDM), "self-interacting dark matter" (SIDM) and "fuzzy dark matter" (FDM).

**WDM** refers generically to candidates whose abundance is set in the early Universe when they are relativistic. These particles have a free-streaming scale that is essentially set by their mass ($\gtrsim$ keV). Eamples include low-mass gravitinos and sterile neutrinos, which are examples of dark matter candidates arising from minimal models of particle physics. The free-streaming scale introduces a cut-off in the linear power spectrum. The lack of power on small scales implies structure grows later and the minimum halo mass can be as large as that allowed by the data (around $10^8 \mathrm{M}_\odot$). There is no concrete guidance from theory for why the WDM mass would be in the regime that impacts structure formation; the minimum halo mass could be much smaller and the model would be indistinguishable from CDM in terms of its gravitational imprints.

**SIDM** refers to dark matter candidates that have appreciable non-gravitational self interactions (Spergel & Steinhardt, 2000; Vogelsberger et al., 2012; Rocha et al., 2013). As an example, imagine a hidden sector with stable neutrons. In this case, the dark matter particles (hidden neutrons) have velocity-dependent self-interactions set by the mass of the hidden sector pion. If the hidden pion and neutron masses are similar to that in the Standard Model, then the self-interaction cross section over mass $\sigma/m$ would of order $10\,\mathrm{cm}^2/\mathrm{g}$. Cross sections over mass larger than about $0.1\,\mathrm{cm}^2/\mathrm{g}$ are detectable (Kaplinghat et al., 2016) because they change the density profile of halos and subhalos. The reduced central densities of subhalos can also impact their survival in the tidal field of the parent halos. Other examples of SIDM include hidden H-atoms. In this case, we have the additional process in the early Universe of dark matter particles scattering off of *massless* hidden photons (Feng et al., 2009) and this will lead to dark acoustic oscillations and a cut-off in the linear power spectrum (Cyr-Racine & Sigurdson, 2013), mimicking WDM behavior but with the additional late-time phenomenology due to self scattering (Buckley et al., 2014; Boddy et al., 2016).

**FDM** is another late time effect that has recently been explored through simulations (Hu et al., 2000; Hui et al., 2016). In this scenario, dark matter particles have masses below about $10^{-22}$eV and the effective de Broglie wavelength in galaxies is kpc-sized. Thus the wave nature of dark matter becomes important. The pressure due to the wave nature of dark matter leads to the creation of dense solitonic cores (denser than CDM). The wave nature of dark matter in fuzzy dark matter models also implies a lack of structure on scales below the Jeans length, which is set by the mass $m$ of the (fuzzy) dark matter particle, and the corresponding Jeans mass is roughly $4 \times 10^7 M_\odot (10^{-22}\mathrm{eV/m})^{3/2}$ (Hui et al. 2017).

| Halo property | WDM | SIDM (massive mediators) | SIDM (massless mediators) | FDM |
|---|---|---|---|---|
| Slope of halo density profile | N | Y | Y | Y |
| Central density of subhalos and dwarf halos | Y | Y | Y | Y |
| Central density of more massive halos | N | Y | Y | N |
| Subhalo mass function | Y | N | Y | Y |

*Table 6: A partial list of how different models of dark matter can impact the halo and subhalo properties in comparison to the CDM predictions. For SIDM, two different examples are included; one in which the dark matter particles interact with a massive* $O(\mathrm{MeV})$ *mediator (e.g., hidden stable neutrons) and one in which the dark matter particles interact with a massless mediator (e.g., atomic dark matter).*

#### 6.2.2 Observables

Broadly speaking, there are three aspects of dark matter halos that we would like to infer from observations to constrain or measure dark matter particle properties: the mass function of subhalos, the phase-space distribution of subhalos, and the internal density profiles of field halos and subhalos. A particle physics model can be mapped on to these observables that are amenable to constraints from structure formation.

The **mass function of subhalos** is set, after non-linear processing, by a variety of processes. The linear power spectrum has a direct impact, as discussed before, through the suppression of structure below a threshold halo mass. Processes such as dark matter decay and mass-loss due to self interactions, in conjunction with tidal interactions with the disk and halo of the Milky Way will impact the number of subhalos that survive (D'Onghia et al., 2010; Errani et al., 2017). The survival probability will be a function of the orbital properties such as the pericenter distance and the number of pericenter passages. Thus, the **phase space distribution** of the subhalos (which includes the radial distribution) will be impacted by dark matter physics.

The internal **density profile** of field halos and subhalos has been the other main avenue for constraining dark matter physics. Viable models of WDM do not create cores (the profile retains the $1/r$ cusp) on observable scales (Kuzio de Naray et al., 2010; Villaescusa-Navarro & Dalal, 2011; Macciò et al., 2012). However, the concentration of the halos (and hence the inner density) is lower due to the delayed structure formation (Lovell et al., 2014). In SIDM, the density profile is shallower in the center due to heat transfer from the outer to inner parts of the halo (Davé et al., 2001), but this statement assumes that the core has not started contracting (which is the generic late-time behavior). In FDM, the outer profile is expected to be similar to the CDM profile but in the inner parts a dense solitonic core forms (Schive et al., 2014), a feature that is disfavored by rotation curve data (Bar et al., 2018; Robles et al., 2019).

A few examples will help further elucidate the influence of particle physics on these observable quantities. Table 6 provides a concise summary of some of the major effects.

- FDM cuts off halo formation below a mass scale determined by the FDM mass. This could be tested with substructure detections using strong lensing or gaps in stellar streams;

- SIDM transports kinetic energy in halos and subhalos and changes the density profile from the CDM predictions. In halos with small $M_\star/M_{\rm halo}$ (like the Local Group dwarf spheroidal galaxies) this leads to constant density cores (for moderate cross sections), which could be tested with resolved stellar velocities and rotation curves;

- At the other extreme end, dark matter density profile of the halos of clusters of galaxies provide a sensitive probe of the self interaction cross section at velocities of order $1000\,{\rm km\,s^{-1}}$ or larger;

- SIDM interactions (if large enough) could also evaporate subhalos and change the number of subhalos. The survival of subhalos and their radial distribution in the halo is sensitive to self-interaction strength;

- Both FDM and SIDM (when it forms constant density cores) change the strength of dynamical friction and this could be a testable prediction. For example, in SIDM models, it is expected that the BCG will slosh about the center of the halo on the scale of the constant density core size.

### 6.2.3 The impact of baryons

The presence of baryons can change the above description in dramatic ways. Sufficiently vigorous and bursty star formation, strong winds from massive stars and supernova-driven outflows that remove gas rapidly compared with local dynamical timescales can significantly alter the structure of dark matter halos (Navarro et al., 1996; Read et al., 2005; Governato et al., 2010; Pontzen & Governato, 2012, 2014; Tollet et al., 2016; Fitts et al., 2017). As a result of these influences, the orbits of dark matter particles expand non-adiabatically, potentially transforming central cusps into cores and lowering masses within the half-light radii of dwarf galaxies. If the feedback processes are also important in satellite galaxies (for example, the Fornax dSph), this would lower the binding energies of subhalos, leaving them more vulnerable to tidal disruption analogous to effect of self interactions.

The impact of feedback on WDM and SIDM has been studied for dwarf galaxies (Fitts et al., 2018). For SIDM models in which the cross section is large – $\sigma/m$ of few ${\rm cm}^2/{\rm g}$ – the drive to thermalization renders the final dark matter density profile insensitive to the star formation history (Robles et al., 2017), but it does depend sensitively on the final distribution of stars and gas (Kaplinghat et al., 2014).

The presence of a disk in the parent halo can have a marked effect on the survival of the subhalos that venture close to the centers of galaxies (e.g., Brooks et al., 2013; Wetzel et al., 2016). At present, however, it is not clear if the disk leads to a divergence in the predictions of these models for the radial distribution of the subhalos or makes the models similar. The answer may depend on the observable of interest, for example, ultrafaint dwarf galaxies or stellar streams may be affected quite differently.

The promise of precision data across a wide range of scales, from ultrafaint dwarf galaxies deep within the galactic potential to low-surface brightness galaxies in the field to galaxies in the cores of clusters, holds out hope that we can disentangle signatures of the particle nature of dark matter from signatures of feedback from star formation. For example, when feedback is efficient at producing cores in isolated field dwarfs, does it simultaneously also produce high-surface brightness galaxies? When self interactions are efficient at creating cores in isolated low-surface brightness galaxies, do they also match the properties of the dwarf spheroidals of the Milky Way and Andromeda (Collins et al., 2014; Tollerud et al., 2014)?

In the remainder of this chapter, we explore the possible ways in which MSE can help to reveal the nature of dark matter. We have grouped the science cases into four sections based on the distance of the stars being targeted: stars and streams in the Milky Way (Section 6.3); dwarf galaxies in the Milky Way and beyond (Section 6.4); galaxies in the low redshift ($z < 0.05$) Universe (Section 6.5); galaxies beyond the low redshift Universe (Section 6.6).

## 6.3 Stars and stellar streams in the Milky Way

The Milky Way is a great laboratory for studying the distribution of dark matter on small scales. The ability to measure 3D positions and 3D velocities for individual stars in the Milky Way implies that we can gravitationally map the dark matter distribution in great detail. The Milky Way's mass distribution is already well constrained in its inner part (e.g., Bovy, 2015; McMillan, 2017) where much of the mass is baryonic, but constraints on the total mass, radial profile, and shape of the smooth dark-matter distribution remain weak even though these are key observables when comparing against the predictions from different dark matter models. Understanding the smooth dark matter density and velocity distribution better is also necessary for the interpretation of laboratory direct-detection experiments and for indirect-detection probes. Finally, tidal stellar streams in the Milky Way are one of a small number of methods known today for measuring the clustering of dark matter on very small scales ($M \lesssim 10^8\,M_\odot$) by looking for the impact of small dark-matter subhalos on the structure of cold stellar streams. Detecting dark subhaloes that do not have any detectable gas or stars would be a stunning confirmation of the dark matter paradigm. Detection or constraints on the presence of these subhalos would also provide information about the particle properties of dark matter.

We now describe how, in conjunction with data from Gaia, LSST, and WFIRST, MSE has the ability to transform our knowledge of dark matter in the Milky Way. The essential contribution of MSE to this science is the measurement of line-of-sight velocities to a precision of 1 to 5 $\mathrm{km\,s^{-1}}$ for extremely large numbers of stellar tracers.

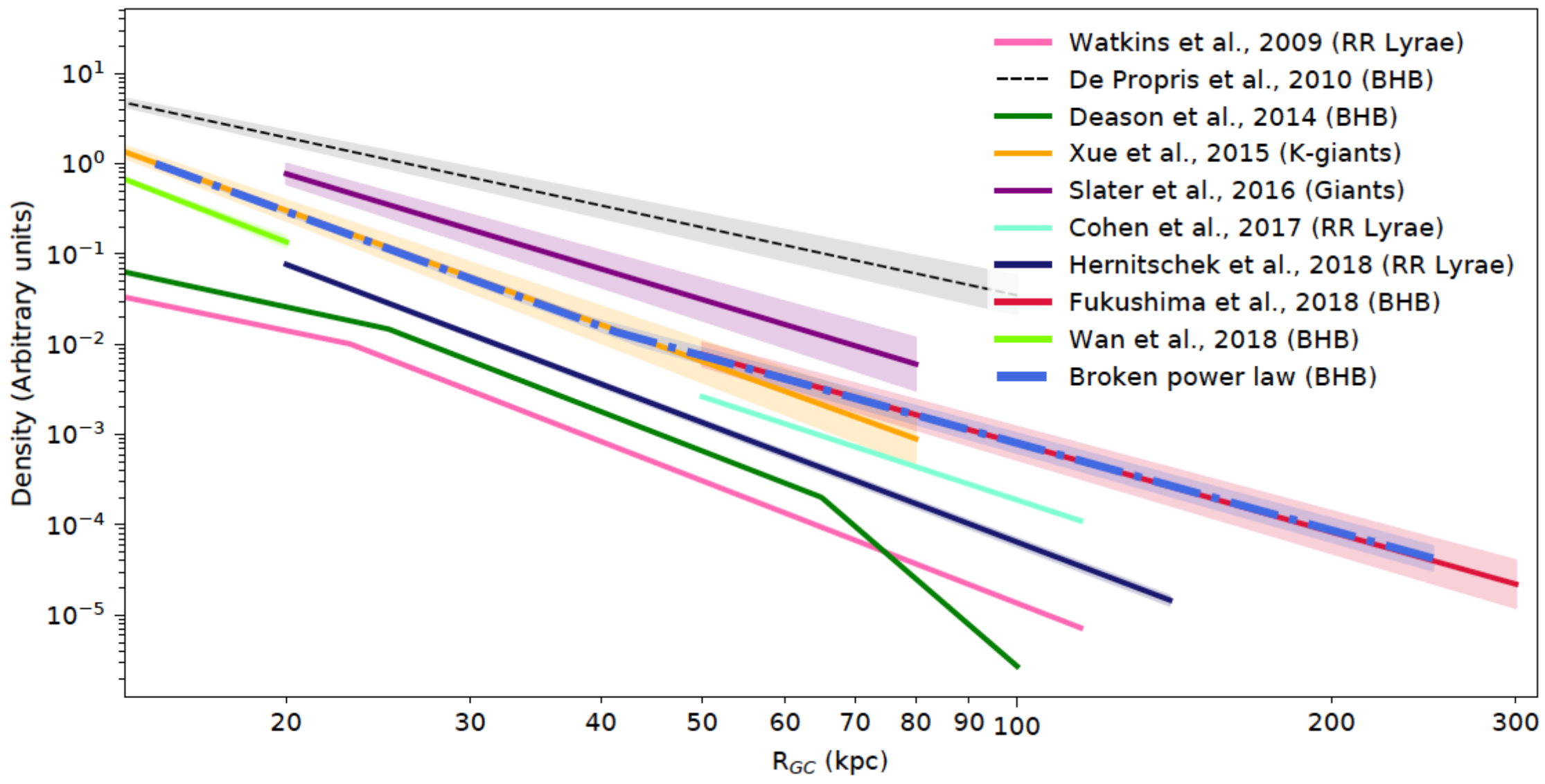

*Figure 55: Profile of the outer stellar halo ($R > 15kpc$) using different tracers. The typical power slope of the halo is in the range $3-4$. However, even for a similar population, such as the Blue Horizontal Branch (BHB) stars, the profile beyond 80 kpc is still very uncertain. Figure from Thomas et al. (2018).*

### 6.3.1 Mapping the Milky Way's gravitational potential with stars, dwarf galaxies, and stellar streams

Our understanding of the mass, shape, and mass profile of the Milky Way underlies every effort to test theories of dark matter using the galaxy we know best. The *mass* of the Milky Way's halo determines which simulated galaxies we should compare to when assessing consistency of our observations with predictions sensitive to the dark matter model, such as the number and structure of satellite galaxies. The concentration and *radial profile* of the Galaxy's dark matter set constraints on its accretion history and formation time (Wechsler et al., 2002), which are responsible for some of the remaining scatter in comparisons with simulations (Mao et al., 2015). The *shape* of the Galactic halo could potentially differentiate between different dark matter models, for example between CDM and SIDM (Sameie et al., 2018; Tulin & Yu, 2018) or superfluid dark matter (Khoury, 2015). It can also help constrain the effect of baryons on Galactic dark matter (e.g. Butsky et al., 2016), and at large distances can test predictions from simulations about the memory of the direction of filamentary dark matter accretion onto the halo (e.g. Vera-Ciro et al., 2011). Understanding the global Milky Way potential is also a necessary first ingredient to setting limits on substructure through its interactions with e.g. tidal streams (Section 6.3.3) since it is needed to construct a model of the unperturbed stream, to understand the contribution of non-regular orbits to stream structure, and to determine whether interaction rates are consistent with one dark matter theory or another (since the expected number of interactions varies with galactocentric distance; e.g. Yoon et al. 2011; Carlberg 2012).

Many of the possible tests of dark matter in the Milky Way thus require constraints on the halo's properties on scales comparable to its virial radius, or at least to its scale length.

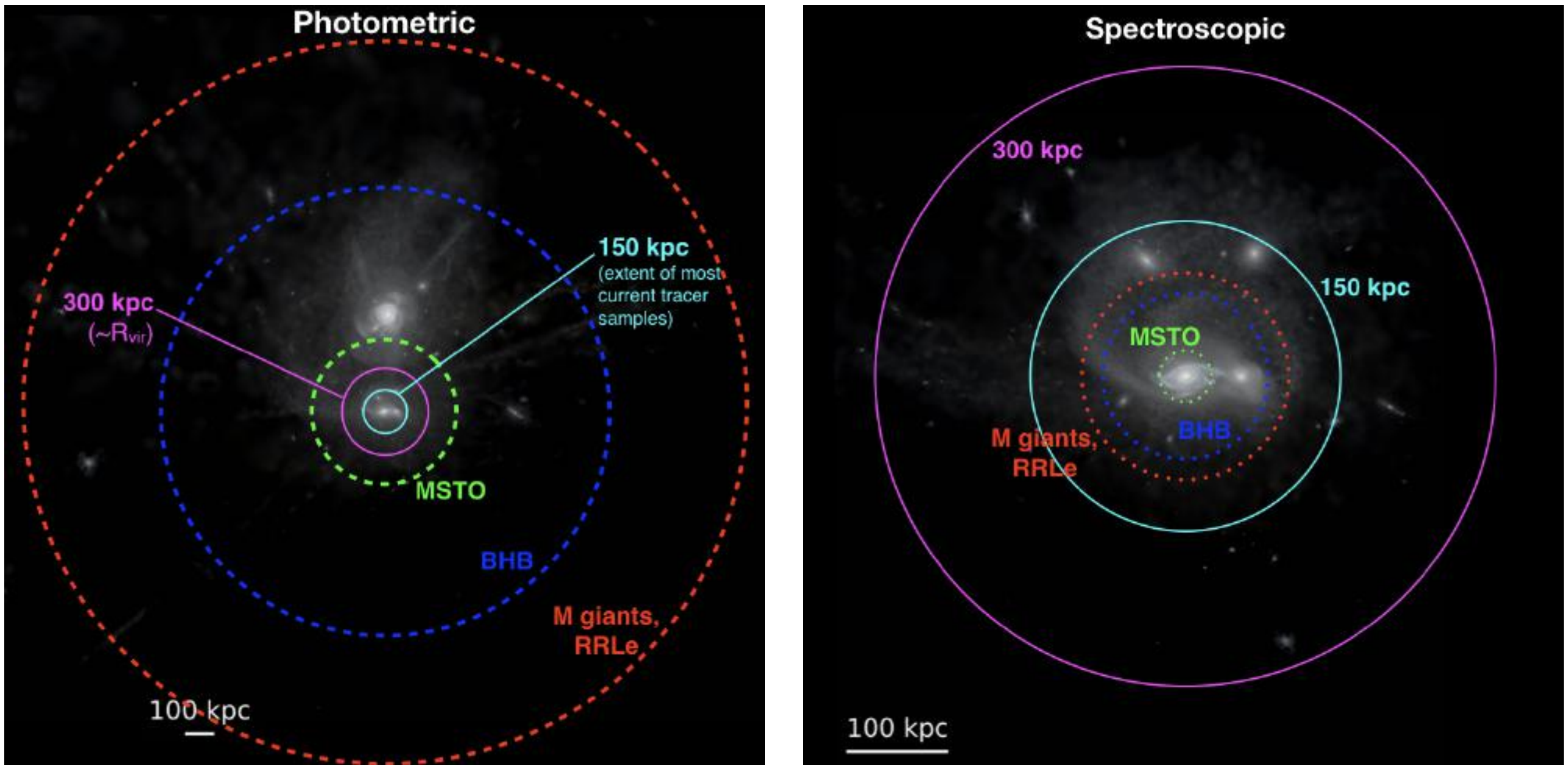

*Figure 56: Left: Distances to which LSST will detect various stellar tracers in coadded fields (limiting magnitude* $r = 26.7$*), superimposed on an image from a cosmological-hydrodynamical simulation of a Local Group-like system (Garrison-Kimmel et al., 2018). Right: Distances to which planned 4-meter multiplexed spectroscopic instruments will be able to observe the same tracers, superimposed on a zoomed view of the Milky Way-like galaxy in the same simulation. To realize the promise of new, deep photometric surveys to map the Milky Way halo requires the deep, multiplexed spectroscopic capabilities of MSE.*

However, the region where we have good dynamical constraints is set by where we have data: that is, where we can reliably measure distances and velocities to tracer populations. Satellite galaxies and globular clusters provide a good start, and many already have well-measured six-dimensional (6D) positions and velocities (Sohn et al., 2018; Fritz et al., 2018), but are limited in number. It is also unclear if they represent an equilibrium population, given that in external galaxies many globular clusters appear to trace tidal features and thus could have been contributed by accreted galaxies (e.g. Veljanoski et al., 2013), and given that based on cosmological simulations, we expect at least some satellite galaxies to arrive in hierarchical groups (e.g. Wetzel et al., 2015). With less than 6D data, equilibrium analysis of tracers is also subject to the mass-anisotropy degeneracy.

Thus, to build an accurate map of the dark matter halo we will also need to make use of more abundant stellar tracers, and to account for the fact that at large distances these tracers are often not in dynamical equilibrium (and thus not directly suitable for, e.g., Jeans analyses). Tidal streams, which are one example of a non-equilibrium population, should be more sensitive to the halo's shape in particular (via, e.g., the precession of the orbital plane) and can extend to very large distances (the Sagittarius stream is now mapped to $\gtrsim$ 200 kpc; Hernitschek et al. 2017). The large radial range explored by a single stream can also help break degeneracies between the scale radius and total mass of the Galaxy that arise from stream modeling (e.g., Bovy et al., 2016; Sanderson, 2016; Bonaca & Hogg, 2018). Gaia is providing data that will allow us to construct maps of the Galactic potential with increased accuracy using various equilibrium stellar tracers in six-dimensional phase space, to about $20-25$ kpc so far (e.g., Wegg et al., 2018). For a dark matter halo of $\sim 10^{12}\,M_{\odot}$ (estimates place the Milky Way's mass in this range to within a factor of two to three), this is comparable to or a little larger than the scale radius, which is probably 1/10 – 1/20 of $R_{\rm vir}$ depending on the halo concentration. Indeed, the majority of known stellar tracers in the Galaxy are strongly concentrated near its center, with only a few exceptions (Figure 55). This is partially a function of their steep radial profile compared to what we predict for the dark matter, but also of the limiting magnitudes of current photometric and spectroscopic surveys. Crucially, while future deep photometric surveys like LSST (coadded limiting magnitude of $g = 26.7$) will easily be able to identify stellar tracers all the way to $R_{\rm vir}$ (Figure 56 left), MSE is the only high multiplexed spectroscopic facility under development that is capabile of matching LSST's limiting photometric depth.

Most new spectroscopic surveys with a significant stellar halo component, planned for 4-meter-class telescopes, have a limiting magnitude matched to the depth of Gaia's proper-motion survey, $g \simeq 21$ (Figure 56 right), or even shallower. For comparison, to measure velocities for BHB stars ($M_g \simeq 0.5$) near the Milky Way's estimated virial radius would require spectroscopy down to $g \simeq 23$. To obtain RVs near $R_{\rm vir}$ for the extremely valuable RR Lyrae standard candles, which by the LSST era will likely be calibrated to yield distances with 2% accuracy, will require reaching $g \sim 22-23$ in an integration time of $\lesssim 15$ minutes, given their typical pulsation period of $6-12$ hours. To reach the main sequence turnoff (MSTO), and the huge increase in tracer density that it offers, would require a depth of $g \sim 25$. Fortunately, the necessary velocity precision at these distances is not large; even at the virial radius stars should have orbital speeds of $100-150\,{\rm km\,s^{-1}}$, and their motion should be primarily radial since they are mostly expected to be in tidal tails from accreted satellites, so a precision of $\sim 5\,{\rm km\,s^{-1}}$ is sufficient for this application.

Besides achieving sufficient depth, the other challenge in observing distant halo tracers is their extremely low density on the sky. We expect that LSST will discover hundreds of new satellites (we discuss other science with new dwarfs in Section 6.4.1, and see also Hargis et al., 2014; Kim et al., 2018; Nadler et al., 2018), as well as tens of thousands of bright stellar tracers in distant tidal streams (Sanderson et al., 2017). Therefore, *multiplexing and multithreading* (i.e. opportunistic scheduling of individual spare fibers for distant stars in streams) will be a necessary component of these (and many other) science programs with MSE.

### 6.3.2 Dark matter halo distortions from the LMC in the MW halo

The fundamental process through which galaxies merge in any dark matter model is through dynamical friction. Using linear perturbation theory in spherical systems, Tremaine & Weinberg (1984); Weinberg (1989) showed that dynamical friction is a resonant process between the orbit of a satellite galaxy and the host galaxy causing distortions of the underlying host dark matter density field from the outer to the inner region of the halo. This leads to a shift in the center of mass of the system but also higher order distortion terms which can be quantified in terms of spherical harmonics. In the Milky Way, Weinberg (1998) used the Kalnaj matrix method (Kalnajs, 1977) to demonstrated that these distortions can also lead to torques on the disc leading to warping in the case of the LMC-MW interaction, for which the direct tides of the LMC are too weak otherwise to lead to a warp. This has not been appreciated in numerical N-body simulations for years leading to conflicting results, mainly due to the too poor resolution to resolve the resonant interactions at place (e.g. see Weinberg & Katz, 2007, for a discussion). However, more recently this has been possible due to the recent advances in computing power and increase in particle numbers able to resolve appropriately enough the phase-space structure of dark matter halos, and the effect of halo distortions on the stability of discs has been investigated in full hydrodynamical N-body cosmological settings (Gómez et al., 2016).

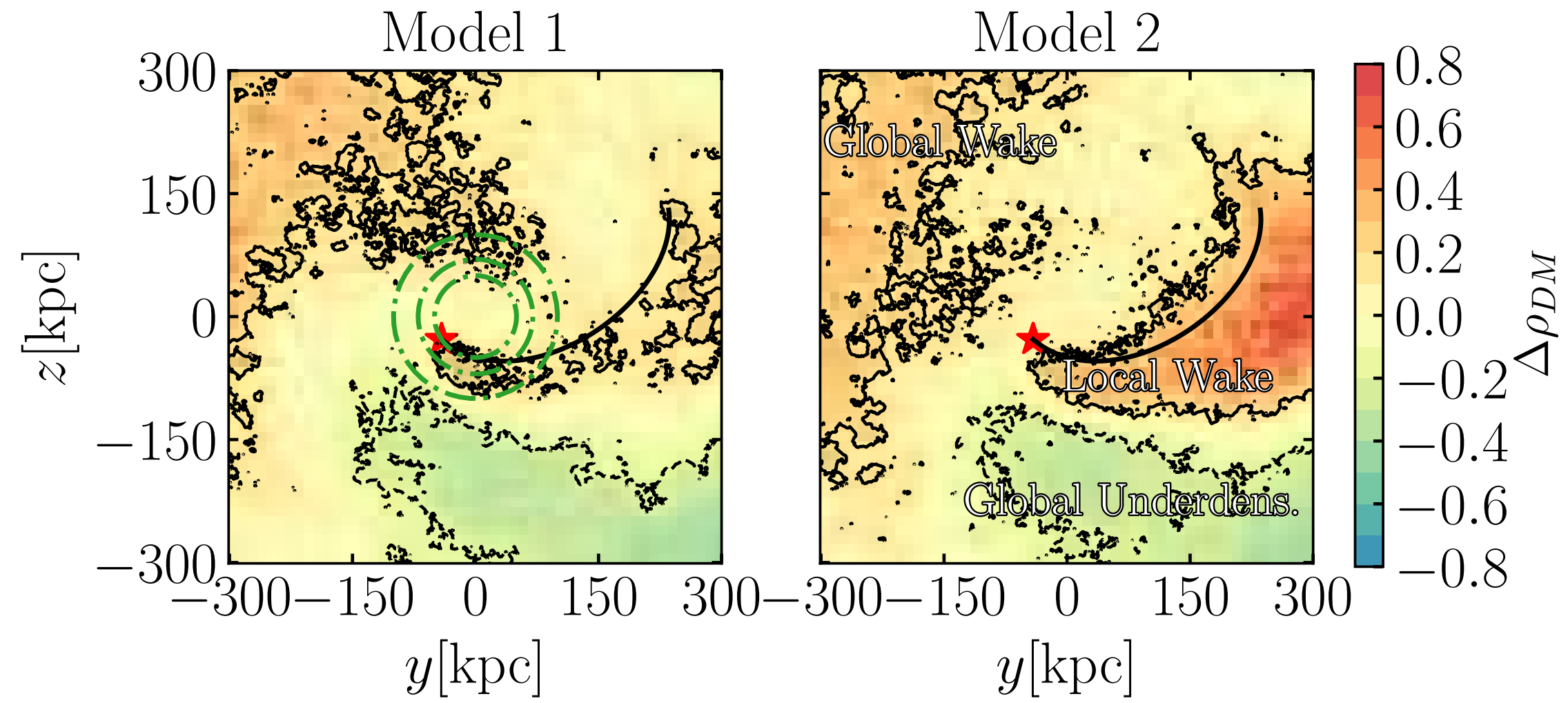

*Figure 57: Milky Way's dark matter density distortions, revealing the DM halo response induced by the LMC in the y-z plane (through a slab of 10 kpc thickness in the x-direction). The black line represents the LMC's past orbit. The disc is confined to the x-y plane and the Sun is at $x = -8.3$ kpc. The green circles delineate Galactocentric distances of 45, 70 and 100 kpc. Three features are defined: (1) The local wake as the DM over-density trailing the LMC, tracing its past orbit, (2) The Global Wake as the over-density that appears in the North and (3) the Global Under-density, which are the regions that surround the Local Wake. One notices that the strength of the signal varies depending on the kinetic structure of the dark matter halo (isotropic for Model 1 and anisotropic for Model 2) which is expected due to the resonant nature of the interaction. However, the general shape of the signal is generic between models. Figure from Garavito-Camargo et al. (2019).*

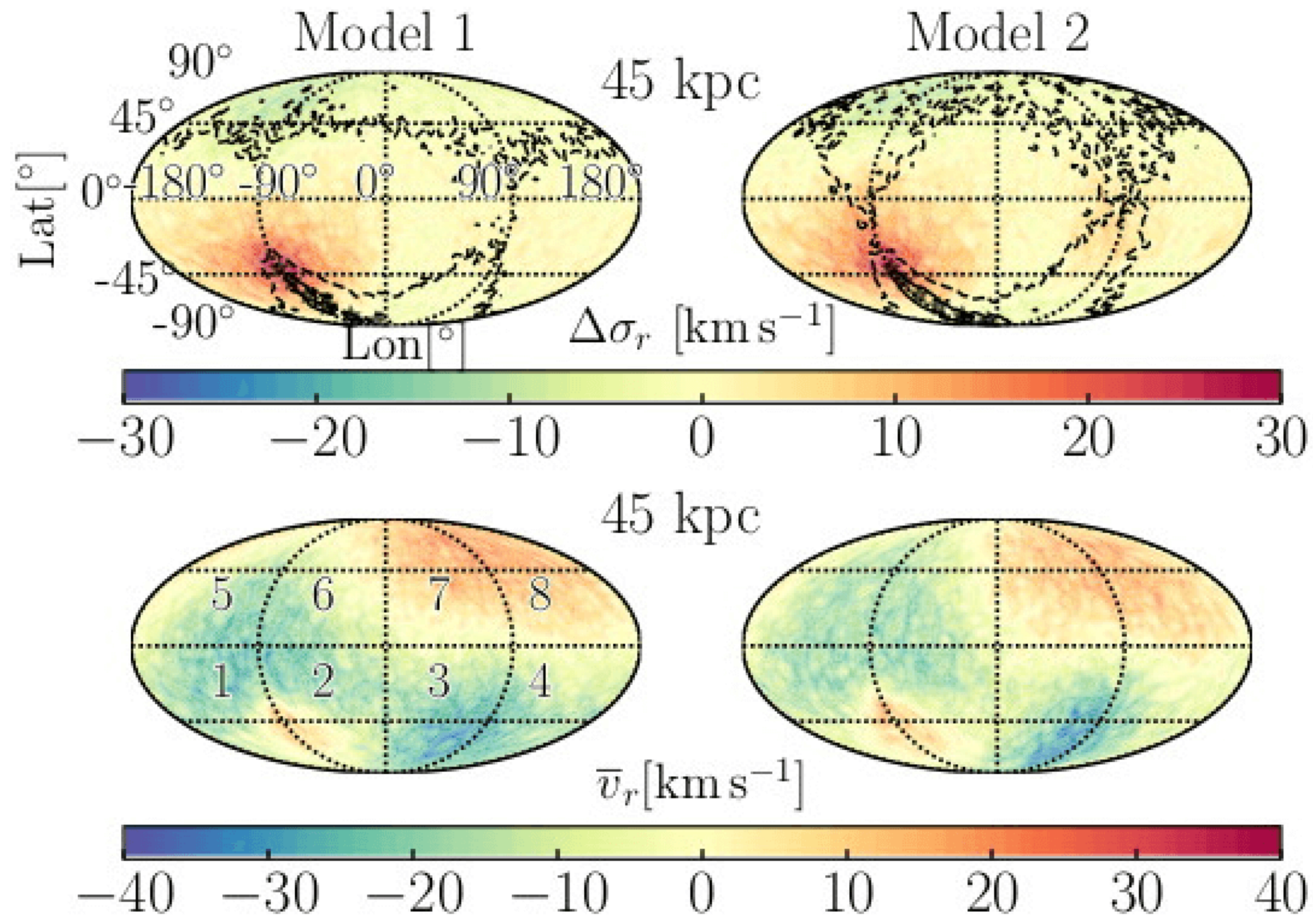

*Figure 58: Mollweide projections in Galactocentric coordinates of radial motions of the stellar halo induced by the LMC. Relative changes to the average 1D radial velocity dispersion are presented in the top panels, while the bottom ones show the local mean radial velocity at* $r = 45\,\mathrm{kpc}$*. The contours denote the overdensities in the North and South. The past trajectory of the LMC are represented by the grey stars.* $\sigma_r$ *increases by* $\sim 25\,\mathrm{km/s}$ *near the LMC in Octants 1 and 2, while in Octants 4 and 5 it decreases by* $\sim 14\mathrm{km/s}$ *to the North of the LMC, giving rise to a kinematically cold spot. The velocity dispersion in Octants 7 and 8, however remain largely unaffected. In the space of radial motions,* $v_r$*, the region in Octant 2 closest to the LMC is moving away from the Galactic Center, while the Local Wake behind the LMC in Octants 3 and 4 follows the LMC's the past orbital motion towards the MW. Figure from Garavito-Camargo et al. (2019).*

Since the recently revised proper motions of the LMC (Kallivayalil et al., 2013), which suggest that it is on a first infall orbit (Besla et al., 2007), interest in the interaction of our most massive satellite with the Milky Way has been renewed. Firstly, this has led to the possibility that the LMC may be more massive than previously thought, putting it in halos ranging from $10^{10}\,\mathrm{M}_{\odot}$ to $3 \times 10^{11}\,\mathrm{M}_{\odot}$. There are various arguments suggesting for a more massive LMC, ranging from abundance matching arguments (e.g. Moster et al., 2013), its association with the SMC and other satellites (e.g. D'Onghia & Lake, 2008; Jethwa et al., 2016; Kallivayalil et al., 2018) to timing argument constraints favouring a mass of $M_{LMC} = 2.5^{+0.9}_{-0.8} \times 10^{1}1\,\mathrm{M}_{\odot}$ (Peñarrubia et al., 2016), to name a few.

A large mass for the LMC would have dramatic consequences on the Galaxy and its subsequent modeling. In the stellar halo, this would lead to systematic biases in the modeling of tracers in the halo (globular clusters, streams, satellites, stellar halo) and as a result the inference of the orbits and associations of satellites bodies. A direct consequence of the tides from the LMC acting on stellar streams has been observed in a recent analysis of the southern part of the Orphan (Koposov et al., 2019) showing proper motions perpendicularly offset from the stream track. Erkal et al. (2018b) demonstrated that in order to fit the full stream a perturbation from the LMC needs to be invoked favouring a mass of $1.3^{+0.27}_{-0.24} \times 10^{11}\,\mathrm{M}_{\odot}$. It is expected that other streams in the close vicinity of the LMC should also exhibit proper motions that are misaligned with their stream tracks, which should further help constrain the mass of the LMC (Erkal et al., 2018a).

In addition to tides, given its much larger inferred mass than typically assumed, it is expected that the whole halo of the MW is also reacting. Gómez et al. (2015) predict a shift in the center of mass of the MW due to the interaction with the MW, which would result in an upward bulk motion of order $v \sim 40\,\mathrm{km/s}$ in the stellar halo beyond a radius of $r \sim 30\,\mathrm{kpc}$ (Erkal et al., 2018b). Indeed, Laporte et al. (2018a) presented live N-body models of the interaction between the LMC and the Galaxy on a first infall orbit, showing that the resulting warp followed the lines of node of the HI warp, producing similar asymmetric distortions. This N-body experiment showed that these could result in density perturbations in the Milky Way's dark matter halo of order 40% around $r \sim 40\,\mathrm{kpc}$ thus confirming earlier studies of the impact of halo distortions on the disc through linear perturbation theory pioneered by Weinberg (1998); Weinberg & Blitz (2006).

In particular, the effect of these distortions on the halo and stellar halo were studied in some depth in Garavito-Camargo et al. (2019) where they varied the internal kinematics of the MW dark matter halo and observed generic trends between realizations. Figure 57 shows the resulting response of MW dark matter halo in those models. By generating a smooth stellar halo tracer population in the halo, they demonstrated how the signal from the halo distortions would be imprinted on the density and kinematics profiles of the stellar halo. As an example, the radial velocity signals at $r = 45$ kpc for the explored models are presented in figure 58. For a set of reasonable assumptions on the number K-giants as a function of radius in the halo, Garavito-Camargo et al. (2019) estimated that measuring an overdensity in the halo due to the response of the MW halo to the LMC would require about $100 - 1000$ tracers in 20 square degree fields. Thus mapping such a signal will require a large number of tracers to probe density contrasts and bulk velocity motions making the well populated MSTO stars a prime and valuable tracer at intermediate distances in the stellar halo $r < 50$ kpc.

Measuring the wake of the dark matter halo would allow us to directly study dynamical friction in 6-D phase space in the Milky Way. Such a measurement would set strong limits on the particle nature of the dark matter (its cross-section in particular) and use the Milky Way as a complementary probe to similar analyses in galaxy clusters that examine the "wobble" of the brightest cluster galaxy (see details in Section 6.6.3). Thus MSE could provide the necessary data for three-quarters of the entire sky to map the dark matter halo's response to the LMC (e.g. Garavito-Camargo et al., 2019).

### 6.3.3 Identifying the dark sub-halo population with stellar streams

Tidal streams are a promising tool to detect the presence of the dark subhaloes predicted by a wide range of dark matter models as summarized in Figure 54 (Ibata et al., 2002; Johnston et al., 2002). These streams form as globular clusters or dwarf galaxies are disrupted by the tidal field of the Milky Way (Bovy, 2014, and references therein). To date, about 50 streams have been discovered in our Galaxy, with many recent discoveries aided by Gaia (Grillmair & Carlin, 2016; Shipp et al., 2018; Malhan et al., 2018; Ibata et al., 2019). Although these streams appear as coherent bands on the sky (e.g. Belokurov et al., 2006b), they are extremely fragile and the nearby passage of a subhalo can induce relative changes to the orbits of stream stars which causes gaps and wiggles to form (Siegal-Gaskins & Valluri, 2008; Yoon et al., 2011; Carlberg, 2013; Erkal & Belokurov, 2015a). Indeed, signatures consistent with such flybys have already been claimed in the Palomar 5 (Carlberg et al., 2012; Bovy et al., 2017; Erkal et al., 2017) and the GD-1 stream (Carlberg & Grillmair, 2013; de Boer et al., 2018; Price-Whelan & Bonaca, 2018). In addition to dark matter subhaloes, baryonic substructure like giant molecular clouds (Amorisco et al., 2016; Banik & Bovy, 2018), the Milky Way bar (Erkal et al., 2017; Pearson et al., 2017), spiral arms (Banik & Bovy, 2018), as well as the disruption of the progenitor (Webb & Bovy, 2018) can also produce perturbations in streams. Fortunately, most of these effects are mitigated for streams on retrograde orbits like GD-1 (Banik et al., 2018; Amorisco et al., 2016).

Two independent techniques have been proposed for how to quantitatively extract the properties of the subhaloes which created these gaps, both of which rely on multiple dimensions of observables for each stream. First, Erkal & Belokurov (2015b) demonstrated that each subhalo flyby produces a unique signature which can be used to almost uniquely determine the subhalo's properties, i.e. its mass, scale radius, flyby velocity, and point of impact. The inference is not quite unique since there is a one-dimensional degeneracy between the sub-halo mass and its velocity relative to the stream. This degeneracy can be broken, in the first instance, by placing a prior on the subhalo's velocity, e.g. that it is bound to the Milky Way. This inference requires at least three observables of the stream, e.g. stream density, stream track, and radial velocities along the stream. This technique can be used to determine the properties of individual subhaloes and build up a catalogue of impacts. Since it determines the properties of the subhalo, it can also be used on compact baryonic substructure like giant molecular clouds. Second, Bovy et al. (2017) developed a statistical technique which determines the amount of substructure required to reproduce the statistical properties of the stream, e.g. the power spectrum of its density. While this technique can be used with just the stream density, it is more powerful when used with multiple observables since the gaps have correlated features in all of the observables. This technique has already been used on the Palomar 5 stream, which was found to have density variations consistent with ΛCDM (Bovy et al., 2017). Both of these techniques would benefit from radial velocities along the stream to faint magnitudes.

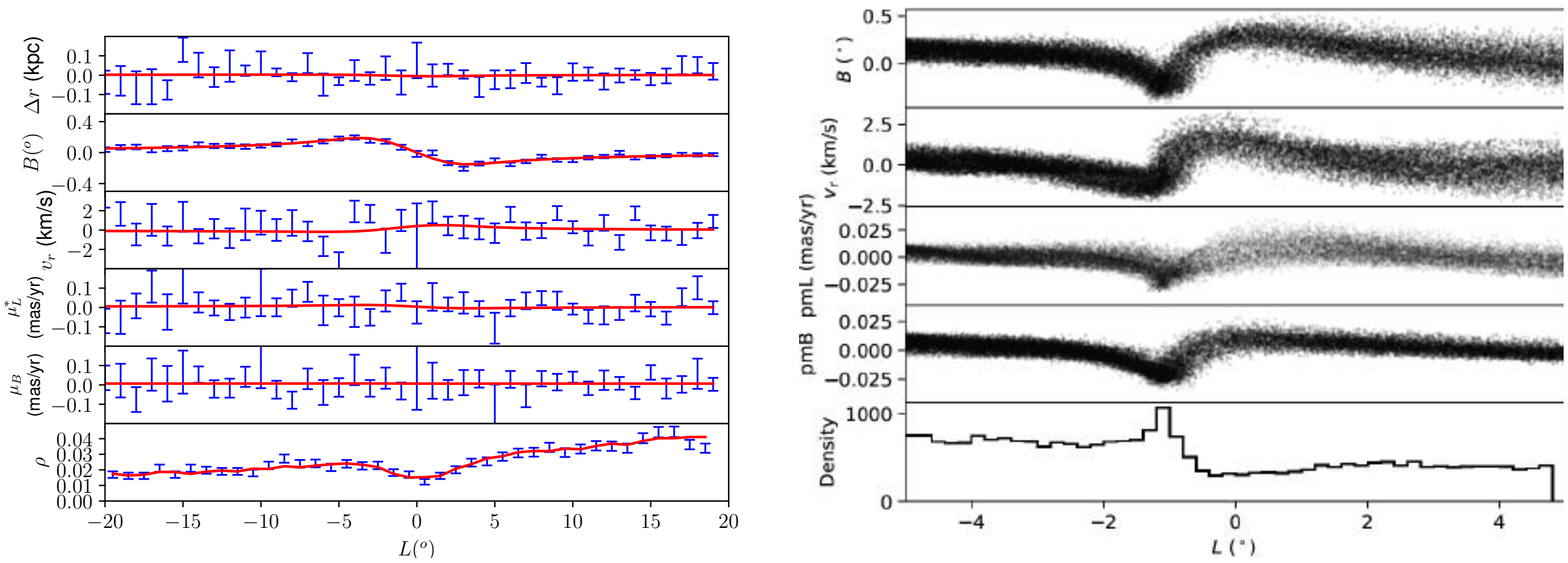

*Figure 59: Left: Fit of a* $10^7 M_\odot$ *subhalo impact adapted from Erkal & Belokurov (2015b). The blue error bars show mock observations of an N-body stream impacted by a* $10^7 M_\odot$ *subhalo 450 Myr ago and the red line shows an analytic model which is fit to the mock data. These mock observations are made with observational errors which will be available in the near future, e.g. Gaia proper motions, radial velocities from spectroscopic surveys like WEAVE, and DES-quality photometry. Even with these errors, the fits at* $10^7 M_\odot$ *are accurate and precise. Right: Gap in a simulated GD-1 like stream from a* $10^6 M_\odot$ *subhalo. This signal would be readily detectable in the density (bottom panel), on the sky (top panel), and in the radial velocities (second panel). However, the proper motions (third and fourth panel) would be undetectable even with Gaia DR2.*

In order to show how these gaps look in practice, Figure 59 shows two gaps produced from the nearby passage of a subhalo. The left panel shows an adapted figure from Erkal & Belokurov (2015b) showcasing how an individual gap can be fit. The gap in this example is caused by a $10^7 M_\odot$ subhalo that impacted the stream 450 Myr ago and the properties of the subhalo can be fit up to the degeneracy described above. The right panel shows a gap in a GD-1 like stream from a $10^6 M_\odot$ subhalo. As can be seen from both panels, the signatures in the different observables are correlated which is what makes both techniques so powerful.

In order to quantitatively assess what precision from MSE is needed to constrain these subhaloes, we use the results of Erkal et al. (2016) who derived the distribution of impact properties from a distribution of subhaloes. We model the GD-1 stream which is one of the best candidates for detecting the presence of dark matter. These impact properties can then be used to determine the distribution of velocity kicks on the stream. Figure 60 shows the maximum velocity kick imparted on a stream from the expected distribution of ΛCDM subhaloes over a period of 5 Gyr. This figure shows that if MSE can measure the radial velocity of stream stars down to $100 - 300$ m s$^{-1}$, we will be able to probe subhaloes down to $10^5 - 10^7 M_\odot$. Note that at the distance of GD-1, $\sim 1$ km s$^{-1}$ would correspond to $\sim 0.02$ mas/yr in proper motion which would only be measurable for the brightest stars in Gaia DR2. Thus, while Gaia is an excellent tool with which to measure the overall motion of the stream, spectroscopic surveys like MSE are crucial for measuring the effect of low mass substructure in velocity space.

Finally, we note that the effectiveness of MSE will be even better than suggested in Figure 60 since the change in radial velocity occurs over a scale given by the size of the gap (see Figure 59). The typical size of a gap is related to the subhalo mass (Erkal et al., 2016; Bovy et al., 2017) with a size of a few degrees expected for a $10^6 M_\odot$ subhalo (although this gap size stretches and compresses along the orbit of the stream). Thus, instead of needing a precision of $\sim 100$ m/s per star, we actually need this precision when averaging over roughly a degree. In this context, the Palomar 5 stream has $\sim 100$ stars/degree brighter than $r \sim 23.5$ (Erkal et al., 2017) and the GD-1 stream has $\sim 10$ stars/degree brighter than $r \sim 23$ (de Boer et al., 2018). A velocity precision per star of $\sim 1$ km s$^{-1}$ would allow measurements of stream perturbations at the $\sim 100$ m s$^{-1}$ level for an average of 100 stars. The main limiting factor in the science described here will likely be the systematic uncertainty in MSE and not the statistical uncertainty for each star. Given an expected systematic uncertainty in the high resolution mode of $\sim 100\ \mathrm{m\,s^{-1}}$, MSE should be sensitive to subhaloes down to $10^5 - 10^6 M_\odot$.

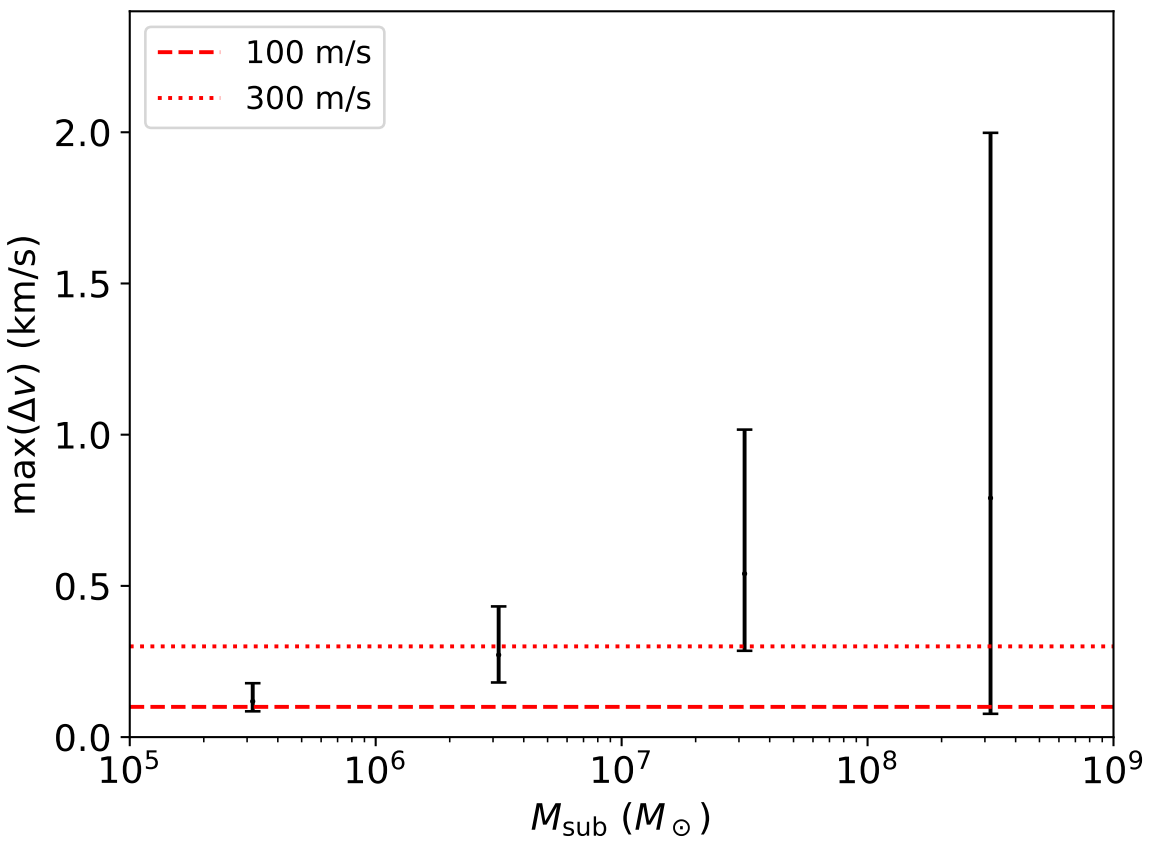

*Figure 60: Typical maximum velocity kick in a GD-1 like stream over a 5 Gyr duration. The black error bars show the scatter (median with 1 −σ spread) from 1000 realizations of ΛCDM subhaloes in 4 decades of subhalo mass from* $10^5 - 10^6 M_\odot$ *up to* $10^8 - 10^9 M_\odot$*. The red dotted and dashed lines show a 300 m/s and 100 m/s uncertainty. Given an expected systematic uncertainty in the high resolution mode of* $\sim 100\ m\,s^{-1}$*, MSE should be sensitive to subhaloes down to* $10^5 - 10^7 M_\odot$*.*

#### 6.3.4 Local dark matter distribution and kinematics for direct detection

One of the possible detection mechanisms of dark matter is direct detection (Goodman & Witten, 1985), the process in which dark matter scatters off a heavy nucleus, where the recoil of the latter emits a detectable signal. Experiments have excluded large parts of parameter space of one of the most popular dark matter scenarios for WIMPs (the most constraining limits have been performed by the Xenon1T collaboration; Aprile et al. 2018). The rate $R$ of this process depends on both the velocity distribution of dark matter as well as the local density of dark matter:

$$R \propto \rho_{\mathrm{DM}} \times \int_{v_{\mathrm{min}}}^{\infty} \frac{f(v)}{v} dv, \tag{6.1}$$

where $\rho_{\mathrm{DM}}$ is the local dark matter density, and $f(v)$ is the local velocity distribution of dark matter. $v_{\mathrm{min}}$ is the minimum velocity for a particular dark matter mass that could produce a signal, and is related to the experimental threshold as

$$v_{\mathrm{min}} = \sqrt{\frac{Q m_N}{2\mu^2}}, \tag{6.2}$$

where $Q$ is the recoil energy, $m_N$ the mass of the nucleus (Xenon for example), and $\mu = m_\chi m_N/(m_\chi + m_N)$ is the reduced mass of the dark matter $m_\chi$ and the heavy nucleus.

Astrophysical errors on the local density of dark matter can change current limits from $\sim 30\%$ to a factor of two, depending on the method used (see Read 2014 for a review). Using local stars as tracers, it is possible to reduce the errors on the local measurement of dark matter local density (Bovy & Tremaine, 2012; Piffl et al., 2014). A particular systematic in the measurement of the dark matter density is the uncertainty in the density and distribution of the baryonic component (stars and gas). With radial velocities from MSE for a large set of stars, we will be able to improve on the existing measurements. Coupled with the proper motions of Gaia, we expect to resolve smaller structures and improve our understanding of the baryonic components, leading to a more accurate measurement of the local density of dark matter.

Another potential reducible systematic is the local velocity distribution. A new strategy to empirically obtain the velocity distribution of dark matter from metal poor stars has been introduced in Herzog-Arbeitman et al. (2018a) for the case of the metal-poor relaxed component, and in Necib et al. (2018) for the case of more recent mergers. These metal poor stars have mostly been accreted, like dark matter, and hence the most metal poor stars should trace the velocity distribution of the oldest dark matter component. Such a correlation has been used to determine the local velocity distribution from Gaia in Herzog-Arbeitman et al. (2018b) and Necib et al. (2018). Here, a new structure in velocity space called Gaia Enceleadus (Belokurov et al., 2018a,b; Helmi et al., 2018b; Myeong et al., 2018a) has been modeled for its dark matter content (see discussion in Chapter 5).

In order to get the most detailed velocity distribution of dark matter locally, an accurate measurement of the metallicity as well as the 3D velocities of a large number of nearby stars is required. Gaia DR2 has already shown the capability of finding nearby velocity substructure (Myeong et al., 2018a). MSE will be able to provide the missing radial velocity component of all Gaia stars across its full magnitude range, and will extend even further to 23rd magnitude to match with future space mission such as WFIRST. At the bright end, the estimated errors on the radial velocities from MSE stars will be of order hundreds of $\mathrm{m\,s^{-1}}$ or better. When coupled with excellent distance measurements from Gaia, this will provide the best set of 3D velocity measurements that exist. The dominating errors in that case will be systematics of the strength of the correlation between the dark matter and the stars (e.g., Bozorgnia et al., 2018).

### 6.3.5 Dark matter distribution in the Galactic Center for indirect detection

Dark matter could also be detected indirectly through its self-annihilation or decay into Standard Model particles, such as gamma rays, neutrinos, electrons and positrons. The likely sources for this search are places known to have a dense concentration of dark matter, including the Galactic Center and the satellites of the Milky Way. The self-annihilation or decay processes depend on the density of dark matter at the source, and the rate is therefore condensed into a parameter called the J-factor, defined as

$$J = \int_{\text{l.o.s}} (\rho(l))^p dl, \tag{6.3}$$

where $\rho$ is the density of dark matter, $p$ is the number of dark matter particles participating in the interaction, $p = 1$ for decay and $p = 2$ for annihilation, and the integral is set along the line of sight from the object to the experiment.

Getting accurate measurement of the velocities of stars closer to the center is crucial in obtaining the correct density profile of the dark matter. Although the field at the Galactic center is crowded, MSE will be able to get radial velocity measurements for a large number of stars (with better than $10\,\mathrm{km\,s^{-1}}$ precision) within a few kiloparsecs of the Galactic Center. We can couple radial velocity measurements of MSE with Gaia to find the escape velocity at different Galactocentric distances (similarly to the analysis in Monari et al. 2018), or work using the radial velocity alone but explore a larger range of distances surpassing Gaia measurements (e.g., Williams et al. 2017). This leads to a better determination of the slope of the density of dark matter, and therefore would constrain the J-factor for indirect detection.

## 6.4 Dwarf galaxies in the Milky Way and beyond with resolved stars

Galaxies like the Milky Way are expected to contain a plethora of dark matter subhaloes (Klypin et al., 1999). The largest of these subhaloes, above $\sim 10^8 M_\odot$, are large enough to have hosted star formation in the early Universe and are thus visible as satellite galaxies (Jethwa et al., 2018; Kim et al., 2018; Norman et al., 2018; Wheeler et al., 2018).

Local Group dwarf galaxies are attractive targets for investigating the nature of dark matter due to their proximity, large dynamical mass-to-light ratios, and early formation times (see also Chapter 5). The standard cosmology model predicts the abundance and internal structure of the dark matter halos that host dwarf galaxies. The particle physics governing dark matter could lead to observable consequences including reducing the number of dwarf galaxies, flattening the density profiles of their dark matter halos, or producing energetic Standard Model particles through annihilation or decay. MSE's ability to gather large stellar-kinematic samples for faint dwarf galaxies, combined with data from X-ray and gamma-ray observatories, will be crucial for testing these predictions and illuminating the physical nature of dark matter.

The theoretical landscape of dark matter models described in Section 6.2 clearly provides support for searches of deviations from the cold collisionless dark matter idea, using Local Group dwarf galaxies. Several apparent discrepancies between CDM-simulated and the observed Universe, particularly on the smallest galactic scales, also motivate such searches.

Around galaxies with $\mathrm{M_{vir}} \sim 10^{12}\mathrm{M_\odot}$, cosmological N-body simulations that consider only gravitational interactions among CDM particles typically form $\sim 10$ times more subhalos with $\mathrm{M_{vir}} \sim 10^8\mathrm{M_\odot}$ than have been detected as luminous dwarf-galactic satellites of either the Milky Way or M31. This has been dubbed the "missing satellites" problem (Klypin et al., 1999; Moore et al., 1999). However, the ultrafaint dwarfs discovered recently seem to bring the census of dwarf galaxies into agreement with predictions of the CDM model, after taking into account detectability and the impact of reionization (Kim et al., 2018).

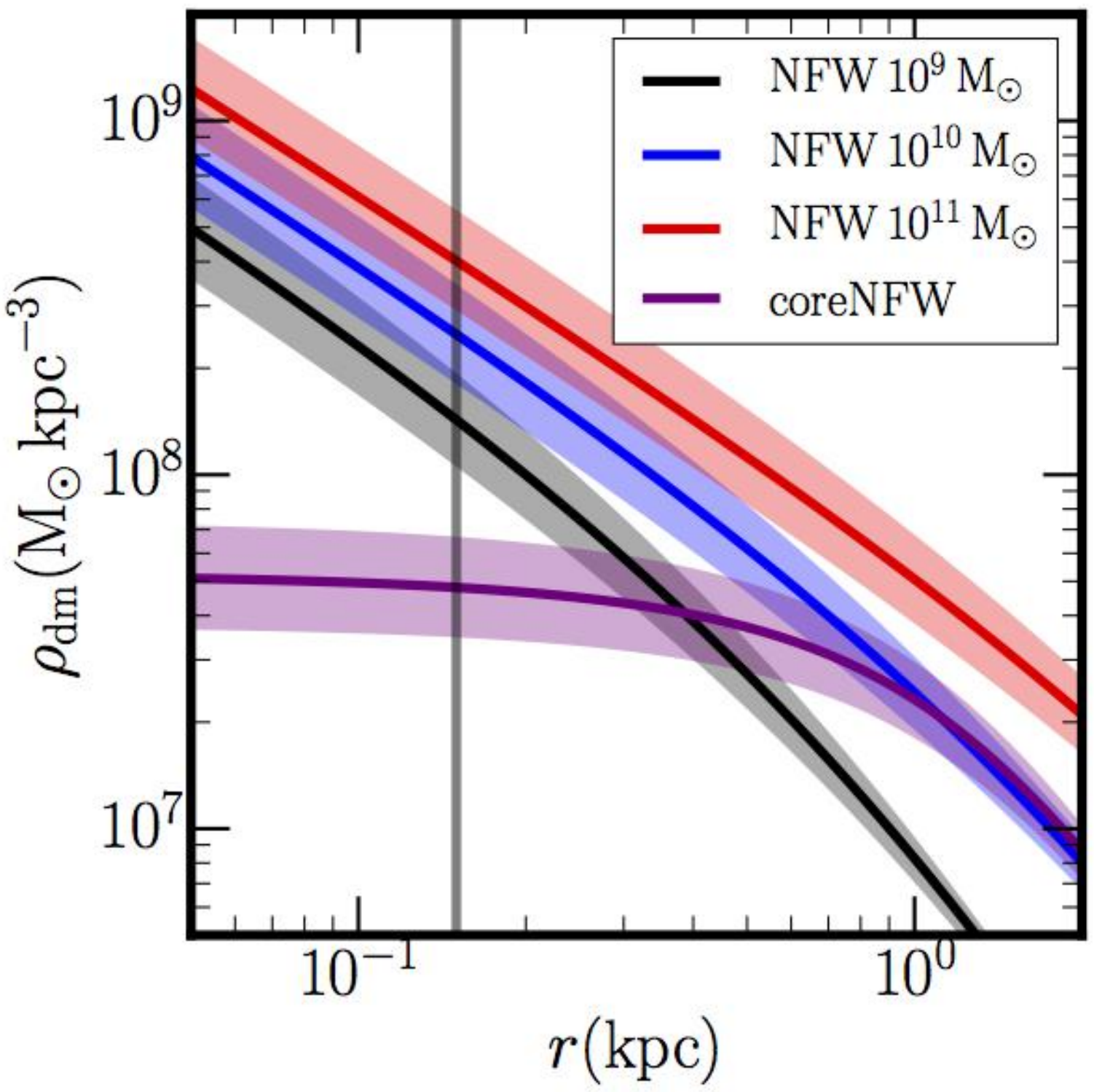

*Figure 61: Density profiles for dark matter halos having virial masses in the range expected for dwarf galaxies. Black, blue and red curves correspond to the pure 'NFW' profiles that characterize standard cold dark matter halos. The purple curve shows how the halo with* $M_{200} \sim 10^{10} M_\odot$ *might evolve in response to energetic feedback from star formation, lowering its central density and flattening the inner density gradient. Figure from Read et al. (2019).*

The dynamical masses of observed satellites (estimated within their half-light radii) are systematically smaller than the masses (evaluated at the same radii) of halos in simulations (Boylan-Kolchin et al., 2011, 2012a). This "too big to fail" problem is perhaps a symptom of a further discrepancy between the shapes of simulated and observationally-inferred mass-density profiles, $\rho(r)$ – the "cusp-core" problem alluded to previously. This discrepancy could be pointing to a deviation from the CDM paradigm, but it could also be correlated with "baryonic physics" including feedback from star formation, as discussed in Section 6.2.

Current N-body+hydro simulations agree with simple expectations that feedback processes become inefficient in the least luminous, most dark-matter-dominated galaxies, such that dwarf galaxies with $L < 10^6 L_\odot$ should retain their primordial CDM cusps (Peñarrubia et al., 2012; Garrison-Kimmel et al., 2013, Figure 61). This implies that dark matter physics can be separated from the astrophysics of galaxy formation, provided that observations can constrain the density profile of ultrafaint dwarfs.

The current census of Local Group dwarf galaxies includes ~ 50 such systems, ~10 of which host more than $10^3$ stars brighter than $V \sim 23$. For these galaxies, an MSE survey will deliver stellar-kinematic samples as large as those currently used to distinguish dark matter cores from cusps in dwarf galaxies with $L > 10^6 L_\odot$ (Walker et al., 2011; Read et al., 2019, 2018). Thus, MSE will have unprecedented power to constrain the nature of dark matter by measuring the density profiles of dwarf galaxies.

### 6.4.1 Luminosity function of Milky Way satellites in the era of LSST

If the dark matter distribution is well-described by numerical simulations of cold, collisionless dark matter, LSST is expected to discover another ~ 200 dwarf galaxies out to the virial radius of the Milky Way (Tollerud et al., 2008; Hargis et al., 2014; Kim et al., 2018; Newton et al., 2018; Nadler et al., 2018; Kelley et al., 2018). Spectroscopic followup of these objects will require a large amount of telescope time on a 10+ meter class telescope. Such observations are necessary to provide stellar chemo-dynamic samples with sufficient precision (~ 1 km $s^{-1}$ velocities, ~ 0.1 dex metallicities) to distinguish dark-matter-dominated dwarf galaxies from outer halo star clusters, to estimate dynamical masses and metallicity distributions, and to measure systemic line-of-sight velocities.

In order to quantify the contributions that MSE can make towards understanding the dark matter content of the Galactic satellite population, we first consider the spectroscopic sample sizes that would be achievable in an MSE survey of Local Group galaxies. For *known* systems less luminous than M32 ($M_V \gtrsim 16.5$), we estimate the number of stars brighter than a fiducial magnitude limit of $V \leq 23$ by integrating a log-normal stellar luminosity function Dotter et al. (2008) of a 10-Gyr-old stellar population with the metallicity reported by McConnachie (2012). Figure 62 compares these numbers to the largest spectroscopic sample sizes that are currently available in the literature for each observed galaxy. In most cases, an MSE survey would increase the available spectroscopic sample by more than an order of magnitude. For the 'ultrafaints' with $M_V \gtrsim -5$, this would imply stellar samples of several hundred to a thousand member stars. For the Milky Way's 'classical' dwarf spheroidals ($-7 \gtrsim M_V \gtrsim -13$) and more luminous distant objects (e.g., NGC 185, NGC 205 and NGC 6822), it means samples reaching into the tens of thousands of member stars. For each class of object, the precision with which we can infer dark matter densities from line-of-sight velocities alone increases by more than an order of magnitude. Used in combination with proper motion data, from the final Gaia release, 30m-class telescopes, or future space missions (e.g., WFIRST, Theia), MSE data will provide definitive constraints on the inner density profiles that distinguish various particle physics models.

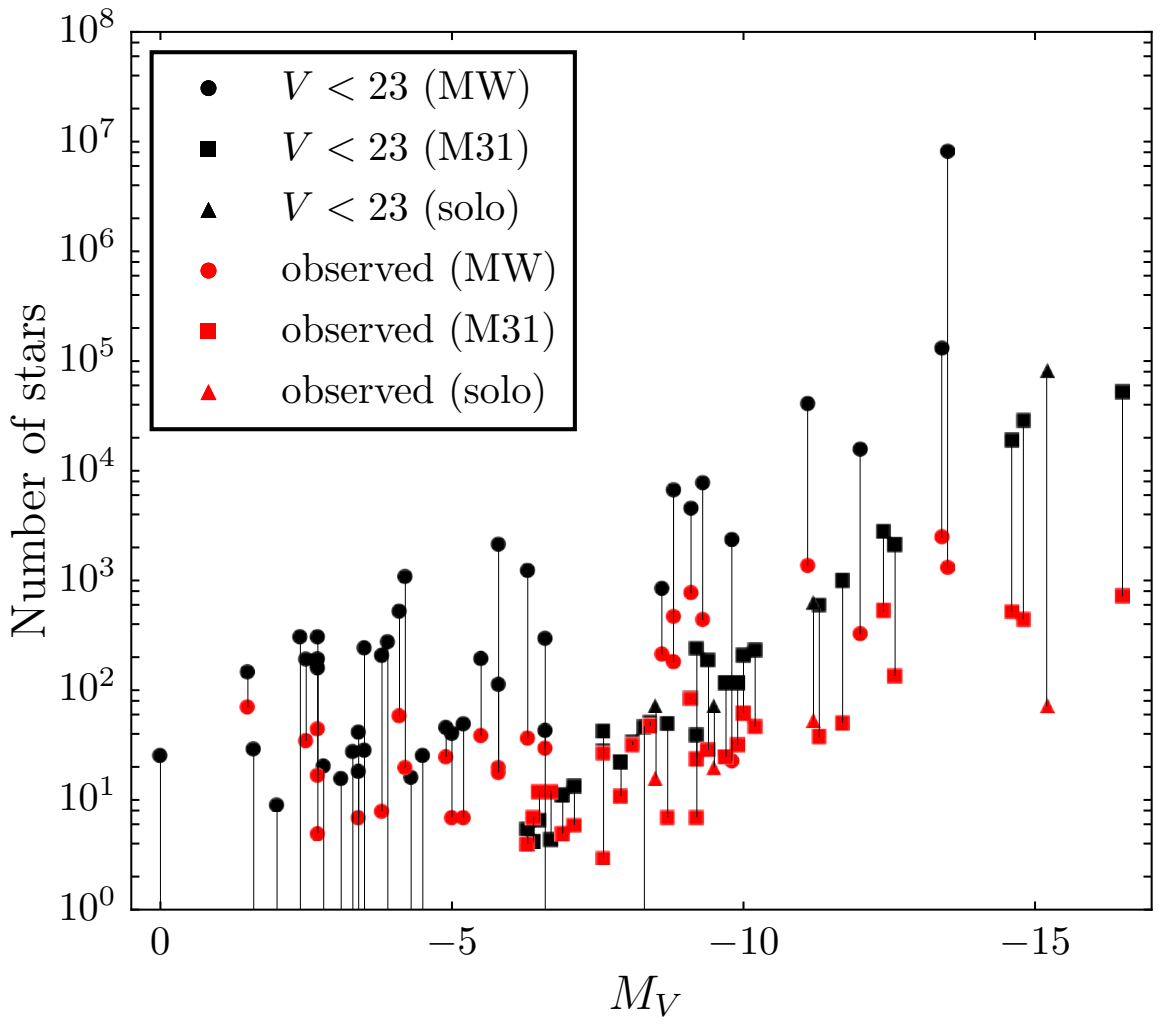

*Figure 62: Spectroscopic sample sizes for known Local Group dwarf galaxies with* $M_V \gtrsim -16.5$*. Black points indicate number of stars stars brighter than a fiducial magnitude limit of* $V < 23$ *that would be observable with MSE. Connected red points indicate the current sample size available in the literature. Marker types specify whether the dwarf galaxy is a satellite of the Milky Way (circles), a satellite of M31 (squares), or an isolated system (triangles). Note that the sample size is displayed on a logarithmic scale.*

To gauge the effects of stellar sample size on inferences about dark matter, we compare results from dynamical analyses of three mock data sets consisting of $N = 10^2$, $N = 10^3$ and $N = 10^4$ stars. In each case, the artificial sample is drawn from a phase-space distribution function describing a stellar population that follows a Plummer surface brightness profile and traces a gravitational potential dominated by a NFW dark matter halo. The analysis uses standard Bayesian procedures to fit simultaneously for the velocity anisotropy, surface brightness and dark matter density profiles (a by-product is specification of the line-of-sight velocity dispersion profile), similar to the procedures described by Geringer-Sameth et al. (2015); Bonnivard et al. (2015).

Figure 63 compares the resulting inferences for velocity dispersion and dark matter density profiles, displaying bands that enclose 95% credible intervals in each case. Given the samples that MSE can provide, inferences about kinematics and dark matter content of dwarf galaxies will become dramatically more precise. We will infer the dark matter densities of ultra-faint satellites with precision similar to what is achieved today only for the most luminous classical dwarfs, for which we will infer density profiles with unprecedented precision from pc to kpc scales. This improvement will render dynamical analyses limited by systematics (e.g., triaxiality, non- equilibrium kinematics, unresolved binary stars) instead of statistics. Moreover, since the smallest ultra-faints have half-light radii of just a few tens of parsecs, large MSE samples for these objects will provide strong constraints on dark matter densities at these smallest galactic scales.

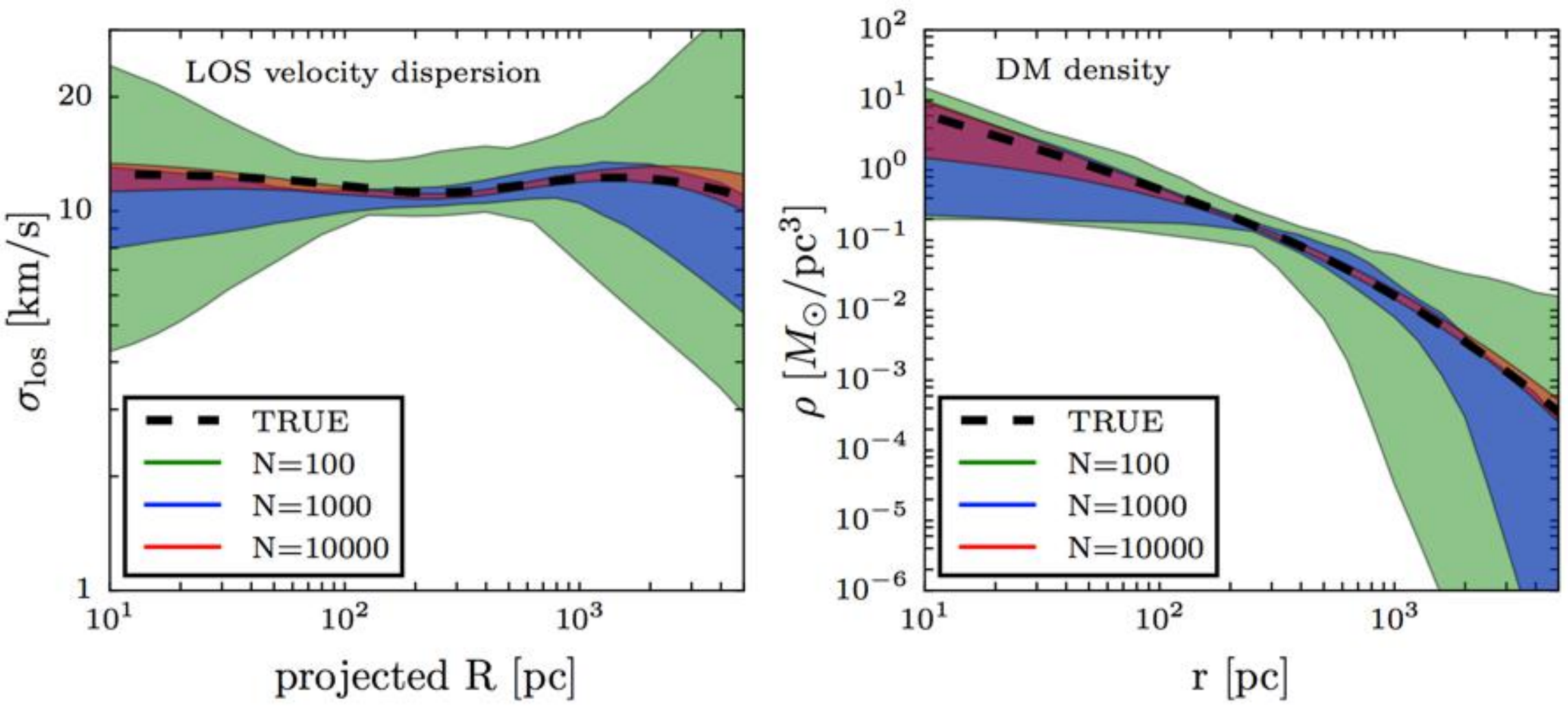

*Figure 63: Recovery of intrinsic line-of-sight velocity dispersion (left) and inferred dark matter density (right) profiles as a function of spectroscopic sample size. Shaded regions represent 95% credible intervals from a standard analysis (based on the Jeans equation) of mock data sets consisting of line of sight (LOS) velocities for $N = 10^2$, $10^3$ and $10^4$ stars (median velocity error 2 km$^{-1}$), generated from an equilibrium dynamical model for which true profiles are known (thick black lines, which correspond to a model having a cuspy NFW halo with $\rho(r) \propto r^{-1}$ at small radii).*

#### 6.4.2 Precise determination of the J-factor of nearby ultra-faint dwarf galaxies

Due to the high dark matter densities and the lack of astrophysical backgrounds (e.g., pulsars, scattering of cosmic rays off ISM, etc.) that contaminate searches near the Galactic center (Section 6.3.5), dwarf galaxies represent the cleanest available targets in searches for annihilation and decay signals (Gunn et al., 1978; Lake et al., 1990).

The flux of photons received from annihilation and decay of dark matter is proportional to the $J$-factor given by Equation 6.3. Thus, given a measurement of photon flux, or even a non-detection, one can use stellar-kinematic estimates of the dark matter density profile to infer or constrain relevant particle physics properties. For example, Figure 64 shows constraints on the dark matter self-annihilation cross section as a function of particle mass (Ackermann et al., 2015; Geringer-Sameth et al., 2015). These upper limits are derived by combining non-detections of gamma-rays from the Fermi-LAT with density profiles estimated from stellar-kinematics of fifteen of the Milky Way's dwarf satellites. For particle masses $M_\chi < 100$ GeV, these limits begin to constrain the cross section that would naturally give the cosmologically-required $\Omega_{\rm DM} \sim 0.2$ in the case of thermally-produced WIMPs (Steigman et al., 2012). Stellar-kinematic data have also been used to evaluate the significance of reported decay signals in X-ray observations of individual dwarf spheroidals (Loewenstein &

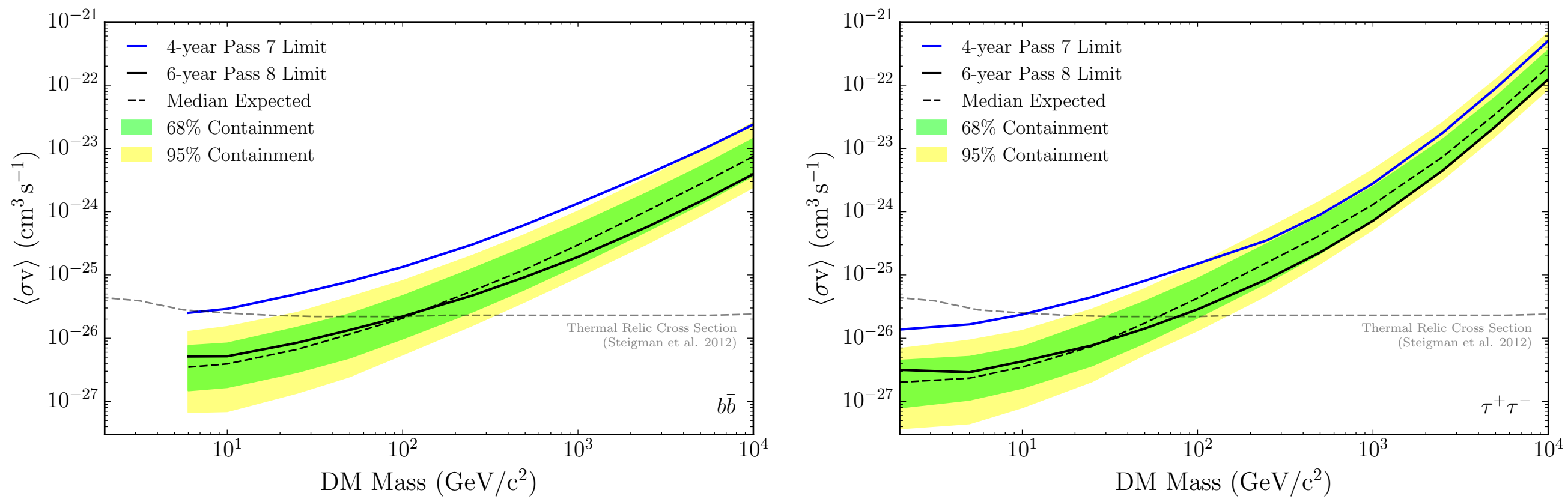

*Figure 64: Upper limits on the cross section for dark matter annihilation into final states of quarks (left) and leptons (right), derived from analysis of gamma-ray data and stellar spectroscopy of 15 Galactic satellites. Dashed lines show the median expected sensitivity, while bands indicate 68% and 95% quantiles. Dashed gray lines indicate the cross section expected for a thermally-produced weakly interacting massive particle (Steigman et al., 2012). Figure from Ackermann et al. (2015).*

Kusenko, 2010; Boyarsky et al., 2010; Jeltema & Profumo, 2015; Ruchayskiy et al., 2016). Regardless of whether an unambiguous photon signal is ultimately detected, the resulting inferences about particle properties are only as good as estimates of dark matter densities (via the $J$-factor) derived from stellar kinematics. As shown in Albert et al. (2017) decreasing the uncertainty in J-factor from 0.6 dex to 0.2 dex, can result a factor of $2-3$ improvement in the sensitivity of constraints on the annihilation cross-section. This dependence highlights the impact MSE will have on efforts to determine dark matter's particle nature. Among the known dwarf galaxies, the most attractive targets for annihilation/decay searches are also the least luminous ($L_V < 10^3 L_\odot$), primarily because these happen also to be the nearest. However, while the dynamical masses of these ultrafaint systems are also consistent with extremely large dark matter densities (Martin et al., 2007; Simon & Geha, 2007; Martinez et al., 2011b; Simon et al., 2011), the small number of spectroscopic measurements and the possibility of binary orbital motion inflation the velocity dispersion (McConnachie & Côté, 2010; Minor et al., 2010) have been serious hurdles. Improving upon existing samples by more than an order of magnitude, MSE will have a major impact in this area, which represents one of the best opportunities to resolve dark matter's particle nature.

#### 6.4.3 Controlling systematics with spatial and temporal completeness at high resolution

Nearly all methods for inferring the amount and distribution of dark matter within dwarf galaxies rest on the assumption that the observed stellar kinematics faithfully trace the underlying gravitational potential. However, the degree to which tidal effects and orbital velocity of binary stars invalidate this assumption are poorly constrained. Moreover, the ability to resolve the internal velocity dispersion $\lesssim 4\,\mathrm{km\,s^{-1}}$ of the ultra-faint galaxies is limited by the spectral resolution of existing instrumentation. MSE will solve these problems by observing thousands of targets simultaneously over a wide field at a resolution capable of measuring sub-km $s^{-1}$ dispersion, enabling inexpensive repeat observations. As a result, MSE will deliver spectroscopic samples with unprecedented completeness in both the spatial and time domains at high resolution. These capabilities are crucial for obtaining accurate dynamical masses for the nearest ultrafaint dwarf galaxies, which are the most important targets for indirect detection searches.

Nearly all published spectroscopic data sets for Galactic satellites suffer from spatial incompleteness; more specifically, selection biases that leave outer regions either under-sampled or neglected altogether. The outer regions of dwarf galaxies are important for investigating the outer structure of dark matter halos (Walker et al., 2007), identifying kinematic signatures of tidal disruption (Muñoz et al., 2006), mapping the evolution of internal halo structure (El-Badry et al., 2016), measuring systemic proper motions (Kaplinghat & Strigari, 2008; Walker et al., 2009), and tracing stellar population gradients (Harbeck et al., 2001; McConnachie et al., 2007). To date, the relatively low fractions of bona fide members at large radii has made thorough observations of these regions prohibitively expensive.

Existing spectroscopic samples also suffer from *temporal* incompleteness. For the vast majority of measured stars, published velocities are based on a single observation. However, in cases where repeat observations exist, there is evidence for velocity variability of individual stars (Olszewski et al., 1996; Minor et al., 2010; Simon et al., 2011; Koposov et al., 2011; Spencer et al., 2017), most likely due to the internal motions of unresolved binary systems. Radial velocity surveys of the Galactic halo indicate that stellar multiplicity increases toward lower metallicity, suggesting that the binary fractions in dwarf galaxies may be large. Indeed, recent studies of the limited multi-epoch data sets available for dwarf galaxies estimate binary fractions in the range of $50 - 75\%$ (Minor, 2013; Spencer et al., 2018). Since binary motions alone can contribute velocity dispersions of $\sim 2 - 3$ km $s^{-1}$ (McConnachie & Côté, 2010; Minor et al., 2010), such contamination can potentially dominate the dynamical masses estimated for the coldest ultra-faint systems, which typically have intrinsic dispersions estimated to be $\lesssim 3$ km $s^{-1}$ (Martin et al., 2007; Simon & Geha, 2007; Caldwell et al., 2017). The possible inflation of dynamical mass measurements of ultra-faint dwarfs presents a major caveat for conclusions about dark matter physics that rely on ultra-faint dwarfs, including constraints on indirect detections with gamma-rays (Section 6.4.2).

The relatively low spectral resolution at which most of the coldest ultra-faint systems have been observed compounds the sampling problems described above. The multi-slit DEIMOS spectrograph at the Keck telescopes has resolution $R \sim 5000$, with a systematic error floor estimated at $\sim 2 - 3$ km $s^{-1}$ (Simon & Geha, 2007). Thus, the variability introduced by existing instrumental effects, as well as astrophysical effects like binarism, is similar to the intrinsic velocity dispersions measured for the coldest systems. Using mock data sets, Figure 65 shows the effects of both spectral resolution and unresolved binaries on measurements of velocity dispersion and dark matter density. Clearly we require velocity errors on single stars that are no larger than the intrinsic velocity dispersions to be resolved (top panels); the coldest known ultra-faint dwarfs exhibit dispersions near $\sim 2$ km $s^{-1}$ (Caldwell et al., 2017). Furthermore, for such cold systems, failure to identify (and/or to model) binary stars generally leads to overestimation of velocity dispersions and dynamical masses. Fortunately, given its unprecedented ability to observe faint targets across a wide field with high resolution at multiple epochs, MSE will not only provide samples of unprecedented size, but also unprecedented constraints on the above sources of systematic error.

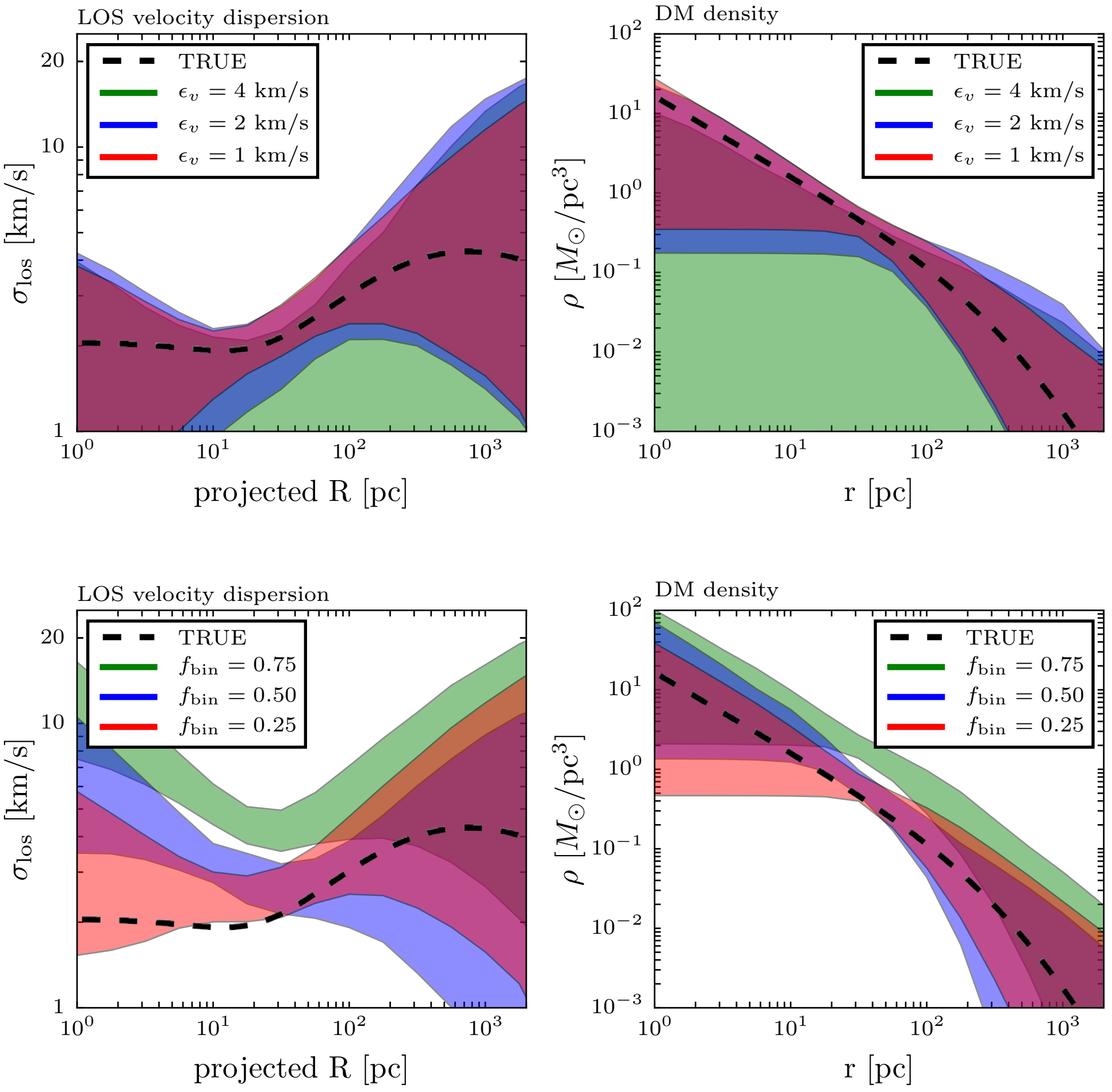

*Figure 65: Recovery of intrinsic line-of-sight velocity dispersion (left) and inferred dark matter density (right) profiles as a function of individual-star velocity precision (top) and the fraction of stars that belong to unresolved binary systems (bottom, with orbital parameters drawn from Galactic distributions), for an input model representative of low-mass ultra-faint satellites ($M_{200} = 10^8 M_\odot$, $R_{\rm half} = 20$ pc).*

## 6.5 Galaxies in the low redshift Universe

The CDM paradigm predicts the number, spatial distribution, and properties of galaxies over a wide range of masses. While CDM predictions are in remarkably good agreement with observations of high-mass galaxies ($M_\star > 10^{10}$M$_\odot$), studies of low-mass galaxies have raised several open questions (e.g., Weinberg et al., 2013). In particular, the low-mass end of the halo mass function and the profiles of low-mass dark matter halos are both especially sensitive to deviations from CDM. Current deep studies of dwarf galaxies either target the Local Group (a unique environment, see the previous section), or target relatively massive dwarf galaxies ($M_\star \gtrsim 10^9$ M$_\odot$). Spectroscopy is an essential ingredient to mapping the low-redshift ($z < 0.05$) dwarf galaxy population, linking galaxies to halos, and thus making inferences on the nature of dark matter.

New imaging surveys focused on the low-surface-brightness Universe are identifying large populations of dwarf and ultradiffuse galaxies outside the Local Group (van Dokkum et al., 2015; Muñoz et al., 2015; Carlin et al., 2016; Yagi et al., 2016; Geha et al., 2017; Crnojević et al., 2018; Greco et al., 2018; Smercina et al., 2018), a discovery trend that will only accelerate with LSST. However, their interpretation is severely hampered by a lack of distance measurements (van Dokkum et al., 2018; Trujillo et al., 2018). Methods such as resolved stars and low surface brightness detections can identify dwarf galaxies out to $\sim 3$ kpc and several tens of Mpc, respectively, with existing facilities (Danieli et al., 2018). Within $z < 0.05$, dwarf galaxies have photometric properties that are very similar to background galaxies, and photometric redshifts alone are uninformative in this regime (Figure 68). Spectroscopic redshifts or tip-of-the-red-giant-branch distance estimates are critical for establishing distances for faint dwarf and ultradiffuse galaxies. At a minimum, a subset of spectroscopic-confirmed low-redshift galaxies are needed to calibrate photometric distance measures (e.g., using surface brightness fluctuations or training photometric redshifts).

### 6.5.1 The faint end of the galaxy luminosity function

A key prediction of ΛCDM is the hierarchy of halos down to small halo masses. Because galaxies are similarly hierarchical (large galaxies live in large halos, and small galaxies live in small halos), a comparison of the predicted dark matter halo distribution to the luminosity function of galaxies is an important test of ΛCDM. Since there is some expected scatter in the galaxy – halo connection, this test requires a large observed volume. For brighter galaxies in the local Universe this has been possible with SDSS and GAMA (Driver et al., 2011), however, it is below these masses where uncertainties arise.

For galaxies fainter than $M_r \sim -14$, the uncertainty in the global low-redshift luminosity function is large, due to the incompleteness in both available photometric and spectroscopic surveys. This uncertainty in return results in large uncertainty in the galaxy–halo connection for halo masses below $10^{10}M_\odot$. Consequently, many inconsistencies between small-scale observations and model predictions (Bullock & Boylan-Kolchin, 2017) have not crystallized into concrete evidence for the need for non-cold dark matter due to the degeneracy between the galaxy – halo connection models and dark matter models. In addition, a more complete sample of low-z galaxies also opens up the possibility of utilizing galaxy clustering statistics to further constrain galaxy – halo connection models (Buckley & Peter, 2018).

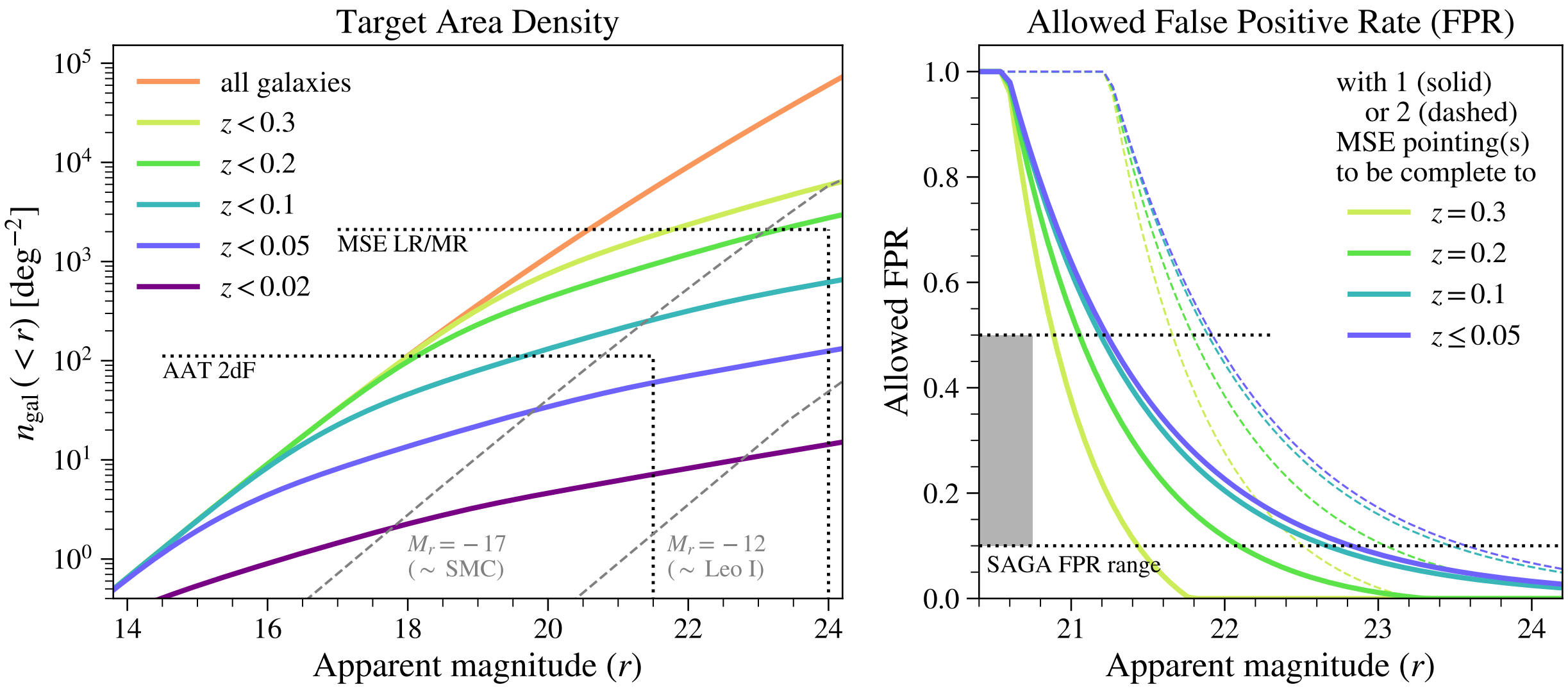

*Figure 66: Left: Target area density (number of galaxies per squared degree; y-axis) as a function of apparent magnitude (x-axis) and redshifts (different solid color lines; redshift values shown in legend). The black dotted lines show the fiber area densities (horizontal) and magnitude limits (vertical) for AAT 2dF and MSE LR/HR. The grey dashed lines are isolines of equal absolute magnitudes (−17 and −12, roughly corresponding to the absolute magnitudes of SMC and Leo I, respectively). Right: Maximal allowed false positive rate (FPR) in target selection (the ratio of the number of targeted high-redshift galaxies to the number of fibers) to complete all galaxies below certain redshifts (different solid color lines; redshift values shown in legend) within only 1 (solid lines) or 2 (dashed lines) MSE pointing(s), as a function of the apparent magnitude of the galaxies. The effect fiber collision is not considered in this plot. FPR = 1 means that all high-redshift galaxies can be targeted and the low-redshift galaxy sample will still be complete; in other words, no extra photometric selection is needed in completing the low-redshift sample. FPR = 0.5 means that the photometric or morphological selection is needed so that only half of the fibers are targeted at high-redshift galaxies in order to complete the low-redshift sample. FPR = 0 means that the area density of low-redshift galaxies is equal to or higher than the fiber area density; therefore even if all fibers are used for the low-redshift galaxies, the sample is still not complete. The horizontal black dotted lines (at FPR = $0.5, 0.1$) are the range of SAGA FPR, i.e. in SAGA the high-redshift galaxy contamination is about 10%-50% after photometric or morphological selection (note that SAGA targets only go down to $r = 20.75$).*

Tightening the global low-redshift luminosity function is a critical, yet challenging, task. In a given patch of sky, low-redshift galaxies are scarce due to the limited volume. To obtain a luminosity function that is as complete as possible, we need to measure redshifts of as many galaxies as possible. Figure 66 demonstrates the areal density as a function of redshift and apparent magnitude limits. In the left panel, we see that for $r < 20.5$, the low resolution fibers of MSE can obtain spectra for all galaxy targets in the field of view with just a single pointing (ignoring fiber collision). For $r < 24$, there are about 20 times more galaxy targets than the available fibers. In the right panel, we see that with an efficient target selection, MSE will be able to obtain a complete sample of low-z galaxies up to $z \sim 0.1$ at $r < 24$ (c.f., the SAGA survey, Geha et al. 2017).

### 6.5.2 Satellite populations in Milky Way analogs

Populations of satellite galaxies are particularly important probes of hierarchical formation models. The Milky Way is the most well-studied galaxy, and so far we have identified its $\sim 50$ dwarf galaxy satellites, including the Magellanic Clouds, the classical dwarf spheroidals, and more recently discovered ultra-faint dwarfs from SDSS and DES (Bechtol et al., 2015; Drlica-Wagner et al., 2015). These satellites provide a unique probe of galaxy and star formation at early times in low mass objects, and also a powerful tool for distinguish different dark matter models.

However, the Milky Way satellite population constitutes a small, and perhaps biased, sample from which it is difficult to extrapolate. For example, do host galaxies with similar luminosity, morphology, and mass as the Milky Way harbor a similar population of satellites? Applying our detailed knowledge of the Milky Way satellites to broader questions of dark matter properties requires an improved understanding of satellite populations in the context of cosmology.

In fact, we know little about dwarf satellite galaxies outside the Milky Way and M31: the faintest detectable satellite galaxies around Milky Way analogs in SDSS (spectra to $r < 17.7$) are similar to the Magellanic Clouds (MC). In SDSS, Milky Way analogs on average have only $\sim 0.3$ MC-like satellites, v.s. two for the Milky Way (e.g., Busha et al., 2011). The recent SAGA Survey (Satellites Around Galactic Analogs, Geha et al., 2017) has taken on the exploration of finding satellite systems around Milky Way analogs. So far, the SAGA has constructed complete satellite luminosity function down to $M_r \sim 12$ (corresponding to Leo I), around 8 Milky Way analogs. This helps to constrain the intrinsic distribution of satellites around Milky Way mass galaxies, of which the Milky Way itself is a single realization. However, a much larger sample is needed to reach any concrete conclusions. SAGA aims to obtain the satellite luminosity function for Milky Way mass galaxies between 20–40 Mpc, which amounts to about 100–200 hosts (depending on image availability). MSE will be able to push the boundary much further beyond 40 Mpc, or to push the satellite luminosity function to a fainter limit. As Figure 66 demonstrates, MSE can easily obtain a satellite luminosity function down to $M_r \sim -12$ for any host galaxy within $z \sim 0.02$, provided a good target selection strategy, i.e. a false positive rate in target selection similar to SAGA.

Comprehensive analyses of the satellite systems of Milky Way analogues should provide particularly important insights into recent results suggesting that planes of satellites exist around the Milky Way and other galaxies, possibly in conflict with predictions of CDM+simple galaxy formation models (see Figure 67 and references in Pawlowski, 2018). Indeed, testing whether such structures exist using a larger sample of satellites than currently exists offers a unique test of CDM. Satellite kinematics are less affected by baryonic effects: the positions and motions of satellite galaxies on scales of 100s of kpc is not strongly affected by their internal dynamics. While this makes the issue particularly challenging to address within the ΛCDM framework, it holds potential to provide clues on the formation of dwarf galaxies and their accretion patterns that are not strongly dependent on the implementation of baryonic physics in simulations, but dominated by the overall dynamics governed by the dark matter distribution and properties.

Planes of Satellite Galaxies have two characteristic properties: (1) a spatial flattening in the positions of satellites around their host, and (2) a kinematic coherence, either seen in a preferred orbital direction (for the Milky Way where proper motions are available) or in line-of-sight velocities that are indicative of a rotating plane (i.e. satellites on one side are blueshifted, those on the other side are redshifted relative to the host). Obtaining spectroscopic systemic velocities for potential satellite galaxies around a larger sample of hosts is necessary to study both characteristics:

1. The spatial analysis is improved by rejecting fore- and background contamination which can be assumed to contribute an isotropic signal that dilutes any spatial flattening. This is particular important since studying the spatial arrangement of satellite galaxies at distances beyond $\sim 5$ Mpc loses one of three spatial dimensions; it is only possible in projection. This is because even small uncertainties of $\sim 5\%$, as achievable with the tip of the red-giant branch method, correspond to the whole virial volume of a Milky Way-like host;

2. While full 3D velocities would be required to confirm or refute a rotational support of found satellite planes, line-of-sight velocities alone can already give a statistical answer to the prevalence of kinematically coherent structures. For random sight-lines, a plane of satellite galaxies is seen to higher than $60°$ inclination in 50% of the cases, which implies that the line-of-sight velocity is dominated by the in-plane component if the satellite plane is indeed rotating.

More generally, a larger sample of satellite systems with spectroscopically confirmed satellite galaxies allows studies of other, potentially related, phase-space correlations among satellites. One example is the apparent over-abundance of pairs of satellites in the Local Group compared to ΛCDM simulation (Fattahi et al., 2013), another is the abundance and properties of groups and associations of dwarf galaxies pre- and post-infall onto a host galaxy (e.g., Sales et al. 2011, 2017 and references therein). A third example is the alignment of satellite planes with their surrounding large scale structure, which could give clues on their origin. Current studies (Libeskind et al., 2015) give mixed results – the planes around M31 and Centaurus A appear to align with the ambient shear field, that around the Milky Way does not – but these studies are severely limited by the small sample size of known satellite planes. MSE can provide a much larger sample to investigate these questions in a statistically meaningful way.

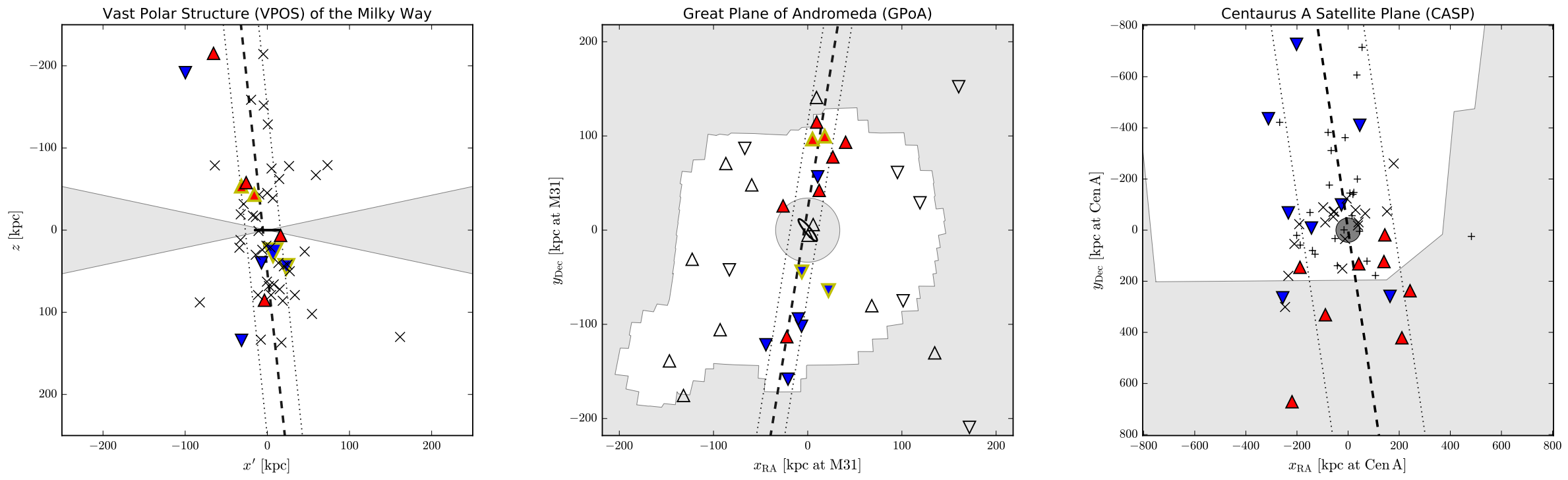

*Figure 67: Edge-on views of three planes of satellites around the Milky Way (left), M31 (middle) and Centaurus A (right). The orientation of the flattened structures and their width are indicated with dashed and dotted lines, respectively. Satellites that are part of a planar structure and for which line-of-sight velocities are known (in the case of the 11 classical satellites of the Milky Way these are calculated from their proper motions) are shown as filled symbols, with upward red triangles indicating receding and downward blue triangles indicating approaching motion in these views. The two close pairs of satellites in the Milky Way and two in M31 identified by Fattahi et al. (2013) are highlighted with yellow outlines. Regions obscured by the Galactic disk, or outside of survey footprints, are indicated in grey. Figure from Pawlowski (2018).*

### 6.5.3 Local galaxies as gravitational lenses

Above halo masses of $10^{11}M_\odot$ (corresponding roughly to an LMC-mass galaxy), there is good agreement among various probes of the relationship between the stellar mass and halo mass (e.g., Wechsler & Tinker, 2018). Below this mass scale, measurements of the relation based on dynamical masses within the optical radius of galaxies (e.g., with rotation curves) and based on abundance matching diverge, with dynamical measurements preferring smaller halo masses (e.g., Klypin et al., 2015; Papastergis et al., 2015; Buckley & Peter, 2018). This mismatched is sometimes called the "Too Big to Fail in the Field" problem. The fundamental problem is that neither of these measurements of halo mass is direct, but instead rely on extrapolations. In the case of dynamical measurements, the match between the stellar mass and halo mass depends on strong priors about the density profile to connect the optical radius to the virial radius (a factor of $\sim 10-100$ in scale). In the case of abundance matching, the key assumptions are that the halo mass function is the CDM halo mass function, and that all halos at these masses should have galaxies inside of them. The field needs a direct measurement of the halo mass of objects of Magellanic Cloud-mass or below in order to tell if the mismatch in the indirect measurements is a result of a density profile problem, a halo count problem, or a star-formation stochasticity problem. To this end, weak (gravitational) lensing provides a direct unbiased measurement of the total mass, and of the density profile of the halo outside the optical radius.

Weak lensing refers to the subtle distortion of galaxy shapes due to light being deflected from the large-scale matter distribution in the line-of-sight (Bartelmann & Schneider, 2001). The weak lensing signal is typically detected statistically as the average distortion of the galaxies are very weak (sub-percent compared to average intrinsic galaxy shapes at about 30 percent). Galaxy-galaxy lensing refers to a specific statistics used to extract this signal, specifically a correlation of the foreground (lens) galaxy position and the background (source) galaxy shape distortion, or shear. This effectively gives a measure of the average mass distribution around the foreground galaxy sample. With sufficient signal-to-noise, one can use weak lensing to constrain the total mass of the low-mass galaxies described above. Sifón et al. (2018) showed a proof of concept using 784 ultra-diffuse galaxies (UDGs, see more discussion in Section 6.5.4) around 18 clusters at $z < 0.09$ and obtained an upper bound for the average mass of the UDGs. Similar measurements can be done with dwarf galaxies or other specific low-mass lens samples.

One further possibility is to use the full weak lensing profile shape (in addition to the amplitude) to constrain different dark matter models. Baryonic and dark matter physics can both alter density profiles from their NFW-like baryon-free CDM form (Davé et al., 2001; Colin et al., 2000; Governato et al., 2010; Fitts et al., 2018). These effects may be separable (although baryons lead to "convergent evolution" of halo properties relative to their baryon-free predictions), and so measuring deviation of the profiles from NFW could provide constraints on e.g. SIDM models (Ren et al., 2018). Moreover, a measurement of the density profile can help determine the physical origin of the discrepancy between dynamics and abundance-matching-based inferences of the stellar-mass–halo-mass relation.

The two main challenges of these measurements are (1) the number and average redshift of these galaxies are low, and so we expect the signal to be relatively weak; (2) contamination of high-redshift galaxies in the lens sample could introduce spurious signal and result in a bias in the weak lensing mass/profile inferred from the data. Having accurate spectroscopic redshifts from MSE will help for both of these aspects compared to the case where only photometric redshift (photo-z) estimates are available. First, even assuming optimistically that the photo-z estimates are unbiased, the scatter would smear out the signal and lower the detection significance. Second, having spectroscopic redshift ensures a much cleaner lens sample without catastrophic photo-z outliers that could bias the inferred weak lensing mass. Figure 68 demonstrates the potential problems of using photo-z estimates for the weak lensing measurements of the low-redshift galaxies. At low redshift ($z < 0.015$) and faint magnitudes ($r > 17.7$), where the population of interest lies, one can see a significant fraction of outliers in the photo-z estimates and large scatter.

### 6.5.4 Ultra diffuse galaxies

The recent research focus on very low surface-brightness galaxies (so-called "Ultra-Diffuse Galaxies" or UDGs) in nearby galaxy clusters has opened a surprising new avenue to investigate the galaxy-halo connection. While such objects have been known to exist for decades (Sandage & Binggeli, 1984; Dalcanton et al., 1997), their abundance in clusters was not fully appreciated; hundreds are now known to live in the most massive clusters (Koda et al., 2015; van der Burg et al., 2017). Notably, the few UDGs for which spectroscopic studies have been conducted exhibit unusual properties associated with their dark-matter halos, such as large globular cluster populations (van Dokkum et al., 2016) or very high (van Dokkum et al., 2016; Beasley et al., 2016) or low velocity dispersions (van Dokkum et al., 2018). Although many explanations have been put forth to explain these unusual galaxies, none are able to completely reproduce the observed properties (e.g. Amorisco & Loeb, 2016; Chan et al., 2018; Carleton et al., 2018). Further spectroscopic observations are necessary to both directly probe their dark-matter halos and develop a more complete understanding of their formation and evolution. As these unusual galaxies primarily live in group and cluster environments (van der Burg et al., 2017), they are natural targets for the high multiplex of MSE.

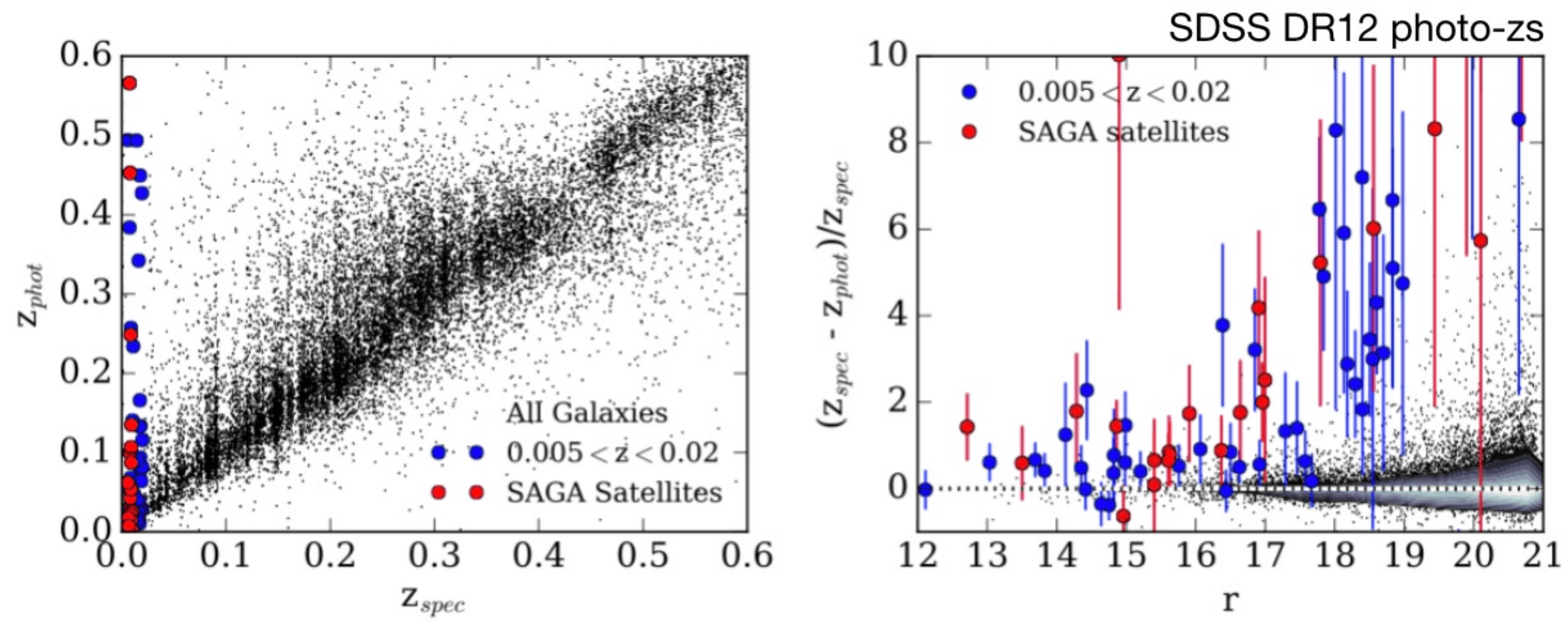

*Figure 68: (Left) Spectroscopic redshift plotted against the SDSS DR12 photometric redshifts from Beck et al. (2016). (Right) Apparent r-band magnitude versus the fractional difference between the spetroscopic and photometric redshifts. In both panels, we plot SAGA satellite galaxies as red circles and a larger number of field galaxies over a similar redshift range (*$0.005 < z < 0.015$*) as blue squares. For the majority of galaxies with redshift* $z < 0.015$*, particularly for galaxies fainter than* $r_o > 17.7$*, photometric redshifts are neither accurate nor precise. Figures from Geha et al. (2017).*

Low-surface-brightness galaxies have long been understood as excellent laboratories for understanding the galaxy – halo connection, as their dynamics are governed by their dark matter halos both at small and large radii. However, current studies arrive at vastly divergent conclusions regarding the dark-matter content of UDGs, with reported halo masses spanning 4 orders of magnitude (van Dokkum et al., 2016; Sifón et al., 2018; van Dokkum et al., 2018). While indirect measurements suggest that UDGs typically live in dwarf halos (e.g. Amorisco et al., 2018), direct spectroscopic measurements persistently find unusually high (van Dokkum et al., 2016; Beasley et al., 2016) or low (van Dokkum et al., 2018) halo masses. In the latter case, follow-up observations paint an even more complex picture; globular clusters and planetary nebulae are both consistent with a $10\,\mathrm{km\,s^{-1}}$ dispersion (Laporte et al., 2019a; Martin et al., 2018; Emsellem et al., 2018), whereas the more concentrated stellar component has a slightly higher dispersion at $16\,\mathrm{km\,s^{-1}}$ (Emsellem et al., 2018). A larger sample of UDGs with direct mass estimates from a variety of tracers is crucial to resolving this apparent discrepancy as only four galaxies have been spectroscopically followed-up (in terms of their globular cluster abundance or stellar velocity dispersion). Regardless, these unexpected results indicate that further study of the dark-matter content of UDGs will be particularly fruitful for our understanding of the physics of dark matter and the role it plays in galaxy evolution.

These measurements will be possible with the large field-of-view and multiplexing capabilities of MSE. While direct stellar velocity dispersion measurements will only be possible for a few UDGs with dozens of hours of observations, MSE will measure the velocity dispersion of the associated globular cluster population for all UDGs out to the distance of Virgo. Additionally, spectroscopic observations of UDGs in clusters as far out as Coma with MSE will provide spectroscopic distance measurements necessary to confirm the large sizes and cluster membership of UDGs, paving the way for deeper follow up studies. Current samples of UDGs suffer from significant foreground contamination (van der Burg et al., 2016), and simple spectroscopic redshifts obtained with MSE will greatly improve the sample of UDGs that are spectroscopically confirmed. For example, this spectroscopic identification will greatly improve weak lensing constraints on UDG masses (Sifón et al., 2018), in a similar way that it will improve weak lensing constraints on dwarf galaxies in general (see previous section). Furthermore, stellar metallicities and ages of both the UDGs and their globular clusters will provide valuable checks on theories for UDG formation and evolution (e.g. Gu et al., 2018). In particular, robust identification and measurement of the rich UDG globular cluster population will allow for a better understanding of this least-understood aspect of UDGs.

The capabilities of MSE place it in an excellent position to address these questions regarding UDGs. MSE's large field of view allow it to cover the $\sim 300$ UDGs in Coma in less than 10 pointings. While the central surface brightness of UDGs, ranging from $23 - 25$ mag/arcsec$^2$, push the limit of MSE, a spectroscopic census of UDGs down to 25 mag/arcsec$^2$ is possible. Deeper observations will provide high-quality spectra necessary to measure metallicities and ages of the stellar populations of UDGs, as well as direct stellar dispersions for a sample of brighter UDGs. Given the wealth of data obtained through spectroscopic observations of individual UDGs, this type of survey will transform our understanding of the low surface brightness Universe. Additionally, MSE is a particularly powerful instrument to target globular clusters around UDGs in Virgo, as it will be able to identify and the velocity dispersion of $10 - 20$ globular clusters (with $r$-band magnitudes spanning $20 - 24$) around the very extended (half light radii $> 1$ arcmin) UDGs in Virgo in a single pointing. Spectra of these objects with velocity resolution better than several km s$^{-1}$ will provide a robust constraint on the dark-matter content of these galaxies, provided enough tracers are present (Laporte et al., 2019a) or at least spectroscopically confirm globular cluster candidates.

### 6.6 Galaxies beyond the low redshift Universe

MSE can measure the dark matter distribution and the halo mass function in galaxies beyond the low-redshift ($z > 0.05$) Universe. We consider three different probes in the following sections that show promise for constraining the particle nature of dark matter. Two of these are based on gravitational lensing effects of small halos and they will both measure the halo mass function to scales smaller than $10^8\,M_\odot$. The third is based on kinematics of the bright cluster galaxies close to the center, with the power to constrain the smooth halo profiles of clusters of galaxies.

One of the most robust predictions of the ΛCDM model is the ubiquity of hierarchical mass substructure at all scales down to the free-streaming cutoff length of the dark matter particle (e.g. Springel et al., 2008, Figure 69, left). On the scale of individual massive galaxies, the absence of dwarf satellite galaxies in comparable abundance could be considered a challenge to the ΛCDM paradigm, or alternatively could just be a reflection of mass-dependent star-formation efficiency and observational selection processes (e.g., Klypin et al., 1999). Strong gravitational lensing on the scale of individual galaxies provides a method to constrain substructure directly in the dark sector and beyond the local Universe, since lensing is sensitive to all gravitating mass independent of luminosity. For unresolved sources such as lensed quasars, this method operates through the detection of flux-ratio anomalies: differences between the relative magnifications of lensed images as compared to the predictions of smooth mass models (e.g., Dalal & Kochanek, 2002). For resolved sources such as lensed normal galaxies, the method operates through the detection of surface brightness perturbations (e.g., Vegetti et al., 2010b) associated with the presence of gravitating substructure in the lens galaxy. The right panel of Figure 69 illustrates both these cases.

The large number of new galaxy-scale strong lenses that will be delivered by MSE (both alone and in conjunction with imaging facilities) will enable strong-lensing tests of dark-matter substructure at high significance, providing a fundamental test of the ΛCDM hypothesis through constraints on the parameters of the dark-matter halo and subhalo mass functions. We consider the prospects for these experiments through the flux-ratio anomaly and surface-brightness perturbation channels in the following sections. We then discuss a novel technique to infer the dark matter distribution in the inner regions of clusters via the wobbling of the brightest central galaxies (BCGs).

### 6.6.1 Quasar lensing: flux ratio anomalies due to low mass dark matter halos

In strong gravitational lensing, multiple images of a background source appear due to the distortions in space-time caused by one or more intervening massive objects along the line of sight to the observer. The positions and relative brightnesses of these multiple images depend on the first and second derivatives of the gravitational potential of the deflector, respectively. Due to this dependency, the image positions provide a strong constraint on the smooth, larger-scale mass distribution of the deflector, while the relative image magnifications are extremely sensitive to low-mass halos. The current limit is $M_{200} \sim 10^{6.5}\mathrm{M}_\odot$ with current technology (e.g. Nierenberg et al., 2014, 2017).

This method was first applied to strong lenses two decades ago (Mao & Schneider, 1998), but it has not yet reached its full potential, because the number of suitable quasar lenses has been too small. In particular, useful quasar lenses must have *four* images to provide an accurate constraint on the smooth mass distribution. Additionally, the source must be at least milli-arcseconds in size to avoid significant perturbations by stars in the plane of the deflector. These two requirements have heretofore limited the field to the small number of currently known radio-loud quasar lenses (Dalal & Kochanek, 2002). New technologies have recently made it possible to extend this analysis to more systems by measuring strongly lensed quasar narrow-line emission, which is observed in virtually all quasars (unlike radio). This is extremely promising: nearly all optically selected quasars have significant narrow-line emission, making it possible to extend the strong-lensing measurement of the dark matter mass function to many more systems (Moustakas & Metcalf, 2003).

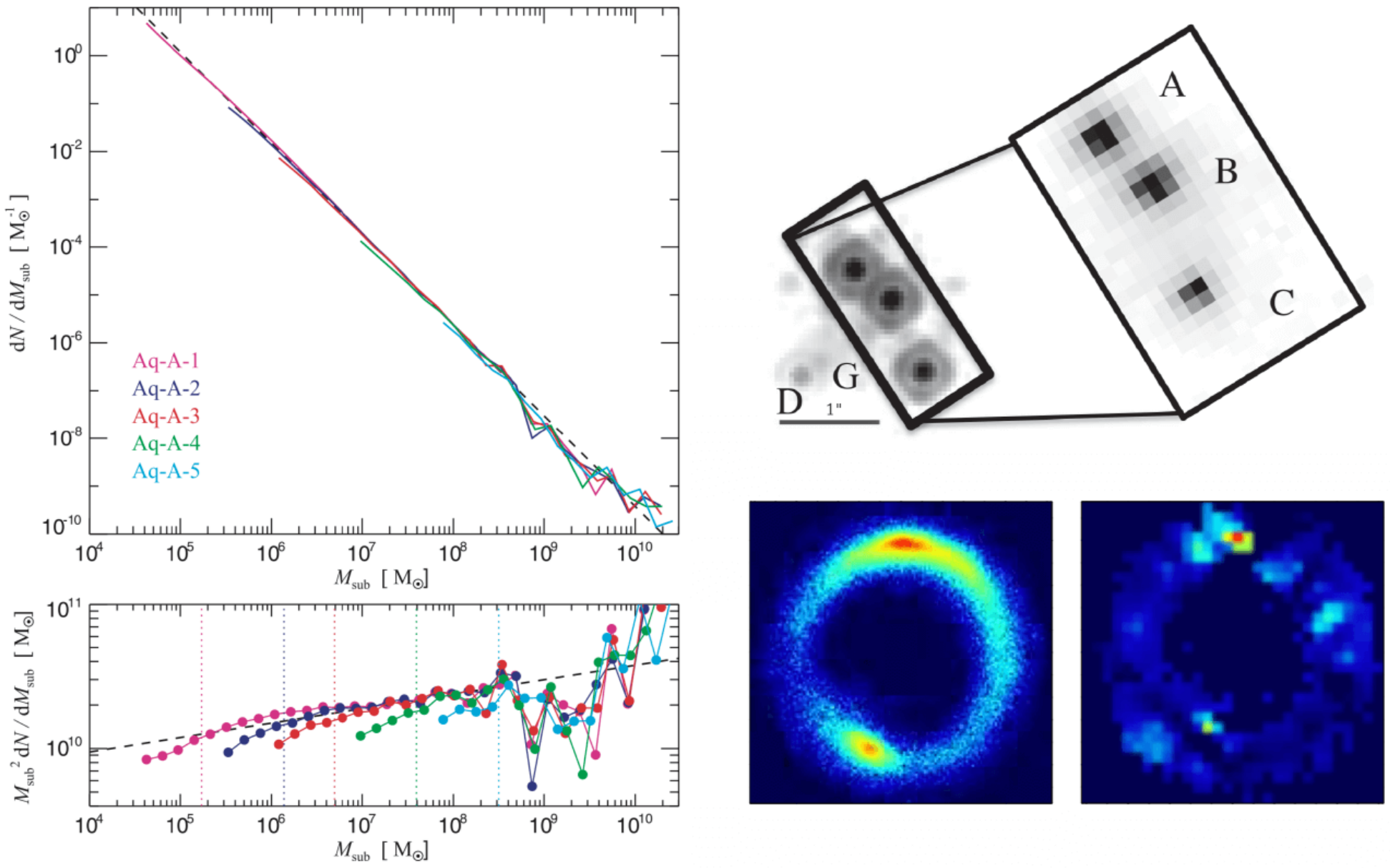

*Figure 69: Left: subhalo abundance within simulated Milky Way-scale halos from Springel et al. (2008). Integrated abundances are of the order 1,000 subhalos in the decade between $10^7$ and $10^8$ $M_\odot$, and of the order 100 subhalos in the decade between $10^8$ and $10^9$ $M_\odot$. Gravitational lensing is sensitive to perturbations from these subhalos as well as all halos along the line-of-sight (e.g Despali et al., 2019; Gilman et al., 2018). Upper right: HST-NICMOS imaging and Keck-OSIRIS integral-field data for the quadruply lensed quasar B1422+231 from Nierenberg et al. (2014), used to detect the presence of substructure in the lens galaxy through flux-ratio anomalies. Lower right: Keck NIRC2 adaptive optics-assisted image of the lensed galaxy B1938+666 from Vegetti et al. (2012a), along with reconstructed density perturbation from a dark-matter dominated satellite to the lens galaxy detected via surface-brightness perturbations to the Einstein ring image.*

Based on recent, state-of-the-art simulations Gilman et al. (2018), it is estimated that, with approximately 100 lenses (depending on flux precision), it will be possible to place a more stringent constraint on the 'warmth' of dark matter — compared to the constraints provided by the Ly-$\alpha$ forest (Viel et al., 2013). With several hundreds of lenses it will be possible to rule out even lower-mass cutoffs in the power spectrum. Such a measurement will have completely independent systematic uncertainties and therefore provide a crucial probe of dark matter below the mass scale at which dark matter halos are currently known to reliably form galaxies.

LSST will contain $\sim 1000$ four-image quasar lenses, in which three images will be brighter than the survey magnitude limit (Oguri & Marshall, 2010). With current gravitational lensing techniques and IFU sensitivity alone, this number will be sufficient to provide stringent new constraints on a turnover in the dark matter power spectrum (Gilman et al., 2018). MSE will play two crucial roles in this constraint both by confirming many of the quasar lenses, and by selecting ideal candidates for follow-up with the next generation of 30-meter class telescopes. MSE will provide an essential step in reaching this goal of measuring microlensing-free fluxes for hundreds of quasar lenses.

Deep, high-resolution imaging will enable morphological and color selection of quasar lens candidates. However, based on results from Agnello et al. (e.g. 2015) and Agnello & Spiniello (e.g. 2018), color and morphological information alone is insufficient to separate quasar lenses from the "blue cloud" of galaxies. For example, lens searches in DES have relied on WISE infrared photometry to isolate objects with quasar-like colors. As is shown in Figure 70, the number of identified four-image lenses is incomplete relative to theoretical predictions at magnitudes fainter than an $i$-band magnitude of 18, which is far below the survey depth of $i \sim 24$. This is due to the limit imposed by requiring WISE photometry, as shown in the right hand panel of Figure 70. In contrast, in SDSS, matching spectroscopy enabled the discovery of quasar lenses down to the limiting survey magnitude of $i \sim 21$ (Inada et al., 2012). MSE can provide critical spectroscopy for this science in conjunction with current and future imaging surveys.

For the goal of measuring a turnover in the low mass end of the halo mass function, we conservatively require approximately 200 lenses (Gilman et al., 2018). Over the area of LSST, this can be achieved by reaching a limiting spectroscopic depth of $i \sim 22$ (Oguri & Marshall, 2010). The number of quasar candidates can be estimated with a purely optical color selection using the results from Richards et al. (2002), which identified approximately 18 candidates per square degree to a limiting $i$-band magnitude of 19 with SDSS photometry. Given a true number at this magnitude of approximately one quasar per square degree, and making the coarse assumption that the purity remains constant with magnitude, we therefore expect about 600 quasar candidates based on optical color selection for the 30 true quasars with $i < 22$ per MSE pointing. This is certainly feasible given the $> 3000$ low resolution fibers for MSE. However, the number can likely be further reduced significantly with the addition of morphological cuts.

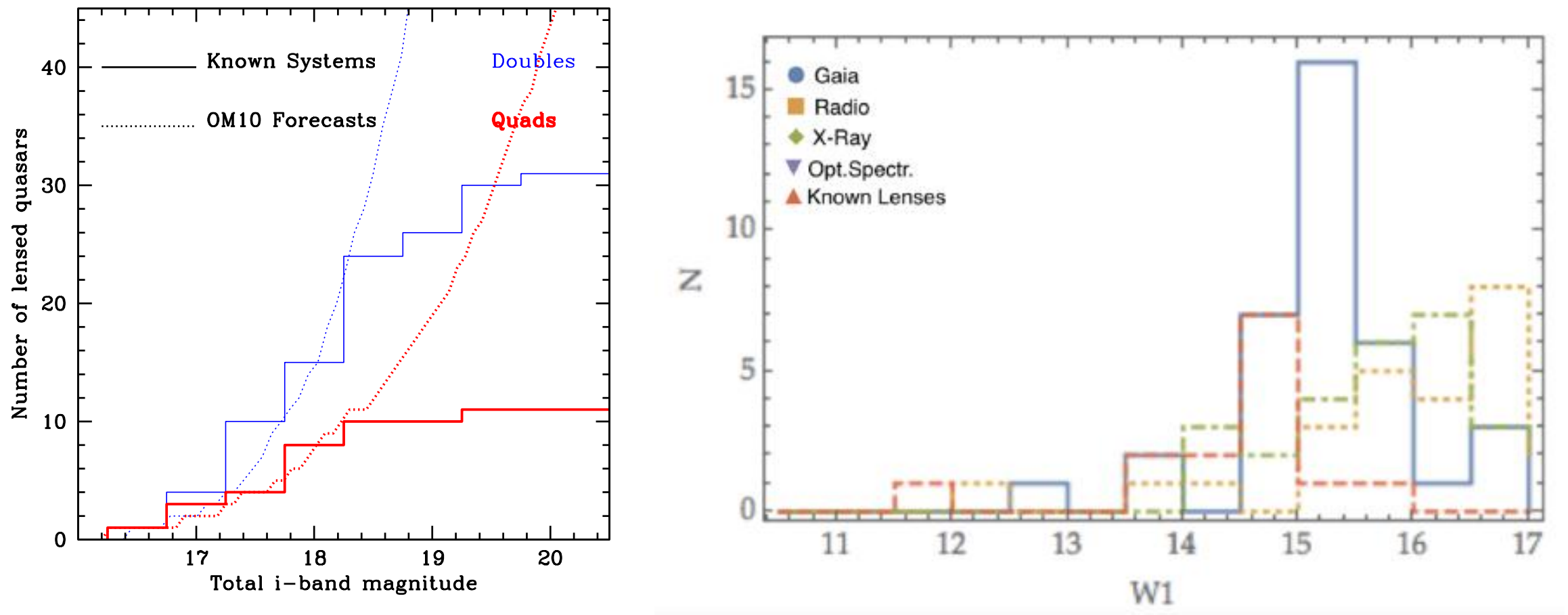

*Figure 70: Detection of quasar lenses in optical data relies heavily on ancillary data. In the case of DES it relies particularly on infrared data from WISE. Left: The number of quasar lenses from Oguri & Marshall (2010) predicted to be found in the DES footprint (dotted lines) compared with the observed numbers (solid histogram). Lens finding algorithms become incomplete well below the depth of the DES survey (~ 24 i-band). Figure from Treu et al. (2018). Right: Distribution of WISE W1 magnitudes of quasar lenses and lens candidates. The candidate numbers peak at the limiting magnitude WISE W1 filter. In order for optical surveys to reach their full potential for gravitational lensing, deep ancillary data is crucial. Figure from Agnello (2017).*

In addition to confirming quasar lenses and to measuring their source redshifts, MSE will enable the measurement of spatially-blended, narrow-line fluxes. Given the fiber size, spectra from all four images of a lensed quasar will be blended into a single measured spectrum. This narrow-line flux measurement will enable accurate planning for follow-up observations with higher spatial resolution facilities such as TMT, and enable the elimination of targets with significant spectral contamination from broad iron or hydrogen lines.

#### 6.6.2 Galaxy-galaxy lensing: image perturbations by low mass dark matter halos

Galaxy redshift survey spectra have proven to be an unparalleled resource for the discovery of new strong galaxy-galaxy lensing systems in large numbers. Through analysis of multiple generations of the SDSS, it has been found that $\sim 0.1 - 1\%$ of galaxy spectra show evidence for another galaxy (or quasar) along the same line of sight. Follow-up imaging reveals a substantial fraction of these 'lens candidates' to be bona fide strong gravitational lenses (e.g., Bolton et al. 2006, 2008; Brownstein et al. 2012; Shu et al. 2017, 2018). This technique is illustrated in Figure 71.

The sensitivity and multiplexing capability of MSE, combined with its dedicated survey operations mission, can enable flux-limited galaxy surveys ten times larger than the original SDSS. These surveys can be expected to deliver significant new samples of strong galaxy-galaxy lenses, which can in turn enable high precision lensing tests of the low mass end of the CDM halo mass function.

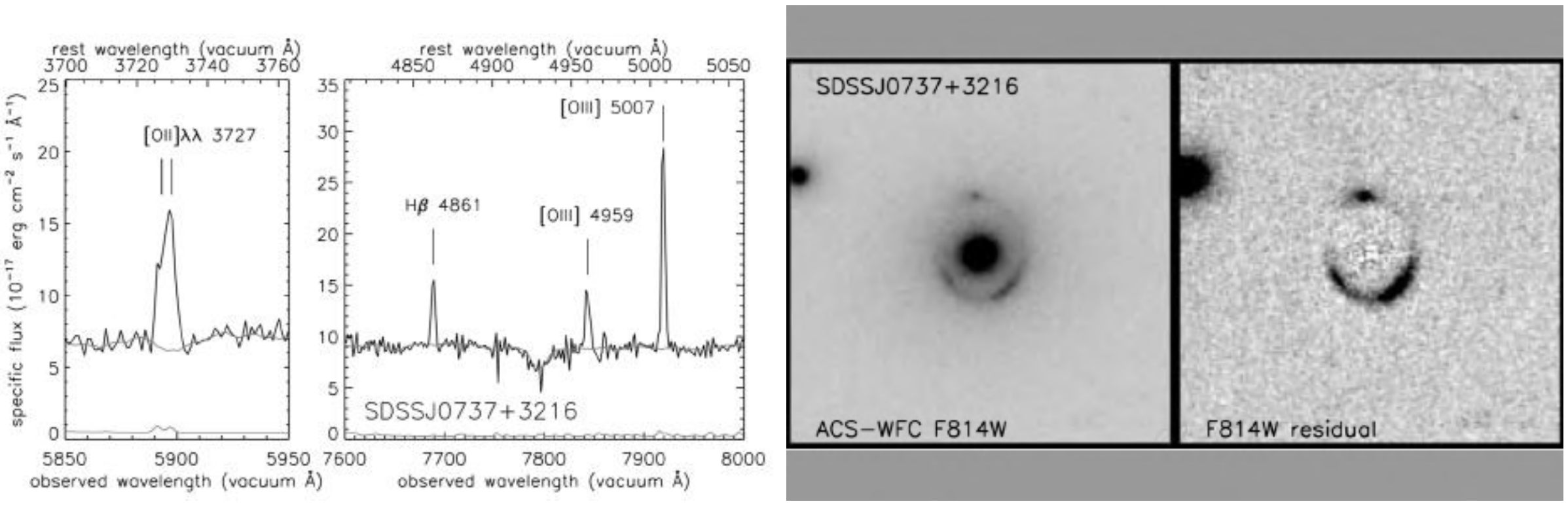

*Figure 71: Spectroscopic selection of strong galaxy-galaxy gravitational lenses in the Sloan Lens ACS (SLACS) survey. Left: [O*II*], Hβ, and [O*III*] emission lines superimposed on the spectrum of a lower-redshift quiescent galaxy. Right: HST imaging of this same system, confirming it to be a gravitational lens. Figure from Bolton et al. (2006).*

In a strong galaxy-galaxy lens system, dark matter perturbations to the smooth mass distribution of the lensing galaxy can be detected via astrometric perturbations to lensed background sources, which appear as distortions in lensed arcs (Vegetti & Koopmans, 2009; Vegetti et al., 2010a,c, 2012b; Hezaveh et al., 2016). Both the subhalos of the main lens halo and the halos along the line-of-sight lead to these distortions, with the latter dominating at higher redshifts (Despali et al., 2019). The mass sensitivity of this method depends on the depth and spatial resolution of the available imaging data, an accurate model of the intrinsic unperturbed source light distribution, and the redshifts of the lens and source galaxies. Currently, the method is not as sensitive to the small-scale power spectrum (Vegetti et al., 2018; Ritondale et al., 2019) as Ly-$\alpha$ forest measurements Viel et al. (2013) but it shows great promise for future applications given the expected sensitivity to dark matter halos of virial masses as low as $\sim 10^8 M_\odot$ (Vegetti et al., 2014).

In order to improve the constraints on the small-scale dark matter power spectrum with 'gravitational imaging' of lensed galaxies at optical and IR wavelengths, a combination of improved imaging precision, larger sample sizes, and selection of a set of 'ideal' galaxy-galaxy lenses is required. While the first requirement can only be met by deep, high-resolution imaging facilities such as HST, JWST, and AO-assisted instruments on large-aperture ground-based telescopes, the latter two requirements are directly addressed by the aperture and spectroscopic multiplex capacity of MSE. The third requirement is especially relevant in consideration of the depth that MSE will achieve relative to SDSS, since higher redshift sources probe a longer path through the Universe and thus provide a tighter constraint on CDM (Despali et al., 2019). The discovery of many new galaxy-galaxy lenses with MSE, so that the best possible subset can be studied in detail, is a key factor in pushing this field forward.

MSE will detect many lenses that are also identified in imaging surveys such as LSST. However, galaxy-scale strong gravitational lenses discovered by spectroscopic survey facilities such as MSE afford significant advantages over those identified in imaging surveys alone. First, the spectroscopic evidence of two objects along one line of sight eliminates the ambiguity of interpretation that can be associated with many imaging-selected lens candidates. Second, lenses that are selected through their spectra have known foreground and background redshifts immediately upon discovery, which is necessary to translate angular observables into mass measurements, and to quantify the integrated optical depth to low-mass CDM halos.

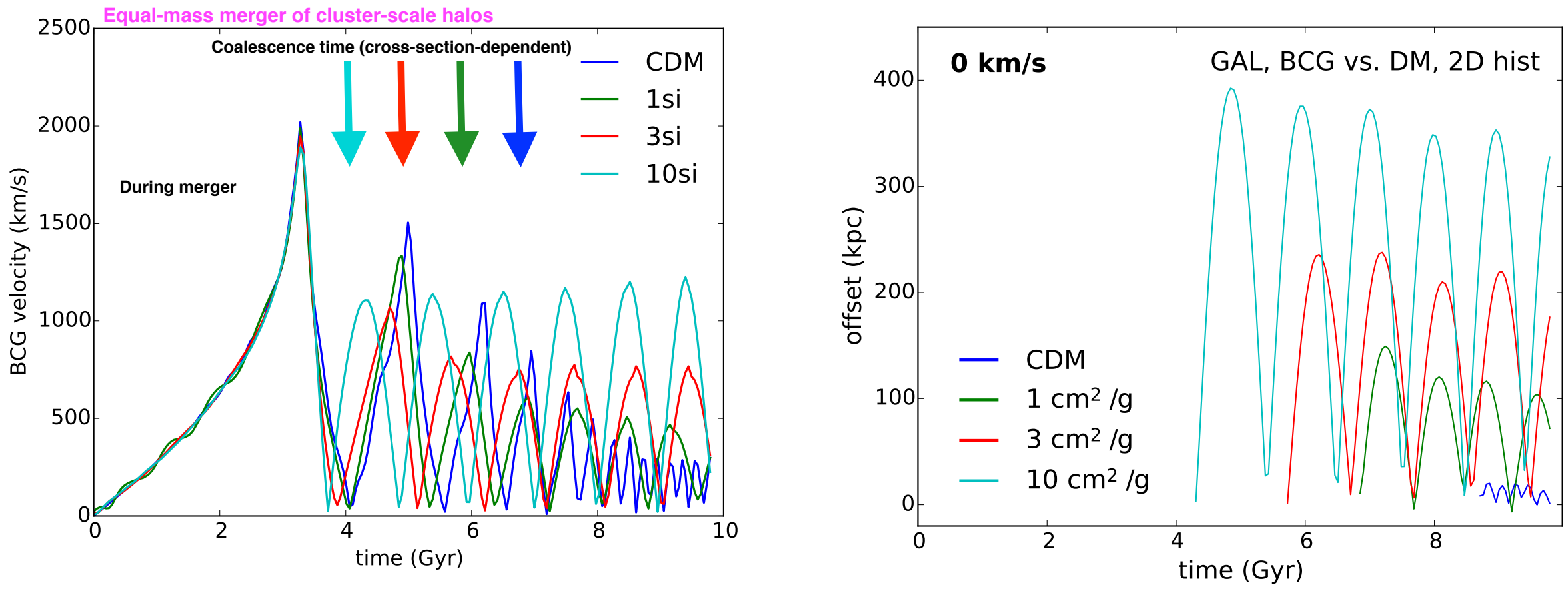

*Figure 72: Left: Relative velocities of the two BCGs from an equal-mass merger of two* $M_h = 10^{15} M_\odot$ *cluster halos. Simulations start at* $t = 0$ *when the virial radii of the two halos touch, and free-fall on a radial trajectory toward each other. Right: Separation of the BCGs with the center of the merged cluster. Clusters are simulated with cross sections* $\sigma/m = 0$ *(CDM),* $1, 3,$ *and* $10$ *cm*$^2$*/g. The time at which the clusters coalesce (the halo finder finds only one, not two, distinct halo centers) is marked with an arrow for each simulation. The equal-mass merger simulations come from Kim et al. (2017). Figure from Kim & Peter, in preparation.*

### 6.6.3 Wobbling of the brightest cluster galaxies

The hierarchical assembly of galaxy clusters leads to another probe of dark matter self-interactions, distinct from the density profiles illuminated by gravitational lensing. Excitingly, strong SIDM cross section constraints may be obtained after the merger between two cluster-scale objects is complete, possibly stronger than those arising from phenomenology during the merger process. In both cases, spectroscopy is essential for assessing the dynamical state of the system, a key ingredient in mapping cluster observables to SIDM cross section measurements. Furthermore, in both cases, robust SIDM constraints will require a careful observational analysis of many systems matched with sophisticated, observationally-focuse, simulations. The era of “easy” SIDM constraints is over.

Consider the case where we catch a cluster during an accretion event. Clusters consist of dark matter (halo of the host and subhalos containing galaxies), hot gas, and galaxies, in decreasing order of contribution to the total mass. Each of these three components behaves differently during a merger. Galaxies are effectively collisionless particles. The hot gas is a highly collisional fluid. In the CDM model, dark matter behaves like the galaxies, collisionlessly, although its dynamical evolution differs from that of galaxies on account of its different mass distribution throughout the object. If dark matter is collisional, it will behave neither as galaxies nor as gas. During the passage of a smaller halo (a cluster-, group- or galaxy-sized halo) through a cluster, the smaller halo will feel an extra force beyond gravitational if the self-scatter cross section is non-zero. Non-gravitational interactions between host and subhalo particles can lead to a loss of specific momentum of the subhalo, either because of a "drag"-like force (typical for small-angle scattering, but it can also happen for large-angle scattering for close to equal-mass mergers; Markevitch et al., 2004; Kahlhoefer et al., 2014; Kim et al., 2017; Kummer et al., 2018) or because the tail of particles ejected from the subhalo can pull gravitationally on the subhalo (Kim et al., 2017). The latter will affect the stellar component of the subhalo as well. The former, though, is a force solely on the dark matter and not the baryonic components. This may lead to a separation ("offset") between galaxies and dark-matter halos.

Spectroscopy is required to measure the kinematics of the merger as well as to obtain high-fidelity strong lensing maps. This is because offsets are expected to be smaller than typical strong-lensing uncertainty on the peaks of the dark matter density field and are velocity-dependent. Historically, the spectacular Bullet Cluster merger has been the focus of SIDM constraints (e.g., Markevitch et al., 2004; Clowe et al., 2006; Randall et al., 2008; Robertson et al., 2017b,a), although more merging clusters with a variety of configurations are also being discovered and analyzed in the context of SIDM (e.g., Bradač et al., 2008; Dawson et al., 2012; Golovich et al., 2017). Constraints based on simple analytic models of SIDM-induced offsets were typically of order $\sigma/m \lesssim 5\ \mathrm{cm}^2/\mathrm{g}$. However, recent simulations show that offsets are much smaller than these simple analytic models suggest, are transient (largest offset just after pericenter passage and approaching the next pericenter), and depend on both the dynamics of the merger and the microphysical scattering model (Kim et al., 2017; Robertson et al., 2017b,a). Offsets are typically no greater than $\sim 20$ kpc for a hard-sphere cross section of $\sigma/m = 1\ \mathrm{cm}^2/\mathrm{g}$, which is of order or smaller than typical uncertainties in the centroid of the subcluster galaxy distribution (a problem intrinsic to the small number of confirmed galaxies even for relaxed clusters; Ng et al., 2017) and lensing peak positions. More spectroscopy of more member galaxies and background lensed galaxies are critical to better constraining the merger dynamics, the centroid of the galaxy distribution, and the dark matter mass map from lensing.

A recent analysis highlighted the importance of spectroscopy for measurements of "bulleticity", the ensemble measurement of the offsets of many galaxy-mass subhalos infalling on cluster-scale halos (Massey et al., 2011; Harvey et al., 2014, 2015). An initial detection of offsets in Abell 3827 (and a measurement of a cross section around $\sigma/m = 1\ \mathrm{cm}^2\mathrm{g}$ for large-angle scattering) was recently excluded on account of improved spectroscopy (both optical and mm; Massey et al., 2015; Kahlhoefer et al., 2015; Massey et al., 2018).

Interestingly, competitive constraints on SIDM may arise from the positions and kinematics of galaxies in relaxed clusters. In staged simulations of equal-mass mergers of cluster-sized halos, Kim et al. (2017) found that the orbits of galaxies at the centers of halos—notably the central galaxy, typically but not always the Brightest Cluster Galaxy (BCG)—continue to “slosh” or “wobble” about the relaxed cluster center after the merger (Figure 72) for SIDM halos, but not CDM. The origin of this effect is the inefficiency of dynamical friction for damping orbits in shallow gravitational potentials, and as such the amplitude and frequency of the oscillations about the relaxed cluster center (in position and velocity space) are expected to depend on the size of the SIDM core region. This in turn depends on the cross section. Harvey et al. (2017) claim a detection of BCG wobbling with respect to halo centers in position space. However, the offsets have so far only been theoretically explored in detail in the context of equal-mass mergers (Kim et al., 2017), and are currently being investigated for more complex merger histories (Harvey et al., 2018).

An interesting approach is to look for relaxed remnants of equal-mass mergers (where we expect the wobbling to be greatest) by looking for systems with two bright central galaxies, and comparing the ensemble of position and velocity differences between the two central galaxies against simulations. Observations of cluster gas and the kinematics of member galaxies can provide information on how relaxed the cluster is. Exploring the separations in space and velocity of systems with two bright central galaxies has the advantage of not requiring a $\sim 1-10$ kpc-scale measurement of the halo center (George et al., 2012; Ng et al., 2017).

# Chapter 7
# Galaxy formation and evolution

**Abstract**

MSE will allow the types of revolutionary extragalactic surveys that have been conducted at $z = 0$ to be conducted as a function of redshift out to the peak of cosmic star formation. At low redshift, MSE will probe a representative volume of the local Universe to lower stellar and halo masses then is achievable with current and other upcoming surveys. These surveys will allow a diverse array of science topics from dwarf galaxies, to galaxy interactions in the low stellar mass regime, the environmental impact on galaxy evolution and the extension of large-scale structure analyses to low mass groups. A fundamental measurement for MSE will be the extension of the stellar mass function to masses below $10^8 M_{\odot}$, for a cosmologically representative, unbiased, spatially complete spectroscopic sample. High redshift extragalactic surveys with MSE will provide a high-completeness, magnitude limited sample of galaxy redshifts spanning the epoch of peak cosmic star-formation ($1.5 < z < 3.0$). They will cover the diverse range of environments probed by surveys such as SDSS and GAMA (groups, pairs, mergers, filaments, voids), but at an epoch when the Universe was under half its current age. The scale of insights available from such surveys are difficult to predict given the relatively scarce amount of information we currently have in these regimes (thousands as opposed to millions of targets). As such, MSE will likely have a generation-defining impact on our understanding of how galaxies evolve over 12 billion years of cosmic time.

**Science Reference Observations** (appendices to the *Detailed Science Case, V1*):
**DSC − SRO − 06** Nearby Galaxies and their Environments
**DSC − SRO − 07** Baryonic structures and the dark matter distribution in Virgo and Coma
**DSC − SRO − 08** Evolution of galaxies, halos, and structure over 12 Gyrs

**DSC − SRO − 09** The chemical evolution of galaxies and AGN over the past 10 billion years ($z < 2$)
**DSC − SRO − 10** Connecting high redshift galaxies to their local environment: 3D tomography mapping of the structure and composition of the IGM, and galaxies embedded within it

## 7.1 Extragalactic surveys with MSE

MSE has the unique ability to survey both wide and deep volumes of the Universe. Extragalactic surveys with MSE will be truly boundary pushing in the science outcomes given its simultaneous spectral coverage from the optical blue ($\sim 400$ nm) to the far H-band ($\sim 1,800$ nm). Most compellingly, this opens up the prospect for the type of far reaching galaxy-evolution-focussed science that has been possible at $z \sim 0$ (e.g. SDSS, GAMA), but at the peak of cosmic star formation ($z \sim 1.5$).

Figure 73 puts the baseline MSE galaxy evolution surveys discussed in this chapter in the context of current and near future confirmed surveys. In terms of pure parameter space, it is clear that MSE can push into a new domain in all dimensions displayed in these figures. Another major advantage is that MSE has access to an expanded spectral window that is partly invisible to all current and future competitor facilities.

Figure 74 demonstrates the main practical observing constraints that must be considered when constructing an MSE survey. The number counts for the $i$-band present the maximum possible sources available before any sort of photo-z selection has been applied. Given the typical galaxies and the fiber density of MSE, it is then straightforward to calculate the limiting factor to achieve a certain area for a given source depth or vice-versa. Any pragmatic survey would need to be designed such that these two characteristics are well balanced — e.g. there is little point in conducting an overly shallow wide area survey that leaves many spare fibers. In practice it is clear there are many more targets available on sky (assuming one hour is spent observing each field), and the limiting factor is the repeat visits to achieve a target completeness. This also allows for a large scope to create photo-z (or colour etc) sub-selections, as there is an excess of sources available for a given depth. This enables a variety of galaxy evolution surveys, making use of high quality next generation photo-z.

A major, fortunate, advantage MSE has over existing and planned spectroscopic surveys is the use of next generation photometric surveys from the ground and space for target selection. These catalogues will only become available post 2020, and offer a paradigm shift in optical imaging quality. The resulting homogeneity and precise calibration will improve the robustness of almost every type of analysis made.

In this chapter we explore potential local wide-area and high redshift surveys which will both maximise the scientific potential of MSE, while staying within a baseline of a $2-3$ year dark-time survey. Following this, we discuss breakthrough science topics that can be addressed with these surveys.

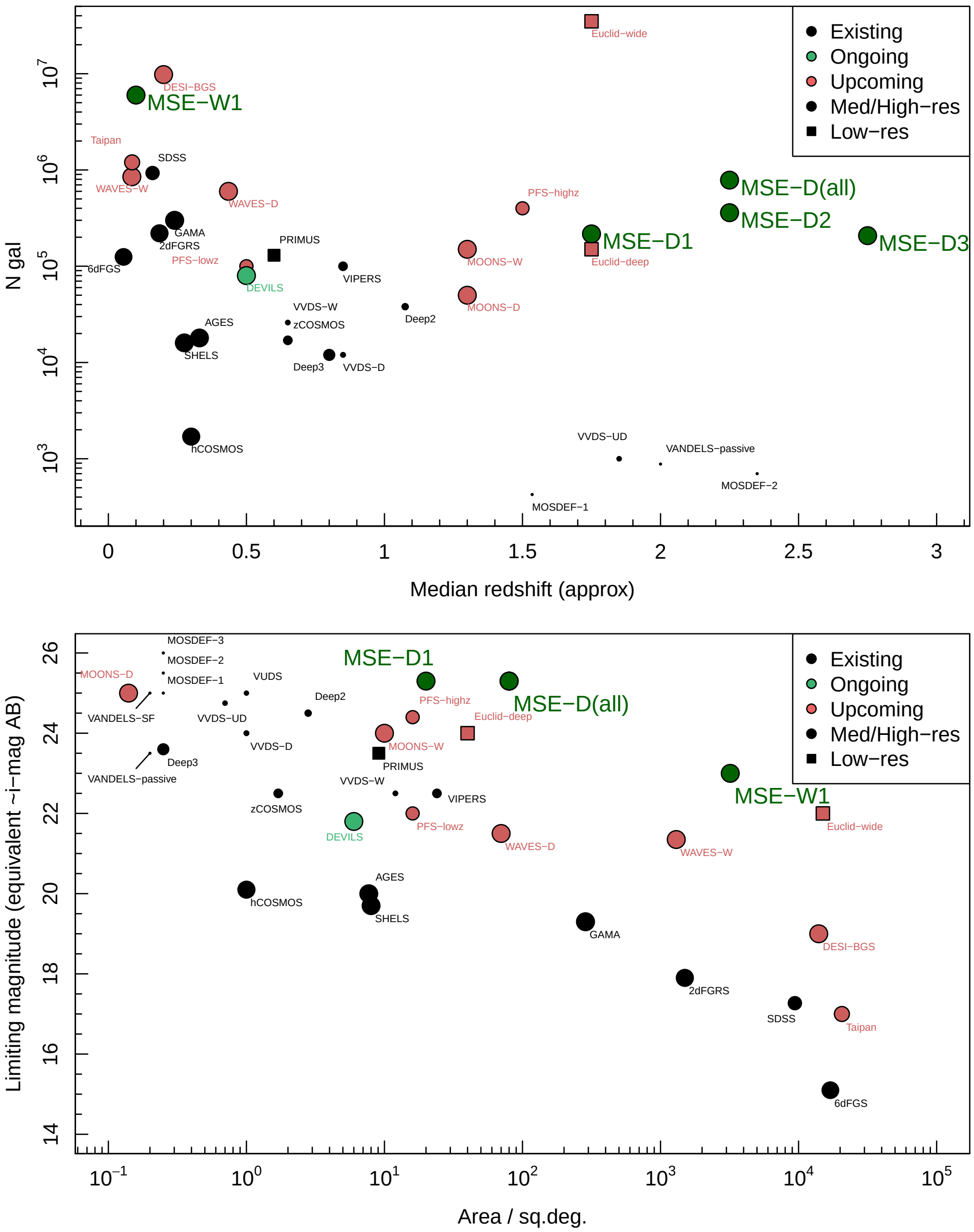

*Figure 73: Comparison between proposed MSE wide and deep surveys (dark green points) with existing, ongoing and upcoming spectroscopic surveys. Point size approximately scales with survey completeness down to a fixed magnitude limit.*

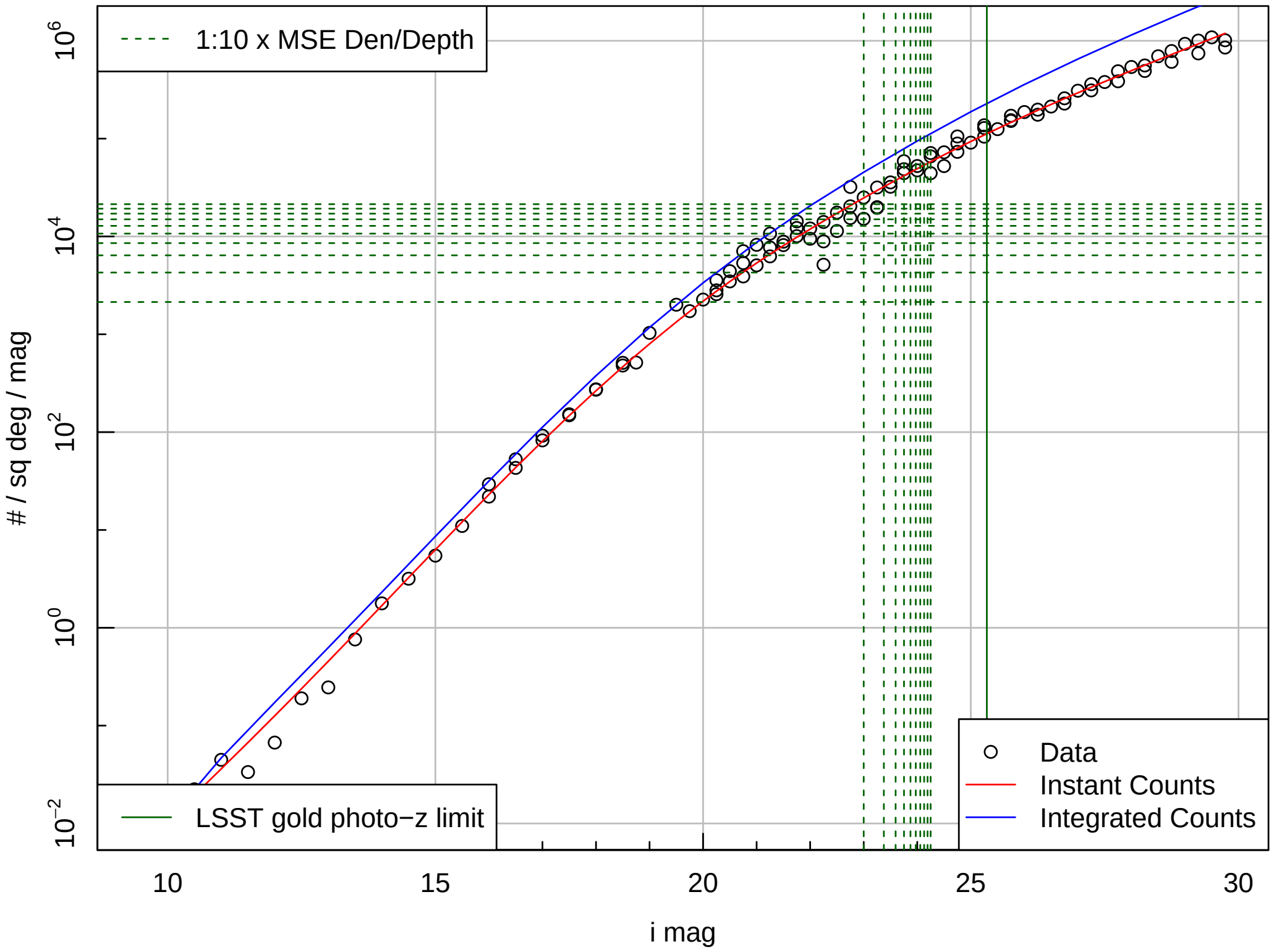

*Figure 74: Deep i-band counts matched against MSE low resolution fibre density. Note we expect to get to* $\mathbf{i} \sim 24$ *in one hour with MSE (low resolution SNR* $= 2$ *for a point source), so* $\mathbf{i} \sim 23$ *for galaxies. Given the multiplexing, a single pass survey would be* $\mathbf{i} < 19.5$ *(2137.5 fibres per sq. deg. in low resolution). To achieve* $\mathbf{i} < 23$ *requires* $\sim 20$ *passes and* $43\mathbf{k}$ *fibre hours per sq. deg. assuming perfectly efficient tiling. A 100 sq. deg. survey will be in excess of 4 million fiber hours, which is reasonably close to the amount of dark time we can expect in 2 years of survey operation.*

### 7.1.1 Local Universe wide-field surveys

The focus of potential local Universe MSE surveys is to probe a representative volume of the local Universe to lower stellar and halo masses then is achievable with current and other upcoming surveys. These surveys will allow a diverse array of science topics from dwarf galaxies, to galaxy interactions in the low stellar mass regime, the environmental impact on galaxy evolution and the extension of large-scale structure analyses to low mass groups. For these science goals, the key survey requirements are low-to-moderate spectral resolution to measure group-scale velocity dispersions and highly-complete samples to faint magnitudes.

We baseline the MSE low redshift strategy to consist of two surveys (referred to as S1-W and S1-D), both targeting galaxies selected to lie at $z < 0.2$. S1-W aims to cover a cosmologically representative, contiguous volume, while still reaching substantially deeper than previous surveys. S1-D targets specific environments over a smaller area, to exceptionally deep limits.

For both surveys we require the full spectral coverage up to 800nm, and there is significant advantage in extending to 1500nm in order to obtain Pa$\beta$ and IMF-sensitive absorption features at $z < 0.2$. Moderate spectral resolution improves the radial velocity measurements of all galaxies, important for association within small groups and pairs; moreover it enables precision stellar population measurements, including of IMF-sensitive features, for high signal-to-noise spectra. It is expected that the faintest-galaxies in our sample will only be read-noise limited at the bluest wavelengths, < 4300Å; it is only there that there will be a small but significant cost to rebinning the low SNR spectra.

Targets will be selected based on $i-$band magnitude and photometric redshift. The choice of $i-$band is made to obtain a sample that is close to stellar mass-limited. In order to obtain good redshifts and stellar population constraints for this low redshift sample it will be important to obtain sufficiently high SNR at blue wavelengths, ~ 400nm. Given the lower instrument throughput at these wavelengths, and the intrinsically red colours of galaxies, the attainable depth of a complete $i-$band selected sample is shallower than would be the case for a higher redshift sample where the features of interest lie at > 600nm.

**S1-W:** The volume of this wide-area survey is set by the desire to minimally contain a 300×300×300 Mpc/h co-moving cube, so that the volume is large enough to ensure Universal homogeneity in all three dimensions (Driver & Robotham, 2010). With an upper redshift limit of $z = 0.2$ this drives a requirement of > 3200 square degrees, in a contiguous region with comparable angular extents in RA and Dec. The total volume is about 0.18 Gpc$^3$, and the cosmologically representative box lies at $z > 0.1$. In a blind 3000 sq degree survey, we expect to cover ~ 10 massive galaxy clusters ($M_{\rm halo} > 10^{15} M_\odot$), and > 200 low-mass clusters ($M_{\rm halo} > 5 \times 10^{14} M_\odot$). For these systems we will be able to study the properties of galaxies out to arbitrarily large distances from the cluster centre. This is critical for many of the remaining questions about galaxy transformations in dense environments (Bahé et al., 2013). Since galaxies are highly clustered on the sky, and fibers cannot be placed arbitrarily close together, a minimum of two fiber configurations are required, whatever the source density of targets. The minimum integration time should be 1 hour; anything less than this can be currently undertaken with a 4-m class telescopes. With the 1.5 deg$^2$ MSE field of view, and assuming 20% overheads, it will therefore take at least 5100 hours to complete the survey.

Given the equatorial location of MSE, a good option for imaging would be UNIONS), covering the northern Euclid footprint. This covers 5000 square degrees in $u(g)riz$, where the $g$ photometry is still being negotiated. In addition, the LSST survey single-visit depth of $i = 24$ is also sufficient for target identification, and a survey area of 3200 square degrees is easily accessible by MSE (more than half of the LSST footprint is visible from Hawaii at good airmass). The WFIRST High Latitude Survey will cover about 2000 square degrees, but may be located in the south, and it is currently unclear how much of that area will be accessible to MSE. It is expected that WFIRST will ask LSST to prioritize observations in such a way as to reach the full 10-year depth within this region as soon as possible. It would be advantageous to target as much of this area as possible, spectroscopically.

| Survey | Area (sq. deg) | Depth (Selection band) | Depth (equivalent $i$) | Sample size |
|---|---|---|---|---|
| S1-W | 3200 | $i < 23$ | $i < 23$ | 6M |
| S1-D | 100 | $i < 24.5$ | $i < 24.5$ | 800k |
| SDSS-Legacy | 8032 | $r < 17.8$ | $i < 16.8 - 17.8$ | 928k |
| 6dF | 17046 | $K < 12.75$ | $i < 15.6$ | 150k |
| GAMA | 300 | $r < 19.8$ | $i < 18.8 - 19.8$ | 238k |
| DESI | 14000 | $r < 19.5$ | $i < 18.5 - 19.5$ | 9.8M |
| WAVES-Wide | 1500 | $i < 21$ | $i < 21$ | 1.0M |

*Table 7: The baseline MSE extragalactic surveys discussed in this chapter, S1-W and S1-D, are compared with other relevant spectroscopic surveys in terms of their area, depth and sample size. To compare $r-$band selected surveys with the proposed $i-$band selection, we assume a colour range of $(r - i) = 0 - 1$, typically of galaxies at $0 < z < 0.2$. This is appropriate to determine the faintest $i$-band magnitude for which a r-selected sample would be complete.*

**S1-D:** The deep component will be used to probe down to very low stellar masses in the most over-dense clusters in the local Universe to explore the enviromental impact of dwarf/low-mass galaxies. This component requires 100 square degrees of deep ($i > 25$) multicolour imaging; not necessarily contiguous but in fact including rare, massive clusters. Here the LSST survey would be ideal, as it will reach the required depth in about one year. Alternatively, some of the HSC survey fields might be appropriate, at least for the lower-density environments. The galaxy cluster fields should be selected based on their X-ray emission and bright-galaxy spectroscopy. Suitable X-ray catalogues already exist, and will be further improved by eROSITA, though access to the latter remains uncertain. In order to cover the virialized region in a single MSE pointing, clusters should be at $z > 0.1$; lower-redshift clusters, for which the lowest-mass galaxies can be studies, should be the focus of a separate survey.

### 7.1.2 High redshift surveys

The main focus of our baseline MSE high redshift extragalactic survey is to provide a high-completeness, magnitude limited sample of galaxy redshifts spanning the epoch of peak cosmic star-formation (cosmic noon, $1.5 < z < 3.0$). Such a survey would cover the diverse range of environments probed by surveys such as SDSS and GAMA (groups, pairs, mergers, filaments, voids), but at an epoch when the Universe was under half its current age. This survey would also provide an extensive sample of intermediate/high redshift galaxies with which to finely sub-divide on properties and undertake a diverse array of galaxy evolution projects.

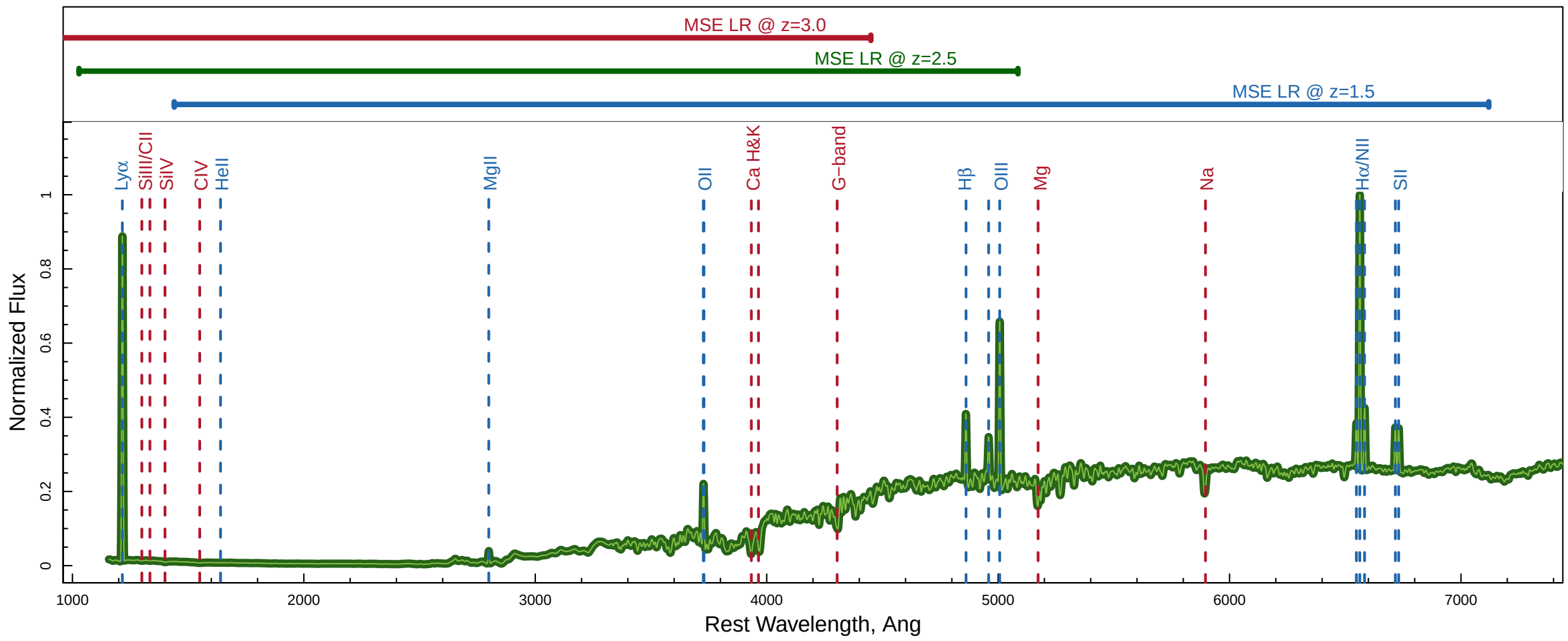

*Figure 75: A typical galaxy spectrum showing key emission and absorption lines features. The top panel displays the MSE low/mid-resolution wavelength range at* $1.5 < z < 2.5$*. At* $z \sim 1.5$*, MSE will observe all key optical features; at* $z \sim 2.5$*, it simultaneously links Lyman-α and UV-absorption lines with optical emission lines; MSE retains the ability to observe OII and Ca H&K features to* $z \sim 3$*.*

The primary focus of this surveys will be securing redshifts with the minimum required SNR. Due to the need for multiple passes when target density is high, we will also obtain higher SNR spectra for a significant fraction of targets ($\sim 30\%$). The NIR capabilities of MSE make it ideal for obtaining robust redshifts for sources at this epoch, where other high-multiplex large FOV facilities largely require the use of single line redshift and photometric priors. Figure 75 displays the spectral coverage of MSE at $1.5 < z < 2.5$ in comparison to key spectral features.

To cover the range of environments explored by local surveys requires that we probe samples (i) to similar stellar masses, (ii) to similar completeness, and (iii) over similar cosmological volumes. Here we propose three sub-surveys spanning $1.5 < z < 3.0$ (Table 8). The lower of these is designed to explore galaxies and their environments at $1.5 < z < 2.0$, while the higher redshift volumes will combine similar studies at earlier times with a tomographical study of the IGM and the galaxies embedded within it. Hence, each sub-survey volume has slightly varying characteristics.

Firstly, our sub-surveys must cover enough volume to reach Universal homogeneity in all three dimensions (as in S1-W, a $\sim 300 \times 300 \times 300$ Mpc/h co-moving cube, or $> 0.03\text{Gpc}^3$, Driver & Robotham 2010). At $z > 1.5$, and in $\Delta z = 0.5$ bins, this equates to a survey area of at least $\sim 20\,\text{deg}^2$. The two higher redshift survey windows have a suggested survey area of ~80 deg$^2$. This is driven by the area and depth required to produce a competitive IGM tomographic mapping survey in parallel with the main galaxy evolution surveys (refer to Section 7.2.1).

*Table 8: Example MSE surveys covering the peak of star formation. Collectively these three surveys would require most of the dark time for a few years on MSE (ignoring RA constraints), but this number would increase if higher redshift targets (e.g. for tomographic mapping) are desired, or higher $S/N$ is desired for sub populations.*

| Area (sq.deg) | $z_{lo}$ | $z_{hi}$ | Vol / Gpc$^3$ | Selection | N ($10^3$) | Density ($10^3$/sq.deg) | Texp (m.hrs) |
|---|---|---|---|---|---|---|---|
| 20 | 1.5 | 2.0 | 0.04 | $i < 25.3$ | 220 | 11.0 | 1.8 |
| 80 | 2.0 | 2.5 | 0.16 | $i < 25.3$ | 360 | 4.5 | 3.2 |
| 80 | 2.5 | 3.0 | 0.16 | $i < 25.3$ | 200 | 2.6 | 2.0 |

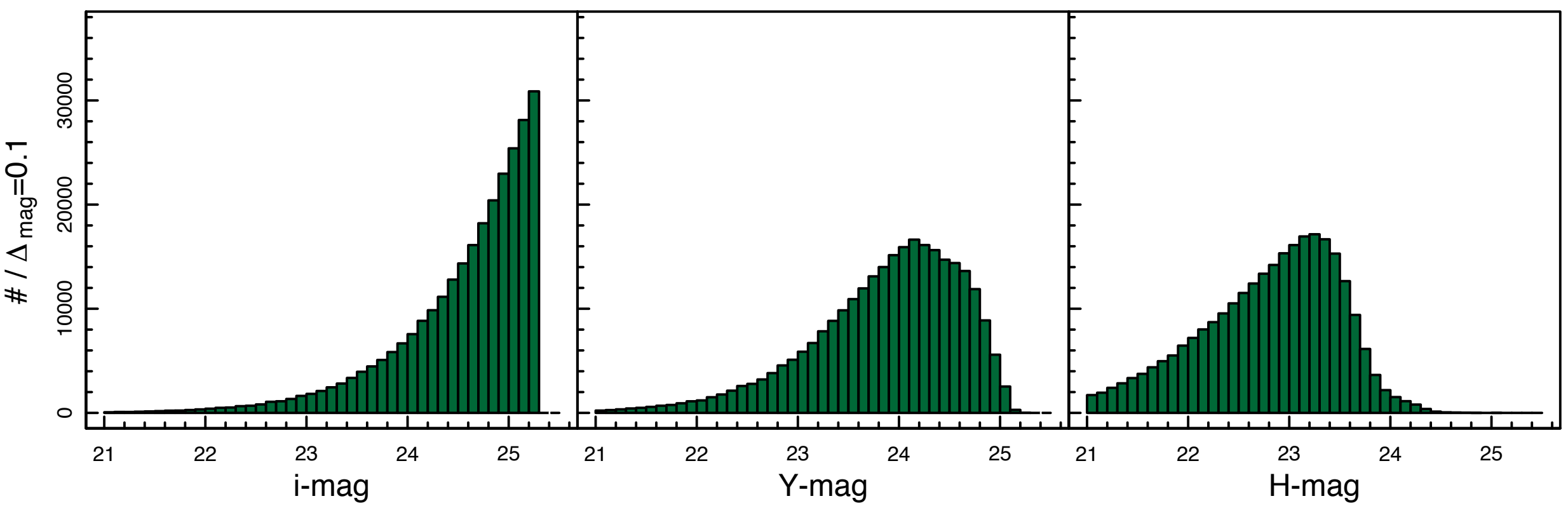

*Figure 76: The resultant magnitude distribution of galaxies in $i$, $Y$ and $H$ for the survey described in Table 8. Simulated using SURFS+Shark mocks as described in the text.*

The stellar mass limits for which we can hope to reach with MSE will be largely constrained by the amount of available dark time, and the presence of deep imaging with robust photo-z measurements (required for target selection). For this survey, we assume that we will have available robust photo-z measurementss for our target selection ($i < 25.3$).

To predict the likely survey depth we can hope to achieve with these constraints, we use mocks generated using the SURFS N-body simulations (Elahi et al., 2018) coupled with the Shark mocks Lagos et al. (2018) and Viperfish SED generation (Lagos et al in prep). We then predict the number density of sources as a function of i-band magnitude over a ~20 deg$^2$ region and at $1.5 < z < 3.0$ (Table 8). We use the predicted exposure times required to obtain a redshift as a function of i-mag, taken from the analytic form derived from VVDS: $T = 0.5 \times 4.5^{(i-23)}$. Using both the number density of objects and predicted exposure times, we can estimate the depth to which we can reach in ~ 3 millions fiber hours with MSE (Figure 74). These are also given in Table 8. We find that for an $i < 25.3$ selection, we can cover a ~20 deg$^2$ region with LSST photo-z pre-selection at $1.5 < z < 3.0$, to > 90% completeness in ~ 3 million fiber hours with MSE. The resultant likely magnitude distribution of sources in given in Figure 76.

To explore that galaxy population that would be targeted using these selections, the top panel of Figure 77 shows the redshift-stellar mass distribution from the Shark mocks, in comparison to a number of current and upcoming galaxy-evolution-focused surveys (GAMA, WAVES-deep, DEVILS, zCOSMOS). The sample will detect $10^{9.5-10.5}$M$_\odot$ galaxies $1.5 < z < 3.0$; at $1.5 < z < 2.0$, it will contain > 1000 galaxies per bin [$\Delta z = 0.05$, $\Delta$Log[M$_*$/M$_\odot$]=0.1].

We can also predict the distribution of haloes which we will probe with this selection. To do this, we assume that haloes which have $N > 2$ members selected in our sample will have their halo mass parameterised (through velocity dispersions, see Robotham et al., 2011). We then identify all haloes within the Shark mock for which $N > 2$ galaxies are at $i <$ 25.3. The resultant group $M - z$ relation is shown in the bottom panel of Figure 77. MSE will identify $\sim$ 800 M$_h$ = $10^{12-14}$M$_\odot$ groups at $1.5 < z < 3$. This will probe a range of environments that have previously been unexplored outside of the local Universe. MSE will also spectroscopically identify over 8000 $N = 2$ (close pair) systems, allowing the first detailed parameterisation of the major merger rate at cosmic noon.

### 7.1.3 Ancillary surveys

Given the high target density of the low and high redshift surveys described previously, there will be tthe need for multiple passes per field with MSE. This allows many compelling science cases to be naturally pursued in parallel with the main extra-galactic surveys. The only strong requirement is that these are restricted to the same areas of sky, but this is rarely an issue for blind targeted surveys.

**High SNR Science:** Within the above mentioned photo-z surveys, there will naturally be a range of exposure times required for sources of different magnitude. Some fainter sources will only be suitable for measuring redshifts and will allow for a large range of structural and evolutionary focussed science cases. Brighter sources have the potential for longer exposures, well beyond that required for pure redshifts. This opens the door to a host of ancillary science that requires higher SNR spectra, such as gas phase metallicity science (continuum SNR $\sim$ 10) and even stellar phase metallicity science (continuum SNR $\sim$ 50).

$z > 4$ **science:** The majority of the currently known galaxies (galaxy candidates) at $z > 4$ are Lyman Break Galaxies (LBGs) selected using the drop-out technique in several deep fields. A tiny fraction of these LBGs (usually the most luminous ones) have been spectroscopically confirmed. In addition, deep ground-based narrow-band surveys have obtained a few small samples of Lyman-$\alpha$ Emitters (LAEs) at particular redshift slices $z \sim 4.5$, 4.8, 5.7, 6.5, 7.0, and 7.3. Many of these have subsequently been spectroscopically confirmed. However, current studies of galaxy evolution at $z > 4$ have been largely based on photometrically-selected LBG samples without secure redshifts. Therefore, a large (and statistically complete) sample of galaxies at $z > 4$ will be very important for studies of early galaxy evolution.

The area and depth of $z > 4$ targets will naturally depend on the characteristics of the primary survey, given $z > 4$ science will be making use of naturally occurring spare fibers. It is not necessary to cover all redshifts at $z > 4$. Rather, such a survey could target several redshift slices that have the highest target selection efficiency. A main survey area of $20 - 80$ sq. deg. is acceptable for these goals, if a sufficient number of fibers are assigned to these targets. A depth of $\sim 25-26$ mag would be required to perform a competitive survey of $z > 4$ galaxies, with the selection band varying from $r$ (z$\sim$ 4) to $z$ ($z \sim 6$). These targets will be pre-selected using traditional LBG selection techniques.

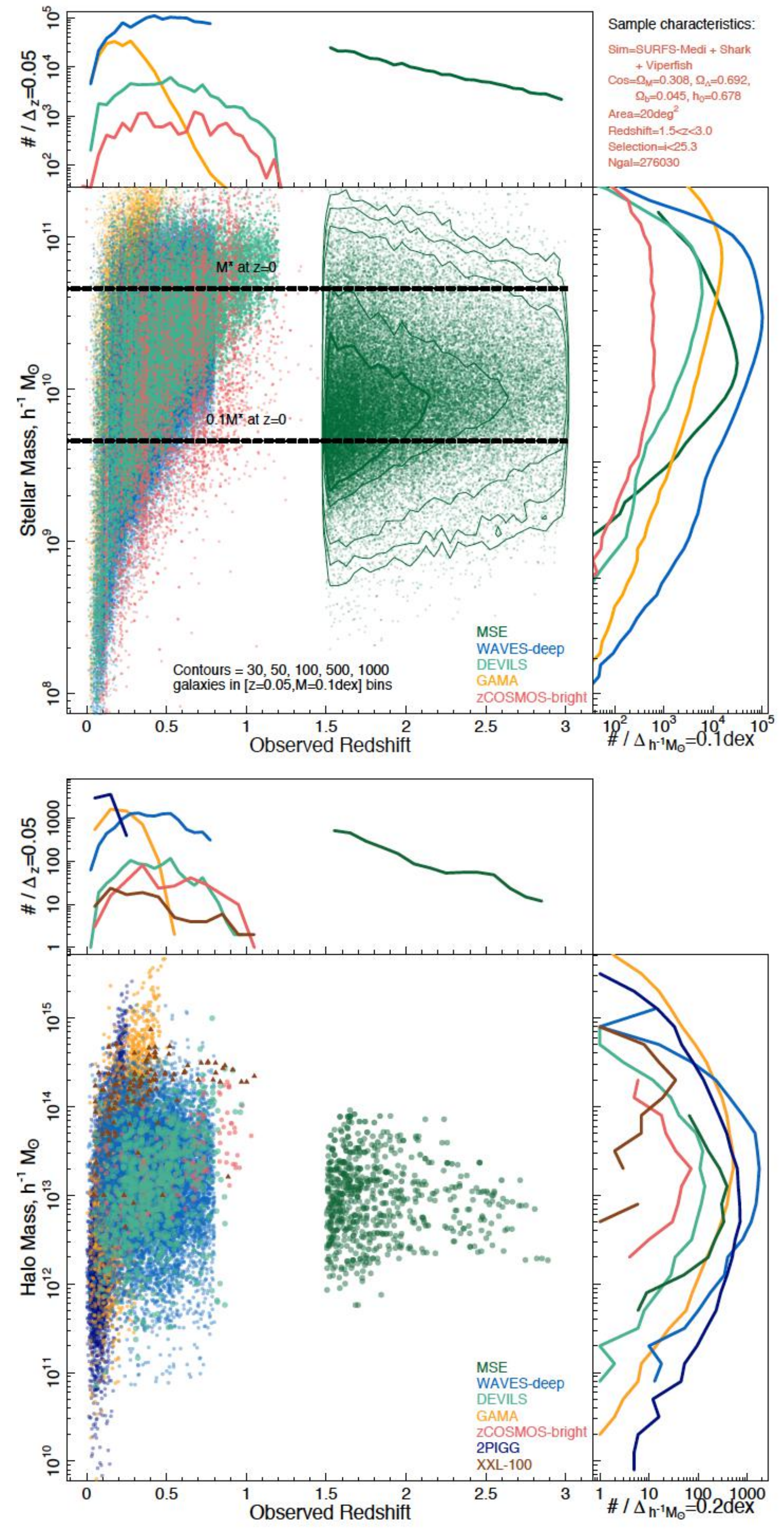

*Figure 77: The predicted galaxy (top) and group (bottom) M-z distribution from MSE for a 20 sq. deg. survey described in Table 8, in comparison to other galaxy-evolution-focused surveys. Simulated using SURFS+Shark mocks. Data points show only a representative sub-sample of the selection.*

For galaxies at $z < 6.5$, LSST imaging data are sufficient. However, if we wish to probe to earlier epochs, we would require deep near-IR imaging data. The Euclid deep-field data in the YJH bands reach 26 AB mag, and are deep enough for the selection of LBG candidates.

The potential observing strategy depends of the main survey area and depth, and the number of fibers assigned to $z > 4$ targets. We will likely prioritise targets with photo-z greater than 4 and brighter than 25.3 mag. The density of bright targets is low, and as such will be easily targeted with unused main survey fibres. For fainter targets that have a much higher number density, we will not be able to identify them if they do not have Ly$\alpha$ emission lines. As we mention above, we only need to focus on several redshift slices that have the highest target selection efficiency, for example, redshifts at which Ly$\alpha$ and/or the Lyman break moves from one filter to the next filter: $z \sim 4.7$ (from $r$ to $i$), 5.7 (from$i$ to $z$), 6.6 (from $z$ to $y$), and so on. Higher-redshift candidates like $y-$dropouts and $J-$dropouts are much rarer, and as such, will also be more easily assigned fibres.

Such a survey will aim to build the largest sample of spectroscopically confirmed galaxies at $z > 4$. This unique sample will be used to probe a variety of properties of high-redshift galaxies and their implications to cosmology, such as luminosity functions, mass functions, physical properties, morphology, stellar population, galaxy clusters/proto-clusters, cosmic reionization, etc.

## 7.2 Large scale structure and galaxy halos

### 7.2.1 IGM tomographic mapping

Inter-galactic medium (IGM) tomographic mapping (TM) uses high source densities of HI Lyman-$\alpha$ forest background sources to tomographically reconstruct the 3D absorption on scales comparable to the mean sightline separation. Most recently, the CLAMATO Survey (Lee et al., 2014) has covered 0.17 deg$^2$ with background sources at $2.3 < z < 3.0$ to probe the $2.0 < z < 2.6$ Ly$\alpha$ forest with sightline densities of $dN/dz \sim 900/\text{deg}^2$, or average transverse separation of 3.4$c$ Mpc. CLAMATO was carried out on Keck-I/LRIS with $2-3$ hour integrations on $r < 24.7$ background sources. Subaru/PFS is currently planning on incorporating an IGM TM program into their Galaxy Evolution Survey which will run from $2021-2026$. Subaru/PFS will cover 15 deg$^2$ but with a slightly coarser sightline sampling than CLAMATO (4$c$ Mpc sightline separation). For MSE, it therefore makes sense to cover a significantly larger footprint than PFS (such as the 80 square degree $z > 2$ photo-z surveys mentioned above) but with comparable sightline sampling as CLAMATO (or better).

The lowest-SNR background spectra in CLAMATO are SNR$\sim$ 1.5 per angstrom within the Ly$\alpha$ forest at around 4000Åin the observed frame, on $r = 24.7$ LBGs at the faint end. Using the MSE exposure time calculator[1] (ETC), this requires 2.5 hour integration times assuming 0.5 arcsec seeing on a $r = 24.7$ point source at 1.2 airmass and sky brightness of

[1] https://mse.cfht.hawaii.edu/?page_id$=$17

20.7 mag/arcsec$^2$. The projected background source density is 1400 deg$^{-2}$ in CLAMATO covering the $2.0 < z < 2.5$ Ly$\alpha$ forest, but with MSE we aim for an extended redshift range of $2.0 < z < 3.0$ (with background sources at $2.3 < z < 3.5$), that requires a factor of ∼ 50% more background sources to maintain the same transverse sightline separation as CLAMATO. Factoring in redshift failures, this would probably require a *target* density of 3000 deg$^{-2}$ assuming a 65% redshift success rate using 6 − 7 photometric bands. The observing requirement is thus 7500 fiber hours per sq deg, and 600$K$ fiber hours to cover 80 sq. deg.

Much of the scientific yield from the IGM tomographic mapping will arise from synergy with a sample of foreground galaxies coeval with the $2 < z < 3$ tomography map volume defined above. A strawman sample would be 320k galaxy redshifts over the 80 sq deg tomography volume, or 4k per sq deg. This sample size is driven by two particular science cases:

1. Constraining intrinsic alignments between galaxy spin or morphology with the cosmic web defined by the IGM tomographic map at $z \sim 2.5$ (Krolewski et al., 2017). A sample of $> 10^5$ galaxies should be able to detect or rule out any such alignments at high statistical significance.

2. Providing a large sample of galaxy protoclusters (with descendant masses of $M_{z=0} > 10^{14.5} M_\odot$), which require comparatively high galaxy number densities ($n_g \sim 7.5 \times 10^{-4} h^3$Mpc$^3$ in this strawman sample) to identify protoclusters with high purity and completeness. These protoclusters, detected in both HI absorption and galaxy redshifts, will allow us to study the evolution of the intra-cluster medium and its effect on member galaxies.

Depending on how the selection and integration times are chosen, this foreground $2.0 < z < 3.0$ sample would probably require 1 million fiber hours (this assumes 2 hour integrations and a redshift success rate of 65%).

### 7.2.2 Halo occupation modelling in the Local Universe

The MSE low-z surveys can help establish how the evolution of the lowest-mass satellite galaxies is influenced by their environments. Analysis of satellites in the Local Group has demonstrated that quenching of star formation has a complex dependence on stellar mass that is not understood. Fillingham et al. (2015) shows that while the effectiveness with which star formation is quenched in satellite galaxies steadily declines with decreasing stellar mass in the mass regime probed by all large spectroscopic surveys, at lower masses the trend appears to reverse (see Figure 78). While the effect is subtle, this differential measurement (comparing satellite galaxies with central galaxies) is a potentially very powerful way to constrain feedback parameters (e.g. McGee et al., 2014). It is important to make this measurement over a cosmologically relevant volume, and with a homogeneous selection of galaxies over the full mass range.

In Figure 79, we show the theoretical halo mass function, with coloured lines indicating how many of those haloes in the S1-W survey are hosted by multiple galaxies with $i < 23$. With a high sampling completeness, the proposed survey will identify the haloes of all galaxies at the peak of the star formation efficiency function within the $z < 0.2$ survey volume. As been shown in (e.g.,) Han et al. (2015) and Viola et al. (2015), weak lensing calibration of the masses of spectroscopically identified groups can be extremely valuable, both in terms of acquiring a more robust and statistically better calibrated estimate of the group mass, but also in terms of constraints on halo assembly bias (e.g. Brouwer et al., 2016; Tojeiro et al., 2017; Dvornik et al., 2017). Hence, there is a large science value for MSE to overlap with complementary weak lensing datasets.

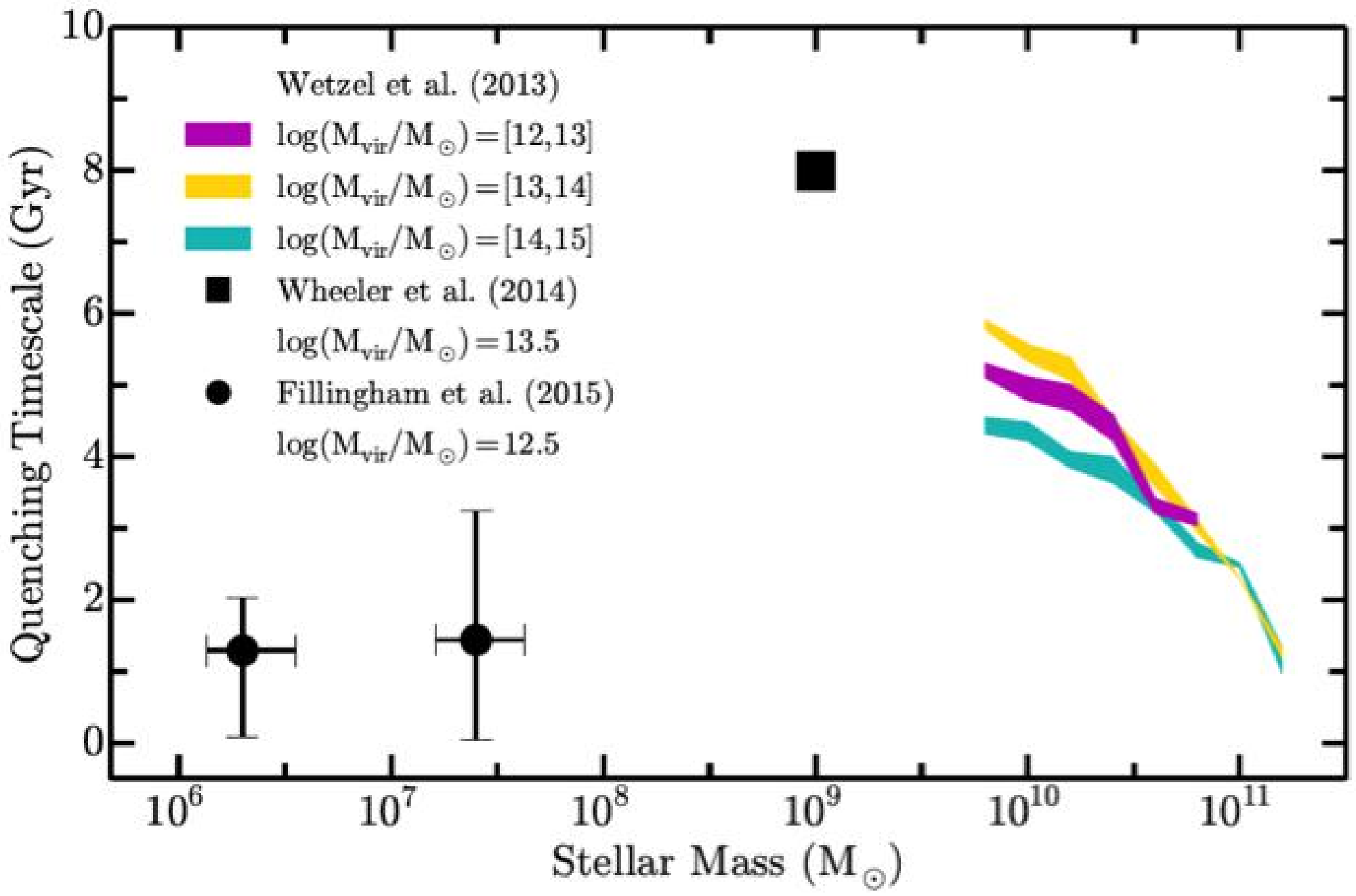

*Figure 78: The inferred star formation quenching timescale of satellite galaxies as a function of their stellar mass. Short quenching timescales lead to high fractions of galaxies with little or no star formation at the epoch of observation. Large spectroscopic surveys like SDSS and GAMA have been instrumental in uncovering the trend in decreasing timescale with increasing stellar mass shown as the coloured lines at the high-mass end. However, this and other analyses of the Local Group show that the lowest-mass satellites may respond very differently to environment.. Figure from Fillingham et al. (2015).*

The galaxy pairwise velocity dispersion (PVD, the line of sight dispersion in peculiar velocities between galaxy pairs) can provide a sensitive test of halo occupation distribution (HOD) model predictions, (e.g. Loveday et al. 2018), as can the mean infall of satellite galaxies into overdense regions. Moreover, comparing dynamically-inferred halo masses with lensing estimates provides a sensitive test of modified gravity (Zu et al., 2014). 4MOST/WAVES will do an excellent job of measuring the PVD and infall kinematics at low redshifts; MSE will allow us to observe the evolution of these statistics to significantly higher redshift and lower luminosities. The main requirement for this science is complete spectroscopic sampling in high-density regions, requiring multi-pass spectroscopy. Ideally, one wants a sample that is complete to at least 3 magnitudes fainter than $M^*$. The proposed wide ($i < 23$ mag) survey (S1-W), about 2 mag deeper than WAVES-Wide, would allow one to probe the dynamics of dwarf galaxies ($M_i \approx -15$ mag) to redshift $z \sim 0.1$ and $\sim M_\star$ galaxies to $z \sim 0.8$. The deep ($i < 24.5$ mag) survey would allow one to push 1.5 mag fainter or correspondingly higher in redshift. The volume of both surveys would ensure that a representative range of environments are probed, from voids to rich clusters.

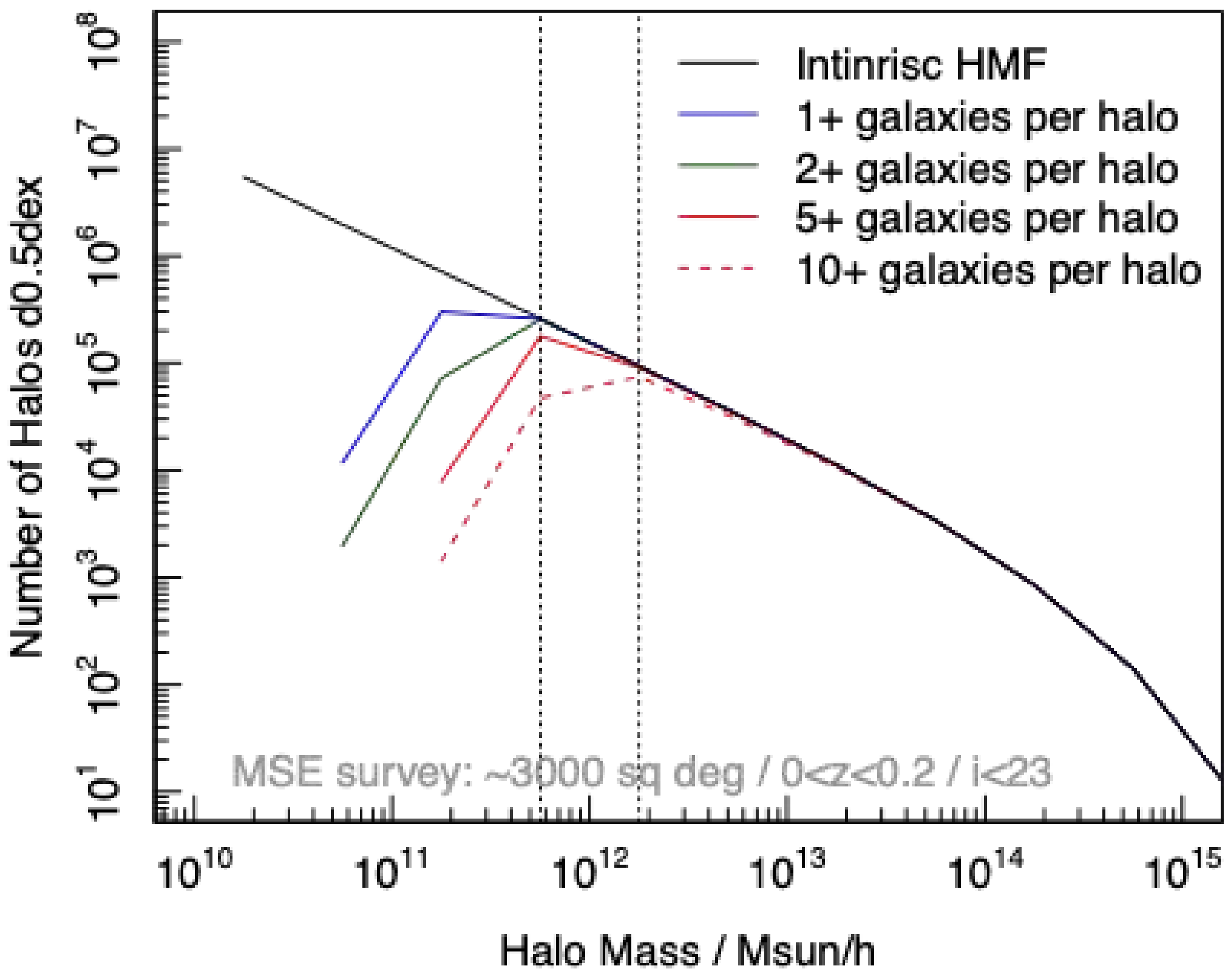

*Figure 79: The expected number of dark matter haloes in the S1-W survey that are populated by 1 or more galaxies with $i < 23$ are shown by the coloured lines. If the survey is 100% efficient in obtaining redshifts for all galaxies to this limit, this is the expected performance. Galaxies at the peak of the galaxy formation efficiency curve, with halo masses $M_h = 10^{12} M_{\odot}$, are all identified with at least one galaxy throughout the survey volume, with most populated by 5 or more galaxies.*

#### 7.2.3 Galaxy groups/clusters at $z \sim 2 - 3$

The advent of sensitive infrared imaging surveys has provided a powerful method for identifying galaxy clusters at $z > 1.5$ (Spitler et al., 2012). We are now able to track how clusters build up their galaxy populations over $\sim 80\%$ of cosmic time ($0 < z < 2.5$). Because the $z \sim 1.5 - 2$ clusters are still assembling, we can better disentangle evolution driven by environment versus galaxy mass (Tran et al., 2015, 2017).

The primary challenge to studying cluster galaxy populations at $z > 1.5$ is the need for NIR spectroscopy (Fig. 80; Yuan et al., 2014). With the installation of efficient multi-object near-IR spectrographs like MSE, we can now obtain key rest-frame features from [O II]$\lambda 3727, 3729$Å to [S II]$\lambda 6717, 6731$Å for galaxy clusters at $z > 1.5$ (Kacprzak et al., 2015). These spectral features have been used to establish empirical scaling relations at $z \sim 0$ (Figure 81) that we can now test at $z \sim 2$ (Kewley et al., 2016; Alcorn et al., 2016). Currently results on whether environment already plays a role at $z \sim 2$ are mixed with measurements in only a handful of galaxy clusters. Only by increasing the number of overdensities at $z > 1$ with measurements of, e.g. gas-phase metallicities, can we better track how environment affects galaxy evolution at this pivotal epoch.

#### 7.2.4 Satellite planes in the nearby universe

Recent years have seen exciting and perplexing discoveries of coherent rotating planes of dwarf galaxy satellites around nearby giant galaxies including the Milky Way, M31, and NGC5128 (Pawlowski et al., 2012, 2013; Ibata et al., 2013; Tully et al., 2015; Müller et al., 2018). Figure 82 shows the kinematics of the NGC5128 plane, where the velocities of dwarf galaxies (squares) are seen to have a clear 1 Mpc-scale rotational signature (left panel), with a rotation axis that broadly aligns with the known kinematics of low-mass GC and PNe tracers within $\sim 50$ kpc of the host giant (right zoom-in panels). The observation that the rotation axes of the "inner" and "outer" satellites align so closely is intriguing, but by no means definitive. While there are $\sim 2000$ GC tracers in the intermediate $50 - 200$ kpc region (black points; Taylor et al., 2017), efforts to investigate whether the two populations are truly kinematically connected are hampered by the lack of a wide-field, highly multiplexed spectroscopic instrument mounted on an $8 - 10$ m class telescope. Regardless, the confirmation of a coherent kinematical signature reaching out to Mpc scales around a giant galaxy would be a stunning and unexpected result, which would present a significant challenge to the ΛCDM paradigm (e.g. Pawlowski et al., 2014; Ibata et al., 2014). Even if this particular alignment is coincidental, the capabilities of MSE would open the door to mapping the kinematics of giant galaxies out to an unprecedented $\sim 25 - 30$ Mpc, and would reveal the frequency of such satellite planes, as traced by systems of CSSs, around hundreds of giant galaxies in the nearby Universe. Further discussion of satellite planes can be found in Chapter 6.

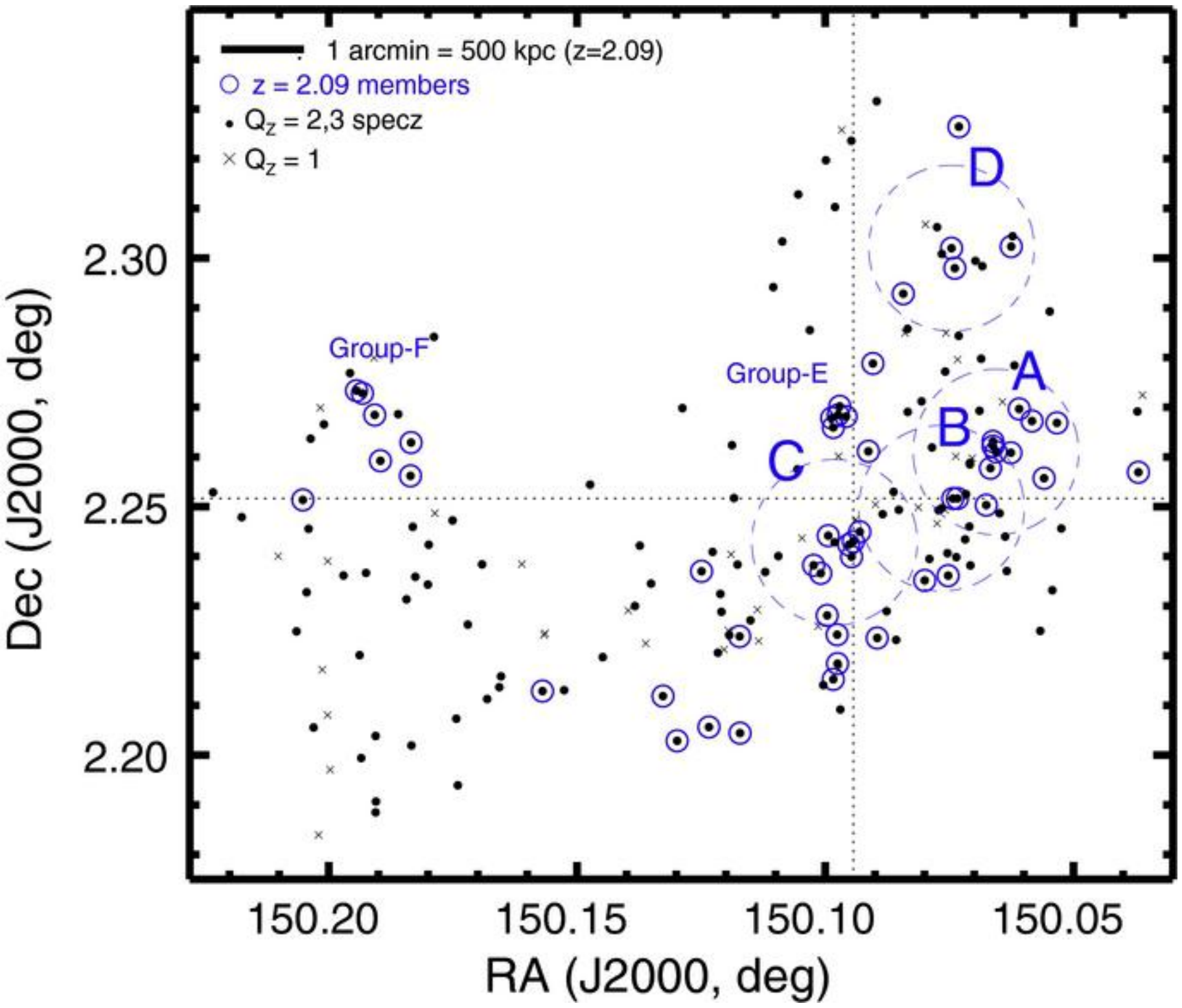

*Figure 80: 180 galaxies at $z \sim 2$ with reliable spectroscopic redshift identifications obtained with Keck/MOSFIRE. With the increasing number of star-forming cluster members at $z \sim 1.5-2$ (Tran et al., 2010; Brodwin et al., 2013), MSE can measure key rest-frame features such as [O II]$\lambda 3727, 3729 Å$ and H$\beta \lambda 4861 Å$. Figure from Yuan et al. (2014).*

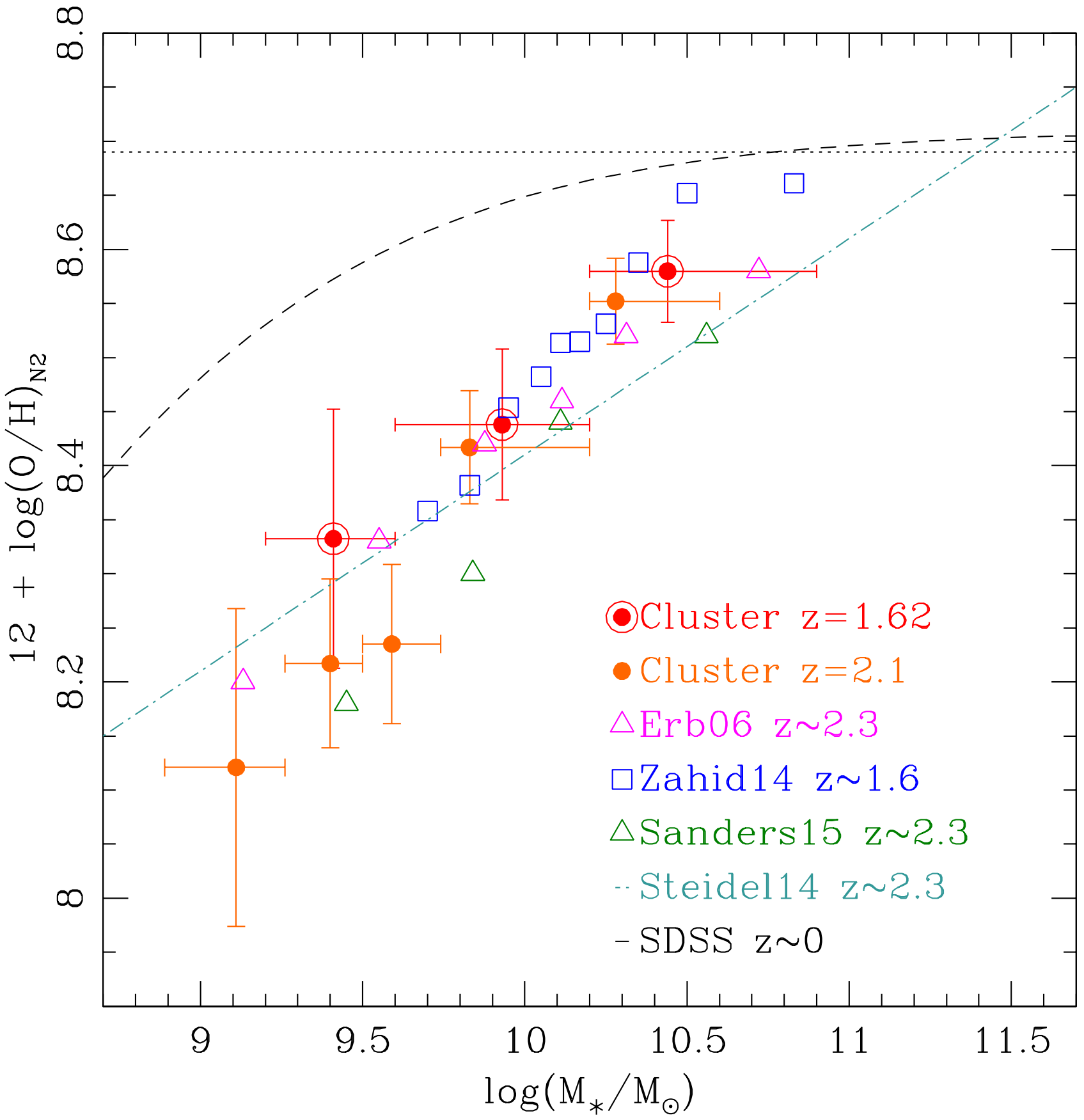

*Figure 81: A spectroscopic survey of the galaxy cluster XMM-LSS J02182-05102 at* $z_{\rm spec} = 1.6233 \pm 0.003$ *with cluster velocity dispersion of* $254 \pm 50$ *km s*$^{-1}$*. The galaxies in IRC 0218 show a similar trend between gas-phase metallicities and stellar masses (MZR) as the field at* $z \sim 2$*, i.e. there is no environmental imprint on the MZR at this epoch. However, measurements of gas-phase metallicities at* $z \sim 2$ *show large scatter and surveys of more overdensities at* $z > 1$ *are needed. Figure from Tran et al. (2015).*

## 7.3 Massive galaxies

### 7.3.1 Mapping giant galaxy assembly at $z \sim 0$ with compact stellar systems

In the standard Λ-cold dark matter (ΛCDM) paradigm, giant galaxies are progressively built up through the accretion of lower mass dwarf galaxies over cosmological time. However, the question of where giant galaxy halos transition from material formed in-situ to where dwarf galaxy disruption takes place is difficult to address for distant galaxies due to the faintness of diffuse outer-halo light. Meanwhile, nearby giant galaxies are equally challenging due to the large angular extents that their halos subtend on the sky. For this reason, comprehensive studies to map the outer halos of nearby giants require instruments with very wide fields of view, and even then are hampered by the fact that resolved stellar population studies are mostly restricted to the Local Volume within $\sim 10\,$Mpc. Happily, rich systems of compact stellar systems (CSSs) that are ubiquitous around giant galaxies throughout the universe have the potential to trace halo chemodynamics to large radii (tens to hundreds of kpc).

Globular clusters (GCs) are CSSs with luminosities and sizes of $M_V \sim -7.5\,$mag and $r_{\mathrm{eff}} \lesssim 10\,$pc, whose birth in the early universe and evolution around their host galaxies make them excellent probes of galaxy formation processes (e.g. Brodie & Strader, 2006). GC colour distributions are typically bi-modal with metal-poor (blue) GCs forming from pristine gas around primordial dwarf galaxies, and metal-rich (red) GCs from material enriched by their giant galaxy hosts (Côté et al., 1998, 2000). The contributions of each to a giant galaxy's GC system thus hints at the relative importance of major and minor mergers to its mass assembly history.

Ultra-compact dwarfs (UCDs) were discovered ~20 years ago (Hilker et al., 1999; Drinkwater et al., 2000). These are massive ($\mathcal{M}_\star \sim 10^{6-9}\,M_\odot$) cousins to GCs with $-13.5 \lesssim M_V \lesssim -11.5\,$mag and $r_{\mathrm{eff}} \lesssim 50\,$pc. These properties place UCDs intermediate between dwarf galaxies and GCs, and blur their distinction. Their discovery elicited an ongoing debate on whether they are galactic in origin (e.g., the remnant nuclei of dwarf galaxies threshed in giant galaxy tidal fields; Bekki & Couch 2001; Pfeffer & Baumgardt 2013), or are simply very bright GCs. The discovery of a $\sim 10^6 M_\odot$ black hole in a UCD by Seth et al. (2014), combined with elevated dynamical mass-to-light ratios for UCDs above $\gtrsim 10^6\,M_\odot$ (Taylor et al., 2010; Mieske et al., 2013; Forbes et al., 2014; Taylor et al., 2015) indicate that many UCDs have origins as stripped nucleated dwarf galaxies. Meanwhile, recent results from the Virgo cluster by Zhang et al. (2018a) show UCDs to have a kinematical signature distinct from metal-rich GCs, further implying that the growth of UCD systems is linked to disrupted dwarf galaxies, thus demonstrating the utility of CSS in exploring the assembly of giant galaxy halos.

Figure 82 shows seven fields of the 2dF instrument on the 3.9m AAT, tiling the outer halo of the giant elliptical galaxy NGC 5128, at a distance of ~4 Mpc. These observations sample a rich population of GCs. For comparison, the orange shading shows the same seven fields sampled by MSE. While the overall spatial sampling per pointing is somewhat less than 2dF, this is more than made up for by the significantly larger MSE aperture, that would cut the required exposure time per field by an order of magnitude. In fact, at $\sim 4\,\mathrm{Mpc}$ distance, to reach $SNE \gtrsim 15$ at $R \sim 2000$ - thus enabling meaningful chemodynamical studies - for an average GC requires several hours of exposing on a 4m-class telescope. This limits us to only a small handful of galaxies for large-scale kinematical studies of giant galaxy halos. The advent of MSE would not only improve observing efficiency in the very nearby universe by an order of magnitude, but would open the door to sampling myriad galaxy environments out to $\sim 25-30\,\mathrm{Mpc}$.

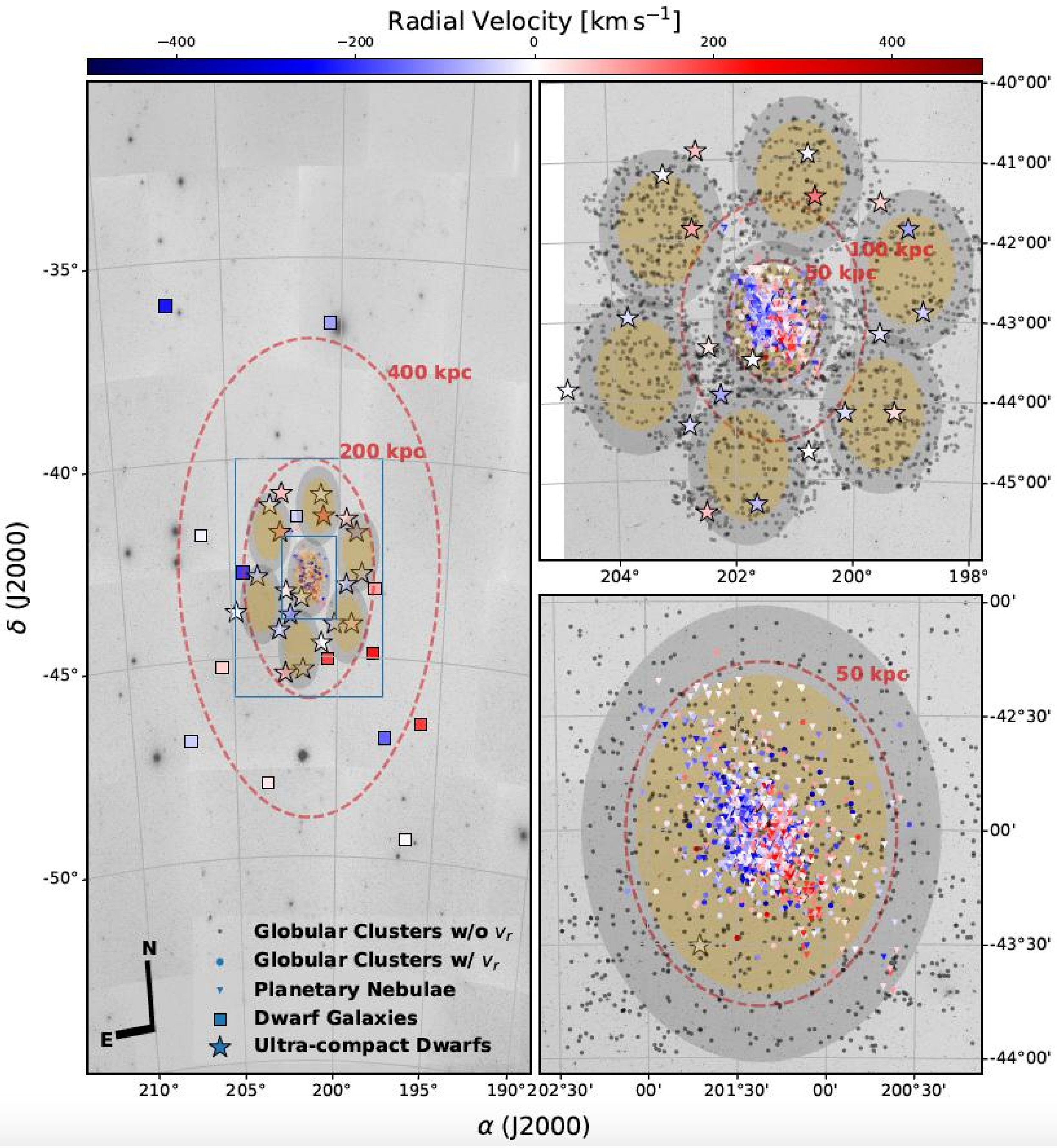

*Figure 82: Low-mass satellite kinematics for the nearby giant elliptical galaxy NGC 5128 superposed on an archival DSS image. Left: a 1 Mpc-scale view shows dwarf galaxies (squares) with $v_r$ parameterized by color. Rotation is evident with northern satellites approaching along the line of sight, and southern objects mostly receding. Colored stars indicate UCDs and show no obvious sign of coherent rotation. Grey shaded regions indicate the seven fields-of-view required to cover the known CSSs system by the AAOmega/2dF facility. Yellow shading indicates the corresponding MSE patrol fields that would achieve the same science while extending the limits of reachable targets from ∼ 5 Mpc to ≲ 30 Mpc. Top-right: an intermediate view of NGC 5128 showing the extent of the known system of CSSs with UCDs and GCs (dark grey dots) surrounding the inner satellites. Bottom-right: a ∼ 60×60 kpc$^2$ cut-out illustrating the kinematics of the inner satellite GCs (colored dots) and PNe (triangles). Note the similar satellite rotational signatures between the inner and outer regions, but lack of detailed kinematical data for intermediate regions where the UCDs and newly discovered GCs reside.*

### 7.3.2 A census of massive galaxies at cosmic noon

MSE's broad wavelength coverage up to $\lambda = 1.78\mu$m allows us to probe $\lambda_{\rm res} = 5400$Å out to $z = 2.3$. We can therefore think of a deep spectroscopic survey of massive galaxies ($M_{\rm star} > 5\times10^{10}\,\mathrm{M}_\odot$ or $> 10^{11}\,\mathrm{M}_\odot$, depending on the redshift) to study their stellar populations through stellar continuum studies. Out to $z = 2.3$, we can get the full set of absorption lines/feautures (CaII(H+K), D4000, H$\delta$, H$\gamma$, G-band[4300], Mgb[5175], Fe[5270], Fe[5330]) to measure stellar metallicity, abundance ratios, age, and velocity dispersion. The combination of these measurements will provide constraints on the duration of the star formation and the quenching, defining their SFHs, the dynamical mass (combined with size measurements), and the mass density.

At the same time, we will also measure faint emission lines, and the gas-phase properties (this is particularly true for the star-forming galaxies). In comparison to similar surveys that currently exist at lower redshift, this extends LEGA-C (van der Wel et al., 2016) from $0.6 < z < 1$ to $z = 2.3$. If we limit ourself to $\lambda_{\rm res}$ <4200Å (hence losing the ability to do abundances, but still able to measure velocity dispersion and stellar ages), this can be extended out to $z = 3.2$. A SNR > 10 per resolution element is probably the minimum needed (but we may want to aim for more). For MSE, we can expect to get SNR = 10 per resolution element in 4 hrs at $m_{\rm H} = 23$ (Table 1).

From Figure 83, $m_{\rm H} < 23$ should allow us to build a stellar mass complete sample to $z \sim 2-2.5$, and still representative out to $z \sim 3$. At $z > 3$, the rest-frame optical break rapidly shifts out of the H-band, and we no longer have a representative sample of massive galaxies if H-band selected down to reasonable magnitudes. Compared to VLT/MOONS, MSE has 1.5× larger mirror, 3× more fibers, and 10× wider field-of-view, so we can exploit this to sample a wide range of environments (from voids to filaments, to proto-clusters). Thus to $z \sim 2.3$, we can do detailed studies of many thousands (parent survey dependent, and dependent on the ultimate targeting and tiling efficiency achieved with MSE) of massive galaxies as a function of environment, star-formation activity, etc, i.e. we can extend the LEGA-C type of surveys out to $z \sim 3$.

In terms of potential competition in this domain, Subaru/PFS goes out to $J$ band ($\lambda = 1.26\mu$m), so it is limited to lower redshift for the aforementioned stellar continuum studies. JWST will do some of these targets too, but it will be limited to very small samples. VLT/MOONS will also target some of this sample, but it has a much smaller field of view on a smaller telescope (although there is clear overlap in the planned surveys). 30m telescopes will also cover aspects of this science (including with resolved sources), but for very small numbers of objects. A critical aspect for achieving these science goals is the quality of the NIR sky subtraction, which is an area of active development within MSE (McConnachie et al. 2018a).

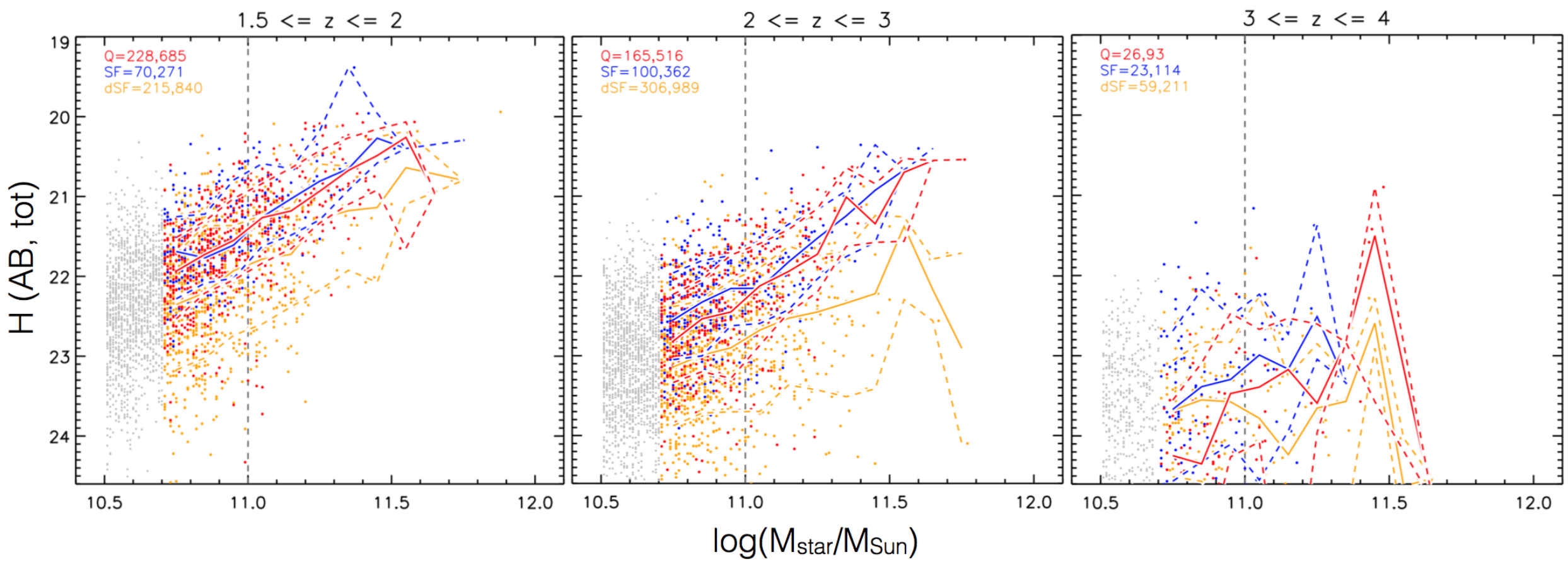

*Figure 83: Total H-band magnitude versus stellar mass from 0.7 sq. deg. (COSMOS); red, orange, and blue points are UVJ quiescent, dusty star-forming, and relatively unobscured star-forming galaxies, respectively (definition from Martis et al., 2016). Solid curves show the median trends, whereas dashed curves are the 15$^{\rm th}$ and 85$^{\rm th}$ percentiles. Listed are the numbers of galaxies with $M_\star > 10^{11}$ $M_\odot$ (first number) and with $M_\star > 5\times10^{10}$ $M_\odot$ (second number). Out to $z\sim2$ we could aim for a stellar mass complete sample at $H<23$; at $2<z<3$, we could still construct a fairly representative sample of massive galaxies with $H<23$; at $z>3$, we would have to target the brightest galaxies, and no longer close to a mass-selected sample (unless we go for lower SNR spectra). One expects ~2,300 and ~8,000 galaxies with $M_{\rm star} > 5\times10^{10}$ $M_\odot$ and $> 10^{11}$ $M_\odot$, respectively, over 1.5 sq. deg.*

Tens of thousands of galaxies up to $z\sim3$ have been already targeted with 10-meter class telescopes (DEEP2, zCOSMOS, VIPERS, MOSDEF, KBSS). Knowledge about more detailed physical properties was gained from relatively small samples or from stacks. If MSE has the potential of deriving an unprecedented sample of massive galaxies out to $z\sim3$ and will create a unique opportunity to study the full range of physical properties of galaxies that dominate the stellar mass budget at cosmic noon ($z\sim2$).

#### 7.3.3 Star formation & stellar assembly histories

The formation of massive galaxies has long been a challenge for galaxy formation theory, as extreme feedback efficiencies are required to quench them early on. Recent discoveries of massive (stellar masses of $10^{11}\,M_\odot$), quenched galaxies at $z\sim3.7$ (Glazebrook et al., 2017) pose even more challenge: few physical processes are capable of quenching galaxies in the short timescales that are required to reproduce the strong H$\delta$, H$\gamma$ and H$\beta$ absorption reported by Glazebrook et al. (2017) (see Figure 84). The stellar mass of these galaxies and their star formation histories imply that $\sim50\%$ of all the baryons in the halo are locked up in stellar mass. The current problem we face is that it is unclear what is the number density of the massive/passive galaxies and how common they are among the overall sample of massive galaxies. The latter is essential to understand in order to pin down whether our current galaxy formation theory needs further development to reproduce such a population as the mainstream, or whether they can be explained as rare events within the current model.

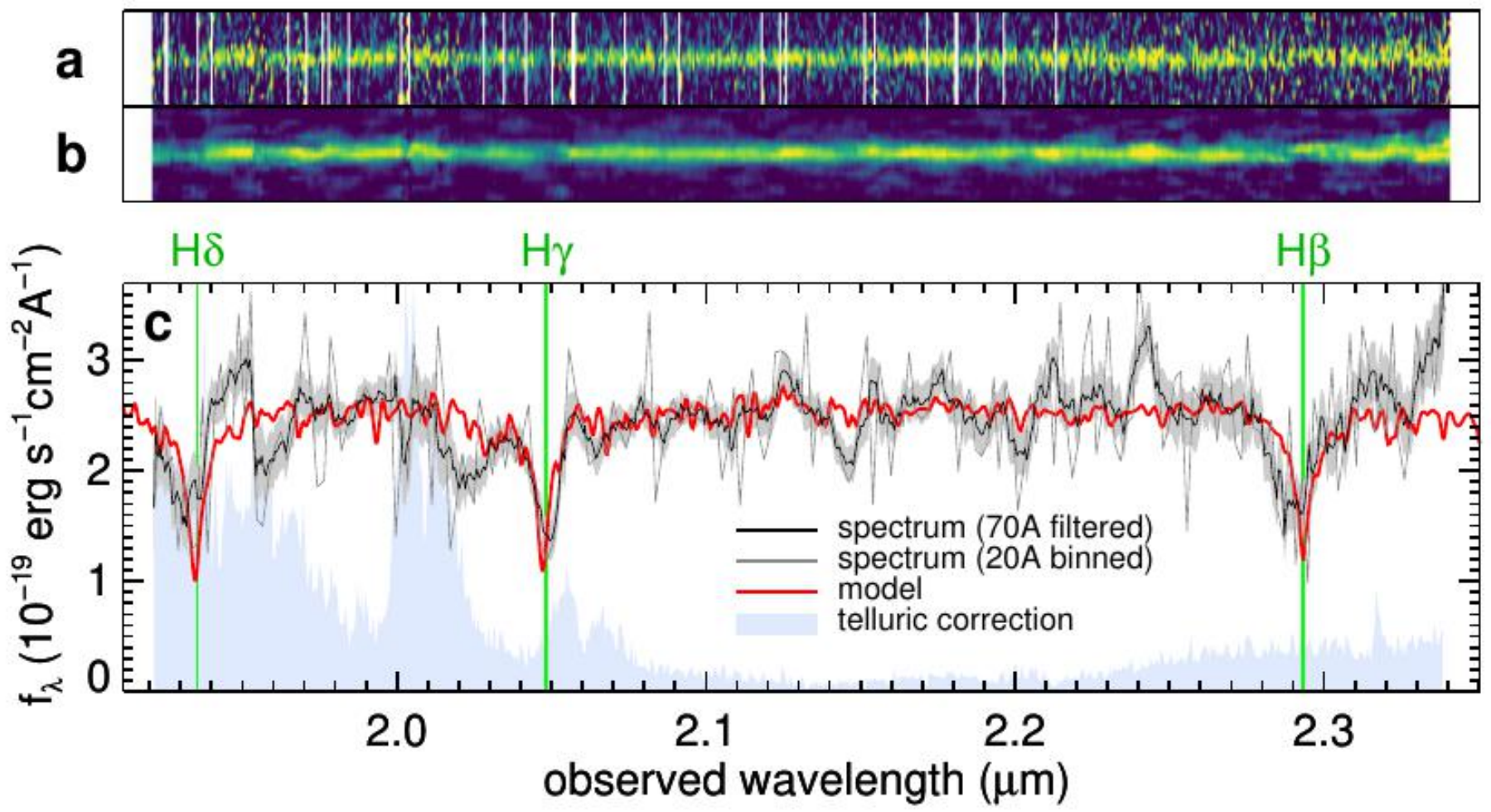

*Figure 84: Spectrum of ZF-COSMOS-20115 at $z \sim 3.7$ in the near-infrared $\boldsymbol{K}-$band showing the deep absorption line features indicative of a post-starburst spectrum. The lack of emission lines suggest lack of current star formation. This galaxy has a stellar mass of* $10^{11}$ $M_\odot$. *Figure from Glazebrook et al. (2017).*

MSE should be able to observe hundreds of galaxies with stellar masses $> 10^{11}\,M_\odot$ out to $z \approx 3$, which would serve as a parent sample for deeper spectroscopic follow up to obtain accurate stellar continuum and emission/absorption (CaII(H+K), D4000, H$\delta$, H$\gamma$, G-band[4300], Mgb[5175], Fe[5270], Fe[5330]). Constraining the SFH of these galaxies would also lead to understanding the baryon collapse and star formation efficiency of halos in the early Universe.

### 7.3.4 Environment & mergers

Mergers between galaxies are thought to be closely linked to the evolution in galaxy properties: they drive the morphological transformation of galaxies (e.g., Toomre & Toomre, 1972), trigger starburst and active galactic nucleus (AGN) episodes (e.g., Barnes & Hernquist, 1991; Mihos & Hernquist, 1994; Wuyts et al., 2010; Davies et al., 2015), and lead to quenching of star formation and the buildup of massive quiescent systems (e.g., Robotham et al. 2013, 2014; Toft et al. 2014; Davies et al. 2016b). Major galaxy mergers produce the most luminous AGNs and ultra-luminous infrared galaxies (ULIRGs; Kartaltepe et al. 2010; Ellison et al. 2013). In comparison with isolated galaxies, interacting and merging systems are characterized by enhanced star formation activity (Patton et al., 2011; Yuan et al., 2012).

Furthermore, massive quiescent galaxies at $z \sim 2$ are on average a factor of $5-6$ smaller than their massive counterparts in the Local Volume (Daddi et al., 2005; Trujillo et al., 2007; Damjanov et al., 2011; van der Wel et al., 2014). A fraction of this average size growth is attributed to the growth of individual quiescent systems (Newman et al., 2012; Fagioli et al., 2016; Damjanov et al., 2018). Theoretical models of individual galaxy size growth include both major mergers between gas-poor galaxies of similar stellar mass and minor merger or accretion of low surface brightness objects (e.g., Naab et al., 2009; Hopkins et al., 2010). As galaxy merging is suggested to have a profound influence on galaxy evolution, it is essential to measure merger fractions and infer merger rates over a broad redshift range in order to quantify their effects on the observed evolutionary trends.

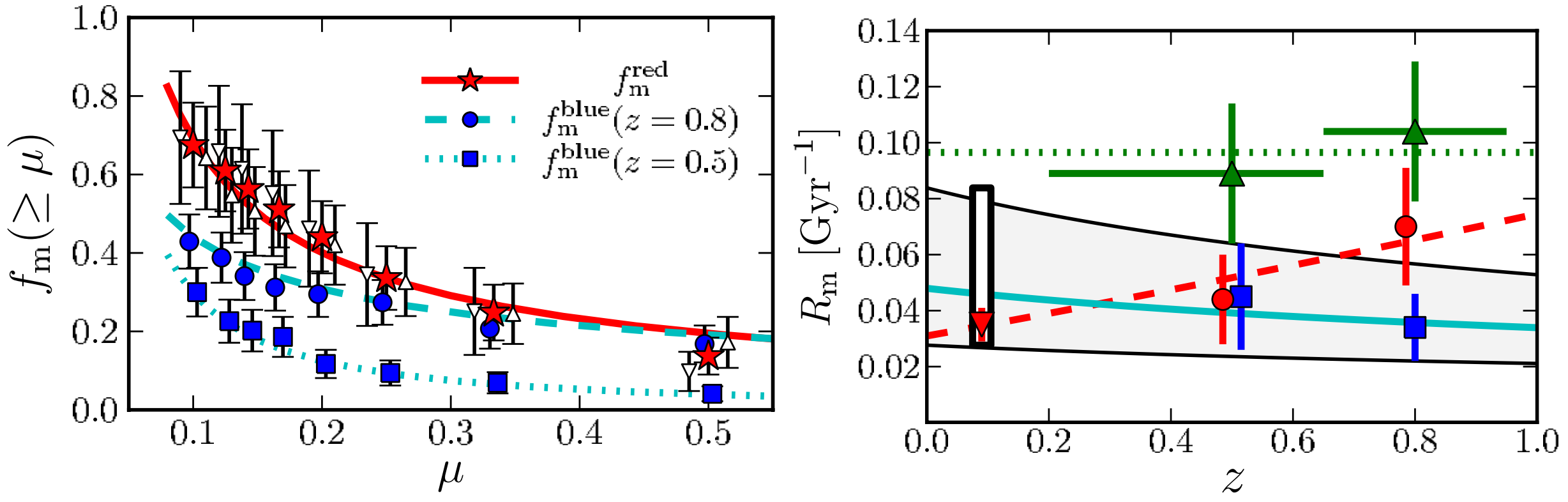

*Figure 85: Left: Merger fraction versus luminosity ratio in B-band ($\mu$). Stars, triangles and inverted triangles are the merger fraction of red primaries at $0.2 \geq z \geq 0.95$, $z = 0.8$, and $z = 0.5$, respectively. Dots and squares are the merger fraction of blue primaries at $z = 0.8$ and $z = 0.5$, respectively. The lines are the best fits power-law functions ($f_m [\geqslant \mu] \propto \mu^s$). Right: Merger rate of galaxies versus redshift. Dots are the major merger rate ($\mu \geq 1/4$), squares are the minor merger rate ($1/10 \leqslant \mu < 1/4$), and triangles are the total (major + minor, $\mu \geqslant 1/10$) merger rate. The error bars in the total merger rate mark the redshift range spanned by VVDS-Deep data. The inverted triangle is the major merger rate at $z = 0.09$. The white rectangle identifies the $z = 0.09$ minor merger fraction estimated from the total and the major merger fractions. The gray area marks the most probable minor merger rate values in the range $0 < z < 1$. The solid and dotted lines are the best-fit power-law function for the evolution in minor merger and major merger rates, respectively. The dotted line represents a constant major + minor merger rate. Figure adapted from López-Sanjuan et al. (2011).*

Merger fraction measurements at $z > 1$ produce conflicting results (Man et al., 2016) that can be partially attributed to the difference in parent galaxy selection criteria and mass ratio limits (Lotz et al., 2011), as well as to the cosmic variance due to small survey areas (López-Sanjuan et al., 2012). These studies of close galaxy pairs are also influenced by the size of spectroscopically confirmed galaxy pair samples that decreases drastically with increasing redshift and increasing galaxy mass ratio (López-Sanjuan et al., 2011). With magnitude (and surface brightness) limited samples of galaxies covering large areas on the sky, future MSE surveys will provide pure samples of galaxy pairs at $0.5 < z < 4$ based on their kinematic properties. These samples will generate merger fraction measurements that are minimally affected by cosmic variance over the full range of surveyed redshift interval.

The complete spectroscopic surveys with MSE that cover large area and, due to its near-IR component (thorough $H$-band), include broad redshift range will probe the evolution in merger fractions and merger rates as a function of mass ratio for star-forming and quiescent $L \gtrsim L^*$ systems over > 10 Gyr of cosmic time. Kinematically confirmed galaxy pair samples will follow the trends that so far have been explored with spectroscopic redshifts only to $z \lesssim 1$ (Figure 85). Furthermore, the number of galaxies in these large spectroscopic samples and their large survey area will enable the first investigations of the evolution in merger rate and accreted mass as a function of both galaxy intrinsic properties (e.g., stellar mass from SED fitting with known spectroscopic redshift, star formation rate from nebular rest-frame optical lines, average age of stellar population from $D_n4000$ index, velocity dispersion from absorption line modelling) and large scale structure distribution. By covering more than $\sim 80\%$ of cosmic history MSE galaxy pair investigations will provide robust observational constraints on the effects that gas-rich and gas-poor mergers with various mass ratios have on the mass assembly of galaxies in a range of different environments, from voids to galaxy superclusters.

## 7.4 M$^\star$galaxies (Milky Way analogues)

### 7.4.1 Star formation/stellar assembly histories

It is now clear that, in order to quantify the growth of galaxies across the Hubble time, matching galaxy populations at different redshift by their stellar mass alone is not enough. Indeed, a $10^{10}$ $M_\odot$ main-sequence galaxy at $z \sim 1$ or 2 has gone (and will go) through a significantly different evolutionary path than a $10^{10}M_\odot$ main-sequence galaxy at $z \sim 0$. In other words, we need to be able to link a progeny to its correct progenitors. To reach this goal, there are several key requirements relevant for MSE surveys.

First, it is critical to be complete (or at least representative) across a stellar mass range that includes all potential progenitors of interest for Milky-Way like galaxies. For example, as shown in the right panel of Figure 86, to include all the potential progenitors of MW-size galaxies at $z \sim 0$, it is important to reach at least $10^9 M_\odot$ at $z \sim 2$ (ideally, $10^{8.5}$ $M_\odot$ would provide a buffer to make sure we observe the MW-like stellar mass range). In terms of survey volume, it should be such that we can have, at the very least, three redshift bins in the $1 < z < 2$ range and, ideally, $\geq 200$ galaxies per $0.25 - 0.3$ dex stellar mass bin in each bin at each redshift (for example, >4000 galaxies in the $1 < z < 2$ range). This may be too optimistic given the constraints of the 2-3 yr dark-time survey. Pushing in this direction will enable SDSS Legacy Surveys at $1 < z < 2$.

A selection like the one described above would significantly improve our reconstruction of the redshift evolution of the main-sequence of star-forming galaxies. This is shown in the left panel of Figure 86, which compares the evolution of the main-sequence as tabulated by Whitaker et al. (2012, solid lines) and Schreiber et al. (2015, dashed lines). Here we extrapolate outside the stellar mass range within which the analytic formalism has been calibrated, but this is done intentionally to highlight our still very poor knowledge of the evolution of the main sequence, particularly for stellar masses $M_\star < 10^{10.5}M_\odot$ at $z > 1$. This has huge implications in our ability to reconstruct the evolution of galaxies of different masses across the star formation rate versus stellar mass plane (see right panel of Figure 86).

Pushing further still, it is highly desirable to go a step further and push for getting spectra that provide more than just redshifts. In particular, reaching a high SNR across the optical rest-frame continuuum will make it possible to reconstruct star formation histories for individual galaxies, which can then be used to accurately link each galaxy to their potential progenitor population. To make this unique with respect to other surveys, we would likely need to push to low stellar masses (a baseline of typically SNR = 10 at rest-frame 5000Å for galaxies with stellar mass $10^{9.5}$ $M_\odot$ at $z \sim 1.5$).

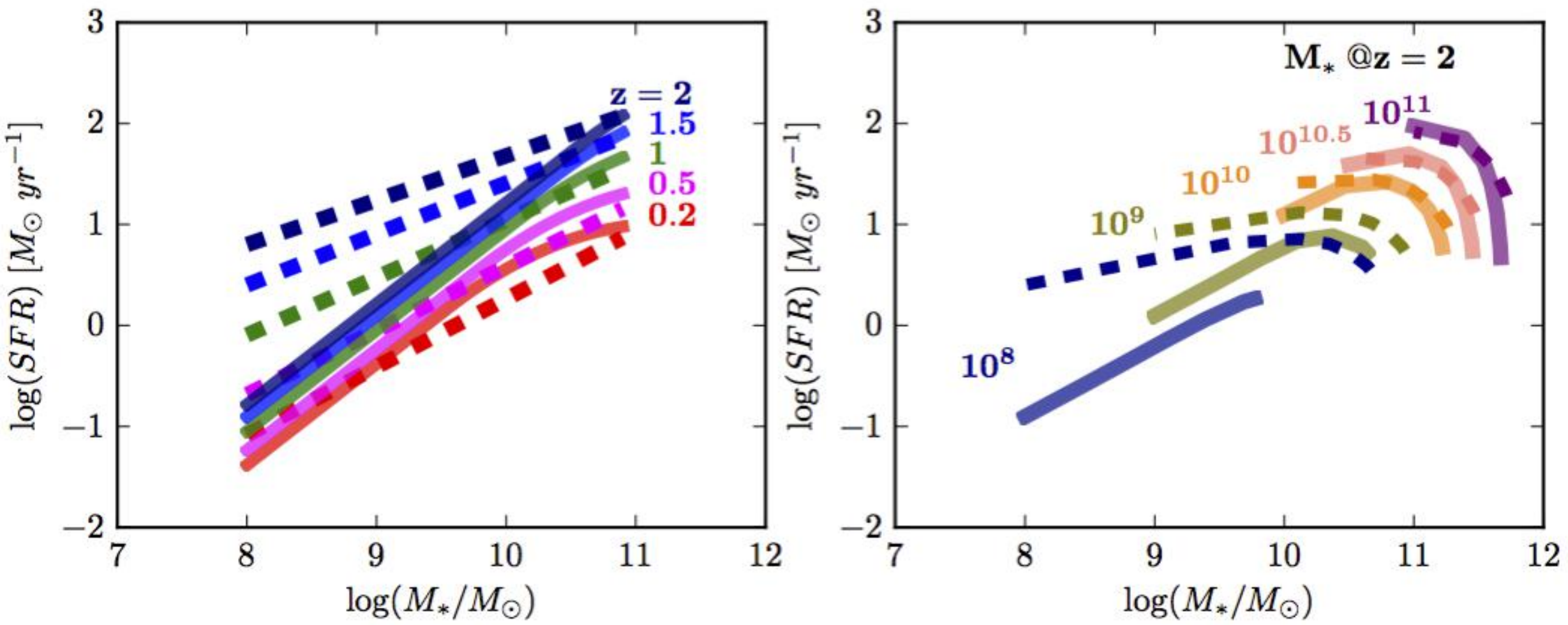

*Figure 86: Left: the evolution of the main-sequence of star-forming galaxies from $z = 2$ to 0.2 derived using the analytic equations presented in Whitaker et al. (2012, solid lines) and Schreiber et al. (2015, dashed lines), respectively. Note that the main sequences have been intentionally extrapolated outside the stellar mass range of calibration to highlight our poor knowledge for z >1 and stellar masses < $10^{10}$ $M_\odot$. Right: the evolution in the stellar mass vs. star formation rate plane for galaxies having stellar masses at $z = 2$ between $10^8$ and $10^{11} M_\odot$. Solid and dashed lines show the result obtained by assuming the Whitaker (solid) or Schreiber (dashed) prescription, respectively. This clearly highlights our inability to link progenitors and progeny between $z \sim 0$ and $z \sim 2$.*

### 7.4.2 Co-evolution of AGN and galaxies

There is an interesting synergy between MSE and the latest suites of radio continuum surveys, e.g., the Very Large Array Sky Survey (VLASS, Lacy et al. in prep) in the north and square kilometre array projects such as the Evolutionary Map of the Universe (EMU, Norris et al., 2011) in the south (see discussion in Chapter 2). The science cases for these state-of-the-art radio observations cover several key questions related to both AGN and galaxy physics, from the efficiency of star-formation to the impact of galaxy environment on AGN accretion mode. Many of these questions will require complimentary spectroscopy to address, and, given the sensitivity of the radio observations (see Figure 87), to a greater depth than previous spectroscopic surveys. The source number densities in these next generation radio surveys is expected to be relatively low ($\sim 2,000\,\mathrm{deg}^{-2}$ for EMU, Norris et al., 2011, and $\sim 100\,\mathrm{deg}^{-2}$ for VLASS, Mark Lacy, private communication) and could potentially exploit spare fibres.

In combination with neutral hydrogen observations, star-formation rates can be used to determine the efficiency of a galaxy in converting baryonic building blocks into stellar matter. The LADUMA survey (Looking At the Distant Universe with the MeerKAT Array, Holwerda et al., 2012) will provide HI observations for galaxies out to $z \sim 1$ in the Chandra deep field south. Obtaining spectra with MSE for these galaxies will allow for a measurement of the star-forming efficiency at high-$z$. Over the next few years, the Wide-field ASKAP L-band Legacy All-sky Blind surveY (WALLABY, Koribalski, 2012) will provide H I measurements for 75 % of the sky out to $z > 0.2$ providing a large local-Universe benchmark against which such high-redshift measurements can be compared.

The sensitivity of surveys such as VLASS and EMU will enable the detection of star-forming galaxies based on their radio-emission out to $z > 0.5$ (see Figure 87). In the local-Universe many star-forming galaxies detected in the radio are relatively low-mass, with the median stellar mass of Best & Heckman (2012) star-forming galaxy sample being $10^{10.5}$ $M_\odot$, and more than 10 % of that sample having $M_* < 10^{10}$ $M_\odot$. Consequently, these radio observations have the potential to act as an input catalogue with which to target low-mass star-forming galaxies for spectroscopy.

Whilst high-luminosity AGN are frequently studied in depth at $z > 1$, fewer such observations exist for low- to intermediate luminosity AGN at this epoch - a natural result of the Malmquist bias. Any reasonably complete survey targeting galaxies at $z \sim 1$ might expect $\sim 20$ % of those targets observed to host an AGN (Wang et al., 2017). The ability to construct an AGN luminosity function at $z \sim 1$ is thus a convenient ancillary result from the MSE galaxy evolution survey. Additionally, the classification of Seyfert galaxies based on their optical spectra at high-redshift in comparison to their MIR inferred SFRs provides a compelling opportunity to assess the AGN - star-formation connection at earlier epochs.

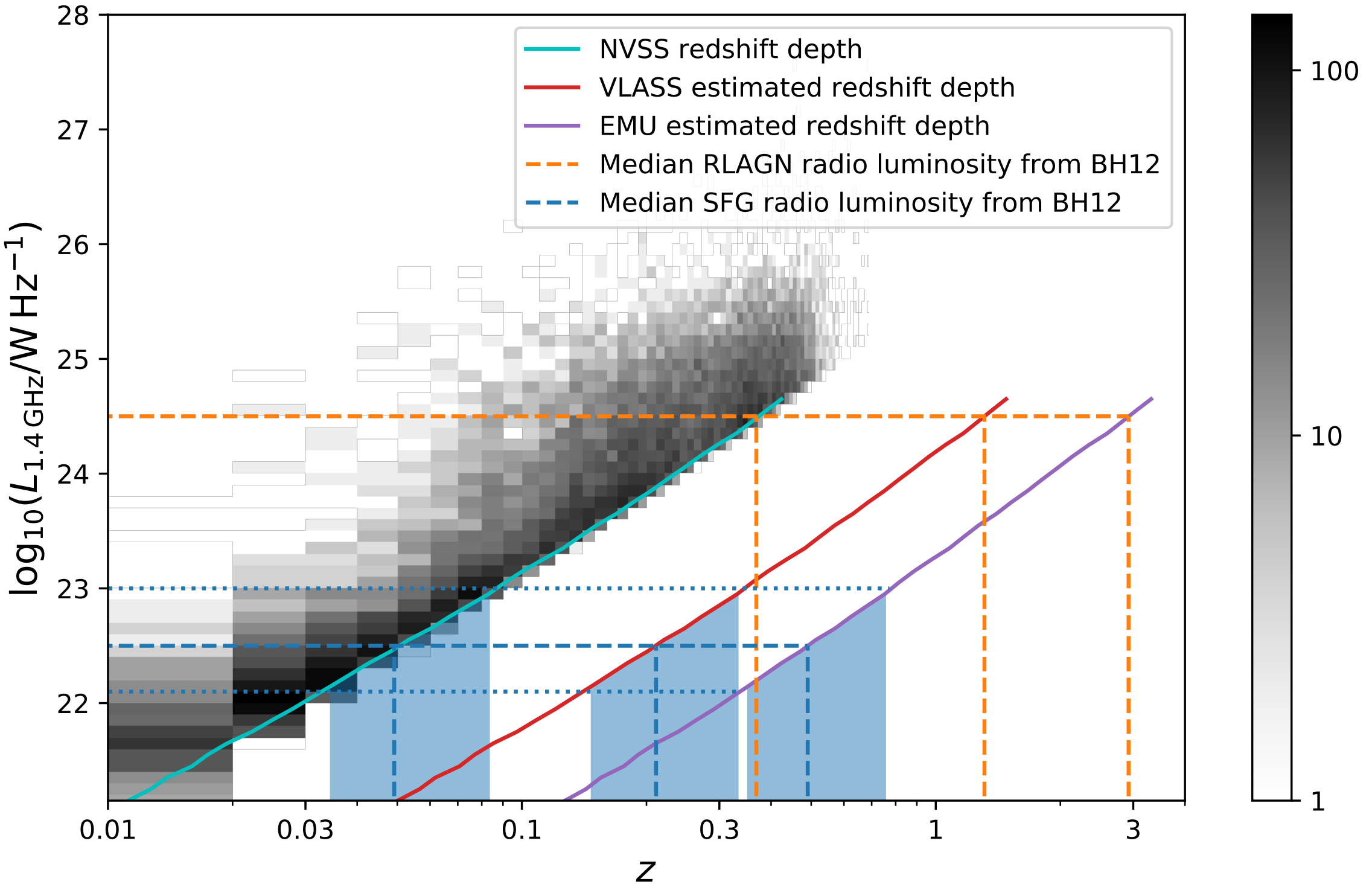

*Figure 87: Estimated redshift depth of VLASS and EMU surveys extrapolated from the radio luminosities of the Best & Heckman (2012) catalogue of radio galaxies using the FIRST/NVSS sensitivity limit. The 2D histogram shows the radio luminosity and redshifts of the Best & Heckman (2012) catalogue. The VLASS (red) and EMU (violet) z-depth curves account for both the changes in sensitivity and frequency (assuming $\alpha = -0.7$) of the respective surveys relative to NVSS (cyan). The blue dashed lines show the median 1.4 GHz radio luminosity of star-forming galaxies in the Best & Heckman (2012) sample, while the dotted blue lines represent the 25th and 75th percentiles of this distribution. Blue shaded regions show the redshift range across which star-forming galaxies are detected in the radio using NVSS/FIRST, as well as estimated ranges for VLASS and EMU. With the increased sensitivity of VLASS and EMU surveys, star forming galaxies will be detected by their radio emission out to $z \sim 0.8$ and beyond. For reference, the median AGN 1.4 GHz luminosity of the Best & Heckman (2012) catalogue is also shown (orange dashed line), demonstrating that that next-generation continuum surveys will observe RLAGN that were active during, and prior to, the cosmic noon.*

Diagnostics of spectroscopically-obtained emission line measurements for radio-loud AGN (RLAGN) can be used to determine the accretion mode of the AGN (e.g., Hardcastle et al., 2007; Buttiglione et al., 2010; Best & Heckman, 2012). A complete spectroscopic census of the sky around the target AGN will allow for accurate mapping of the host galaxy environment, providing not only reliable positional data for nearby galaxies, but information on their stellar populations, SFRs, and nuclear activity. The capabilities of MSE and next-generation radio continuum surveys are ideally positioned to conduct such an environmental analysis of low and high excitation radio galaxies at $z \sim 1$. The observations of MSE additionally have the potential to clarify our understanding of the evolution of the AGN feedback processes required to account for the $M - \sigma$ relation (Magorrian et al., 1998; Ferrarese & Merritt, 2000) and the shape of the high-mass end of the galaxy stellar mass function (Croton et al., 2006). Specifically, the complimentarity of MSE and SKA-era radio observations provide an opportunity to constrain the evolution of radio-mode feedback. This can be achieved both by constructing a radio-luminosity function at high redshift, and thus providing the numbers of galaxies capable of producing radio-mode feedback, and by comparing the kinetic jet powers of high- and low-redshift RLAGN.

## 7.5 Dwarf galaxies

A fundamental measurement for MSE will be the extension of the stellar mass function to masses below $10^8 M_\odot$, for a cosmologically representative, unbiased, spatially complete spectroscopic sample, as shown in Figure 88. Through halo modelling techniques it is possible to associate the galaxies in this stellar mass function to dark matter haloes, and thus measure the efficiency of galaxy formation as a function of halo mass and environment. Extending existing work to observations of galaxies more than an order of magnitude lower in mass provides leverage on decoupling the effects of many heating mechanisms (e.g. photoionization, supernovae feedback, supermassive black hole accretion, and stellar winds) from one another. Specifically, in the Local Group, the abundance of such low mass galaxies is orders of magnitude less than expected by simply extrapolating the dark matter halo occupation of more massive galaxies (e.g. Moore et al., 1999; Boylan-Kolchin et al., 2012a). This is at least partly due to the important but poorly understood heating, feedback and disruption processes; it may also be telling us about the nature of dark matter itself. However, characterization of this “missing satellites” problem is currently limited to galaxies in the Local Group, which is the only volume over which such low-mass galaxies are observable (see also discussion in Chapter 6). MSE will allow the crucial step of measuring the universality, or environmental dependence, of this remarkably low efficiency of star formation.

In the past decade, imaging and spectroscopic surveys of large, unbiased samples of local galaxies, both in clusters and in the field, have revealed the existence of important scaling relations, such as those between stellar mass, SFR and metallicity, and how these depend on other parameters, such as morphology, AGN activity and environment (Brinchmann & Ellis, 2000; Kauffmann et al., 2003; Tremonti et al., 2004; Balogh et al., 2004), These, in turn, inform and constrain theoretical models (e.g., Lilly et al. 2013; Peng et al. 2015). However, the extension of large statistical studies into the dwarf galaxy regime is largely uncharted, due to the flux-limited nature of most surveys. At low masses, we expect chemical enrichment to become increasingly stochastic and sensitive to factors such as winds, infall and environment (Kirby et al. 2013), potentially with a metallicity floor where self-enrichment is driven by a few generations of stars (Sweet et al. 2014). Here, again, the environmental dependence of these scaling relations is mass-dependent (Ellison et al. 2009; Cole et al. 2014) and provides key evidence for the physical drivers of both satellite and central galaxies.

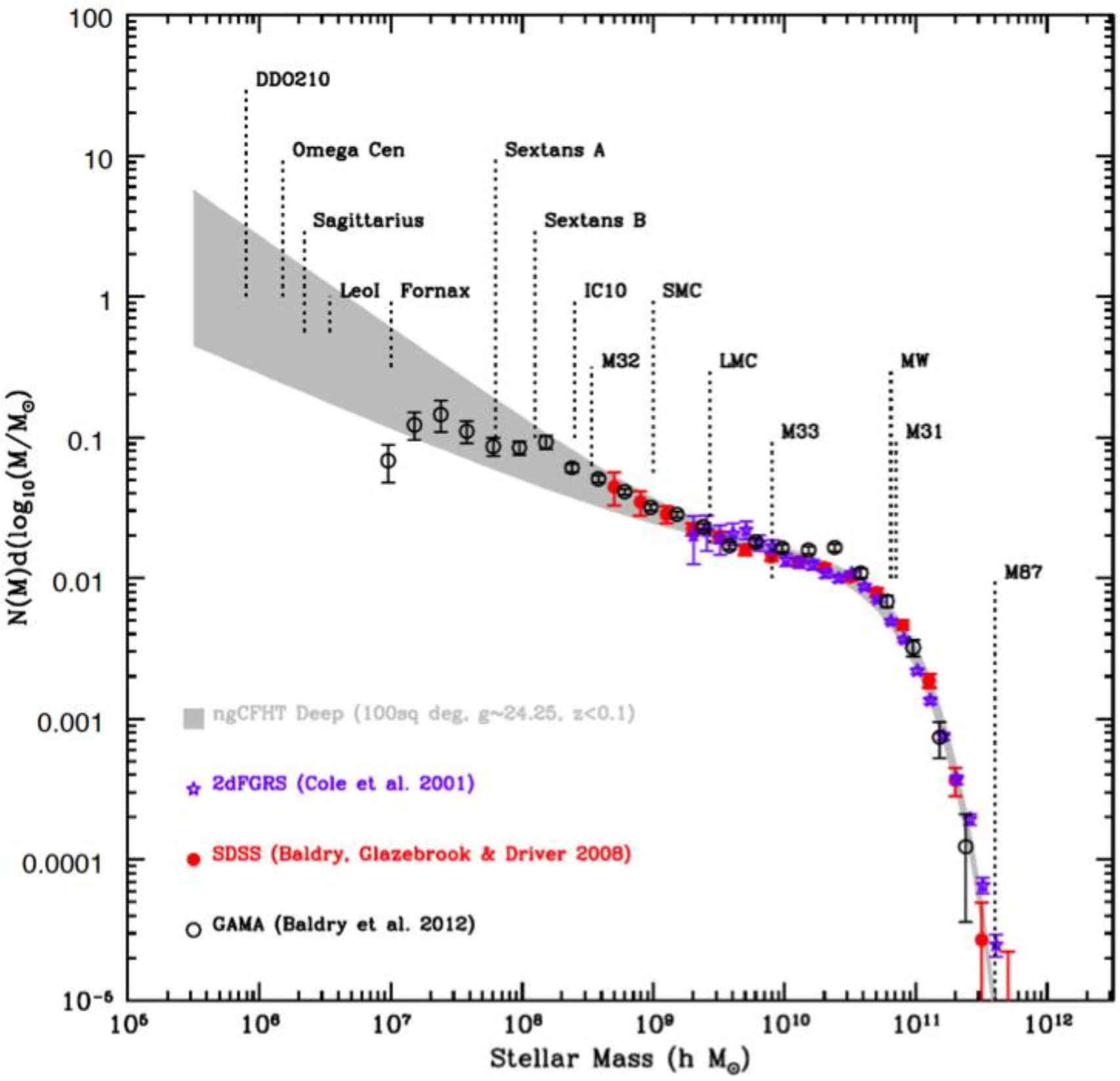

*Figure 88: The local stellar mass function as measured by SDSS, 2dFGRS and GAMA. GAMA is the deepest of these surveys, and probes robustly down to the scale of M32; lower mass measurements are actually lower-limits due to surface brightness limitations. The grey, shaded region shows a proposed MSE survey in the feasibility study, comparable to the S1-D survey described in Section 7.1.1.*

Galaxies today have assembled their mass today through a combination of in-situ star formation and mergers with other galaxies. For massive galaxies we have developed a good picture of how the star formation rate depends on stellar mass, halo mass and epoch (Behroozi et al., 2013). At low redshift it is clear, for example, that low mass galaxies are moderately more efficient at forming stars (Gilbank et al., 2010; Bauer et al., 2013). There is evidence that the relationship becomes much steeper for dwarf galaxies, in a way that is not predicted by current state-of-the-art simulations (Figure 89). MSE will extend the analysis shown in Figure89 by at least an order of magnitude in stellar mass, over a much larger area. Furthermore it will be possible to measure any dependence on halo mass, which is a prediction of most satellite quenching models but remains controversial observationally (McGee et al., 2009; Wetzel et al., 2013). Finally, from analysis of emission line ratios is will be possible to determine the occupation fraction of AGN in low mass galaxies, and determine whether supermassive black holes relate more closely to the spheroidal (bulge) component or other properties of the host (e.g. total gravitational mass). This science is discussed further in Chapter 8.

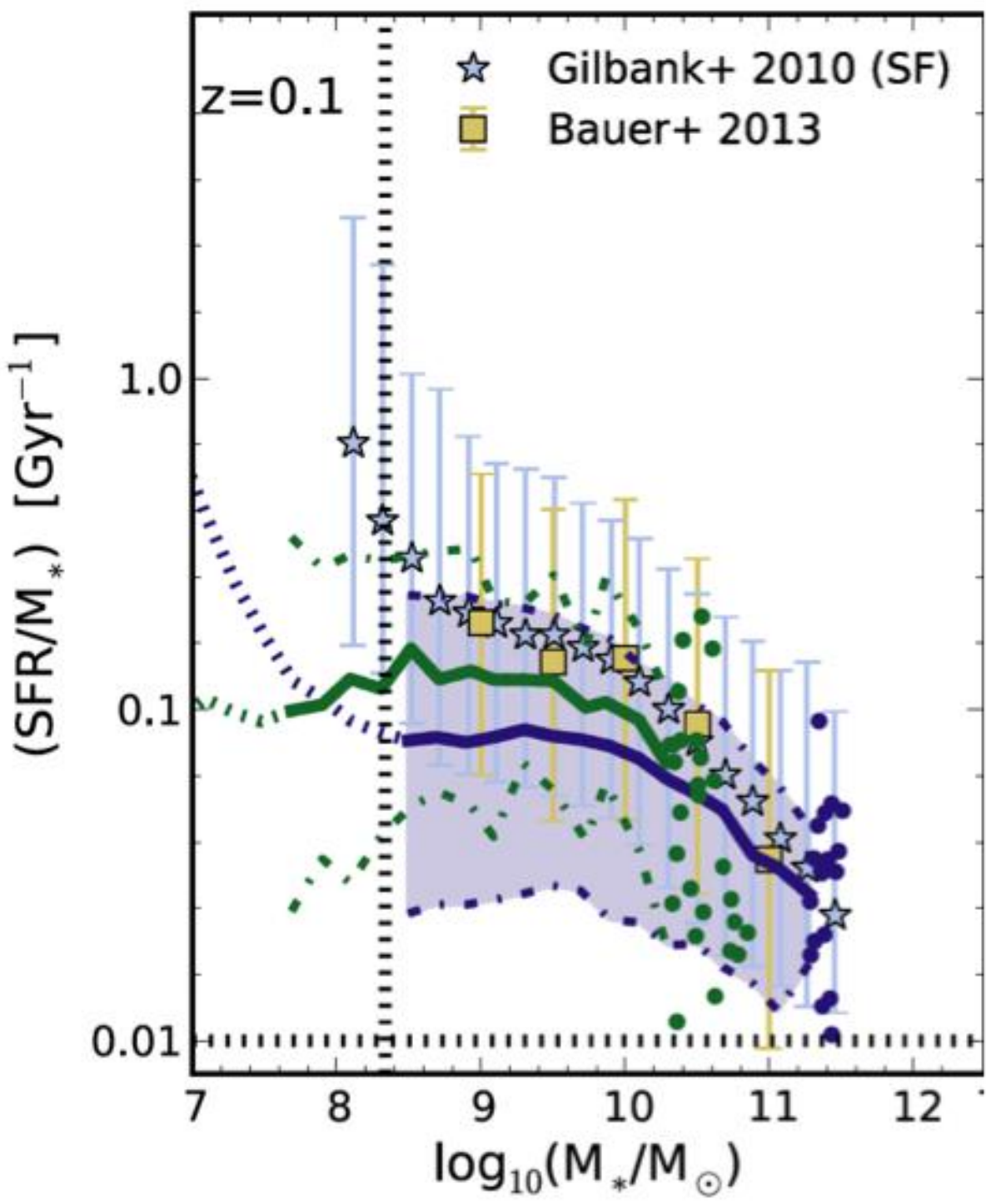

*Figure 89: The prediction of the specific SFR – mass relationship from the EAGLE hydrodynamic simulations, compared with observations at low redshift. The purple and green solid lines are low and high-resolution simulation results, respectively; the dashed lines indicate where resolution effects may be important. There is indication of a discrepancy with the SDSS Stripe 82 analysis of Gilbank et al. (2010) at the lowest stellar masses probed. Figure from Furlong et al. (2015).*

# Chapter 8

# Active Galactic Nuclei and Supermassive Black Holes

**Abstract**

MSE will probe the growth of supermassive black holes (SMBHs), and will characterize their relationship with host galaxies near and far, by measuring luminosity functions, clustering, outflows, variability and mergers. A multi-epoch reverberation mapping campaign with MSE will yield $2000-3000$ robust time lags of the quasar broad-line region over a broad range of redshift and luminosity. This is an order of magnitude more than the expected yields from current campaigns, and enables accurate SMBH mass measurements for the largest sample of quasars to date and unprecedented mapping of their central regions. MSE will provide large, statistical samples of growing SMBHs with sufficient areal coverage, depth, and temporal character to cover the AGN zoo at $z = 0-3$. It will also build a large sample of very high-$z$ ($z > 7.5$) quasars, and so probe the most distant SMBHs. MSE will simultaneously study the radiation environment close to growing SMBHs and the star formation histories of their host galaxies. MSE will provide better determination of the cosmological density of galaxies that host a binary SMBHs and will constrain the rate of SMBH mergers. Further, MSE will also allow us to better constrain how the cluster environment evolves from one that is conducive to the triggering of efficiently accreting AGN at high-$z$, to one that inhibits (efficient) AGN activity at low-$z$.

**Science Reference Observations** (appendices to the *Detailed Science Case, V1*):
**DSC – SRO – 08** Evolution of galaxies, halos, and structure over 12 Gyrs

**DSC – SRO – 09** The chemical evolution of galaxies and AGN over the past 10 billion years ($z < 2$)
**DSC – SRO – 11** Mapping the Inner Parsec of Quasars with MSE

## 8.1 Context

In conjunction with the stars, gas and dark matter, the central supermassive black hole (SMBH) is one of the fundamental components of galaxies. It has been known for sometime that these SMBHs seem to grow in lock-step with other galaxy components (Kormendy & Richstone, 1995), implying that they are a key component in the story of galaxy assembly and growth. However, there are many aspects of SMBHs that remain mysterious: how are they first seeded into galaxies? How do they grow? What are their size and mass distributions? Beyond their fundamental nature, SMBHs also have the potential to influence their host galaxies, and are thus an important ingredient in galaxy evolution models (e.g., Kormendy & Ho, 2013; Heckman & Best, 2014). Finally, accreting SMBHs shine as some of the most luminous objects in the universe, providing beacons for cosmological studies.

To describe the multi-faceted research that MSE will contribute to the broad field of AGN, we have divided the science presented here into hierarchical layers. First, we consider the SMBH itself (Section 8.2), then its role within the galaxy, and finally as a cosmological probe.

## 8.2 The central engine: supermassive black holes and accretion

### 8.2.1 How are SMBHs seeded in galaxies?

The mechanisms by which SMBHs form are still largely unknown. At high redshifts, the extremely rapid growth experienced by luminous quasars makes it difficult, if not impossible, to reconstruct the mass function of the initial seeds, or even shed light on their formation mechanisms (Volonteri & Gnedin, 2009). More direct clues can be found in the local Universe: largely unperturbed SMBH seeds are expected to persist to the present day in the nuclei of galaxies in which SMBH growth, be it by gas accretion or merging, is limited. This implies that one of the best diagnostics of 'seed' formation mechanisms is to measure the masses of SMBHs in local dwarf galaxies (van Wassenhove et al., 2010). As an example, Figure 90 (taken from Greene, 2012) shows how measuring the fraction of low-mass galaxies that contain SMBH with $M > 3 \times 10^5$ $M_\odot$ can help distinguish between two of the most credited formation mechanisms (Volonteri et al., 2008; Volonteri, 2010) – seeds originating from the death of Pop III stars (dashed green line), or from direct collapse of gas clouds though by gas-dynamical instabilities (solid magenta line).

MSE's contribution to this field is two-fold:

1. In quiescent galaxies, SMBH detections require stellar or gas dynamical studies based on high SNR observations that spatially resolve the SMBH "sphere of influence", i.e. the region of space within which the kinematics are dominated by the SMBH gravitational potential. In all but the most nearby dwarf galaxies, such observations can only be performed with AO-assisted IFU or long-slit spectroscopy at 30m-class facilities. For reference, the sphere of influence of the SMBH expected in NGC 205 (at $M_{gal} \sim 10^{10}$ $M_\odot$ one of the most massive galaxies of interest for these studies) is only 0.2 pc, barely resolvable by HST (Valluri et al., 2005). MSE, however, can assist in the target selection: medium resolution fiber spectroscopy of the nuclei of nearby dwarf galaxies can reveal the presence of emission lines [e.g., H$\alpha$, [NII]$\lambda$, $\lambda$6548, 6584, [SII]$\lambda$, $\lambda$6717, 6731) associated with gas that could potentially be in Keplerian rotation around a central SMBH (Ferrarese et al., 1996); the presence of a broad symmetric component in the forbidden lines could be further indication of rotational broadening. By targeting a large sample of such galaxies (thousands exist out to the distance of Virgo and Fornax), MSE can pre-select the most promising candidates for follow-up resolved spectroscopy.

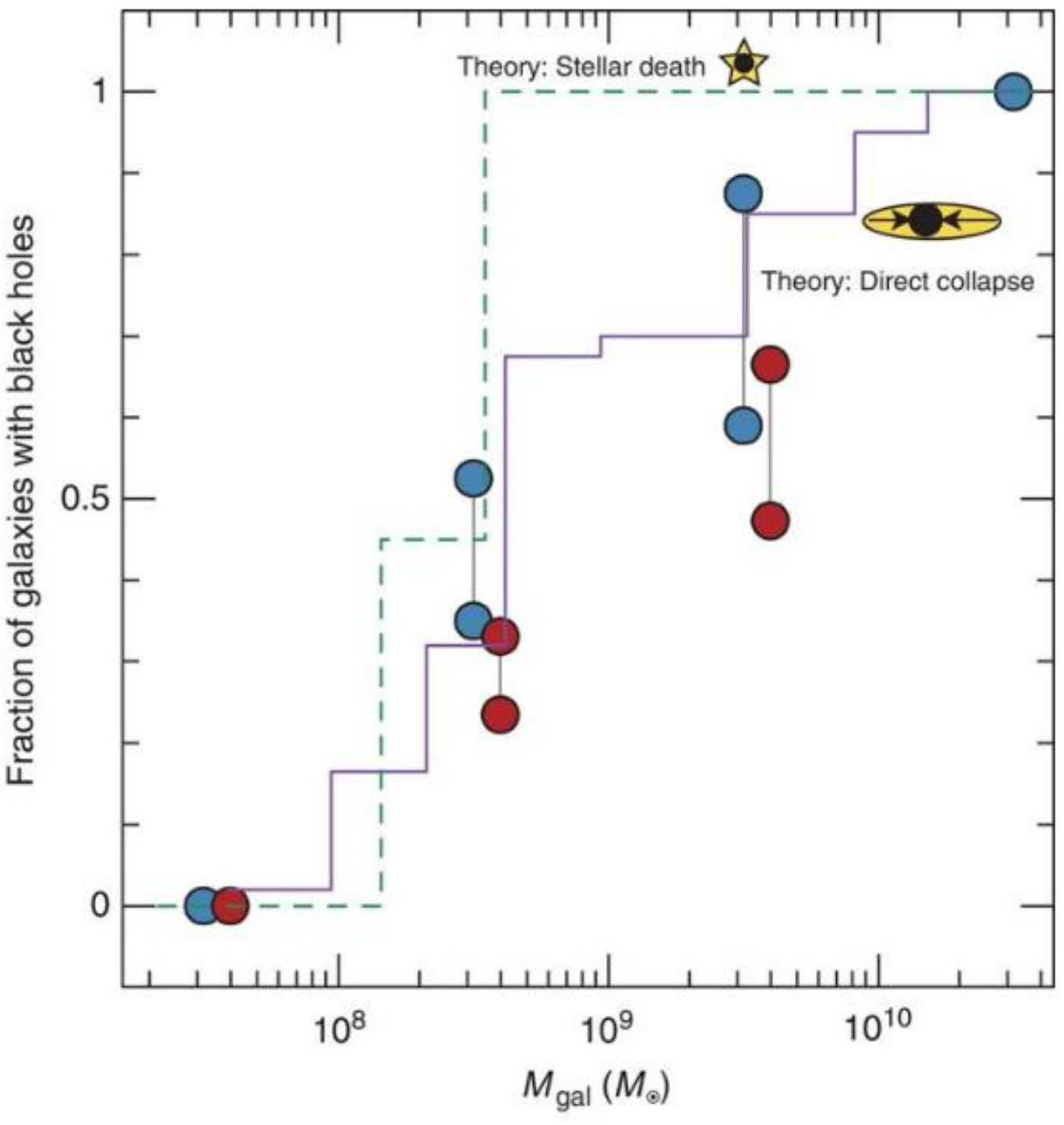

*Figure 90: Expected fraction of galaxies with $M_{gal} < 10^{10}$ $M_\odot$ that contain black holes with $M_{BH} > 3\times10^5$ $M_\odot$, for high efficiency massive seed formation (solid purple line), as well as stellar deaths (green dashed line). Figure from Greene et al. (2012).*

2. AGN samples can be selected from optical spectroscopy using the narrow-line [OIII]/H$\beta$ vs [NII]/H$\alpha$ diagram as a diagnostic. The same data then yields virial SMBH mass estimates by combining the width of the H$\beta$ emission line (under the assumption that the Broad Line Region is virialized) and the AGN continuum luminosity at $\lambda$5100 (which time resolved reverberation mapping studies show to be tightly correlated with the size of the Broad Line Region, e.g. Bentz et al., 2009). The methodology has been proven for dwarf galaxies out to $z \sim 0.055$ using SDSS data (Reines et al., 2013). However, while SDSS is sensitive to AGNs with bolometric luminosity $L_{bol} > 10^{42}$ erg s$^{-1}$ (translating into a SMBH mass of $\sim 10^5$ M$_\odot$ assuming 10 percent of the Eddington limit), MSE is expected to push this limit by a factor ~50, and therefore probe SMBH down to $\sim 10^{3.3}$ M$_\odot$. At this limit, the FWHM of the broad H$\beta$ emission is expected to be in the few hundred km s$^{-1}$ range, requiring MSE's medium resolution mode to resolve the lines. MSE follow-up observations (~5h integration assuming a central stellar surface brightness of $g \sim 25$ mag arcsec$^{-2}$) of those galaxies for which a SMBH mass can be estimated, will allow the placement of low mass galaxies in the M$-\sigma$ relation and place further constraints on the formation of SMBH seeds.

### 8.2.2 Measuring SMBH masses: Reverberation Mapping the Inner Regions of Quasars

At the present epoch SMBH are ubiquitous in the centres of massive galaxies, but they grew as luminous quasars when the Universe was a fraction of its present age. Though quasars have been studied in radio through X-ray wavelengths for decades, there remain fundamental, open questions about accretion disk physics, and about the geometry and kinematics of the region generating the broad emission lines. Quasar light is generated in a region a few light years across, and yet it can outshine the stars in its host galaxy by a thousand times. These distant, cosmic powerhouses have such small angular sizes that they cannot be resolved with existing or near-term technologies. Our only access to constraining their structure empirically is through time-domain astrophysics. A ground-breaking MSE campaign of ~100 observations of ~5000 quasars over a period of several years (totaling ~600 hours on-sky) would accurately measure SMBH masses of high-$z$ quasars and map their inner regions. In addition, a well-calibrated radius-luminosity relation for quasars would enable the construction of a high-$z$ Hubble diagram to constrain the expansion history of the Universe.

The powerful technique of reverberation mapping (RM) takes advantage of quasar variability to measure the physical sizes of line-emitting regions (e.g., Blandford & McKee, 1982). In particular, repeat spectroscopy enables measurements of the rest-frame time lag of the response of a broad emission line's flux to changes in the continuum to provide a characteristic distance, $R_{\rm line}$, between the broad-line region gas and the accretion disk. With this characteristic distance (from the time lag) and velocity (from the line width), the SMBH mass is derived (e.g., Peterson, 2011, and references therein). Furthermore, for an individual quasar, comparison of the characteristic time lags from different lines puts powerful constraints on the structure, kinematics, and physical conditions (e.g., gas density and ionization parameter) of the broad-line region gas (e.g., Korista & Goad, 2004).

The essential requirements for a successful RM campaign are to accurately measure the time lag between continuum variability and the response of each emission line, and to obtain a root-mean-square (RMS) flux spectrum of the quasar. The time-lags provide the size scale of the emission-line region, while the RMS spectrum shows the velocity structure of the material responding to the continuum variability. Both of these are required to map the structure of the broad-line region and to obtain accurate black-hole masses.

To detect low-amplitude, short rest-frame timescale continuum and emission-line variations requires high SNR spectra with accurate ($\leq$ 4%) relative flux calibration. A minimum resolution of 200 km s$^{-1}$ ($R \sim 1500$), and ideally of 100 km s$^{-1}$ ($R \sim 3000$), is required to adequately sample highly structured and often blended broad emission lines, and to resolve narrow-emission line widths of hundreds of km s$^{-1}$. Continuous wavelength coverage in the regions from 360 nm to 1.8 $\mu$m maximizes the science return for quasars up to $z \sim 3$, and in addition provides simultaneous coverage of CIV and H$\beta$ for $z = 1.3 - 2.7$ quasars, enabling $R_{\rm line}$ measurements from the innermost regions to the dust-sublimation radius. Furthermore, including rest-frame optical emission lines is essential for tying the typical high-redshift quasar to the extensively calibrated local reverberation-mapped AGN.

Repeat observations to monitor continuum and emission-line variability are at the heart of our proposed science program. The number of required spectroscopic epochs is based on the science goal of using time delay versus velocity information to construct images of the broad-line region. To sample both short lags (days) and long lags (years), we envision a multi-year program with a cadence of several days in the first year, and reduced cadence in each successive year, totaling ~ 100 epochs per field over 3 – 5 years (Horne et al., 2004). Based on experience with previous RM campaigns, we anticipate that such a program will yield 2000 to 3000 robust time lags. This is an order of magnitude more than the expected yields from current multi-object spectroscopic RM programs (Shen et al., 2015b). Such broad-band spectral coverage with the proposed time cadence would enable accurate black hole mass measurements for the largest sample of quasars to date and unprecedented mapping of the central regions.

Though the demographics of the quasar population have changed remarkably since $z \sim 3$, the structure of luminous quasars shows surprisingly little evolution in fundamentals such as metallicity and spectral energy distribution. They are thus promising objects for constructing a high-$z$ Hubble diagram given an appropriate independent estimate of luminosity such as a well-calibrated $R_{\rm line} - L_{\rm cont}$ relation (e.g., Bentz et al., 2013). The size of a line-emitting region can be measured from reverberation; this then yields the average quasar luminosity. From the measured flux and the redshift, a Hubble diagram to $z \sim 3$ can be made from the proposed MSE high-$z$ quasar reverberation-mapping campaign, thus providing important constraints on general cosmological models (e.g., King et al., 2014).

### 8.2.3 How do SMBHs grow? AGN triggering mechanisms and different fuelling/accretion modes

A key unsolved puzzle in extragalactic astronomy is how SMBHs grow. Random infall of gas and stars, major and minor mergers, and cluster environments can send fuel toward central super-massive black holes transforming the center of galaxies into AGN. The relative significance of these AGN triggering mechanisms is yet unclear. To help solve this puzzle, MSE can provide spectroscopic, optical to NIR observations of large, statistical samples of growing SMBHs with sufficient areal coverage, depth, and temporal character to cover the AGN zoo at $z = 0 - 3$. MSE will enable the simultaneous study of the radiation environment close to the growing SMBHs and the star formation histories of their host galaxies.

**Major and minor mergers:** Gravitational interactions between gas rich galaxies and associated starburst activity stir up the gas and dust near the black hole, triggering quasar activity (e.g. Sanders et al., 1988; Hopkins et al., 2008). The quasar emerges only when its winds and UV emission clear the obscuring debris from the nuclear region. Hopkins et al. (2016) show that in some cases winds can clear out material while creating an obscuring geometry similar to that of a clumpy torus. This theory predicts that (1) QSO2s (narrow-line QSOs) and QSO1s (broad-line QSOs) have different Eddington ratios and (2) the hosts QSO2s have younger host galaxies and higher star-formation rates (SFR) than QSO1s. MSE will allow us to measure, stellar masses (from H band and stellar population modeling), star-formation histories, and star-forming properties to systematically study the host galaxies of QSO1s and QSO2s.

In recent years deep-wide imaging surveys such as the Kilo-Degree Survey (de Jong et al., 2013) and DECaLS (Blum et al., 2016) have allowed in-depth studies of the role of mergers in galaxy evolution (Kaviraj et al., 2014). Over the next decade projects such as UNIONS, Subaru/HSC and LSST will probe even deeper ($\mu_r \lesssim 30\,\mathrm{mag\,arcsec^{-2}}$). Such deep imaging surveys will naturally provide targets for follow up spectroscopy with MSE, allowing detailed study of their role in the AGN phenomenon at earlier epochs. MSE's simultaneous optical to NIR spectral coverage will give access to emission and absorption lines from which to measure the impact of dust extinction on the observed line emission, and have access to excitation/ionization and kinematic diagnostics needed to relate the interactional history of the host galaxy to (e.g. van Dokkum & Conroy, 2010; Heavens et al., 2004; Schiavon, 2007).

**AGN triggering in cluster environments:** X-ray studies of AGN suggest that the AGN fraction in clusters is significantly rising as a function of redshift (e.g. Fassbender et al., 2012). With MSE, we will be able to study galaxy clusters spanning a wide range of mass/red-shift/richness range selected from SZ-radio/millimeter samples. MSE will thus be able to trace the number of AGN as a function of cluster properties (richness, mass, density profile). MSE will allow also us to better constrain how the cluster environment evolves from one which is conducive to the triggering of efficiently accreting AGN at high-z (Krishnan et al., 2017), to one that inhibits (efficient) AGN activity at low-z (e.g., Haines et al. 2012; Pimbblet et al. 2013).

At at $1 < z < 2$, $\sim 20\%$ of massive galaxies host AGN (based on photometric data from COSMOS; Wang et al., 2017), and $5 - 10\%$ of galaxies are involved in mergers (Conselice, 2006). This equates to $\sim 960$ AGN and up to $\sim 480$ merging systems per square degree at $1 < z < 2$ that are observable by MSE. To select higher redshift mergers, we will use sample of MIR selected AGN from wide field studies with Spitzer and WISE (Lacy et al., 2007, 2015, e.g.). To study the role of mergers in AGN triggering at cosmically relevant epochs, we will need to survey an area of the order a hundred square degrees to a limiting magnitude of $i \sim 25$. The list of spectral features required for this analysis include Hydrogen recombination lines:;Mg lines at 517, 880, and 1574 nm; Ca at 422.7, 849.8, 854.2,866.2, and 1620.5 nm; O at 630 nm; Ba at 649.9 nm; Al at 1671.9; Li at 670.7 nm; Na at 818.5,819.3, and 1639.5 nm; molecular features due to CH at 430.0 nm; CN at 417, 800 nm, and 1,100 nm; MgH at 517.0 nm; C2 at 466.8 nm; Fe lines across the optical and NIR.

### 8.2.4 Measuring accretion in real-time: extreme variability, tidal disruption events and changing-look AGN

Extreme variations (e.g., magnitude changes > 1 mag) of quasars over multi-year timescales are observationally rare events, and defy our current understanding of steady-state accretion in AGN (Graham et al., 2017; Rumbaugh et al., 2018). In some cases, spectroscopy at the bright and faint states reveals dramatically different emission line properties. This is referred to as the changing-look phenomenon[1] that blurs the traditional division between type-1 and type-2 AGN (e.g., LaMassa et al., 2015; Runnoe et al., 2016). MSE will work in synergy with future time domain imaging surveys to identify and monitor large samples of AGN. LAMOST has successfully employed this technique (Yang et al., 2018) to find 21 new changing look AGN. Deep MSE spectroscopy for candidate changing-look AGN selected from pre-imaging surveys, or from spectroscopic monitoring of the MSE reverberation mapping sample, will confirm a large number of spectroscopic changing-look AGN given the more than decade-long pre-MSE light curves for large samples of quasars from existing surveys. The new MSE spectrum, when combined with earlier spectra, can be used to study the responses in the broad-emission lines to the extreme continuum variations over multi-year timescales, providing further constraints on the properties of the broad-line region. In addition, deep MSE spectra taken at the faint state can be used to measure host galaxy properties, and can be correlated with the black hole mass measured from the spectrum taken at the bright state. This will allow the study of the SMBH mass and host scaling relations in broad-line AGN up to $z \sim 1.5$.

Dramatic absorption variability events (Hall et al., 2011; Rafiee et al., 2016; Stern et al., 2017) are another observationally rare phenomena. These are also cases of brightening by a magnitude or more at rest wavelengths < 3000 Å. However, they are caused by extremely strong low-ionization broad absorption line troughs nearly disappearing, leaving behind quasars with relatively weak absorption troughs; the reverse has also been seen to happen. Only in a few cases does the underlying continuum flux level appear to rise as the absorption disappears, as expected if an increase in emission from the inner disk is responsible for ionizing away the absorbing gas (He et al., 2017). The origin of these objects may instead lie in rapid changes in the geometry of shielding gas that exposes our line of sight to an increase in extreme-UV ionizing photons without a large increase in longer-wavelength photons.

Tidal disruption events (TDEs) are cases where a star wanders too close to a supermassive black hole and is torn apart by the gravity differential across the diameter of the star (e.g., Piran et al., 2015). As some of the stellar matter spirals into the black hole, it forms a short-lived accretion disk that can in some cases launch jets detectable by their radio emission. TDEs are more likely to occur when the black hole mass is relatively low, due to the stronger tidal force of lower-mass black holes. Thus, TDEs offer a way to probe intermediate-mass black holes, including quiescent ones.

Targeting transient events like TDEs and extreme-variability quasars will not drive the choice of MSE observing locations on the sky, but can drive targeting of single fibers within those chosen MSE observing locations. Observation planning that integrates alerts from LSST and other time-domain imaging surveys and enables the targeting of the most interesting known transient in each MSE field will enable excellent science at the cost of only a single fiber per pointing.

[1] The term "Changing-look" was first introduced in observations of AGN X-ray variability (Matt et al., 2003; Puccetti et al., 2006; Bianchi et al., 2009; Risaliti et al., 2009; Marchese et al., 2012).

#### 8.2.5 Binary SMBHs

In addition to growth through accretion, SMBHs can experience rapid growth when they coalesce with another SMBH in the late stages of a major merger. Such systems are additionally important for finding prime targets for low-frequency gravitational wave signatures by space-based observatory, such as LISA. Since SMBH mergers are very rare, the key to a successful observational program is in large number statistics and strategic follow-up. MSE will provide a large and deep sample of quasars with measured emission line properties, with accompanying spectroscopic monitoring of a $\sim$ 5000 quasars over a period of several years with $\sim$ 100 observations for each quasar. This will provide the largest homogeneous sample of quasar spectral light curves with high cadence, an ideal dataset from which to detect new candidate SMBH systems. We discuss below two science cases for binary SMBH detection.

**Unresolved binaries:** Direct imaging of binary SMBHs on (kilo)parsec scales can only be carried out in the local universe (see e.g. Komossa, 2006; Popović, 2012). For SMBH binary detection on sub-pc and those which are at larger cosmological redshifts, spectroscopic methods can be applied. The SMBH binaries on sub-pc scales can be detected through radial velocity curves of complex line-profiles of strong emission lines or through asymmetries and shifts in line profiles (e.g. Popović et al., 2000; Shen & Loeb, 2010; Tsalmantza et al., 2011; Popović, 2012). Additionally, one can expect that in a relatively long period, oscillation patterns can be detected in the continuum and broad-line light curves (e.g., Kovačević et al., 2018).

The best candidates, showing opposite motions of red and blue peaks are NGC 4151 and NGC 5548, monitored for several decades (e.g., Bon et al., 2012; Li et al., 2016). Approximately 100 other candidates have been identified and monitored (Runnoe et al., 2017; Guo et al., 2018). The line profile velocity shift is larger and easier to detect in large mass binaries, however they have longer dynamical time-scale, whereas small mass binaries may have ordinary line profiles, typical for a single quasar (Simić & Popović, 2016), but spectral monitoring could reveal and confirm their binary nature (Figure 91). In addition, the velocity field from RM data can reveal sub-pc binary SMBH if homogeneous, high-cadence, reasonable spectral resolution, and well calibrated spectral data are available (Wang et al., 2018). The estimated probability to find a sub-pc SMBH binary is of the order of few percents (Shen et al., 2013; Pflueger et al., 2018) but the detection rate can be as high as ~30% (Guo et al., 2018). Therefore, dedicated spectroscopic monitoring with MSE will provide radial velocity curves with enough cadence to sample the orbit well, exclude the red-noise (i.e., more power at longer timescales) variability, test more complex physical models, and detect more sub-pc SMBH binaries.

**Resolved binaries:** Dual quasars with separations of ~kpc have been found to be common (for reviews, see Komossa, 2006; Popović, 2012). However, only a handful of unambiguous cases of dual quasars are confirmed with multiwavelength observations (e.g., Liu et al., 2013). They are detected through spectroscopic surveys, like SDSS, which reveal objects with two-component narrow line profiles (e.g., Wang et al., 2009b; Comerford et al., 2009; Smith et al., 2010; Liu et al., 2010) or with higher angular resolution observations with space-based observatories and very long baseline radio interferometers (e.g., Liu et al., 2013). The estimated fraction of low-redshift dual quasars is a few percent in the nearby Universe and ~ 0.1% at intermediate-redshifts. MSE will enable the identification of large numbers of galaxies with spectroscopic signatures of binary quasars on kpc and sub-kpc scales at different redshifts, including at high redshift thanks to the extensive spectral coverage of MSE. Thus, MSE will provide better determination of the cosmological density of galaxies that host a binary SMBHs and will constrain the rate of galaxy mergers. We note that complex broad line profiles as well as double narrow line profiles can be caused by other effects (e.g., the bi-conical motion in the narrow line region) and other spectral characteristics should be explored to confirm the duality in high-redshift quasars.

The moderate resolution capabilities of MSE are required across the optical and NIR wavelength range, with typical SNR of ~ 20 near H$\alpha$. This will provide reliable spectral measurements and estimations of line profiles, radial velocity shift and asymmetries with a precision of $50-70$ km s$^{-1}$). This translates to a limiting magnitude of $i \sim 23.5$ assuming 1hr exposure with typical observing conditions.

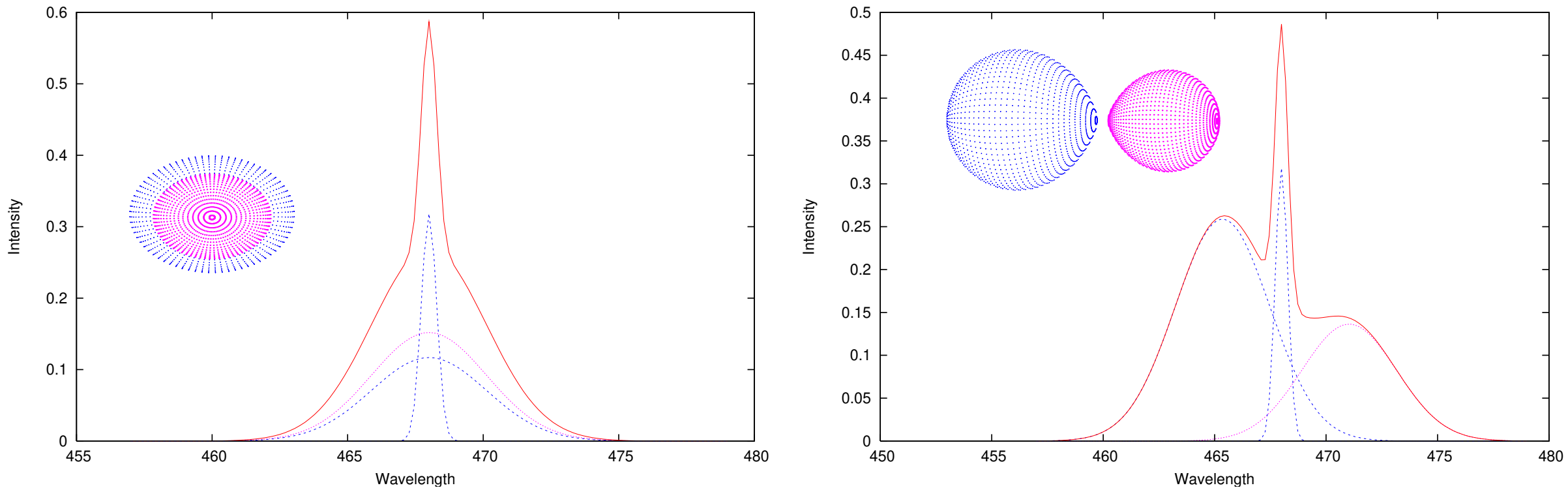

*Figure 91: Modeled H$\beta$ line profiles in the case of the SMBH binary (and binary broad line region) in different phases for the case where the broad line profile is symmetric (left) and where the broad line profile shows a large asymmetry (right). The central narrow component is assumed to be emitted from the narrow line region. Figures from Popović et al. (2000); Popović (2012).*

#### 8.2.6 Outflows from SMBHs as traced through BALs and intrinsic NALs

A substantial fraction of all AGN show absorption signatures in their spectra from gas associated with the central engine. These absorption systems span intrinsic narrow absorption-line systems (NALs) to broad absorption line systems (BALs).

BALs are seen in ~15% of optically selected quasars at redshifts $z > 1.65$, though the true percentage is higher after correcting for selection effects (e.g., Allen et al. 2011). Outflows separated by over 10,000 $\mathrm{km\,s^{-1}}$ have equivalent widths (EWs) that strengthen or weaken together much more than they vary in opposite directions (e.g., Filiz Ak et al. 2012). We now know that the main driver of BAL variability is ionization state changes in the outflows in response to changing flux levels of the ionizing source (Wang et al., 2015; He et al., 2017).

BALs and NALs must be launched in wind(s) from the disk and/or torus. As such, they offer a probe of the accretion physics near black holes at the centers of distant galaxies. Also, the masses and distances of BAL and NAL outflows are potentially of great significance for understanding quasar feedback on galaxy formation. Up to half of BALs may be located more than 100 pc from the black hole (Arav et al., 2018), though the average distance may be closer to 25 pc (Hamann et al. 2019).

What science can be done with tens of thousands of MSE BAL and NAL quasar spectra?

- Link AGN and BAL variability (e.g., He et al. 2015), especially in MSE-RM fields (and LSST-MSE overlap). Note that better spectrophotometric calibration produces cleaner results.

- Study physical trends of BAL outflows and their host quasars as a function of key observables and physical parameters by using composite spectra (e.g., Hamann et al. arXiv:1810.03686).

- Study BAL variability in MSE-RM fields, especially down to short timescales (e.g., Hemler et al. arXiv:1811.00010). What fraction of BAL troughs show short-term variability? How well can we constrain the structure and properties of the absorbers from their time variability?

- MSE will extend the time baseline for BAL quasar spectroscopy into the 2030s, enabling a statistical search for the deceleration expected as outflows sweep up their host galaxy's ISM.

- Predict BAL (re-)appearances. BAL host quasars appear to be those which have weaker He II 1640 emission, although that line can be difficult to measure in individual spectra.

## 8.3 AGN host galaxies

### 8.3.1 AGN host galaxies and type-2 AGN

For about two decades there have been two leading theories for the relationship between broad-line (AGN1) and narrow-line (AGN2) AGNs. The orientation theory (Antonucci, 1993) postulates that QSO1 and QSO2 objects both contain a dusty torus, but are viewed at different angles. Elitzur (2012) proposed a more sophisticated version of this: all AGNs may indeed have a torus, this torus is clumpy, and both the viewing geometry and the number and properties of clumps along the line of sight dictate the observable characteristics of the majority of AGN. This is supported by spectro-polarimetric observations of AGN2s (e.g. Zakamska et al., 2005), where the spectra in scattered light reveal the broad-line emission, the defining characteristic of AGN1s, suggesting that the broad-line region is obscured from line-of-sight in AGN2s by the dusty torus. A complementary evolutionary theory centers on gravitational interactions between gas rich galaxies (e.g. Sanders et al., 1988; Hopkins et al., 2008).

The merger and associated starburst activity stir up the gas and dust near the black hole, triggering quasar activity. The quasar emerges only when its winds and UV emission clear the obscuring debris from the nuclear region. Hopkins et al. (2016) show that in some cases winds can clear out material while creating an obscuring geometry similar to that of a clumpy torus. This theory predicts that (1) QSO2s and QSO1s have different Eddington ratios and (2) the hosts QSO2s have younger host galaxies and higher SFRs than QSO1s. Observations in the far-infrared that trace the total cold ISM content of their host galaxies, and can be used together with radio observations to estimate star-formation rates, suggests that at least for optically luminous, nearby samples, AGN2s appear to have higher star-formation rates (Petric et al., 2015; Zakamska et al., 2016). However the implications of the FIR results are uncertain: a full decomposition of the sources of IR emission (Lani et al., 2017) is needed, and fits using NIR and optical spectroscopy to estimate the ages of the stellar populations in the host galaxies are necessary. MSE will allow us to measure, stellar masses (from H-band and stellar population modeling), star formation histories, and star-forming properties to systematically study the host galaxies of QSO1s and QSO2s.

Demographic studies of AGN populations show that the number density of MIR-selected AGN2s peak at a higher redshift than that of unobscured counterparts $z \sim 2-3$ (Lacy et al., 2015; Mauduit et al., 2012). Lacy et al. (2015) suggest that there are evolutionary differences between obscured and unobscured sources, and speculate that these may be driven by the increased frequency of major mergers of gas rich galaxies at high redshift (see Figure 92). Lacy et al. (2015) reach those conclusions from optical and NIR spectroscopic follow up of MIR-selected AGN from wide-field micro-Jansky level surveys with Spitzer IRAC. MSE can push the sensitivity limit to probe the higher redshift regime of $z \sim 3-4$.

Amarantidis et al. (2019) studied a wide range of cosmological hydrodynamical simulations and semi-analytic models to estimate the frequency of AGN that are growing via QSO mode (starburst drive, typically Eddington ratios $> 0.01$), radio mode (hot-halo mode driven; Eddington ratio of $< 0.01$) and super-Eddington (typically associated to QSOs as well). The largest differences in the predictions seem to be concentrated in the range $z = 2-4$. Most models invoke AGN as the primary way of quenching star formation in massive galaxies, and predict similar stellar mass functions and stellar mass densities versus redshift. This suggest a degeneracy between our understanding of AGN and the effect on massive stars. Detailed studies of gas metallicities, ionization conditions, and stellar populations in the hosts of AGN can help break this degeneracy.

It is notoriously difficult to measure host properties (such as host luminosity, stellar mass, and stellar velocity dispersion) in broad-line quasars, where the nuclear light can easily overwhelm the host light. However, with sufficient signal-to-noise ratio, it is possible to decompose the spectrum of a broad-line quasar into nuclear and host light (e.g., Vanden Berk et al., 2006), and measure host properties (such as stellar population and velocity dispersion) from the decomposed host spectrum. This exercise has been demonstrated with the deep, co-added spectroscopy from the SDSS-RM project (Shen et al., 2015a; Matsuoka et al., 2015), where the $\sim 60$ hr total exposure time provided sufficient SNR to apply this spectral decomposition technique for quasars up to $z \sim 1$. With the reverberation mapping program within MSE, it is possible to measure host properties from decomposed spectroscopy for quasars that are at least one magnitude fainter than those monitored by SDSS-RM. MSE's near-infrared coverage will also extend the redshift range of host measurements in unobscured broad-line quasars.

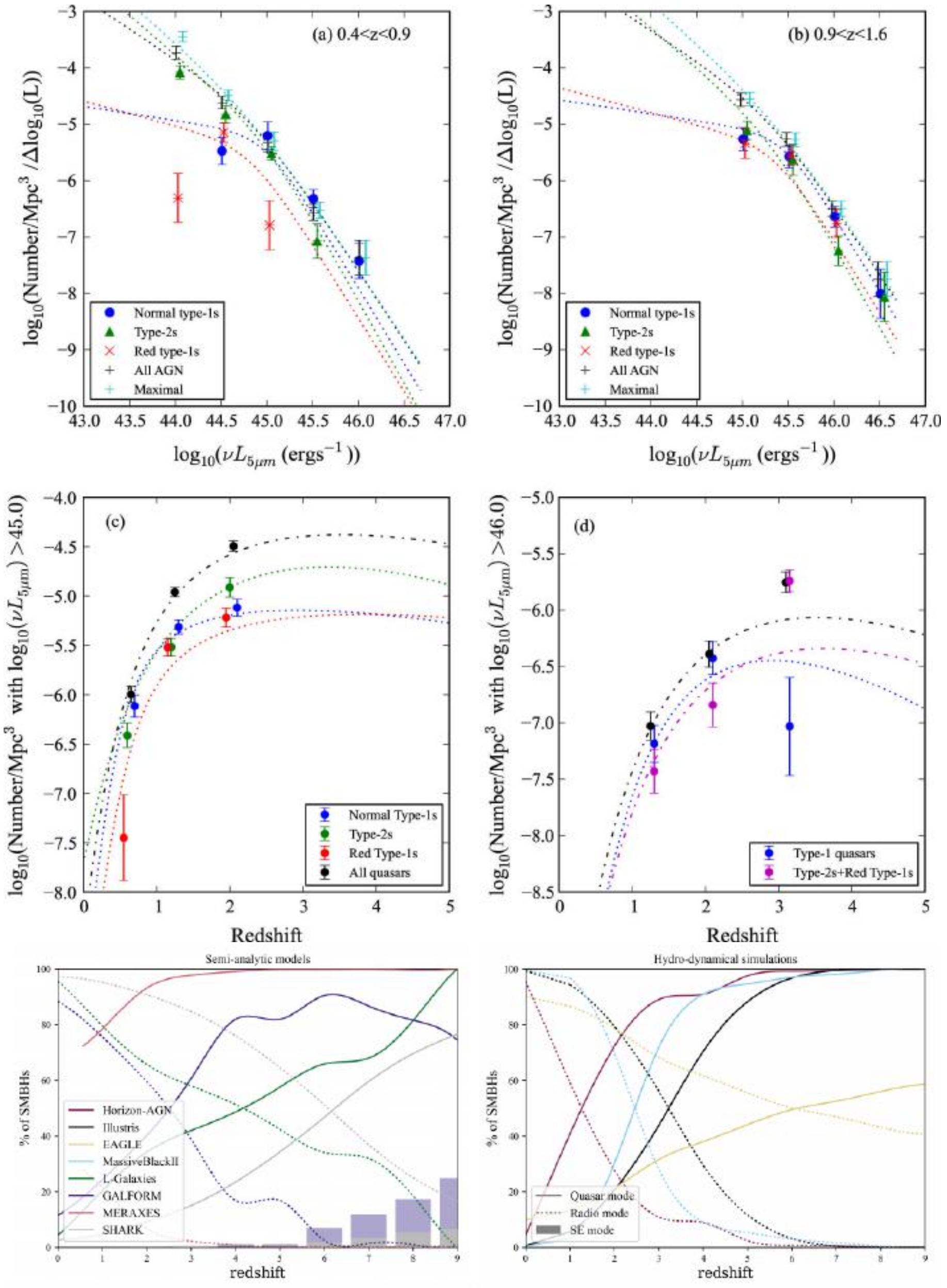

*Figure 92: Top: the evolution by type of AGN from a survey of MIR-selected AGN. Figure from Lacy et al. (2015). Bottom: the frequency of AGN growing via several modes in various cosmological semi-analytic models of galaxy formation (left) and hydrodynamical simulations (right). Figure from Amarantidis et al. (2019).*

### 8.3.2 From the epoch of re-ionization – high-$z$ AGN and their hosts

High-redshift ($z \geq 6$) quasars represent powerful probes of the structure formation during the epoch of the cosmic reionization. These objects are in fact bright enough to be studied in detail up to very high redshifts ($z \geq 10$), while the characterization of normal galaxies at similar redshifts is practically unfeasible by current facilities due their intrinsic faintness. Deep, medium resolution observations of the most distant quasars can provide a comprehensive view of their central engine and of their surrounding environment, allowing us to address a wide range of questions, such as the birth and growth of the first SMBHs, the early co-evolution of BHs and galaxies, the history of cosmic reionization, etc.

**MSE survey of high-$z$ quasars:** In the past 20 years, more than 200 quasars at $z > 5.7$ have been discovered (e.g. Fan et al., 2001, 2006; Jiang et al., 2008, 2016; Willott et al., 2007, 2010; Mortlock et al., 2011; Bañados et al., 2018; Matsuoka et al., 2016). They have played a key role in our understanding of the early quasar population, quasar host galaxies, SMBH formation, and cosmic reionization. However, current studies are still limited by the scarcity of very high-$z$ ($z > 7$) sources and low-luminosity sources. MSE will allow us to identify large samples of quasars, breaking the existing barriers both in redshift and luminosity, and allow us to really characterize the quasar population in the early Universe.

**The earliest SMBH formation:** How the early luminous quasars formed and evolved is still a mystery. The analysis of the optical and NIR spectra of the highest-$z$ luminous quasars known to date ($z \leq 7.5$) has shown that they harbour SMBHs with masses $\geq 10^9\ M_\odot$ when the Universe is less than one Gyr old (e.g. Mortlock et al., 2011; De Rosa et al., 2014; Mazzucchelli et al., 2017). How black holes can grow their substantial masses in the short time available poses a challenge to the current theories of formation and evolution of SMBHs. MSE will help us build a large sample of very high-$z$ ($z > 7.5$) quasars (owing to its NIR capability), and probe the most distant SMBHs. In particular, it will likely solve the mystery, because the time available to grow SMBHs is getting even shorter towards higher redshift.

**The quasar population in the early epoch:** The spectra and SEDs of high-$z$ quasars show little to no evolution from $z \sim 0$ to 7.5 (e.g. Jiang et al., 2007; Kurk et al., 2007; De Rosa et al., 2011; Mazzucchelli et al., 2017; Shen et al., 2018). How the AGN structure assembled from the highly structured environment of the early galaxies and how the metal content built up so rapidly remains a puzzle. With the large sample of $z > 7.5$ quasars from MSE, we will probe a cosmic time during which the SMBH in the brightest and most massive quasars are in their major growth phase. On the other end, MSE will allow us to build statistically significant samples that extend to lower and more typical luminosities at $z > 6$, in order to probe the structure of objects that are representative of the quasar population at these early cosmic times.

**Reionization history:** Detection of Gunn-Peterson absorption troughs in the absorption spectra of $z > 6$ quasars (e.g. Fan et al., 2006)and recent Cosmic Microwave Background polarization measurements by the Planck mission (Planck Collaboration et al., 2016b) constrain the peak of reionization to $6 < z < 10$. How the reionization unraveled and the nature of the sources that played a major role in the process are still big unknowns. The key observable is the Ly$\alpha$ emission line. At $z > 7$ this requires near-IR spectra covering up to 1700 nm to allow an accurate modeling of the intrinsic spectrum. MSE will help us build a larger sample of bright high-$z$ ($z > 7$) quasars, allowing us to obtain improved measurement of the Hydrogen neutral fraction along several lines of sight.

### 8.3.3 AGN feedback

AGN feedback is the prime candidate to provide the energy required to cease star formation in massive galaxies and produce the steep slope at the high-mass end in the stellar mass function of galaxies (Croton et al., 2006; Bower et al., 2006; Heckman & Best, 2014), and to establish the $M_{BH} - \sigma$ relation (e.g., King & Pounds, 2015). The reason behind this is that plenty of energy is available from accreting SMBHs: one solar mass of gas which is accreted by a black hole produces $10,000$ times more energy than that released by supernova for the same mass of stars formed. AGN feedback therefore encompasses the general concept of the (self-regulating) process which links the energy released by the AGN to the surrounding gaseous medium, impacting on the evolution of the host galaxy. The energy injected by the AGN can therefore provide the mechanism to quench star formation and limit the SMBH's growth, either by preventing the cooling of gas or by expelling gas from the galaxy. The latter are generally seen as two broad, distinct mechanisms of AGN feedback, generally referred to as "radio-mode" and "QSO-mode", respectively. The QSO-mode feedback (also referred to as "radiative mode") is linked to the AGN bolometric luminosity and the capability of SMBHs of launching large scale outflows from the galaxy (e.g. Fabian 1999, 2012; Murray et al. 2005), while radio-mode feedback (also referred to as "jet mode") is expected to be more efficient in AGN that are capable of generating mechanical energy (i.e. in the form of jets) and acts as a maintenance mode (e.g., McNamara & Nulsen, 2012). Because of these properties, QSO-mode feedback is expected to be more prominent at the peak of AGN activity ($z \sim 2$; Croom et al. 2009), while radio-mode feedback should dominate at lower redshifts. With MSE, we will investigate these two types of AGN feedback and distinguish their respective effect on galaxies during the cosmic time around $1 \lesssim z \lesssim 3$, covering the relevant epoch for both phenomena.

**Impact of QSO-mode Feedback:** QSO-mode feedback operates through AGN-driven winds that eject and/or heat the gas of the surrounding host galaxy. In order to quantify the impact of the feedback, we will select AGN spanning a wide range of luminosities, and control for both host galaxy properties (stellar mass, star formation rates), and their environment. The latter is important to isolate AGN feedback from environment quenching (e.g., Peng et al., 2012). Reliable measurements of environment will require a dense spectroscopic coverage, high-completeness strategy. MSE spectra will reveal AGN-driven outflows from the profiles of emission lines, tracing the ionized gas (e.g., [OIII]; see Fig. 93) and/or blueshifted absorption lines, tracing the neutral gas (e.g., Na D). We will study the relationship between AGN luminosity, outflow velocities and outflow rates to understand the energy budget, and put the findings in context of host galaxies properties, focusing on specific star formation rates (SFR/$M_\star$), and comparing to control samples. H-band sensitivity allowa us to reach [OIII] out to $z = 2 - 2.5$, the peak epoch of star formation and black hole activity (Figure 94).

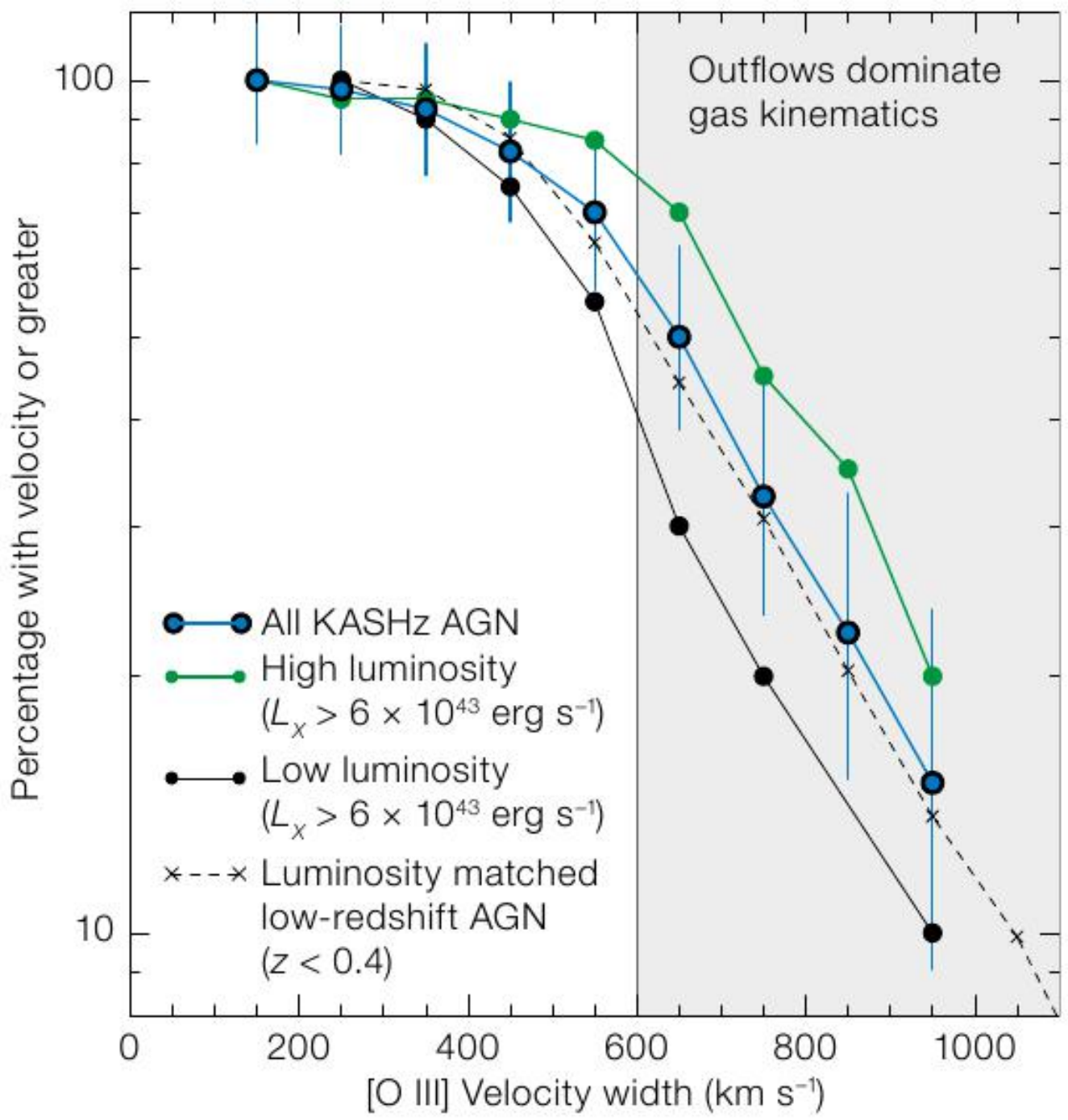

*Figure 93: Cumulative histogram of the [OIII] emission line velocity width for the KMOS AGN survey KASHz at* $1 \lesssim z \lesssim 1.7$*. Also shown are the subsamples of X-ray bright and faint AGN and a sample of low redshift AGN. If the [OIII] velocity widths exceed* $\sim 600\,\mathrm{km\,s^{-1}}$*, the kinematics of the gas is expected to be dominated by the outflowing component. Such a feature is a "smoking" gun of AGN-driven outflows and the QSO-mode feedback in action. Figure from Harrison et al. (2016).*

**Impact of Radio Mode Feedback:** Radio mode feedback is provided by Low Excitation Radio Galaxies (LERGs, Heckman & Best, 2014). LERGS are the most common radio galaxy at the current epoch, but are expected to decline at higher redshift (e.g., Amarantidis et al. 2019). We need to determine the space density over a range of radio power (i.e., luminosity function) of LERGS during cosmic noon in order to establish (1) the amount of kinetic power provided by radio jets and (2) the relative contribution to feedback from quasar mode and radio mode at this critical time for galaxy and AGN evolution. The interplay between MSE and the next generation of radio surveys will be key. We will select radio sources from the EMU and VLASS surveys for which photometric redshifts give $z > 0.8$. Another important resource is the JVLA surveys of the COSMOS regions (Smolčić et al., 2017). With MSE spectra, we will in turn determine host galaxy properties as well as signatures of neutral and ionized gas outflows from absorption, and emission line profiles.

**Measuring Black Hole Mass:** The SMBH mass is the product of the integrated accretion history (Soltan, 1982). But in quasar mode, the accretion rate is related to the energy released which can power feedback. So by looking at the distribution of BH mass as a function of epoch, we can see the evolution in the feedback over that time. For broad-line AGN the H$\beta$ and Mg II lines can be used to determine the mass of the SMBH (Wang et al., 2009a). Obtaining these masses for high-$z$ quasars will allow the evolution of black hole masses to be quantified when compared to low-$z$ observations and help constrain simulations. After $z \sim 1.5$, H$\beta$ will be beyond the reach of the low-resolution spectrograph on MSE, but Mg II can be used out to $z \sim 3.5$. Exploiting the H-band window will permit black hole masses to be obtained out to $z \sim 6$ using the magnesium line (Figure 94). Improved recipes for quasar black hole mass estimation will also be available from the MSE reverberation mapping program (see Section 8.2.2).

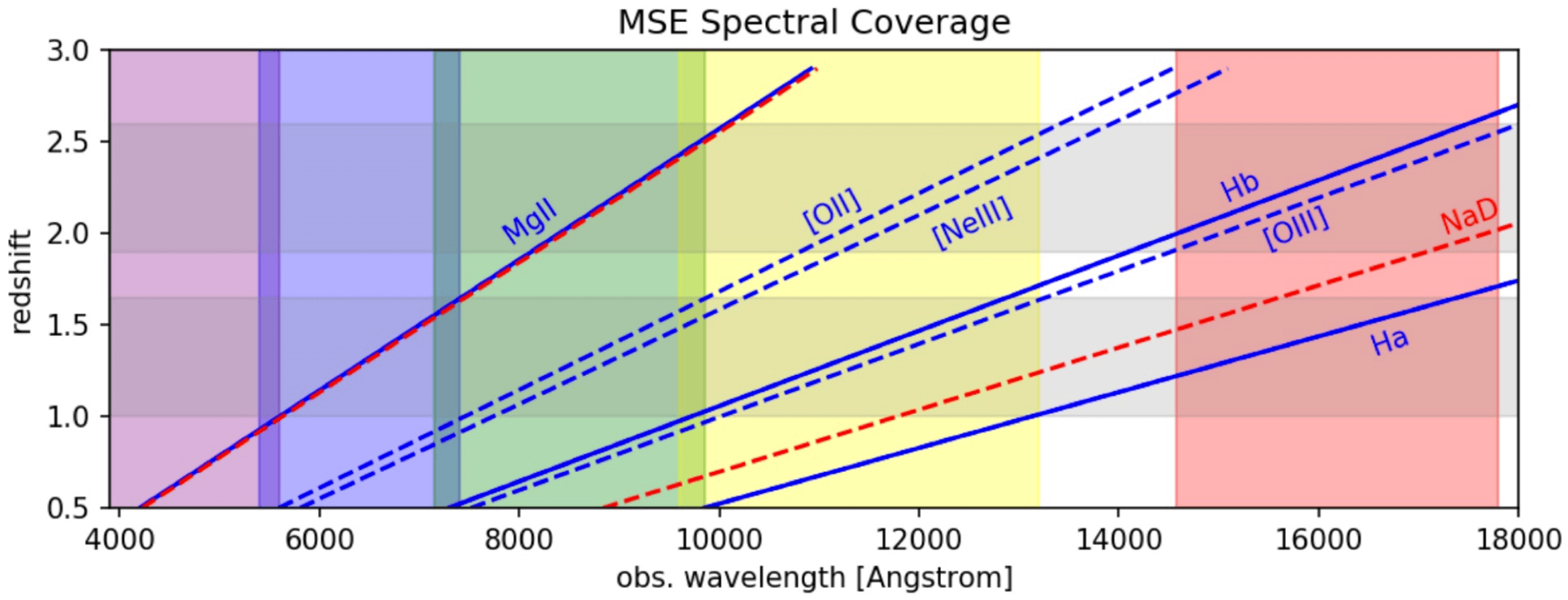

*Figure 94: Observability of spectral features as a function of redshift and observed wavelength for MSE. Vertical colored bands show the available spectral windows. Dashed blue lines mark narrow emission lines, solid blue lines mark emission lines that are potentially broad (in the case of Type 1 AGNs), and red dashed lines mark absorption lines.*

## 8.4 Beyond the host galaxy – cosmological aspects and applications of AGN and SMBHs

### 8.4.1 AGN clustering and demography

MSE will provide a spectroscopic quasar sample that will dwarf any earlier quasar samples in terms of the limiting magnitude and the sensitivity to measure emission line properties of low-luminosity AGN at high redshift from optical/near-IR spectroscopy. The latter measurements are necessary to derive physical quantities, such as black hole mass and Eddington ratios, for these distant quasars. Such a sample will enable the best measurements on the clustering and demography of these quasars at high redshifts, providing critical constraints on cosmological models of SMBHs. We highlight two science cases on quasar demographics and clustering below.

**Quasar demographics** The demography of quasars, usually measured in terms of the quasar luminosity function (LF, e.g., Hopkins et al., 2007), contains critical information about the evolution of the global SMBH accretion activity. In addition to luminosity, an equally important physical quantity of quasars is SMBH mass. SMBH mass is directly related to growth, and determines how efficiently the quasar is accreting mass (with an assumed radiative efficiency). Thus by jointly studying quasar demography in terms of luminosity (instantaneous accretion activity) and BH mass (integrated accretion history), one can obtain a more complete understanding of the evolution of quasars. The MSE quasar sample will enable the measurements of both the LF and the black hole mass function (BHMF), with SMBH masses measured from single-epoch spectroscopy (e.g., Shen, 2013), to the very faint end of the quasar population over a broad redshift range ($0 < z \lesssim 6$).

**Quasar clustering** The spatial clustering of quasars (e.g., the two-point correlation function) contains information about the host dark matter halos where these quasars reside in. Earlier measurements of quasar clustering have revealed the typical halo mass of quasars and constraints on their lifetime (or duty cycle) (e.g., Shen et al., 2007; Ross et al., 2009). Clustering measurements thus provides a critical piece to understanding the cosmic formation of SMBHs at galaxy centers. The MSE quasar sample can be used to measure quasar clustering at the very faint end of the LF and at high redshifts with unprecedented statistics, and to test models of quasar light curves with the luminosity (and mass) dependency of quasar clustering (right panel of Figure 95).

In a 1 hour exposure with typical observing conditions, we expect to reach a limiting magnitude of $i \sim 23.5$ (at a target density of $\sim 250 - 300\,\mathrm{deg}^{-2}$). This is $\sim 3.5$ mags fainter than SDSS-DR7 ($\sim 1$ mags fainter than DESI). To robustly detect a clustering signal in each bin of magnitude and redshift we require $> 1000$ pairs at separations $< 20\,h^{-1}$Mpc (White et al., 2012), which requires a minimal sky coverage of $\sim 5000\,\mathrm{deg}^2$ in order to measure the clustering at $M_i(z = 2) = -24$ at $z = 2$. Reducing the sky area reduces the pair counts linearly. Fixing the sky area but reducing the number of quasars reduces the pair counts quadratically. To boost the clustering SNR one could also cross-correlate with a much larger spectroscopic (from MSE) or photometric galaxy sample. The MSE quasar sample will supersede the DESI sample in terms of fainter limiting magnitude and the near-IR spectral coverage (providing more reliable single-epoch SMBH masses based on H$\beta$and Mg IIto higher redshifts than DESI.)

### 8.4.2 Gravitational lensing applications

**Quasar strong lensing** Even though Zwicky's prescient prediction in the 1930's was for gravitational lensing by galaxy clusters, the first gravitational lens discovered by Walsh et al. (1979) was the doubly lensed quasar, Q0957+561. The extreme intrinsic brightness of quasars ($\sim 4 \times 10^{12} L_\odot$) and their point source nature, make them easier to detect than the typically low surface brightness, extended lensed arcs of background galaxies (sources), formed by galaxy clusters or even massive field galaxies (lenses). With the advent of large all-sky photometric surveys, the predicted discovery numbers of lens candidates will skyrocket from the current few hundreds to the tens of thousands (Oguri & Marshall, 2010) in the era of MSE. This discovery rate would be augmented further by data science and machine learning based efficient gravitational lens finding techniques, e.g., *LensFlow* (Pourrahmani et al., 2018).

The purity and utility of these large lensed quasars datasets rely entirely on spectroscopic follow-up, for the confirmation and characterization of each lens system, especially the red-shifts of the lenses and sources, and the velocity dispersions of the lensing galaxies and clusters. Without this spectroscopic confirmation, a lens candidate is of little use. MSE will arguably be the only platform capable of effectively handling this data flood, and is designed to use its mulitplexing capabilities to simultaneously obtain the redshifts and all other characteristics of the lenses and sources.

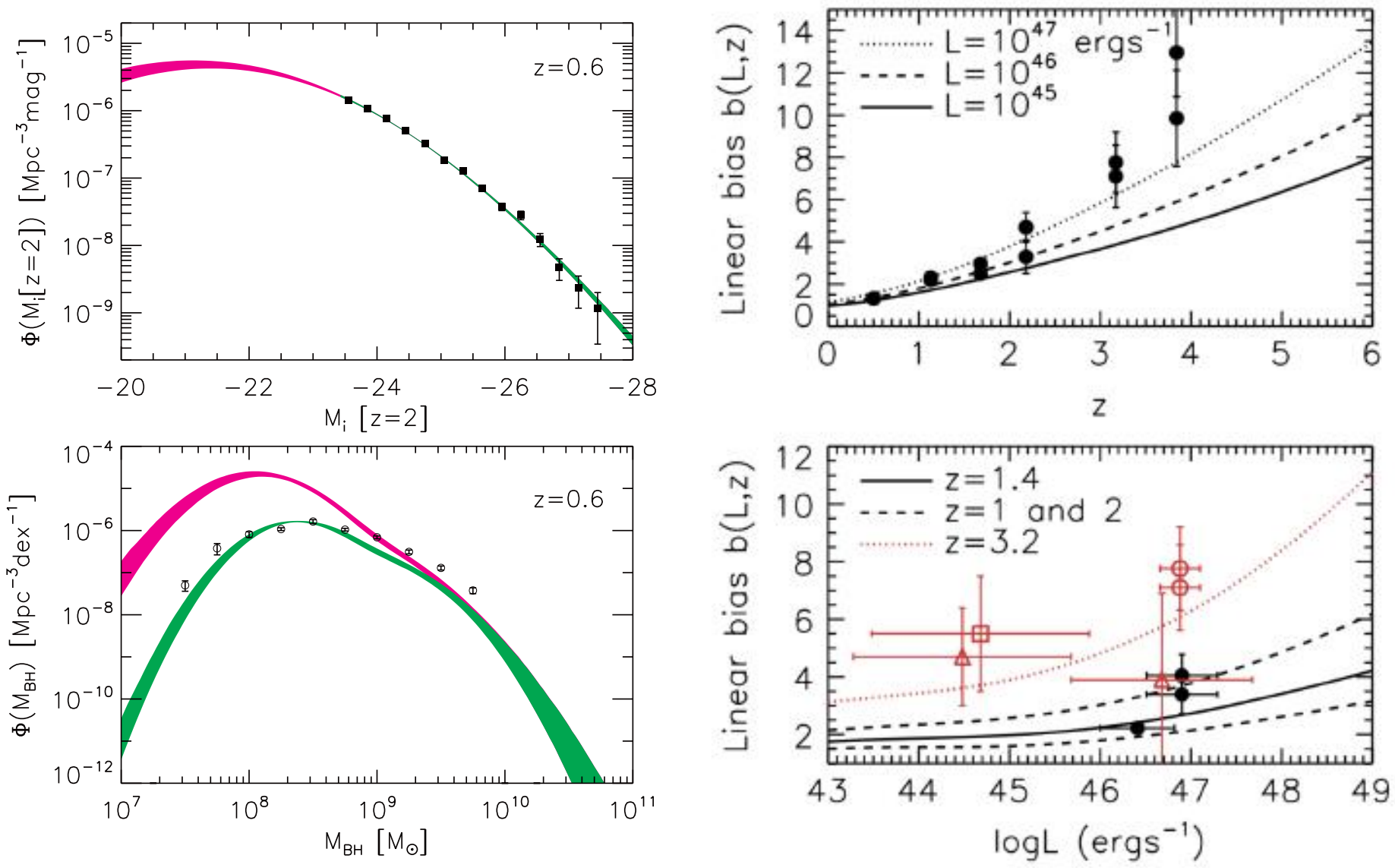

*Figure 95: Left: Quasar LF (top) and BHMF (bottom) at $z = 0.6$. The black points are data from the SDSS DR7 quasar sample and the shaded bands are models from Shen & Kelly (2012). The arrows indicate the approximate depth of MSE, which is several magnitudes fainter than SDSS. Right: The quasar linear bias (from clustering measurements) as functions of redshift and luminosity. The data points are from various surveys and the lines are the model in Shen (2009). The MSE quasar sample will much improve the measurement precision of high-z quasar clustering at faint luminosities, and differentiate quasar light curve models.*

There are multifold astronomical applications of these strong gravitational lenses, from determining $H_0$ in a model independent way (Refsdal, 1964; Treu & Marshall, 2016), detecting dark matter substructures in the lensing galaxies (Keeton & Zabludoff 2004; Kochanek 2006; see also Chapter 6), to studying the chemical make up of the inter-galactic medium (Smette et al., 1999). Here, we focus on only three of these many aspects, (i) the demography of lensed quasars, (ii) time domain observations of multiply imaged lensed quasars, and (iii) the unique capability of MSE to estimate the size of the accretion disk and the innermost stable circular orbit (ISCO) in the accretion disk around the central engines of these quasars.

**Lensed quasar demography:** The excellent image quality of MSE means that the typically multiply lensed quasar images (separation distances of $1-2$arcsecs), and the lensing galaxy can all be targeted individually. The total collecting area of MSE coupled with the low-to-moderate spectral resolution modes extending from the optical to the NIR, will permit us to obtain high SNR spectra of these lensed quasars, whose images, even with the lensing boost, are normally low luminosity due to their high redshift. The key advantage offered by these multi-fiber observations of the multiply imaged lensed quasars (with $n$ images), shown in the left panel of Figure 96, is that we get a $\sqrt{n}$ boost in the SNR of the combined quasar spectrum (from $n$ images of the same object) for free (i.e., without having to stack multiple exposures in single-object-spectroscopy mode). Adopting a model of the radiation efficiency, these measurements will permit us to estimate the Eddington ratios, and the mass accretion rates, and thereby the masses of the black holes powering these quasars. Gravitationally lensed quasars will permit us to extend these measurements to $z \sim 6$ and beyond.

**Lensed quasars in 4D:** Acting as a small integral field unit (IFU), the multi-fiber observations described above will place individual fibers on each of the images of a multiply lensed quasar. Repeating such observations at a regular cadence will then add an entirely new spectral dimension to the time domain observations of lensed quasars, which have already been extensively used for the determination of $H_0$ (see review by Treu & Marshall (2016) and references therein). By temporally monitoring the continuum and broad line properties of the different lensed images simultaneously, we obtain a 4D datacube and look for temporal variations in the spectral lines in these quasar spectra (see Figure 96 for an example). With this simultaneous monitoring of the spectral lines in each lensed image, we can alleviate the systematic uncertainties in the $H_0$ measurements due to microlensing effects in the lens galaxy. Mass clumps in the lensing galaxy lead to microlensing and therefor differential magnification of the temperature fluctuations in the accretion disk, exacerbated further by any inclination of the disk to observer's line of sight. These lead to systematic, aperiodic variations and associated uncertainties in the observed light curves of the lensed images, which are used for the determination of $H_0$. Tie & Kochanek (2018) argue that such uncertainties impose a hard systematic limit on the overall uncertainty of the $H_0$ determination, which cannot be improved simply by addressing only the statistical uncertainties. In other words, unless this systematic uncertainty introduced by microlensing is addressed, the benefit of increased sample sizes of lensed quasars in the era of mega-surveys will be of little use. In other words, these spectral + time domain 4D observations made feasible by MSE may be the only viable option to improve the precision in the $H_0$ estimate from time delay measurements of lensed quasars.

**Lensed quasar structure** Tie & Kochanek (2018) also point out that the microlensing effect of accretion disk temperature variations, despite posing a serious systematic uncertainty in $H_0$ measurement, may at the same time provide a unique way to probe the physical structure of the accretion disks of even these high redshift quasars. As they have pointed out, microlensing studies may be used to estimate the radius of the accretion disk if the masses of clumps causing the microlensing, and the peculiar velocities of the lensing clumps relative to the source may be independently estimated. At the same time, as shown in the right panel of Figure 96, the variability of the time delay measured in each lensed image in the presence of microlensing is dependent on the masses of the clumps in the lens causing the microlensing, and the relative velocity of the lensed source and these clumps. Using a suitable prior, typically the peculiar velocities of the source and clumps, this degeneracy may be broken, providing an estimate of the size of the accretion disk (see Tie & Kochanek (2018) for details).

Gravitational lensing is in principle achromatic, i.e. the deflection angle of the light rays does not depend on its wavelength. However, according the unified model of AGNs, an AGN consists of a super-massive black hole surrounded by an accretion disc that is emitting in the continuum. It is expected that microlensing is wavelength-dependent since the size of a continuum emission region of the accretion disc depends on the wavelength (Popović & Chartas, 2005). Moreover, the accretion disc is surrounded by an emitting region that emits the broad emission lines, the so-called broad line region (BLR). The BLR can be different in sizes different emission lines. Thus, the wavelength-dependent geometry of the different emission regions in lensed quasars may result in chromatic effects. Investigation of the influence of microlensing on the spectra of lensed quasars needs to account for the complex quasar emitting structure and it can be used to find the sizes of different emitting quasar regions (e.g. the size of accretion disc in different wavelength and complex BLR, see Fian et al. 2018).

### 8.5 QSO (intervening) absorption line applications

Quasars provide very sensitive probes of the gas in the Universe. The intergalactic medium provides by far the largest cross-section on the sky and produces the Lyman-$\alpha$ forest seen in quasar spectra. As the line-of-sight gets closer to galaxies, the gas density increases and several phases can be observed, starting from warm diffuse phases in the circumgalactic medium up to a mix of warm gas with cold, dense and possibly molecular-rich clouds in the interstellar medium. Detecting the later phases is challenging because of their small total cross-section on the sky, hence requiring a large number of sightlines to be surveyed. Moreover, the associated dust extinction from the cold and dense gas phase requires high sensitivity. MSE will be a unique tool to overcome these challenges (see Figure 97. Indeed, the spectroscopic quasar sample that MSE will be able to provide will by far outperform the previous large-sky quasar sample from SDSS in terms of limiting magnitude (up to 4 magnitudes deeper) while simultaneously providing higher spectral resolution (by roughly a factor of 2) and extended wavelength coverage (out to 1.3$\mu$m). The deeper quasar survey will be crucial in order to study intervening absorption systems causing larger amounts of optical extinction, and the higher resolution will allow an improved deblending of the Lyman-forest lines together with easier identification of different types of absorption classes.

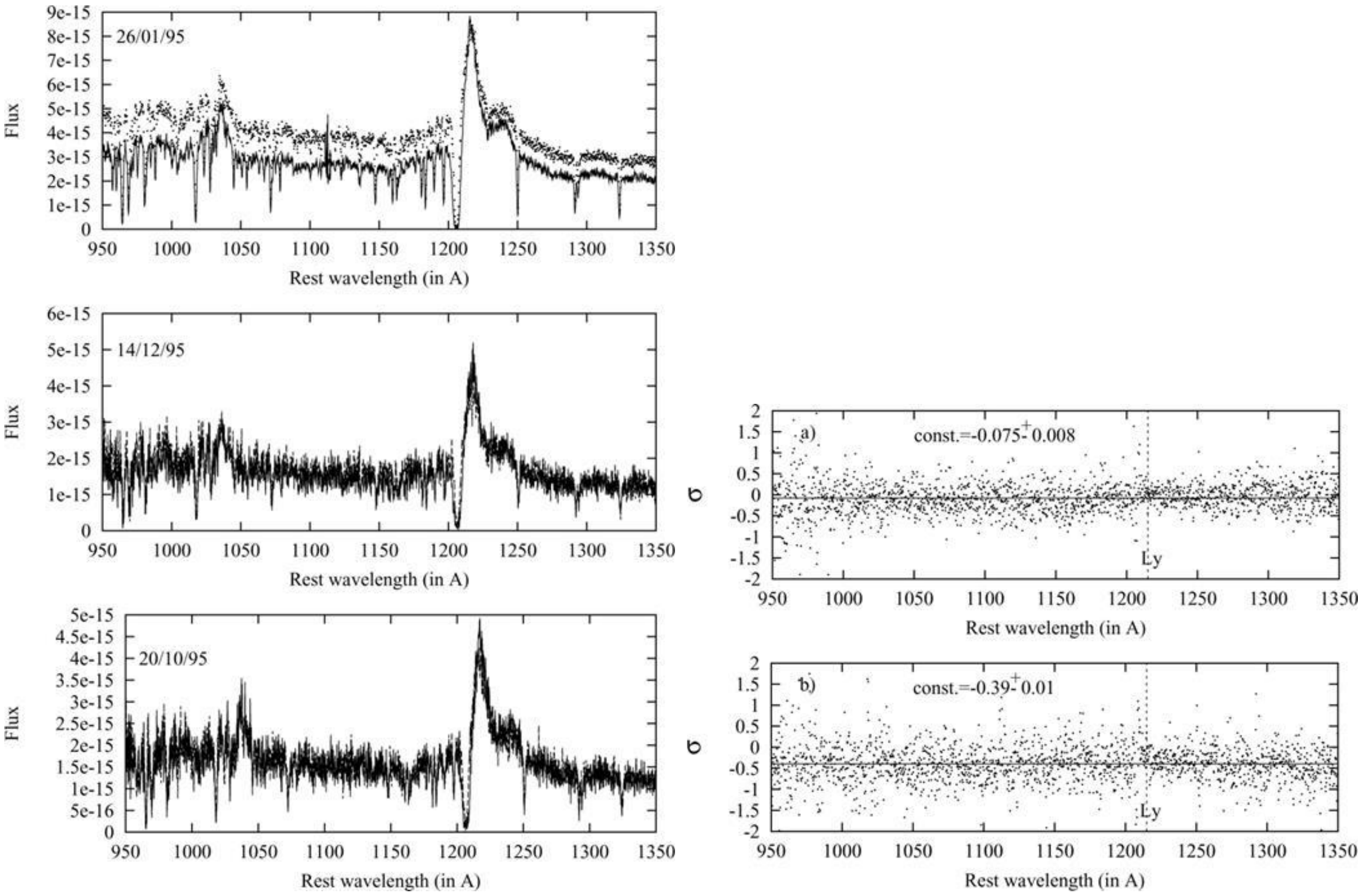

*Figure 96: Left: The spectra of Q0957+561 observed with HST for three different epochs. The solid line and dots indicate the spectra of images A and B, respectively. The flux is given in units of* $erg.cm^{-2}.s^{-1}.^{-1}$*. Right: The flux ratio between the spectra of image A of Q0957+561 for the epochs: (a) 1995 December 14 and 1995 October 20; and (b) 1995 January 26 and 1995 December 14. Figures Popović & Chartas (2005).*

We highlight a few science cases for intervening quasar absorption line systems that will see significant advance thanks to MSE.

**Statistical studies of diffuse neutral gas:** The current samples of neutral hydrogen absorbers (damped Lyman-$\alpha$ absorbers; DLAs) are limited to column densities above $\log(N/\mathrm{cm}^{-2}) > 20$ due to the low spectral resolution of SDSS which makes deblending of the Lyman-$\alpha$ forest difficult. This blending even hampers robust detection of DLAs at $z > 3.5$ (Noterdaeme et al., 2012). With an increase in spectral resolution by a factor of 2, not only we will be able to identify absorption systems down to $\log(N/\mathrm{cm}^{-2}) > 19$, but also detect them more robustly at $z > 3.5$, a capability not available for previous large surveys such as the SDSS. These measurements will provide cutting-edge constraints on the neutral hydrogen distribution function.

We note that the MSE will resolve the C IV doublet at $\sim 1550$ Å. A first large and systematic survey of these lines tracing ionised gas will then be possible, leading to a measurement of $\Omega_{\rm CIV}(z)$ with unprecedented accuracy.

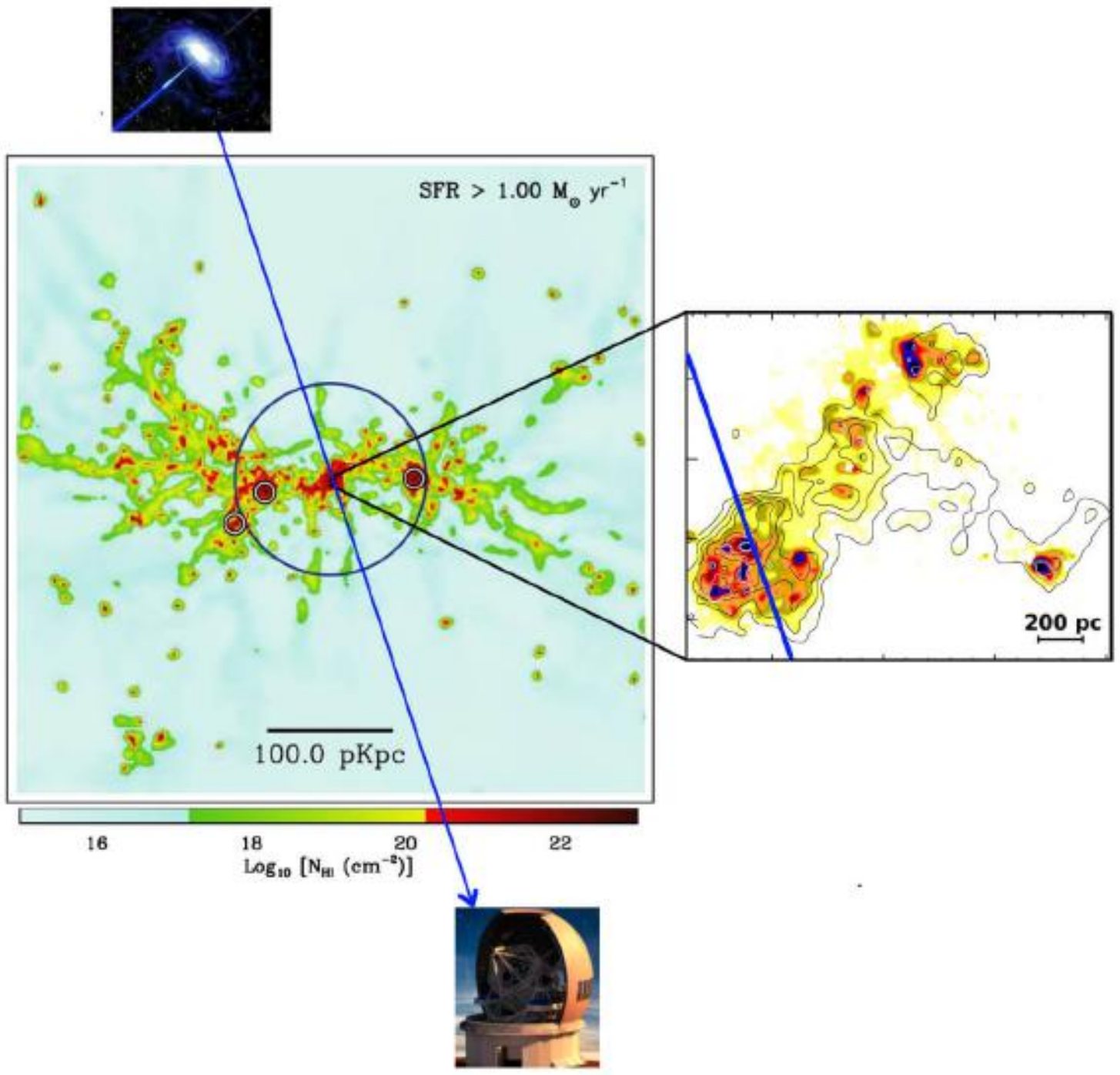

*Figure 97: Illustration of quasar absorption spectroscopy. Only a small fraction of random lines of sight probe the central regions of galaxies, where the associated dust also extinguishes the background quasar light. MSE will overcome these two challenges by allowing not only the observation of a large number of quasars but also down to faint magnitudes.*

**Absorption systems with high extinction:** While the global H I census has long been recognised as being little affected by dust obscuration biases (e.g., Ellison et al., 2005) the same is likely not true for the census of metals, dust and molecules. Indeed, SDSS-IV has only permitted the detection of cold gas with extinction typically less than 0.5 mag (Heintz et al., 2016) and the highest $H_2$ columns detected so far have $\log N(H_2) \sim 21$ (Balashev et al., 2017; Ranjan et al., 2018) with moderate metallicities. Molecular-rich, high-metallicity systems have been found (e.g., Krogager et al., 2016; Noterdaeme et al., 2018), but with lower total column densities. With MSE, we will be able to make a step forward in characterising absorbers with extinction more similar to what is studied in the solar neighbourhood towards nearby stars. This will allow us to quantify better the dust bias and provide quantitative measurements of the cosmological density of metals and dust, all three important constraints for simulations of galaxy formation and evolution. Furthermore, the higher resolution of MSE will allow us to statistically detect the $H_2$ signal more easily than currently done with SDSS (Balashev & Noterdaeme, 2018) and derive its frequency distribution function.

**Associated emission to metallic absorbers (MgII/CIV):** The line emission from galaxies responsible for quasar absorbers, falling into the same fibre centred on the background quasar, have been proposed as a tool to study the cosmic star formation history (Rahmani et al., 2010; Ménard et al., 2011). However, as highlighted by López & Chen (2012), there is a degeneracy between the actual galaxy luminosity and its impact parameter to the quasar sightline. Such a degeneracy has been studied by Joshi et al. (2017) thanks to the different fiber sizes of SDSS-II (3 arcsec) and IV (2 arcsec). MSE will be able to to narrow the constraint further with 1 arcsec fibers, collecting the light only close to the gas that makes the absorption system. Moreover, the extended wavelength coverage of MSE will allow us to study the [OII]$\lambda$3727 emission out to redshift $z \sim 2.5$ corresponding to the peak in cosmic star formation history.

**The cosmological metallicity evolution of absorbers:** With a SNR of at least 6 (per pixel), an MSE quasar survey would enable the detailed study of the cosmological metal enrichment of the Universe, as traced by DLAs (Rafelski et al., 2014; De Cia et al., 2018). This SNR criterion would allow us to probe metallicities down to [Fe/H] $\sim -3$. A sample of $z > 3.5$ quasars would probe an epoch which is currently not well constrained. Additionally, we also expect to uncover some of the most metal-poor DLAs in the Universe which could harbour the signatures of Population III nucleosynthesis and pinpoint the nature of the first stars (Cooke et al., 2015). Such low-metallicity systems provide excellent conditions for measuring the primordial deuterium-to-hydrogen ratio, putting a strong constraint on the cosmological density of baryons ($\Omega_b$).

Lastly, we can study metallicities in greater detail using the MSE high resolution mode, however, this is only possible for very specific redshift bins due to the narrow observable wavelength ranges. We can use various elements to cover different redshift ranges and different metallicity ranges based on the strength of the absorption lines. The targets for the high-resolution mode will be pre-selected from other spectroscopic surveys (e.g., SDSS and DESI) from which the absorption redshift and a rough estimate of the metal strength is known. As an example, ZnII has two transitions at 2026 Åand 2062 Åboth of which fit within the three narrow wavelength ranges nominally defined as 401 – 415 nm, 472 – 485.5 nm, and 626.5 – 672 nm, hereafter referred to as band 1, band 2 and band 3, respectively. Using the ZnII lines will allow us to probe metallicities down to 1/10 $Z_{\odot}$ (for SNR of 10 per pixel and resolution $R \sim 20K$).

Since all three bands will be observed simultaneously we will be able to study several lines for any given redshift range. For the reddest band, we can cover ZnII 2026 and 2062 from $z = 2.092 - 2.259$. At this redshift, we simultaneously cover fine-structure complexes of CI at 1277 and 1280 Å and OI at 1302 Å in band 1 together with the 1509 Å band of CO in band 2. This will allow us to put constraints on the cold gas properties of a large sample of $z \sim 2$ absorbers. Since the cold gas typically exhibits narrow features of $\sim 5\,\mathrm{km\,s^{-1}}$, we need high resolution to resolve such intrinsically narrow lines. As the optical elements controlling the wavelength ranges can be changed during the survey lifetime, it will be possible to cover targets over larger redshift ranges in the high-resolution mode.

# Chapter 9

# Cosmology

**Abstract**

MSE can answer two of the most important remaining questions within physics, namely determining the masses of neutrinos and providing insight into the physics of inflation. It can do this by undertaking a cosmological redshift survey that will probe a large volume of the Universe with a high galaxy density. With such a survey, we expect a measurement of the level of non-Gaussianity as parameterized by the local parameter $f_{NL}$ to a precision $\sigma(f_{NL}) = 1.8$. Combining these data with data from a next generation CMB stage 4 experiment and existing DESI data will provide the first $5\sigma$ confirmation of the neutrino mass hierarchy from astronomical observations. In addition, the Baryonic Acoustic Oscillations observed within the sample will provide measurements of the distance-redshift relationship in six different redshift bins between $z = 1.6$ and 4.0, each with an accuracy of $\sim 0.6\%$. These high-redshift measurements will provide a probe of the dark matter dominated era and test exotic models where dark energy properties vary at high redshift. The simultaneous measurements of Redshift Space Distortions at redshifts where dark energy has not yet become important directly constrain the amplitude of the fluctuations parameterized by $\sigma_8$, at a level ranging from 1.9% to 3.6% for the same redshift bins. In addition to this major program, MSE is able to address many other areas of cosmological interest, for example a deep survey for LSST photometric redshift training and pointed observations of galaxy clusters to $z = 1$.

## 9.1 Motivation

### 9.1.1 Background cosmology

Observations over the last 50 years have provided a tremendous insight into the Universe, with many of the key parameters including the age of the Universe and the split of energy-density components determined with high precision (e.g. Planck Collaboration et al. 2018b). The standard cosmological model, also known as ΛCDM, postulates that the late-time evolution of the Universe is driven by dark energy in the form of a cosmological constant (Λ) (Riess et al., 1998; Perlmutter et al., 1999). Although this model is a tremendous success, matching measurements of ever-increasing precision, this has thrown up some big questions: the origin of a Λ term is difficult to understand physically, suggesting that the ΛCDM might be an approximation to a more complicated theory (e.g. Mortonson et al. 2013. Our knowledge of the very early Universe is also limited, except for the evidence that a period of rapid acceleration (called inflation) is needed to solve a number of problems with the standard cosmological model; we do not know exactly what drives this inflation, however (e.g. Liddle 1999). We also do not know how structure growth affects the large-scale cosmological model (a concept often called back-reaction, e.g. Buchert & Räsänen 2012). Finally, there are a number of unknown physical parameters, such as the summed neutrino particle mass, whose influence spans the fields of cosmology and particle physics (e.g. Lesgourgues & Pastor 2012).

At least two neutrino species are known to have non-negligible mass thanks to flavor oscillation experiments (Ahmad et al., 2001; Hosaka et al., 2006). However, current observations are consistent with many neutrino mass models, and determining the absolute mass scale is an obvious goal in the field of particle physics. It is the target of terrestrial experiments such as the searches for neutrinoless double beta decay (Dell'Oro et al., 2016) or tritium beta decay experiments (Otten & Weinheimer, 2008), but can also be measured through astronomical observations (Lesgourgues & Pastor, 2012). The best current constraints on the summed neutrino mass are $\sum m_\nu < 0.12$ eV (95% confidence), combining low-redshift BAO data with the 2018 CMB data from Planck (Planck Collaboration et al., 2018b). The normal hierarchy with one particle of negligible mass has $\sum m_\nu = 0.057$ eV, while the inverted hierarchy with one negligible mass neutrino has $\sum m_\nu = 0.097$ eV. Thus, for example, we need to measure the neutrino mass with an error of $\sigma = 0.008$ eV in order to rule out the inverted hierarchy at $5\sigma$ if neutrino masses are distributed in the normal hierarchy and $\sum m_\nu = 0.057$ eV. As we will see below, MSE will be a vital component in enabling such a measurement, which is achievable when the MSE data is combined with other available cosmological data.

Measuring the level of primordial non-Gaussianity is one of the most powerful ways to test inflation, and the high-energy early Universe (e.g. Bartolo et al. 2004). The simplest slow-roll models of inflation predict the generation of primordial fluctuations that are almost Gaussian distributed, with only a tiny deviation from Gaussianity. If we consider fluctuations in the potential arising after inflation, then one can denote the portion of the potential that can be described as a Gaussian random field as $\varphi$ and assume that the level of primordial non-Gaussianity is a local function of the potential. To 2nd order, this approach yields $\Phi = \varphi + f_{NL}(\varphi^2 - \langle\varphi^2\rangle)$, where $f_{NL}$ is the parameter that we want to measure. A local designator is often ascribed to the parameter, as defined in this way, but for brevity we do not do this. The best current constraints on $f_{NL}$ come from Planck, who in 2015 (the 2018 update has yet to be published) used Bispectrum measurements of the CMB to measure $f_{NL} = 2.5 \pm 5.7$, consistent with no signal (Planck Collaboration et al., 2016a). A detection of $f_{NL}>1$ would rule out the class of slow-role models. The proposed cosmology survey with MSE will provide a measurement with error ±1.8 from the power spectrum alone. Including constraints from the Bispectrum would further improve these predictions Karagiannis et al. (2018).

Although the standard cosmological model has been an incredible success, there are tensions between measurements at the $2-3\,\sigma$ level. The most significant of these (at $3.5\sigma$) is the mismatch between Hubble parameter measurements made locally with supernovae (Riess et al., 2018b), and those obtained from the combination of galaxy survey and Cosmic Microwave Background (CMB) measurements (Planck Collaboration et al., 2018a). In addition, locally measured structure growth rates from weak lensing (Shan et al., 2018; Abbott et al., 2018a) differ from the ΛCDM extrapolation of CMB measurements, while RSD measurements are more in agreement (Alam et al., 2017). In the past, $2-3\sigma$ tensions between data sets such as these have been revealed as unknown systematic experimental errors, and so it pays to be skeptical. However, these observations might be a hint of new physics so, given the physical issues with the standard model, there are many reasons to investigate further. The proposed MSE survey will push consistency tests to redshifts where there are no current constraints, offering tremendous discovery space.

In what follows, we propose a large-volume galaxy survey designed to probe the inflationary Universe through measurement of primordial non-Gaussianity, and to measure the summed neutrino mass both to physically interesting levels. A by-product of this survey will be more standard measurements of the distance-redshift relation and rate of structure growth, testing the standard model over a redshift range not studied prior to this survey. These measurements will test the ΛCDM model in ways (particularly the redshift range) not previously tested, offering the potential to discover deviations.

### 9.1.2 The role of galaxy redshift surveys

Galaxy surveys provide several complementary mechanisms by which the cosmological measurements described above can be made.

Massive neutrinos alter the observed distribution of galaxies by changing the matter-radiation equality scale compared with the same cosmological model without massive neutrinos: as they are still relativistic at decoupling they hence effectively act as radiation instead of matter around the time of equality. This produces an enhancement of small-scale perturbations in the CMB, especially near the first acoustic peak, and slightly alters the linear matter power spectrum as traced by galaxies. Their effect on structure growth in the matter-dominated era leaves a clearer imprint on galaxy redshift survey measurements due to their relativistic to non-relativistic transition. Massive neutrinos suppress growth on scales below the Hubble radius at the time of the non-relativistic transition, because free-streaming leaves the density contrast for neutrinos lower than that of CDM and they can only slowly catch up through growth. On scales larger than the Hubble radius at the transition time, fluctuations in the distribution of neutrinos are of the same order as those in CDM, as free-streaming of neutrinos out of overdensities is ineffective on these scales and so growth is not as suppressed. A review of these effects is given in Lesgourgues & Pastor (2012). As a result, the summed neutrino mass can be measured from the scale-dependent growth of matter fluctuations as traced by galaxies. This is the primary mechanism that MSE will exploit in order to measure the summed neutrino mass.

Primordial non-Gaussianity alters the expected power spectrum of galaxies in a complementary way to the effects of massive neutrinos. The change in the relative number of high mass haloes that the non-Gaussianity creates alters the expected mass function of dark matter halos in such a way as to produce a scale dependent bias for galaxies that diverges as $1/k^2$, where $k$ is the Fourier wave number of the clustering modes (Dalal et al., 2008). The best $f_{NL}$ measurements to date from a spectroscopic survey used the Baryon Oscillation Spectroscopic Survey (BOSS), finding $f_{NL}$ consistent with zero and an error of order ±100 (Ross et al., 2013). This will soon be supplanted by the extended-BOSS survey, which is expected to provide constraints of order ±10 (Zhao et al., 2016). As this scale-dependent bias arises on very large scales, we need large survey volumes in order to beat down sample/cosmic variance. The eBOSS survey achieves this by observing quasars out to redshift 2.2 over of order 2000 deg$^2$ (for the DR14 sample; the final sample should total 4500 deg$^2$). By pushing to a larger sky area and to higher redshifts, MSE can significantly improve on these measurements. Here, we adopt a conservative strategy and only consider $f_{NL}$ constraints coming from the power spectrum. Measurements made from the Bispectrum have the potential to improve these measurements by a factor $\sim 1.5$ (see Table 9 of Karagiannis et al. 2018) for local $f_{NL}$ measurements, and a larger factor for other types of primordial non-Gaussianity to which the Bispectrum is more sensitive than the Power Spectrum. However these measurements come at the cost of an increased reliance on being able to model non-linear effects, and we do not present any explicit predictions here.

More standard cosmological measurements come from features in the clustering pattern of galaxies arising due to Baryon Acoustic Oscillations (BAO). BAO are caused by acoustic waves in the early Universe, and act as a standard ruler of fixed comoving length, whose apparent size when observed can be used to constrain the distance-redshift relationship and the geometry of the Universe. Information about the growth rate of large-scale structure is obtained through Redshift-Space Distortions (RSD). RSD arise because we do not observe true galaxy positions, but instead infer distances from measured redshifts, which include coherent flows due to the growth of structure. A review of the physics of BAO and RSD is given in Percival (2013). The combination of geometrical and structure-growth measurements has incredible power to constrain theories of acceleration based on modifications to GR (e.g. Joyce et al. 2016). If GR is correct, the growth of large-scale-structure fluctuations can be predicted directly from the expansion history; deviations from this relationship indicate that new physics is needed. Conversely, non-GR theories of gravity cannot be easily tuned to match both the expansion rate and structure growth rate simultaneously.

In addition to the probes discussed above, once a galaxy redshift survey has been produced it acts as a powerful resource for other cosmological measurements and tests, such as those made using the bispectrum (Gil-Marín et al., 2015) or from identifying clusters of galaxies or voids (Nadathur, 2016; Mao et al., 2017). One also obtains significant information by cross-correlating galaxy surveys with other data. For MSE, one interesting avenue will be the cross-correlation of the galaxy survey with CMB data. In particular, the proposed survey will include coverage over $2 < z < 3$, which corresponds to the peak of the CMB lensing effectiveness (Planck Collaboration et al., 2018c). Although these science interests are not the focus of the design of the survey described here, it is important to remember that this survey will offer significant legacy both within the field of cosmology and beyond.

## 9.2 A high redshift cosmology survey with MSE

### 9.2.1 Survey baseline

The MSE High-z Cosmology Survey is designed to probe a large volume of the Universe with a density of galaxies sufficient to measure the extremely-large-scale density fluctuations required to explore primordial non-Gaussianity. By pushing to high redshifts, the growth of structure in the Universe is closer to linear dynamics on average, potentially simplifying the modeling, although the strong biasing of the galaxy samples to be observed will limit the extent of this improvement. The proposed strategy is not optimized for BAO and RSD measurements at high redshift where dark energy does not dominate the cosmological energy budget. However, measurements of these quantities still form a secondary goal of the survey. RSD measurements at redshifts where dark energy has not yet become important, directly constrain the amplitude of the fluctuations parameterised by $\sigma_8$. BAO will measure the angular diameter distance to high redshift, providing a probe of the dark matter-dominated era and tests of exotic models where dark energy properties vary at high redshift.

Our proposed survey covers 10,000 deg$^2$, measuring redshifts for three classes of target objects: Emission Line Galaxies (ELGs), Lyman Break Galaxies (LBGs), and quasars. The ELGs and LBGs will be used as direct tracers of the underlying density field, while the Lyman-$\alpha$ (Ly$\alpha$) forests of the quasars will be used to probe structure along their lines of sight. Details of our proposed target selection and estimated redshift efficiencies are given below. Figure 98 shows how the proposed MSE survey compares to other galaxy redshift surveys.

We propose to undertake exposures of duration 1,800 seconds. Each exposure covers 1.52 square degrees and we expect to pack the observations onto the sky to cover the maximum area possible without gaps, such that each exposure provides approximately 1.25 square degrees of extra coverage. Thus the survey will be comprised of 8,000 pointings, and will take 4,000 hours on target in total. With the expected MSE efficiency of 2,336 hours on target per year, the survey will take approximately 1.7 years of telescope time in total to complete. As we will only use Dark-Time observations, the survey will need to be spread over a number of years, for example taking ∼100 nights per year over 5 years, interspersed with other programs.

The area to be covered by the survey is limited by the availability of imaging data from which to target, while the number of galaxies to be targeted and redshifts obtained will be limited by the number of fibers and the total exposure time available for the survey. We note that having more fibers on MSE would improve the cosmological return, although the return would diminish with each fibre added as we will always preferentially target the galaxies most likely to provide cosmological information. Similarly, increasing the exposure times would improve the efficiency of successful redshift measurement, and would be particularly useful given an increase in the number of fibres. We also note that increasing the field of view while keeping the number of fibres fixed does not significantly help given that we wish to increase the density of observations rather than increase the survey area within a fixed total observing time. Techniques developed for the dark energy Spectroscopic Instrument (DESI) will be used to mitigate selection effects given the patrol radii of individual fibres (Burden et al., 2017; Pinol et al., 2017; Bianchi & Percival, 2017; Percival & Bianchi, 2017), and thus we do not expect any problems with accurately measuring clustering due to sampling targets for observation.

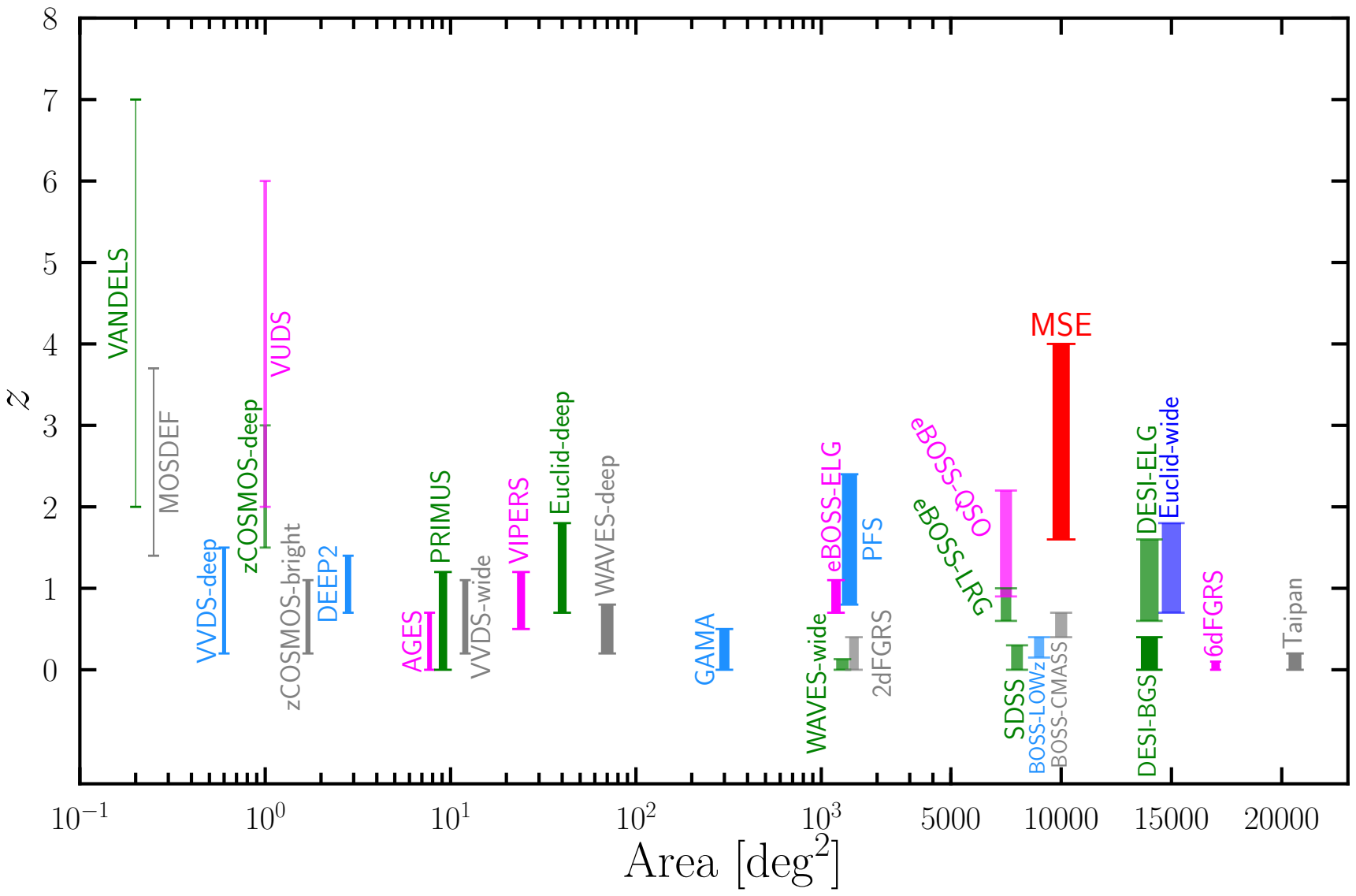

*Figure 98: Recent galaxy redshift surveys as a function of their area and redshift range, compared with the proposed MSE survey. The thickness of each bar is proportional to the total number of galaxies. Notice the transition from logarithmic to linear scale on x-axis at* 5000 $\text{deg}^2$.

### 9.2.2 Detailed predictions for targeting and exposure times

We will use three tracers:

1.6<z<2.4: Emission Line Galaxies (ELGs)

2.4<z<4.0: Lyman Break Galaxies (LBGs) and Ly$\alpha$ emitters (LAEs)

2.1<z<4.0: Ly$\alpha$ forests of QSOs (Ly$\alpha$ QSOs)

We want to optimize the tiling for the survey outlined above. With 3249 low-res fibres per 1.25 square degree gained for the survey, we have a budget of 2600 fibres per square degree on average. Assuming that $\sim 10\%$ of the fibers are used for calibration (sky and reference star fibers), the available density for science targets is $\sim 2340$ fiber per deg$^2$. We will share the same pointings between the three tracers (ELGs, LBGs and Ly$\alpha$ QSOs); hence all will receive the same exposure time. The fiber assignment priorities will be as follows: first the ELGs, then Ly$\alpha$ QSOs and finally the LBGs.

First, we observe observe the ELGs with a density adequate for BAO measurements, i.e, $nP(k = 0.1\text{h.Mpc}^{-1}) = 1$. Assuming a redshift efficiency of order 90%, the density required is 600 deg$^{-2}$ targets. To achieve such an efficiency for ELGs with $r < 24$, as explained below we calculate that this requires an exposure time of 1800s. In addition this exposure time allows us to observe the Ly$\alpha$ forest in QSOs with $r < 24$ or even $r < 24.5$, with a SNR per resolution element of the order of 2-3 in the forest, which is optimal for studies of the cross-correlation between the HI absorption and the positions of the other tracers. Assuming the quasar luminosity function of Palanque-Delabrouille et al. (2013) we can expect there to be 150 to 170 deg$^{-2}$ quasars with $z > 2.1$. The rest of the fibres will be filled with LBGs at the level of approximately 1400 deg$^{-2}$ targets. As the exposure time of 1800s is driven by getting a high redshift efficiency for the ELGs, as estimated below we expect for the LBGs a lower redshift efficiency at the order of 50%. As a result the LBG density is not optimized for BAO measurements but it is still sufficient for the measurement of the primordial non-Gaussianity, i.e, $f_{NL}$. Note that the combination of all three samples together with the sky fibres, fits within the budget for our survey set at 2600 fibres/deg$^2$, with a small amount of margin remaining.

**Emission Line Galaxies:** ELGs are being intensively used to trace the matter within the redshift range $0.6 < z < 1.6$ in the recent survey eBOSS and the upcoming DESI survey. These galaxies are characterized by high star formation rates, and therefore exhibit strong emission lines from ionized H II regions around massive stars, as well as spectral energy distributions with a relatively blue continuum, thanks to which they be selected from optical $ugriz$-band photometric surveys such as LSST in the southern hemisphere or UNIONS in the northern hemisphere. The prominent [OII] (3727 Å) doublet in ELG spectra consists of a pair of emission lines separated in rest-frame wavelength by 2.78 Å. The wavelength separation of the doublet provides a unique signature if spectral resolution is sufficiently high, allowing definitive line identification and secure redshift measurements.

In contrast to DESI, the NIR coverage of the LR spectrograph ($940\,\text{nm} < \lambda < 1320\,\text{nm}$) allows us to determine ELG redshifts up to $z = 2.5$, opening a new research window for RSD and BAO studies. Considering an $r$-band magnitude limit of $r < 24$, the next generation of photometric surveys such as LSST will provide a photometry deep enough to select ELGs based on their optical colors ($u-r$, $g-r$ and $r-z$). A $(u-r)$ or $(g-r)$ color cut combined with a r-z color cut is ideal for selecting galaxies in the desired redshift range, $1.6 < z < 2.4$. The strategy is to measure the UV excess which has a strong correlation with the star-formation activity of ELGs by limiting the $(u - r)$ or $(g - r)$ colors. In addition, as the OII line and the Balmer break are redshifted beyond the $r$-band and the Lyman break is not in the g band, the ELGs are expected to be blue in $(g - r)$ color. Following the strategy proposed in the PFS proposal (Takada et al., 2014b) for a similar ELG survey, a pure optical bands is sufficient to achieve the required ELG density. This selection was successfully tested over the COSMOS field. However it may suffer from a small contamination by low-z galaxies.

To avoid this possible contamination, as the ELGs are redshifted compared to DESI ELGs, a selection based on NIR can improve its purity. The future LSST project with its $y$-band, can provide such photometry. Currently, we can consider KiDS + VIKING in the south hemisphere and Subaru/HSC in the north hemisphere, as test-beds for developing an optical and NIR approach.

It is therefore reasonable to assume that we will be able to select a minimum of 600 deg$^{-2}$ ELGs in the desired redshift range. Based on comparison with Subaru/PFS (Takada et al., 2014b), we can expect a redshift efficiency of order of 90% for ELGs.

**Lyman Break Galaxies:** The LBGs are strongly star-forming galaxies with blue spectra longward of the Lyman break (at 912 Å). Shortwards of the break almost all light is absorbed by neutral HI along the line of sight. At $2.4 < z < 3.5$ they are thus selected by having a very red $(u-g)$ color combined with very blue $(g-r)$ color. For more distant quasars, $z \sim 4$, we can use a similar strategy based on $(g-r)$ and $(r-i)$ colors. For more than a decade, these selections using the Lyman break color techniques have been used extensively. The primary uncertainty for this tracer is on our ability to determine the redshift for samples with a magnitude limit $r < 24$ or even $r < 24.5$.

Recent observations with MUSE on the VLT (Caruana et al., 2018) have shown that at the redshifts of interest more than half of LBGs exhibit a detectable Ly$\alpha$ emission line, which can help to facilitate the determination of redshifts, as we demonstrate below. For the LBGs without Ly$\alpha$ emission, the redshift can be determined thanks to the position of the Ly$\alpha$ and Ly$\beta$ absorption features and a number of interstellar medium absorption lines.

Given an exposure time of 1800 seconds, we wish to estimate the efficiency to determine redshifts for LBGs at 0.1% precision $(0.001 \times (1+z))$. To compute the redshift efficiency we adopt a conservative approach by using two different sets of templates for LBG spectra; one is used to create "simulated spectra" and the other is used to determine the redshifts of the simulated spectra.

The three templates from Figure 7 of Hathi et al. (2016) are implemented in the MSE Exposure Time Calculator (ETC)[1] and thus we obtain the SNR per resolution element for each of them, given a set of parameters (sky brightness, airmass, exposure time, seeing, source type, magnitude of the source and its redshift). In order to obtain the simulated spectra we just redshift the templates and then add gaussian noise given by the SNR.

These spectra are given to a redshift finder, called PandoraEZ (Garilli et al., 2010), which measured their redshifts using the seven spectral templates from Figure 15 of Bielby et al. (2011) and Figure 14−15 of Bielby et al. (2013). We then count the number of LBGs having a redshift which fulfills the condition that $|z_{meas} - z_{real}| < 0.001 \times (1 + z_{real})$. In each set of templates, the difference between the spectra is the strength of the Ly$\alpha$ emission line, expressed through the value of the Equivalent Width ($EW$).

In Figure 99, we present the redshift efficiency as a function of the r magnitude for the three different templates ($EW < 0$, $0 < EW < 20$, and $EW > 20$, the last case corresponding to Ly$\alpha$ emitters (LAE)) and for 4 redshift bins. One can see that the redshift efficiency is higher for the LBGs with stronger Ly$\alpha$ emission line, demonstrating that the determination of the redshift is facilitated by the presence of this emission line. The next step is to make an average over the three templates assuming the Ly$\alpha$ emitting fraction of LBGs from Caruana et al. (2018) (i.e. 40% of LBGs having $EW < 0$; 30% having $0 < EW < 20$; 30% having $EW > 20$).

[1] `http://etc-dev.cfht.hawaii.edu/mse/`

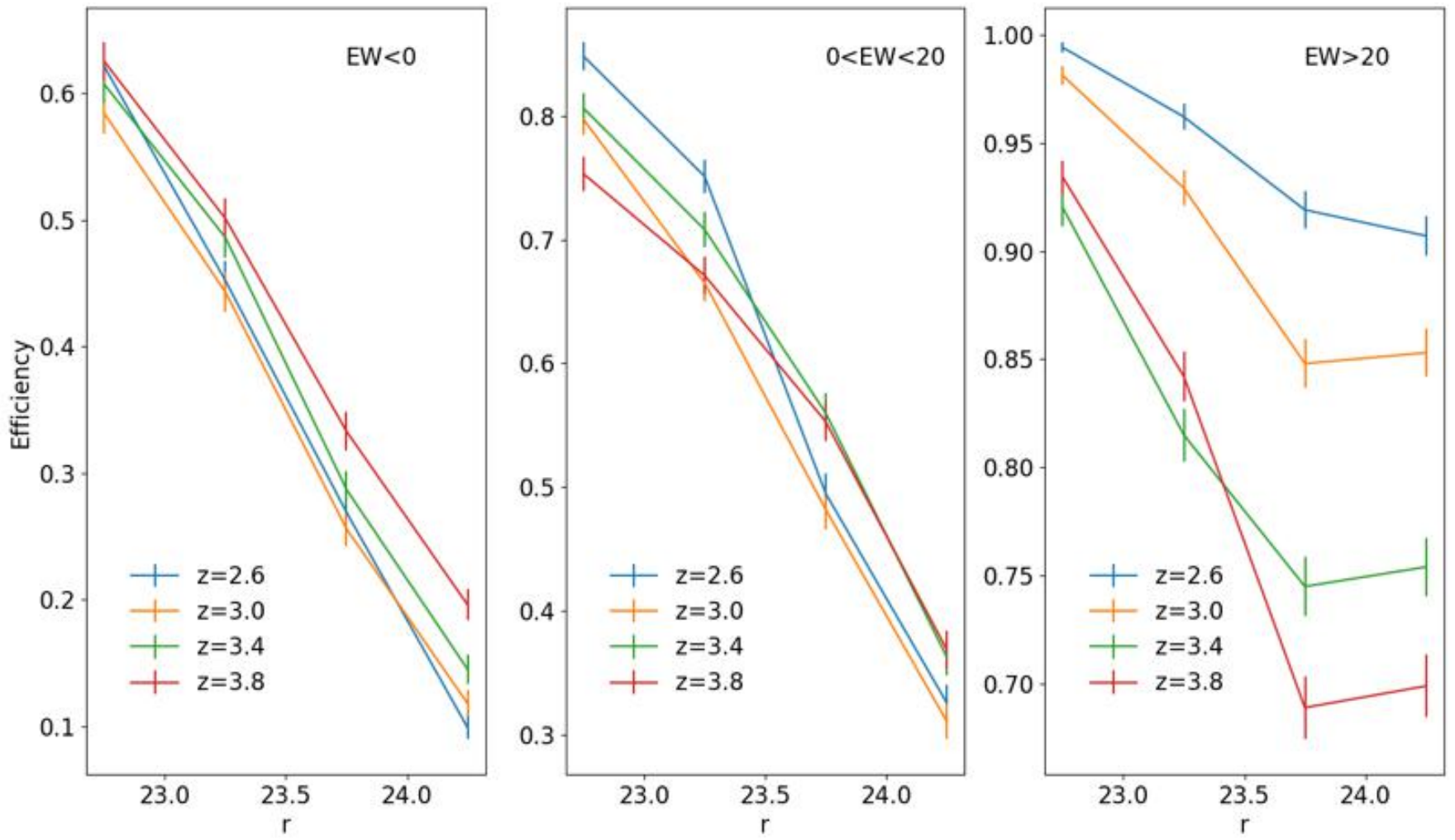

*Figure 99: Redshift efficiency as function of r magnitude for the 3 templates from Hathi et al. (2016) and for 4 redshift bins. The efficiency does depend strongly on the strength of the* Ly$\alpha$ *emission line and on continuum magnitude.*

Using the luminosity function of LBGs from Steidel et al. (1999), the average redshift efficiency and the average density of LBGs with a good redshift measurement is computed. For $z = 3$ and $r < 24.0$ we obtain an efficiency of 56% resulting in a density of 218 LBGs with good z per deg$^2$ , while for $r < 24.5$ the efficiency was 44% with a number density of 481 LBGs with good z per deg$^2$. Consequently, we demonstrate that we have enough LBGs and we have at least 50% redshift efficiency for $r < 24$ to ensure the required precision used by the cosmological forecasts.

**Ly-$\alpha$ forest:** During the last few years, the BOSS collaboration has shown that it is possible to study the large scale structure of the Universe using the Ly$\alpha$ forest of absorption features in the spectra of high-z quasars. In their last analysis, the combination of Ly$\alpha$ auto-correlation and its cross-correlation with quasars provided a BAO measurement at $z = 2.4$ with a 2% accuracy.

While low-z galaxy BAO measurements are dominated by sample/cosmic variance, this is not the case for high-z Ly$\alpha$ BAO, where the error bars can be dramatically reduced by increasing the density of lines of sight. MSE can increase the density by a factor of 10, bringing Ly$\alpha$ BAO analyses to sub-percent precision. MSE will dedicate $\sim 170$ deg$^{-2}$ fibers to Ly$\alpha$ quasar targets. Assuming a targeting efficiency of 90% for $r < 24$ targets, this translates into 150 deg$^{-2}$ forests, close to the cosmic variance limit of $nP = 1$ (see McQuinn & White (2011) for a discussion of Ly$\alpha$ forecasts). In addition to a color selection similar to what is used in DESI, for the footprint overlapping LSST, the quasar can be selected by their intrinsic variability with LSST.

In addition, cross-correlations between the Ly$\alpha$ forest and the other galaxy tracers (ELGs and LBGs) will allow multiple internal cross-checks. The different catalogs will be affected by very different systematics, and consistency between the different measurements will make our results more robust.

### 9.2.3 Cosmological measurements

Given the uncertainties in targeting, we adopt rounded numbers for our baseline predictions, assuming a 10,000 deg$^2$ survey consisting of two samples:

- ELG sample: 1.6<z<2.4, 5.4M galaxies (with a 540 deg$^{-2}$ density corresponding to 600 deg$^{-2}$ targets)
- LBG sample: 2.4<z<4, 7.0M galaxies (with a 700 deg$^{-2}$ density corresponding to 1400 deg$^{-2}$ targets)

For both ELGs and LBGs, we assume a large-scale structure bias $b(z) = G(0)/G(z)$, where G is the linear growth rate of overdensities in the matter. This assumption was adopted for eBOSS ELGs (Zhao et al., 2016) and is based on Comparat et al. (2013a) and Comparat et al. (2013b). For LBGs, this matches the results of Jose et al. (2013); empirically, the overall clustering of luminous galaxies on linear scales changes little with redshift, even as the underlying dark matter overdensities change greatly in amplitude.

We predict the constraints on the cosmological parameters obtainable with MSE using the Fisher Information Matrix. This quantifies the amount of information available about a particular parameter within some observable and accounts for the first-order correlations between different parameters. In our case the observable of interest is the two-point clustering (the power spectrum) of ELGs and LBGs measured with MSE, whilst the cosmological parameters are the standard parameters of the ΛCDM cosmological model, the sum of neutrino masses, and $f_{NL}$ as a measure of primordial non-Gaussianity. The inverse of the Fisher matrix provides a lower limit on the expected variance of these parameters and can be calculated using only a model of the galaxy power spectrum that we might expect to measure with MSE and the baseline numbers for the ELG and LBG samples given above.

Hence, for our given baseline survey, we can provide simple estimates for how well a high-redshift survey of galaxies undertaken with the MSE will allow us to constrain both primordial non-Gaussianity and the sum of neutrino masses. This is done in the following sections. In both cases we anticipate being able to measure the galaxy power spectrum across a range of scales between $k_{min} = 2\pi/V^{1/3}$ and $k_{max} = 0.2h$/Mpc where $V$ is the comoving volume of the Universe covered by our proposed survey; for a 10,000 deg$^2$ sky area between 1.6<z<4.0, $V \simeq 90(\mathrm{Gpc}/h)^3$. We choose our value of $k_{max}$ based on what is achievable with our current state-of-the-art theoretical modelling of the non-linear clustering of galaxies. However, in the era of MSE it is likely we will be able to model even further into the non-linear regime and so also present some scenarios with a slightly higher $k_{max} = 0.25h/\mathrm{Mpc}$. In this scenario we find that our proposed MSE survey will produce tighter constraints on the sum of neutrino masses than any other galaxy survey and when combined with other data (namely from the highly complementary DESI survey and future CMB experiments) will offer the first $5\sigma$ confirmation of the neutrino mass hierarchy, a fundamental measurement for particle physics.

**Predictions for neutrino mass:** For ease of comparison with other upcoming surveys, we use the method of Font-Ribera et al. (2014) for our neutrino mass error estimates, which was also adopted by DESI Collaboration et al. (2016). We consider an MSE survey where we measure the anisotropic power spectrum of ELGs and LBGs in narrow redshift bins of $\Delta z = 0.1$ between the $k_{min}$ and $k_{max}$ given previously. In each redshift bin we marginalise over the unknown effects of galaxy bias. Our model for the clustering of the ELGs and LBGs is sensitive to the sum of neutrino masses in a number of ways. Different neutrino masses change the expansion history of the Universe, changing the length scales associated with a measurement of the power spectrum. Neutrinos also affect the way in which structures grow, adding a scale dependence to an otherwise scale-independent growth rate. Different neutrino masses will also change the shape of the CDM+baryon power spectrum and its normalisation. Hence, a measurement of the clustering of ELGs and LBGs using the MSE is rich in information regarding the neutrino masses.

For the full MSE ELG+LBG sample from $z = 1.6$–$4.0$ and combining with Planck constraints on the standard ΛCDM parameters, we obtain an error on the sum of neutrino masses of 0.018 eV. This is better than any other current or planned survey and would enable a $3\sigma$ constraint on the sum of neutrino masses in either hierarchy. For comparison, the DESI survey between $z = 0.0 - 1.6$ forecasts an error of 0.020 eV, however realising such a measurement from DESI will require careful cross-correlation of several overlapping galaxy samples. Our proposed survey is somewhat less complex with samples separated in redshift, and benefits from the increased cosmological volume available with the MSE.

In addition to its ability as a stand-alone survey, a strength of our proposed survey is in its complementarity with other planned projects. In the era of MSE, the DESI data will be publicly available and can be combined with the survey proposed here. The two surveys can be combined easily given their lack of overlap, but together will form an enormous survey stretching from $z = 0.0 - 4.0$, with a combined volume of over $130(\mathrm{Gpc}/h)^3$. We forecast a combined error on the sum of neutrino masses, including current CMB data from Planck, of 0.013 eV (0.012 eV), for $k_{max} = 0.2\,h\,\mathrm{Mpc}^{-1}(0.25\,h\,\mathrm{Mpc}^{-1})$, which corresponds to a $4\sigma$ constraint on the sum of neutrino masses, *and* a $3\sigma$ detection of the difference between hierarchies.

Finally, adding in information from potential future CMB experiments (i.e., a 'Stage 4' CMB experiment following Abazajian et al. 2016)[2] will enable the constraint on the sum of neutrino masses of $0.008 - 0.009$eV depending on the value of $k_{max}$ used. This will be the first $5\sigma$ constraint on the neutrino mass estimate in either hierarchy, *and* of the difference between hierarchies. Such a measurement is only possible with the inclusion of data from the MSE cosmology survey of ELGs and LBGs.

[2]Constraints on ΛCDM parameters for such an experiment provided by Renée Hložek (private communication) using the OxFish code (Allison et al., 2015) after representative foreground cleaning.

**Predictions for $f_{NL}$:** In addition to neutrino mass estimates, the extremely large cosmological volume available to us with the baseline MSE cosmological survey compared to other planned surveys will enable exquisite constraints on the amount of primordial non-Gaussianity in the early universe and will allow us to rule out a number of inflationary models. Using the same Fisher matrix method as before, we can predict the error on the primordial non-Gaussianity parameter $f_{NL}$ that we can achieve with our baseline survey.

Unlike neutrinos, the effect of primordial non-Gaussianity is to add a scale-dependence to the large scale linear bias of galaxies with respect to the underlying dark matter. The unique $1/k^2$ dependence of the galaxy bias introduced by primordial non-Gaussianity also means that the lower values of $k$ that can be reliably measured (i.e., the larger the cosmological 'baseline' available to us), the stronger constraints we can obtain. The fact that a cosmology survey with MSE will allow us to probe higher redshifts and larger cosmological volumes than other surveys is a strength that will result in superior constraints

To demonstrate this, we take the approach used for the eBOSS survey (Zhao et al., 2016), and again envision measuring the clustering of ELGs and LBGs observed by MSE in narrow redshift bins. For primordial non-Gaussianity forecasts, we take the bias and $f_{NL}$ parameters as free parameters, and report the precision of $f_{NL}$ with the bias parameter marginalised over. The minimum wave-vector we use in the Fisher matrix calculation is determined by the volume of the survey, i.e., $k_{\rm min} = 2\pi/V^{1/3}$. MSE ELGs can constrain primordial non-Gaussianity of the local form to a precision $\sigma(f_{NL}) = 4.1$. MSE LBGs at higher redshifts can reach a tighter constraining precision of $\sigma(f_{NL}) = 2.0$. The individual QSOs that give rise to our proposed Ly$\alpha$ forest sample can also be used as tracers of the density field and reach a precision of $\sigma(f_{NL}) = 5.7$, which itself is already at the level of current CMB constraints.

Overall, the combination of MSE ELGs and LBGs will provide a constraint on primordial non-Gaussianity (the local ansatz) to a precision $\sigma(f_{NL}) = 1.8$. This will be further improved by including the QSOs measured in the same volume; measurements of $f_{NL}$ have been shown to be greatly improved by the use of multiple overlapping tracers of the same density field with different galaxy bias (Seljak, 2009). This measurement is better than the constraint from CMB temperature and polarisation data of Planck 2015, reaching $\sigma(f_{NL}) = 5.7$ (Planck Collaboration et al., 2016a), and also more stringent than the predictions of $\sigma(f_{NL}) = 15$ by the ongoing eBOSS survey (Zhao et al., 2016), $\sigma(f_{NL}) = 5$ by DESI (DESI Collaboration et al., 2016), and $\sigma(f_{NL}) = 11$ by PFS (Takada et al., 2014b).

**Predictions for BAO & RSD:** We present the predictions on BAO RSD precision by MSE in Table 9. For BAO forecasts, we follow the method of Seo & Eisenstein (2007). Using both ELGs and LBGs, we expect to obtain 6 measurements in different redshift bins, each with accuracy $\sim 0.6\,\%$. Binned in a slightly different way, we would obtain approximately 17 1% BAO measurements. These measurements would provide an exquisite determination of the distance-redshift relationship during the matter dominate era over the redshift range $1.6 < z < 4.0$. They would set an incredible benchmark to compare with the low redshift data, and test exotic early dark energy models.

For the RSD forecasts, we follow the Fisher matrix calculation in White et al. (2009), and conservatively use modes within the scale range of $k < 0.1$[h/Mpc]. The RSD parameter can be constrained to 2.1 per cent by MSE ELGs in two redshift bins. At higher redshifts, MSE LBGs can measure RSD precision at 3.6 per cent level. These measurements rely on smaller scale observations than the BAO constraints and so the low density of LBGs in particular limits the precision achievable. However, the measurements will still test gravitational growth over a range of redshifts not previously probed in this way. Note that at $z = 4$, for standard ΛCDM models, $f \sim 0.99$, and we see that the RSD constraint is only weakly dependent on the gradient of the growth rate, and gives instead a strong measurement of $\sigma_8$. These measurements would therefore help to understand any remaining discrepancies between probes of structure growth, such as the current mismatch between weak lensing and CMB predictions.

| Sample | $z$ | $\bar{n}$ [$10^{-4}$h$^3$/Mpc$^3$] | $V$ [Gpc$^3$/h$^3$] | $\sigma_{D_A}/D_A$ [%] | $\sigma_H/H$ [%] | $\sigma_{D_V}/D_V$ [%] | $\sigma_{f\sigma_8}/f\sigma_8$ [%] $k_{\rm max} = 0.1$[h/Mpc] |
|---|---|---|---|---|---|---|---|
| ELGs | 1.6 – 2.0 | 1.8 | 15.56 | 0.81 | 1.43 | 0.56 | 1.86 |
| | 2.0 – 2.4 | 1.8 | 16.20 | 0.74 | 1.30 | 0.51 | 2.05 |
| LBGs | 2.4 – 2.8 | 1.1 | 16.27 | 0.96 | 1.59 | 0.64 | 2.68 |
| | 2.8 – 3.2 | 1.1 | 16.00 | 0.94 | 1.54 | 0.63 | 2.94 |
| | 3.2 – 3.6 | 1.1 | 15.54 | 0.93 | 1.52 | 0.62 | 3.23 |
| | 3.6 – 4.0 | 1.1 | 14.99 | 0.94 | 1.52 | 0.62 | 3.59 |

*Table 9: Forecast constraints on BAO distance precision and growth of structure precision by MSE.*

#### 9.2.4 Discussion

We have shown that MSE can answer two of the most important remaining questions within physics, namely determining the masses of neutrinos and providing insight into the physics of inflation. It will do this by targeting the high redshift Universe, measuring cosmological density fluctuations over an enormous volume (approximately 280 Gpc$^3$). The large collecting area of MSE allows us to measure galaxy redshifts out to $z \sim 4$ with exposure times that allow a large-area survey to be undertaken within a reasonable amount of time. The multiplexed spectroscopic capability matches that required to observe a population of galaxies with sufficient density to measure the large-scale overdensity modes required to understand inflation. At the equatorial location of MSE, UNIONS and LSST will likely provide the major targeting datasets (with both supported by Euclid).

The baseline survey clearly pushes beyond the capabilities of DESI and Euclid into a new regime for galaxy surveys. The focus of the predictions and design of the survey we have presented have been measuring neutrino mass and primordial non-Gaussianity, which we do with theoretically interesting precision. A summary of the improvement achievable by MSE over other projects is shown in Figure 100. We have also shown that the MSE High-z Cosmology Survey will provide unique BAO and RSD measurements over an untested redshift range, offering significant discovery space.

Our cosmological predictions focused on using galaxies as point-tracers of large-scale struc-ture, as these measurements are the key driver of the survey design. The measurements made possible by observations of the Ly$\alpha$ forest of absorption features in the spectra of high-z quasars, and their cross-correlation with the galaxy samples at the same redshifts, are also very exciting, even though they only represent a small fraction of the targets. These observations represent excellent "value" in terms of the information content obtained about the large-scale structure for each observation.

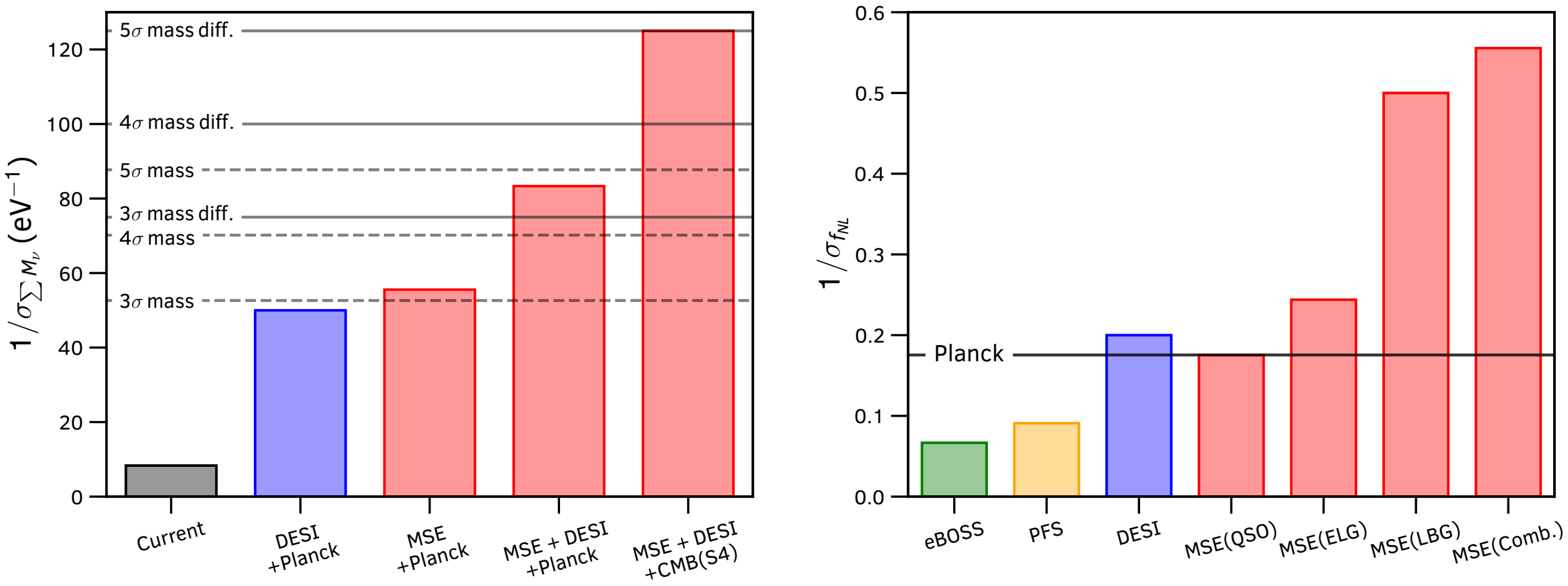

*Figure 100: A summary of the neutrino mass (left) and primordial non-Gaussianity (right) constraints achievable with the MSE compared to other surveys. The dashed and solid horizontal lines in the left panel show the requirements for 3, 4 and $5\sigma$ constraints on the sum of neutrino masses (either hierarchy) and on the mass difference between hierarchies respectively. The horizontal line on the right panel shows the current constraints on $f_{NL}$ from Planck CMB data.*

Once obtained, a survey such as the MSE High-z Cosmology Survey will offer a goldmine of data to be used for many different astrophysical applications, including providing insights into galaxy formation and evolution, cross-correlations science with the Cosmic Microwave Background, calibration of photometric redshift errors from other surveys, analyses of voids, clusters, filaments and other large-scale structures. The public release of the galaxy catalogues will provide the definitive high-redshift spectroscopic survey database, enabling serendipitous science similar to that enabled by the SDSS public database, and will have broad impact in many areas, probably in entirely unanticipated ways, as has also been the experience with the SDSS.

### 9.3 A deep survey for LSST photometric redshift training

All of the probes of dark energy to be employed by LSST will rely on estimates of photometric redshifts for galaxies, either directly or indirectly. In simulations with perfect knowledge of galaxy spectral templates (or perfect spectroscopic training sets for machine-learning based algorithms) and expected LSST measurement errors, it is found that LSST is capable of delivering photometric redshifts with 2% uncertainties; however, existing datasets of compa-rable depth yield uncertainties closer to 5%.

Enabling optimal LSST photo-z's – i.e., reducing uncertainties from 5% to 2% – would require a better understanding of the relationship between galaxy colors and redshifts. This requires spectroscopy of galaxies spanning the range of the properties of the photometric samples to be studied (e.g., objects with $i < 25.3$ for the LSST weak lensing sample, going significantly fainter than the spectroscopic samples being obtained for Euclid training) with sufficient signal-to-noise ratio to achieve highly-secure redshift measurements for a large fraction of targets. Such a sample would greatly improve LSST dark energy constraints, improving the figure of merit from BAO and weak lensing alone by almost $\sim 50\,\%$ (large-scale structure studies especially benefit due to the sharper maps provided by higher-fidelity photo-$z$'s). Cosmology with galaxy cluster counts and photometric supernovae (i.e., those with redshifts assigned based on their host photo-$z$ rather than spectroscopy) would likely benefit more.

Newman et al. (2015) lays out a basic strategy for photometric redshift training spectroscopy for LSST. It is estimated that spectroscopy of $20-30K$ galaxies to the weak lensing magnitude limit of LSST ($i < 25.3$) is necessary. In order to limit the impact of sample/cosmic variance, it is optimal to distribute those objects over a number of widely-separated fields; fifteen 20-arcminute diameter fields (the minimum considered viable in Newman et al. (2015), as smaller fields would not allow the characterization of galaxy clustering) would have similar variance to five widely-separated single MSE pointings, or one 10 square degree field. Given that $\sim$ 3000 objects can be targeted per MSE pointing, seven widely-separated pointings would be the minimum viable program. More pointings would enable better characterization of the impact of cosmic variance (estimates of standard deviations from seven samples are unstable) and the rejection of fields which are outliers at a particular redshift.

This work is extremely synergistic with galaxy evolution observations of interest for MSE (see Chapter 7). Many objects that would be targeted for galaxy evolution surveys would be relevant for photo-z training, so programs could be pursued simultaneously (purely targeting limited redshift ranges would be highly undesirable for this work, however). Furthermore, photometric redshift training is fundamentally equivalent to the problem of determining the range of galaxy spectral energy distributions at a given luminosity and redshift; as a result, the proposed survey would provide strong constraints on models of galaxy evolution.

Additionally, shrinking LSST photo-z measurement errors from 5% per-object errors to 2% will improve the sharpness of maps of the large-scale structure proportionally. This will improve SNR in studies of relationships between galaxy properties and environment, measurements of galaxy clustering, and photometric identification of galaxy clusters and groups; LSST can help constrain the broader context of the galaxies whose spectra MSE would obtain. Star-galaxy separation for studies of local dwarf galaxies and the Milky Way halo would also benefit from improved knowledge of galaxy SEDs provided by this sample. For galaxy evolution studies, having more fields versus fewer (at fixed area) improves measurements of galaxy clustering on the scales relevant for galaxy evolution studies, reduces sample/cosmic variance in count-based statistics (e.g., luminosity and mass functions), and enables variance to be well quantified.

For photometric redshift training, in general we wish to obtain spectroscopic redshifts for as broad a range of galaxies within the magnitude range of LSST weak lensing and galaxy evolution samples. For weak lensing, this represents a limit of $i = 25.3$; for galaxy evolution samples are currently less defined but going fainter than this spectroscopically for highly-complete samples is infeasible. We then aim to obtain a signal-to-noise per angstrom equivalent to that obtained by the DEEP2 Galaxy Redshift Survey at $i = 22.5$ (which yielded a $> 80\%$ redshift success rate at that magnitude) and comparable spectral resolution, but covering a broader wavelength range. This leads to an estimated exposure time of 135 dark hours per pointing, for a total of 1000 (1500) hours for a 20,000 (30,000) object survey. [3]

If secure redshifts can be obtained for $> 99\%$ of the galaxies targeted for spectroscopy, this sample would provide a direct calibration of the redshift distributions of photo-z samples accurate enough for LSST dark energy uncertainties to be dominated by random errors rather than photo-z systematics. However, given past redshift failure rates for deep samples of $20-40\%$, it is quite likely that this high threshold will not be reached. In that case, cross-correlations between LSST photo-z objects with galaxies with precision redshifts from wide-area spectroscopic surveys (such as the one proposed above) present the most promising route for this calibration, exploiting the fact that both sets of objects trace the same underlying large-scale-structure (Newman 2008). Even if $> 99\%$ redshift success rates are not achieved, however, the proposed sample would reduce uncertainties on individual objects' photo-z measurements from $\sim 5\%$ to $\sim 2\%$ in those regions of parameter space with good redshift success, greatly enhancing the value of the LSST dataset for both galaxy evolution and cosmology studies.

We note that although we have focused on LSST here, it is worth noting that WFIRST will have similar, overlapping (but not identical) spectroscopic training needs in order to enable core parts of its cosmology program.

### 9.4 Pointed observations of galaxy clusters to z = 1

Galaxy clusters may be detected using observational techniques sensitive to each main cluster component: weak-lensing is sensitive to the total cluster mass; the Sunyaev-Zeld'ovich (SZ) decrement and X-ray imaging are sensitive to the cluster ICM; galaxy overdensity studies are sensitive to the member galaxy population. MSE is poised to be an essential tool for cosmological studies based on galaxy clusters through high completeness targeted observations that will provide detailed information on their dynamical state for comparison to SZ, X-ray, and galaxy overdensity surveys.

The landscape of galaxy cluster studies over the next two decades will be dominated by large samples of clusters identified using multiple, complementary techniques from observations executed over many thousands of square degrees. The aim of such studies is to obtain precise and accurate knowledge of our cosmological model, particularly the dark energy equation of state, in addition to compiling a detailed picture of how the mass and physical history of galaxy clusters in turn affects the evolutionary history of their member galaxies. There are two critical issues that affect the extent to which such galaxy cluster samples can be used to test cosmological models:

1. The relationship between the observable used to identify each cluster and its true mass (in this case measured with respect to a common overdensity scale, e.g., M500,c).
2. The relationship between a sample of clusters identified using a given observational method (e.g., SZ, X-ray, galaxy overdensity) and the true population of clusters existing in the universe.

[3] Our estimate of 135 hours per pointing is based upon the scaling of photon statistics assuming the same throughput for MSE as for Keck/DEIMOS and that redshift success will be determined purely by SNR. We have also estimated the time required to reach the same SNR per angstrom as DEEP2 using the MSE exposure time calculator, and obtain results within a factor of two of this. However, the ETC-based estimate is more uncertain, as the DEEP2 redshift success rate at $i = 22.5$ can be measured directly from catalog-level DEEP2 data products, but SNRs for direct comparison to ETC results can only be determined from a new pixel-level analysis. As a result, the ETC-based calculation relies on an uncertain (and spectrum-dependent) conversion from SNR at $r = 23.5$ (as tabulated by DEEP2) to SNR at $i = 22.5$, which is not an issue for the photon statistics conversion.

Cluster mass — the key parameter whose evolution is predicted by theory and N-body simulations — is a problematic parameter because it cannot be observed directly. Traditionally, cluster masses for large samples of clusters could only be inferred statistically, via various "observable"-mass scaling relations. Popular observables include X-ray temperature/luminosity, SZ decrement, and cluster richness, $N_{gal}$. However, there is currently no mass proxy that is simultaneously accurate (unbiased) and precise. Cluster cosmology benefits hugely from knowing the average cluster mass accurately (for the amplitude of the mass function) and the relative masses of clusters precisely (to get the shape of the mass function). For the absolute calibration, weak lensing currently achieves the highest accuracy. To get relative masses, X-ray gas mass and/or temperature are more useful because they have a smaller intrinsic scatter than WL.

So what can MSE bring to this field of study? Put simply, redshifts. Spectroscopic redshifts can be used to fix individual galaxies in space and thus determine cluster velocity distributions to a few percent, translating into a $\sim 10\%$ accuracy in mass. This would provide a highly competitive mass calibration among the various cluster mass proxies. The statistical accuracy could be further improved by "stacking" clusters of similar values of observables (e.g., richness, X-ray temperature). MSE could also be employed to determine the recent star formation history of member galaxies viable the observation of narrow emission and absorption features generated within the photospheres of stars or in the interstellar gas.

# Appendix A
# MSE science team membership

*(Current as of March 2019)*

Carlos Abia
Vardan Adibekyan
João Alves
George Angelou
Borja Anguiano
Nobuo Arimoto
Martin Asplund
Herve Aussel
Carine Babusiaux
Eduardo Balbinot
Michael Balogh
Michele Bannister
Daniela Barria Diaz
Nate Bastian
Giuseppina Battaglia
Chiara Battistini
Chetan Bavdhankar
Amelia Bayo
Rachael Beaton
Paul Beck
Megan Bedell
Tim Beers
Earl Bellinger
Thomas Bensby
Trystyn Berg
Maria Bergemann
Lemasle Bertrand
Joachim Bestenlehner
Maciej Bilicki
Jessica Birky
Bertram Bitsch
John Blakeslee
Joss Bland-Hawthorn
Adam Bolton
Alessandro Boselli
Herve Bouy
Jo Bovy
Angela Bragaglia
Niel Brandt
Terry Bridges
James Bullock
Denis Burgarella
Adam Burgasser
Etienne Burtin
Derek Buzasi
Claudio Cáceres
Elisabetta Caffau
Jan Cami
Luca Casagrande
Giada Casali
Viviana Casosola
Santi Cassisi
Denise Castro
Marcio Catelan
Ken Chambers
Chihway Chang
William Chaplin
Scott Chapman
Corinne Charbonnel
Yanmei Chen
Norbert Christlieb
Andrew Cole
Remo Collet
Johan Comparat
Jeff Cooke
Andrew Cooper
Luca Cortese
Pat Cote
Helene Courtois
Nicolas Cowan
Scott Croom
Chris D'Andrea
Ivana Damjanov
Luke Davies
Tamara Davis
Peter De Cat
Richard de Grijs
Roelof De Jong
Gisella de Rosa
Gayandhi De Silva
Alis Deason
Eric Depagne
Darren DePoy
Arjun Dey
Marcella Di Criscienzo
Paola Di Matteo
Simon Driver
Alex Drlica-Wager
Maria Drout
Pierre-Alain Duc

Patrick Durrell
Gwendolyn Eadie
Sara Ellison
Prahlad Epili
Dawn Erb
Denis Erkal
Ana Escorza
Benoit Famaey
Ginevra Favole
Annette Ferguson
Laura Ferrarese
Nicolas Flagey
Scott Fleming
Hector Flores
Andreu Font
Morgan Fouesneau
Wes Fraser
Ken Freeman
Xiaoting Fu
Maksim Gabdeev
Boris Gaensicke
Bryan Gaensler
Sarah Gallagher
Rafael Garcia
Peter Garnavich
Patrick Gaulme
Marla Geha
Mario Gennaro
Karoline Gilbert
Andreja Gomboc
Anais Gonneau
Rosa Gonzalez-Delgado
Yjan Gordon
Aruna Goswami
John Graham
Catherine Grier
Richard Griffiths
Carl Grillmair
Sam Grunblatt
Puragra Guhathakurta
Guillaume Guiglion
Szabó Gyula
Daryl Haggard
Pat Hall
Gerald Handler
Terese Hansen
Lei Hao
Nimish Hathi
Despina Hatzidimitriou
Misha Haywood
Vincent Henault-Brunet
Gregory Herczeg
JJ Hermes
Juan Hernandez
Falk Herwig
Vanessa Hill
Lynne Hillenbrand
Derek Homeier
Andrew Hopkins
Anna Hourihane
Cullan Howlett
Daniel Huber
Mike Hudson
Narae Hwang
Rodrigo Ibata
Dragana Ilic
Pascale Jablonka
Matthew Jarvis
Nada Jevtic
Linhua Jiang
Ning Jiang
David Jones
Stephanie Juneau
Umanath Kamath
Devika Kamath
Manoj Kaplinghat
Amanda Karakas
Lisa Kewley
Stacy Kim
Xu Kong
Georges Kordopatis
Janet Kos
George Koshy
Marina Kounkel
Jens-Kristian Krogager
Mark Lacy
Claudia Lagos
Rosine Lallement
Chervin Laporte
Mahy Laurent
Damien Le Borgne
Yveline Lebreton
K.G. Lee
Chien-Hsiu Lee
Claus Leitherer
Geraint Lewis
Ting Li
Sophia Lianou
Chien-Cheng Lin
Karin Lind
Jiren Liu
Chengze Liu
Guilin Liu
Xin Liu
Nicolas Lodieu
Giuseppe Longo
Alessia Longobardi
Jonathan Loveday
Mikkel Lund
Robert Lupton
Carla Maceroni
Ted Mackereth
Dougal Mackey
Laura Magrini
Martin Makler
Katarzyna Malek
Danilo Marchesini
Jennifer Marshall
Sarah Martell
Nicolas Martin
Mario Mateo
Sean McGee
Carl Melis
Olga Mena
Thibault Merle
Szabolcs Mészáros
Andres Meza
Andrea Miglio
Faizan Mohammad
Karan Molaverdikhani
Richard Monier
Thierry Morel
Ulisse Munari
Muthu Muthumariappan
Adam Muzzin

David Nataf
Lina Necib
Ignacio Negueruela
Hilding Neilson
James Nemec
Jeffrey Newman
David Nidever
Ewa Niemczura
Anna Nierenberg
Peder Norberg
Pasquier Noterdaeme
Chris O'Dea
Mahmoudreza Oshagh
Andrew Pace
N. Palanque-Delabrouille
Gajendra Pandey
Casey Papovich
Cirino Pappalardo
Laura Parker
Javier Pascual-Granado
David Patton
Marcel Pawlowski
Jorge Penarrubia
Eric Peng
Yingjie Peng
Will Percival
Enrique Pérez
Celine Peroux
Annika Peter
Patrick Petitjean
Andreea Petric
Vinicius Placco
Bianca Poggianti
Agnieszka Pollo
Luka Popovic
Abhishek Prakash
Adrian Price-Whelan
Andrej Prša
Mathieu Puech
Swara Ravindranath
Alejandra Recio-Blanco
Céline Reylé
Michael Rich
Johan Richard
Stephen Ridgway
Annie Robin
Aaron Robotham
Leslie Rogers
Martino Romaniello
Laurie Rousseau-Nepton
Jan Rybizki
Giueseppe Sacco
Charli Sakari
Anya Samadi
Robyn Sanderson
Will Saunders
Ricardo Schiavon
Carlo Schimd
Mathias Schultheis
George Seabroke
Arnaud Seibert
Aldo Serenelli
Prajval Shastri
Yue Shen
Jingjing Shi
Yong Shi
Jianrong Shi
Cristóbal Sifón
Victor Silva Aquirre
Sergio Simon-Diaz
Jeffrey Simpson
Raghubar Singh
Małgorzata Siudek
Gregory Sivakoff
Ása Skúladóttir
Rodolfo Smiljanic
Daniel Smith
Marcelle Soares-Santos
Jennifer Sobeck
Sergio Sousa
C.S. Stalin
Else Starkenburg
Dennis Stello
Louis Strigari
Guy Stringfellow
Mark Sullivan
Mouyuan Sun
Firoza Sutaria
Róbert Szabó
Jamie Tayar
Edward Taylor
Matthew Taylor
Elmo Tempel
Karun Thanjuvur
Sivarani Thirupathi
Guillaume Thomas
Yuan-sen Ting
Jeremy Tinker
Erik Tollerud
Silvia Toonen
Kim-vy Tran
Pier-Emmanuel Tremblay
Laurence Tresse
Jonathan Trump
Maria Tsantaki
Brad Tucker
Brent Tully
David Tytler
Meg Urry
Murat Uzundag
Marica Valentini
Mathieu van der Swaelmen
Sophie Van Eck
Jennifer van Saders
Ludo van Waerbeke
Anna Lisa Varri
Kim Venn
Eva Villaver
Matthew Walker
Jian-Min Wang
Jing Wang
Yuting Wang
Yiping Wang
Huiyuan Wang
Tinggui Wang
Tracy Webb
Daniel Weisz
Vivienne Wild
Jon Willis
Mike Wilson
Joanna Woo
Clare Worley
Nick Wright
Xufen Wu
Yanxia Xie

Siyi Xu
Yongquan Xue
Hong-Liang Yan
Ming Yang
Christophe Yeche
Mutlu Yildiz
David Yong
Olga Zamora
Gail Zasowski
Huawei Zhang
Binbin Zhang
Hongxin Zhang
Gongbo Zhao
Ying Zu
Konstanze Zwintz